\documentclass[12pt]{report}
\usepackage{amssymb,amsmath}
\usepackage{comment}
\usepackage{graphicx}
\usepackage{bm}
\usepackage{enumerate}
\usepackage{subfigure}
\usepackage{here}
\usepackage{hyperref}

\numberwithin{equation}{section}

\begin{document}

\thispagestyle{empty}
\begin{flushright}
CALT-TH-2015-011\\
OU-HET 852\\

\end{flushright}
\vskip3.5cm
\begin{center}
{\LARGE \bf Superconformal Quantum Mechanics\\
\vskip5mm
from M2-branes}

\vskip1.5cm
{\large Tadashi Okazaki\footnote{tadashi@theory.caltech.edu}}

\bigskip\bigskip
\vskip3mm
{\it 
Walter Burke Institute for Theoretical Physics\\
California Institute of Technology \\
Pasadena, CA 91125, USA}
\vskip2mm
{\it
and}
\vskip3mm
{\it Department of Physics, Graduate School of Science \\
Osaka University \\
1-1 Machikaneyama-cho Toyonaka, Osaka 560-0043, Japan\\
}

\end{center}

\begin{abstract}
We discuss 
the superconformal quantum mechanics arising from the M2-branes. 
We begin with a comprehensive review 
on the superconformal quantum mechanics and emphasize that 
conformal symmetry and supersymmetry in quantum mechanics 
contain a number of exotic and enlightening properties 
which do not occur in higher dimensional field theories. 
We see that superfield and superspace formalism is available 
for $\mathcal{N}\le 8$ superconformal mechanical models. 
We then discuss the M2-branes 
with a focus on the world-volume descriptions of the 
multiple M2-branes which are superconformal 
three-dimensional Chern-Simons matter theories. 
Finally we argue that 
the two topics are connected in M-theoretical construction 
by considering the multiple M2-branes wrapped around a compact Riemann surface 
and study the emerging IR quantum mechanics. 
We establish that 
the resulting quantum mechanics realizes 
a set of novel $\mathcal{N}\ge 8$ superconformal quantum mechanical models  
which have not been reached so far. 
Also we discuss possible applications of 
the superconformal quantum mechanics to mathematical physics. 
\end{abstract}


\newpage
\setcounter{tocdepth}{2}
\tableofcontents

\newpage
\begin{center}
\section*{Acknowledgments}
\end{center}
I would like to acknowledge kind support and help 
with deep appreciation and gratitude to many people. 
I owe a debt of appreciation to Hirosi Ooguri 
for continuous guidance during my days at Caltech 
over the past two and a half years. 
It goes without saying that 
my study and experiences at Caltech would not be possible 
if were not for his arrangement. 
Moreover this thesis would not have been completed without 
his suggestion and assistance. 
I cannot thank him enough for all that he has done for me. 
I was really happy to study at Caltech guided by whom I respect. 
I would like to express my gratitude to 
Yutaka Hosotani for always helping me to study at Caltech 
and consulting me when I am in trouble. 
His advice was always a useful resource for me. 
I am profoundly grateful to 
Satoshi Yamaguchi for continuing encouragement 
and innumerable discussions. 
Over the past five years, 
particularly in the first two and a half years 
I could learn many things through collaboration works 
and countless times of long hours of discussions with him. 
I am glad that 
I was able to start my study of string theory 
with him at Osaka University. 
It was my indelible memory to talk about physics and life 
with brilliant members and alumni at Caltech including,  
Hee-Joong Chung, Abhijit Gadde, Siqi He, Koji Ishiwata, 
Hyungrok Kim, Kazunobu Maruyoshi, Elena Murchikova, 
Yu Nakayama, Chan Youn Park, Ke Ye, Du Pei, Pavel Putrov and Wenbin Yan. 
I extend my attitude to 
my colleagues at Osaka University including 
Tetsuya Enomoto, Tomohiro Horita, 
Takuya Shimotani, Akinori Tanaka, Akio Tomiya. 
The study meeting continuing for 
a long stretch of time with them is a good memory now. 
I wish to thank Greg Marlowe, 
the house owner for sharing with me a nice place to live at Pasadena 
and for always encouraging me. 
I was often relaxed myself by talking with him.  
I would like to thank 
Carol Silberstein, the secretary at Caltech 
and Kumiko Nakayama, Keiko Takeda, 
the secretaries at Osaka University for arranging 
a lot of procedures so that I could do my research at Caltech. 
They are always kind and I was helped many times. 
Finally I am grateful to my parents 
for giving constant support to me.

\newpage
\chapter{Introduction and Overview}
\section{Historical background}
The pursue of Theory of Everything that describes our nature 
and achieves unification of all fundamental interactions, 
i.e. electroweak, strong and gravitational 
interactions has been a mission agitating theoretical physicists. 
Historically much of significant developments in theoretical physics 
were achieved by overcoming the inconsistency between 
existing concepts. 
The special relativity was established 
from the crisis between 
classical mechanics and electrodynamics, 
the general relativity was proposed 
by reconciling the special relativity and Newtonian gravity 
and quantum field theory was acquired 
by combining quantum mechanics and the special relativity. 
However, we are now confronting another contradiction 
between quantum field theory and the general relativity. 
Although a quantum field theory
as the standard model 
successfully describes and predicts 
almost all phenomena controlled by electroweak and strong forces, 
the general relativity describing the gravity 
that is the remaining fundamental interaction 
seriously disagrees with the quantum field theory. 
This indicates that 
quantum field theory cannot lead to 
the correct quantization of gravity.  
Therefore it is expected that the standard model is regarded as 
the low-energy effective theory of a more fundamental theory.

%
String theory has been proposed as the promising candidate 
for such a fundamental theory 
since it can naturally describe all fundamental interactions. 
In string theory all particles are recognized as 
various vibrational modes of only two different types of fundamental
strings; 
the open strings which have two endpoints and 
the the closed strings which have no endpoint. 
One of the most beautiful structures in string theory is that 
Yang-Mills gauge theories 
which govern the electroweak and strong interactions 
as the standard model arise from the vibrations of the open strings 
while the general relativity that describes the gravitational
interaction appears from the vibrations of the closed string.  
Among the massless states of the open string 
there are spin-one particles which behave as gauge bosons 
while among the massless states of the closed string 
there is a spin-two particle which can be identified with graviton 
\cite{Scherk:1974ca,Yoneya:1974jg}.

%
However, the bosonic string theory is not realistic 
since the vibrations of the bosonic strings yield only bosonic particles. 
The lack of fermionic particles can be resolved by 
introducing supersymmetry in string theory, 
i.e. the superstring theory. 
The spectrums of superstrings contain both bosonic and fermionic particles. 
Hence string theory supports the existence of the supersymmetry.

%
One of the most fascinating predictions in superstring theory 
is the existence of the extra dimension in space-time.  
It turns out that 
the unitarity and the Lorentz invariance of space-time in which 
the superstrings live are guaranteed 
only for ten-dimensional space-time. 
In other words, flat space superstrings can only exist in ten dimensions.  
In order to reconcile the difference between 
the ten-dimensional space-time in string theory and 
the four-dimensional space-time in 
our instinctive knowing physics, 
the notion of compactification has been proposed. 
The idea is that 
since the six extra dimensional compact spatial directions are 
much smaller than the four-dimensional space-time, 
the original $(1+9)$-dimensional space-time effectively 
looks like $(1+3)$-dimensional space-time. 
For the six-dimensional spaces 
Calabi-Yau manifolds are known 
to possess phenomenologically promising properties.

Ten-dimensional superstring theory is not a 
single theory but rather a set of possible five theories; 
(i) type IIA, (ii) type IIB, (iii) type I 
(iv) $SO(32)$ heterotic (v) $E_{8}\times E_{8}$ heterotic. 
When the both left-moving and right-moving modes 
are taken as superstrings, there are 
two possibilities; opposite handedness or the same handedness. 
The former theory is called type IIA 
while the latter is called type IIB. 
Type I superstring theory is obtained by 
the orientifold projection 
that mods out left-right symmetry of type IIB superstring theory. 
When the left-moving mode is chosen as the bosonic string 
and the right-moving mode is taken as the superstring, 
consistency allows only two different theories; 
$SO(32)$ heterotic and $E_{8}\times E_{8}$ heterotic 
superstring theories. 
The type II superstring theories have 
$d=10$, $\mathcal{N}=2$ supersymmetry 
and the type I and heterotic superstring theories 
possess $d=10$, $\mathcal{N}=1$ supersymmetry 
\footnote{For $d=10$, $\mathcal{N}=1$ supersymmetry 
a consistent local gauge symmetry group 
is characterized 
by the Lie algebras $\mathfrak{so}(32)$ and $E_{8}\times E_{8}$.}.

%
It has been argued that 
the five superstring theories are connected with each other. 
T-duality relates a pair of the two type II superstring theories 
and also a pair of the two heterotic superstring theories. 
S-duality relates the type I superstring theory 
to the $SO(32)$ heterotic superstring theory 
and the type IIB superstring theory to itself. 
T-dualities and S-dualities 
generates a discrete non-abelian group, 
the so-called U-duality group \cite{Hull:1994ys,Witten:1995ex}. 
Remarkably the U-duality groups are recognized as 
discretization of global symmetry groups of supergravity. 
In fact it is known that 
type IIA and IIB superstring theories are 
the ultraviolet (UV) completions of 
$d=10$ type IIA and IIB supergravities 
\footnote{Originally type IIB supergravity was discovered
\cite{Green:1981yb} and constructed \cite{Schwarz:1983wa,Howe:1983sra} as 
the low-energy limit of type IIB superstring
theory.} 
whereas type I and heterotic superstring theories 
are the UV completions of $d=10$ type I supergravities.

%
%
From the supergravity point of view, 
it is interesting to note that 
$d=10$ type IIA supergravity arises by dimensional reduction 
of $d=11$ supergravity \cite{Giani:1984wc,Huq:1983im,Campbell:1984zc}. 
$d=11$ supergravity \cite{Cremmer:1978km} is furnished 
with a particular interest 
since $d=11$ is the highest space-time dimension 
which realizes a consistent supersymmetric theory 
containing particles with spins $\ge 2$ \cite{Nahm:1977tg}. 
$d=11$ supergravity possesses a single 32-component spinor supercharge 
and its Lagrangian is unique if we require that 
the theory contains at most two-derivative interactions. 
Therefore the relation between superstring theory and 
$d=10$ supergravity indicates the existence 
of the UV completion of $d=11$ supergravity. 
It has been argued that 
in the strong string coupling limit 
an eleventh direction arises in type IIA superstring theory 
and the resulting eleven-dimensional theory is referred to as M-theory 
\cite{Townsend:1995kk,Witten:1995ex,Schwarz:1995jq} 
\footnote{The letter ``M'' proposed by 
E. Witten embodies several possible meanings; 
membrane, matrix, magic and mother.}. 
Conversely M-theory reduces to type IIA superstring theory 
upon the compactification on a spatial circle. 
Up until now M-theory is the most prospective candidate 
for the fundamental theory 
in that it may explain the origin of strings 
and unify the five superstring theory. 
However, a familiar perturbative method in string theory 
is not applicable because M-theory describes the 
strong coupling region of string theory.

%
As the fundamental string is a fundamental object 
in ten-dimensional superstring theory, 
the membrane appears to play a fundamental role in M-theory. 
This membrane is called M2-brane. 
Indeed $d=11$ supergravity contains a three-form gauge field, 
which leads to two stable extended objects 
as solitonic solutions; electric membrane and 
magnetic five-brane. 
Moreover it has been pointed out \cite{Duff:1987bx} that 
the M2-brane is identified with the fundamental string 
when M-theory is compactified on a circle 
and reduces to type IIA superstring theory. 
In spite of the prospective importance for the membranes in M-theory 
a number of attempts for the quantization of the membranes 
does not work well hitherto.

%
Although M-theory is much less understood than string theory 
due to the difficulty of the quantization of the membranes,  
we can still obtain several insights and clues from string theory 
and supergravity. 
In addition to the fundamental string, 
string theory also contains extended objects, 
the so-called D-branes on which open strings can end 
\cite{Polchinski:1995mt}. 
In fact ten-dimensional supergravities possess the solutions 
describing the geometries around such branes. 
There is a remarkable conjecture, 
the so-called AdS/CFT correspondence 
\cite{Maldacena:1997re,Gubser:1998bc,Witten:1998qj} 
which states that there is the correspondence between 
string/M-theory on certain supergravity geometries 
with anti-de-Sitter (AdS) factors and 
certain conformally invariant quantum field theories. 
The most basic evidence for the AdS/CFT correspondence 
is the equivalence between 
type IIB superstring theory 
on the $\textrm{AdS}_{5}\times S^{5}$ supergravity geometry 
constructed as a set of $N$ coincident D3-branes 
and the $d=4$, $\mathcal{N}=4$ 
superconformal $U(N)$ Yang-Mills gauge theory. 
Namely the low-energy dynamics for  
the world-volume of the planar $N$ D3-branes in flat space-time 
can be effectively described 
by $(1+3)$-dimensional maximally supersymmetric 
$U(N)$ Yang-Mills gauge theory \cite{Witten:1995im}. 

%
Similarly, in the near-horizon limit 
$d=11$ supergravity solutions describing 
planar M2-branes in flat space-time contain the AdS$_{4}$ factors 
and the low-energy dynamics of the M2-branes are expected to be 
described by the $(1+2)$-dimensional conformal field theories. 
As the eleven-dimensional 
flat background geometry can possess 32 space-time supercharges, 
the world-volume effective field theory of planar M2-branes 
should preserve half of the supersymmetry. 
Also the gauge degrees of freedom are needed to 
describe the internal degrees of freedom of multiple M2-branes.  
Thus the low-energy effective field theories 
of planar M2-branes are $d=3$, $\mathcal{N}=8$ 
superconformal gauge theories. 
The candidates for such effective field theories of 
world-volume dynamics of multiple M2-branes 
have been proposed as 
three-dimensional superconformal Chern-Simons matter theories, 
the so-called BLG-model 
\cite{Bagger:2006sk,Bagger:2007jr,Bagger:2007vi,
Gustavsson:2007vu,Gustavsson:2008dy} and the ABJM model \cite{Aharony:2008ug}.

%
In order to obtain new AdS/CFT examples 
it is desirable to consider more general supergravity solutions
describing the wrapped branes around certain cycles which may be curved. 
However, in the generic setup where 
the branes are wrapping an arbitrary manifold, 
all of the supersymmetries are destined to break down. 
In other words, 
specific background geometries of branes and 
specific cycles wrapped by branes 
must be chosen to preserve supersymmetry. 
Mathematically the supersymmetric cycles 
are characterized by the calibration \cite{MR666108}. 
There is a remarkable observation \cite{Bershadsky:1995qy} 
that topologically twisted field theories may 
give rise to the world-volume theories of wrapped branes. 
For the Euclidean D3-branes wrapping four-manifold 
there are three calibrated cycles embedded in 
special holonomy manifolds; 
(i) special Lagrangian submanifold in Calabi-Yau four-fold, 
(ii) coassociative submanifold in $G_{2}$ manifold and 
(iii) Cayley submanifold in $Spin(7)$ manifold. 
Each of them corresponds to 
three distinct topological twisting procedures 
\footnote{Also see \cite{Lozano:1999ji,Kato:2005fj,Mikhaylov:2012jm,Ito:2013eva} for 
interesting applications of the topological twisted 
$\mathcal{N}=4$ super Yang-Mills theories.}; 
(i) geometric Langlands (GL) twist 
\cite{Marcus:1995mq,Blau:1996bx,Labastida:1997vq,Kapustin:2006pk}, 
(ii) Vafa-Witten twist \cite{Vafa:1994tf} and 
(iii) Donaldson-Witten twist \cite{Yamron:1988qc}. 
The world-volume of D3-branes can be put on 
the product of two Riemann surfaces $\mathcal{C}\times \Sigma_{g}$. 
For the compact Riemann surface $\Sigma_{g}$ of genus $g>1$  
the field theories on the D3-branes are partially twisted 
on the curved Riemann surface to preserve supersymmetry. 
Since the compact manifold wrapped 
by branes introduces into the theory the typical energy scale as its volume, 
one can consider an additional limit 
where the energy is much smaller than the inverse size of the cycles. 
The resulting effective field theories 
then reduce to the two-dimensional topological sigma-models 
whose target space is specified by the BPS equations 
\cite{Bershadsky:1995qy,Yau:1995mv,Maldacena:2000mw,
Kapustin:2006pk,Benini:2012cz,
Benini:2013cda,Karndumri:2013iqa,Nagasaki:2014xya}.

When the Euclidean M2-branes wrap a compact curved three-manifold, 
the three-dimensional effective theories 
on the branes are fully twisted \cite{Lee:2008cr}. 
The $SO(3)$ Euclidean symmetry on the world-volume 
is topologically twisted in terms of the $SO(3)$ subgroup of the R-symmetry. 
For the M2-branes 
wrapping compact Riemann surface $\Sigma_{g}$ of genus $g$ 
supersymmetry is unbroken 
if the Riemann surface is chosen as holomorphic curve 
in Calabi-Yau manifold, 
which are the only known supersymmetric two-cycles. 
From the supergravity point of view, 
the solutions which describe the M2-branes 
wrapping compact Riemann surfaces have been studied 
\cite{Gauntlett:2001qs,Gauntlett:2003di,Kim:2013xza}
by using the gauged supergravity method \cite{Maldacena:2000mw}. 
The basic observation \cite{deWit:1984nz,deWit:1986iy} is that 
the dimensional reduction of $d=11$ supergravity on a seven-sphere 
can be truncated to give rise to 
the four-dimensional $SO(8)$ gauged supergravity 
where $SO(8)$ gauge symmetry corresponds to 
the isometry of the seven-sphere. 
Since the planar M2-branes take the form of 
$\mathrm{AdS}_{4}\times S^{7}$,  
the non-trivial coupling of the external $SO(8)$ gauge field 
which is nothing but an R-symmetry of the world-volume theory 
of the planar M2-branes may realize the curved geometries of the form 
$\mathrm{AdS}_{2}\times \Sigma_{g}$ instead of $\mathrm{AdS}_{4}$. 
Thus the uplift of the four-dimensional $SO(8)$ supergravity solutions 
can be used to construct the $d=11$ supergravity solutions 
describing the M2-branes wrapped on holomorphic curves. 
Correspondingly the three-dimensional 
effective superconformal field theories 
are partially twisted for $g\neq 0$ \cite{Okazaki:2014sga}. 
The $SO(2)$ Euclidean symmetry on the curved Riemann surface 
is topologically twisted in terms 
of the $SO(2)$ subgroup of the R-symmetry. 
Furthermore 
it is shown \cite{Okazaki:2014sga} that 
in the limit where the size of the Riemann surface goes to zero, 
superconformal quantum mechanical models arise as 
the low-energy effective theories.  
This thesis explores the new connection 
between the M2-branes 
and the superconformal quantum mechanics.

%
We should note that 
conformal symmetry and supersymmetry in one-dimensional field theory, 
i.e. quantum mechanics, contain a bunch of intriguing properties which 
do not appear in higher dimensional field theories, 
as we will discuss in this thesis. 

%
Although supersymmetric quantum mechanics 
was originally studied as the simple testing model 
for non-perturbative breaking of supersymmetry 
\cite{Witten:1981nf,Cooper:1982dm}, 
supersymmetric quantum mechanics is much more interesting itself. 
Supersymmetry is closely related to the translational symmetry 
as the square of the supercharges generates the momentum. 
However, in one dimension there are no spatial directions 
and the translational symmety generator is just the Hamiltonian, 
which reflects the reduced Poincar\'{e} symmetry in one-dimension. 
The reduced Poincar\'{e} symmetry 
looses the constraints for supersymmetry in one dimension 
and leads to richer structures than higher dimensional field theories.  
Indeed there exist supersymmetric quantum mechanical models which 
cannot reached via naive dimensional reductions 
from higher dimensional field theories. 
In parallel with that, 
there may be a large number of supermultiplets in one-dimension 
(see Table \ref{susynumber001})
and there is no relationship between 
the physical bosonic degrees of freedom and fermionic degrees of freedom 
in supersymmetric quantum mechanics. 
These properties are special in one dimension. 

%
Conformal symmetry in one-dimension also exhibits unique features. 
The reduced Poincar\'{e} symmetry identifies  
the generator of a translation with the Hamiltonian $H$ 
and does not allow for the generator of a rotational symmetry. 
Therefore the one-dimensional conformal symmetry 
is generated by the Hamiltonian $H$, 
the dilatation generator $D$ 
and the conformal boost generator $K$, 
all of which together form the $\mathfrak{sl}(2,\mathbb{R})$ algebra. 
Therefore the one-dimensional conformal group 
is $SL(2,\mathbb{R})\cong SO(1,2)$. 
%
The first detailed analysis of conformal mechanics  
appeared in \cite{deAlfaro:1976je}. 
The conformal mechanical models are typically 
characterized by the inverse-square potential 
 \footnote{The treatment of the
inverse-square potential in quantum mechanics was discussed in \cite{
shortley1931inverse,Case:1950an,
landau2009quantum,
tietz1959discrete,
bastai1963treatment,
meetz1964singular,
englefield1964scattering,
guggenheim1966inverse,
Ferreira:1970bx,
Jackiw:1972cb}.}. 
Inverse-square potential in quantum mechanics 
is a jewellery box in theoretical physics and mathematics containing 
black hole physics \cite{Claus:1998ts,Gibbons:1998fa,deAzcarraga:1998ni,
Michelson:1999dx,Michelson:1999zf,Maloney:1999dv,Kallosh:1999mi,
Papadopoulos:2000ka,BrittoPacumio:2000sv,Ivanov:2002tb,
Bellucci:2002va,Galajinsky:2008ce,Bellucci:2008xu}, 
AdS$_{2}$/CFT$_{1}$ correspondence 
\cite{Strominger:1998yg,Townsend:1998qp,Nakatsu:1998st,
Spradlin:1999bn,Blum:1999pc,Cadoni:1999ja,
Strominger:2003tm,Leiva:2003kd,Chamon:2011xk,Jackiw:2012ur}, 
QCD \cite{Dosch:2014wxa,deTeramond:2014asa}, 
quantum Hall effect \cite{Azuma:1993ra,kane1991general}, 
Tomonaga-Luttinger liquid \cite{Kawakami:1991yr}, 
string theory 
\cite{Dabholkar:1991te,
McGreevy:2003dn,Verlinde:2004gt,Gates:2004di,
Agarwal:2006nv,Wen:2008hi}, 
spin chains
\cite{Haldane:1987gg,SriramShastry:1987gh,Minahan:1992ht,Ahn:1995jt,hikami2000supersymmetric,BasuMallick:2007nb,
barba2008polychronakos,basu2009exactly}, 
Efimov effect 
\cite{Bawin:2003dm,Camblong:2003mb,Braaten:2004pg,Barford:2004fz}, 
mesoscopic physics 
\cite{beenakker1994exact,caselle1995distribution}, 
quantum dot 
\cite{date1998classical}, 
quantum chaos 
\cite{simons1994matrix}, 
fractional exclusion statistics 
\cite{murthy1994thermodynamics,Ha:431407}, 
random matrix model 
\cite{feigel1997supersymmetric,
khare1997quantum,oas1997universal,Bardek:2010jg}, 
Seiberg-Witten theory \cite{DHoker:1999ft,Gorsky:2000px}, 
Jack polynomial \cite{Desrosiers:2001ri,Desrosiers:2002sv,
Desrosiers:2002ww,Desrosiers:2003me,MR2259017,MR2587340,MR2753265}, 
and relevant algebraic and integrable structures 
\cite{Hikami:1994du,Awata:1994xd,Bordner:1999us,Khastgir:2000ig}.  
%
One of the well-known such quantum mechanical models 
is the Calogero model \cite{Calogero:1969xj,Calogero:1969b} 
which is the multi-particle system 
with the pairwise inverse-square interaction 
\footnote{See \cite{BrittoPacumio:1999ax,Fedoruk:2011aa,Ghosh:2011tu} for the 
enlightening reviews on (super)conformal mechanics 
and also see 
\cite{Olshanetsky:1981dk,Olshanetsky:1983wh,askol1990integrable,
Pasquier:1994cs,DHoker:1998yg,Polychronakos:1999sx,Polychronakos:2006nz,
Milekovic:2006kj,kuramoto2009dynamics}
for excellent reviews on the Calogero model. 
}. 
It was firstly proven in
\cite{Barucchi:1976am,Wojciechowski:1977uj} that 
the Calogero model has the $SL(2,\mathbb{R})$ conformal symmetry. 
The Calogero model and its generalizations can be viewed as 
a system of free indistinguishable particles 
\cite{Polychronakos:2006nz}. 
The indistinguishableness implies that 
the permutation group acting on the configuration of the particles 
is treated as a discrete gauge symmetry in the system and therefore 
the Calogero model and its generalized models can be obtained  
from gauged matrix models 
\cite{MR0478225,Polychronakos:1991bx}. 
This observation is used to find new conformal mechanical systems 
by starting gauged matrix models or gauged quantum mechanical models 
and reducing the systems via Hamiltonian reduction 
\cite{Fedoruk:2008hk,Fedoruk:2011aa}.

%
Since the appearance of the seminal works 
\cite{Akulov:1984uh,Fubini:1984hf} on 
superconformal quantum mechanics (SCQM),  
there has been a great deal of efforts to 
construct superconformal quantum mechanics. 
The superconformal quantum mechanical models are  
characterized by the superconformal group, 
i.e. the Lie supergroup which contains 
one-dimensional conformal group $SL(2,\mathbb{R})$ 
and R-symmetry group as factored bosonic subgroups. 
One of the most powerful methods to build up superconformal mechanical
systems is the superspace and superfield formalism. 
In fact for $\mathcal{N}\le 8$ supersymmetric cases it does work and 
several superconformal quantum mechanical systems are constructed.  
For $\mathcal{N}=1$ supersymmetric case,  
the superconformal group is $OSp(1|2)$ and 
there is no non-trivial one particle superconformal quantum mechanics.  
For $\mathcal{N}=2$ supersymmetric case,  
the superconformal group is $OSp(2|2)\cong SU(1,1|1)$ 
and the simplest one particle model is the pioneering work of 
\cite{Akulov:1984uh,Fubini:1984hf}. 
For $\mathcal{N}=4$ supersymmetric case, 
the generic superconformal group is $D(2,1;\alpha)$ 
which is a one-parameter family of supergroup.  
The superspace and superfield formalism 
keeping track of the exceptional supergroup 
can be derived by the non-linearlizations technique
\cite{Coleman:1969sm,Callan:1969sn,Volkov:1973vd} 
and several models have been constructed. 
In the case of $\mathcal{N}=8$ 
there exist four different superconformal groups; 
$SU(1,1|4)$, $OSp(8|2)$, $OSp(4^{*}|4)$ and $F(4)$. 
Such several choices of superconformal group 
cannot occur in higher dimensional field theories 
and thus present a various families 
of $\mathcal{N}=8$ superconformal quantum mechanics. 
However, for $\mathcal{N}>8$ 
the superspace and superfield formalism is not 
unrealistic and unsuccessful. 
One of the signals for such difficulty is that 
the number of bosonic and fermionic component fields 
in the supermultiplets typically becomes greater than 
the number of supersymmetry when $\mathcal{N}$ is larger than eight 
(see Table \ref{susynumber001}). 
In spite of the depression of the superspace and superfield formalism, 
several $\mathcal{N}>8$ superconformal 
quantum mechanical models have been constructed 
via reduction of the three-dimensional 
quiver type superconformal Chern-Simons matter theories
\cite{Okazaki:2014sga}. 
As mentioned before, 
these superconformal quantum mechanical models 
may capture the low-energy dynamics of 
the multiple M2-braens wrapped on a compact Riemann surface. 
We will spell out the details of 
these superconformal quantum mechanical models in this thesis.

\section{What I did}
The organization of this thesis consists of three parts. 
In  part I and II we will review two main subjects; 
the superconformal quantum mechanics and the M2-branes. 
The original part of the author's work based on \cite{Okazaki:2014sga} 
is part III, in which we will discuss the new connection 
between the two subjects, 
that is the superconformal quantum mechanics 
emerging from M2-branes.

Part I contains two chapters; 
chapter \ref{chapcqm} and \ref{chsc1m01}. 
In chapter \ref{chapcqm} 
we will discuss various aspects of conformal quantum mechanics. 
We will start with section \ref{seccqm1} 
by studying the DFF-model \cite{deAlfaro:1976je} 
and its $SL(2,\mathbb{R})$ conformal symmetry 
and then in section \ref{secana0} 
we will explore the quantum properties of the system. 
In section \ref{gcqmsec1} 
we will see that 
the conformal mechanical models 
can be derived from the gauged mechanical system 
via Hamiltonian reduction or Routh reduction. 
In section \ref{secbh1} 
we will review the observation \cite{Claus:1998ts} 
that 
in the near horizon of the extreme Reissner-Nordstr\"{o}m 
black hole the motion of the charged particle 
can be described by the conformal mechanics (\ref{bhcqm10}).
In section \ref{secnonlin1} 
we will present the non-linear realization 
technique which is useful 
to construct (super)conformal quantum mechanical models 
and then review the statement in \cite{Ivanov:2002tb} 
that DFF-model (\ref{conflag1}) 
is equivalent to the black hole conformal mechanics (\ref{bhcqm10}) 
in \cite{Claus:1998ts}. 
We will extend the analysis to the multi-particle models, 
i.e. the sigma-models in section \ref{mulcqm0}. 
We will review the discussion in \cite{Michelson:1999zf} that 
the target space of the conformal sigma-model 
possesses a homothety vector field whose 
associated one-form is closed. 
We will argue that the gauging procedure for the 
multi-particle model, the matrix model yield the Calogero model in 
section \ref{mulcqm2}. 
In chapter \ref{chsc1m01} 
we will turn to the discussion on 
the superconformal quantum mechanics. 
We will recall the Lie superalgebra 
and Lie supergroup and then 
discuss the one-dimensional 
superconformal group (see in Table \ref{listsup1}) 
in section \ref{secsuperalg01}.  
In section \ref{1dsusysec001} 
we will explain the exisotic structures of 
supersymmetry in one-dimension, 
which allows us to construct various supermultiplets. 

Part II, which is comprised of two chapters; 
chapter\ref{blgch1} and \ref{abjmch1}, 
is devoted to the low-energy effective field theories of the M2-branes. 
We will review the BLG-model 
in chapter \ref{blgch1} 
and the ABJM-model in chapter \ref{abjmch1}. 
We will present our notations and conventions 
and also the several conjectural statements for 
the BLG-model and the ABJM-model in these chapters. 

Part III is the most important part of this thesis. 
It is based on the author's work of \cite{Okazaki:2014sga},  
in which we will engage in the superconformal quantum mechanical models 
arising from the M2-branes. 
We consider the multiple membranes 
on a compact Riemann surface and study the IR quantum mechanics 
by taking the limit where the energy scale is much lower than the 
inverse size of the Riemann surface. 
In chapter \ref{secflat1} we will demonstrate that 
the resulting quantum mechanics from the BLG-model 
compactified on a torus is the $\mathcal{N}=16$ 
superconformal gauged quantum mechanics. 
Furthermore we will find the $OSp(16|2)$ superconformal quantum mechanics 
from the reduced system. 
Similarly in chapter \ref{abjmscqm} 
we will investigate the 
IR quantum mechanics from the ABJM-model 
compactified on a torus, 
which turns out to be the $\mathcal{N}=12$ superconformal gauged 
quantum mechanics. 
By the Hamiltonian reduction, or the Routh reduction 
we will also find the $SU(1,1|6)$ superconformal quantum mechanics 
from the gauged quantum mechanics. 
In chapter \ref{curvedtwistingcg01} 
we will present various examples of the topological twisting, 
which is the important concept to describe curved branes 
in string theory and M-theory. 
In chapter \ref{seccurv} 
we will survey the M2-branes wrapped on a curved 
Riemann surface which is taken as a holomorphic curve 
in a Calabi-Yau manifold to preserve supersymmetry. 
We will present a prescription of the topological twisting for the case 
where the Calabi-Yau space is constructed as the direct sum of the 
line bundles over the Riemann surface. 
In chapter \ref{cy2sec} 
we will complete the analysis of the 
M2-branes wrapped around the holomorphic Riemann surface 
in a K3 surface. 
We will find the $\mathcal{N}=8$ superconformal gauged quantum 
mechanics which may describe the motion of the two M2-branes 
wrapping holomorphic curve in a K3 surface. 
Finally in chapter \ref{secconc} we will present conclusion 
and discuss the future directions.

\part{Superconformal Mechanics}

\chapter{Conformal Mechanics}
\label{chapcqm}
In this chapter 
we will review the conformal quantum mechanics. 
The simplest model is the so-called DFF model \cite{deAlfaro:1976je}. 
In section \ref{seccqm1}, \ref{secana0}, \ref{seccqm2} and \ref{seccqm3}  
we will learn from the DFF model several remarkable features 
of the conformal symmetry in one-dimension, 
which cannot occur in higher dimensional field theories. 
Then in section \ref{gcqmsec1} we will argue the alternative formulation 
of the conformal mechanical models as the gauged mechanical models. 
As an interesting application of 
the conformal quantum mechanics 
we will discuss the relationship between 
the conformal mechanics and black hole in section \ref{secbh1} 
and introduce the non-linear realization method 
to construct (super)conformal quantum mechanics in section
\ref{secnonlin1}. 
Finally we will extend the analysis to 
the multi-particle conformal mechanical models in section \ref{mulcqm0}
and \ref{mulcqm2}. 

\section{$SL(2,\mathbb{R})$ conformal symmetry}
\label{seccqm1}
In $d$-dimensions 
a scale invariant Lagrangian for a scalar field $\phi$ has the form 
\begin{align}
\label{scaleinvlag1}
L=\frac12 \partial_{\mu}\phi \partial^{\mu}\phi -\gamma\phi^{\frac{2d}{d-2}}
\end{align}
where $\gamma$ is a dimensionless coupling constant. 
In one-dimensional case we get the Lagrangian
\begin{align}
\label{conflag1}
L=\frac12\left(
\dot{x}^{2}-\frac{\gamma}{x^{2}}
\right).
\end{align}
This simple quantum mechanical model is the so-called DFF-model 
\cite{deAlfaro:1976je}. 
To keep particles from falling into the origin, 
the coupling constant $\gamma$ should be positive classically. 
As we will see in the following discussion, 
quantum mechanically the uncertain principle 
gives rise to the minimum value $\gamma=-\frac14$, however, 
the normalizability of the wavefunction of the ground state requires
that $\gamma$ is positive. 
So we will denote $\gamma=g^{2}$ for convenience. 
The Lagrangian (\ref{conflag1}) leads to the equation of motion
\begin{align}
\label{dffeom1}
\ddot{x}=\frac{g^{2}}{x^{3}}.
\end{align}

The action
\begin{align}
\label{confac1}
S=\int L \ dt = \frac12 \int dt \left(
\dot{x}^{2}-\frac{g^{2}}{x^{2}}
\right) 
\end{align}
is invariant under the transformations
\begin{align}
\label{finconf01}
t'&=\frac{\alpha t+\beta}{\gamma t+\delta},\\
\label{finconf02}
x'(t')&=\frac{1}{\gamma t+\delta}x(t)
\end{align}
where the real numbers 
$\alpha$, $\beta$, $\gamma$ and $\delta$ 
form a real $2\times 2$ matrix with determinant one
\begin{align}
\label{finconfmt1}
A=\left(
\begin{array}{cc}
\alpha&\gamma\\
\beta&\delta\\
\end{array}
\right), \ \ \ \ \ 
\det A=1.
\end{align}

\begin{enumerate}
 \item \textbf{translation}

The subgroup of the matrix (\ref{finconfmt1}) 
\begin{align}
\label{finconfmt2}
\left(
\begin{array}{cc}
1&0\\
a&1\\
\end{array}
\right)
\end{align}
with 
$\alpha=1$, $\beta=a$, $\gamma=0$, $\delta=1$ 
yields 
\begin{align}
\label{finconf03}
t'&=t+a,\nonumber\\
x'(t')&=x(t).
\end{align}
This corresponds to the translation. 

\item \textbf{dilatation}

The subgroup of the matrix (\ref{finconfmt1}) 
\begin{align}
\label{finconfmt3}
\left(
\begin{array}{cc}
e^{\frac{b}{2}}&0\\
0&e^{-\frac{b}{2}}\\
\end{array}
\right)
\end{align} 
with 
$\alpha=e^{\frac{b}{2}}$, $\beta=0$, $\gamma=0$, 
$\delta=e^{-\frac{b}{2}}$ 
generates the transformations 
\begin{align}
\label{finconf04}
t'&=e^{b}t,\nonumber\\
x'(t')&=e^{\frac{b}{2}}x(t).
\end{align}
This is the dilatation.

\item \textbf{conformal boost}

The subgroup of the matrix (\ref{finconfmt1}) 
\begin{align}
\label{finconfmt4}
\left(
\begin{array}{cc}
1&-c\\
0&1\\
\end{array}
\right)
\end{align}
with $\alpha=1$, 
$\beta=0$, $\gamma=-c$, $\delta=1$ 
corresponds to the transformations
\begin{align}
\label{finconf05}
t'&=\frac{t}{-ct+1},\nonumber\\
x'(t')&=\frac{x(t)}{-ct+1}.
\end{align}
This is the conformal boost transformation. 
\end{enumerate}
From a set of finite transformations 
(\ref{finconf03}), (\ref{finconf04}) and (\ref{finconf05}) 
we see that the action (\ref{confac1}) is invariant 
under the infinitesimal one-dimensional conformal transformations
\begin{align}
\label{inconf1a}
\delta t&=f(t)=a+bt+ct^{2},\\
\label{inconf1b}
\delta x&=\frac12 \dot{f}x
=\frac12(b+2ct)x.
\end{align}
The passive transformations 
(\ref{inconf1a}) and (\ref{inconf1b}) lead to 
the active transformations
\begin{align}
\label{inconf2a}
\delta t&=0,\\
\label{inconf2b}
\delta x&=\frac12 \dot{f}x-f\dot{x}.
\end{align}
Noting that $a$, $b$ and $c$ are 
the infinitesimal parameters of the translation, 
the dilatation and the conformal boost, 
we can compute the Noether charges, i.e. 
the Hamiltonian $H$, 
the dilatation operator $D$ 
and the conformal boost operator $K$
\begin{align}
\label{hktbos1}
H&=\frac{p^{2}}{2}+\frac{g^{2}}{2x^{2}},\\
\label{hktbos2}
D&=tH-\frac14\left(xp+px\right),\\
\label{hktbos3}
K&=t^{2}H-\frac12 t(xp+px)+\frac12 x^{2}
\end{align}
where $p=\dot{x}$ is the canonical momentum. 
The operators $D$ and $K$ 
are the constants of motion in the sense that 
\begin{align}
\label{dkconst1}
\frac{\partial D}{\partial t}+[H,D]&=0,&
\frac{\partial K}{\partial t}+[H,K]&=0.
\end{align}

One can carry out the canonical quantization by 
establishing the equal time commutation relation
\begin{align}
\label{canonical00}
[x,p]=i.
\end{align}
Using the commutation relation (\ref{canonical00}), 
we can show that 
\begin{align}
\label{hkd00}
[H,D]&=iH,\\
\label{hkd01}
[K,D]&=-iK,\\
\label{hkd02}
[H,K]&=2iD
\end{align}
and 
\begin{align}
\label{hkd00a1}
i[H,x(t)]&=\dot{x}(t),\\
\label{hkd00a2}
i[D,x(t)]&=t\dot{x}(t)-\frac12 x(t),\\
\label{hkd00a3}
i[K,x(t)]&=t^{2}\dot{x}(t)-tx(t).
\end{align}
If we express the time independent part of $D$ and $K$ as 
\begin{align}
\label{hkdconf2a}
D_{0}&:=-\frac14 
\left(xp+px\right),\\
\label{hkdconf3a}
K_{0}&:=\frac12 x^{2},
\end{align}
then the equations 
(\ref{hkd00a1}), (\ref{hkd00a2}) and (\ref{hkd00a3}) are rewritten as
\begin{align}
\label{hkd0b1}
i[H,x(t)]&=\dot{x}(t),\\
\label{hkd0b2}
i[D_{0},x(t)]&=-\frac12 x(t),\\
\label{hkd0b3}
i[K_{0},x(t)]&=0.
\end{align}
These equations are regarded as the Heisenberg equations.  
The equation (\ref{hkd0b1}) is familiar for general quantum mechanical
systems and yields the variation of the operator with respect to time 
while the equation (\ref{hkd0b2}) 
gives rise to the scale dimension of the operator. 

Note that the explicit time dependence of $D$ and $K$ 
can be absorbed into the similarity transformations
\begin{align}
\label{simdk0}
D&=e^{itH}D_{0}e^{-itH},&
K&=e^{itH}K_{0}e^{-itH}
\end{align}
So we will use the time independent parts as the explicit 
expressions for $D$ and $K$ and drop off the subscripts.

Defining
\begin{align}
\label{so12a}
T_{0}&=\frac12\left(
\frac{K}{a}+aH
\right),\\
T_{1}&=D,\\
T_{2}&=\frac12\left(
\frac{K}{a}-aH
\right)
\end{align}
where $a$ is a constant with dimension of length, 
we find from (\ref{hkd00})-(\ref{hkd02}) the explicit representation 
of the $\mathfrak{so}(1,2)$ algebra
\begin{align}
\label{so12b}
[T_{i},T_{j}]=i\epsilon_{ijk}T^{k}
\end{align}
where $\epsilon_{ijk}$ is a three-index anti-symmetric tensor with 
$\epsilon_{012}=1$ and 
$g_{ij}=\mathrm{diag}(1,-1,-1)$. 
If we introduce
\begin{align}
\label{vira1}
L_{0}&=\frac12 \left(\frac{K}{a}+aH\right)=T_{0},
\\
\label{vira1a}
L_{\pm}&=
\frac12\left(
\frac{K}{a}-aH\pm 2iD
\right)=T_{2}\pm iT_{1},
\end{align}
then we get the explicit representation 
of the $\mathfrak{sl}(2,\mathbb{R})$ algebra in the Virasoro form
\begin{align}
\label{vira2}
[L_{n},L_{m}]=(m-n)L_{m+n}
\end{align}
with $m,n=0,\pm$. 
Note that 
\begin{align}
\label{vira2a1}
H&=\frac{1}{a}\left[
L_{0}-\frac12 \left(L_{+}+L_{-}\right)
\right],\\
\label{vira2a2}
D&=\frac{1}{2i}\left(L_{+}-L_{-}\right),\\
\label{vira2a3}
K&=a\left[
L_{0}+\frac12\left(L_{+}+L_{-}\right)
\right].
\end{align}

Recall that 
in the representation theory 
the Casimir invariants play an important role 
since their eigenvalues may characterize the representations.  
The one-dimensional conformal group $SL(2,\mathbb{R})$ 
is of rank one and therefore possesses 
one independent second-order Casimir invariant. 
The second-order Casimir operator $\mathcal{C}_{2}$ 
of the $\mathfrak{sl}(2,\mathbb{R})$ algebra is given by
\begin{align}
\label{casimir2}
\mathcal{C}_{2}
&=T_{0}^{2}-T_{1}^{2}-T_{2}^{2}\nonumber\\
&=L_{0}(L_{0}-1)-L_{+}L_{-}\nonumber\\
&=\frac12 (HK+KH)-D^{2}\nonumber\\
&=\frac{g^{2}}{4}-\frac{3}{16}.
\end{align}

\section{Spectrum}
\label{secana0}
\begin{figure}
\begin{center}
\includegraphics[width=8.5cm]{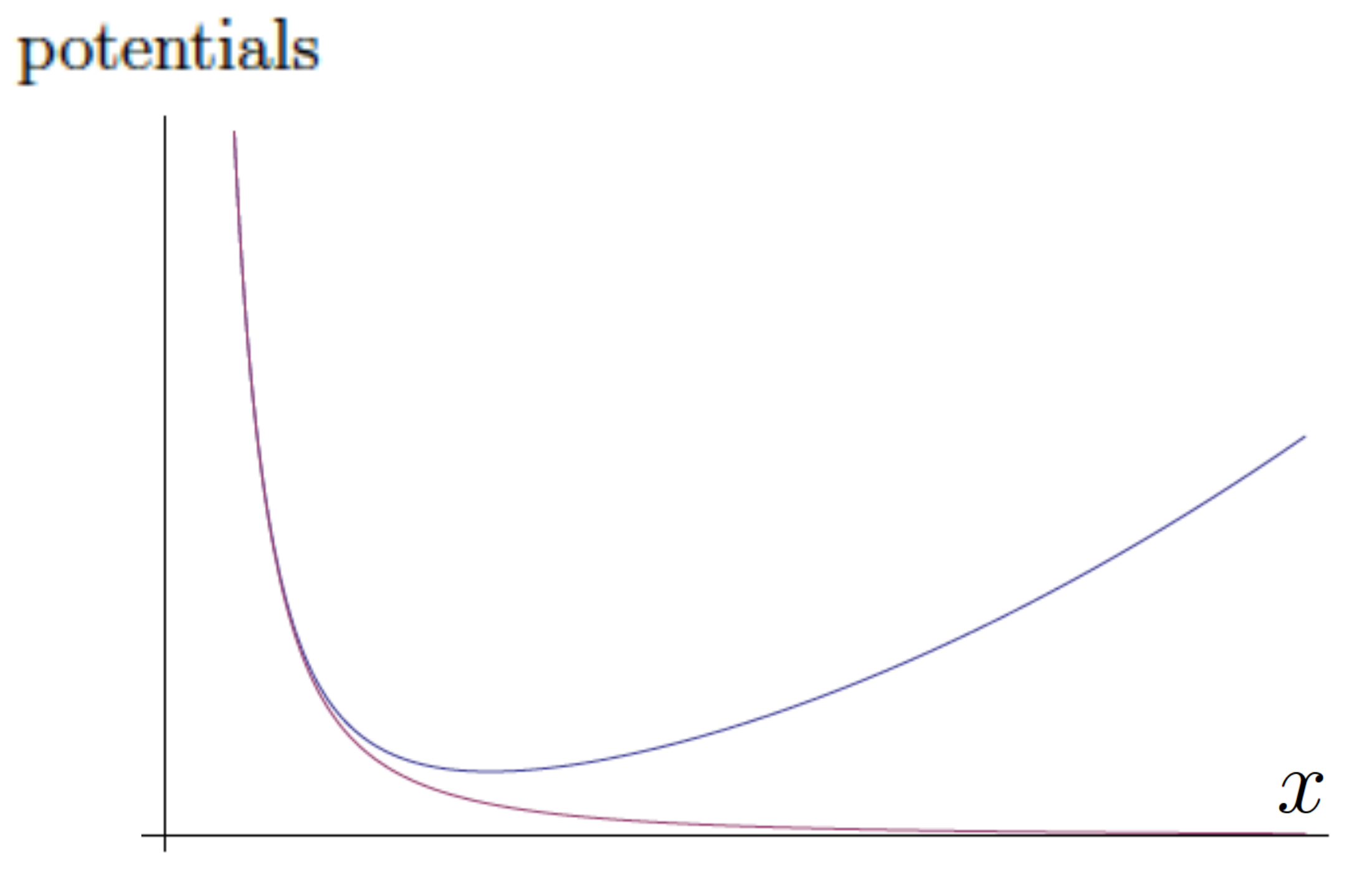}
\caption{
The potentials for the original Hamiltonian $H$ and 
the new Hamiltonian $L_{0}$. 
The red line is the potential for $H$ 
and the blue one is for $L_{0}$.  
}
\label{potfig1}
\end{center}
\end{figure}

It is known that 
the quantum formalism based on the Hamiltonian $H$  
is awkward to describe the conformal quantum mechanics. 
The spectrum of $H$ is continuous 
due to the existence of $D$ \footnote{
For the DFF-model (\ref{conflag1}) this can be readily seen 
from the behavior of the inverse-square potential as $x\rightarrow
\infty$ (Figure \ref{potfig1}). 
}. 
This is because if $|E\rangle$ is an eigenstate of energy $E$, 
then $e^{i\alpha D}|E\rangle$ is that of energy $e^{2\alpha} E$ 
with $\alpha$ being an arbitrary real parameter. 
Thus the spectrum contains all $E>0$ eigenvalues of $H$. 

The corresponding wave functions are given by 
\begin{align}
\label{ewave1}
\psi_{E}(x)=C\sqrt{x}J_{\sqrt{g^{2}+\frac14}}\left(\sqrt{2E}x\right)
\end{align}
where $C$ is a normalization factor and  
$J_{\alpha}$ is the Bessel function of the first kind. 
For each of the eigenstates with the eigenvalues $E>0$ 
there exists a normalizable plane wave.

On the other hand, the wavefunction of the zero energy state is given by 
\footnote{
Due to the infinite repulsive potential at the origin, 
we here consider the solution $\psi_{0}(x)$ satisfying 
the boundary condition $\psi_{0}(x)|_{x=0}=0$. 
}
\begin{align}
\label{ewave2}
\psi_{0}(x)=Cx^{-\frac12 +\frac{\sqrt{1+4g^{2}}}{2}}
\end{align}
where $C$ is a constant value. 
To make matters worse, 
this eigenfunction is not even plane wave normalizable and 
this makes it difficult 
for us to regard the state with $E=0$ as the ground state. 

However. it is important to note that \cite{deAlfaro:1976je} 
any combination 
\begin{align}
\label{ghdk01}
G=uH+vD+wK
\end{align}
of the three conformal generators is a constant of motion 
in the sense that 
\begin{align}
\label{ghdk02}
\frac{\partial G}{\partial t}+i[H,G]=0.
\end{align}
This implies that 
the transformations generated by $G$ leave the action invariant. 
Hence we may use the operator $G$ as 
the new Hamiltonian to study the evolution of the system. 

\begin{figure}
\begin{center}
\includegraphics[width=12.5cm]{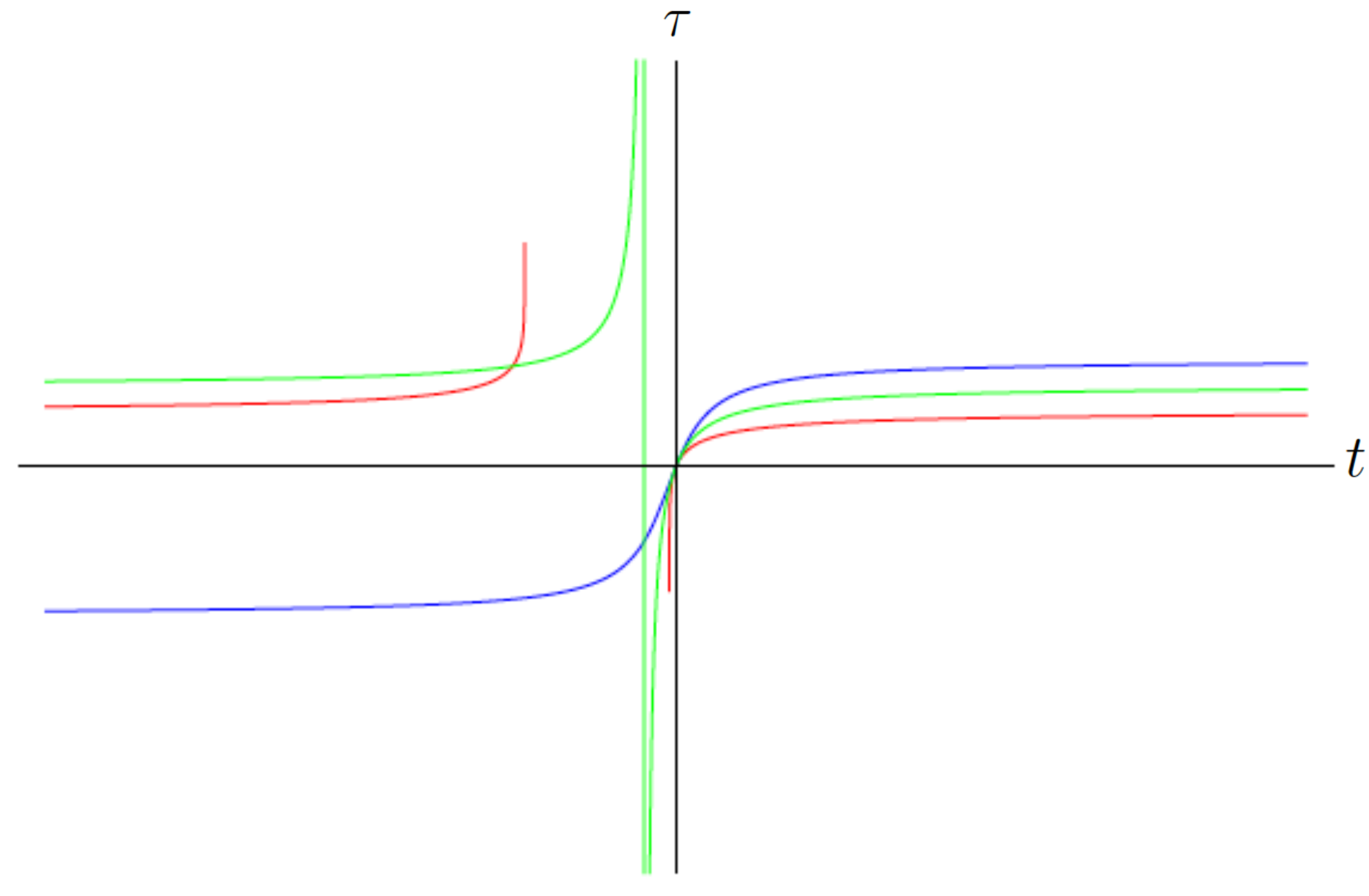}
\caption{The new time parameter $\tau$ as a function of the original time $t$. 
The red curve represents the non-compact case $\Delta >0$, 
which cannot sweep over the whole time $t$.  
The blue curve corresponds to the compact case $\Delta <0$ covering all
 the time region. 
Green curve denotes the case $\Delta =0$, 
which contains one singular point in $\tau$.}
\label{taufig}
\end{center}
\end{figure}

The switching from $H$ 
to the new evolution operator $G$ 
can be interpreted as a redefinition of the time and the coordinate. 
Let us introduce a new time parameter
\begin{align}
\label{newtime1}
d\tau=\frac{dt}{u+vt+wt^{2}}
\end{align}
and a new variable 
\begin{align}
\label{newcoord1}
q(\tau)=\frac{x(t)}{\sqrt{u+vt+wt^{2}}}.
\end{align}
Then we find the action of $G$ on the operator and 
on the state given by
\begin{align}
\label{newcoord2}
\frac{dq(\tau)}{d\tau}&=i[G,q(\tau)],\\
G|\Psi(\tau)\rangle&=i\frac{d}{d\tau}|\Psi(\tau)\rangle
\end{align}
as required. 
Although the operator $G$ may describe the evolution 
in the new time $\tau$, 
it is not yet complete to justify the passing to the new description. 
We need to examine whether the new Hamiltonian and 
the new coordinates cover the whole evolution in time 
from $t=-\infty$ to $+\infty$. 
From (\ref{newtime1}) we can express the new time parameter as
\begin{align}
\label{newtime2}
\tau=\int_{t_{0}}^{t}\frac{dt'}{u+vt'+w{t'}^{2}}+\tau_{0}.
\end{align}
This integral depends on the zeros of the denominater 
and the result is classified by the discriminant
\begin{align}
\label{tau000}
\Delta=v^{2}-4uw.
\end{align}
We find
\begin{align}
\label{tau001}
\tau=
\begin{cases}
\frac{1}{\sqrt{\Delta}}\left(
\ln \left|
\frac{2wt+v-\sqrt{\Delta}}{2wt+v+\sqrt{\Delta}}
\right|-
\ln \left|
\frac{v-\sqrt{\Delta}}{v+\sqrt{\Delta}}
\right|
\right)&\textrm{for $\Delta>0$}\cr
&\cr
\frac{2}{\sqrt{|\Delta|}}
\left(
\tan^{-1}\frac{2wt+v}{\sqrt{|\Delta|}}
-\tan^{-1}\frac{v}{\sqrt{|\Delta|}}
\right)&\textrm{for $\Delta<0$}\cr
&\cr
-\left(
\frac{2}{2wt+v}-\frac{2}{v}
\right)&\textrm{for $\Delta=0$}\cr
\end{cases}
\end{align}
where we normalize as $\tau_{0}=0$. 
For $\Delta >0$, 
the parameter $\tau$ 
cannot sweep the whole time region $-\infty\le t\le \infty$. 
This unpleasant feature is associated with the fact that 
the corresponding operators are non-compact 
and their spectrums are physically unacceptable. 
The dilatation operator $D$ belongs to this class. 
When $\Delta <0$, 
$\tau$ can be defined over the whole time interval $-\infty\le t\le \infty$. 
The corresponding operators in this case generate 
a compact rotation and their spectrums have physically satisfactory
characteristics. 
In the case of $\Delta=0$, 
the whole time interval $-\infty\le t\le \infty$ 
can be described over $-\infty\le \tau\le \infty$, 
however, there exists one singular point in $\tau$ at 
$t=-\frac{v}{2w}$. 
This is the case for the Hamiltonian $H$ and the 
conformal boost operator $K$. 
These three cases are illustrated in Figure \ref{taufig}.

In terms of the new set of coordinates 
(\ref{newtime1}) and (\ref{newcoord1}), 
we can rewrite the action (\ref{confac1}) as
\begin{align}
\label{confac2}
S&=\int d\tau 
\left[
\frac12\dot{q}^{2}
-\frac{g^{2}}{2q^{2}}
+\frac12\left(
\frac{v}{2}
+wt\right)^{2}q^{2}
+\frac12(v+2wt)q\dot{q}
\right]\nonumber\\
&=\int d\tau
\left[
\frac12\dot{q}^{2}
-\frac{g^{2}}{2q^{2}}
+\frac{\Delta}{8}q^{2}
+\frac12\frac{d}{d\tau}\left\{
\left(\frac{v}{2}+wt\right)q^{2}
\right\}
\right]\nonumber\\
&=\int d\tau
\left[
\frac12\dot{q}^{2}
-\frac{g^{2}}{2q^{2}}
+\frac{\Delta}{8}q^{2}
\right]\nonumber\\
&=\int d\tau L_{\tau}
\end{align}
up to the total $\tau$ derivative. 
Note that the dot denotes $\tau$ derivative in (\ref{confac2}). 
The new Lagrangian $L_{\tau}$ leads to the new Hamiltonian 
\begin{align}
\label{newham001}
H_{\tau}
&=\dot{q}\frac{\partial L_{\tau}}{\partial\dot{q}}-L_{\tau}
\nonumber\\
&=\frac12 \left(
\dot{q}^{2}+\frac{g^{2}}{q^{2}}-\frac{\Delta}{4}q^{2}
\right)
\end{align}
with 
\begin{align}
\label{newham002}
G(x(t),\dot{x}(t))=H_{\tau}(q(\tau,)\dot{q}(\tau)).
\end{align}

Note that $L_{0}=T_{0}$ is the compact generator satisfying
$\Delta=-1<0$. 
Qualitatively one can see that 
the potential energy of this new Hamiltonian $L_{0}$ acquires 
the minimum and assymptotes to infinity 
(Figure \ref{potfig1}). 
The new time coordinate $\tau$ and variable $q(\tau)$ associated with 
the generator $L_{0}$ are given by
\begin{align}
\label{newtime1a}
\tau&=2\tan^{-1}\left(\frac{t}{a}\right),\\
\label{newcoord1a}
q(\tau)&=\sqrt{\frac{2}{a}}\frac{x(t)}{\sqrt{1+\left(\frac{t}{a}\right)^{2}}}.
\end{align}
As we will discuss in section \ref{secbh1}, 
in the black hole interpretation 
$\tau$ can be identified with the proper time of the test particle 
near the horizon of the extremal black hole \cite{Claus:1998ts}.

The fact that the operator $L_{0}$ is regarded as the new Hamiltonian 
of the system can be paraphrased as 
the group theoretical statement that 
infinite dimensional unitary representations 
in terms of Hermitian operators 
of the non-compact group $SL(2,\mathbb{R})$ 
are characterized by the discrete eigenvalues of 
the Casimir operator $\mathcal{C}_{2}$ 
and of the compact generator $L_{0}$  
\footnote{The diagonalization of the non-compact operator 
requires the continuous basis 
\cite{MR0223489,MR0242438,MR0226912,MR0273920}.}.

\begin{figure}
\begin{center}
\includegraphics[width=6cm]{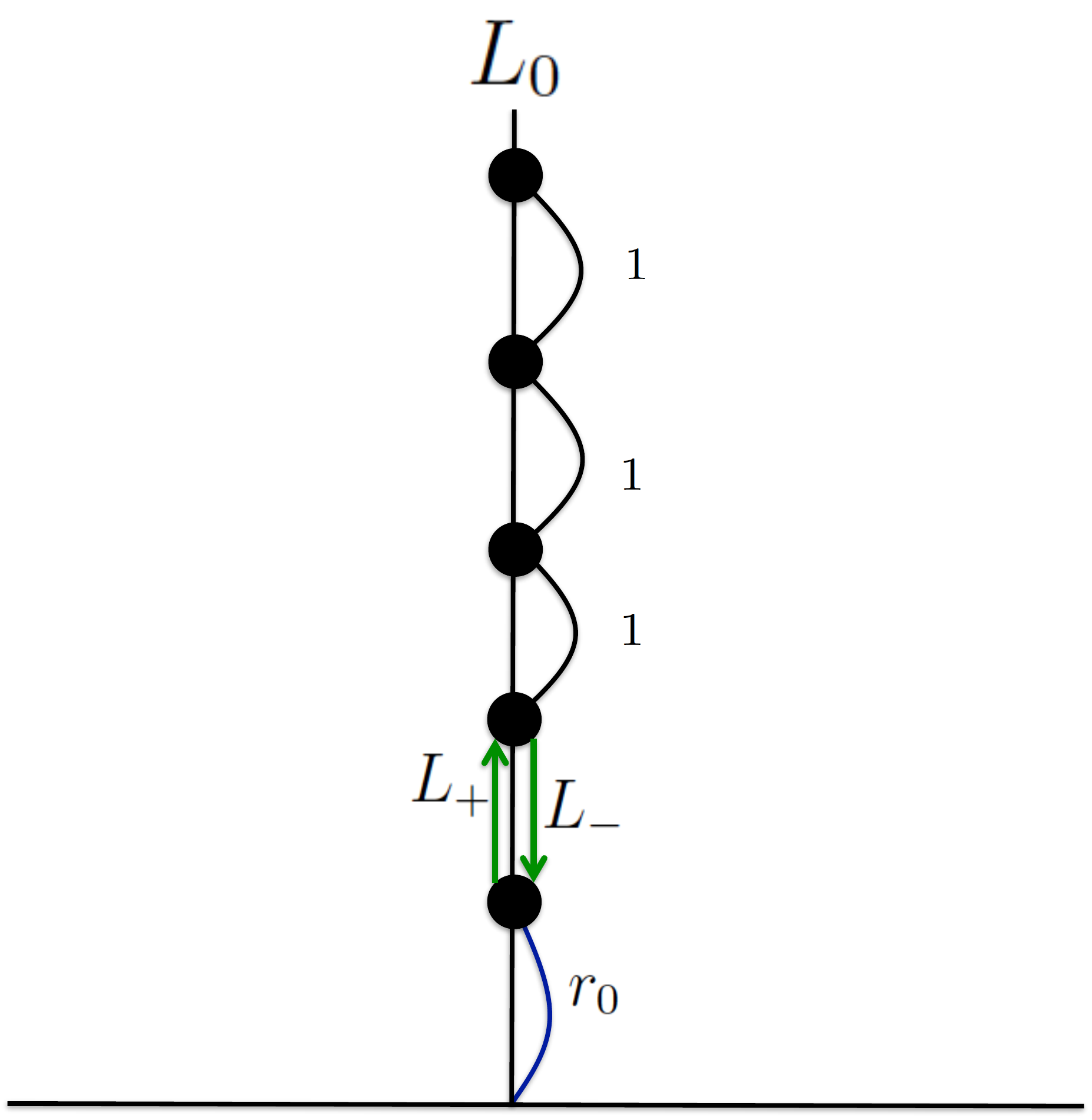}
\caption{The $L_{0}$ spectrum. The ground state has eigenvalue $r_{0}$
 and the excited states generated by $L_{+}$ are equally spaced with
 unit one.}
\label{l0spe1}
\end{center}
\end{figure}

We now look for the eigenvalues and eigenstates 
of $L_{0}$. 
Let us denote the eigenvalues and eigenstates of $L_{0}$ by $r_{n}$ 
and $|n\rangle$
\begin{align}
\label{slalg001}
L_{0}|n\rangle =r_{n}|n\rangle.
\end{align}
From the $\mathfrak{sl}(2,\mathbb{R})$ algebra (\ref{vira2}) 
one can show that
\begin{align}
\label{slalg002}
L_{0}L_{-}|r_{n}\rangle&=(r_{n}-1)L_{-}|r_{n}\rangle,\\
\label{slalg003}
L_{0}L_{+}|r_{n}\rangle&=(r_{n}+1)L_{+}|r_{n}\rangle,\\
\label{slalg004}
L_{-}L_{+}&=-\mathcal{C}_{2}+L_{0}(L_{0}+1),\\
\label{slalg005}
L_{+}L_{-}&=-\mathcal{C}_{2}+L_{0}(L_{0}-1).
\end{align}
The relations (\ref{slalg002}) and (\ref{slalg003}) imply that 
the operators $L_{-}$ and $L_{+}$ play the role of 
the annihilation and creation operators respectively. 
Since the norm of the states $L_{\pm}|r_{n}\rangle$ 
must be positive or zero, 
we require that 
\begin{align}
\label{slalg006}
0&\le \left|
L_{\pm}|r_{n}\rangle
\right|^{2}\nonumber\\
&=-\mathcal{C}_{2}+r_{n}(r_{n}\pm1).
\end{align}
Assuming that 
there exists one positive eigenvalue $r_{n}$ 
among the allowed eigenvalues, 
the creation operator $L_{+}$ yields the infinite 
chain of states
\begin{align}
\label{slalg007}
|r_{n}\rangle, \ \ \ |r_{n}+1\rangle, \ \ \ |r_{n}+2\rangle, \ \ \ \cdots.
\end{align}
If we require the existence of the ground state, 
the spectrum need to be bounded below 
and the chain must terminate. 
This means that 
\begin{align}
\label{slalg008}
L_{-}|r_{0}\rangle=0
\end{align}
and 
\begin{align}
\label{slalg009}
L_{+}L_{-}|r_{0}\rangle 
=\left[
-\mathcal{C}_{2}+r_{0}(r_{0}-1)
\right]|r_{0}\rangle=0.
\end{align}
Therefore the eigenvalues of $L_{0}$ are given by a discrete series 
(see Figure \ref{l0spe1}) 
\begin{align}
\label{vira3}
r_{n}&=r_{0}+n,\ \ \  n=0,1,2,\cdots\\
\label{vira3a}
\mathcal{C}_{2}&=r_{0}(r_{0}-1).
\end{align}
In the following discussion 
we will thus simplify the expression as $|n\rangle=|r_{n}\rangle$.

\begin{figure}
\begin{center}
\includegraphics[width=10cm]{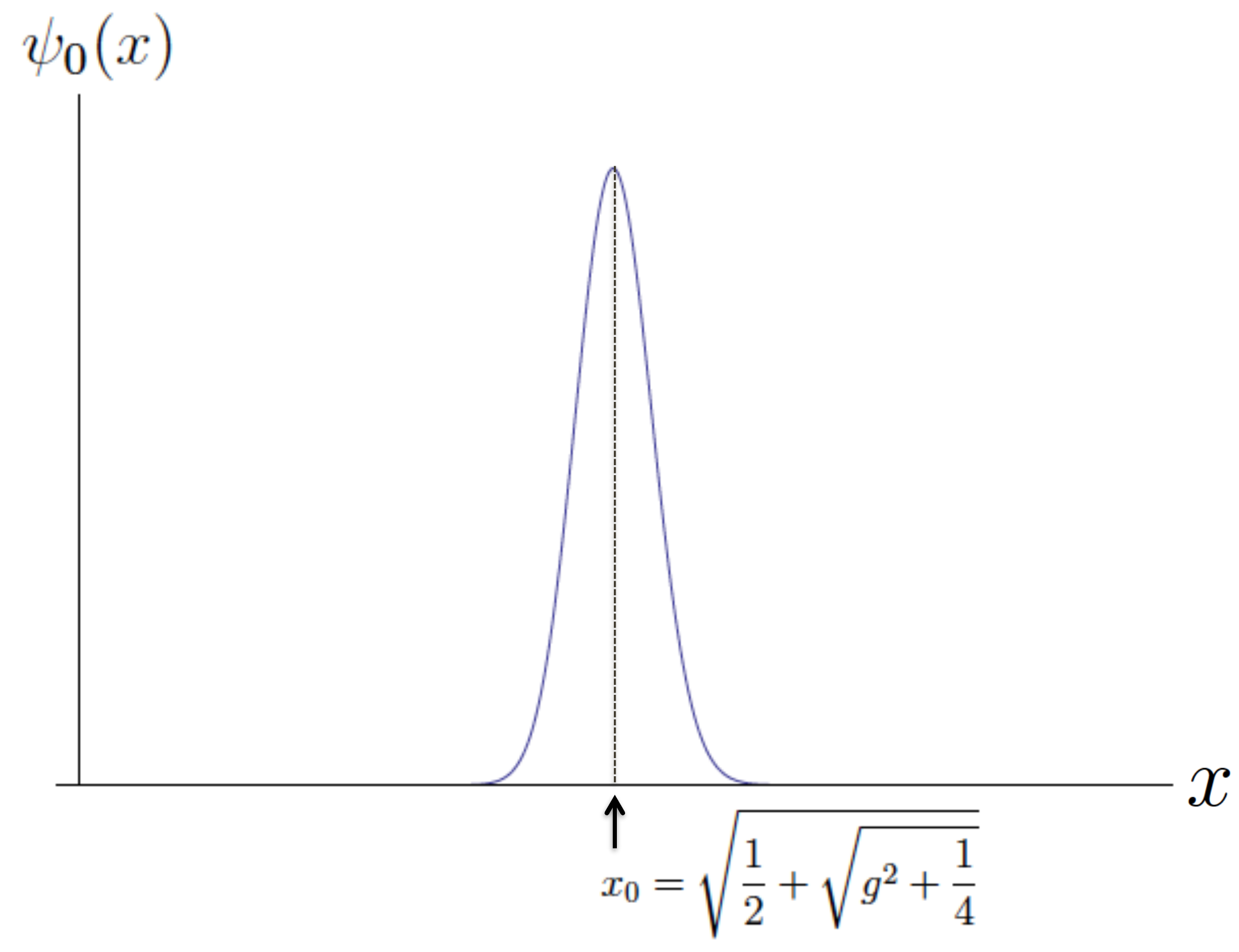}
\caption{Wavefunction $\psi_{0}(x)$ of the $L_{0}$ ground state. }
\label{wfctfig1}
\end{center}
\end{figure}

Combining the relation (\ref{vira3a}) and 
the expression (\ref{casimir2}), 
we find two possible values for $r_{0}$ 
as the choice of the positive or negative signs for the square root. 
However, it turns out that only the positive sign for the square root 
should be selected. 
\begin{align}
\label{vira4}
r_{0}=\frac12 \left(
1+\sqrt{g^{2}+\frac14}
\right).
\end{align}
To see this let us determine 
the lowest eigenfunction $\psi_{0}(x)$. 
From the equation (\ref{slalg008}) and 
the explicit expressions for $L_{\pm}$
\begin{align}
\label{vira4a1}
L_{-}&=L_{0}-\frac{x^{2}}{2a}-\frac12 x\frac{d}{dx}-\frac14,\\
\label{vira4a2}
L_{+}&=L_{0}-\frac{x^{2}}{2a}+\frac12 x\frac{d}{dx}+\frac14,
\end{align}
we see that $\psi_{0}(x)$ satisfies the equation 
\begin{align}
\label{vira5}
\left[
x\frac{d}{dx}+\frac{x^{2}}{a}-\left(
2r_{0}-\frac12
\right)
\right]\psi_{0}(x)=0.
\end{align}
Let us choose our units so that $a=1$. 
The generic solution is given by
\begin{align}
\label{vira6}
\psi_{0}(x)
=Ce^{-\frac{x^{2}}{2}}x^{2r_{0}-\frac12}
\end{align}
where $C$ is the constant of integration. 
The presence of the infinitely repulsive potential barrier 
at the origin 
and the confinement property of the wavefunction requires that 
\begin{align}
\label{vira7}
\lim_{x\rightarrow 0}\psi_{0}(x)&=0,\\
\label{vira7b}
\lim_{x\rightarrow 0}\psi_{0}'(x)&=0.
\end{align}
These conditions lead to
\begin{align}
\label{vira8}
r_{0}>\frac34
\end{align}
which is only satisfied by the positive root solution. 
Note that (\ref{vira8}) is equivalent to the condition $\gamma>0$ 
for the coupling constant as we mentioned. 
Also one can determine $C$ by the normalization 
condition $\int_{0}^{\infty}\left|\psi_{0}(x)\right|^{2}dx=1$ as
\begin{align}
\label{vira9}
C=\sqrt{\frac{2}{\Gamma(2r_{0})}}.
\end{align}
Therefore the eigenfunction of the ground state is given by
\begin{align}
\label{vira10}
\psi_{0}(x)
=\sqrt{\frac{2}{\Gamma(2r_{0})}}e^{-\frac{x^{2}}{2}}
x^{\frac12+\sqrt{g^{2}+\frac14}}.
\end{align}
This is illustrated in Figure \ref{wfctfig1}. 
Curiously a particle in the $L_{0}$ ground state 
has zero probability of existing at $x=0$. 

From (\ref{slalg006}) and (\ref{vira3a}) 
one can see that the raising and lowering operators $L_{\pm}$ 
act as 
\begin{align}
\label{viara11}
L_{\pm}|n\rangle 
=\sqrt{r_{n}(r_{n}\pm1)-r_{0}(r_{0}-1)}|n\pm1\rangle,
\end{align}
which leads to the relation
\begin{align}
\label{vira12}
|n\rangle=
\sqrt{\frac{\Gamma(2r_{0})}
{n!\Gamma(2r_{0}+n)}}(L_{+})^{n}|0\rangle.
\end{align}
Upon the repeated application of the creation operator $L_{+}$ 
on the ground state $|0\rangle$, 
the eigenfunctions of the excited states found to be
\cite{deAlfaro:1976je}
\begin{align}
\psi_{n}(x)
=\sqrt{\frac{\Gamma(n+1)}{2\Gamma(n+2r_{0})}}
x^{-\frac12}\left(\frac{x^{2}}{a}\right)^{r_{0}}
e^{-\frac{x^{2}}{2a}}L_{n}^{2r_{0}-1}\left(\frac{x^{2}}{a}\right)
\end{align}
where $L_{n}^{2r_{0}-1}$ is the associated Laguerre polynomial.

\begin{figure}
\begin{center}
\subfigure[$L_{0}=10,50,100; g=1,a=1$.]
{\includegraphics*[width=.35\linewidth]{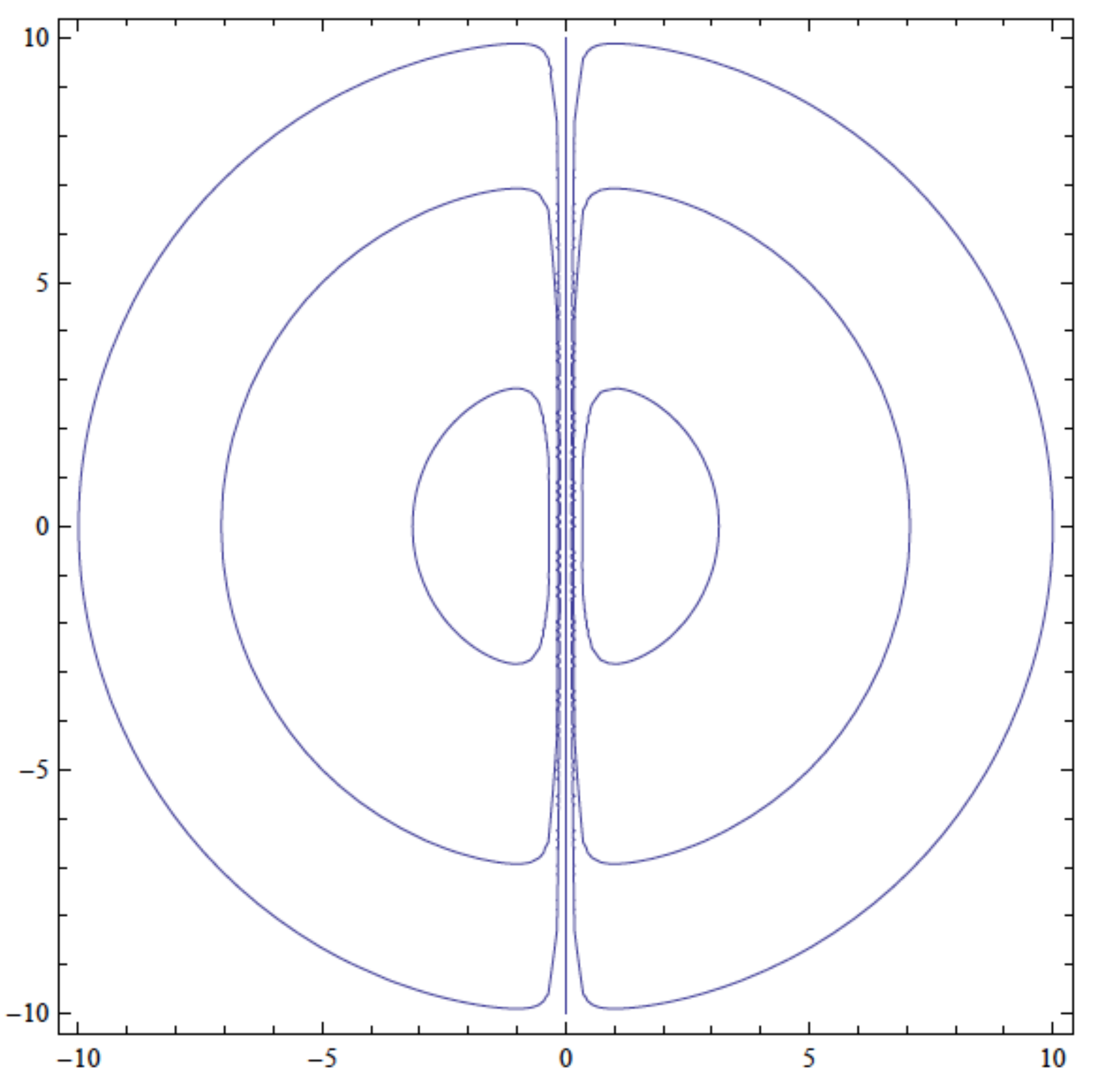}}\qquad\qquad
\subfigure[$g^{2}=1,100,500; L_{0}=50,a=1$.]
{\includegraphics*[width=.36\linewidth]{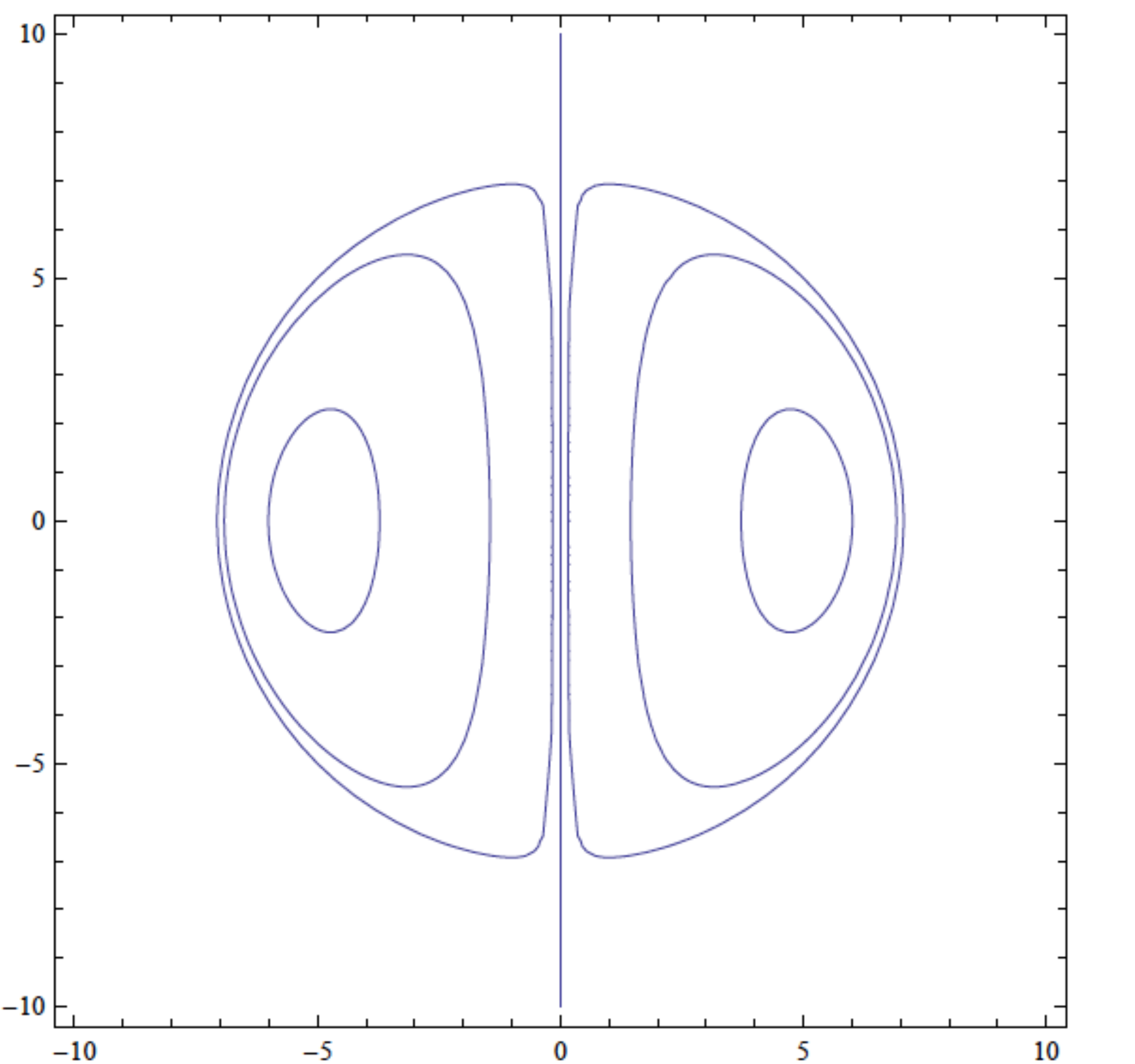}}\qquad\qquad
\subfigure[$a^{2}=1,5,25; L_{0}=50,g=1$.]
{\includegraphics*[width=.35\linewidth]{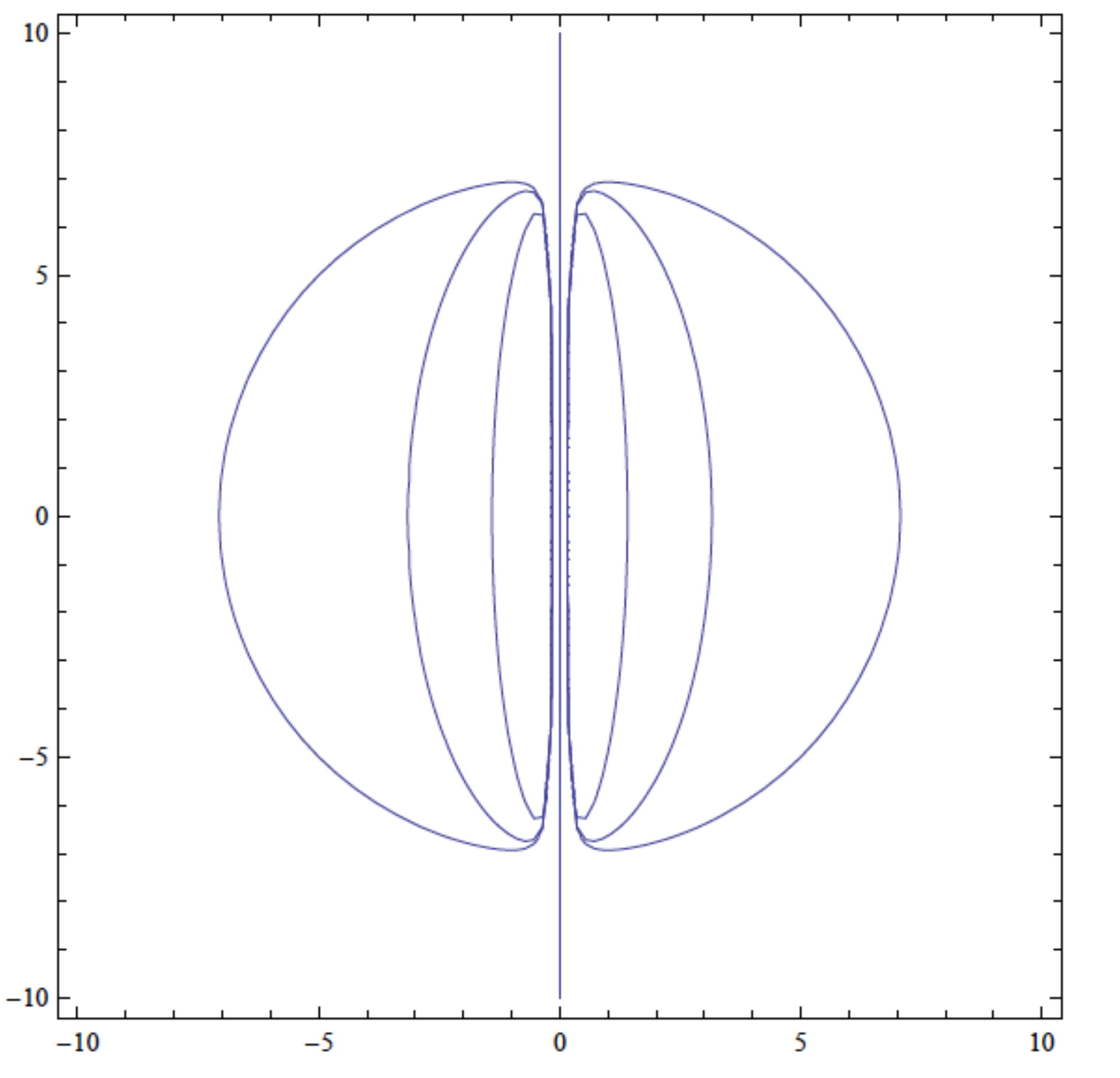}}
\end{center}
\caption{The surface of the constant value of $L_{0}$ 
in the classical phase space. 
The horizontal axis denotes the canonical variable $x$ 
while the vertical axis represents the canonical momentum $p$. 
The volume of the phase space 
with ``the energy'' below $L_{0}$ decrease with increase in the coupling constant $g$ 
and the deformation parameter $a$. 
Qualitatively $g$ keeps a particle away from the origin 
whereas $a$ sucks it into the origin.}
\label{xpfig1}
\end{figure}

Now consider the thermodynamical aspect of the DFF-model. 
As we have been discussing, 
it has been proposed that 
$L_{0}=\frac12 (aH+\frac{K}{a})$ is treated as the new Hamiltonian 
instead of the original Hamiltonian $H$ 
in the DFF-model. 
The surface of the constant value of $L_{0}$ 
in the classical phase space is given by 
\begin{align}
\label{thdy01}
p=\pm\sqrt{2L_{0}-\frac{g}{x^{2}}-a^{2}x^{2}}
\end{align}
and illustrated in Figure \ref{xpfig1} \footnote{
Note that the phase space is restricted to 
either $x>0$ or $x<0$ region due to the infinite potential at the origin.}.

Thus the volume of the phase space 
with the ``energy'' below $L_{0}$ 
can be evaluated to be
\begin{align}
\label{thdy02}
\Gamma(L_{0})
&=2\int_{0}^{\infty}dx
\sqrt{2L_{0}-\frac{g}{x^{2}}-a^{2}x^{2}}\nonumber\\
&=\pi\left(\frac{L_{0}}{a}-g\right).
\end{align}
According to the additional term $-\pi g$, 
the result is slightly modified from a 
simple harmonic oscillator. 
This corresponds to the fact 
that the $L_{0}$-ground state 
of the DFF-model is raised 
by the increase of the coupling constant $g$ as in (\ref{vira4}). 
As seen form Figure \ref{xpfig1}, 
the volume of the phase space with 
``the energy'' below $L_{0}$ decrease 
with increase in the coupling constant $g$ 
and the deformation parameter $a$. 
Therefore qualitatively $g$ keeps a particle 
away from the origin whereas $a$ sucks it into the origin. 
These features are in accord with the behavior of the 
wavefunction $\psi_{0}(x)$ 
of the ground state given in (\ref{vira10})
(see also Figure \ref{wfctfig1}).

In quantum mechanics 
the $L_{0}$-spectrum is the discrete value given in  
(\ref{vira3}). 
By summing over the spectrum one obtains the partition function 
\begin{align}
\label{thdy05}
Z&=\sum_{n=0}^{\infty}e^{-\beta L_{0}}
=\frac{e^{-\beta r_{0}}}{1-e^{-\beta}}.
\end{align}

\section{Time evolution}
\label{seccqm2}
So far the DFF-model (\ref{conflag1}) 
has been studied in the $x$ space, 
i.e. the stationary problem at $t=0$. 
Now let us consider the state $|t\rangle$ which is characterized by the
time $t$. 
Let us define the time-dependent function
\begin{align}
\label{timecqm5}
\beta_{n}:=\langle t|n\rangle,
\end{align}
on which the action of the Hamiltonian is realized with 
the time derivative 
\begin{align}
\label{timecqm1}
H =i\frac{d}{dt}.
\end{align}
Combining the expression (\ref{timecqm1}) 
with the $\mathfrak{sl}(2,\mathbb{R})$ algebra (\ref{hkd00}) 
and the form of the Casimir operator (\ref{casimir2}), 
one finds the action of the dilatation operator $D$ and 
the conformal boost operator $K$ on $\beta_{n}$ as 
\begin{align}
\label{timecqm2}
D&=
\left(it\frac{d}{dt}+ir_{0}\right),\\
\label{timecqm3}
K&=
\left(it^{2}\frac{d}{dt}+2ir_{0}t\right).
\end{align}
Thus the compact operator $L_{0}$ acts on $\beta_{n}(t)$ as
\begin{align}
\label{timecqm4}
L_{0}=\frac{i}{2}\left[
\left(
a+\frac{t^{2}}{a}
\right)\frac{d}{dt}
+2r_{0}\frac{t}{a}
\right].
\end{align}
From the expressions (\ref{timecqm1})-(\ref{timecqm4}) 
we can write the actions of the corresponding operators on 
the state $|t\rangle$ as
\begin{align}
\label{timecqm1a}
H|t\rangle&=-i\frac{d}{dt}|t\rangle,\\
D|t\rangle&=-i\left(t\frac{d}{dt}+r_{0}\right)|t\rangle,\\
K|t\rangle&=-i\left(t^{2}\frac{d}{dt}+2r_{0}t\right)|t\rangle,\\
L_{0}|t\rangle&=-\frac{i}{2}
\left[
\left(a+\frac{t^{2}}{a}\right)\frac{d}{dt}
+2r_{0}\frac{t}{a}
\right]|t\rangle.
\end{align}
Then the explicit expression (\ref{timecqm4}) 
for the operator $L_{0}$ leads to the differential equation 
\begin{align}
\label{timecqm6}
\frac{i}{2}
\left[
\left(a+\frac{t^{2}}{a}\right)\frac{d}{dt}
+2r_{0}\frac{t}{a}
\right]\beta_{n}=r_{n}\beta_{n}.
\end{align}
and its solution is given by
\begin{align}
\label{timecqm7}
\beta_{n}(t)
=(-1)^{n}\left[
\frac{\Gamma(2r_{0}+n)}{n!}
\right]^{\frac12}
\left(
\frac{a-it}{a+it}
\right)^{r_{n}}\frac{1}
{\left(1+\frac{t^{2}}{a^{2}}\right)^{r_{0}}}.
\end{align}
Using the above solution (\ref{timecqm7}) 
one finds 2-point function \cite{deAlfaro:1976je}
\begin{align}
\label{timecqm8}
F_{2}(t_{1},t_{2})
&=\langle t_{1}|t_{2}\rangle\nonumber\\
&=\sum_{n}\beta_{n}(t_{1})\beta^{*}_{n}(t_{2})\nonumber\\
&=\frac{\Gamma(2r_{0})a^{2r_{0}}}
{[2i(t_{1}-t_{2})]^{2r_{0}}}
\propto 
\frac{1}{\left(t_{1}-t_{2}\right)^{2r_{0}}}.
\end{align}
The expression (\ref{timecqm8}) indicates that 
the 2-point function is the value 
of two operators whose effective dimensions are $r_{0}$. 
Note that the 2-point function satisfies the set of conditions 
\begin{align}
\label{timecqm9a}
\left(
\frac{\partial}{\partial t_{1}}+
\frac{\partial}{\partial t_{2}}
\right)F_{2}(t_{1},t_{2})&=0,\\
\label{timecqm9b}
\left(
t_{1}\frac{\partial}{\partial t_{1}}
+t_{2}\frac{\partial}{\partial t_{2}}
+2r_{0}\right)F_{2}(t_{1},t_{2})&=0,\\
\label{timecqm9c}
\left(
t_{1}^{2}\frac{\partial}{\partial t_{1}}
+t_{2}^{2}\frac{\partial}{\partial t_{2}}
+2r_{0}(t_{1}+t_{2})
\right)F_{2}(t_{1},t_{2})&=0.
\end{align}

Now we want to consider the $E$ space. 
The eigenstate $|E\rangle$ is defined by 
\begin{align}
\label{ecqm1}
H|E\rangle=E|E\rangle
\end{align}
and we can expand the eigenstate $|n\rangle$ of 
the compact operator $L_{0}$ as
\begin{align}
\label{ecqm2}
|n\rangle=\int dE C_{n}(E)|E\rangle
\end{align}
where we have defined
\begin{align}
\label{ecqm3}
C_{n}(E):=\langle E|n\rangle.
\end{align}
Note that the eigenvalue $E$ of the original Hamiltonian $H$ 
is continuous as we have already mentioned. 
Requiring the $\mathfrak{sl}(2,\mathbb{R})$ algebra (\ref{hkd00}) 
and the realization of the Casimir operator (\ref{casimir2}), 
we get 
\begin{align}
\label{ecqm4}
D|E\rangle&=-i\left(E\frac{d}{dE}+\frac12\right)|E\rangle,\\
\label{ecqm5}
K|E\rangle&=\left[
-E\frac{d^{2}}{dE^{2}}-\frac{d}{dE}+
\left(r_{0}-\frac12\right)^{2}\frac{1}{E}
\right]|E\rangle.
\end{align}
Then we can write the compact operator $L_{0}$ as
\begin{align}
\label{ecqm6}
L_{0}=\frac12\left[
-E\frac{d^{2}}{dE^{2}}
-\frac{d}{dE}+E
+\left(r_{0}-\frac12\right)^{2}
\frac{1}{E}
\right]
\end{align}
and the explicit expression for $C_{n}(E)$ can be found 
by solving the equation
\begin{align}
\label{ecqm7}
\frac12 
\left[
-E\frac{d^{2}}{dE^{2}}
-\frac{d}{dE}
+E
+\left(r_{0}-\frac12\right)^{2}\frac{1}{E}
\right]C_{n}(E)
=r_{n}C_{n}(E).
\end{align}
If we set 
\begin{align}
\label{ecqm8}
C_{n}=2^{r_{0}}E^{r_{0}-\frac12}e^{-E}\varphi_{n}(E),
\end{align}
then we see that the function $\varphi_{n}$ satisfies the differential
equation 
\footnote{The associated Laguerre polynomials $L_{n}^{k}(x)$ are defined by the 
solution of the differential equation
\begin{align}
\left[
x\frac{d^{k+2}}{dx^{k+2}}
+(k+1-x)\frac{d^{k+1}}{dx^{k+1}}
+n\frac{d^{k}}{dx^{k}}
\right]L_{n}^{k}(x)=0\nonumber
\end{align}
with $0\le k\le n$. 
}
\begin{align}
\label{ecqm9}
\eta \varphi_{n}''+(2r_{0}-\eta)\varphi_{n}'
+n\varphi_{n}=0
\end{align}
of the associated Laguerre polynomial $L_{n}^{2r_{0}-1}$. 
Putting all together we obtain 
\begin{align}
\label{ecqm10}
C_{n}(E)
=2^{r_{0}}
\sqrt{\frac{\Gamma(n+1)}{\Gamma(n+2r_{0})}}
\frac{(aE)^{r_{0}}}{\sqrt{E}}e^{-aE}L_{n}^{2r_{0}-1}(2aE).
\end{align}
Note that the function $C_{n}(E)$ has the following properties:
\begin{align}
\label{ecqm11}
&\sum_{n}C_{n}(E)C_{n}^{*}(E')=\delta(E-E'),\\
\label{ecqm12}
&\int_{0}^{\infty} dE C_{n}(E)C_{n'}^{*}(E)=\delta_{nn'},\\
\label{ecqm13}
&C_{n}(E)=2^{r_{0}}E^{\frac12-r_{0}}
\int_{-\infty}^{\infty}
\frac{dt}{2\pi}e^{iEt}\beta_{n}(t).
\end{align}

\begin{figure}
\begin{center}
\includegraphics[width=10cm]{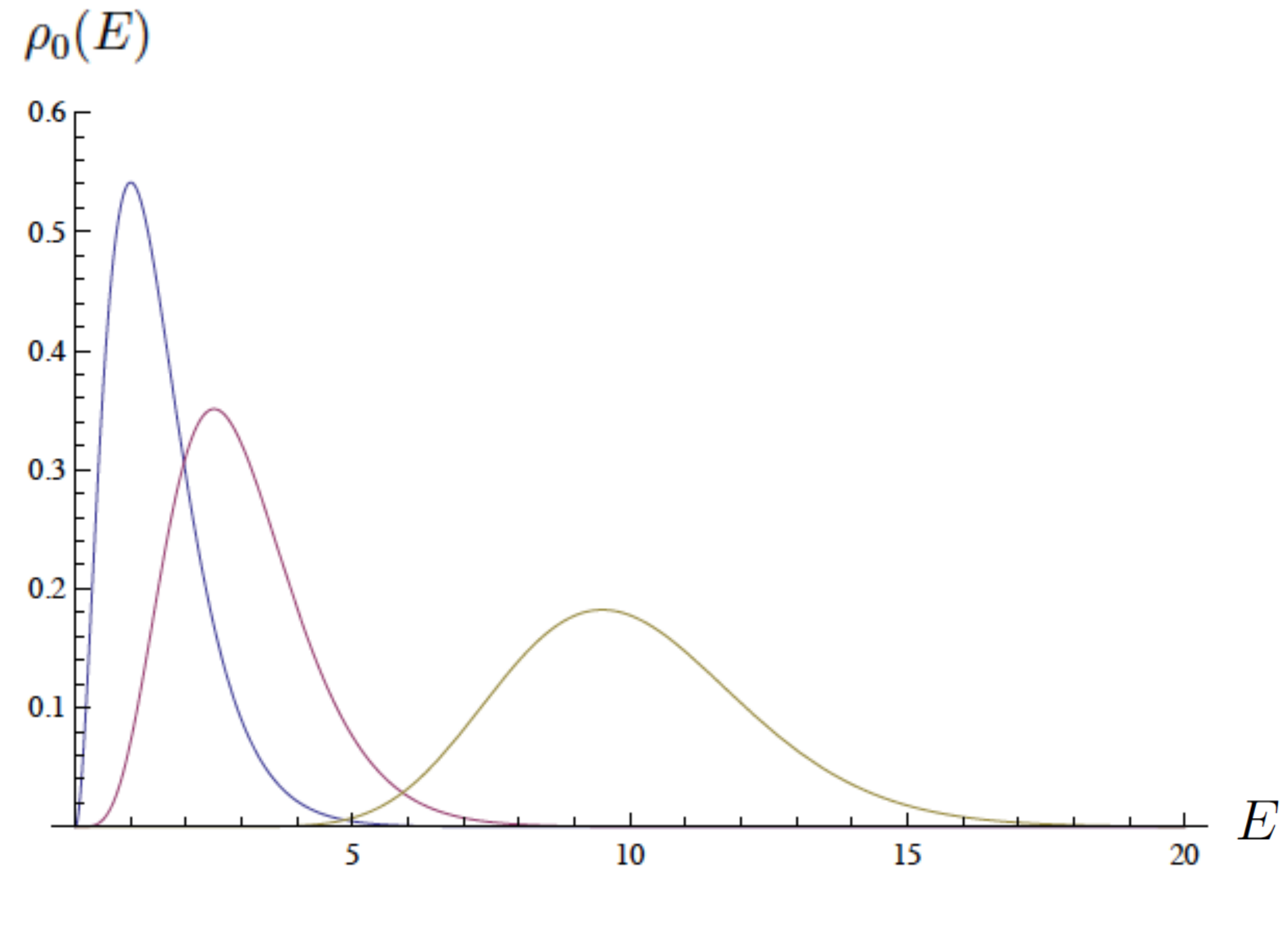}
\caption{The probability density $\rho_{0}(E)$ 
of the ground state in  $E$ space. 
The blue, red and brown lines denote the cases with 
$r_{0}=\frac{3}{2}$, $3$ and $10$. 
They have the maximum values at $E=r_{0}-\frac12$.}
\label{wfctfig2}
\end{center}
\end{figure}

Let us discuss the probability density $\rho_{n}(E)$ in $E$ space 
defined by
\begin{align}
\label{ecqmdens1}
\rho_{n}(E):=\left|C_{n}(E)\right|^{2}.
\end{align}
For $n=0$, i.e. for the ground state of $L_{0}$, 
the probability density $\rho_{0}(E)$ is given by
\begin{align}
\label{ecqmdens2}
\rho_{0}(E)
=\frac{4r^{r_{0}}}{\Gamma(2r_{0})}
E^{2r_{0}-1}
e^{-2E}\left[
L^{2r_{0}-1}(2E)
\right]^{2}
\end{align}
with $a=1$. 
This is shown in Figure \ref{wfctfig2}. 

The distribution of $E$ of the ground state is peaked at
\begin{align}
\label{ecqmdens3}
E_{0}=r_{0}-\frac12
\end{align}
and its effective width is  
\footnote{The effective width $\Gamma$ is defined by 
$\frac{8}{\Gamma}:=
-\frac{\rho_{0}(E_{0})''}{\rho_{0}(E_{0})}\nonumber$.}
\begin{align}
\label{ecqmdens4}
\Gamma=2\sqrt{E_{0}}
=2\sqrt{r_{0}-\frac12}.
\end{align}
(\ref{ecqmdens3}) shows that 
the peak of the distribution of $E$ increases with 
the increase of $r_{0}$ or $g$. 
(\ref{ecqmdens4}) implies that 
the width grow and the probability dense spread in $E$ space with
$r_{0}$.

The expectation values for the ground state $|0\rangle$ 
can be evaluated and we find
\begin{align}
\label{ecqmdens5}
&\langle H\rangle_{0}
=\langle 0|H|0\rangle=r_{0}\\
\label{ecqmdens6}
&\left(\Delta E\right)^{2}
:=\langle H^{2}\rangle_{0}
-\langle H\rangle_{0}^{2}
=\frac{r_{0}^{2}}{2}.
\end{align}

\begin{figure}
\begin{center}
\includegraphics[width=10cm]{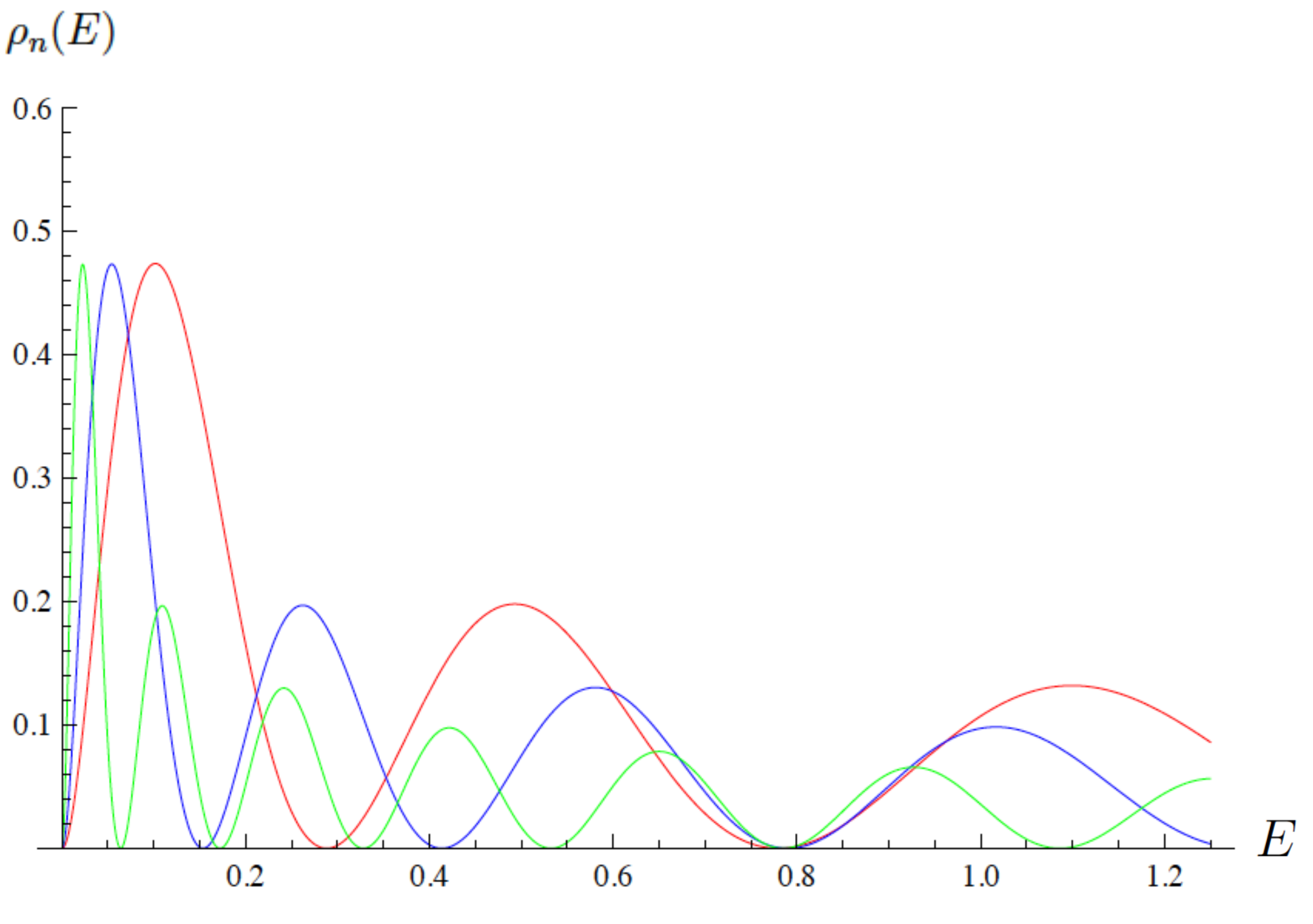}
\caption{The probability densities $\rho_{n}(E)$ of the excited states
 for $r_{0}=\frac32$ in $E$ space. 
The red, blue and green lines correspond to 
the densities with $n=10$, $20$ and $30$. 
The number of peaks increases with the increase of the excited levels.}
\label{wfctfig3}
\end{center}
\end{figure}

For $n>0$ the probability density $\rho_{n}(E)$ with $r_{0}=\frac32$ 
is illustrated in Figure \ref{wfctfig3}.
The red, blue and green lines represent the case with 
$n=10$, $20$ and $30$ respectively. 
In this case several peaks appear due to the 
presence of the $n$-th order polynomial. 
The expectation value of $H$ in the excited state $|n\rangle$ 
is calculated to be \cite{deAlfaro:1976je}
\begin{align}
\label{ecqmdens7}
\langle H\rangle_{n}=\langle n|H|n\rangle =
r_{n}.
\end{align}

The state $|E\rangle$ provides us with the further properties of the 
state $|t\rangle$. 
From the expression (\ref{timecqm8}) ant the relation (\ref{ecqm13}) 
with the use of the Hankel integral representation of the gamma
function
\begin{align}
\label{hankelgamma1}
\frac{1}{\Gamma(z)}
=\frac{i}{2\pi}\int_{C}dt (-t)^{-z}e^{-t},
\end{align}
we can show that 
\begin{align}
\label{ecqm14}
\langle t|E\rangle
&=2^{-r_{0}}
\frac{(aE)^{r_{0}}}{\sqrt{E}}e^{-iEt}.
\end{align}
Using the two relations 
(\ref{ecqm13}) and (\ref{ecqm14})
we see that 
\begin{align}
\label{ecqm15}
\int_{-\infty}^{\infty}\frac{dt}{2\pi}
|t\rangle \langle t|
&=\frac{1}{H}\left(\frac{aH}{2}\right)^{2r_{0}}.
\end{align}
This indicates the incompleteness of 
the state $|t\rangle$.

\section{Operator}
\label{seccqm3}
Now consider the tensor operator 
$B(t)$ with the mass dimension $\Delta$. 
As seen from the set of the Heisenberg equations 
(\ref{hkd0b1})-(\ref{hkd0b3}), 
the $\mathfrak{sl}(2,\mathbb{R})$ invariance of the operator implies that 
\begin{align}
\label{bcqm1}
i[H,B(t)]&=\frac{d}{dt}B(t),\\
\label{bcqm2}
i[D,B(t)]&=t\frac{d}{dt}B(t)+\Delta B(t),\\
\label{bcqm3}
i[K,B(t)]&=t^{2}\frac{d}{dt}B(t)+2t\Delta B(t)
\end{align}
where $D=tH+D_{0}$ and $K=t^{2}H-\frac12\left\{x,p\right\}+K_{0}$. 
This is the $SO(1,2)$ Wigner-Eckart theorem. 
We now want to compute the 3-point function 
\begin{align}
\label{bcqm4}
F_{3}(t;t_{2},t_{1})
&=\langle t_{2}|B(t)|t_{1}\rangle\nonumber\\
&=\sum_{n_{1},n_{2}}\beta_{n_{2}}(t_{2})
\beta_{n_{1}}^{*}(t_{1})\langle n_{2}|B(t)|n_{1}\rangle.
\end{align}
In analogy with (\ref{newtime1a}) and (\ref{newcoord1a}) 
it is convenient to introduce the new variables
\begin{align}
\label{bcqm5}
\tau&=2\tan^{-1}t,\\
b(\tau)&=B(t)(1+t^{2})^{\Delta}
\end{align}
with the relations
\begin{align}
\label{bcqm6}
\frac{db(\tau)}{d\tau}
&=i[L_{0},b(\tau)],\\
\label{bcqm7}
b(\tau)&=e^{iL_{0}\tau}b(0)e^{-iL_{0}\tau}.
\end{align}
By using the above expressions we can show that 
\begin{align}
\label{bcqm8}
\langle n_{2}|B(t)|n_{1}\rangle
&=\frac{1}{(1+t^{2})^{\Delta}}
\left(
\frac{1-it}{1+it}
\right)^{n_{1}-n_{2}}\langle n_{2}|B(0)|n_{1}\rangle.
\end{align}
The equation (\ref{bcqm8}) enables us 
to rewrite the 3-point function (\ref{bcqm4}) as
\begin{align}
\label{bcqm9}
F_{3}(t;t_{2},t_{1})
&=\sum_{n_{1},n_{2}}(-1)^{n_{1}+n_{2}}
\sqrt{\frac{\Gamma(n_{1}+2r_{0})\Gamma(n_{2}+2r_{0})}{n_{1}!n_{2}!}}
\nonumber\\
&\times 
z_{1}^{-n_{1}}z_{2}^{n_{2}}
\frac{(1+z_{1})^{2r_{0}}(1+z_{2})^{2r_{0}}}
{2^{4r_{0}}z_{1}^{2r_{0}}}
\nonumber\\
&\times 
2^{-2\Delta}\left|
\frac{1+z}{z}
\right|^{2\Delta}z^{n_{1}-n_{2}}
\langle n_{2}|B(0)|n_{1}\rangle
\end{align}
where we have defined 
\begin{align}
\label{bcqm10}
z&:=\frac{1-it}{1+it}=e^{-i\tau},&
z_{i}&:=\frac{1-it_{i}}{1+it_{i}}=e^{-i\tau_{i}}.
\end{align}

On the other hand, 
the one-dimensional conformal $\mathfrak{sl}(2,\mathbb{R})$ covariance 
of the 3-point function implies that 
\begin{align}
\label{bcqm11}
\left(
\frac{\partial}{\partial t}
+\frac{\partial}{\partial t_{1}}
+\frac{\partial}{\partial t_{2}}
\right)F_{3}(t;t_{2},t_{1})&=0,\\
\label{bcqm12}
\left(
t\frac{\partial}{\partial t}
+t_{1}\frac{\partial}{\partial t_{1}}
+t_{2}\frac{\partial}{\partial t_{2}}
+2\Delta t+2r_{0}(t_{1}+t_{2})
\right)F_{3}(t;t_{2},t_{1})&=0,\\
\label{bcqm13}
\left(
t^{2}\frac{\partial}{\partial t}
+t_{1}^{2}\frac{\partial}{\partial t_{1}}
+t_{2}^{2}\frac{\partial}{\partial t_{2}}
+2\Delta t+2r_{0}(t_{1}+t_{2})
\right)F_{3}(t;t_{2},t_{1})&=0.
\end{align}
As the first condition (\ref{bcqm11}) 
says that $F_{3}(t;t_{2},t_{1})=F_{3}(t-t_{1},t-t_{2},t_{1}-t_{2})$, 
we impose the ansatz 
$F_{3}
=f(t-t_{1})^{\alpha_{1}}(t-t_{2})^{\alpha_{2}}(t_{1}-t_{2})^{\alpha_{3}}$
with $f$ being an arbitrary constant value. 
Then the remaining two conditions (\ref{bcqm12}) and (\ref{bcqm13}) 
restrict to the form of the 3-point function as
\begin{align}
\label{bcqm14} 
F_{3}(t;t_{2},t_{1})
&=
fi^{2r_{0}+\Delta}
\frac{1}{(t-t_{1})^{\Delta}(t_{2}-t)^{\Delta}(t_{1}-t_{2})^{-\Delta+2r_{0}}}
\nonumber\\
&=2^{-\Delta -2r_{0}}f
\frac{(1+z)^{2\Delta}(1+z_{1})^{2r_{0}}(1+z_{2})^{2r_{0}}}
{(z-z_{1})^{\Delta}(z_{2}-z)^{\Delta}(z_{1}-z_{2})^{-\Delta+2r_{0}}}.
\end{align}

Combining the two expressions 
(\ref{bcqm9}) and (\ref{bcqm14})
for the 3-point function, 
we obtain the relation
\begin{align}
\label{bcqm15}
\sum_{n_{1},n_{2}}
(-1)^{n_{1}+n_{2}}
\sqrt{\frac{\Gamma(n_{1}+2r_{0})\Gamma(n_{2}+2r_{0})}
{n_{1}!n_{2}!}
}z_{1}^{-n_{1}}z_{2}^{n_{2}}
2^{-\Delta}
\left|
\frac{1+z}{z}
\right|^{2\Delta}
z^{n_{1}-n_{2}}\langle n_{2}|B(0)|n_{1}\rangle\nonumber\\
=2^{r_{0}}fz_{1}^{2r_{0}}
\frac{(1+z)^{2\Delta}}{(z-z_{1})^{\Delta}
(z_{2}-z)^{\Delta}(z_{1}-z_{2})^{-\Delta+2r_{0}}}.
\end{align}
By redefining the variables $z_{i}\rightarrow zz_{i}$, 
we can factor out the time $t$ dependence and 
thus we shall take $t=0$ or equivalently $z=1$ in the following
discussion. 
Then the left hand side can be regarded as the 
double Taylor series expansions and 
one can write the quantity $\langle n_{2}|B(0)|n_{1}\rangle$ 
as the coefficients of the expansions as
\begin{align}
\label{bcqm16}
\langle n_{2}|B(0)|n_{1}\rangle
&=\frac{f}{(2\pi i)^{2}}
\sqrt{\frac{n_{1}!n_{2}!}{\Gamma(n_{1}+2r_{0})\Gamma(n_{2}+2r_{0})}}
(-1)^{n_{1}+n_{2}}2^{2r_{0}+\Delta}\nonumber\\
&\times \oint_{C_{1}}dz_{1}
\oint_{C_{2}}dz_{2}
\frac{z_{1}^{2r_{0}+n_{1}-1}z_{2}^{-(n_{2}+1)}}
{(1-z_{1})^{\Delta}(z_{2}-1)^{\Delta}(z_{1}-z_{2})^{-\Delta+2r_{0}}}
\end{align}
where $C_{1}$ is the suitable anti-clockwise 
contour for the coordinates $z_{1}=\infty$ 
and $C_{0}$ is for $z_{2}$. 
Changing the pair of the variables 
$z_{1}=\frac{1}{w_{1}}$ and $z_{2}=w_{2}$, 
we find the expression
\begin{align}
\label{bcqm17}
\langle n_{2}|B(0)|n_{1}\rangle
&=\frac{f}{(2\pi i)^{2}}
\sqrt{\frac{n_{1}!n_{2}!}{\Gamma(n_{1}+2r_{0})\Gamma(n_{2}+2r_{0})}}
(-1)^{n_{1}+n_{2}}2^{2r_{0}+\Delta}\nonumber\\
&\times \oint_{C_{2}}\frac{dw_{2}}{w_{2}^{n_{2}+1}}
\oint_{C_{1}}\frac{dw_{1}}{w_{1}^{n_{1}+1}}
\frac{(1-w_{1}w_{2})^{\Delta-2r_{0}}}
{(1-w_{1})^{\Delta}(w_{2}-1)^{\Delta}}
\end{align}
where the integral are carried out around the 
contours $C_{1}$ for $w_{1}$ 
and $C_{2}$ for $w_{2}$. 
Applying the Cauchy theorem, 
the integration (\ref{bcqm17}) can be calculated to be \cite{deAlfaro:1976je}
\begin{align}
\label{bcqm17a}
\langle n_{2}|B(0)|n_{1}\rangle 
&=2^{2r_{0}+\Delta}f\sqrt{
\frac{n_{1}!n_{2}!}{\Gamma(2r_{0}+n_{1})\Gamma(2r_{0}+n_{2})}
}\nonumber\\
&\times \sum_{k=0}^{\min[n_{1},n_{2}]}
(-1)^{k}\left(
\begin{array}{c}
\Delta-2r_{0}\\
k\\
\end{array}
\right)
\left(
\begin{array}{c}
-\Delta\\
n_{1}-k\\
\end{array}
\right)
\left(
\begin{array}{c}
-\Delta\\
n_{2}-k\\
\end{array}
\right).
\end{align}
For $n_{1}=n_{2}=0$ 
we can read off the explicit formula 
for the constant $f$ as
\begin{align}
\label{bcqm18}
f=2^{-\Delta-2r_{0}}\Gamma(2r_{0})
\langle 0|B(0)|0\rangle.
\end{align}
Inserting (\ref{bcqm18}) 
into (\ref{bcqm14}) and 
reviving the constant factor $a$, 
we finally get the formula for the 3-point function 
\begin{align}
\label{bcqm19}
F_{3}(t,;t_{2},t_{1})
&=\langle 0|B(0)|0\rangle 
\left(\frac{i}{2}\right)^{2r_{0}+\Delta}
\frac{\Gamma(2r_{0})a^{2r_{0}}}
{(t-t_{1})^{\Delta}(t_{2}-t)^{\Delta}(t_{1}-t_{2})^{-\Delta+2r_{0}}}
\nonumber\\
&\propto 
\frac{1}
{(t-t_{1})^{\Delta}(t_{2}-t)^{\Delta}(t_{1}-t_{2})^{-\Delta+2r_{0}}}.
\end{align}
From the above form (\ref{bcqm19}) we see that 
the 3-point function $F_{3}$ consists of 
the two operators with the same mass dimension $r_{0}$ 
and the third operator $B$ with the mass dimension $\Delta$.

As seen from (\ref{timecqm8}) and 
(\ref{bcqm19}), the structures of the 2- and 3-point functions 
suggest that there exists the averaging state and the 
corresponding operator with the mass dimension $r_{0}$. 
Let us define the operator $\mathcal{O}(t)$ 
which acts on the $L_{0}$-vacuum to create the state $|t\rangle$:
\begin{align}
\label{opcqm1}
|t\rangle =\mathcal{O}(t)|0\rangle.
\end{align}
Making use of the 
formulae (\ref{timecqm7}) and (\ref{vira12}), 
we can write 
\footnote{Such construction of the state $|t\rangle$ 
has also been considered in \cite{Nakayama:2011qh,Freivogel:2011xc,Anninos:2011af}}
\begin{align}
\label{opcqm2}
\mathcal{O}(t)
&=N(t)\exp\left[-\overline{z}(t)L_{+}\right]
\end{align}
where 
\begin{align}
\label{opcqm3}
N(t)&=\sqrt{\Gamma(2r_{0})}\left(
\frac{\overline{z}(t)+1}{2}
\right)^{2r_{0}}.
\end{align}
Then the 2- and 3-point functions are given by
\begin{align}
\label{opcqm4a}
F_{2}(t_{1},t_{2})
&=\langle t_{1}|t_{2}\rangle 
=\langle 0|\mathcal{O}^{\dag}(t_{1})\mathcal{O}(t_{2})|0\rangle,\\
\label{opcqm4b}
F_{3}(t;t_{2},t_{1})
&=\langle t_{2}|B(t)|t_{1}\rangle
=\langle 0|\mathcal{O}^{\dag}(t_{1})B(t)\mathcal{O}(t_{2})|0\rangle.
\end{align}
Therefore the averaging state is 
the $L_{0}$ ground state $|0\rangle$ 
and the corresponding operators are 
$\mathcal{O}^{\dag}(t)$ and $\mathcal{O}(t)$. 
We should note that 
the conformal invariant correlation functions 
can be built up although 
the averaging state $|0\rangle$ is not conformally invariant 
and the operators $\mathcal{O}(t)$ and $\mathcal{O}^{\dag}(t)$ 
are not primary operators. 
This is the significant difference from 
other higher dimensional conformal field theories 
where one can assume the existence of the 
normalizable and invariant vacuum states. 
For quantum field theories we generally treat with the Fock spaces 
which are underlying on the empty no-particle vacuum states. 
However, in quantum mechanics 
we deal with the Hilbert space which is the 
subspace of the Fock space with the fixed number of the particle. 
This fact prevents us from constructing the 
normalizable and invariant empty vacuum state 
in conformal quantum mechanics. 

Noting that 
\begin{align}
\label{opcqm5a}
[L_{-},e^{-\overline{z}(t) L_{+}}]&=
e^{-\overline{z}(t) L_{+}}
\left(-2\overline{z}(t) L_{0}+\overline{z}^{2}(t)L_{+}\right),\\
\label{opcqm5b}
[L_{0},e^{-\overline{z}(t) L_{+}}]&=
-e^{-\overline{z} L_{+}}\overline{z}(t)L_{+},
\end{align}
we see that the state $|t\rangle$ is 
the eigenstate of the operator $L_{-}+\overline{z}(t)L_{0}$
\begin{align}
\label{opcqm6}
\left(
L_{-}+\overline{z}(t)L_{0}
\right)|t\rangle
=-r_{0}\overline{z}(t)|t\rangle
\end{align}
with the eigenvalue $-r_{0}\overline{z}(t)$. 
We see that 
the state $|t\rangle$ is similar to the coherent state $|a\rangle$ 
which satisfies $L_{-}|a\rangle=a|a\rangle$, 
however, 
the additional term $\overline{z}(t)L_{0}$ deviates from it. 
In fact the coherent state $|a\rangle$ can be constructed as
\begin{align}
\label{opcqm6a}
|a\rangle&=\sqrt{\Gamma(2r_{0})}
\sum_{n}\frac{a^{n}}{\sqrt{n!\Gamma(2r_{0}+n)}}|n\rangle\nonumber\\
&=\Gamma(2r_{0})
\sum_{n}\frac{a^{n}}{n!\Gamma(2r_{0}+n)}
(L_{+})^{n}|0\rangle.
\end{align}

Let us define the state 
\begin{align}
\label{opcqm7}
|\Psi\rangle&:=e^{-Ha}|t=0\rangle
=e^{-Ha}e^{-L_{+}}|0\rangle.
\end{align}
By using the relations
\begin{align}
\label{opcqm8}
L_{0}e^{-Ha}&=e^{-Ha}\left(\frac{K}{2a}+iD\right),\\
\label{opcqm9}
\left(\frac{K}{2a}+iD\right)e^{-L_{+}}
&=e^{-L_{+}}
\left(L_{0}-\frac14 L_{-}\right),
\end{align}
one finds that
\begin{align}
\label{opcqm10}
L_{0}|\Psi\rangle=r_{0}|\Psi\rangle.
\end{align}
Thus the state $|\Psi\rangle$ defined by 
(\ref{opcqm7}) is proportional to the $L_{0}$ vacuum state $|0\rangle$ 
\begin{align}
\label{opcqm11}
|\Psi\rangle = C|0\rangle
\end{align}
where the proportional constant $C$ 
can be determined by noting the relation (\ref{ecqm14}) as
\begin{align}
\label{opcqm12}
C=\frac{\sqrt{\Gamma(2r_{0})}}{2^{2r_{0}}}
\end{align}
up to a phase factor. 
Then we obtain the alternative 
description for the state $|t\rangle$ 
\begin{align}
\label{opcqm13}
|t\rangle
&=e^{iHt}|t=0\rangle\nonumber\\
&=e^{iHt}e^{Ha}\left(C|0\rangle\right)\nonumber\\
&=\sqrt{\Gamma(2r_{0})}2^{-2r_{0}}
e^{(a+it)H}|0\rangle.
\end{align}

Let us consider the 4-point function 
\begin{align}
\label{blcqm1}
F_{4}(t_{1},t_{2},t_{3},t_{4})
&=\langle t_{1}|B(t_{2})\tilde{B}(t_{3})|t_{4}\rangle\nonumber\\
&=\langle 0|
\mathcal{O}^{\dag}(t_{1})B(t_{2})B(t_{3})\mathcal{O}(t_{4})|
0\rangle
\end{align}
where the two different fields $B(t)$ and $\tilde{B}(t)$ 
carry the mass dimension $\Delta$ and $\tilde{\Delta}$ respectively. 
It is calculated to be \cite{Jackiw:2012ur}
\begin{align}
\label{blcqm2}
&F_{4}(t_{1},t_{2},t_{3},t_{4})
=\langle 0|B(0)|0\rangle 
\langle 0|\tilde{B}(0)|0\rangle
\frac{\Gamma(2r_{0})}{2^{\Delta+\tilde{\Delta}+2r_{0}}}\nonumber\\
&\times 
\frac{1}
{(t_{13})^{\Delta-r_{0}}(t_{24})^{\tilde{\Delta}-r_{0}}
(t_{12})^{\tilde{\Delta}+r_{0}}(t_{34})^{\Delta+r_{0}}
(t_{14})^{2r_{0}-\Delta-\tilde{\Delta}}}
x^{r_{0}}\ _{2}F_{1}(\Delta,\tilde{\Delta};2r_{0};x)\nonumber\\
&=p(t_{1},t_{2},t_{3},t_{4})x^{r_{0}}\ _{2}F_{1}(\Delta,\tilde{\Delta};2r_{0};x)
\end{align}
where the parameter $a$ set to be one and we have introduced 
the expressions $t_{ij}:=t_{i}-t_{j}$ 
and $x:=\frac{t_{12}t_{34}}{t_{13}t_{24}}$. 
$_{2}F_{1}(\Delta,\tilde{\Delta};2r_{0};x)$ 
is the hypergeometric function that 
possesses the Mellin-Barnes representation
\begin{align}
\label{blcqm3}
_{2}F_{1}(a,b;c;x)=
\frac{\Gamma(c)}{\Gamma(a)\Gamma(b)}
\frac{1}{2\pi i}
\int^{+i\infty}_{-i\infty}
ds \frac{\Gamma(a+s)\Gamma(b+s)\Gamma(-s)}
{\Gamma(c+s)}(-x)^{s}.
\end{align}
Note that the single Mellin integral 
appears in the formula of the 4-point function (\ref{blcqm2}). 
This reflects the fact that 
four points lead to a single invariant in one-dimension 
in contrast to the two invariants in higher dimensions.

It is known that 
4-point functions $F_{4}$ can be 
expressed by the superposition of the conformal blocks, 
or the conformal partial waves $G$ in higher dimensional conformal field
theories \cite{Dolan:2011dv}. 
The conformal block $G$ 
can be determined by requiring 
that it is the eigenfunction of the quadratic 
Casimir of the conformal group. 
As seen from the formula (\ref{blcqm2}), 
we can easily read off a single conformal block as \cite{Jackiw:2012ur}
\footnote{
For higher dimensional field theories, 
the conformal block can be obtained through the operator product
expansion.}
\begin{align}
\label{blcqm4}
G=x^{r_{0}} _{2}F_{1}(
\Delta,\tilde{\Delta};2r_{0};x),
\end{align}
which satisfies 
the differential equation
\begin{align}
\label{blcqm5}
\mathcal{C}_{2}\left[
p(t_{1},t_{2},t_{3},t_{4})G
\right]
=r_{0}(r_{0}-1)
p(t_{1},t_{2},t_{3},t_{4})G.
\end{align}

\section{Gauged conformal mechanics}
\label{gcqmsec1}
It has been pointed out 
\cite{
Fedoruk:2011aa} 
that the DFF-model (\ref{conflag1}) 
can be obtained by the gauged quantum mechanics. 
Let us consider a simple complex free particle Lagrangian
\begin{align}
\label{conflag2}
L=\frac12\dot{z}\dot{\overline{z}}
\end{align}
where $z$ is a complex one-dimensional field. 
The system (\ref{conflag2}) is invariant under the following 
$U(1)$ transformations
\begin{align}
\label{confg1}
z'&=e^{-i\lambda}z,&
\overline{z}'&=e^{i\lambda}\overline{z}
\end{align}
where $\lambda$ is a real parameter. 
Let us gauge this symmetry by promoting 
$\lambda\rightarrow\lambda(t)$. 
Then the gauge invariant Lagrangian is given by
\begin{align}
\label{confg2}
L&=\frac12 D_{0}zD_{0}\overline{z}
+cA_{0}
\nonumber\\
&=\frac12 \left(\dot{z}+iA_{0}z\right)
\left(\dot{\overline{z}}-iA_{0}\overline{z}\right)
+cA_{0}
\end{align}
where $A_{0}(t)$ is the one-dimensional $U(1)$ gauge field. 
The term $cA_{0}$ is a Fayet-Iliopoulos (FI) term with $c$ being a constant. 
This term is gauge invariant itself up to total derivative. 

The action (\ref{confg2}) is invariant under the one-dimensional
conformal transformations
\begin{align}
\label{confg2a}
\delta t&=f(t)=a+bt+ct^{2},\\
\label{confg2b}
\delta x&=\frac12 \dot{f}x,\\
\label{confg2c}
\delta A_{0}&=-\dot{f}A_{0}.
\end{align}
Here the transformation of the gauge field $A_{0}(t)$ 
is the same as that of the time derivative $\partial_{0}$. 

Note that the Lagrangian (\ref{confg2}) 
is quadratic in the $U(1)$ gauge field $A_{0}$ 
and contains no time derivative of $A_{0}$. 
This immediately implies that 
the gauge field $A_{0}$ is identified with the auxiliary gauge field. 
Hence we attempt to integrate out the auxiliary gauge field. 
However, we should be careful of the exclusion of the auxiliary field 
because it is a gauge field. 
We need to integrate out the auxiliary gauge field in two steps; 
firstly we fix a gauge to eliminate residual degrees of freedom 
and then solve the equation of motion of the auxiliary gauge field 
or impose the resulting Gauss law constraint to 
ensure the consistency of the gauge fixing. 
Let us choose the gauge such that 
\begin{align}
\label{confg3}
z(t)=\overline{z}(t)=x(t).
\end{align}
Then the Lagrangian (\ref{confg2}) becomes
\begin{align}
\label{confg4}
L=\frac12 \dot{x}^{2}
+\frac12 A_{0}^{2}x^{2}+cA_{0}.
\end{align}
Using the algebraic equation of motion 
for the auxiliary gauge field $A_{0}$
\begin{align}
\label{confg5}
A_{0}=-\frac{c}{x^{2}},
\end{align}
we can integrate out gauge field and obtain the reduce Lagrangian
\begin{align}
\label{confg6}
L=\frac12\left(
\dot{x}^{2}-\frac{c^{2}}{x^{2}}
\right).
\end{align}
This is nothing but (\ref{confg1}), the DFF-model Lagrangian. 
Thus the conformal invariance is preserved under the gauging procedure. 
This procedure, 
i.e. the integration of the auxiliary gauge field 
can be interpreted as the reduction process 
for the mechanical systems with symmetry. 
Let us summarize the basic concepts of the classical theory 
of Hamiltonian dynamical systems. 

A manifold $\mathcal{M}$ is said to be 
endowed with a Poisson structure 
if there is an operation 
assigning to every pair of functions 
$F,G\in \mathcal{F}(\mathcal{M})$ 
a new function $\left\{F,G\right\}\in \mathcal{F}(\mathcal{M})$ 
which is linear in $F$ and $G$ and 
has the following properties
\begin{enumerate}
 \item skew symmetry 
\begin{align}
\label{hm1}
\left\{F,G\right\}=-\left\{G,F\right\}
\end{align}
\item Jacobi identity
\begin{align}
\label{hm2}
\left\{F,\left\{G,H\right\}\right\}
+\left\{G,\left\{H,F\right\}\right\}
+\left\{H,\left\{F,G\right\}\right\}=0
\end{align}
\item Leibniz rule
\begin{align}
\label{hm3}
\left\{F,GH\right\}
=\left\{F,G\right\}H
+\left\{F,H\right\}G.
\end{align} 
\end{enumerate}
As the above three identities 
(\ref{hm1})-(\ref{hm3}) 
are the axioms of the Lie algebra, 
the space $\mathcal{F}(\mathcal{M})$ is nothing but 
an infinite dimensional Lie algebra. 
Then a dynamical system on $\mathcal{M}$, 
the so-called Hamiltonian dynamical system 
can be introduced as
\begin{align}
\label{hm4}
\dot{x}^{i}=\left\{H(x),x^{i}\right\}=X_{H}^{i}
\end{align}
where 
$x^{i}$ are local coordinates on $\mathcal{M}$,  
$H(x)$ is the Hamiltonian of the dynamical system 
and the vector field $X_{H}^{i}$ is referred to as 
a Hamiltonian vector field. 
For such system we have 
\begin{align}
\label{hm5}
\dot{F}=\left\{H,x\right\}
\end{align}
and the functions which satisfy
$\left\{H,F\right\}=0$ 
are conserved quantities, i.e. the integrals of motion. 

The important class of phase spaces is known as 
a symplectic manifold $(\mathcal{M},\omega)$, 
which possesses closed nondegenerate differential two-form $\omega$, 
i.e. symplectic forms and their Poisson structures are given by
\begin{align}
\label{hm5}
\left\{F(x),G(x)\right\}
&=\omega^{ij}\partial_{i}F\partial_{j}G\nonumber\\
&=\omega(X_{F},X_{G}).
\end{align}

Suppose that a Lie group 
$G$ acts on $\mathcal{M}$. 
Then one can represent the 
corresponding Lie algebra $\mathfrak{g}$ of $G$ 
in the Lie algebra of vector fields on $\mathcal{M}$. 
In other words there is a vector field $X_{\xi}$ on $\mathcal{M}$ 
to each $\xi\in \mathfrak{g}$. 
If one can associate a function $H_{\xi}$ on $\mathcal{M}$ 
to each $X_{\xi}$ obeying the conditions
\begin{align}
\label{hm6}
X_{\xi}^{i}&=\left\{H_{\xi},x^{i}\right\},\\
H_{\xi+\eta}&=H_{\xi}+H_{\eta},\\
H_{[\xi,\eta]}&=\left\{H_{\xi},H_{\eta}\right\},
\end{align}
the action of $G$ is called Hamiltonian 
and $H_{\xi}$ the Hamiltonian function.  
Namely an action of $G$ on $\mathcal{M}$ is 
called Hamiltonian if the map $\xi\mapsto H_{\xi}$ 
is a homomorphism of the Lie algebra $\mathfrak{g}$ 
into the Lie algebra $\mathcal{F}(\mathcal{M})$. 
It is known that 
any symplectic action of Lie group $G$ is Hamiltonian 
if $H^{2}(\mathfrak{g},\mathbb{R})=0$. 

Since the Hamiltonian function $H_{\xi}$ of $G$ 
depends on $\xi\in \mathfrak{g}$ linearly 
we may write it as
\begin{align}
\label{hm7}
H_{\xi}(x)=\langle \mu(x),\xi\rangle
\end{align}
where the notation 
$\langle f,\xi\rangle$ denotes the value of $f$ at $\xi\in
\mathfrak{g}$ 
and $\mu(x)$ belongs to $\mathfrak{g}^{*}$, 
the dual of the Lie algebra $\mathfrak{g}$. 
Therefore there is a map
\begin{align}
\label{hm8}
 \mu:\mathcal{M}\mapsto \mathfrak{g}^{*}
\end{align}
for any Hamiltonian action of $G$ on $\mathcal{M}$. 
This is called the moment map . 

If we have the Hamiltonian action of a 
Lie group $G$ on $\mathcal{M}$ which 
leaves the Hamiltonian $H(x)$ invariant, 
the quadruple 
\begin{align}
\label{hm9}
\left\{
\mathcal{M}, \left\{\ ,\ \right\}
, H, G
 \right\}
\end{align} 
is called a Hamiltonian system with $G$-symmetry. 
There is an important property 
of Hamiltonian system with $G$-symmetry \cite{MR0690288}
\begin{quote}
If the Hamiltonian $H(x)$ is invariant 
under a Hamiltonian action of a Lie group $G$ on $\mathcal{M}$, 
then the moment map $\mu(x)$ is an integral of motion. 
\end{quote}
This is a generalization 
of the well-known Noether's theorem. 
Since the symmetries give rise to 
the integrals of motion, 
one can reduce the dynamical system 
to one with fewer degrees of freedom 
\cite{MR0402819,MR0690288,abraham1978foundations,moser2011various}. 
Suppose we have a Hamiltonian 
action of $G$ on a symplectic manifold $\mathcal{M}$ 
and the corresponding moment map $\mu:\mathcal{M}\mapsto 
\mathfrak{g}^{*}$. 
We consider the inverse image 
of a point $c\in \mathfrak{g}^{*}$ for $\mu$ 
and represent this set by $\tilde{\mathcal{M}}_{c}$
\begin{align}
\label{hm10}
\tilde{\mathcal{M}}_{c}
=\mu^{-1}(c).
\end{align}
We require that $c$ is a regular value of $\mu$. 
This implies that 
either the differential of $\mu$ 
at every point of $\tilde{M}_{c}$ maps 
the tangent space to $\mathcal{M}$ 
onto $\mathfrak{g}^{*}$ or $\tilde{M}_{c}$ is empty 
\footnote{In \cite{MR766739} it has been discussed 
that almost all $c$ are regular values.}. 
In this case $\tilde{\mathcal{M}}_{c}$ 
is a smooth submanifold of $\mathcal{M}$. 
The isotropy subgroup of $c$ 
consists of the elements $\mathfrak{g}$ of $G$ 
for which 
\begin{align}
\label{hm11}
\mathrm{Ad}^{*}_{\mathfrak{g}}c=c.
\end{align}
Put in another way, 
the isotropy subgroup is the subgroup relative to the 
coadjoint action which leaves $\tilde{\mathcal{M}}_{c}$ invariant. 
Let us denote this isotropy subgroup 
by 
\begin{align}
\label{hm12}
G_{c}=\left\{g:\mathrm{Ad}^{*}_{\mathfrak{g}}c=c\right\}.
\end{align}
Now that 
the space $\tilde{\mathcal{M}}_{c}$ 
decomposes into orbits of the action of $G$, 
we can define the reduced phase space by 
\begin{align}
\label{hm13}
\mathcal{M}_{c}=\tilde{\mathcal{M}}_{c}/G_{c}.
\end{align}
It has been shown in \cite{MR0690288,abraham1978foundations} that 
if the isotropy subgroup $G_{c}$ is compact 
and acts on $\tilde{\mathcal{M}}_{c}$ without 
fixed points, 
the reduced phase space (\ref{hm13}) 
is shown to symplectic manifold 
and that 
the reduced field, the vector field on the reduced phase space 
$\mathcal{M}_{c}$ remains Hamiltonian vector field on it 
and the corresponding Hamiltonian function pulled back to 
$\tilde{\mathcal{M}}_{c}$ coincides with the 
original Hamiltonian function restricted to $\tilde{\mathcal{M}}_{c}$.

An Abelian version of the Lagrangian reduction 
with the integrals of motion was firstly proposed by 
Routh \cite{MR0068369}.  
Recall that there are two formulations 
for the classical dynamical system; 
the Lagrangian formalism and the Hamiltonian formalism. 
The Lagrangian is a functional of coordinates and 
their time derivatives and it leads to the equations of motion 
as a set of second order differential equations  
while the Hamiltonian is a functional of coordinates and 
their canonical momenta and leads to 
the equations of motion as a set of first order differential equations 
at the cost of the twice number of the equations. 

Suppose we have a system whose Lagrangian 
is independent of some subset of coordinates. 
We will refer them as cyclic coordinates and denote by $y^{i}$ 
and the remaining non-cyclic ones by $x^{i}$. 
The Lagrangian can be written as
\begin{align}
\label{1routh1}
L(x^{i},y^{i},\dot{x}^{i},\dot{y}^{i};t)
&=L(x^{i},\dot{x}^{i},\dot{y}^{i};t).
\end{align} 
Note that the canonical momenta of the cyclic coordinates $y^{i}$ 
\begin{align}
\label{1routh2}
p_{y^{i}}&=\frac{\partial L}{\partial \dot{y}^{i}}
\end{align}
are conserved quantities. 
In this case the differential equations associated with 
these momenta are trivial and therefore 
the Hamiltonian formulation is more advantageous. 

The Routhian $R$ is regarded as the new Lagrangian 
, which is the mixture of the Lagrangian with the Hamiltonian. 
More precisely it is defined by setting $p_{y^{i}}=h_{i}=\textrm{constant}$ 
and performing a partial Legendre transformation 
on the cyclic coordinates $y_{i}$
\begin{align}
\label{1routh3}
R(x^{i},\dot{x}^{i},h_{i};t):=
L-\sum_{i}h_{i}\dot{y}^{i}.
\end{align}
Let us consider the Euler-Lagrange expressions 
for the Routhian
\begin{align}
\label{1routh4}
&\frac{d}{dt}
\left(
\frac{\partial R}{\partial \dot{x}^{i}}\right)
-\frac{\partial R}{\partial x^{i}}\nonumber\\
=&\frac{d}{dt}\left(
\frac{\partial L}{\partial \dot{x}^{i}}+
\frac{\partial L}{\partial \dot{y}^{i}}
\frac{\partial \dot{y}^{i}}{\partial \dot{x}^{i}}
\right)
-\frac{d}{dt}\left(
h_{i}\frac{\partial \dot{y}^{i}}{\partial \dot{x}^{i}}
\right)\nonumber\\
&-\left(
\frac{\partial L}{\partial x^{i}}
+\frac{\partial L}{\partial \dot{y}^{i}}
\frac{\dot{y}^{i}}{\partial x^{i}}
\right)
-h^{i}\frac{\partial \dot{y}^{i}}{\partial x^{i}}.
\end{align}
The first and fourth terms cancel 
by the original Euler-Lagrange equations 
and the remaining terms vanish by the definition of the canonical
momenta $h_{i}=\frac{\partial L}{\partial \dot{y}^{i}}$. 

This shows that 
the Euler-Lagrange equations for $L(x,\dot{x},\dot{y})$ 
together with the conserved quantities $h_{i}=p_{y^{i}}$ 
are equivalent to the Euler-Lagrange equations for the Routhian
$R(x,\dot{x})$. 
The Euler-Lagrange equations for the Routhian are 
called the reduced Euler-Lagrange equations 
because the phase space $\mathcal{M}$ with variables $\{x^{i},y^{i}\}$ 
is now reduced to the small phase space $\tilde{\mathcal{M}}$ with 
variables $\{x^{i}\}$ 
\footnote{In other words the naive substitution of the conserved
quantities into the original Lagrangian spoils the role of the
Lagrangian.}.
Note that the Hamilton equations for the cyclic coordinates yield 
the trivial statement; 
the constant property of $h_{i}$ (i.e. $\dot{h}_{i}=0$) 
and the definition of $h_{i}$ 
(i.e. $h_{i}=\frac{\partial L}{\partial \dot{y}^{i}}$).

Now let us go back to the gauged mechanical Lagrangian (\ref{confg2}) 
and apply the Routh reduction 
\footnote{
The application of the Routh reduction 
in the gauged mechanical systems was 
discussed in \cite{Okazaki:2014sga}.}. 
 We will parametrize the complex variable $z$ as $z=qe^{i\varphi}$ 
where $q\ge 0$ and $0\le \varphi< 2\pi$ are real variables. 
We then can write the Lagrangian (\ref{confg2}) as
\begin{align}
\label{confg7}
L=
\frac12\dot{q}^{2}
+\frac12(q\dot{\varphi})^{2}
+q\dot{\varphi}A_{0}
+\frac12 q^{2}A_{0}^{2}
+cA_{0}.
\end{align}
By choosing the temporal gauge $A_{0}=0$, 
we get 
\begin{align}
\label{confg8}
L=\frac12\dot{q}^{2}+\frac12(q\dot{\varphi})^{2}
\end{align}
and the Gauss law constraint
\begin{align}
\label{confg9}
\phi=q^{2}\dot{\varphi}+c=0.
\end{align}
Note that the conserved quantity
$h:=\frac{\partial L}{\partial \dot{\varphi}}=q^{2}\dot{\varphi}$ 
appears in the Gauss law. 
The Gauss law constraint is the result 
of fixing the gauge action on the phase space. 
Thus it is interpreted as the moment map condition. 
Since the variable $\varphi$ is cyclic coordinate, 
we can now apply the Routh reduction (\ref{1routh3}) 
and derive the reduced action. 
We find the new Lagrangian as the Routhian
\begin{align}
\label{confg10}
R=\frac12
\left(
\dot{q}^{2}-\frac{c^{2}}{q^{2}}
\right).
\end{align}
Again this is exactly the DFF-model Lagrangian 
(\ref{confg1}) (or (\ref{confg6})) as expected. 
Therefore upon the reduction procedure of the gauged mechanical model 
we get the conformal mechanics (DFF-model).

\section{Black hole}
\label{secbh1}
An interesting connections between 
black holes and conformal mechanical modelds have been proposed  
in \cite{Claus:1998ts} 
\footnote{Also see \cite{Gibbons:1998fa} for the conjectural 
relation between black holes and the Calogero model. }.
Let us consider the $d=4$ Einstein-Maxwell theory which has the action 
\begin{align}
\label{bhcqm1}
S=\frac{1}{16\pi}\int d^{4}x\sqrt{-g}\left(
R-F^{2}
\right).
\end{align}
The theory admits a single extreme Reissner-Nordstr\"{o}m black hole
solution with the metric in isotropic coordinate
\begin{align}
\label{bhcqm2}
ds^{2}&
=-\left(1+\frac{|Q|l_{p}}{\rho}\right)^{-2}dt^{2}
+\left(1+\frac{|Q|l_{p}}{\rho}\right)^{2}
\left(d\rho^{2}+\rho^{2}d\Omega^{2}\right)
\end{align}
and the gauge field
\begin{align}
\label{bhcqm3}
A&=\left(1+\frac{|Q|l_{p}}{\rho}\right)^{-1}dt
\end{align}
where $Q$ is the black hole charge, 
$l_{p}$ is the Planck length 
with the black hole mass $M=\frac{|Q|}{l_{p}}$, 
and $d\Omega^{2}=d\theta^{2}+\sin^{2}\theta d\varphi^{2}$ 
is the $SO(3)$ invariant metric on $S^{2}$. 
In the near-horizon limit the metric 
(\ref{bhcqm2}) becomes the 
Bertotti-Robinson (BR) metric
\begin{align}
\label{bhcqm4}
ds^{2}=-\left(
\frac{\rho}{|Q|l_{p}}
\right)^{2}dt^{2}
+\left(
\frac{|Q|l_{p}}{\rho}
\right)^{2}d\rho^{2}
+\left(
|Q|l_{p}
\right)^{2}d\Omega^{2},
\end{align}
which is $SO(1,2)\times SO(3)$ invariant conformally flat 
metric on $\textrm{AdS}_{2}\times S^{2}$. 
Defining the horospherical coordinate as $(t,\phi=\frac{\rho}{Ql_{p}})$ 
for $AdS_{2}$ part, 
we can express the 
BR metric (\ref{bhcqm4}) as
\begin{align}
\label{bhcqm5}
ds^{2}=-\phi^{2}dt^{2}
+\frac{\left(|Q|l_{p}\right)^{2}}{\phi^{2}}d\phi^{2}
+\left(|Q|l_{p}\right)^{2}d\Omega^{2}
\end{align}
where the quantity $|Q|l_{p}$ 
is interpreted as the $S^{2}$ radius 
and also as the radius of the curvature 
of the $\textrm{AdS}_{2}$ space. 
To go further, 
let us introduce a new radial coordinate $r$ by 
\begin{align}
\label{bhcqm6}
\phi=\left(
\frac{2M}{r}
\right)^{2}.
\end{align}
Putting together the black hole solutions 
(\ref{bhcqm2}) and (\ref{bcqm3}) now become
\begin{align}
\label{bhcqm7}
ds^{2}&=-\left(
\frac{2M}{r}
\right)^{4}dt^{2}
+\left(
\frac{2M}{r}
\right)^{2}dr^{2}
+M^{2}d\Omega^{2},\\
\label{bhcqm8}
A&=\left(\frac{2M}{r}\right)^{2}dt
\end{align}
where we have chosen the unit so that $l_{p}=1$ and $M=|Q_{p}|$. 

Now we consider the test particle with 
mass $m$ and charge $q$. 
The world-line action of the particle is 
\begin{align}
\label{bhcqm9}
S&=-m\int ds +q\int A.
\end{align}
Putting the black hole solutions 
(\ref{bhcqm7}) and (\ref{bhcqm8}) into (\ref{bhcqm9}), 
we find the action
\begin{align}
\label{bhcqm10}
S=\int dt 
\left(\frac{2M}{r}\right)^{2}
\left[
q-m
\sqrt{
1
-\left(\frac{2M}{r}\right)^{-2}\dot{r}^{2}
-M^{2}\left(\frac{2M}{r}\right)^{-4}
\left(\dot{\theta}^{2}+\sin^{2}\theta\dot{\varphi}^{2}\right)}
\right].
\end{align}
The action is invariant under the conformal transformations 
\cite{Ivanov:2002tb}
\begin{align}
\label{bhcqm11a}
\delta t&=f(t)+c\left(\frac{1}{M^{2}}\right)r^{4}
=a+bt+ct^{2}+c\left(\frac{1}{M^{2}}\right)r^{4},\\
\label{bhcqm11b}
\delta r&=\frac12 \dot{f}r
=\frac12 (b+2ct)r,\\
\label{bhcqm11c}
\delta\theta&=\delta\varphi=0.
\end{align}
The corresponding conformal generators, 
the Hamiltonian $H$, the dilatation operator $D$ 
and the conformal boost operator $K$ are given by 
\begin{align}
\label{bhcqm12a}
H&=\left(
\frac{2M}{r}
\right)^{2}\left[
\sqrt{m^{2}+\frac{r^{2}p_{r}^{2}+4L^{2}}{4M^{2}}}-q
\right]\nonumber\\
&=\frac{p_{r}^{2}}{2f}
+\frac{m\gamma}{2r^{2}f},\\
\label{bhcqm12b}
D&=-\frac 14\left(rp_{r}+p_{r}r
\right),\\
\label{bhcqm12c}
K&=\frac12 fr^{2}
\end{align}
where we have introduced 
\begin{align}
\label{bhcqm13a}
L^{2}&=p_{\theta}^{2}
+\frac{p_{\varphi}^{2}}{\sin^{2}\theta},\\
\label{bhcqm13b}
f&=\frac12 \left[
\sqrt{m^{2}+\frac{1}{4M^{2}}\left(r^{2}p_{r}^{2}+4L^{2}\right)}
+q
\right],\\
\label{bhcqm13c}
\gamma&=4M^{2}\frac{m^{2}-q^{2}}{m}
+\frac{4L^{2}}{m}.
\end{align}
It can be shown that 
three generators $H$, $D$ and $K$ form the one-dimensional 
conformal $\mathfrak{sl}(2,\mathbb{R})$ 
algebra under the Poisson brackets. 

It has been pointed out \cite{Claus:1998ts} that 
this conformal mechanical model (\ref{bhcqm10}) give rise to 
the DFF-model (\ref{conflag1}) in the specific limit 
\footnote{However, the physical meaning of this particular limit is 
not clear and we will see that 
the mechanical model (\ref{bhcqm10}) and 
the DFF-model (\ref{conflag1}) can be realized as 
two different non-linear realizations of the 
one-dimensional conformal group $SL(2,\mathbb{R})$. }. 
Considering the limit 
\begin{align}
\label{bhcqm14}
M\rightarrow \infty,\ \ \ \ \ 
(m-q)\rightarrow 0,\ \ \ \ \ 
M^{2}(m-q)=\mathrm{fixed}
\end{align}
and noting that $f\rightarrow m$ in this limit, 
we obtain the DFF Hamiltonian 
\begin{align}
\label{bhcqm15a}
H=\frac{p_{r}^{2}}{2m}
+\frac{\gamma}{2r^{2}}
\end{align}
with the coupling constant
\begin{align}
\label{bhcqm15b}
\gamma=8M^{2}(m-q)+\frac{4l(l+1)}{m}.
\end{align}
Here $l(l+1), l\in \mathbb{Z}$ is the quantum number of the operator
$L^{2}$. 
Note that this quantization corresponds to the freezing of the 
$S^{2}$ angles, $\theta$, $\varphi$, i.e. 
$\theta=\textrm{const.}$, 
$\varphi=\textrm{const.}$ 
Therefore the DFF-model (\ref{bhcqm15a}) 
describes the radial motion 
of the $\textrm{AdS}_{2}\times S^{2}$ particle, 
i.e. the particle 
near the horizon of the extreme Reissner-Nordstr\"{o}m 
black hole in the limit (\ref{bhcqm14}).

\begin{figure}
\begin{center}
\includegraphics[width=10cm]{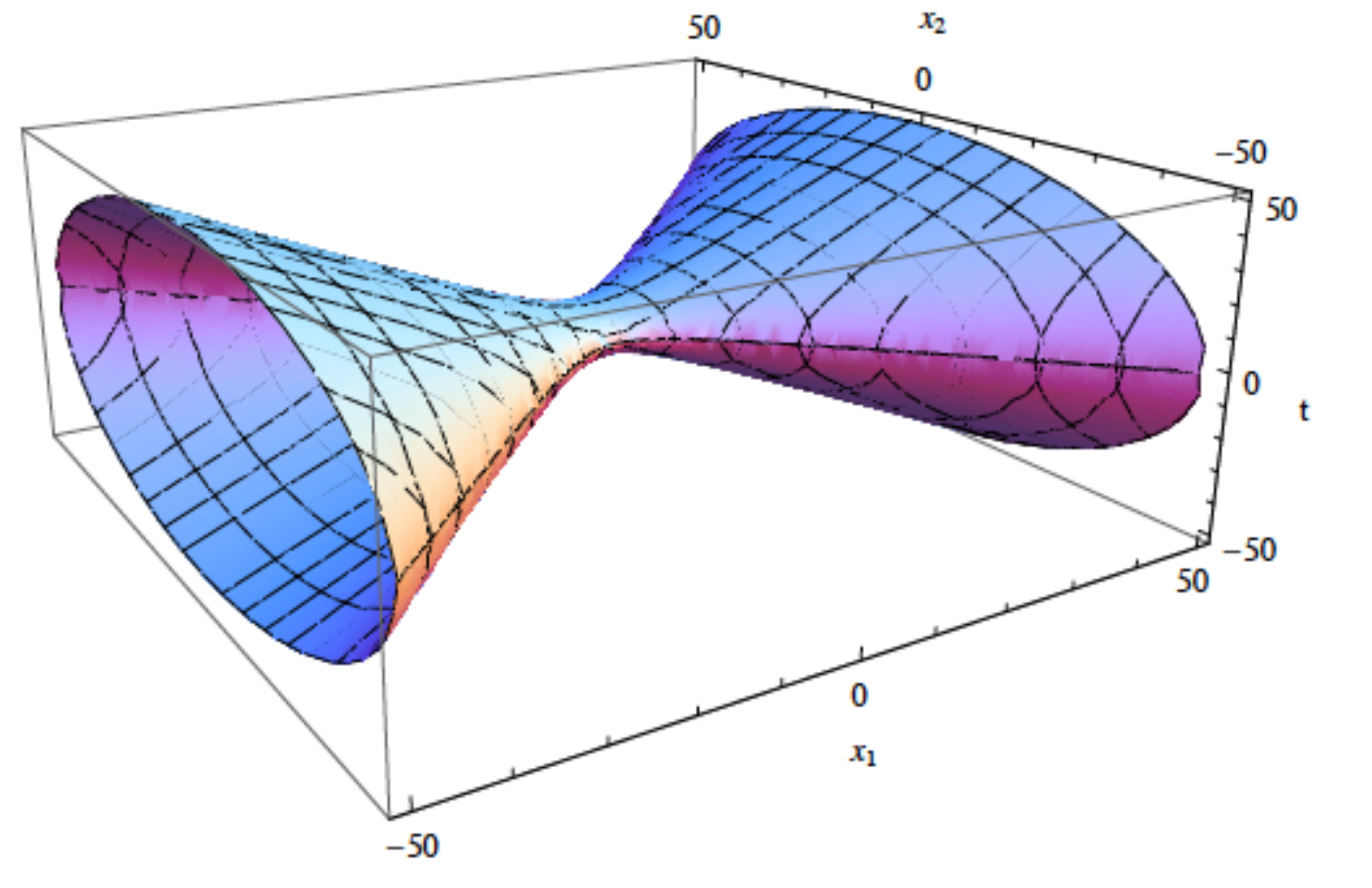}
\caption{$\textrm{AdS}_{2}$ space viewed as a hyperboloid
 of one sheet in a three dimensional Minkowski space.}
\label{figads2}
\end{center}
\end{figure}

Let us discuss the 
procedure proposed by DFF to cure the problem of the absence of the ground state for the Hamiltonian $H$ 
from the perspective of the particle motion near the black hole
horizon. 
Firstly we see that 
the metric (\ref{bhcqm5}) is singular at $\phi=0$, 
however, this is just a coordinate singularity 
and $\phi=0$ is a non-singular degenerate Killing horizon of the 
time-like Killing vector field $\frac{\partial}{\partial t}$. 
To see this we recall the definition 
of the $\textrm{AdS}_{2}$ space as a Lobachevski-like embedded 
surface in a three dimensional 
Minkowski space (see Figure \ref{figads2})
\begin{align}
\label{bhcqm16}
-(x^{0})^{2}+(x^{1})^{2}-(x^{2})^{2}=-R^{2}.
\end{align}
Using the hypersurface coordinate $(\phi,t)$ defined by
\begin{align}
\label{bhcqm17a}
x^{0}&=t\phi,\\
\label{bhcqm17b}
x^{+}&=x^{2}+x^{1}=\frac{R^{2}-t^{2}\phi^{2}}{R\phi},\\
\label{bhcqm17c}
x^{-}&=x^{2}-x^{1}=R\phi,
\end{align}
we obtain the $\textrm{AdS}_{2}$ factor of 
the BR metric (\ref{bhcqm5}) with $|Q|l_{p}=R$. 
The horospherical coordinates $(t,\phi)$ can only cover 
the half of the $\textrm{AdS}_{2}$ region. 
At $\phi=0$ the metric (\ref{bhcqm5}) is singular 
and $\phi>0$ or $\phi<0$ should be chosen. 
Correspondingly the coordinate $x^{-}$ is restricted to 
$x^{-}>0$ or $x^{-}<0$. 
Since the coordinate $x^{0}, x^{+}$ and $x^{-}$ 
are smooth on the hypersurface, 
at the horizon $\phi=0$ 
the time coordinate $t$ is ill-defined. 
To avoid such situation, 
let us define new coordinates 
\begin{align}
\label{bhcqm18a}
&t_{1}=\frac12 (x^{+}+x^{-}),\ \ \ \ \ 
t_{2}=x^{0},\ \ \ \ \ 
\mathfrak{t}=t_{1}+it_{2},\\
\label{bhcqm18b}
&r=\frac12 (x^{+}-x^{-}).
\end{align}
Then the equation (\ref{bhcqm16}) becomes 
\begin{align}
\label{bhcqm19}
-|\mathfrak{t}|^{2}+r^{2}=-R^{2}
\end{align}
and thus we can write 
\begin{align}
\label{bhcqm20}
\mathfrak{t}=e^{\frac{i\tau}{R}}
\sqrt{R^{2}+r^{2}}
\end{align}
with $\tau\in \mathbb{R}$ being a new coordinate. 
In terms of these coordinates the BR metric (\ref{bhcqm5}) 
can be written as
\begin{align}
\label{bhcqm21}
ds^{2}=
-\left(
\frac{R^{2}+r^{2}}{R^{2}}
\right)d\tau^{2}
+\left(
\frac{R^{2}}{R^{2}+r^{2}}
\right)dr^{2}
+R^{2}d\Omega^{2}.
\end{align}
In fact this shows that 
the horizon is not a true singularity as we mentioned.

Now we want to get further insights of conformal mechanics 
from black hole viewpoint. 
As seen from the expression (\ref{bhcqm17a}), 
the classical analog of an eigenstate vector 
of the Hamiltonian $H$ in conformal mechanics is an orbit of a 
time-like Killing vector $k=\frac{\partial}{\partial t}$ 
in the $\mathrm{AdS}_{2}$ region outside the horizon $(\phi\neq 0)$ 
and the energy eigenvalue $E$ is the value of $k^{2}$. 
This implies that the ground state $|E=0\rangle$ 
of $H$ with $E=0$ in conformal mechanics 
corresponds to the orbit of $k$ with $k^{2}=0$ which 
is a null geodesic generator of the event horizon. 
Therefore the absence of the ground state $|E=0\rangle$ 
can be interpreted as the fact that 
the orbit of $k^{2}=0$ cannot be covered by 
the static coordinate $t$ as we discussed. 

In classical general relativity 
it is a general procedure to demonstrate that 
the horizon is not a true singularity by changing the coordinate
system. 
Note that 
the $\textrm{AdS}_{2}$ isometry $SO(1,2)$ is linearly realized 
on the coordinates $(x^{0},x^{1},x^{2})$ as 
rotations $\delta x^{\mu}={\Lambda^{\mu}}_{\nu}x^{\nu}$ 
whose generators $J^{\mu\nu}=ix^{[\mu}\partial^{\nu]}$ 
form the $\mathfrak{so}(1,2)$ algebra
\begin{align}
\label{bhcqm210}
[J^{\mu\nu},J^{\rho\sigma}]
=i\left(\eta^{\mu[\rho}J^{\sigma]\nu}
-\eta^{\nu[\rho}J^{\sigma]\mu}\right)
\end{align}
with $\eta^{\mu\nu}=\textrm{diag}(-1,+1,-1)$. 
Then we can find new generators in our new coordinates 
$(t_{1},t_{2},r)$ and the corresponding operators in the DFF-model as follows: 
\begin{enumerate}
 \item rotation: $t_{1}\leftrightarrow t_{2}$

This rotation is expressed as 
the $U(1)$ rotation of the complex coordinate $\mathfrak{t}$
\begin{align}
\label{bhcqm22}
\mathfrak{t}=e^{\frac{i\tau'}{R}}\sqrt{R^{2}+r^{2}}=e^{i\alpha}\mathfrak{t}
\end{align}
with $\alpha\in \mathbb{R}$ being the infinitesimal parameter. 
and thus yields the time $\tau$ translation
\begin{align}
\label{bhcqm23}
\tau'=\tau+R\alpha.
\end{align}
Since it generates compact rotation in the non-compact $SO(1,2)$
       symmetry group, 
the corresponding generator $J^{t_{1}t_{2}}$ is identified with 
the $L_{0}$ in the conformal mechanics. 

\item rotation: $(t_{1},t_{2})\leftrightarrow r$

In this case the rotations are expressed as two boost operations
\begin{align}
\label{bhcqm24}
\delta \mathfrak{t}&=\beta,&
\delta r=\frac12\left(\beta \mathfrak{t}^{*}+\beta^{*}\mathfrak{t}\right)
\end{align}
where $\beta\in \mathbb{C}$ is the infinitesimal parameter. 
The complexified generator $J^{t_{1}r}\pm iJ^{t_{2}r}$ 
can be regarded as $L_{\pm}$ in the DFF conformal mechanics.

\end{enumerate}

Therefore from the black hole perspective 
the DFF prescription can be thought of 
as the different choice of time coordinates in which 
the world-lines of static particles can pass through the 
black hole horizon.

\section{Non-linear realization}
\label{secnonlin1}
The non-linear realization 
\cite{Coleman:1969sm,Callan:1969sn,Volkov:1973vd,Zumino:1977av} 
is a useful method to 
construct the non-linear invariant Lagrangian. 
The basic idea is the following:
\begin{enumerate}
 \item Start from the 
Lie (super)group $G$ that reflects the symmetry in the theory.  
\item Find the invariants under $G$ 
from the Cartan forms $\omega$ belonging to the (super)coset $G/H$ 
where $H$ is the stability subgroup of $G$. 
\item Construct the invariant Lagrangian under $G$ 
in terms of the Goldstone fields associated 
with the (super)coset Cartan forms $\omega$.
\end{enumerate}
By making use of the non-linear realization, 
it has been showed \cite{Ivanov:2002tb} that 
the DFF-model (\ref{conflag1}) 
and the black hole conformal mechanics (\ref{bhcqm10}) 
are essentially equivalent modulo redefinition 
of the time coordinates and the variables at classical level. 
Moreover the non-linear realization approach 
provides us with a powerful method 
to construct the irreducible supermultiplets 
for superconformal mechanical models. 
Much of the irreducible constraints and 
transformation laws can be automatically obtained 
from the non-linear realization technique. 

Let $G$ be a Lie (super)group 
and $H$ be its subgroup. 
We call $Y_{i}$ the generator of $H$. 
and $X_{i}$ the generator of the remaining generators. 
We assume that the commutator $[X_{i},Y_{j}]$ 
is a linear combination of $X_{i}$ alone
\begin{align}
\label{nonl1a}
[X_{i},Y_{j}]=f_{ij}^{k}X_{k}
\end{align}
where $f_{ij}^{k}$ are the structure constants. 
(\ref{nonl1a}) implies that 
the remaining generators $X_{i}$ form 
the representation of the subgroup $H$, 
which we will call the stability subgroup. 
Then a group element $g$ of $G$ 
can be represented uniquely by 
\cite{Coleman:1969sm,Callan:1969sn,Volkov:1973vd,Zumino:1977av}
\begin{align}
\label{nonl1b}
g&=e^{x\cdot X}h
\nonumber\\
&=\tilde{g}h
\end{align}
where $h$ is an element of $H$, $x\cdot X:=\sum_{i}x^{i}X_{i}$ and 
$x^{i}$ are the coordinates parametrizing the coset space $G/H$. 
The actions of the group $G$ can be realized by left multiplications 
on the coset $G/H$. 
This fact is the key statement of the 
non-linear realization method. 

Now we want to apply the basic statement (\ref{nonl1b}) to 
find the non-linear realization of $G$ symmetry group   
and to construct the $G$-invariant Lagrangian. 
In order to achieve this, 
we classify the parameters $x^{i}$ in two classes
\begin{align}
\label{nonl1c}
x^{i}=\begin{cases}
\textrm{(super)space coordinates}&
\textrm{if $X_{i}$ is (super)translation} \cr
\textrm{Goldstone (super)fields}&
\textrm{otherwise}. \cr
\end{cases}
\end{align}
In other words, 
the (super)space and time coordinates are the 
parameters of the (super)translation generators 
while the remaining coset parameters are treated as the 
(super)fields. 
We should note that 
the number of the Goldstone (super)fields 
is not always same as the number of the coset generators. 
In fact some of the Goldstone (super)fields may be 
expressed by other Goldstone (super)fields. 
This phenomenon is known 
as the inverse Higgs effect \cite{Ivanov:1975zq}.

For one-dimensional conformal algebra $\mathfrak{sl}(2,\mathbb{R})$ 
given by (\ref{hkd00})-(\ref{hkd02}), 
the stability subgroup $H$ is trivial 
and thus the coset is parametrized by the coordinates for all generators
\begin{align}
\label{nonl1d}
\tilde{g}=e^{itH}e^{iz(t)K}e^{iu(t)D}.
\end{align}
Since we are now considering one-dimensional field theory, i.e. 
quantum mechanics, 
we introduce time coordinate $t$ for the Hamiltonian $H$. 
The remaining two coordinates $z(t)$ 
for $K$ and $u(t)$ for $D$ are Goldstone fields. 

Then we can find the realization 
of the conformal group in our coset (\ref{nonl1d}). 
The translation $H$ 
is realized by acting on $g$ by $g_{0}=e^{iaH}$ from 
the left
\begin{align}
\label{nonl1e}
g_{0}\cdot \tilde{g}&=e^{iaH}\cdot e^{itH}e^{iz(t)K}e^{iu(t)D}.
\end{align}
We thus obtain the translations 
\begin{align}
\label{nonl1f}
\delta t&=a,\\
\delta u&=0,&
\delta z&=0.
\end{align}
The dilatation $D$ 
is realized by acting on $g$ by $g_{0}=e^{ibD}$ from 
the left
\begin{align}
\label{nonl1g}
g_{0}\cdot \tilde{g}&=e^{ibD}\cdot e^{itH}e^{iz(t)K}e^{iu(t)D}
\nonumber\\
&=(e^{ibD}e^{itH}e^{-ibD})
(e^{ibD}e^{izK}e^{-ibD})e^{ibD}e^{iuD}\nonumber\\
&=e^{i(t+bt)H}e^{i(z+bz)K}e^{i(u+b)D}.
\end{align}
One finds the dilatations
\begin{align}
\label{nonl1h}
\delta t&=bt,\\
\delta u&=b,&
\delta z&=-bz.
\end{align}
The conformal boost $K$ 
is realized by acting on $g$ by $g_{0}=e^{icK}$ from 
the left
\begin{align}
\label{nonl1i}
g_{0}\cdot \tilde{g}&=e^{icK}\cdot e^{itH}e^{izK}e^{iuD}
\nonumber\\
&=e^{itH}
(e^{-itH}e^{icK}e^{itH})e^{izK}e^{iuD}
\nonumber\\
&=e^{i(t+ct^{2})H}e^{i(z+c-2ctz)K}e^{i(u+2ct)D}
.
\end{align}
We get the conformal boost transformations
\begin{align}
\label{nonl1j}
\delta t&=ct^{2},\\
\delta u&=2ct,&
\delta z&=c-2ctz.
\end{align}

Let us discuss the construction of the $G$-invariant expressions. 
To this end we introduce the Maurer-Cartan form $\Omega$ 
for the coset $G/H$ defined by
\begin{align}
\label{nonl2a}
\Omega
&=\tilde{g}^{-1}d\tilde{g}\nonumber\\
&=e^{-x\cdot X}d(e^{x\cdot X})\nonumber\\
&=i\omega^{i}X_{i}+i\tilde{\omega}^{i}Y_{i}.
\end{align}
Then one can show 
\cite{Coleman:1969sm,Callan:1969sn,Volkov:1973vd,Zumino:1977av} that 
the forms $\omega^{i}$ on the coset 
transform homogeneously 
and therefore any expression constructed with 
$\omega^{i}$ is invariant under $G$. 
On the other hand it turns out 
\cite{Coleman:1969sm,Callan:1969sn,Volkov:1973vd,Zumino:1977av} 
that the forms $\tilde{\omega}^{i}$ 
on the stability subgroup $H$ transform like connections 
and can be used to construct covariant derivatives.

The Maurer-Cartan forms for the coset 
(\ref{nonl1d}) is 
\begin{align}
\label{nonl2a1}
\Omega&=\tilde{g}^{-1}d\tilde{g}
=i\left(\omega_{H}H+\omega_{K}K+\omega_{D}D\right)
\end{align}
where 
\begin{align}
\label{nonl2b}
\omega_{H}&
=e^{-u},\\
\label{nonl2c}
\omega_{D}&
=du-2zdt,\\
\label{nonl2d}
\omega_{K}&
=e^{u}\left(dz+z^{2}dt\right).
\end{align} 
Alternatively 
the Maurer-Cartan forms associated with 
the generators $T_{i}$, $i=0,1,2$ defined (\ref{so12a}) 
are given by
\begin{align}
\label{nonl2e1}
\omega_{0}&=\frac{1}{m}\omega_{K}
+m\omega_{K},\\
\label{nonl2e2}
\omega_{1}&=\frac{1}{m}\omega_{K}
-m\omega_{H},\\
\label{nonl2e3}
\omega_{2}&=\omega_{D}
\end{align}
where $m$ is a constant parameter. 
Since the form $\omega_{1}$, $\omega_{2}$ 
are the coset forms, 
they transform homogeneously. 
we can impose the following $SL(2,\mathbb{R})$ invariant conditions 
\footnote{
Although the choice of $\omega_{0}=0$ also yields 
the $SL(2,\mathbb{R})$ invariant constraint, 
it does not lead to the good dynamical systems.}
\begin{align}
\label{nonl2e4}
\omega_{1}&=0,\\
\label{nonl2e5}
\omega_{2}&=0.
\end{align}
The first condition (\ref{nonl2e4}) turns out to be 
the equation of motion for the system and 
the second condition (\ref{nonl2e5}) leads to the relation 
\begin{align}
\label{nonl2e6}
z=\frac12 \dot{u},
\end{align}
which implies that 
the Goldstone field $z(t)$ can be represented 
by the other Goldstone field $u(t)$.  
This is the inverse Higgs effect \cite{Ivanov:1975zq}. 
In terms of the remaining Maurer-Cartan forms $\omega_{0}$, 
one can construct the $SL(2,\mathbb{R})$ action \cite{Ivanov:1988vw}
\begin{align}
\label{nonl2e7}
S&=-c\int \omega_{0}\nonumber\\
&=-c\int dt \left[
\frac{c}{m}e^{u}\left(
\dot{z}+z^{2}
\right)+mce^{-u}
\right]\nonumber\\
&=\int dt \left[
x^{2}-\frac{c^{2}}{x^{2}}
\right]
\end{align}
where we have used the relation (\ref{nonl2e6}) 
and introduced
\begin{align}
\label{nonl2e8}
x&:=\mu^{-\frac12}e^{\frac{u}{2}},\\
\label{nonl2e9}
\mu&=\frac{m}{c}
\end{align}
The action (\ref{nonl2e7}) is just the DFF-model (\ref{confac1}).

Let us introduce 
\begin{align}
\label{nonl2h}
\hat{K}&=mK-\frac{1}{m}H,&
\hat{D}&=mD
\end{align}
where $m$ is a constant parameter. 
Then the one-dimensional conformal $\mathfrak{sl}(2,\mathbb{R})$ algebra 
can be written as
\begin{align}
\label{nonl2i1}
[H,\hat{D}]&=imH,\\
\label{nonl2i2}
[\hat{K},\hat{D}]&=
-2iH-im\hat{K},\\
[H,\hat{K}]&=
2i\hat{K}.
\end{align}
Defining the corresponding coset by 
\begin{align}
\label{nonl2i3}
\tilde{g}=
e^{i\tau H}
e^{i\phi(\tau)\hat{D}}
e^{i\Omega(\tau)\hat{K}}
\end{align}
and acting the corresponding elements on the coset (\ref{nonl2i3}) from
the left, 
one can find the $SL(2,\mathbb{R})$ transformations for 
the new coordianates
\begin{align}
\label{nonl2i4}
\delta \tau&=a+b+c\tau+\frac{1}{m^{2}}ce^{2m\phi},\\
\label{nonl2i5}
\delta \phi&=\frac{1}{m}\left(b+2c\tau\right),\\
\label{nonl2i6}
\delta \Omega&=\frac{1}{m}ce^{m\phi}
\end{align}
where $a,b,c$ are infinitesimal constant parameters. 
Note that the transformation of the new time coordinate $\tau$ 
contains the additional term $\frac{1}{m}ce^{2m\phi}$, 
which is similar to (\ref{bhcqm11a}). 
We can read the Maurer-Cartan forms for the coset (\ref{nonl2i3}) 
\cite{Ivanov:2002tb}
\begin{align}
\label{nonl2j1}
\hat{\omega}_{H}
&=\frac{1+\Lambda^{2}}{1-\Lambda^{2}}
e^{-m\phi}d\tau-2\frac{\Lambda}{1-\Lambda^{2}}d\phi,\\
\label{nonl2j2}
\hat{\omega}_{D}
&=\frac{1+\Lambda^{2}}{1-\Lambda^{2}}
d\phi-2\frac{\Lambda}{1-\Lambda^{2}}e^{-m\phi}d\tau,\\
\label{nonl2j3}
\hat{\omega}_{K}
&=m\frac{\Lambda}{1-\Lambda^{2}}
\left(\Lambda e^{-m\phi}d\tau-d\phi\right)
+\frac{d\Lambda}{1-\Lambda^{2}}
\end{align}
where 
\begin{align}
\label{nonl2j4}
\Lambda=\tanh\Omega.
\end{align}
Let us impose the $SL(2,\mathbb{R})$ invariant conditions as
\begin{align}
\label{nonl2j5}
\hat{\omega}_{D}=0,
\end{align}
which results in the inverse Higgs effect \cite{Ivanov:1975zq}
\begin{align}
\label{nonl2j6}
\partial_{\tau}\phi=2e^{-m\phi}\frac{\Lambda}{1+\Lambda^{2}}.
\end{align}
So the Goldstone field $\Lambda$ or $\Omega$ 
can be expressed by $\phi$ 
\begin{align}
\label{nonl2j7}
\Lambda&=
\partial_{\tau}\phi
e^{m\phi}
\frac{1}{1+\sqrt{1-e^{2m\phi}(\partial_{\tau}\phi)^{2}}}.
\end{align}

Using the non-vanishing Maurer-Cartan forms, 
one can construct the $SL(2,\mathbb{R})$ invariant action 
\cite{Ivanov:2002tb}
\begin{align}
\label{nonl2j8}
S&=\int 
\left[
(q-\tilde{\mu})\hat{\omega}_{H}
-\frac{2}{m}q\hat{\omega}_{K}
\right]\nonumber\\
&=-\int d\tau 
e^{-m\phi}
\left[
\tilde{\mu}
\sqrt{1-e^{2m\phi}(\partial_{\tau}\phi)^{2}}-q
\right].
\end{align}
We see that 
the action (\ref{nonl2j8}) is the 
conformal mechanical model (\ref{bhcqm10}) which 
describes the radial motion of the $\textrm{AdS}_{2}\times S^{2}$ 
particle \cite{Claus:1998ts}.

Therefore we see that 
the two mechanical model (\ref{bhcqm10}) 
and the DFF-model (\ref{conflag1}) 
can be realized as two different 
non-linear realizations of 
the one-dimensional conformal group $SL(2,\mathbb{R})$. 
From this point of view, 
we can conclude that 
the two conformal mechanical models 
are equivalent up to 
the redefinition of the time coordinate and 
the physical variable.

\section{Multi-particle conformal mechanics}
\label{mulcqm0}
Let us study the conformal mechanical models 
with many degrees of freedom for different particles. 
Generically $n$-particle quantum mechanics 
can be viewed as a sigma-model 
with an $n$-dimensional target space $\mathcal{M}$. 
So we will see the conditions 
\cite{Michelson:1999zf} 
for the target space $\mathcal{M}$ 
for the existence of conformal operators $D$ and $K$.

Consider the Hamiltonian 
\begin{align}
\label{mul1a1}
H&=\frac12 p^{\dag}_{a}g^{ab}p_{b}+V(x).
\end{align}
Here $g_{ab}(x)$ is the metric of the target space $\mathcal{M}$ 
where the indices $a,b=1,\cdots,n$ label the particles. 
\begin{align}
\label{mul1a2}
p_{a}&=g_{ab}\dot{x}^{b}
\end{align}
are the canonical momenta obeying
\begin{align}
\label{mul1a2a} 
[x^{a},p_{b}]&=i\delta_{ab},& 
[x^{a},x^{b}]&=0,&
[p_{a},p_{b}]&=0.
\end{align} 
The Hermitian conjugate of $p_{a}$ are
\begin{align}
\label{mul1a3}
p_{a}^{\dag}
&=\frac{1}{\sqrt{g}}p_{a}\sqrt{g}\nonumber\\
&=p_{a}-i\Gamma^{b}_{ba}
\end{align}
where $\Gamma_{ab}^{c}$ is the Christoffel symbol 
constructed from $g_{ab}$. 

Let us assume that 
the theory has a dilatational invariance of the form
\begin{align}
\label{mul1a4}
\delta t&=bt,&
\delta {x}^{a}&=\frac12 D^{a}(x)b,
\end{align}
which is 
a generalization of (\ref{inconf1a}) and (\ref{inconf1b}) 
with $b$ being an infinitesimal parameter for the dilatation. 
Then the dilatation generator $D$ is given by
\begin{align}
\label{mul1a5}
D&=\frac14 \left(
D^{a}p_{a}+p_{a}^{\dag}D^{a\dag}
\right).
\end{align}
Under the canonical relations (\ref{mul1a2a}) 
we find the commutation relation of 
the Hamiltonian (\ref{mul1a1}) 
and the dilatation generator (\ref{mul1a5}) as \cite{Michelson:1999zf}
\begin{align}
\label{mul1a6}
[H,D]&=
\frac{i}{4}
p_{a}^{\dag}\left(
\mathcal{L}_{D}g^{ab}
\right)p_{b}
+\frac{i}{2}\mathcal{L}_{D}V
+\frac{i}{8}\nabla^{2}\nabla_{a}D^{a}
\end{align}
where $\mathcal{L}_{d}$ is the Lie derivative
\begin{align}
\label{mul1a7}
\mathcal{L}_{D}g_{ab}
&=D^{c}g_{ab,c}+{D^{c}}_{,a}g_{cb}
+{D^{c}}_{,b}g_{ac}.
\end{align}
From the $\mathfrak{sl}(2,\mathbb{R})$ algebra 
(\ref{hkd00}) and the expressions (\ref{mul1a1}), (\ref{mul1a6}), 
the existence of the dilatation generator $D$ requires that 
\begin{align}
\label{mul1a8}
\mathcal{L}_{D}g_{ab}&=2g_{ab},\\
\label{mul1a9}
\mathcal{L}_{D}V(x)&=-2V(x),\\
\label{mul1a10}
\nabla^{2}\nabla_{a}D^{a}&=0.
\end{align}
A vector field $D$ 
is called homothetic vector field or 
similarity vector field on $\mathcal{M}$ 
\footnote{Note that
\begin{align}
\label{mul1a9a}
X
=\begin{cases}
\textrm{conformal Killing field}&\textrm{if\ \ } \mathcal{L}_{X}g_{ab}=\rho(x) g_{ab}\cr
\textrm{homothetic vector field}&\textrm{if\ \ }\mathcal{L}_{X}g_{ab}=cg_{ab}\cr
\textrm{Killing vector field}&\textrm{if\ \ }\mathcal{L}_{X}g_{ab}=0\cr
\end{cases}
\end{align}
where $\rho(x)$ is a function on $\mathcal{M}$ 
and $c$ is a constant on $\mathcal{M}$.  
}. 
A homothetic vector field generates 
a similarity transformation group. 
It is shown that 
along any integral curve of a homothetic vector field 
the space-like, time-like or null character of the 
tangent vector does not change 
and that there is necessarilly a singularity in 
each orbit of the similarity transformation group 
\cite{MR868722,MR940051,MR1045498}. 
 
Furthermore 
the remaining commutation relations 
(\ref{hkd01}) and (\ref{hkd02}) lead to \cite{Michelson:1999zf}
\begin{align}
\label{mul1b1}
\mathcal{L}_{D}K&=2K,\\
\label{mul1b2}
D_{a}dx^{a}&=dK
\end{align}
respectively. 
As the solutions to the equations (\ref{mul1b1}) and (\ref{mul1b2}), 
one can express the conformal boost generator $K$ 
as the norm of $D^{a}$
\begin{align}
\label{mul1b3}
K=\frac12 g_{ab}D^{a}D^{b}.
\end{align}
The equation (\ref{mul1b2}) means that 
the one-form $D=D_{a}dx^{a}$ is exact, 
however, it is shown \cite{Michelson:1999zf} that 
closed homothety vector field $D$ is always exact. 
So it is enough to impose the closeness condition for 
the homothetic vector field $D$ 
\begin{align}
\label{mul1b3a}
d\left(
D_{a}dx^{a}
\right)=0.
\end{align}

Therefore we can conclude that 
in order to obtain conformal quantum mechanical sigma-models, 
\begin{itemize}
 \item the target space $\mathcal{M}$ must admit 
a homothety vector field $D$ whose associated one-form $D_{a}dx^{a}$ 
is closed \footnote{The vector field $D$ with the required properties 
for conformal mechanical sigma-model is referred to as a closed homothety
       vector field in \cite{Michelson:1999zf}}; 
(\ref{mul1a8}) and (\ref{mul1b3a})
\item the potential $V(x)$ must satisfy (\ref{mul1a9})
\item $D^{a}$ must obey the vanishing condition (\ref{mul1a10}).
\end{itemize}

\section{Calogero model}
\label{mulcqm2}
One of the most celebrated multi-particle 
conformal mechanical models is the Calogero model, 
which is the system of multi-particles 
scattering on the line with inverse-square potentials 
\cite{Calogero:1969xj,Calogero:1969b}. 
The Hamiltonian is given by
\begin{align}
\label{calogero1}
H=\sum_{i=1}^{n}\frac12 p_{i}^{2}+\sum_{i<j}\frac{\gamma}{(x_{i}-x_{j})^{2}}
\end{align}
where the indices $i=1,\cdots,n$ label the particles 
and $\gamma$ is a coupling constant. 

In what follows we will discuss 
that the Calogero model and its generalization 
can be obtained from gauged matrix models. 
This formulation not only indicates the intimite relationship between 
the conformal quantum mechanical models and the gauged quantum
mechanical models but also provides us 
with non-trivial (super)conformal mechanical models. 

Let us start with the gauged matrix model action 
\begin{align}
\label{calogero1a1}
S=\int dt \left[
\mathrm{Tr}(DXDX)+\frac{i}{2}\left(
\overline{Z}DZ-D\overline{Z}Z
\right)+c\mathrm{Tr}A
\right].
\end{align}
Here $X_{a}^{b}(t)$, $\overline{X}_{a}^{b}=X_{b}^{a}$, $a=1,\cdots, n$ 
are the bosonic Hermitian $(n\times n)$ matrices, 
$Z_{a}(t)$, $\overline{Z}^{a}=(\overline{Z}_{a})$ are 
bosonic complex matrices and 
$A_{a}^{b}(t)$, $(\overline{A}_{a}^{b})=A_{b}^{a}$ 
are the $U(n)$ gauge fields with $n^{2}$ component fields. 
$c$ is a real constant parameter. 
In the action (\ref{calogero1a1}) the covariant derivatives are defined as
\begin{align}
\label{calogero1a2}
DX&:=\dot{X}+i[A,X],&
DZ&:=\dot{Z}+iAZ,&
D\overline{Z}&:=\dot{\overline{Z}}-i\overline{Z}A.
\end{align}
Note that in the third term, the Fayet-Iliopoulos term 
the non-abelian traceless part of the gauge field $A$ drops out and 
only the $U(1)$ part has contributions.

The action (\ref{calogero1a1}) is 
invariant under the one-dimensional $SL(2,\mathbb{R})$ 
conformal transformations
\begin{align}
\label{calogero1a3}
\delta t&=f(t),&
\delta \partial_{0}&=-\dot{f}\partial_{0},\\
\label{calogero1a4}
\delta X&=\frac12\dot{f}X,&
\delta Z&=0,&
\delta A&=-\dot{f}A
\end{align}
where $f(t)=a+bt+ct^{2}$ with $a,b,c$ being infinitesimal real
parameters. 
The action (\ref{calogero1a2}) is 
invariant under the $U(n)$ gauge transformations 
\begin{align}
\label{calogero1a5}
X&\rightarrow gXg^{-1},\\
\label{calogero1a6}
Z&\rightarrow gZ,&
\overline{Z}&\rightarrow \overline{Z}g^{-1},\\
\label{calogero1a7}
A&\rightarrow gAg^{-1}+i\dot{g}g^{-1}
\end{align}
where $g\in U(n)$. 
Let us impose a partial gauge fixing condition
\begin{align}
\label{calogero1a8}
X_{a}^{b}=x_{a}\delta_{a}^{b}
\end{align}
where $x_{a}$ are real component fields since 
$X$ are Hermitian matrices. 
Then the action (\ref{calogero1a2}) becomes 
\begin{align}
\label{calogero1a9}
S=\int dt 
\sum_{a,b}
\Biggl[
&\dot{x}_{a}\dot{x}_{a}
+\frac{i}{2}\left(
\overline{Z}^{a}\dot{Z}_{a}
-\dot{\overline{Z}}^{a}Z_{a}
\right)\nonumber\\
&+(x_{a}-x_{b})^{2}A_{a}^{b}A_{b}^{a}
-\overline{Z}^{a}A_{a}^{b}Z_{b}+cA_{a}^{a}
\Biggr].
\end{align}
By noting that the action (\ref{calogero1a9}) is invariant 
under the $U(1)^{n}$ gauge transformations 
\begin{align}
\label{calogero1b1}
x_{a}&\rightarrow x_{a},\\
\label{calogero1b2}
Z_{a}&\rightarrow e^{i\lambda_{a}}Z_{a},&
\overline{Z}^{a}&\rightarrow e^{-i\lambda_{a}}\overline{Z}^{a},\\
\label{calogero1b3}
A_{a}^{a}&\rightarrow A_{a}^{a}-\dot{\lambda}_{a}
\end{align}
where $\lambda_{a}(t)$ are local parameters, 
we further impose the gauge fixing condition as
\begin{align}
\label{calogero1b4}
Z_{a}=\overline{Z}^{a}.
\end{align}
Then the action (\ref{calogero1a9}) reduces to 
\begin{align}
\label{calogero1b5}
S=\int dt \sum_{a,b}
\left[
\dot{x}_{a}^{2}
+(x_{a}-x_{b})^{2}A_{a}^{b}A_{b}^{a}
-Z_{a}Z_{b}A_{a}^{b}+cA_{a}^{a}
\right].
\end{align}
At this stage we attempt to integrate out the gauge field $A$. 
From the action (\ref{calogero1b5}) 
we obtain the equations of motion for $A_{a}^{a}$ and 
for $A_{a}^{b}$,$a\neq b$ as
\begin{align}
\label{calogero1b6}
(Z_{a})^{2}&=c\\
\label{calogero1b7}
A_{b}^{a}&=\frac{Z_{a}Z_{b}}{2(x_{a}-x_{b})^{2}}.
\end{align}
Substituting 
the equations (\ref{calogero1b6}) and (\ref{calogero1b7}) 
into the action (\ref{calogero1b5}) and rescaling $x_{a}$ appropriately, 
we obtain the Calogero model action
\begin{align}
\label{calogero1b8}
S=\frac12 \int dt \left[
\sum_{a}\dot{x}_{a}^{2}
-\sum_{a\neq b}\frac{c^{2}}{(x_{a}-x_{b})^{2}}
\right].
\end{align}

\chapter{Superconformal Mechanics}
\label{chsc1m01}
In this chapter we will proceed to the superextension 
of the conformal quantum mechanics; 
the superconformal quantum mechanics. 
Firstly in section \ref{secsuperalg01} we will recall the basic facts 
about Lie superalgebra and Lie supergroup 
and will clarify the one-dimensional superconformal group. 
Then in section \ref{1dsusysec001} we will stress that 
supersymmety in one-dimension possesses 
many peculiar properties. 
In section \ref{secscqm1}, \ref{secscqm2}, \ref{secscqm4} and
\ref{secscqm8} 
we will review the persistent efforts to construct 
$\mathcal{N}=1$, $2$, $4$ and $8$ 
superconformal quantum mechanics 
by using the superspace and superfield formalism 
and also review the interesting topics which 
are relevant to those superconformal mechanical models. 

\section{Superalgebra and supergroup}
\label{secsuperalg01}
In $d$-dimensional superconformal field theories 
the ordinary supersymmetry and the conformal symmetry 
lead to a second supersymmetry. 
The corresponding generator $S_{\alpha}$ 
with $\alpha,\beta,\cdots$ being spinor indices 
can be found by taking the commutator 
of the conformal boost operator $K_{\mu}$ 
with space-time indices $\mu,\nu,\cdots=0,1,\cdots,d-1$ 
and the original supersymmetry $Q_{\alpha}$ 
\begin{align}
\label{spg1}
[K_{\mu},Q_{\alpha}]={(\Gamma_{\mu})^{\beta}}_{\alpha}S_{\beta}
\end{align}
where $\Gamma_{\mu}$ is a $d$-dimensional gamma matrix. 
Additionally the ani-commutator of supersymmetries 
$Q_{\alpha}$ and $S_{\alpha}$ 
generates the bosonic symmetry, the so-called R-symmetry. 
In general these generators form the superconformal algebras 
which are isomorphic to the simple Lie superalgebras. 
Hence it is expected that one can specify 
the corresponding Lie superalgebras, i.e. the superconformal algebras  
which characterize the superconformal field theories.

\subsection{Lie superalgebra}
\label{secliesupalg}
A superalgebra is a $\mathbb{Z}_{2}$-graded algebra 
$\mathfrak{g}=\mathfrak{g}_{\overline{0}}\oplus
\mathfrak{g}_{\overline{1}}$. 
This means that 
if $a\in \mathfrak{g}_{\alpha}$, $b\in \mathfrak{g}_{\beta}$, 
$\alpha,\beta\in\mathbb{Z}_{2}=\left\{\overline{0},\overline{1}\right\}$, 
then $ab\in \mathfrak{g}_{\alpha+\beta}$. 
We say that $a$ is of degree $\alpha$ and 
 write $\textrm{deg}a=\alpha$. 
$\mathfrak{g}_{\overline{0}}$ is a Lie algebra, 
which is called the even or bosonic part of $\mathfrak{g}$ 
while $\mathfrak{g}_{\overline{1}}$ is 
called the odd or fermionic part of $\mathfrak{g}$, 
which is not an algebra.

A Lie superalgebra is the superalgebra endowed with 
the product operation $[\ ,\ ]$ possessing the following axioms:
\begin{enumerate}
 \item graded anticommutativity
\begin{align}
\label{spg2a}
[a,b]=-(-1)^{\alpha\beta}[b,a]
\end{align}
\item generalized Jacobi identity
\begin{align}
\label{spg2b}
[a,[b,c]]=[[a,b],c]+(-1)^{\alpha\beta}[b,[a,c]]
\end{align}
\end{enumerate}
where $a\in \mathfrak{g}_{\alpha}$, $b\in \mathfrak{g}_{\beta}$. 
The product $[a,b]$ is referred to as 
the Lie superbracket or supercommutator for 
two elements $a,b\in\mathfrak{g}$.

Let $V=V_{\overline{0}}\oplus V_{\overline{1}}$ be 
 $\mathbb{Z}_{2}$-graded vector space where 
$\dim V_{\overline{0}}=m$ and $\dim V_{\overline{1}}=n$.  
Then the algebra $\mathrm{End}V$ is endowed with a
$\mathbb{Z}_{2}$-graded superalgebra structure. 
Hence the Lie superbracket $[\ ,\ ]$ satisfying (\ref{spg2a}) and
(\ref{spg2b}) turns $\textrm{End}V$ into 
a Lie superalgebra $\mathfrak{l}(m,n)$. 
The Lie superalgebra $l(V)$ 
plays the same role as the general linear Lie algebra 
in the theory of Lie algebra. 
Let $e_{1},\cdots, e_{m};e_{m+1},\cdots,e_{m+n}$ 
be a basis of $V$, formed by the bases 
of $V_{\overline{0}}$ and $V_{\overline{1}}$.  
In this basis the matrices of an element $a$ from 
the Lie superalgebra $\mathfrak{l}(m,n)$ can be written in the form 
\begin{align}
\label{spg2b1}
a=\left(
\begin{array}{cc}
\alpha&\beta\\
\gamma&\delta\\
\end{array}
\right)
\end{align}
where $\alpha$ and $\delta$ are $\mathfrak{gl}(m)$ 
and $\mathfrak{gl}(n)$ matrices and 
$\beta$ and $\gamma$ are $m\times n$ and $n\times m$ rectangular matrices. 
On the Lie superalgebra $\mathfrak{gl}(m,n)$ the supertrace is defined by 
\begin{align}
\label{spg2b2}
\mathrm{str}(a)=\mathrm{tr}\alpha-\mathrm{tr}\delta.
\end{align}
In terms of the supertrace (\ref{spg2b2}), 
we can define the bilinear form $B_{R}$ associated with 
the representation $R$ of $\mathfrak{g}$ by
\begin{align}
\label{spg2b3}
B_{R}(a,b)=\textrm{str}(R(a),R(b)),\ \ \ \ 
\forall a,b\in \mathfrak{g}
\end{align}
where $R(a)$ is the matrix of the elements $a\in \mathfrak{g}$ 
in the representation $R$. 
As a special case the Killing form $K$ can be defined 
as the bilinear form on $\mathfrak{g}$ 
associated with the adjoint representation 
\begin{align}
\label{spg2b4}
K(a,b)=\textrm{str}(\textrm{ad}(a),\textrm{ad}(b)),\ \ \ \ \ 
\forall a,b \in \mathfrak{g}.
\end{align}

The Lie superalgebra $\mathfrak{g}$ is called simple 
if it contains no non-trivial ideal. 
The Lie superalgebra $\mathfrak{g}$ is called semi-simple 
if it contains no non-trivial solvable ideal. 
If a Lie superalgebra $\mathfrak{g}=\mathfrak{g}_{\overline{0}}\oplus
\mathfrak{g}_{\overline{1}}$ is simple, 
the representation of $\mathfrak{g}_{\overline{0}}$ 
on $\mathfrak{g}_{\overline{1}}$ is faithful 
and
$\left\{\mathfrak{g}_{\overline{1}},\mathfrak{g}_{\overline{1}}\right\}
=\mathfrak{g}_{\overline{0}}$. 
If the representation of $\mathfrak{g}_{\overline{0}}$ 
on $\mathfrak{g}_{\overline{1}}$ is irreducible, 
then $\mathfrak{g}$ is simple. 
Unlike the Lie algebras, 
semi-simple Lie superalgebra cannot be written 
as the direct sum of simple Lie superalgebras. 
However, there is a construction which allows us to 
build finite-dimensional semi-simple Lie superalgebras 
in terms of simple ones \cite{MR0486011}.

It is known that simple Lie superalgebras are 
classified into two families; the classical Lie superalgebras 
and (non-classical) Cartan type superalgebras. 
The simple Lie superalgebra is said to be classical 
the representation of the Lie algebra $\mathfrak{g}_{\overline{0}}$ 
on $\mathfrak{g}_{\overline{1}}$ is completely reducible.

For the classical Lie superalgebras there are further classifications. 
Firstly the representation of $\mathfrak{g}_{\overline{0}}$ 
on $\mathfrak{g}_{\overline{1}}$ can be either 
(i) irreducible or (ii) the direct sum of two irreducible
representations of $\mathfrak{g}_{\overline{0}}$. 
The superalgebra of the case (i) is called the type I 
and that of the case (ii) is called type II. 
In addition, the Lie superalgebra $\mathfrak{g}$ is called basic 
if there is a non-degenerate invariant bilinear form, 
the Killing form $K$ on $\mathfrak{g}$ while strange if it is not basic. 
The basic Lie superalgebras is divided into 
(a) four infinite series: $A(m,n)$, $B(m,n)$, $C(n)$ and $D(m,n)$, 
that is $\mathfrak{sl}(m+1|n+1)$, $\mathfrak{osp}(2m+1|2n)$, 
$\mathfrak{osp}(2|2n)$ and $\mathfrak{osp}(2m|2n)$; 
(b) three exceptional series: 40-dimensional $F(4)$, 
31-dimensional $G(3)$ and 17-dimensional $D(2,1;\alpha)$ which is 
a one-parameter family of superalgebras. 
The strange algebras split into 
two infinite families $P(n)$ and $Q(n)$. 

For the Cartan type superalgebras 
there are four infinite families $W(n)$, 
$S(n)$, $H(n)$ $\tilde{S}(n)$, where the 
first three series are analogous to the corresponding series 
of simple infinite-dimensional Lie algebra of Cartan type 
and $\tilde{S}(n)$ is a deformation of $S(n)$. 

\begin{figure}
\begin{center}
\includegraphics[width=15cm]{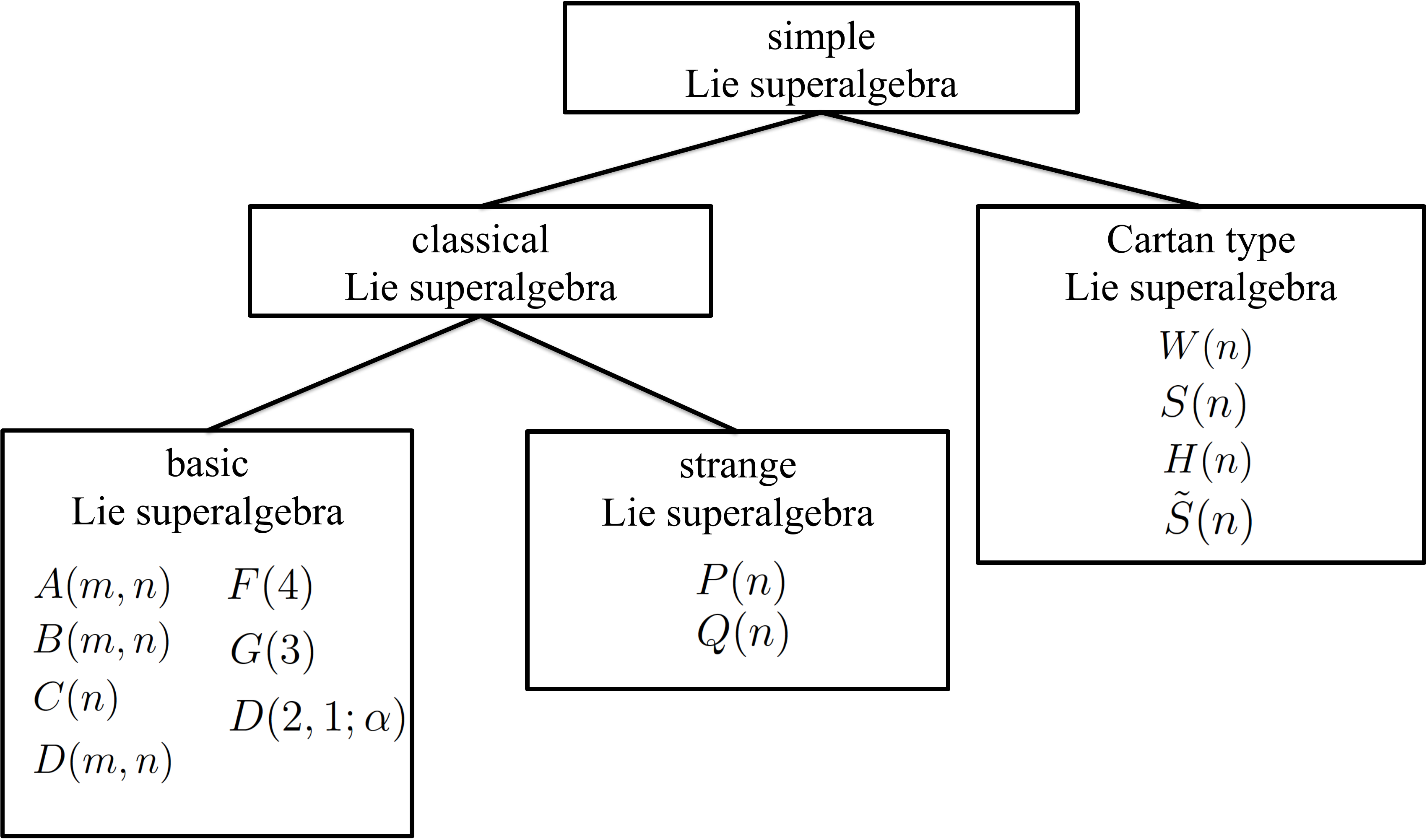}
\caption{The classification of simple Lie superalgebra.}
\label{liesuper1}
\end{center}
\end{figure}

Summarizing the above, 
the classification of simple Lie superalgebra is illustrated in 
Figure \ref{liesuper1}. 

\subsection{Lie supergroup}
\label{secsupalg01b}
To begin with, let us introduce a supermatrix. 
A supermatrix $M$ is defined as 
the matrix whose entries valued in a Grassmann algebra 
$\Gamma
=\Gamma_{\overline{0}}\oplus \Gamma_{\overline{1}}$ 
of the form 
\begin{align}
\label{spg2c}
M=\left(
\begin{array}{cc}
A&B\\
C&D\\
\end{array}
\right)
\end{align}
where $A$,$B$,$C$ and $D$ are 
$m\times p$, $m\times q$, $n\times p$ and $n\times q$ matrices
respectively. 
The supermatrix $M$ is said to be even and of degree 0
if $A,D\in \Gamma_{\overline{0}}$ 
and $B,C\in \Gamma_{\overline{1}}$ 
whereas it is called odd and of degree 1 if 
$A,D\in \Gamma_{\overline{1}}$ 
and $B,C\in \Gamma_{\overline{0}}$.

The general linear supergroup $GL(m|n)$ 
consists of even invertible supermatrices $M$ 
and its product is defined by the multiplication rule 
of the supermatrices:
\begin{align}
\label{apg2c2}
(MN)_{ij}
=\sum_{k=1}^{p+q}M_{ik}N_{kj}
\end{align}
where $M$ and $N$ are 
two $(m+n)\times (p+q)$ and 
$(p+q)\times (r+s)$ supermatrices 
and $(MN)_{ij}$ denotes the $(i,j)$ 
entry of the $(m+n)\times (r+s)$ supermatrix $MN$.

The operations for the supermatrices are defined as follows:
\begin{enumerate}
 \item transpose $M^{t}$ and supertranspose $M^{st}$
\begin{align}
\label{spg2d}
M^{t}&=\left(
\begin{array}{cc}
A^{t}&C^{t}\\
B^{t}&D^{t}\\
\end{array}
\right),\\
\label{spg2e}
M^{st}&=\left(
\begin{array}{cc}
A^{t}&(-1)^{\deg M}C^{t}\\
-(-1)^{\deg M}B^{t}&D^{t}\\
\end{array}
\right)
=
\begin{cases}
\left(
\begin{array}{cc}
A^{t}&C^{t}\\
-B^{t}&D^{t}\\
\end{array}
\right)&\textrm{if $M$ is even}\cr
\left(
\begin{array}{cc}
A^{t}&-C^{t}\\
B^{t}&D^{t}\\
\end{array}
\right)&\textrm{if $M$ is odd}\cr
\end{cases}
\end{align}

\item supertrace $\textrm{str}(M)$
\begin{align}
\label{spg2f}
\textrm{str}(M)
=\textrm{tr}(A)-(-1)^{\deg M}\textrm{tr}(D)
=\begin{cases}
\textrm{tr}(A)-\textrm{D}&\textrm{if $M$ is even}\cr
\textrm{tr}(A)+\textrm{D}&\textrm{if $M$ is odd}\cr
\end{cases}
\end{align}
\item superdeterminant $\textrm{sdet}(M)$
\begin{align}
\label{spg2g}
\textrm{sdet}(M)
=\frac{\det (A-BD^{-1}C)}{\det (D)}
=\frac{\det (A)}{\det (D-CA^{-1}B)}
\end{align}
\item adjoint $M^{\dag}$ and superadjoint $M^{\ddag}$
\begin{align}
\label{spg2h}
M^{\dag}&=(M^{t})^{*},\\
\label{spg2i}
M^{\ddag}&=(M^{st})^{*}.
\end{align}
\end{enumerate}

The relation between the Lie superalgebra $\mathfrak{g}$
and the corresponding Lie supergroup $G$ 
is analogous to the theory of the Lie algebra. 
Consider the complex Grassmann algebra $\Gamma(n)$ 
of order $n$ with $n$ generators 
$1,\theta_{1},\cdots, \theta_{n}$ obeying the anti-commutation relations 
$\left\{\theta_{i},\theta_{j}\right\}=0$. 
If in the element $\eta=\sum_{0\le m}\sum_{i_{1}<\cdots
i_{m}}\eta_{i_{1}\cdots i_{m}}\theta_{i_{1}}\cdots \theta_{i_{m}}$ 
each complex coefficient $\eta_{i_{1}\cdots i_{m}}$ is an even (odd)
value of $m$, 
the corresponding element is called even (odd). 
In general $\Gamma(n)$ can be decomposed into 
even and odd parts as a vector space;
$\Gamma(n)=\Gamma(n)_{\overline{0}}\oplus \Gamma(n)_{\overline{1}}$.
The Grassmann envelope $G(\Gamma)$ of the Lie superalgebra
$\mathfrak{g}$ is constructed as a formal linear combinations 
$\sum_{i}\eta_{i}a_{i}$ where $a_{i}$ is a basis of $\mathfrak{g}$ 
and $\eta_{i}\in \Gamma$ such that the elements $a_{i}$ 
and $\eta_{i}$ are both even or odd. 
The Lie supergroup $G$ associated with the superalgebra $\mathfrak{g}$ 
is realized as the exponential mapping of the Grassmann envelope
$G(\Gamma)$ of $\mathfrak{g}$; 
the even generators of the superalgebra $\mathfrak{g}$ 
corresponds to commuting parameters, i.e. even elements of the Grassmann
algebra and the odd generators of the superalgebra $\mathfrak{g}$ 
to anti-commuting parameters, i.e. odd elements of the Grassmann algebra 
\cite{MR0265520}.

\subsection{Superconformal algebra}
\label{secsupalg01a}
The requirements for the corresponding superconformal algebra 
have been proposed in \cite{Nahm:1977tg}:
\begin{enumerate}
 \item The $d$-dimensional conformal algebra $\mathfrak{so}(d,2)$
       should appear as a bosonic factored subgroup. 
\item The fermionic generators should be spinor representations of 
the conformal algebra $\mathfrak{so}(d,2)$. 
\end{enumerate}
First of all we can see that 
these conditions can be satisfied for the simple classical Lie 
superalgebras. 
The detail list of the classical Lie superalgebras is 
given in Table \ref{liesup1table} 
\cite{Scheunert:1976uf,Scheunert:1976ug,Kac:1977qb,Frappat:1996pb,Claus:1998us}.
\begin{table}
\begin{center}
\begin{tabular}{|c|c|c|c|c|} \hline
$\mathfrak{g}$&$\mathfrak{g_{\overline{0}}}$&$\mathfrak{g}_{\overline{1}}$
&$K$&type\\ \hline\hline
$A(m-1,n-1)$&$A_{m-1}\oplus A_{n-1}\oplus \mathfrak{u}(1)\delta_{m,n}$&
$(\bm{m},\overline{\bm{n}})$&basic&I \\
 & & $\oplus 
(\overline{\bm{m}},\bm{n})$& & \\
$\mathfrak{su}(m-p,p|n-q,q)$&$\mathfrak{su}(m-p,p)\oplus
     \mathfrak{su}(n-q,q)\oplus \mathfrak{u}(1)\delta_{m,n}$&
&basic&I\\
$\mathfrak{su}^{*}(2m|2n)$&$\mathfrak{su}^{*}(2m)\oplus
     \mathfrak{su}^{*}(2n)\oplus \mathfrak{so}(1,1)\delta_{m,n}$&
 &basic&I\\
$\mathfrak{sl}'(n|n)$&$\mathfrak{sl}(n,\mathbb{C})$&
&basic&I\\ 
$B(m,n)$&$B_{m}\oplus C_{n}$&
$(\bm{2m+1},\bm{2n})$ &basic&II\\
&$m\ge 0, n\ge1$ & & & \\
$C(n+1)$&$C_{m}\oplus \mathfrak{u}(1)$& 
$\bm{2n}\oplus\bm{2n}$ &basic&I\\
&$n\ge1$ & & & \\
$D(m,n)$&$D_{m}\oplus C_{n}$&
$(\bm{2m},\bm{2n})$ &basic&II\\
&$m\ge2,n\ge1,m\neq n+1$ & & & \\
$\mathfrak{osp}(m-p,p|n)$&$\mathfrak{so}(m-p,p)\oplus \mathfrak{sp}(n)$&
 &basic&II\\
$\mathfrak{osp}(m^{*}|n-q,q)$&$\mathfrak{so}^{*}(m)\oplus
     \mathfrak{usp}(n-q,q)$& &basic&II\\ \hline
$D(2,1;\alpha)$&$A_{1}\oplus A_{1}\oplus A_{1}$&
$(\bm{2},\bm{2},\bm{2})$&basic&II\\
&$0<\alpha\le 1$& & & \\
$D^{p}(2,1;\alpha)$&$\mathfrak{so}(4-p,p)\oplus \mathfrak{sl}(2)$&
&basic&II\\ 
$F(4)$&$A_{1}\oplus B_{3}$&
$(\bm{2},\bm{8})$&basic&I\\
$F^{p}(4)$&$\mathfrak{so}(7-p,p)\oplus \mathfrak{sl}(2)$&
&basic&I\\
&$p=0,3$& & & \\
$F^{p}(4)$&$\mathfrak{so}(7-p,p)\oplus \mathfrak{su}(2)$&
&basic&I\\
&$p=1,2$& & & \\
$G(3)$&$A_{1}\oplus G_{2}$&
$(\bm{2},\bm{7})$&basic&I\\
$G_{p}(3)$&$\mathfrak{g}_{2,p}\oplus \mathfrak{sl}(2)$&
&basic&I\\
&$p=-14,2$& & & \\ \hline
$P(m-1)$&$\mathfrak{sl}(m)$&
$(\bm{m}\otimes\bm{m})$&strange&I\\
&$m\ge3$& & & \\
$Q(m-1)$&$\mathfrak{su}(m)$&
$\textrm{adjoint}$&strange&II\\
&$m\ge3$& & & \\
$Q(m-1)$&$\mathfrak{sl}(m)$&
&strange&II\\
$Q((m-1)^{*})$&$\mathfrak{su}^{*}(m)$&
&strange&II\\
$UQ(p,m-1-p)$&$\mathfrak{su}(p,m-p)$&
&strange&II\\ \hline
\end{tabular}
\caption{The list of the classical Lie superalgebras
 $\mathfrak{g}=\mathfrak{g}_{\overline{0}}\oplus
 \mathfrak{g}_{\overline{1}}$ with Killing forms $K$.}
\label{liesup1table}
\end{center}
\end{table}

The unitary superalgebra $A(m-1,n-1)$ or $\mathfrak{sl}(m,n)$ 
with $m>n>0$ possesses 
an even part $\mathfrak{sl}(m)\oplus \mathfrak{sl}(n)\oplus
\mathfrak{u}(1)$ and an odd part 
$(\bm{m},\overline{\bm{n}})\oplus (\overline{\bm{m}},\bm{n})$ as 
a representation of the even part. 
The unitary superalgbra $A(n-1,n-1)$ with $n>1$ has 
an even part $\mathfrak{sl}(n)\oplus \mathfrak{sl}(n)$ 
and an odd part $(\bm{n},\overline{\bm{n}})\oplus 
(\overline{\bm{n}},\bm{n})$. 

The orthosymplectic superalgebras 
consist of three infinite series 
$B(m,n)$, $C(n+1)$ and $D(m,n)$. 
The superalgebra $B(m,n)$ or 
$\mathfrak{osp}(2m+1|2n)$ with $m\ge 0$, $n\ge 1$ 
possesses an even part $\mathfrak{so}(2m+1)\oplus \mathfrak{sp}(2n)$ 
and an odd part $(\bm{2m+1},\bm{2n})$. 
The superalgebra $C(n+1)$ or 
$\mathfrak{osp}(2|2n)$ with $n\ge 1$ 
contains an even part $\mathfrak{so}(2)\oplus \mathfrak{sp}(2n)$ 
and an odd part $\bm{2n}\oplus \bm{2n}$ as 
twice the fundamental representation $\bm{2n}$ of $\mathfrak{sp}(2n)$. 
The superalgebra $D(m,n)$ or 
$\mathfrak{osp}(2m|2n)$ with $m\ge2, n\ge 1$ has 
an even part $\mathfrak{so}(2m)\oplus \mathfrak{sp}(2n)$ 
and an odd part $(\bm{2m},\bm{2n})$.

The superalgebras $D(2,1;\alpha)$ with $\alpha\neq 0,-1,\infty$ 
is a one-parameter family of superalgebras 
of rank 3 and dimension 17. 
It is a deformation of the superalgebra $D(2,1)$ 
that corresponds to the case of $\alpha=1$. 
It has an even part $\mathfrak{sl}(2)\oplus \mathfrak{sl}(2)\oplus
\mathfrak{sl}(2)$ 
and an odd part $(\bm{2},\bm{2},\bm{2})$ 
as the spinor representations of 
$\mathfrak{sl}(2)\oplus \mathfrak{sl}(2)\oplus \mathfrak{sl}(2)$. 
The three $\mathfrak{sl}(2)$ factors appear as 
the anticommutator of the fermionic generators with 
the relative weights $1,\alpha$ and $1-\alpha$.

The superalgebra $F(4)$ is 40-dimensional algebra of rank 4 
and possesses an even part $\mathfrak{sl}(2)\oplus \mathfrak{o}(7)$ 
and an odd part $(\bm{2},\bm{8})$ as the spinor representations 
of $\mathfrak{sl}(2)\oplus \mathfrak{o}(7)$.

The superalgebra $G(3)$ is 31-dimensional algebra of rank 3 
and has an even part $\mathfrak{sl}(2)\oplus G_{2}$ 
and an odd part $(\bm{2},\bm{7})$ as the representations 
of $\mathfrak{sl}(2)\oplus G_{2}$.

By scanning through the list in Table \ref{liesup1table}, 
we can find the superconformal algebras which satisfy the 
required conditions. 
For $d=1$ superconformal field theory, 
that is superconformal quantum mechanics, 
the bosonic conformal algebra is 
$\mathfrak{so}(1,2)=\mathfrak{sl}(2,\mathbb{R})
=\mathfrak{su}(1,1)=\mathfrak{sp}(2)$ 
and richer superconformal structures are allowed 
due to the small conformal group.  
Note that $\mathfrak{so}(1,2)$ may be contained 
as an even part $\mathfrak{g}_{\overline{0}}$ 
for the series of the Lie superalgebra 
$\mathfrak{g}=\mathfrak{osp}(m-p,p|n)$ 
and thus the corresponding R-symmetry algebras are 
the series of the non-compact $\mathfrak{sp}(n)$.  
Therefore if we consider the classical Lie superalgebras 
with compact R-symmetry algebras, 
the corresponding supergroups can be represented 
in terms of supermatrices as 
\begin{align}
\label{scalg1a}
\left(
\begin{array}{cc}
SL(2,\mathbb{R})&B\\
C&\textrm{R-symmetry}\\
\end{array}
\right),\\
\label{scalg1b}
\left(
\begin{array}{cc}
SU(1,1)&B\\
C&\textrm{R-symmetry}\\
\end{array}
\right),\\
\label{scalg1c}
\left(
\begin{array}{cc}
Sp(2)&B\\
C&\textrm{R-symmetry}\\
\end{array}
\right)
\end{align}
where $B$ and $C$ are fermionic matrices. 
Two supermatrices (\ref{scalg1b}) and (\ref{scalg1c}) 
correspond to the infinite series of the Lie superalgebra 
and provide us chains of the one-dimensional superconformal groups. 
The remaining supermatrices (\ref{scalg1a}) may cover 
the exceptional Lie superalgebras and 
other special cases. 
The one-dimensional superconformal groups are 
tabulated in Table \ref{listsup1} 
\cite{Frappat:1996pb,Claus:1998us,BrittoPacumio:1999ax}.
\begin{table}
\begin{center}
\begin{tabular}{|c|c|c|} \hline
supersymmetry&supergroup&R-symmetry \\ \hline\hline
$\mathcal{N}=1$&$OSp(1|2)$&$1$ \\ \hline
$\mathcal{N}=2$&$SU(1,1|1)$&$U(1)$ \\ \hline
$\mathcal{N}=3$&$OSp(3|2)$&$SU(2)$ \\ \hline
$\mathcal{N}=4$&$SU(1,1|2)$&$SU(2)$ \\ 
 &$D(2,1;\alpha), \alpha\neq -1,0,
$&$SU(2)\times SU(2)$ \\ \hline
$\mathcal{N}=5$&$OSp(5|2)$&$SO(5)$ \\ \hline
$\mathcal{N}=6$&$SU(1,1|3)$&$SU(3)\times U(1)$ \\
 &$OSp(6|2)$&$SO(6)$ \\ \hline
$\mathcal{N}=7$&$OSp(7|2)$&$SO(7)$ \\
 &$G(3)$&$G_{2}$ \\ \hline
$\mathcal{N}=8$&$OSp(8|2)$&$SO(8)$ \\
 &$SU(1,1|4)$&$SU(4)\times U(1)$ \\
 &$OSp(4^{*}|4)$&$SU(2)\times SO(5)$ \\
 &$F(4)$&$SO(7)$ \\ \hline
$\mathcal{N}>8$&$OSp(\mathcal{N}|2)$&$SO(\mathcal{N})$ \\
 &$SU(1,1|\frac{\mathcal{N}}{2})$&$SU(\frac{\mathcal{N}}{2})\times U(1)$ \\
 &$OSp(4^{*}|\frac{\mathcal{N}}{2})$&$SU(2)\times Sp(\frac{\mathcal{N}}{2})$ \\ \hline
\end{tabular}
\caption{The simple classical Lie supergroups that contain 
the one-dimensional conformal group $SL(2,\mathbb{R})$ 
as a factored bosonic subgroup. 
For $\mathcal{N}>8$ superconformal quantum mechanics 
there are three different superconformal groups.}
\label{listsup1}
\end{center}
\end{table}

In the cases of $\mathcal{N}<4$ supersymmetry 
the superconformal groups are essentially unique series of $OSp(2|\mathcal{N})$ 
as the isomorphism $SU(1,1|1)\cong OSp(2|2)$ is taken into account. 

For $\mathcal{N}=4$ supersymmetry 
the structure of the superconformal group becomes large 
as the exceptional Lie superalgebra $D(2,1;\alpha)$ 
is a one-parameter family. 
Note that $SU(1,1|2)$ for $\mathcal{N}=4$ case is not simple 
as $SU(m,n|m+n)$ is not even semi-simple. 
The quotient $PSU(1,1|2)\cong SU(1,1|2)/U(1)$ 
is simple and we denote it just by $SU(1,1|2)$. 
As $D(2,1;-1)$ and $D(2,1;0)$ are semi-direct product 
$SU(1,1|2)\rtimes SU(2)$ and they are not simple, 
they are excluded in the Table \ref{listsup1}. 

With $\mathcal{N}=8$ supersymmetry 
one-dimensional superconformal groups 
can be realizes as four different supergroups; 
$OSp(8|2)$, $SU(1,1|4)$, $OSp(4^{*}|4)$ and $F(4)$.

When the highly extended supersymmetry with $\mathcal{N}>8$ 
exists in the quantum mechanics, 
one can have three distinct series of one-dimensional superconformal
groups for even $\mathcal{N}$
; $OSp(\mathcal{N}|2)$, $SU(1,1|\frac{\mathcal{N}}{2})$ 
and $OSp(4^{*}|\frac{\mathcal{N}}{2})$. 
The supergroup $OSp(4^{*}|\frac{\mathcal{N}}{2})$ is the exceptional 
series which does not appear in the theories with fewer supersymmetries. 
It has an even part $SO^{*}(4)\times USp(\mathcal{N})$ 
where the non-compact bosonic subgroup 
$SO^{*}(4)\cong SL(2,\mathbb{R})\times SU(2)$ 
contain the one-dimensional conformal group $SL(2,\mathbb{R})$.

\section{One-dimensional supersymmetry}
\label{1dsusysec001}
Now we want to discuss the concrete construction of superconformal quantum
mechanical models. 
To this end we should note that 
in one dimension the supersymmetry is realized 
containing various peculiarities 
which do not appear in higher dimensional cases 
regardless of whether a conformal symmetry exists or not. 
It is known that 
the supersymmetry of a sigma-model 
imposes strong restrictions on its target space. 
However, the restrictions of 
one-dimensional supersymmetric sigma-models 
are generically weaker than 
higher dimensional sigma-models. 
In other words, 
more couplings among the fields are allowed in one dimension. 
This is because 
in higher dimensional cases 
the Lorentz symmetry rules out particular couplings, 
however,  
in one dimension there is no Lorentz symmetry group 
and much more couplings are possible. 
Moreover we cannot expect the relation 
between the number of bosonic and fermionic fields 
as in higher dimensional supersymmetric field theories.

\subsection{Supermultiplet}
\label{1dsusysec001a}
One of the most powerful methods to 
construct supersymmetric quantum mechanics is to appeal the 
superspace and superfield formalism. 
In what follows we will consider a particularly reasonable class 
of supermultiplets 
\cite{Gates:1995ch,Gates:1995pw,Gates:2002bc} 
and discuss how many components we need to realize the 
$\mathcal{N}$-extended superalgebra 
\footnote{Also see \cite{deWit:1992up,Pashnev:2000ij,Kuznetsova:2005cd} 
for the classification of the supermultiplets.} 
which satisfy 
\begin{align}
\label{smlt1}
[\delta_{\epsilon^{A}},\delta_{\epsilon^{B}}]
=-2i\epsilon^{A}\epsilon^{B}\partial_{t}
\end{align}
where $A,B,\cdots=1,\cdots,\mathcal{N}$ denote the R-symmetry indices 
and $\epsilon^{A}$ are a set of real anti-commuting supersymmetry
parameters. 

Now consider the scalar multiplets $\Phi$ 
which consist of a set of $d$ physical bosons $x_{i}(t)$ 
and a set of $d$ fermions $\psi_{\hat{i}}(t)$ 
where $i=1,\cdots,d$ and $\hat{i}=1,\cdots,d$ 
denote the multiplicities, 
i.e. the numbers of the bosons and the fermions respectively 
and suppose that their supersymmetric transformations 
are given by
\begin{align}
\label{smlt2}
\delta_{\epsilon^{A}}x_{i}
&=-i\epsilon^{A}{(L_{A})_{i}}^{\hat{j}}\psi_{\hat{j}},\\
\label{smlt3}
\delta_{\epsilon^{A}}\psi_{\hat{i}}
&=\epsilon^{A}{(R_{A})_{\hat{i}}}^{j}\dot{x}_{j}
\end{align}
where ${(L_{A})_{i}}^{\hat{j}}$ 
and ${(R_{A})_{\hat{i}}}^{j}$ 
are real $d\times d$ matrices. 
Then the algebra (\ref{smlt1}) imposes constraints 
on the matrices $L_{A}$ and $R_{A}$ as
\begin{align}
\label{smlt4}
{\left(L_{A}R_{B}+L_{B}R_{A}\right)_{i}}^{j}
&=-2\delta_{AB}\delta_{j}^{j},\\
\label{smlt5}
{\left(R_{A}L_{B}+R_{B}L_{A}
\right)_{\hat{i}}}^{\hat{j}}
&=-2\delta_{AB}\delta_{\hat{i}}^{\hat{j}}.
\end{align}
From the algebraic point of view there is no relationship 
between two matrices $L_{A}$ and $R_{A}$, 
however, if we require that the kinetic action for 
the scalar multiplet $\Phi$ with the form
\begin{align}
\label{smlt6}
S=\int dt 
\left[
\frac12 \dot{x}_{i}^{2}
-\frac{i}{2}\psi_{\hat{t}}\dot{\psi}_{\hat{i}}
\right]
\end{align}
is invarinat under the supersymmetric transformations 
(\ref{smlt1}), 
we obtain the relation
\begin{align}
\label{smlt7}
(L_{A}^{T})^{\hat{i}j}=-{(R_{A})}^{\hat{i}j}.
\end{align}

Likewise let us consider the spinor multiplets $\Psi$ 
which are composed of a set of $d$ real fermions $\lambda_{\hat{i}}$ 
and a set of $d$ real bosons $y_{i}$ 
possess the supersymmetry transformations 
\begin{align}
\label{smlt8a}
\delta_{\epsilon^{A}}\lambda_{\hat{i}}
&=\epsilon^{A}{(R_{A})_{\hat{i}}}^{j}y_{j},\\
\label{smlt8b}
\delta_{\epsilon^{A}}y_{i}
&=-i\epsilon^{A}{(L_{A})_{i}}^{\hat{j}}\dot{\lambda}_{\hat{j}}.
\end{align}
Then one finds the same constraints for the 
two matrices $L_{A}$ and $R_{A}$ as (\ref{smlt4}) and (\ref{smlt5}). 
In addition if we require that the quadratic part of the action 
for the spinor multiplet $\Psi$
\begin{align}
\label{smlt9}
S=\int dt \left[
-\frac{i}{2}
\lambda_{\hat{i}}\dot{\lambda}_{\hat{i}}
+\frac12 y_{i}y_{i}
\right]
\end{align}
is invariant under the supersymmetry transformations 
(\ref{smlt8a}) and (\ref{smlt8b}), 
then we find the precisely same relation as (\ref{smlt7}).

Hence the existence of the scalar supermultiplets $\Phi$ 
and $\Psi$ is rooted in the algebra 
\footnote{In \cite{Gates:1995pw,Gates:2002bc,Faux:2004wb} this algebra 
of dimension $d$ and rank $\mathcal{N}$ is called 
$\mathcal{GR}(d,\mathcal{N})$ algebra since the one of the two matrices, 
say $L_{A}$ satisfies a general real ($\mathcal{GR}$) Pauli algebra 
(\ref{smlt4}), (\ref{smlt5}) 
with the other matrix $R_{A}$ determined by the relation (\ref{smlt7}).} 
defined by three conditions 
(\ref{smlt4}), (\ref{smlt5}) and (\ref{smlt7}). 
It is known that 
there is a minimal value of $d$, called $d_{\mathcal{N}}$ 
for which $\mathcal{N}$ linearly independent real $d\times d$ 
matrices $L_{A}$ and $R_{A}$ 
satisfying the relations (\ref{smlt4}), (\ref{smlt5}) and (\ref{smlt7}) exist. 
We see that $d_{\mathcal{N}}$ translates into the minimal number of 
the bosonic or fermionic component fields in the supermultiplets 
for a given the number of supersymmetry $\mathcal{N}$. 
The value of $d_{\mathcal{N}}$ is given by 
\cite{Pashnev:2000ij,Gates:2002bc}
\begin{align}
\label{smlt10a}
d_{\mathcal{N}}=16^{m}\rho(2^{r})
\end{align}
where the number of supersymmetry is 
written as a $\mathrm{mod}8$ decomposition 
\begin{align}
\label{smlt10b}
\mathcal{N}=8m+n.
\end{align}
Here $\rho(2^{r})$ is the so-called Hurwitz-Radon function 
\cite{MR1512117,MR3069384} 
 define by 
\footnote{The Hurwitz-Radon function $\rho(2^{r})$ 
yields the largest integer $\rho$ for
which the square identities can 
\begin{align}
\left(a_{1}^{2}+\cdots +a_{\rho}^{2}\right)
\left(b_{1}^{2}+\cdots +b_{2^{r}}^{2}\right)
=c_{1}^{2}+\cdots c_{2^{r}}^{2}
\end{align}
hold where $a_{1},\cdots, a_{\rho}$ 
and $b_{1},\cdots, b_{2^{r}}$ are the 
independent indeterminates and 
$c_{i}$ is a bilinear form in $a_{1},\cdots, a_{\rho}$ and 
$b_{1},\cdots, b_{2^{r}}$. 
The Hurwitz-Radon function also appears 
in topology \cite{MR0139178} and linear algebra \cite{MR0201460}. 
See also \cite{MR1253071,MR1786291,MR2811573}.}
\begin{align}
\label{smlt10c}
\rho(2^{r})
=\begin{cases}
2r+1&n\equiv 0 \textrm{\ \ mod$4$}\cr
2r&n\equiv 1,2 \textrm{\ \ mod$4$}\cr
2r+2&n\equiv 3 \textrm{\ \ mod$4$}.\cr
\end{cases}
\end{align}
with $r$ being taken as 
the nearest integer greater than or equal to $\log_{2}n$ 
(see Table \ref{rhfct1}). 
\begin{table}
\begin{center}
\begin{tabular}{|c|c|c|c|c|c|c|c|c|} \hline
$n$&$1$&$2$&$3$&$4$&$5$&$6$&$7$&$8$ \\ \hline\hline
$\log_{2}n$&$0$&$1$&$\log_{2}3$&$2$&$\log_{2}5$&$\log_{2}6$&$\log_{2}7$&$3$ 
\\ \hline
$r$&$0$&$1$&$2$&$2$&$3$&$3$&$3$&$3$ \\ \hline 
$\rho(2^{r})$&$1$&$2$&$4$&$4$&$8$&$8$&$8$&$8$ \\ \hline
\end{tabular}
\caption{Hurwitz-Radon function $\rho(2^{r})$ 
where $r$ is the nearest integer greater than or equal to $\log_{2}n$.}
\label{rhfct1}
\end{center}
\end{table}
The results are summarized in Table \ref{susynumber001}
\begin{table}
\begin{center}
\begin{tabular}{|c|c|c|c|c|c|c|c|c|c|c|c|c|c|c|c|c|c|} \hline
$\mathcal{N}$&$1$&$2$&$3$&$4$&$5$&$6$&$7$&$8$
&$9$&$10$&$11$&$12$&$13$&$14$&$15$&$16$ \\ \hline\hline
$d_{\mathcal{N}}$&$1$&$2$&$4$&$4$&$8$&$8$&$8$&$8$
&$16$&$32$&$64$&$64$&$128$&$128$&$128$&$128$ \\ \hline
\end{tabular}
\caption{Then minimal numbers $d_{\mathcal{N}}$ of the component fields 
in the $\mathcal{N}$-extended supermultiplets.}
\label{susynumber001}
\end{center}
\end{table}
From Table \ref{susynumber001} 
one can see that 
when $\mathcal{N}=1,2,4,8$ 
the minimal numbers $d_{\mathcal{N}}$ of the component fields 
coincide with the numbers $\mathcal{N}$ of supersymmetries. 
As we will see in the following, 
the superspace and superfield formalism works well for these four
cases. 
Note that when $\mathcal{N}>8$ 
the minimal numbers $d_{\mathcal{N}}$ of the supermultiplets 
are greater than the numbers $\mathcal{N}$ of supersymmetries 
and the corresponding supermultiplets become much more complicated 
and the superspace and superfield formalism is unsuccessful at present.

\subsection{Automorphic duality}
\label{1dsusy001b}
One of the most significant features in one-dimensional 
supersymmetric field theories, i.e. quantum mechanical models 
is the fact that the the equal number of bosonic and fermionic 
physical degrees of freedom, 
which is valid in higher dimensional field theories, 
does not take place. 
This is because in one dimension 
there is the duality which allows us 
to convert any physical field to auxiliary field 
and vice versa \cite{Gates:1995pw,Gates:2002bc,Faux:2004wb}. 
Consequently even if we consider 
the $\mathcal{N}=1,2,4,8$ supersymmetric cases, 
where $d_{\mathcal{N}}=\mathcal{N}$ is realized, 
a number of supermultiplets can be constructed in one-dimension.

To see this let us take the most 
basic $d=1$ $\mathcal{N}=1$ superalgebra 
\begin{align}
\label{ad1a}
[\delta_{\epsilon_{1}},\delta_{\epsilon_{2}}]
=-2i\epsilon_{1}\epsilon_{2}\partial_{t}.
\end{align}
We introduce $\mathcal{N}=1$ superspace $\mathbb{R}^{(1|1)}$ 
parametrized by 
\begin{align}
\label{n1susy1}
\mathbb{R}^{(1|1)}=(t,\theta)
\end{align}
where $t$ is time and $\theta$ is a real Grassmann coordinate. 
The covariant superderivative $D$ is defined by 
\footnote{This convention yields $\{Q,Q\}=2H$ and leads to simple forms of 
the supersymmetric Lagrangian and its supersymmetric transformation.}
\begin{align}
\label{n1susy1a}
D&=i\frac{\partial}{\partial\theta}-\theta\frac{\partial}{\partial t},&
\left\{D,D\right\}&=-2i\partial_{t}
\end{align}
and the supercharge $Q$ 
is realized as
\begin{align}
\label{n1susy1b}
Q&=i\frac{\partial}{\partial \theta}
+\theta \frac{\partial}{\partial t},& 
\left\{Q,Q\right\}&=2i\partial_{t}
\end{align}
in the superspace. 

In this case there are two irreducible representations of 
(\ref{ad1a}); the scalar multiplet $\Phi$ and 
the spinor multiplet $\Psi$. 
The scalar multiplet contains a real bosonic field $x$ 
as the lowest component and a real fermion $\psi$ as the highest
component while the spinor multiplet $\Psi$ includes a real fermion $\lambda$ 
as the lowest component and a real boson $y$ as the highest component. 
Namely the multiplets can be described by 
\begin{align}
\label{ad1d}
\Phi&=x+i\theta\psi,\\
\Psi&=\lambda+\theta y.
\end{align}

The supersymmetry transformation laws 
for the scalar multiplet $\Phi$ are 
$\delta\Phi=-i[\epsilon Q,\Phi]$, which yield
\begin{align}
\label{ad1b}
\delta_{\epsilon}x&=i\epsilon\psi,\\
\delta_{\epsilon}\psi&=\epsilon\dot{x}
\end{align}
and those for the spinor multiplet $\Psi$ are 
$\delta\Psi=-i[\epsilon Q,\Psi]$, which give rise to 
\begin{align}
\label{ad1c}
\delta_{\epsilon}\lambda&=\epsilon y,\\
\delta_{\epsilon}y&=i\epsilon\dot{\lambda}.
\end{align}
One can write the supersymmetric action for the scalar multiplet $\Phi$ as
\begin{align}
\label{ad1d}
S=-\frac12 \int dt d\theta\ 
D\Phi \dot{\Phi}
\end{align}
and also write that for the spinor multiplet $\Psi$ as
\begin{align}
\label{ad1e}
S=-\frac{i}{2}\int dt d\theta 
\Psi D\Psi.
\end{align}
In component fields the above supersymmetric action 
(\ref{ad1d}) and (\ref{ad1e}) 
can be expressed by 
\begin{align}
\label{ad1f}
S&=\frac12 \int dt\ \left[
\dot{x}^{2}+i\dot{\psi}\psi
\right]
\end{align}
and 
\begin{align}
\label{ad1g}
S&=\frac12 \int dt\ \left[
i\dot{\lambda}\lambda +y^{2}
\right]
\end{align}
respectively. 

As described in \cite{Gates:2002bc}, 
there is a useful operation which maps between the two irreducible 
$\mathcal{N}=1$ multiplets 
\begin{align}
\label{ad2a}
-D\Phi\leftrightarrow \Psi.
\end{align}
In component fields this map is realized by 
performing the following replacements
\begin{align}
\label{ad2b}
(\dot{x},\psi)\leftrightarrow (y,\lambda).
\end{align}
We see that 
the supersymmetry transformations (\ref{ad1b}) 
for the scalar multiplet 
and the transformations (\ref{ad1c}) for the spinor multiplet are 
exchanged under the replacement (\ref{ad2a}) 
and that 
the action (\ref{ad1f}) for the scalar multiplet 
and the action (\ref{ad1g}) for the spinor multiplet 
transform into the other under the operation (\ref{ad2a}). 
Therefore a map (\ref{ad2a}) or (\ref{ad2b}) 
is the operation which replace a 
scalar multiplet $\Phi$ with 
a spinor multiplet $\Psi$ and vice-verse. 
This is called the automorphic duality (AD) map 
because the operation corresponds to 
the automorphism on the space of the representations 
of the superalgebra. 
Intriguingly the AD map (\ref{ad2b}) 
make it possible to 
convert the physical field $x$ into 
the auxiliary field $y$ and vice versa. 
It has been pointed out \cite{Faux:2004wb} that 
this remarkable property in quantum mechanics 
can be interpreted as the Hodge duality in one-dimension. 
In general the Hodge duality maps 
a differential $p$-form $\Omega_{p}$ in $d$-dimension 
into a differential $(d-p-2)$-form $\Omega_{d-p-2}$ in $d$-dimension 
by the Hodge star operation as 
\begin{align}
\label{ad2c}
*:d\Omega_{p}\rightarrow 
d\Omega_{d-p-2}.
\end{align}
If we consider a scalar field, a zero-form  in one dimension, 
then the Hodge duality (\ref{ad2c}) gives rise to a dual $(-1)$-form. 
Formally the exterior derivative of a $0$-form or a scalar $x$ 
is a $(-1)$-form.
Therefore if we denote the component field of the
$(-1)$-form by $y$, we then get the relation 
\begin{align}
\label{ad2d}
\dot{x}=y.
\end{align}
This is just the AD map given in (\ref{ad2b}). 

According to the existence of the AD map in quantum mechanics, 
we will use the notation $(\bm{n},\bm{\mathcal{N}},\bm{\mathcal{N}-n})$ 
for $\mathcal{N}=1,2,4,8$ supermultiplets. 
Here the first entry denoted by $\bm{n}$ 
is the number of physical bosons in the supermultiplet, 
the second number $\bm{\mathcal{N}}$ represents the number of fermions 
which is equal to the number of supersymmetry 
and the last one $\bm{\mathcal{N}-n}$ is the number of bosonic auxiliary
fields. 
Using this notation, 
the $\mathcal{N}=1$ scalar multiplet $\Phi$ is 
$(\bm{1},\bm{1},\bm{0})$ and 
the spinor multiplet $\Psi$ is 
$(\bm{0},\bm{1},\bm{1})$.

\section{$\mathcal{N}=1$ Superconformal mechanics}
\label{secscqm1}
\subsection{One particle free action}
Consider the $\mathcal{N}=1$ $n$ 
particle quantum mechanical system which 
is described by the $n$-dimensional 
scalar superfield $(\bm{1},\bm{1},\bm{0})$. 
In general the $\mathcal{N}=1$ superfield can be thought of 
as a map from the superspace $\mathbb{R}^{(1|1)}$ 
to the target space $\mathcal{M}$. 
In terms of component fields we can write the multiplet as
\begin{align}
\label{n1susy1a1}
\Phi^{i}(t,\theta)=x^{i}(t)+i\theta\psi^{i}(t)
\end{align}
where $i,j,\cdots=1,\cdots,n$. 
Also consider the $(\bm{0},\bm{1},\bm{1})$ spinor superfield $\Psi^{a}$ 
which is a section of the bundle on $\mathcal{M}$ 
with rank $k$ given by 
\begin{align}
\label{n1susy1b1}
\Psi^{a}(t,\theta)=\lambda^{a}(t)+\theta y^{a}(t)
\end{align}
where $a,b,\cdots=1,\cdots,k$. 
We attach the mass dimension as the following:
\begin{align}
\label{n1susy1c}
[t]&=-1,&
[\theta]&=-\frac12,\nonumber\\
[\Phi]&=0,&
[\Psi]&=\frac12,\nonumber\\
[D]&=\frac12,&
[\partial_{t}]&=1.
\end{align}
Then the most general $\mathcal{N}=1$ action 
with dimensionless couplings 
up to cubic terms is given by 
\footnote{See also \cite{Coles:1990hr,Maloney:1999dv} 
for the $\mathcal{N}=1$ superfield action.}
\begin{align}
\label{n1susy1d}
S=\int dt d\theta 
&\Biggl[
-\frac12 g_{ij}D\Phi^{i}\dot{\Phi}^{j}
+\frac{i}{3!}c_{ijk}D\Phi^{i}D\Phi^{j}D\Phi^{k}\nonumber\\
&-\frac{i}{2}h_{ab}\Psi^{a}\nabla \Psi^{b}
+\frac{1}{3!}l_{abc}\Psi^{a}\Psi^{b}\Psi^{c}
+f_{ia}\dot{\Phi}^{i}\Psi^{a}\nonumber\\
&+\frac{i}{2}m_{iab}\Psi^{a}\Psi^{b}D\Phi^{i}
+\frac{i}{2}n_{ija}D\Phi^{i}D\Phi^{j}\Psi^{a}
\Biggr]
\end{align}
where $g_{ij}$ is a metric on $\mathcal{M}$ 
and $h_{ab}$ is a fibre metric on the bundle. 
The covariant derivative for the fermions are defined by 
\begin{align}
\label{n1susy1e}
\nabla\Psi^{a}
=D\Psi^{a}+D\Phi^{i}{(A_{i})^{a}}_{b}\Psi^{b}
\end{align}
with ${(A_{i})^{a}}_{b}$ being the connection on the bundle.

Note that for the one particle case 
where the corresponding target space $\mathcal{M}=\mathbb{R}$ has 
no non-trivial bundle over it, 
the $\mathcal{N}=1$ superspace action 
is described by just a free action (\ref{ad1f}). 
This corresponds to the statement that 
it is not possible to construct 
one-particle $OSp(1|2)$ superconformal quantum mechanics 
with inverse-square type potential 
\cite{Claus:1998ts,deAzcarraga:1998ni,Fedoruk:2011aa}.

\subsection{Multi-particle model}
Let us focus on  
the sigma-model action constructed only from  
the $(\bm{1},\bm{1},\bm{0})$ scalar supermultiplet $\Phi^{i}$ 
\cite{Coles:1990hr,Gibbons:1997iy,Michelson:1999zf,BrittoPacumio:1999ax} 
\footnote{The $(\bm{1},\bm{1},\bm{0})$ supermultiplet 
is also called $\mathcal{N}=1B$ superfield.}
\begin{align}
\label{n1susy2a1}
S&=\int dtd\theta \left[
-\frac12 g_{ij}D\Phi^{i}\dot{\Phi}^{j}
+\frac{i}{3!}c_{ijk}D\Phi^{i}D\Phi^{j}D\Phi^{k}
\right]\nonumber\\
&=\int 
dt \left[
\frac12 g_{ij}\dot{x}^{i}\dot{x}^{j}
+\frac{i}{2}\psi^{i}
\left(
g_{ij}\frac{D\psi^{j}}{dt}
-\dot{x}^{k}c_{ijk}\psi^{j}
\right)
-\frac16 \partial_{l}c_{ijk}\psi^{l}\psi^{i}\psi^{j}\psi^{k}
\right]
\end{align}
where the covariant derivative is defined as
\begin{align}
\label{n1susy2a2}
\frac{D\psi^{i}}{dt}&:=
\dot{\psi}^{i}+\dot{x}^{j}\Gamma^{i}_{jk}\psi^{k}
\end{align}
with $\Gamma^{i}_{jk}$ being the Christoffel symbol on $\mathcal{M}$. 

Instead of the space-time indices $i$ for the fermions $\psi^{i}$ 
we shall introduce the tangent space indices $\alpha=1,\cdots, n$ 
by redefining the fermions $\psi^{\alpha}$ as
\begin{align}
\label{n1susy2a2a}
\psi^{i}={e^{i}}_{\alpha}\psi^{\alpha}.
\end{align}
Note that $\psi^{\alpha}$ commute with $x^{i}$ and $p^{i}$ 
while $\psi^{i}$ does not commute with $x^{i}$ and $p^{i}$ 
\begin{align}
\label{n1susy2a2a1}
[p_{i},\lambda^{j}]=-i\left(
{\omega_{i}}^{j}_{k}-{\Gamma_{ik}^{j}}
\right)\psi^{k}.
\end{align}
Then the action (\ref{n1susy2a1}) can be written as
\begin{align}
\label{n1susy2a2b}
S=\int dt 
\Biggl[
&\frac12 g_{ij}\dot{x}^{i}\dot{x}^{j}
+\frac{i}{2}
\left(
\delta_{\alpha\beta}\psi^{\alpha}\dot{\psi}^{\beta}
+\dot{x}^{i}\omega_{i\alpha\beta}\psi^{\alpha}\psi^{\beta}
\right)
\nonumber\\
&-\frac{i}{2}\dot{x}^{i}c_{i\beta\gamma}\psi^{\alpha}\psi^{\beta}
-\frac16 {e^{l}}_{\delta}\partial_{l}
c_{ijk}{e^{i}}_{\alpha}{e^{j}}_{\beta}{e^{k}}_{\gamma}
\psi^{\delta}\psi^{\alpha}\psi^{\beta}\psi^{\gamma}
\Biggr]
\end{align}
where $\omega$ is the spin connection 
and $c_{i\alpha\beta}:=c_{ijk}{e^{j}}_{\alpha}{e^{k}}_{\beta}$. 
From the fermionic kinetic terms in the action (\ref{n1susy2a2b}) 
we see that the covariant derivatives of the fermions contains 
the connection with torsion $c$. 
Although this is similar to the two-dimensional $(1,0)$ supersymmetric 
sigma models \cite{Howe:1988cj}, the torsion $c$ here is not necessarily closed 
as opposed to two-dimensional case. 
This indicates that there exist new supermultiplets in one dimension 
which have no higher-dimensional ancestors. 
The canonical momenta $p_{i}$ is expressed as
\begin{align}
\label{n1susy2a3}
p_{i}&=g_{ij}\dot{x}^{j}
+\frac{i}{2}\left(\omega_{ijk}-c_{ijk}\right)\psi^{j}\psi^{k}
\end{align}
where 
$\omega_{ijk}:={\omega_{i}}^{\beta}_{\gamma}{e_{j\beta}}{e_{k}}^{\gamma}$. 
The action (\ref{n1susy2a1}) 
is invariant under the supersymmetry transformations
\begin{align}
\label{n1susy2a4}
\delta x^{i}&=-i\epsilon \psi^{i},\\
\label{n1susy2a5}
\delta\psi^{i}&=\epsilon\dot{x}^{i}.
\end{align}
By means of the Noether's method we find the supercharge
\begin{align}
\label{n1susy2a6}
Q=\psi^{i}\Pi_{i}-\frac{i}{3}c_{ijk}\psi^{i}\psi^{j}\psi^{k}
\end{align}
where we have defined
\begin{align}
\label{n1susy2a7}
\Pi_{i}&=g_{ij}\dot{x}^{j}.
\end{align}
Note that the supercharge $Q$ is Hermitian 
though $\Pi_{i}$ is not Hermitian. 
Using the canonical relation for the fermions
\begin{align}
\label{n1susy2a8}
\left\{
\psi^{\alpha},\psi^{\beta}
\right\}=\delta^{\alpha\beta}
\end{align}
and the relations (\ref{mul1a2a}) for bosons, 
one finds 
\begin{align}
\label{n1susy2a9}
\left\{Q,Q\right\}&=2H.
\end{align}
where the Hamiltonian is
\begin{align}
\label{n1susy2b1}
H&=\frac12 p_{a}^{\dag}g^{ab}p_{b},
\end{align}
which agrees with the bosonic sigma-model Hamiltonian (\ref{mul1a1}) with 
the bosonic potential $V(x)$ vanishing. 

At this stage we consider 
the condition so that 
the theory (\ref{n1susy2a2b}) is 
the $OSp(1|2)$ superconformal quantum mechanics. 
The corresponding $\mathfrak{osp}(1|2)$ superalgebra is 
characterized by the following (anti)commutation relations:
\begin{align}
\label{n1susy2c1}
[H,D]&=iH,&
[K,D]&=-iD,&
[H,K]&=2iD,\\
\label{n1susy2c2}
[Q,H]&=0,&
[Q,D]&=-\frac{i}{2}Q,&
[Q,K]&=-iS,\\
\label{n1susy2c3}
[S,H]&=iQ,&
[S,D]&=\frac{i}{2}S,&
[S,K]&=0,\\
\label{n1susy2c4}
\left\{Q,Q\right\}&=2H,&
\left\{Q,S\right\}&=-2D,&
\left\{S,S\right\}&=2K.
\end{align}
From the commutation relation (\ref{n1susy2c2}) 
and the expressions (\ref{n1susy2a6}) and (\ref{mul1b3}) 
we can read the superconformal charge $S$
\begin{align}
\label{n1susy2d1}
S&=\psi^{i}D_{i}
\end{align}
where $D_{i}$ has been introduced in (\ref{mul1a4}) 
as the generalized dilatation. 
From the anti-commutator (\ref{n1susy2c4}) 
we obtain the modified dilatation generator
\begin{align}
\label{n1susy2d2}
D=\frac14 \left(D^{i}\Pi_{i}+\Pi^{\dag}D_{i}^{\dag}\right)
\end{align}
with $p_{i}$ being replaced with $\Pi_{i}$. 
Using the new dilatation generator (\ref{n1susy2d2}), 
$[S,D]=\frac{i}{2}S$ is satisfied, 
however, $[Q,D]$ yields
\begin{align}
\label{n1susy2d3}
[Q,D]&=-\frac{i}{2}Q
-\frac{i}{2}c_{ijk}D^{i}\psi^{j}p^{k}+\mathcal{O}(\psi^{3}).
\end{align}
Thus the $OSp(1|2)$ superconformal symmetry 
imposes the condition so that the second quadratic term in $\psi$ must
vanish
\begin{align}
\label{n1susy2d4}
D^{i}c_{ijk}&=0,
\end{align}
which means that $c$ is orthogonal to $D$. 
With the constraint (\ref{n1susy2d4}), 
the commutator (\ref{n1susy2d3}) becomes
\begin{align}
\label{n1susy2d5}
[Q,D]&=-\frac{i}{2}Q
-\frac{1}{12}\psi^{i}\psi^{j}\psi^{k}
\left(\mathcal{L}_{D}-2\right)c_{ijk},
\end{align}
which implies that 
\begin{align}
\label{n1susy2d6}
\mathcal{L}_{D}c_{ijk}&=2c_{ijk}.
\end{align}
Then one can check that 
the remaining (anti)commutation relations 
(\ref{n1susy2c1})-(\ref{n1susy2c4}) are satisfied 
and there are no further constraints for the $OSp(1|2)$ symmetry 
imposed on the target space $\mathcal{M}$.

Therefore 
the conditions so that 
the $\mathcal{N}=1$ sigma-model action (\ref{n1susy2a2b}) 
realizes the $OSp(1|2)$ superconformal quantum mechanics 
are the conformal condition (\ref{mul1a8}), (\ref{mul1b3a})
and the additional two constraints on the torsion $c$
\begin{align}
\label{n1susy2d7}
D^{i}c_{ijk}&=0,\\
\label{n1susy2d8}
\mathcal{L}_{D}c_{ijk}&=2c_{ijk}.
\end{align}

\subsection{Gauged superconformal mechanics}
As a generalization 
of the gauged mechanics (\ref{confg2}) for the DFF-model 
and the gauged matrix model (\ref{calogero1a1}) for the Calogero model, 
we will discuss the superextension of the 
$\mathcal{N}=1$ supersymmetric gauged mechanical model. 
As we will see this gauging procedure allows for 
the explicit construction of the 
non-trivial $\mathcal{N}=1$ superconformal quantum mechanics 
\cite{Fedoruk:2008hk,Fedoruk:2011aa}. 
Consider the matrix superfield gauged mechanics action 
\begin{align}
\label{n1susy2e1}
S=-i\int dt d\theta 
\left[
\mathrm{Tr}\left(
\nabla_{t}\mathcal{X}\mathcal{D}\mathcal{X}
\right)
+\frac{i}{2}\left(
\overline{\mathcal{Z}}\mathcal{D}\mathcal{Z}
-\mathcal{D}\overline{\mathcal{Z}}\mathcal{Z}
\right)
+c\mathrm{Tr}\mathcal{A}
\right].
\end{align}
Here we have introduced 
\begin{itemize}
 \item 
the $\mathcal{N}=1$ Grassmann-even Hermitian $n\times n$ matrix superfield 
$\mathcal{X}_{a}^{b}(t,\theta)$ 
which satisfies $(\mathcal{X})^{\dag}=\mathcal{X}$ and 
transforms as the adjoint representation of $U(n)$
\item the $\mathcal{N}=1$ Grassmann-even complex superfield 
$\mathcal{Z}_{a}(t,\theta)$
which satisfies  $\overline{\mathcal{Z}}
=\mathcal{Z}^{\dag}$ and 
transform as the fundamental representation of $U(n)$
\item the $\mathcal{N}=1$ Grassmann-odd anti-Hermitian $n\times n$
      matrix superfield $\mathcal{A}_{a}^{b}(t,\theta)$ 
which satisfies $(\mathcal{A})^{\dag}=-\mathcal{A}$ 
and transforms as the adjoint representation of $U(n)$.
\end{itemize}

The covariant derivatives are defined by 
\begin{align}
\label{n1susy2e2a}
\nabla_{t}\mathcal{X}&=D\mathcal{X}+i[\mathcal{A}_{t},\mathcal{X}],\\
\label{n1susy2e2b}
\mathcal{D}\mathcal{X}&=D\mathcal{X}+i[\mathcal{A},\mathcal{X}],\\
\label{n1susy2e2c}
\mathcal{D}\mathcal{Z}&=D\mathcal{Z}+i\mathcal{A}\mathcal{Z}
\end{align}
where \footnote{Note that the notation here is different from
(\ref{n1susy1a}).} 
\begin{align}
\label{n1susy2e2d}
D&=\frac{\partial}{\partial \theta}
+i\theta \frac{\partial}{\partial t},&
\left\{D,D\right\}&=2i\partial_{t},\\
\label{n1susy2e2e}
\mathcal{A}_{t}&=-iD\mathcal{A}-\mathcal{A}\mathcal{A}.
\end{align}

The superconformal boost transformations are found to be
\begin{align}
\label{n1susy2e2e0}
\delta t&=-i\eta \theta t,&
\delta \theta&=\eta t,\\
\label{n1susy2e2e1}
\delta (dtd\theta)&=-i\eta\theta (dtd\theta),&
\delta D&=i\eta \theta D,\\
\label{n1susy2e2e2}
\delta \mathcal{X}&=-i\eta \theta \mathcal{X},&
\delta \mathcal{A}&=i\eta\theta \mathcal{A},\\
\label{n1susy2e2e2}
\delta \mathcal{Z}&=0.
\end{align}

The action (\ref{n1susy2e1}) is invariant under the 
$U(n)$ gauge transformations 
\begin{align}
\label{n1susye2f}
\mathcal{X}&\rightarrow e^{i\Lambda}\mathcal{X}e^{-i\Lambda},\\
\label{n1susye2g}
\mathcal{Z}&\rightarrow e^{i\Lambda}\mathcal{Z},\\
\label{n1susye2h}
\mathcal{A}&\rightarrow e^{i\Lambda}\mathcal{A}e^{-i\Lambda}
-ie^{i\Lambda}\left(
De^{-i\Lambda}
\right),\\
\label{n1susye2i}
\mathcal{A}_{t}&\rightarrow 
e^{i\Lambda}\mathcal{A}_{t}e^{-i\Lambda}
-ie^{i\Lambda}\left(
\partial_{t}e^{-i\Lambda}
\right)
\end{align}
where $\Lambda_{a}^{b}(t,\theta)$ is the 
Hermitian $n\times n$ matrix gauge parameter. 
The $\mathcal{N}=1$ superfields $\mathcal{X}$, $\mathcal{Z}$ and
$\mathcal{A}$  can be expanded in the component fields as
\begin{align}
\label{n1susye3a}
\mathcal{X}_{a}^{b}&=x_{a}^{b}+i\theta\psi_{a}^{b},\\
\label{n1susye3b}
\mathcal{Z}_{a}&=z_{a}+\theta\xi_{a},\\
\label{n1susye3c}
\mathcal{A}_{a}^{b}&=i(\zeta_{a}^{b}+\theta A_{a}^{b}).
\end{align}
From the gauge transformation 
(\ref{n1susye2h}) we can fix the gauge so that 
\begin{align}
\label{n1susye3d}
\mathcal{A}_{a}^{b}&=i\theta(A_{0})_{a}^{b}(t).
\end{align}
Inserting (\ref{n1susye3d}) into the action (\ref{n1susy2e1}), 
performing the integration over $\theta$ 
and integrating out the auxiliary fields 
$\xi,\overline{\xi}$, 
we find the $\mathcal{N}=1$ gauged superconformal matrix model action
\begin{align}
\label{n1susye3e}
S=\int dt 
\left[
\mathrm{Tr}\left(DxDx\right)
-i\mathrm{Tr}\left(\psi D\psi\right)
+\frac{i}{2}\left(
\overline{z}Dz-D\overline{z}z
\right)+c\mathrm{Tr}A_{0}
\right]
\end{align}
where the covariant derivative is defined by
\begin{align}
\label{n1susye3f}
Dx&=\dot{x}+i[A_{0},x],&
D\psi&=\dot{\psi}+[A_{0},\psi].
\end{align}
Note that the action (\ref{n1susye3e}) is 
the supersymmetric generalization of (\ref{calogero1a1}) 
that describes the Calogero model. 
 
Instead of the gauge choice (\ref{n1susye3d}), 
we can fix the gauge as 
\begin{align}
\label{n1susye4a}
\mathcal{X}_{a}^{b}&=\mathcal{X}_{a}\delta_{a}^{b},\\
\label{n1susye4b}
\mathcal{Z}_{a}&=\overline{\mathcal{Z}}^{a}
\end{align}
as we have discussed in (\ref{calogero1a8}) 
and (\ref{calogero1b4}) for the bosonic gauged matrix model. 
In this gauge 
the theory contains $n^{2}$ real $\mathcal{N}=1$ superfields 
$\mathcal{A}_{a}^{b}$, $a\neq b$ and $\mathcal{X}_{a}$ 
while the superfields $\mathcal{Z}_{a}$ and 
$\mathcal{A}_{a}^{a}$ are auxiliary. 
The superfield action (\ref{n1susy2e1}) reads \cite{Fedoruk:2011aa}
\begin{align}
\label{n1susye4c}
S=-i\int dt d\theta 
\Biggl[
&\sum_{a}\dot{\mathcal{X}}_{a}D\mathcal{X}_{a}
+\frac{i}{2}\sum_{a}
\left(
\overline{Z}^{a}D\mathcal{Z}_{a}-D\overline{Z}^{a}\mathcal{Z}_{a}
\right)
-i\sum_{a,b}\left(
\mathcal{X}_{a}-\mathcal{X}_{b}
\right)^{2}D\mathcal{A}_{a}^{b}\mathcal{A}_{b}^{a}\nonumber\\
&-\sum_{a,b}\left(\mathcal{X}_{a}-\mathcal{X}_{b}\right)^{2}
\left(\mathcal{A}\mathcal{A}\right)_{a}^{b}\mathcal{A}_{b}^{a}
+\sum_{a,b}\overline{\mathcal{Z}}^{a}\mathcal{A}_{a}^{b}\mathcal{Z}_{b}
+c\sum_{a}\mathcal{A}_[a]^{a}
\Biggr].
\end{align}

For $n=1$, one particle case,  
the action (\ref{n1susye4c}) becomes free action
\begin{align}
\label{n1susye4d0}
S=-i\int dtd\theta \dot{\mathcal{X}}D\mathcal{X}
\end{align}
and the theory has no bosonic potential in the component action. 

In the case of $n=2$, that is two particles case,  
the action (\ref{n1susye4c}) is written as
\begin{align}
\label{n1susye4d}
S=-i\int dtd\theta 
\Biggl[
&\frac12 \dot{\mathcal{X}}_{+}D\mathcal{X}_{+}
-\frac12 \mathcal{A}_{-}D\mathcal{A}_{-}\nonumber\\
&+\frac12 \dot{\mathcal{X}}_{-}D\mathcal{X}_{-}
-\frac12 \mathcal{A}_{+}D\mathcal{A}_{+}
-c\epsilon_{1}\epsilon_{2}\frac{\mathcal{A}_{+}}{\mathcal{X}_{-}}
\Biggr]
\end{align}
where 
\begin{align}
\label{n1susye4e}
\mathcal{X}_{-}&:=\mathcal{X}_{1}-\mathcal{X}_{2},&
\mathcal{X}_{+}&:=\mathcal{X}_{1}+\mathcal{X}_{2},\\
\label{n1susye4f}
\mathcal{A}_{+}&:=\mathcal{X}(\mathcal{A}_{1}^{2}+\mathcal{A}_{2}^{1}),&
\mathcal{A}_{-}&:=i\mathcal{X}(\mathcal{A}_{1}^{2}-\mathcal{A}_{2}^{1})
\end{align}
and $\epsilon_{1}=\pm 1$, $\epsilon_{2}=\pm 1$ 
are the constans appearing in the constraint 
$\mathcal{Z}_{1}\mathcal{Z}_{2}=-\frac{c\epsilon_{1}\epsilon_{2}}{2}$. 
Note that the superfield action 
(\ref{n1susye4d}) is a sum of 
two free $\mathcal{N}=1$ supermultiplets
$(\mathcal{X}_{+},\mathcal{A}_{-})$ 
and two interacting $\mathcal{N}=1$ supermultiplets 
$(\mathcal{X}_{-},\mathcal{A}_{+})$. 
It has been argued that 
the superfield action (\ref{n1susye4d}) 
is the $\mathcal{N}=1$ superfield form of the 
off-shell $\mathcal{N}=2$ superconformal mechanics 
based on the supermultiplet $(\bm{1},\bm{2},\bm{1})$ 
\cite{Fedoruk:2008hk,Fedoruk:2011aa}.

For $n=3$ it has been shown \cite{Fedoruk:2008hk,Fedoruk:2011aa} that 
the $\mathcal{N}=1$ superfield action (\ref{n1susye4d}) 
cannot be connected to the known $\mathcal{N}=2$ or $\mathcal{N}=3$ 
superconformal mechanical modelds and that 
in the bosonic limit it yields 
the three particle Calogero model for the component fields 
$x_{a}=\mathcal{X}_{a}|$.

\section{$\mathcal{N}=2$ Superconformal mechanics}
\label{secscqm2}
\subsection{One particle model}
The $\mathcal{N}=2$ superspace $\mathbb{R}^{(1|2)}$ 
contains time coordinate $t$ and two Grassmann coordinate
$\theta,\overline{\theta}$
\begin{align}
\label{n2susy1}
\mathbb{R}^{(1|2)}=(t,\theta,\overline{\theta}).
\end{align}
The covariant superderivatives $D$ and $\overline{D}$ are 
\begin{align}
\label{n2susy2a}
D=i\frac{\partial}{\partial \theta}
-\overline{\theta}\frac{\partial}{\partial t},\ \ \ \ \ 
\overline{D}=i\frac{\partial}{\partial\overline{\theta}}
-\theta\frac{\partial}{\partial t},\ \ \ \ \ 
\left\{D,\overline{D}\right\}=-2i\partial_{t}
\end{align}
while the two supercharges $Q$ and $\overline{Q}$ are given by 
\begin{align}
\label{n2susy2b}
Q=i\frac{\partial}{\partial \theta}
+\overline{\theta}\frac{\partial}{\partial t},\ \ \ \ \ 
\overline{Q}=i\frac{\partial}{\partial \overline{\theta}}
+\theta\frac{\partial}{\partial t},\ \ \ \ \ 
\left\{Q,\overline{Q}\right\}=2i\partial_{t}
\end{align}
in the superspace. 

Now consider the $\mathcal{N}=2$ superfield $(\bm{1},\bm{2},\bm{1})$ 
in the superspace (\ref{n2susy1})
\begin{align}
\label{n2susy2c}
\Phi(t,\theta,\overline{\theta})
=x(t)+i\theta\psi(t)+i\overline{\theta}\overline{\psi}(t)
+\theta\overline{\theta}y(t).
\end{align}
The $(\bm{1},\bm{2},\bm{1})$ supermultiplet is also called 
$\mathcal{N}=2A$ multiplet. 
This supermultiplet is related to the two-dimensional $(1,1)$
supersymmetry. 
Making use of the $(\bm{1},\bm{2},\bm{1})$ supermultiplet
(\ref{n2susy2c}), 
we can write $\mathcal{N}=2$ supersymmetric action in the form 
\begin{align}
\label{n2susy2d}
S=\frac12 \int dtd\theta d\overline{\theta}
\ \left[
\overline{D}\Phi D\Phi -W(\Phi)
\right]
\end{align}
where $W(\Phi)$ is a superpotential that is some function of the 
superfield $\Phi$. 
In component the superfield action (\ref{n2susy2d}) 
can be written as
\begin{align}
\label{n2susy2e}
S=\frac12 
\int dt\ 
\left[
\dot{x}^{2}
+i\dot{\overline{\psi}}\psi
-i\overline{\psi}\dot{\psi}+y^{2}
-W'(x)y-W''(x)\psi\overline{\psi}
\right].
\end{align}
To obtain the conformal invariant action, 
let us consider the superpotential in the form
\begin{align}
\label{n2susy3a}
W(\Phi)=f\ln \Phi^{2}.
\end{align}
Then the action (\ref{n2susy2e}) becomes 
\begin{align}
\label{n2susy3b}
S=\frac12 
\int dt\ 
\left[
\dot{x}^{2}
+i\dot{\overline{\psi}}\psi
-i\overline{\psi}\dot{\psi}
+y^{2}
-\frac{2fy}{x}
-\frac{2f\psi\overline{\psi}}{x^{2}}
\right].
\end{align}
By solving the algebraic equation of motion of $y$, 
one can integrate out the auxiliary field $y$. 
Then we find the one-particle $\mathcal{N}=2$ $OSp(2|2)$ 
superconformal mechanical model 
\cite{Akulov:1984uh,Fubini:1984hf}
\begin{align}
\label{n2susy3c}
S=\frac12 
\int dt\ 
\left[
\dot{x}^{2}
+i\dot{\overline{\psi}}\psi
-i\overline{\psi}\dot{\psi}
-\frac{f(f-2\psi\overline{\psi})}{x^{2}}
\right].
\end{align}

In the superspace 
the generators of the superconformal group 
can be realized by the following expressions 
\footnote{Note that 
in \cite{Akulov:1984uh} the
Hamiltonian is expresses by $\left\{Q,\overline{Q}\right\}=-2H$ while in
our notation the additional sign does not appear.}
\begin{align}
\label{n2susy4a}
H&=i\frac{\partial}{\partial t},\\
\label{n2susy4b}
D&=i\left(t\frac{\partial}{\partial t}
+\frac12 \theta \frac{\partial}{\partial \theta}
+\frac12 \overline{\theta} \frac{\partial}{\partial \overline{\theta}}+\Delta
\right),\\
\label{n2susy4c}
K&=i\left(
t^{2}\frac{\partial}{\partial t}
+t\theta\frac{\partial}{\partial \theta}
+t\overline{\theta}\frac{\partial}{\partial \overline{\theta}}
+2t\Delta
\right),\\
\label{n2susy4d}
Q&=i\frac{\partial}{\partial \theta}
+\overline{\theta}\frac{\partial}{\partial t},&
\overline{Q}&=
i\frac{\partial}{\partial \overline{\theta}}
+\theta \frac{\partial}{\partial t},\\
\label{n2susy4f}
S&=tQ-\theta\overline{\theta}\frac{\partial}{\partial \theta}
+2\Delta\overline{\theta},&
\overline{S}&=t\overline{Q}
-\overline{\theta}\theta\frac{\partial}{\partial \overline{\theta}}
+2\Delta \theta,\\
\label{n2susy4h}
B&=-i\theta \frac{\partial}{\partial \theta}
+i\overline{\theta}\frac{\partial}{\partial \overline{\theta}}.
\end{align}
One can show that these generators form 
the $\mathfrak{su}(1,1|1)$ superalgebra
\begin{align}
\label{n2susy5a}
\begin{array}{ccc}
[H,D]=iH,
&[K,D]=-iK,
&[H,K]=2iD,
\cr
\end{array}
\end{align}
\begin{align}
\label{n2susy5b}
\begin{array}{ccc}
[B,H]=0,
&[B,D]=0,
&[B,K]=0,
\end{array}
\end{align}
\begin{align}
\label{n2susy5d}
\begin{array}{ccc}
[H,Q]=0,
&[D,Q]=-\frac{i}{2}Q,
&[K,Q]=-iS,
\cr
[H,\overline{Q}]=0,
&[D,\overline{Q}]
=-\frac{i}{2}\overline{Q},
&[K,\overline{Q}]
=i\overline{S},
\end{array}
\end{align}
\begin{align}
\label{n2susy5e}
\begin{array}{ccc}
[H,S]=iQ,
&[D,S]=\frac{i}{2}S,
&[K,S]=0,
\cr
[H,\overline{S}]
=i\overline{Q}
&[D,\overline{S}]
=\frac{i}{2}\overline{S},
&[K,\overline{S}]=0,
\cr
\end{array}
\end{align}
\begin{align}
\label{n2susy5f}
\begin{array}{ccc}
\{Q,\overline{Q}\}=2H,&
\{S,\overline{S}\}=2K,&
\{Q,\overline{S}\}=2D-B,\\
\end{array}
\end{align}
\begin{align}
\label{n2susy5g}
[B,Q]
&=iQ,&
[B,S]
&=iS,\nonumber\\
[B,\overline{Q}]&=-i\overline{Q},&
[B,\overline{S}]&=-i\overline{S}.
\end{align}

The supersymmetry transformations for the 
$(\bm{1},\bm{2},\bm{1})$ multiplet which follow from 
$
\delta \Phi=-i[\epsilon \overline{Q}+\overline{\epsilon}Q,\Phi]
$
are expressed in the component fields as
\begin{align}
\label{n2susy4a}
\delta x&=i\epsilon\overline{\psi}+i\overline{\epsilon}\psi,\\
\label{n2susy4b}
\delta \psi&=\epsilon\dot{x}-i\epsilon \frac{f}{x},\\
\label{n2susy4c}
\delta \overline{\psi}&=
\overline{\epsilon}\dot{x}+i\overline{\epsilon}\frac{f}{x}.
\end{align}
%
%
Applying the Noether's method, 
we find the explicit expressions for the 
supercharges $Q,\overline{Q}$, 
the three $SL(2,\mathbb{R})$ conformal generators $H,D,K$  
and we also introduce 
the superconformal charges $S,\overline{S}$ 
and the $SO(2)$ R-symmetry generator $B$ as follows:
\begin{align}
\label{n2susy6a}
Q&=\psi\left(-ip+\frac{f}{x}\right), 
&\overline{Q}&=\overline{\psi}\left(ip+\frac{f}{x}\right),\\
\label{n2susy6b}
S&=x\psi,&\overline{S}&=x\overline{\psi},\\
\label{n2susy6c}
H&=\frac12 \left[
p^{2}+\frac{f(f+2B)}{x^{2}}\right]\\
\label{n2susy6d}
D&=-\frac14 (xp+px),\\
\label{n2susy6e}
K&=\frac12 x^{2},\\
\label{n2susy6f}
B&=\frac12 [\psi,\overline{\psi}].
\end{align}
Note that the potential 
in the Hamiltonian $H$ is shifted as a quantum effect. 
Under the canonical relations 
\begin{align}
\label{n2susy7a}
[x,p]&=i,
&\left\{\psi,\overline{\psi}\right\}=1,
\end{align}
the set of operators (\ref{n2susy6a})-(\ref{n2susy6f}) 
form the $\mathfrak{osp}(2|2)$ superalgebra
(\ref{n2susy5a})-(\ref{n2susy5f}).

\begin{figure}
\begin{center}
\includegraphics[width=15cm]{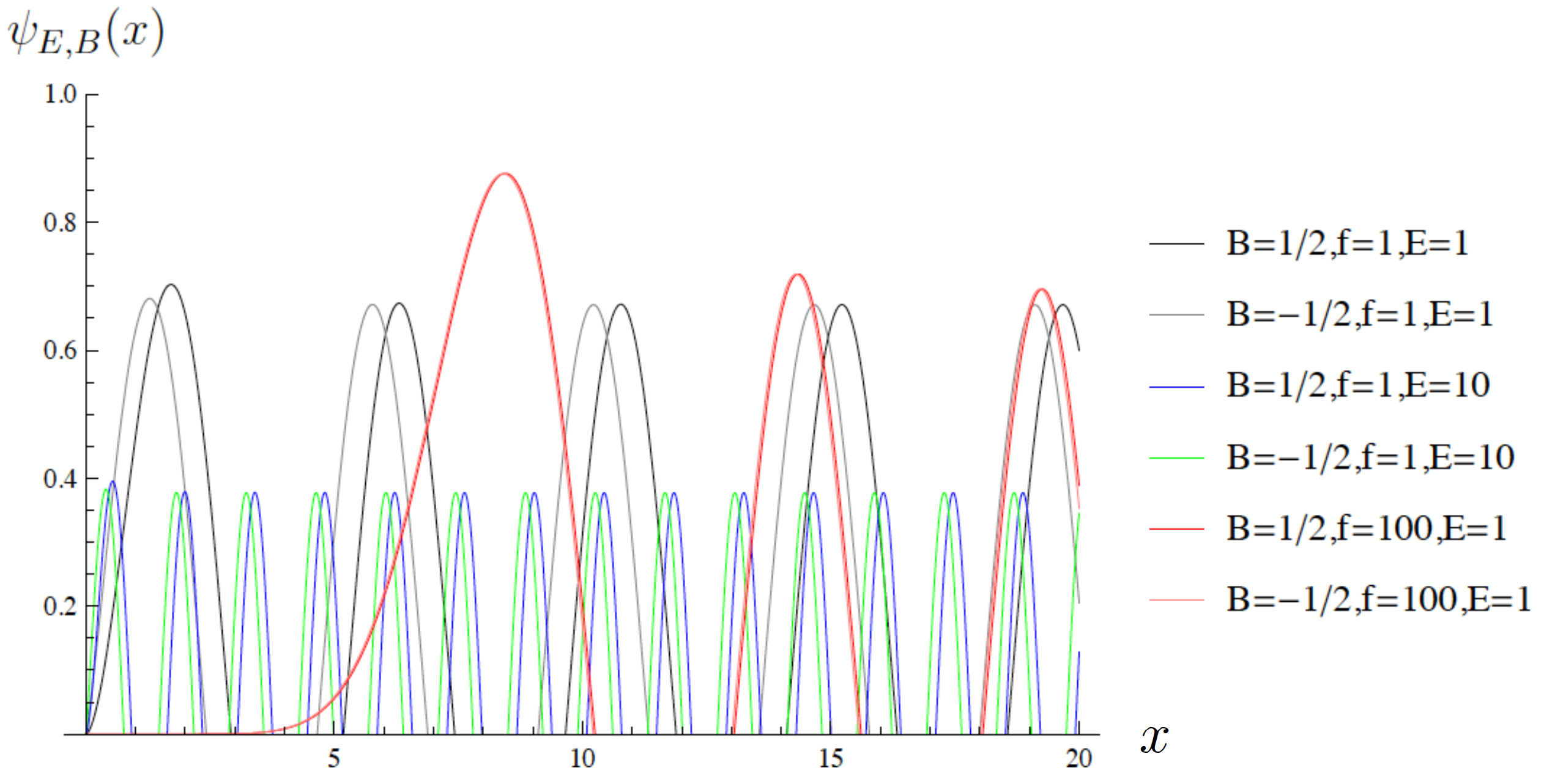}
\caption{The eigenfunctions $\psi_{E,B}(x)$ of the original Hamiltonian
 $H$ with $E\neq 0$. 
There are two sectors labeled by $B=\frac12$ and $B=-\frac12$.}
\label{wavefcts1}
\end{center}
\end{figure}

Let us study the spectrum of the 
one-particle $OSp(2|2)$ superconformal quantum mechanics
(\ref{n2susy3c}). 
In general supersymmetric quantum mechanics 
has the Hamiltonian $H$ which can be written as the sum of squares 
of the Hermitian supercharges $Q_{A}$,$A=1,\cdots,\mathcal{N}$. 
This implies that the energy of any state is positive or zero 
\cite{Iliopoulos:1974zv,Witten:1981nf}. 
If $H|\Omega\rangle=0$, then 
we have $0=\langle\Omega|H|\Omega\rangle
=\sum_{A}\langle\Omega|Q^{2}_{A}|\Omega\rangle 
=\sum_{A}|Q_{A}|\Omega\rangle|^{2}$, 
which is only possible if $Q_{A}|\Omega\rangle$ for any $A$. 
Conversely if a state $|\Omega\rangle$ is annihilated by $Q_{A}$, 
then $H|\Omega\rangle=Q_{A}^{2}|\Omega\rangle=0$, 
i.e. its energy is zero. 
Therefore the supersymmetry generated by $Q_{A}$ 
is broken if the system has no normalizable ground state of $H$. 
Now consider the  equations 
defining the ground state of $H$
\begin{align}
\label{n2susy8a}
Q|\Omega\rangle =\overline{Q}|\Omega\rangle =0.
\end{align}
Using the explicit expressions (\ref{n2susy6a}) and (\ref{n2susy6b}), 
the equation (\ref{n2susy8a}) is written as 
\begin{align}
\label{n2susy8b}
\left(
2iBp-\frac{f}{x^{2}}
\right)|\Omega\rangle=0
\end{align}
which can be interpreted 
as the first order differential equation of $x$. 
Then the generic solution of (\ref{n2susy8b}) 
leads to the $x$-depgroundence of the ground state of $H$ as
\begin{align}
\label{n2susy8c}
|\Omega\rangle=
x^{-2fB}|\textrm{phys}\rangle
\end{align}
where $|\textrm{phys}\rangle$ is any $x$ independent state. 
Noting that the $SO(2)$ R-symmetry operator $B$ 
has eigenvalue $+\frac12$ and $-\frac12$, 
we see that the ground state of $H$ 
may have the two different $x$ dependence
\begin{align}
\label{n2susy8d}
|\Omega\rangle=
\begin{cases}
x^{-f}|\textrm{phys}\rangle&\textrm{for\ \ \ $B=\frac12$}\cr
x^{f}|\textrm{phys}\rangle&\textrm{for\ \ \ $B=-\frac12$}.\cr
\end{cases}
\end{align}
As the wavefunction will blow up for 
either large or small $x$ region, 
there is no normalizable state of $H$ 
and therefore the supersymmetry generated by 
$Q$, $\overline{Q}$ is spontaneously broken. 
Note that the wavefunction with $E>0$ energy 
can be exactly solved by using the result of DFF-model. 
Comparing the quantum Hamiltonian  
(\ref{n2susy6c}) with the DFF-model Hamiltonian 
(\ref{hktbos1}), 
we find the relation 
\begin{align}
\label{n2susy8e}
g^{2}=f(f+2B)=
\begin{cases}
f(f+1)&\textrm{for\ \ \ $B=\frac12$}\cr
f(f-1)&\textrm{for\ \ \ $B=-\frac12$}.\cr
\end{cases}
\end{align}
The appearance of two sectors, 
i.e. the doublet structure of the eigenstates of $H$ 
corresponds to the fact that 
$H$ commutes with two operators $Q$ and $\overline{Q}$. 
From the expression (\ref{ewave1}) 
we find the eigenfunctions 
\begin{align}
\label{n2susy8f}
\psi_{E,B}(x)
=\begin{cases}
C\sqrt{x}J_{\sqrt{f+\frac12}}\left(\sqrt{2E}x\right)
&\textrm{for\ \ \ $B=\frac12$}\cr
C\sqrt{x}J_{\sqrt{f-\frac12}}\left(\sqrt{2E}x\right)
&\textrm{for\ \ \ $B=-\frac12$}.\cr
\end{cases}
\end{align}
These wavefunctions are shown in Figure \ref{wavefcts1}.
From Figure \ref{wavefcts1} we see that 
there are several peaks of the wavefunctions 
with the nearest one from the origin being the maximum value. 
For large coupling constant $f$ 
the relative positions of the particle 
gradually become far from the origin. 
At high energy $E$ the number of peaks increases 
and the probability of the position of the particle is averaged. 

\begin{figure}
\begin{center}
\includegraphics[width=8.5cm]{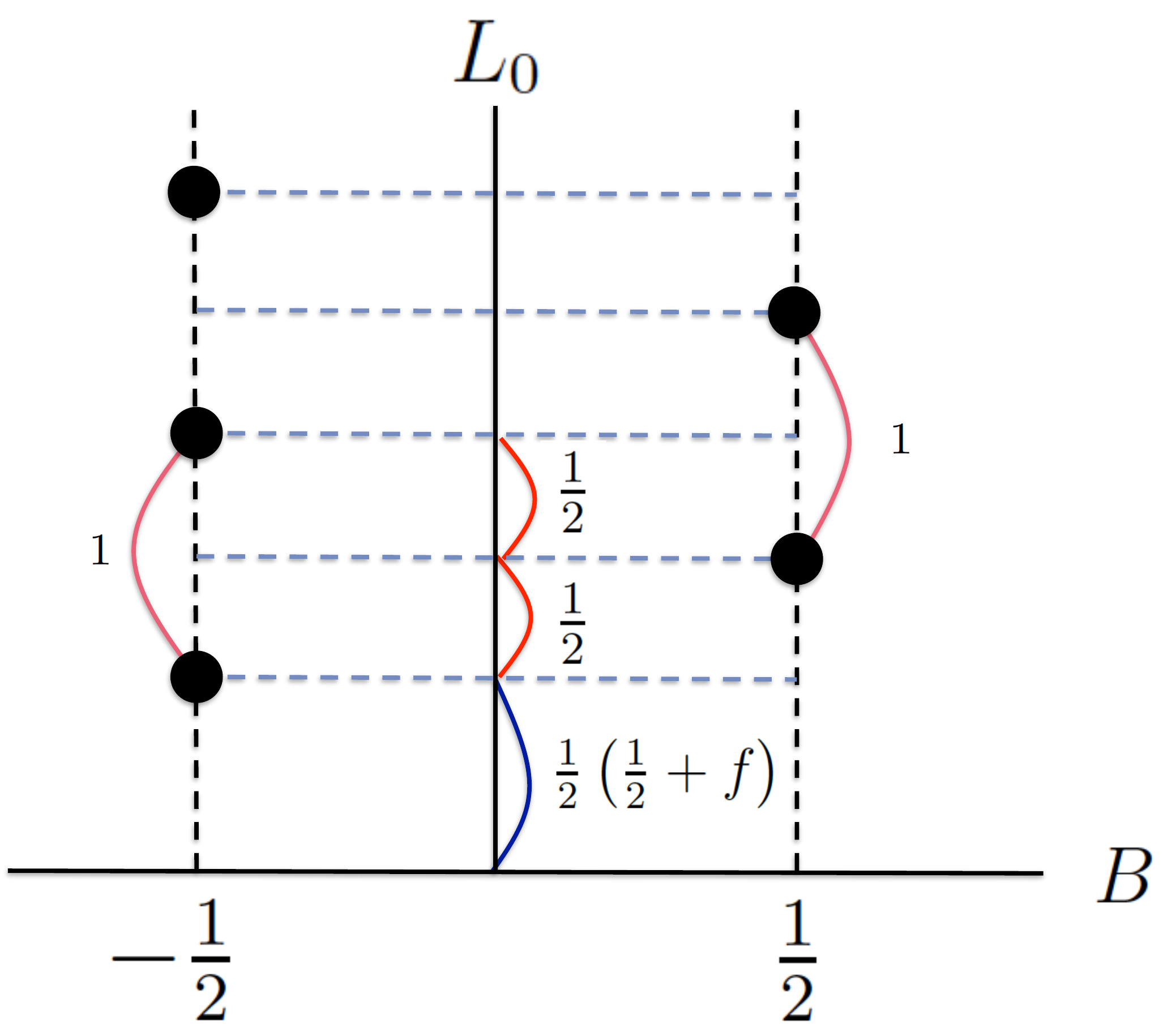}
\caption{The level structure of the spectrum of 
the new Hamiltonian $L_{0}$. 
The spectrum is equally spaced. 
For a fixed $B$ the equal space is $1$ 
while the space with $\Delta B\neq 1$ is $\frac12$.}
\label{l0susy1}
\end{center}
\end{figure}

Then we can follow the previous discussion for the DFF-model 
to solve the problem of the absence of the ground state. 
Instead of the original Hamiltonian 
we now regard the compact operator $L_{0}=\frac12 (H+K)$ as 
the new Hamiltonian. 
Looking at the formulae (\ref{vira3}), (\ref{vira4})
and the relation (\ref{n2susy8e}), 
one finds
\begin{align}
\label{n2susy8g}
r_{n}
=\begin{cases}
\frac12\left(\frac32+f\right)
&\textrm{for\ \ \ $B=\frac12$}\cr
\frac12\left(\frac12+f\right)
&\textrm{for\ \ \ $B=-\frac12$}.\cr
\end{cases}
\end{align}
The level structure of the spectrum of $L_{0}$ 
has two series corresponding to the two different eigenvalues 
$B=-\frac12, \frac12$. 
So it can be represented on the plane of the eigenvalue 
of $B$ and $L_{0}$ (see Figure \ref{l0susy1}).
In order to understand the appearance of 
the half integer shift in an algebraic way, 
let us define the fermionic operators \cite{Fubini:1984hf}
\begin{align}
\label{n2susy8h1}
M&=Q-S=\psi\left(
-ip+\frac{f}{x}-x
\right),\\
\label{n2susy8h2}
\overline{M}&=\overline{Q}-\overline{S}
=\overline{\psi}
\left(
ip+\frac{f}{x}-x
\right),
\\
\label{n2susy8h3}
N&=\overline{Q}+\overline{S}
=\overline{\psi}
\left(
ip+\frac{f}{x}+x
\right),\\
\label{n2susy8h4}
\overline{N}&=Q+S
=\psi\left(
-ip+\frac{f}{x}+x
\right).
\end{align}
Then we find the following anti-commutation relations
\begin{align}
\label{n2susy8g1}
\left\{M,\overline{M}\right\}
&=:4T_{1}
=4L_{0}+2B-2f,\\
\label{n2susy8g2}
\left\{N,\overline{N}\right\}
&=:4T_{2}
=4L_{0}-2B+2f,\\
\label{n2susy8g3}
\left\{M,N\right\}
&=4L_{+}
=2\left(H-K+2iD\right),\\
\label{n2susy8g4}
\left\{\overline{M},\overline{N}\right\}
&=4L_{-}
=2\left(H-K-2iD\right),\\
\label{n2susy8g5}
\left\{M,\overline{N}\right\}&
=\left\{\overline{M},N\right\}=0.
\end{align}
and the commutation relations 
\begin{align}
\label{n2susy8g8}
[L_{0},M]&=-\frac12 M,&
[L_{0},\overline{M}]&=\frac12 \overline{M},\\
\label{n2susy8g9}
[L_{0},N]&=-\frac12 N,&
[L_{0},\overline{N}]&=\frac12 \overline{N},\\
\label{n2susy8g10}
[T_{1},N]&=-N,&
[T_{1},\overline{N}]&=\overline{N},\\
\label{n2susy8g12}
[T_{2},N]&=-N,&
[T_{2},\overline{N}]&=\overline{N},\\
\label{n2susy8g6}
[T_{1},M]&=
[T_{1},\overline{M}]=0,\\
\label{n2susy8g7}
[T_{2},N]&=
[T_{2},\overline{N}]=0,\\
\label{n2susy8g11}
[T_{1},L_{-}]&=-L_{-},&
[T_{1},L_{+}]&=L_{+},\\
\label{n2susy8g13}
[T_{2},L_{-}]&=-L_{-},&
[T_{2},L_{+}]&=L_{+}.
\end{align}
Let us consider the ground states eliminated by the supercharges. 
Since there are now three sets of the supercharges; 
$(Q,\overline{Q})$, $(M,\overline{M})$ and $(N,\overline{N})$, 
we find six possible candidates for the 
$x$ dependence of the ground states $|\Omega\rangle$:
\begin{align}
\label{n2susy9}
|\Omega\rangle =
\begin{cases}
x^{-f}|\textrm{phys}\rangle&\textrm{for $(H,Q,\overline{Q},B=\frac12)$}\cr
 x^{f}|\textrm{phys}\rangle&\textrm{for $(H,Q,\overline{Q},B=-\frac12)$}\cr
x^{-f}e^{\frac{x^{2}}{2}}|\textrm{phys}\rangle
&\textrm{for $(T_{1},M,\overline{M},B=\frac12)$}\cr 
x^{f}e^{-\frac{x^{2}}{2}}|\textrm{phys}\rangle
&\textrm{for $(T_{1},M,\overline{M},B=-\frac12)$}\cr
x^{-f}e^{-\frac{x^{2}}{2}}|\textrm{phys}\rangle
&\textrm{for $(T_{2},N,\overline{N},B=\frac12)$}\cr 
x^{f}e^{\frac{x^{2}}{2}}|\textrm{phys}\rangle
&\textrm{for $(T_{2},N,\overline{N},B=-\frac12)$}\cr
\end{cases}
\end{align}
where $|\textrm{phys}\rangle$ is a $x$ independent state. 
We see that 
only the set of generators $(T_{1},M,\overline{M}$, $B=-\frac12)$ 
can yield the normalizable eigenfunction of the ground state. 
In order to obtain the normalizable ground state, 
$|\textrm{phys}\rangle$ need to be the eigenstate with $B=-\frac12$. 
Let us define a state $|0\rangle$ annihilated 
by the operator $\overline{\psi}$
\begin{align}
\label{n2susy9a}
\overline{\psi}|0\rangle=0.
\end{align}
Then $B|0\rangle=-\frac12 |0\rangle$ and 
we thus can choose the state $|0\rangle$ as $|\textrm{phys}\rangle$. 
Given the state $|0\rangle$, 
one can build up a tower of states by 
multiplying the operator $\psi$. 
Since the square of the Grassmann variable is zero $\psi^{2}=0$, 
the fermionic generators form the two-dimensional space spanned by 
\begin{align}
\label{n2susy9b}
|0\rangle,\ \ \ \psi|0\rangle
\end{align}
and $\overline{\psi}$ and $\psi$ are 
identified with the lowering operator and raising operator 
for fermionic excitation respectively. 
Therefore we obtain the normalizable ground state
\begin{align}
\label{n2susy9c}
|\Omega\rangle =x^{f}e^{-\frac{x^{2}}{2}}|0\rangle
\end{align}
which satisfies 
\begin{align}
\label{n2susy9d}
M|\Omega\rangle&=\overline{M}|\Omega\rangle=0,\\
\label{n2susy9dd}
N|\Omega\rangle&=0,\\
\label{n2susy9e}
\overline{\psi}|\Omega\rangle&=0.
\end{align}

\begin{figure}
\begin{center}
\includegraphics[width=8.5cm]{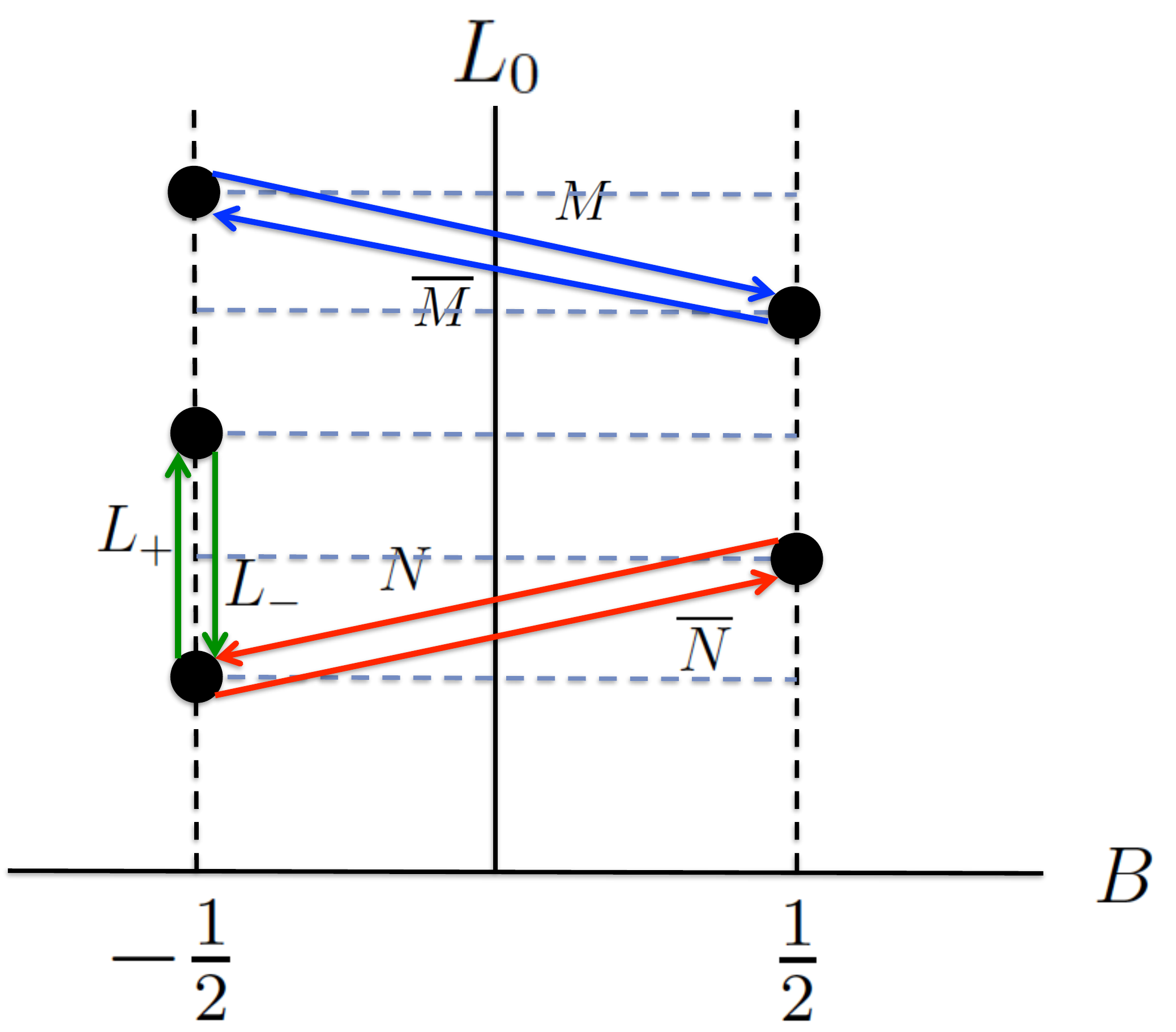}
\caption{The bosonic excitations and the fermionic excitations 
in the $L_{0}$ spectrum. 
For a fixed $B$, i.e. for the bosonic excitation generated by 
$L_{+}$ and $L_{-}$, the space is one unit. 
For a fermionic excitation generated by 
$\overline{N}$, $N$, $\overline{M}$ and $M$ 
the space is half a unit.}
\label{l0susy2}
\end{center}
\end{figure}

Having found the eigenfunction of $L_{0}$, 
we see from (\ref{vira6}) and 
(\ref{n2susy9c}) that the ground state $|\Omega\rangle$ 
is the eigenstate of $L_{0}$ with the eigenvalue 
\begin{align}
\label{n2susy9f}
r_{0}=\frac12\left(f+\frac12\right)
\end{align}
and obtain the two series (\ref{n2susy8g}) labeled by $B$. 
We observe from the commutation relations (\ref{n2susy8g9})
that the fermionic generator $M,N$ 
decreases $L_{0}$ by $\frac12$ 
while $\overline{M},\overline{N}$ increase $L_{0}$ by $\frac12$. 
As seen from the relations (\ref{n2susy9d}), 
the fermionic excitation for 
the ground state $|\Omega\rangle$ can be generated 
by only $\overline{N}$. 
In addition, there are bosonic excitations. 
As in the DFF-model, 
$L_{+}$ increases $L_{0}$ by one 
and $L_{-}$ decreases $L_{0}$ by one. 
While the fermionic excitations shift the eigenvalue of $B$, 
the bosonic excitations does not. 
The excitations in the $L_{0}$ spectrum are drawn 
in Figure \ref{l0susy2}. 

From the relations 
(\ref{n2susy8g1}), (\ref{n2susy8g2}), 
(\ref{n2susy8g6}) and (\ref{n2susy8g7}) 
one can see that 
the two sets of new supercharges $(M,\overline{M})$ 
and $(N,\overline{N})$ yield 
the bosonic operators $T_{1}$ and $T_{2}$ respectively. 
Since the bosonic operators $T_{1}$ and $T_{2}$ are compact, 
one may also use $T_{1}$ or $T_{2}$ as the new Hamiltonian. 
However, unlike the compact operator $L_{0}$, 
$T_{1}$ and $T_{2}$ enjoy the double structures of their spectrums 
according to the commutation relations (\ref{n2susy8g6}) and
(\ref{n2susy8g7}). 

\begin{figure}
\begin{center}
\includegraphics[width=8.5cm]{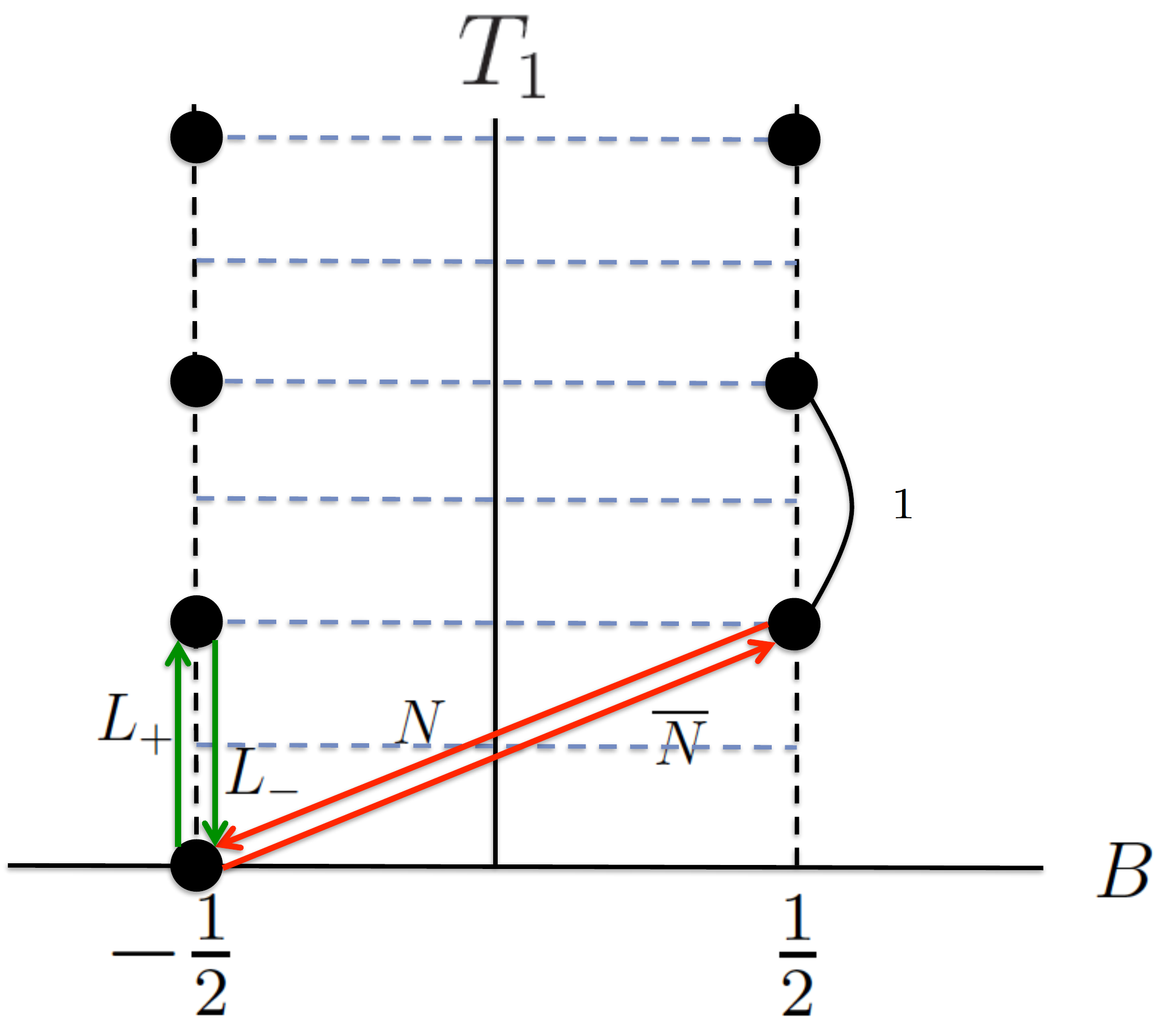}
\caption{The level structure of $T_{1}$ spectrum 
and its bosonic and fermionic excitations. 
Each of the bosonic and fermionic excitations has 
the equal space of one unit. The ground state $|\Omega\rangle$ 
has zero eigenvalue.}
\label{t1susy1}
\end{center}
\end{figure}

Now consider the spectrum of $T_{1}$. 
By noting the relations 
(\ref{n2susy8g1}) and (\ref{n2susy9d}), 
we see that the ground state $|\Omega\rangle$ 
has zero eigenvalue of $T_{1}$. 
According to the 
commutation relations 
(\ref{n2susy8g10}) and 
(\ref{n2susy8g11}), 
one finds that for the $T_{1}$ spectrum 
the bosonic and fermionic excitations have 
the same spacing equal to one, which are 
generated by $L_{+},L_{-}$ and $\overline{N},N$ respectively. 
Note that $\overline{M},M$ commute with $T_{1}$ 
and do not play the role of the raising and lowering operators. 
The $T_{1}$ spectrum is given by the 
two series 
\begin{align}
\label{n2susy10a}
T_{1}=
\begin{cases}
0,1,2,\cdots&\textrm{for $B=-\frac12$}\cr
1,2,\cdots&\textrm{for $B=\frac12$},\cr
\end{cases}
\end{align}
which is illustrated in Figure \ref{t1susy1}. 
For all non-zero $T_{1}$ states, 
there are degenerate structures. 
In other words 
the bosonic and fermionic states are always paired at 
the excited level of $T_{1}$. 
This is due to the relations (\ref{n2susy8g6}), 
which ensure the preservation of the supersymmetry 
generated by $M$ and $\overline{M}$. 
Therefore one can interpret the pairing structure 
of $T_{1}$ spectrum at excited states 
as the consequence of the preserved 
supersymmetry generated by $M$ and $\overline{M}$.

\begin{figure}
\begin{center}
\includegraphics[width=8.5cm]{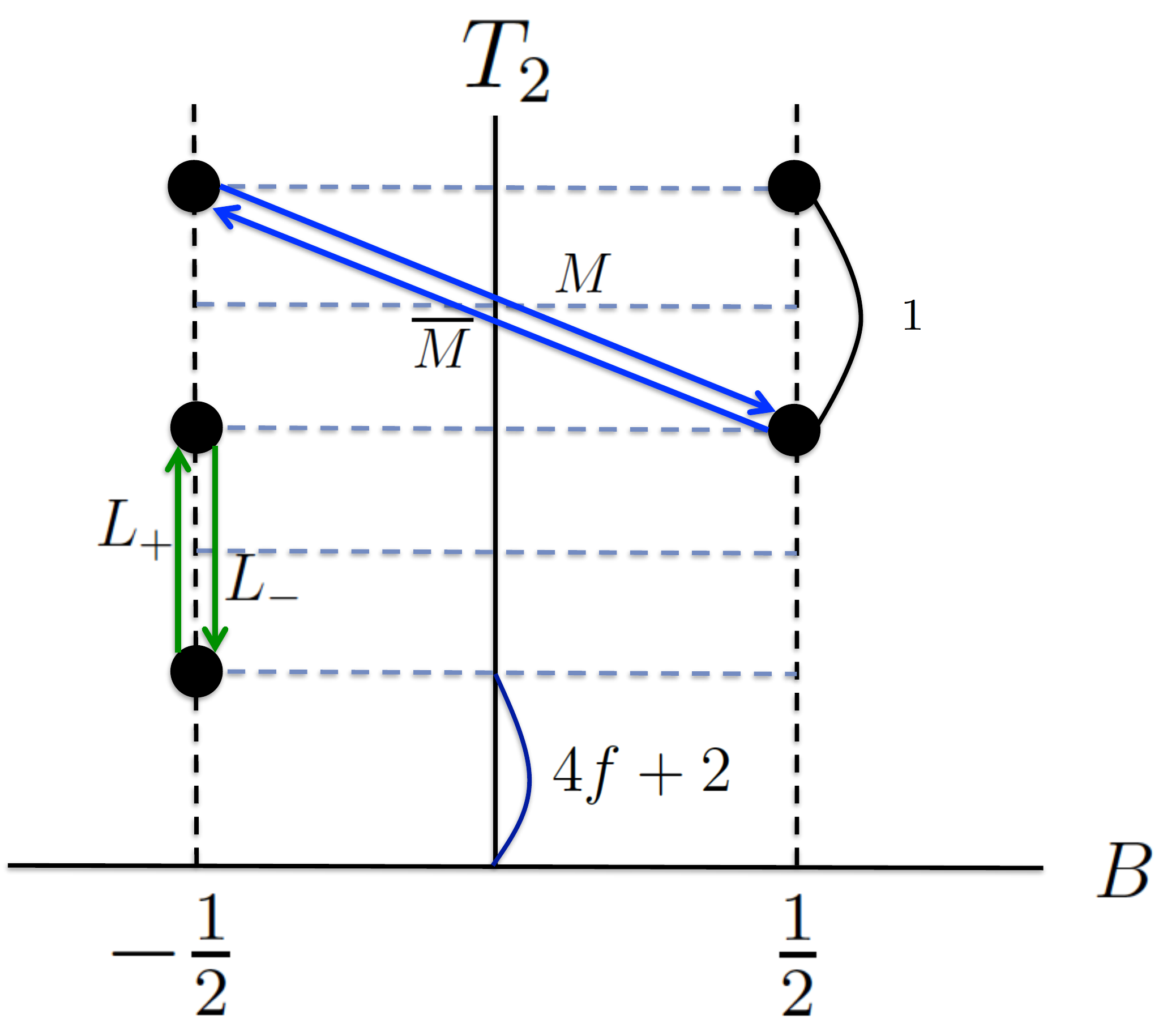}
\caption{The level structure of $T_{2}$ spectrum 
and its bosonic and fermionic excitations. 
Each of the bosonic and fermionic excitations has 
the equal space of one unit. The ground state $|\Omega\rangle$ 
has eigenvalue $(4f+2)$.}
\label{t2susy1}
\end{center}
\end{figure}

Similarly the spectrum $T_{2}$ 
holds the doublet structure because 
$T_{2}$ commute with $N$ and $\overline{N}$ 
and the corresponding supersymmetry is preserved 
as seen from (\ref{n2susy8g7}). 
In this case the bosonic excitation is generated by $L_{+},L_{-}$ 
whereas the fermionic one is generated by $\overline{M},M$. 
Also one can see from (\ref{n2susy8g12}) and (\ref{n2susy8g13}) 
that both bosonic and fermionic excitations 
are produced with equal spacing of one unit. 
In this case, however, 
there is no normalizable zero $T_{2}$ state. 
The normalizable ground state $|\Omega\rangle$ 
has the eigenstate of $T_{2}$ with the eigenvalue $(4f+2)$. 
The $T_{2}$ spectrum is given by
\begin{align}
\label{n2susy10b}
T_{2}=
\begin{cases}
4f+2+n&\textrm{for $B=-\frac12$}\cr
4f+3+n&\textrm{for $B=\frac12$}
\end{cases}
\end{align}
where $n=0,1,2,\cdots$. 
The $T_{2}$ spectrum and its excitation 
are shown in Figure \ref{t2susy1}.

\subsection{Multi-particle model}
\label{mulcqm1}
Now we want to discuss the $\mathcal{N}=2$ superconformal sigma-model. 
Let us start with $n$ $(\bm{1},\bm{2},\bm{1})$ 
supermultiplets $\Phi_{a}$,$a=1,\cdots, n$ 
\footnote{The $(\bm{1},\bm{2},\bm{1})$ supermultiplet is also called 
$\mathcal{N}=2A$ multiplet while $(\bm{2},\bm{2},\bm{0})$ chiral
supermultiplet is also called $\mathcal{N}=2B$ multiplet 
\cite{Gibbons:1997iy}.}. 
The generic action without superpotential terms takes the form 
\cite{Gibbons:1997iy}
\begin{align}
\label{n2mul1a1}
S=\frac12 \int dtd^{2}\theta 
\left[
(g+b)_{ij}\overline{D}\Phi^{i}D\Phi^{j}
+l_{ij}D\Phi^{i}D\Phi^{j}
+m_{ij}\overline{D}\Phi^{i}\overline{D}\Phi^{j}
\right]
\end{align}
where $g_{ij}$ is the metric and $b_{ij}$, $l_{ij}$ and $m_{ij}$ 
are the two-forms on the target space $\mathcal{M}$. 
Note that 
the terms of $l_{ij}$ and $m_{ij}$ 
correspond to the non-Lorentz invariant terms 
in two-dimensions. 
Notice that 
the target space $\mathcal{M}$ of the 
$(\bm{1},\bm{2},\bm{1})$ supermultiplet, 
or $\mathcal{N}=2A$ multiplet is a real manifold.

We have already defined the two covariant derivatives 
$D$ and $\overline{D}$ for $\mathcal{N}=2$ supersymmetry in 
(\ref{n2susy2a}), however, 
more generally in terms of the two $\mathcal{N}=1$ 
covariant derivatives $D_{1}$, $D_{2}$ 
the $\mathcal{N}=2$ two covariant derivatives can be chosen as
\begin{align}
\label{n2mul1a2}
D_{2}\Phi^{i}&={I^{i}}_{j}D_{1}\Phi^{j}
\end{align}
where $I$ is an endomorphism of the tangent bundle of $\mathcal{M}$. 
Then the 
anti-commutation relations $\left\{D_{i},D_{j}\right\}=2i\partial_{t}$ 
impose the conditions 
\begin{align}
\label{n2mul1a3}
I^{2}&=-1,\\
\label{n2mul1a4}
N(I)&=0
\end{align}
where $N(I)$ is the Nijenhuis tensor of the endomorphism $I$. 
The condition (\ref{n2mul1a3}) implies that 
$I$ is the almost complex structure 
and the condition (\ref{n2mul1a4}) further shows that 
the $I$ is an (integrable) complex structure. 
Thus $\mathcal{N}=2$ supersymmetry requires a complex structure 
$I$ on the target space \cite{Coles:1990hr}. 

To go further 
let us follow the strategy in \cite{Howe:1988cj} 
and express the second supersymmetry transformation in terms of 
the $\mathcal{N}=1$ superspace formalism as \cite{Michelson:1999zf}
\begin{align}
\label{n2mul1a5a}
\delta x^{i}&=-i\epsilon {I^{i}}_{j}\psi^{j},\\
\label{n2mul1a5b}
\delta \psi^{i}&=
-\epsilon\left[
{I^{i}}_{j}\dot{x}^{j}
-i\psi^{k}\left(\partial_{k}{I^{i}}_{j}\right)\psi^{j}
\right].
\end{align}
Following the Noether's procedure, 
we obtain the second supercharge
\begin{align}
\label{n2mul1a5c}
Q_{2}&=\psi^{i}{I_{i}}^{j}\Pi_{j}
-\frac{i}{2}\psi^{i}{I^{j}}_{i}c_{jkl}\psi^{j}\psi^{k}
-\frac{i}{6}\psi^{i}\psi^{j}\psi^{k}
{I^{l}}_{i}{I^{m}}_{j}{I^{n}}_{k}c_{lmn}
-\frac{i}{2}\psi^{i}c_{ijk}I^{jk}
\end{align}
where $\Pi_{i}:=g_{ij}\dot{x}^{j}$.  
Then it turns out that the $\mathcal{N}=1$ 
action (\ref{n1susy2a1}) is invariant under the 
$\mathcal{N}=2$ supersymmetry transformations if we have 
\cite{Gibbons:1997iy,DeJonghe:1996fb}
\begin{align}
\label{n2mul1b1}
&g_{ij}={I^{k}}_{i}{I^{l}}_{j}g_{kl},\\
\label{n2mul1b2}
&\nabla_{(i}^{(+)}{I^{k}}_{j)}=0,\\
\label{n2mul1b3}
&\partial_{[i}\left({I^{m}}_{j}c_{|m|kl]}
\right)-2{I^{m}}_{[i}\partial_{[m}c_{jkl]]}=0
\end{align}
where $\nabla^{(+)}_{i}$ is the connection with torsion $c$ 
on $\mathcal{M}$; $\Gamma^{i}_{jk}+{c^{i}}_{jk}$. 
The first constraint (\ref{n2mul1b1}) 
requires that the metric $g$ on $\mathcal{M}$ 
is Hermitian with respect to the complex structure $I$. 
The second condition (\ref{n2mul1b2}) 
is a generalized Yano tensor condition with torsion \footnote{For
vanishing torsion the equation (\ref{n2mul1b2}) coincides with 
the Yano tensor condition as in \cite{Gibbons:1993ap}}. 
This corresponds to the vanishing of $\left\{Q_{1},Q_{2}\right\}$ 
where $Q_{1}$ is the $\mathcal{N}=1$ supercharge given by (\ref{n1susy2a6}). 
The third condition (\ref{n2mul1b3}) yields the restriction 
on torsion and complex structure, however it has no geometrical
interpretation so far. 

It is known that 
the $\mathcal{N}=2$ supermultiplets in one dimension 
are related to the $\mathcal{N}=1$ supersymmetry in two dimensions 
\begin{align}
\label{n2mul1b3a}
\textrm{1d $\mathcal{N}=2A$}
&\Leftrightarrow 
\textrm{2d $\mathcal{N}=(1,1)$},\nonumber\\
\textrm{1d $\mathcal{N}=2B$}
&\Leftrightarrow 
\textrm{2d $\mathcal{N}=(2,0)$}
\end{align}
by the dimensional reduction. 
Note that two-dimensional $(2,0)$ supersymmetry sigma-models 
requires the first condition (\ref{n2mul1b1}), 
however, the last two conditions (\ref{n2mul1b2}) and (\ref{n2mul1b3}) 
do not appear in two-dimensional $(2,0)$ sigma-models. 
Instead of (\ref{n2mul1b2}), 
there appears 
the covariant constant condition of $I$ with respect to the 
connection $\nabla^{(+)}$
\begin{align}
\label{n2mul1b4}
\nabla_{i}^{(+)}{I^{j}}_{k}&=0.
\end{align}

Now we consider the $\mathcal{N}=2$ superconformal condition. 
Promoted from the $\mathfrak{osp}(1|2)$ algebra 
(\ref{n1susy2c1})-(\ref{n1susy2c4}), 
the $\mathfrak{su}(1,1|1)$ algebra (\ref{n2susy5d})-(\ref{n2susy5g}) 
contain the $U(1)$ R-symmetry generator $B$. 
From the commutation of the 
supercharges $Q_{1}$ in (\ref{n1susy2a6}) and $Q_{2}$ 
in (\ref{n2mul1a5c})  
with the conformal boost generator $K$ we can read the 
superconformal charges
\begin{align}
\label{n2mulc1}
S_{1}&=\psi^{i}D_{i},&
S_{2}&=\psi^{i}{I^{j}}_{i}D_{j}.
\end{align}
Then the R-symmetry generator $B$ can be found from 
the commutator of $Q$ and $S_{2}$ as
\begin{align}
\label{n2mulc2}
B=D_{2}^{i}\Pi_{i}-i{I_{ij}}\psi^{i}\psi^{j}
-iD_{2}^{i}c_{ijk}\psi^{j}\psi^{k}.
\end{align}
The constraint can be found from the 
commutation relation $[D,Q_{2}]=\frac{i}{2}Q_{2}$, 
which leads to 
\begin{align}
\label{n2mulc2}
\mathcal{L}_{D}{I^{j}}_{i}&=0.
\end{align}
This implies that 
$D$ preserves the complex structure $I$, 
that is $D$ acts holomorphically. 
Combining the constraint (\ref{n2mulc2}) 
with the other required conditions 
(\ref{n1susy2d7}) and (\ref{n2mul1b2}), 
we also find
\begin{align}
\label{n2mulc3}
\mathcal{L}_{\tilde{D}}{I^{j}}_{i}&=0,&
\mathcal{L}_{\tilde{D}}g_{ij}&=0,
\end{align}
which means that 
$\tilde{D}^{i}:=D^{j}{I^{i}}_{j}$ generates a holomorphic isometry. 

Therefore 
the $SU(1,1|1)$ superconformal quantum sigma-model with vanishing
bosonic potential 
can be realized if the conformal invariant conditions 
(\ref{mul1a8}), (\ref{mul1b3a}), 
the $\mathcal{N}=2$ supersymmetry invariant conditions 
(\ref{n2mul1b1})-(\ref{n2mul1b3}) 
and the $SU(1,1|1)$ superconformal invariant conditions 
(\ref{n2mulc2}), (\ref{n2mulc3}) are satisfied. 
The last additional constraints on the target space $\mathcal{M}$ 
require that $D$ acts holomorphically 
and $\tilde{D}^{i}:D^{j}{I^{i}}_{j}$ generates 
a holomorphic isometry.

\subsection{Freedman-Mende model}
Let us consider $n$ $(\bm{1},\bm{2},\bm{1})$ 
supermultiplets $\Phi_{a}$,$a=1,\cdots, n$ 
and a simple superfield action given by 
\begin{align}
\label{n2susy10c}
S=\frac12 \int dtd^{2}\theta\ 
\left[
\sum_{a=1}^{n}\overline{D}\Phi_{a}D\Phi_{a}
-W(\Phi)
\right]
\end{align}
where $W(\Phi)$ is the superpotential. 
In terms of the component fields 
the action (\ref{n2susy10c}) is expressed as
\begin{align}
\label{n2susy10d}
S=\frac12 
\int dt 
\Biggl[
\sum_{a=1}^{n}
&\left(\dot{x}_{a}^{2}
+i\dot{\overline{\psi}}_{a}\psi_{a}
-i\overline{\psi}_{a}\dot{\psi}_{a}
\right)
\nonumber\\
&-\frac14 \sum_{a=1}^{N}\partial_{a}W\partial_{a}W
-\sum_{a,b}\left(
\partial_{a}\partial_{b}W
\right)
\psi_{a}\overline{\psi}_{b}
\Biggr]
\end{align}
where $\partial_{a}:=\frac{\partial}{\partial x_{a}}$. 
Taking into account the superconformal boost transformation 
on the $(\bm{1},\bm{2},\bm{1})$ multiplet 
\begin{align}
\label{n2susy10e}
\delta \Phi_{a}&=-i\left(
\eta \overline{\theta}+\overline{\eta}\theta
\right)\Phi_{a}
\end{align}
and the invariance of the measure $\delta \left(dt d\theta\right)=0$ 
we find that the action 
(\ref{n2susy10d}) is invariant under the superconformal 
boost transformation only if we have
\begin{align}
\label{n2susy10f}
\Phi_{a}\partial_{a}W(\Phi)=c
\end{align}
with $c$ being a constant. 
It has been shown \cite{Galajinsky:2009iq} that 
$c$ characterizes 
the central charge in $\mathfrak{su}(1,1|1)$ superconformal algebra 
and that 
the superpotential $W(\Phi)$ is a harmonic function 
of $\Phi_{a}$ if quantum Hamiltonian contain 
boson-fermion interaction but no boson-boson interaction.

It is interesting to note that
the superpotential \cite{Bellucci:2003ng}
\begin{align}
\label{n2susy10g}
W(\Phi)=f\sum_{a\neq b}\ln \left(
\Phi_{a}-\Phi_{b}
\right)
\end{align}
where $f$ is a constant 
gives rise to the Freedman-Mende model \cite{Freedman:1990gd}
\begin{align}
\label{n2susy10h}
S=\frac12 \int dt \Biggl[
\sum_{a=1}^{n}
\left(\dot{x}_{a}^{2}+i\dot{\overline{\psi}}_{a}\psi_{a}
-i\overline{\psi}_{a}\dot{\psi}_{a}
\right)
-\sum_{a\neq b}
\frac{f^{2}+4f\psi_{a}\overline{\psi}_{b}}{4(x_{a}-x_{b})^{2}}
\Biggr].
\end{align}
This is the $\mathcal{N}=2$ superconformal generalization 
of the Calogero model. 
For the Freedman-Mende model 
the central charge $Z$ in the $\mathfrak{su}(1,1|1)$ superconformal 
algebra can be identified with
\begin{align}
\label{n2susy10h}
Z=n(n-1)f.
\end{align}
The Freedman-Mende model 
is the supersymmetric rational $A_{n+1}$ Calogero model 
in the sense that 
the original Calogero model is obtained 
by projecting the supersymmetric Hamiltonian 
onto the zero fermion sector.

If we have the superpotential 
\begin{align}
\label{n2susy12a}
W(\Phi)=f\ln \left(\sum_{a}\Phi_{a}\Phi_{a}\right)
\end{align}
with $f$ being a constant, 
then we find \cite{Galajinsky:2006hq}
\begin{align}
\label{n2susy12b}
S=\frac12 \int dt 
\sum_{a=1}^{N} \left[
\dot{x}_{a}^{2}
+i\dot{\overline{\psi}}_{a}\psi_{a}
-i\overline{\psi}_{a}\dot{\psi}_{a}
-\frac{f(f-2\psi_{a}\overline{\psi}_{a})}{x_{a}^{2}}
\right].
\end{align}
Unlike the Freedman-Mende model (\ref{n2susy10h}), 
the interaction terms are not pairwise but 
still possess the inverse square behavior. 
This is the $\mathcal{N}=2$ superconformal mechanics 
describing the motion of the $n$-particle center of mass 
and the corresponding central charge $Z$ in the 
superconformal algebra $\mathfrak{su}(1,1|1)$ is \cite{Galajinsky:2006hq}
\begin{align}
\label{n2susy12c}
Z=2f.
\end{align}

\subsection{Gauged superconformal mechanics}
We start with the $\mathcal{N}=2$ matrix 
superfield gauged mechanical action 
\cite{Fedoruk:2008hk,Fedoruk:2010bt,Fedoruk:2011aa}
\begin{align}
\label{n2susy13a}
S=\int dtd^{2}\theta 
\left[
\mathrm{Tr}\left(
\overline{\mathcal{D}}\mathcal{X}\mathcal{D}\mathcal{X}
\right)
+\frac12\overline{\mathcal{Z}}e^{2V}\mathcal{Z}
-c\mathrm{Tr}V
\right].
\end{align}
Here we have
\begin{itemize}
 \item the $\mathcal{N}=2$ Grassmann-even Hermitian $n\times n$ matrix
       superfield $\mathcal{X}_{a}^{b}(t,\theta,\overline{\theta})$
       which satisfies $(\mathcal{X})^{\dag}=\mathcal{X}$ and 
transforms as the adjoint representation of $U(n)$; the
       $(\bm{1},\bm{2},\bm{1})$ supermultiplet

\item the $\mathcal{N}=2$ Grassmann-even chiral superfield 
$\mathcal{Z}_{a}(t_{L},\theta)$, 
$\overline{\mathcal{Z}}^{a}(t_{R},\overline{\theta})$, 
$t_{L,R}=t\pm i\theta\overline{\theta}$ which transform as 
the fundamental representations of $U(n)$; the 
$(\bm{2},\bm{2},\bm{0})$ supermultiplets

\item the $\mathcal{N}=2$ Grassmann-even complex 
$n\times n$ matrix bridge superfield $b_{a}^{b}(t,\theta,\overline{\theta})$ 
which satisfies $\overline{b}:=b^{\dag}$.
\end{itemize}
Note that gauge superfields are described by the complex $n\times n$ 
matrix bridge superfields $b$ or by the prepotential $V$ defined by 
\begin{align}
\label{n2susy13b}
e^{2V}&=e^{-i\overline{b}}e^{ib}.
\end{align}
The covariant derivatives are defined by
\begin{align}
\label{n2susy13c}
\mathcal{D}\mathcal{X}&=D\mathcal{X}+i[\mathcal{A},\mathcal{X}],&
\overline{\mathcal{D}}\mathcal{X}&=
\overline{D}\mathcal{X}+i[\overline{\mathcal{A}},\mathcal{X}]
\end{align}
where \footnote{The notation here is different from (\ref{n2susy2a}).}
\begin{align}
\label{n2susy13d}
D&=\frac{\partial}{\partial
 \theta}+i\overline{\theta}\frac{\partial}{\partial t},&
\overline{D}&=-
\frac{\partial}{\partial \overline{\theta}}
+i\theta \frac{\partial}{\partial t},&
\left\{D,\overline{D}\right\}&=-2i\partial_{t}
\end{align}
where the connections $\mathcal{A}$ are deduced from the bridge 
superfields 
\begin{align}
\label{n2susy13e}
\mathcal{A}&=-ie^{i\overline{b}}
\left(
De^{-i\overline{b}}
\right),&
\overline{\mathcal{A}}&=
-ie^{ib}\left(
\overline{D}e^{-ib}
\right).
\end{align}

The superconformal boost transformations are \cite{Ivanov:1988it}
\begin{align}
\label{n2susy14a}
\delta t&=-i\left(
\eta\overline{\theta}+\overline{\eta}\theta
\right)t,&
\delta \theta&=
-\eta(t+i\theta\overline{\theta}),\\
\label{n2susy14b}
\delta\overline{\theta}&=
-\eta(t-i\theta\overline{\theta}),&
\delta(dtd^{2}\theta)&=0,\\
\label{n2susy14c}
\delta \mathcal{X}&=-i\left(
\eta\overline{\theta}+\overline{\eta}\theta
\right)\mathcal{X},&
\delta\mathcal{Z}&=0,\\
\label{n2susy14d}
\delta b&=0,&
\delta V&=0.
\end{align}

The action (\ref{n2susy13a}) is invariant under the 
$U(n)$ transformations \cite{Fedoruk:2008hk,Fedoruk:2010bt,Fedoruk:2011aa}
\begin{align}
\label{n2susy14e}
e^{ib}&\rightarrow e^{i\Lambda}e^{ib}e^{-i\lambda},&
e^{i\overline{b}}&\rightarrow e^{i\Lambda}
e^{i\overline{b}}e^{-i\overline{\lambda}},&
e^{2V}&\rightarrow e^{i\overline{\lambda}}e^{2V}
e^{-i\lambda},\\
\label{n2susy14f}
\mathcal{X}&\rightarrow e^{i\Lambda}\mathcal{X}e^{-i\Lambda},&
\mathcal{Z}&\rightarrow e^{i\lambda}\mathcal{Z},&
\overline{\mathcal{Z}}&\rightarrow 
\overline{\mathcal{Z}}e^{-i\overline{\lambda}}.
\end{align}
Here 
$\Lambda$ is the Hermitian $n\times n$ matrix gauge parameter 
and $\lambda$ are the complex $n\times n$ gauge parameters.

Alternatively if we use the $\Lambda$ gauge invariant 
superfields $V$, $\mathcal{Z}$ and the new Hermitian $n\times n$ 
matrix superfield 
\begin{align}
\label{n2susy14g}
\mathcal{Y}=e^{-ib}\mathcal{X}e^{i\overline{b}},
\end{align}
the action (\ref{n2susy13a}) can be written as
\begin{align}
\label{n2susy14h}
S=\int dtd^{2}\theta 
\left[
\mathrm{Tr}\left(
\overline{\mathcal{D}}\mathcal{Y}e^{2V}
\mathcal{D}\mathcal{Y}e^{2V}
\right)
+\frac12 \overline{\mathcal{Z}}e^{2V}
\mathcal{Z}
-c\mathrm{Tr}V
\right]
\end{align}
where the covariant derivatives are 
\begin{align}
\label{n2susy14i}
\mathcal{D}\mathcal{Y}&=
D\mathcal{Y}+e^{-2V}(De^{2V})\mathcal{Y},&
\overline{\mathcal{D}}\mathcal{Y}&=
\overline{D}\mathcal{Y}
-\mathcal{Y}e^{2V}\left(\overline{D}e^{-2V}\right).
\end{align}

The $\mathcal{N}=2$ superfields 
$V$, $\mathcal{Y}$, $\mathcal{Z}$ and $\overline{\mathcal{Z}}$ 
can be expressed in terms of the component fields as
\begin{align}
\label{n2susy15a}
V&=v+\theta \xi-\overline{\theta}\overline{\xi}
+\theta\overline{\theta}A,\\
\label{n2susy15b}
\mathcal{Y}&=x+\theta \psi-\overline{\theta}\overline{\psi}
+\theta\overline{\theta}y,\\
\label{n2susy15c}
\mathcal{Z}&=z+2i\theta\zeta+i\theta\overline{\theta}\dot{z},\\
\label{n2susy15d}
\overline{\mathcal{Z}}&=
\overline{z}+2i\overline{\theta}\overline{\zeta}
-i\theta\overline{\theta}\dot{\overline{z}}.
\end{align}
According to the gauge transformation (\ref{n2susy14e}), 
let us choose the gauge so that
\begin{align}
\label{n2susy15e}
V(t,\theta,\overline{\theta})&=\theta\overline{\theta}A_{0}(t).
\end{align}
After integrating out the auxiliary fields 
$\zeta,\overline{\zeta}$ and performing the Grassmann integrations, 
we get the $\mathcal{N}=2$ superconformal gauged mechanical action
\begin{align}
\label{n2susy15f}
S=\int dt&
\Biggl[
\mathrm{Tr}\left(Dx Dx\right)
+\frac{i}{2}\left(\overline{z}Dz-D\overline{z}z\right)\nonumber\\
&+i\mathrm{Tr}\left(
\overline{\psi}D\psi-D\overline{\psi}\psi
\right)-c\mathrm{Tr}A_{0}
\Biggr]
\end{align}
where the covariant derivatives are
\begin{align}
\label{n2susy15g1}
Dx&=\dot{x}+i[A_{0},x],&
Dz&=\dot{z}+iA_{0}z,\\
\label{n2susy15g2}
D\psi&=\dot{\psi}+i[\psi,A_{0}],&
D\overline{\psi}&=\dot{\overline{\psi}}
+i[\overline{\psi},A_{0}].
\end{align}
The action (\ref{n2susy15f}) is the supersymmetric generalization 
of the Calogero model whose bosonic part agrees with the Calogero model 
(\ref{calogero1a1}). 
The action is invariant with respect to the $U(n)$ gauge transformations 
\begin{align}
\label{n2susy15g3}
x&\rightarrow gXg^{-1},&
z&\rightarrow gz,\\
\label{n2susy15g4}
\psi&\rightarrow g\psi g^{-1},&
A_{0}&\rightarrow gA_{0}g^{-1}+i\dot{g}g^{-1}
\end{align}
where $g\in U(n)$. 
By fixing the gauge as in 
(\ref{calogero1a8}) and (\ref{calogero1b4}), 
$z, \overline{z}$ and the non-diagonal part of $x$ 
are eliminated and thus we have
\begin{align}
\label{n2susy15g5}
&\textrm{$n$ physical bosons $x_{a}^{a}$},\nonumber\\
&\textrm{2$n^{2}$ physical fermions $\psi^{b}_{a},\overline{\psi}_{a}^{b}$}.
\end{align}
This is different from the Freedman-Mende model (\ref{n2susy10h}) 
which possesses $n$ physical bosons and $2n$ physical fermions. 
which can be realized as
\begin{align}
\label{n2susy15g6}
(\bm{n},\bm{2n^{2}},\bm{2n^{2}-n})
=n (\bm{1},\bm{2},\bm{1})\oplus 
(n^{2}-n) (\bm{0},\bm{2},\bm{2}).
\end{align}
It has been pointed out \cite{Delduc:2006yp} 
that the supermultiplet (\ref{n2susy15g6}) can be 
obtained from $n$ $(\bm{1},\bm{2},\bm{1})$ supermultiplets 
by gauging procedure.

\section{$\mathcal{N}=4$ Superconformal mechanics}
\label{secscqm4}
As we have discussed in subsection \ref{secsupalg01a}, 
the most general superconformal algebra of 
$\mathcal{N}=4$ superconformal quantum mechanics is $D(2,1;\alpha)$. 
As opposed to the $\mathcal{N}=1$ superconformal algebra $\mathfrak{osp}(1|2)$ 
and $\mathcal{N}=2$ superconformal algebra $\mathfrak{su}(1,1|1)
\cong \mathfrak{osp}(2|2)$, 
the Lie superalgebra $D(2,1;\alpha)$ is a one-parameter family 
of superalgebra characterized by a real parameter $\alpha$. 
In order to construct the corresponding 
family of $\mathcal{N}=4$ superconformal quantum mechanical modelds  
parametrized by $\alpha$, 
it is desirable to find the 
inequivalent irreducible off-shell supermultiplets 
in a systematic way.

To this end there is the methodical way proposed in
\cite{Ivanov:2003tm} 
by means of the non-linear realizations technique 
\cite{Coleman:1969sm,Callan:1969sn,Volkov:1973vd}. 
We shall start from the superconformal algebra $D(2,1;\alpha)$, 
wihch contains three conformal charges $H,D,K$ which form 
$\mathfrak{sl}(2,\mathbb{R})$, 
four supercharges $Q^{i},\overline{Q}_{i}$, $i=1,2$,  
four superconformal charges $S^{i},\overline{S}_{i}$ 
and two commuting sets of $\mathfrak{su}(2)$ R-symmetry generators 
$J,\overline{J},J_{3}$ and $I,\overline{I},I_{3}$ 
\footnote{
Here we use the notation in
\cite{Frappat:1996pb,Ivanov:2002pc,Ivanov:2003tm}, 
which is slightly different from 
our previous $\mathcal{N}=1$ and $\mathcal{N}=2$
cases in that the signs of the anti-commutators (\ref{d21a1d}) and 
the covariant derivatives (\ref{n4susy1a2}) 
and the supercharges (\ref{n4susy2b}). }
\begin{align}
\label{d21a1a}
\begin{array}{ccc}
[H,D]=iH,
&[K,D]=-iK,
&[H,K]=2iD,
\cr
\end{array}
\end{align}
\begin{align}
\label{d21a1b}
[H,Q^{i}]&=0,&
[D,Q^{i}]&=-\frac{i}{2}Q^{i},&
[K,Q^{i}]&=-iS^{i},\\
\label{d21a1c}
[H,S^{i}]&=iQ^{i},&
[D,S^{i}]&=\frac{i}{2}Q^{i},&
[K,S^{i}]&=0,
\end{align}
\begin{align}
\label{d21a1d}
\{Q^{i},\overline{Q}_{j}\}&=-2\delta^{i}_{j}H,&
\{S^{i},\overline{S}_{j}\}&=-2\delta^{i}_{j}K,&
\{Q^{i},S^{j}\}&=-2(1+\alpha)\epsilon^{ij}\overline{I},\nonumber\\
\{Q^{1},\overline{S}_{2}\}&=2\alpha\overline{J}&
\{Q^{1},\overline{S}_{1}\}&=-2D-2\alpha J_{3}+2(1+\alpha)I_{3},\nonumber\\
\{Q^{2},\overline{S}_{1}\}&=-2\alpha J,&
\{Q^{2},\overline{S}_{2}\}&=-2D+2\alpha J_{3}+2(1+\alpha)I_{3},
\end{align}
\begin{align}
\label{d21a1e}
[J_{3},Q^{1}]&=-\frac{i}{2}Q^{1},&
[J_{3},Q^{2}]&=\frac{i}{2}Q^{2},&
[J,Q^{1}]&=-iQ^{2},&
[J,\overline{Q}_{2}]&=i\overline{Q}_{1},\nonumber\\
[J_{3},S^{1}]&=-\frac{i}{2}S^{1},&
[J_{3},S^{2}]&=\frac{i}{2}S^{2},&
[J,S^{1}]&=-iS^{2},&
[J,\overline{S}_{2}]&=i\overline{S}_{1},\nonumber\\
[I_{3},Q^{i}]&=-\frac{i}{2}Q^{i},&
[I,Q^{i}]&=-i\overline{Q}^{i},&
[I_{3},S^{i}]&=-\frac{i}{2}S^{i},&
[I,S^{i}]&=-i\overline{S}^{i},
\end{align}
\begin{align}
\label{d21a1f}
[J_{3},J]&=iJ,&
[J_{3},\overline{J}]&=-i\overline{J},&
[J,\overline{J}]&=-2iJ_{3},\nonumber\\
[I_{3},I]&=iI,&
[I_{3},\overline{I}]&=-i\overline{J},&
[I,\overline{I}]&=-2iJ_{3}.
\end{align}
The R-symmetry group contains two $SU(2)$ factors 
generated by $J,\overline{J},J_{3}$ and $I,\overline{I},I_{3}$. 
Looking at the commutation relations (\ref{d21a1e}), 
$J$ corresponds to the rotations indices $i$ of $\theta_{i}$ 
while $I$ mixes $\theta_{i}$ with their complex conjugates.

Here we take bosonic conformal generators $H,D,K$  as Hermitian operators 
\begin{align}
\label{d21a1g}
(H)^{\dag}&=H,&
(D)^{\dag}&=D,&
(K)^{\dag}&=K
\end{align}
while the other operators are chosen so that
\begin{align}
\label{d21a1h}
(J)^{\dag}&=\overline{J},&
(J_{3})^{\dag}&=-J_{3},\\
\label{d21a1i}
(I)^{\dag}&=\overline{I},&
(I_{3})^{\dag}&=-I_{3},\\
\label{d21a1j}
\overline{(Q^{i})}&=\overline{Q}_{i},&
\overline{(S^{i})}&=\overline{S}_{i}.
\end{align}

The parameter $\alpha$ only appears  
in the anti-commutation relations (\ref{d21a1d}), 
from which we see that 
two $\mathfrak{su}(2)$ R-symmetry algebras 
appear with relative weights $\alpha$ and $-(1+\alpha)$. 
Note that the conformal algebra $\mathfrak{sl}(2,\mathbb{R})$ 
has relative weight $1$. 
Thus the transformation 
\begin{align}
\label{d21a2a}
\alpha\leftrightarrow -(1+\alpha)
\end{align}
exchanges the role of two R-symmetry algebras; 
$J\leftrightarrow I$. 
On the other hand, the transformation 
\begin{align}
\label{d21a2b}
\alpha\leftrightarrow \frac{1}{\alpha}
\end{align}
is not well-defined for our real $D(2,1;\alpha)$ superalgebra 
because it exchanges the role of the non-compact conformal algebra
$\mathfrak{sl}(2,\mathbb{R})$ and the compact R-symmetry algebra 
$\mathfrak{su}(2)$. 

In particular we have the isomorphism
\begin{align}
\label{d212a2c}
D(2,1;\alpha)
\cong
\begin{cases}
\mathfrak{su}(1,1|2)+ \mathfrak{su}(2)&\textrm{for $\alpha=-1,0$} \cr
\mathfrak{osp}(4^{*}|2)&\textrm{for $\alpha=1,-2$}\cr
\mathfrak{osp}(4|2)&\textrm{for $\alpha=-\frac12$}\cr
\end{cases}
\end{align}
At $\alpha=-1$ and $\alpha=0$ 
one of the R-symmetry $\mathfrak{su}(2)$ algebra is decoupled 
and the superalgebra $D(2,1;-1)$ is isomorphic to the semi-direct sum 
$\mathfrak{su}(1,1|2)+\mathfrak{su}(2)$ 
\footnote{We use 
$\oplus$ for the direct sum and 
$+$ for the semi-direct sum.}. 
In this case one can extend the $\mathfrak{su}(1,1|2)$ 
superalgebra by adding the central charges. 
To see this let us put the $\mathfrak{su}(2)$ generators 
$J$, $\overline{J}$ and $J_{3}$ as
\begin{align}
\label{d212a1f1}
Z&\equiv \alpha J,&
\overline{Z}&\equiv \alpha\overline{J},&
Z_{3}&\equiv \alpha J_{3}
\end{align}
where $Z$, $\overline{Z}$ and $Z_{3}$ 
commute with everything. 
Then the new generators $Z$, $\overline{Z}$ and 
$Z_{3}$ only appear in the anti-commutations (\ref{d21a1d}) 
and they now become
\begin{align}
\label{d21a1d1}
\{Q^{i},\overline{Q}_{j}\}&=-2\delta^{i}_{j}H,&
\{S^{i},\overline{S}_{j}\}&=-2\delta^{i}_{j}K,&
\{Q^{i},S^{j}\}&=-2\epsilon^{ij}\overline{I},\nonumber\\
\{Q^{1},\overline{S}_{2}\}&=2\overline{Z},&
\{Q^{1},\overline{S}_{1}\}&=-2D-2Z_{3}+2I_{3},\nonumber\\
\{Q^{2},\overline{S}_{1}\}&=-2Z,&
\{Q^{2},\overline{S}_{2}\}&=-2D+2Z_{3}+2I_{3}.
\end{align}
Hence the three generators $Z$, $\overline{Z}$ and $Z_{3}$ 
are identified with the central charges. 
Note that we can only have single nonvanishing central charge 
by taking into account the $SU(2)$ transformation on the 
three central charges.

As its name suggests, 
$D(2,1;\alpha)$ is regarded as a deformation of 
the superalgebra $D(2,1)=\mathfrak{osp}(4|2)$ that 
corresponds to the case $\alpha=1$, 
however, 
we are now considering the 
even part of $D(2,1;\alpha)$ 
as $\mathfrak{sl}(2)\oplus \mathfrak{su}(2)\oplus \mathfrak{su}(2)$ 
not as $\mathfrak{sl}(2)\oplus \mathfrak{sl}(2)\oplus
\mathfrak{sl}(2)$. 
The first $\mathfrak{sl}(2)$ factor is 
the conformal algebra and the remaining two 
factors are replaced with the compact algebras $\mathfrak{su}(2)$. 
Consequently $\mathfrak{so}^{*}(4)$, 
the non-compact version of the original factor $\mathfrak{so}(4)$ 
shows up for $\alpha=1$. 
We see that the case of $\alpha=-\frac12$ 
is self-dual under the transformation (\ref{d21a2a}). 
In our case this degenerate case realizes 
the $\mathfrak{so}(4)\cong \mathfrak{su}(2)\oplus \mathfrak{su}(2)$ factor 
and all the other cases can be thought of as 
the deformations of $D(2,1;\frac12)= \mathfrak{osp}(4|2)$.

Using the generators of $D(2,1;\alpha)$, 
let us consider the supercoset of $D(2,1;\alpha)$
\begin{align}
\label{d21a3}
g=e^{itH}
e^{iuD}
e^{izK}
e^{\theta_{i}Q^{i}+\overline{\theta}^{i}\overline{Q}_{i}}
e^{\psi_{i}S^{i}+\overline{\psi}^{i}\overline{S}_{i}}
e^{i\varphi J+i\overline{\varphi}\overline{J}}
e^{\phi J_{3}}
\end{align}
where 
the parameters $t,\theta_{i},\overline{\theta}^{i}$ 
are the coordinates of the $\mathcal{N}=4$ 
superspace $\mathbb{R}^{(1|4)}$ 
and the other parameters 
$u,z,\psi_{i},\overline{\psi}^{i},\varphi,\overline{\varphi}$ and $\phi$ 
are the $\mathcal{N}=4$ Goldstone superfields. 
Note that the R-symmetry group $SU(2)$ generated by 
$(I,\overline{I},I_{3})$, which mixes the fermionic charges 
with their conjugates, is taken into the supercoset 
but considered as the stability subgroup. 
Note that our chice of the supercoset (\ref{d21a3}) 
is allowed for the case of $\alpha\neq 0$ where 
the generators $(J,\overline{J},J_{3})$ exist.

From the supercoset one can extract 
left-covariant Cartan one-form $\Omega$
\begin{align}
\label{d21a4}
\Omega=g^{-1}dg.
\end{align}
Expanding $\Omega$ over the generators, 
we find the the corresponding one-forms \cite{Ivanov:2003tm}
\begin{align}
\label{d21a5a}
\omega_{D}&=idu-2\left(\overline{\psi}^{i}d\theta_{i}
+\psi_{i}d\overline{\theta}^{i}\right)
-2iz d\tilde{t},\\
\label{d21a5b}
\omega_{V}&=
\frac{e^{-i\phi}}{1+\Lambda\overline{\Lambda}}
\left(id\Lambda+\hat{\omega}_{J}
+\Lambda^{2}\hat{\overline{\omega}}_{J}
-\Lambda \hat{\omega}_{J_{3}}\right),\\
\label{d21a5c}
\overline{\omega}_{V}&=
\frac{e^{i\phi}}{1+\Lambda\overline{\Lambda}}
\left(
id\overline{\Lambda}
+\hat{\overline{\omega}}_{J}
+\overline{\Lambda}^{2}\hat{\omega}_{J}
+\overline{\Lambda}\hat{\omega}_{J_{3}}
\right),\\
\label{d21a5d}
\omega_{J_{3}}
&=
d\phi+\frac{1}{1+\Lambda\overline{\Lambda}}
\left[
i\left(
d\Lambda\overline{\Lambda}
-\Lambda d\overline{\Lambda}
\right)
+\left(1-\Lambda\overline{\Lambda}\right)\hat{\omega}_{J_{3}}
-2\left(
\Lambda\hat{\overline{\omega}}_{J}
-\overline{\Lambda}\hat{\omega}_{J}
\right)
\right]
\end{align}
where 
\begin{align}
\label{d21a5d}
\hat{\omega}_{J}
&=2\alpha \left[
\psi_{2}d\overline{\theta}^{1}
-\overline{\psi}^{1}\left(
d\theta_{2}-\psi_{2}d\tilde{t}
\right)
\right],\\
\label{d21a5e}
\hat{\overline{\omega}}_{J}
&=2\alpha \left[
\overline{\psi}^{2}d\theta_{1}
-\psi_{1}\left(d\overline{\theta}^{2}
-\overline{\psi}^{2}d\tilde{t}
\right)
\right],\\
\label{d21a5f}
\hat{\omega}_{J_{3}}
&=2\alpha\left[
\psi_{1}d\overline{\theta}^{1}
-\overline{\psi}^{2}d\theta_{1}
-\psi_{2}d\overline{\theta}^{2}
+\overline{\psi}^{2}d\theta_{2}
+\left(
\overline{\psi}^{1}\psi_{1}
-\overline{\psi}^{2}\psi_{2}
\right)d\tilde{t}
\right],\\
\label{d21a5g}
d\tilde{t}&=dt+i\left(\theta_{i}d\overline{\theta}^{i}
+\overline{\theta}^{i}d\theta_{i}\right),\\
\label{d21a5h}
\Lambda&=\frac{\tan\sqrt{\varphi\overline{\varphi}}}
{\sqrt{\varphi\overline{\varphi}}}.
\end{align}

For the $\mathcal{N}=4$ superspace $\mathbb{R}^{(1|4)}$ 
parametrized by \cite{Ivanov:1988it}
\begin{align}
\label{n4susy1a1}
\mathbb{R}^{(1|4)}
&=(t,\theta_{i},\overline{\theta}^{j}),& 
(\theta_{i})^{\dag}&=\overline{\theta}^{i},&i,j&=1,2
\end{align} 
we will introduce the covariant derivatives 
\begin{align}
\label{n4susy1a2}
D^{i}&=\frac{\partial}{\partial\theta_{i}}
+i\overline{\theta}^{i}\frac{\partial}{\partial t},& 
\overline{D}_{j}&
=\frac{\partial}{\partial\overline{\theta}^{j}}
+i\theta_{j}\frac{\partial}{\partial t},&
\left\{D^{i},\overline{D}_{j}\right\}&=2i\delta^{i}_{j}\partial_{t}.
\end{align}
The supercharges $Q$ and $\overline{Q}$ can be expressed by 
\begin{align}
\label{n4susy2b}
Q^{i}&
=\frac{\partial}{\partial \theta_{i}}
-i\overline{\theta}^{i}\frac{\partial}{\partial t},&
\overline{Q}_{j}&
=\frac{\partial}{\partial \overline{\theta}^{j}}
-i\theta_{j}\frac{\partial}{\partial t},&
\left\{Q^{i},\overline{Q}_{j}\right\}&=-2i\delta^{i}_{j}\partial_{t}
\end{align}
in the superspace. 

By acting a particular element on the 
supercoset element (\ref{d21a3}) 
from the left, 
we can find the corresponding transformations.
\begin{enumerate}
 \item supersymmetry transformations

Acting the element 
\begin{align}
\label{n4susy2c1}
g_{\epsilon}=
e^{\epsilon_{i}Q^{i}+\overline{\epsilon}^{i}\overline{Q}_{i}}
\in D(2,1;\alpha),
\end{align}
we obtain the supersymmetry transformations
\begin{align}
\label{n4susy2c2}
\delta t&=i\left(
\theta\overline{\epsilon}
-\epsilon\overline{\theta}
\right),\\
\label{n4susy2c3}
\delta \theta_{i}
&=\epsilon_{i},\\
\label{n4susy2c4}
\delta \overline{\theta}^{i}
&=\overline{\epsilon}^{i}.
\end{align}

\item superconformal boost transformations

Acting the element 
\begin{align}
\label{n4susy2c5}
g_{\eta}=e^{\eta_{i}S^{i}+\overline{\eta}_{i}\overline{S}_{i}},
\end{align}
one finds the superconformal boost transformations 
\cite{Ivanov:2002pc,Ivanov:2003tm,Ivanov:2003nk}
\begin{align}
\label{n4susy2c6}
\delta t&=
-it\left(\eta\overline{\theta}
+\overline{\eta}\theta
\right)+
(1+2\alpha)
\left(
\theta\overline{\theta}
\right)
\left(
\eta\overline{\theta}
-\overline{\eta}\theta
\right),\\
\label{n4susy2c7}
\delta \theta_{i}
&=\eta_{i}t
-2i\alpha\theta_{i}
(\theta\overline{\eta})
+2i\left(
1+\alpha
\right)\theta_{i}
\left(
\overline{\theta}\eta
\right)
-i(1+2\alpha)
\eta_{i}(\theta\overline{\theta}),\\
\label{n4susy2c8}
\delta u&=
-2i\left(
\eta\overline{\theta}
+\overline{\eta}\theta
\right),\\
\label{n4susy2c9}
\delta\phi&=
2\alpha
\Biggl[
\overline{\eta}^{1}\theta_{1}
-\overline{\eta}^{2}\theta_{2}
-\eta_{1}\overline{\theta}^{1}
+\eta_{2}\overline{\theta}^{2}\nonumber\\
&+\left(
\overline{\eta}^{2}\theta_{1}
-\eta_{1}\overline{\theta}^{2}
\right)\Lambda
+\left(
\overline{\eta}^{1}\theta_{2}
-\eta_{2}\overline{\theta}^{1}
\right)\overline{\Lambda}
\Biggr],\\
\label{n4susy2c10}
\delta\Lambda&
=2i\alpha\Biggl[
\theta_{2}\overline{\eta}^{1}
-\overline{\theta}^{1}\eta_{2}
+\left(
\overline{\theta}^{2}\eta_{1}
-\theta_{1}\overline{\eta}^{2}
\right)\Lambda^{2}\nonumber\\
&+\left(
\overline{\theta}^{1}\eta_{1}
-\theta_{1}\overline{\eta}^{1}
+\theta_{2}\overline{\eta}^{2}
-\overline{\theta}^{2}\eta_{2}
\right)\Lambda
\Biggr]
\end{align}
and
\begin{align}
\label{n4susy2c11}
\delta (dtd^{4}\theta)
&=2i\left(\eta\overline{\theta}+\overline{\eta}\theta\right)
dtd^{4}\theta,\\
\label{n4susy2c12}
\delta D^{i}
&=i
\left[
(2+\alpha)(\eta\overline{\theta})
+\alpha(\theta\overline{\eta})
\right]
D^{i}\nonumber\\
&-2i(1+\alpha)(\overline{\eta}\overline{\theta})\overline{D}^{i}
-2i\alpha \left[
\eta^{(i}\overline{\theta}_{k)}
+\theta^{(i}\overline{\epsilon}_{k)}
\right]
D^{k},\\
\label{n4susy2c13}
\delta\overline{D}_{i}
&=i\left[
(2+\alpha)(\overline{\eta}\theta)
+\alpha(\overline{\theta}\eta)\overline{D}_{i}\right]
\nonumber\\
&-2i(1+\alpha)
(\theta\eta)D_{i}
-2i\alpha 
\left[
\eta_{(i}\overline{\theta}_{k)}
+\theta_{(i}\overline{\epsilon}_{k)}
\right]
\overline{D}^{k}.
\end{align}
\end{enumerate}

At this stage we are ready to derive 
the irreducible off-shell supermultiplets 
which allow us to 
construct the $D(2,1;\alpha)$ superconformal mechanics. 
The strategy is to 
extract the irreducible superfields 
from the Goldstone superfields
$u,z,\psi_{i},\overline{\psi}^{i},\varphi,\overline{\varphi}$ 
by imposing the appropriate constraints. 
Since the number of the fermionic Goldstone superfields is four 
which coincides with the minimal number of the fermionic fields 
in $\mathcal{N}=4$ supermultiplets, 
we attempt to reduce the number of the bosonic Goldstone superfields. 
It has been discussed \cite{Ivanov:2002pc,Ivanov:2003tm} that 
such irreducibility condition can be achieved by 
requiring that 
all spinor derivatives of all bosonic superfields 
are expressed in terms of the fermionic fields 
$\psi_{i}$ and $\overline{\psi}^{i}$. 
From the equations (\ref{d21a5a})-(\ref{d21a5d}), 
we see that this requirement 
corresponds to the constraints on the corresponding Cartan forms 
$\omega_{D}$, $\omega_{J}$, $\overline{\omega}_{V}$ $\omega_{J_{3}}$.

\subsection{$(4,4,0)$ supermultiplet}
Let us begin with the most general case where 
the supercoset (\ref{d21a4}) holds all four bosonic Goldstone 
superfields $u$, $\varphi$ ,$\overline{\varphi}$ and $\phi$. 
If we require that the all spinor covariant derivatives of 
these bosonic superfields can be expressed by
$\psi_{i}$,$\overline{\psi}^{i}$, 
then (\ref{d21a5a})-(\ref{d21a5d}) lead to 
\begin{align}
\label{n4susy2d1}
\omega_{D}=\omega_{J}|=\overline{\omega}_{J}|=\omega_{J_{3}}=0
\end{align}
where $|$ represents the restriction to spinor projection. 
The set of constraints (\ref{n4susy2d1}) can be rewritten as
\begin{align}
\label{n4susy2d2}
D^{(i}q^{j)}&=0,&
\overline{D}^{(i}q^{j)}&=0,&
D^{(i}\overline{q}^{j)}&=0,&
\overline{D}^{(i}\overline{q}^{j)}&=0
\end{align}
where 
\begin{align}
\label{n4susy2d3}
q^{1}&=\frac{e^{\frac12(\alpha u-i\phi)}}
{\sqrt{1+\Lambda\overline{\Lambda}}}\Lambda,&
q^{2}&=\frac{e^{\frac12(\alpha u-i\phi)}}
{\sqrt{1+\Lambda\overline{\Lambda}}},\nonumber\\
\overline{q}_{1}&=
\frac{e^{\frac12(\alpha u+i\phi)}}
{\sqrt{1+\Lambda\overline{\Lambda}}},&
\overline{q}_{2}&=
\frac{e^{\frac12 (\alpha u+i\phi)}}
{\sqrt{1+\Lambda\overline{\Lambda}}}
\end{align}
are identified with four $\mathcal{N}=4$ superfields. 
This multiplet was discussed in 
\cite{Coles:1990hr,Gibbons:1997iy,
Hellerman:1999nr,Hull:1999ng,Ivanov:2003tm,
Bellucci:2005xn,Delduc:2006yp,Delduc:2006pg,Delduc:2007gs} 
and was considered in $\mathcal{N}=4$ harmonic superspace 
\cite{Ivanov:2003nk}. 
The constraints (\ref{n4susy2d2}) lead to the 
following independent fields:
\begin{align}
\label{n4susy2d3a}
\begin{cases}
q^{i}&\textrm{$4$ physical bosons}\cr
D_{i}q^{i},\overline{D}_{i}q^{i},
D_{i}\overline{q}^{i},
\overline{D}_{i}\overline{q}^{i}&
\textrm{$4$ fermions}.\cr
\end{cases}
\end{align}
The superfield $q_{i}$ contains 4 bosonic, 4 fermionic fields 
and no auxiliary fields and is diagnosed as the 
$(\bm{4},\bm{4},\bm{0})$ supermultiplet. 
Since $q^{i}$ and their set of constraints (\ref{n4susy2d2}) are 
similar to the $d=4$ $\mathcal{N}=2$ hypermultiplet, 
it is also called hypermultiplet. 
However, the conditions (\ref{n4susy2d2}) for 
the $(\bm{4},\bm{4},\bm{0})$ multiplet defines the off-shell multiplet 
as opposed to the $d=4$ $\mathcal{N}=2$ hypermultiplet.

Remarkably it has been discussed that 
all other $\mathcal{N}=4$ supermultiplets can be 
obtained from $(\bm{4},\bm{4},\bm{0})$ multiplet 
via reduction process 
either on the component action \cite{Bellucci:2005xn} or on 
the superfield action \cite{Delduc:2006yp,Delduc:2006pg,Delduc:2007gs}. 
Accordingly the $(\bm{4},\bm{4},\bm{0})$ multiplet 
can be viewed as a fundamental multiplet.

Since we know the superconformal boost transformations 
(\ref{n4susy2c6})-(\ref{n4susy2c10})  
for the original Goldstone superfields, 
we can read off the superconformal boost transformations 
for the superfields $q^{i}$, $\overline{q}_{i}$
\begin{align}
\label{n4susy2d4}
\delta q^{i}=2i\alpha 
\left(
\overline{\theta}^{i}\eta_{j}-\theta^{i}\overline{\eta}_{j}
\right)q^{j}.
\end{align}
This leads to the transformations 
$\delta (q\overline{q})=-2i\alpha(\eta\overline{\theta}+\overline{\eta}\theta)
(q\overline{q})$, 
which cancel the transformation (\ref{n4susy2c11}) of the 
integration measure. 
Therefore we can write superconformally invariant superfield action 
\begin{align}
\label{n4susy2d5}
S=\int dtd^{4}\theta\ 
\left(q\overline{q}\right)^{\frac{1}{\alpha}}.
\end{align}
Note that this vanishes when $\alpha=-1$ 
due to the constraints (\ref{n4susy2d2}). 
For $\alpha=-1$ the superconformal superfield action is given by 
\cite{Ivanov:2003nk,Ivanov:2003tm}
\begin{align}
\label{n4susy2d6}
S=\int dtd^{4}\theta\  
\frac{\ln \left(q\overline{q}\right)}{q\overline{q}}.
\end{align}
It is worthwhile to remark that 
these two expressions (\ref{n4susy2d5}) and (\ref{n4susy2d6}) 
can be written uniformly by adding the overall factor as
\begin{align}
\label{n4susy2d7}
\frac{1}{1+\alpha}\int dtd^{4}\theta\ 
\left(q\overline{q}\right)^{\frac{1}{\alpha}}.
\end{align}
One can check that 
(\ref{n4susy2d7}) is regular for 
any $\alpha$ and coincides with (\ref{n4susy2d6}) at $\alpha=-1$.

Although there is a superpotential term for the 
$(\bm{4},\bm{4},\bm{0})$ multiplet \cite{Ivanov:2003nk} 
which is a Wess-Zumino type term 
\footnote{
The superfield Wezz-Zumino type potential term  
for all $\mathcal{N}=4$ multiplets can be represented manifestly  
only in the harmonic superspace \cite{Fedoruk:2011aa}.}  
of first order in time derivative, 
it does not produce any non-trivial potential for physical bosons. 
Therefore 
one cannot construct $\mathcal{N}=4$ superconformal mechanics 
with the non-trivial potential for the physical bosons 
by using the $(\bm{4},\bm{4},\bm{0})$ multiplet only. 
On the other hand, 
it has been discussed \cite{Delduc:2006yp,Delduc:2006pg} 
that the gauged action of the $(\bm{4},\bm{4},\bm{0})$ multiplet 
generates more generic actions.

 \subsection{$({3},{4},{1})$ supermultiplet}
Let us set $\phi=0$ in the supercoset (\ref{d21a3}). 
This enforces us to put the corresponding subgroup $U(1)\subset SU(2)$ 
into the stability subgroup and 
thus the resulting supercoset involves $SL(2,\mathbb{R})\times
SU(2)/U(1)$. 
To realize the spinor covariant derivatives 
of the remaining bosonic superfields $u$, $\Lambda$,
$\overline{\Lambda}$, we should impose the conditions
\begin{align}
\label{n4susy2e1}
\omega_{D}=\omega_{J}|=\overline{\omega}_{J}|=0.
\end{align}
The set of conditions (\ref{n4susy2e1}) 
can be expressed as
\begin{align}
\label{n4susy2e2}
D^{(i}V^{jk)}&=0,&
\overline{D}^{(i}V^{jk)}=0
\end{align}
where 
\begin{align}
\label{n4susy2e3}
V^{11}&=
-i\sqrt{2}e^{\alpha u}\frac{\Lambda}{1+\Lambda\overline{\Lambda}},&
V^{22}&=i\sqrt{2}e^{\alpha u}\frac{\overline{\Lambda}}
{1+\Lambda\overline{\Lambda}},&
V^{12}=\frac{i}{\sqrt{2}}
e^{\alpha u}\frac{1-\Lambda\overline{\Lambda}}
{1+\Lambda\overline{\Lambda}}.
\end{align}
Note that the $\mathcal{N}=4$ superfields $V^{ij}$ is real and 
satisfy the relations
\begin{align}
\label{n4susy2e4}
V^{ij}&=V^{ji},&
\overline{V^{ij}}&=\epsilon_{ik}\epsilon_{jl}V^{kl},&
V^{2}&:=V^{ij}V_{ij}=e^{2\alpha u}.
\end{align}
The superfield $V^{ij}$ obeying the constraints (\ref{n4susy2e2}) 
was firstly introduced in 
\cite{deCrombrugghe:1982un} and later discussed in 
\cite{Ivanov:1990jn,Berezovoj:1991ka,Maloney:1999dv,Ivanov:2002pc,
Ivanov:2003tm,Ivanov:2003nk,Delduc:2006yp,Delduc:2006pg}. 
The constraints (\ref{n4susy2e2}) 
give rise to the independent components 
\begin{align}
\label{n4susy2e5}
\begin{cases}
V^{11},V^{12},V^{12}&\textrm{$3$ physical bosons}\cr
D^{1}V^{12},D^{2}V^{12},\overline{D}^{1}V^{12},\overline{D}^{2}V^{12}
&\textrm{$4$ fermions}\cr
D^{i}\overline{D}^{j}V_{ij}&\textrm{$1$ auxiliary boson}\cr
\end{cases}
\end{align}
Thus we can identify the superfield $V^{ij}$ 
with the $(\bm{3},\bm{4},\bm{1})$ supermultiplet. 
Since the constraints (\ref{n4susy2e2}) are obtained 
by the dimensional reduction from the constraints 
of the $d=4$ $\mathcal{N}=2$ tensor multiplet 
\cite{Wess:1975ns}, 
the $(\bm{3},\bm{4},\bm{1})$ multiplet 
is also called tensor multiplet 
\footnote{
The superfield $V^{ij}$ can also be obtained 
by the dimensional reduction from $d=4$ $\mathcal{N}=1$ 
vector multiplet \cite{Ivanov:1990jn} 
as the spatial component of $d=4$ Abelian gauge vector connection superfield. 
}. 

From (\ref{n4susy2c8})-(\ref{n4susy2c10}), 
one can read the $D(2,1;\alpha)$ 
superconformal boost transformations of $V^{ij}$ 
\begin{align}
\label{n4susy2e6}
\delta V^{ij}
=-2i\alpha 
\left[
\left(
\eta\overline{\theta}+\overline{\eta}\theta
\right)V^{ij}
+
\left(
\eta^{(i}\overline{\theta}_{k}
-\overline{\eta}_{k}\theta^{(i}
\right)V^{j)k}
+\left(
\eta_{k}\overline{\theta}^{(i}
-\overline{\eta}^{(i}\theta_{k}
\right)V^{j)k}
\right]
\end{align} 
The superfield action for the kinetic term is given by 
\cite{Ivanov:2002pc,Ivanov:2003nk,Ivanov:2003tm}
\begin{align}
\label{n4susy2e7}
S_{\textrm{kin}}&=\int dt d\theta\ 
(V^{2})^{\frac{1}{2\alpha}}
\end{align}
where $V^{2}$ is defined in (\ref{n4susy2e4}). 
The action (\ref{n4susy2e7}) vanishes for $\alpha=-1$. 
The superfield action for the kinetic term in the case of $\alpha=-1$ is
\begin{align}
\label{n4susy2e8}
S_{\textrm{kin}}=-\frac12 \int dtd^{4}\theta\ 
(V^{2})^{-\frac12}\ln V^{2}.
\end{align}
It has been pointed out \cite{Ivanov:2002pc} that 
both of the action (\ref{n4susy2e7}) and (\ref{n4susy2e8}) 
can be described in a unified form as
\begin{align}
\label{n4susy2e7a}
S_{\textrm{kin}}&=\frac{1}{1+\alpha}\int dt d\theta\ 
(V^{2})^{\frac{1}{2\alpha}}.
\end{align}

The superconformally invariant 
potential term for the $(\bm{3},\bm{4},\bm{1})$ multiplet 
can be written as \cite{Ivanov:2002pc}
\begin{align}
\label{n4susy2e9}
S_{\textrm{pot}}&=-i\sqrt{2}\int dtd^{4}\theta\ 
\left[
\int_{0}^{1}dy \partial_{y}\mathcal{W}
\frac{1}{\sqrt{V^{2}}}
\right]
\end{align}
where $\mathcal{W}$ is the prepotential satisfying
\begin{align}
\label{n4susy2e10}
V^{ij}&=D^{(i}\overline{D}^{j)}\mathcal{W},&
\overline{W}&=-\mathcal{W}.
\end{align}
Note that the constraints (\ref{n4susy2e2}) 
are solved by an unconstraint prepotential $\mathcal{W}$. 
Alternative way to obtain the potential term 
for the $(\bm{3},\bm{4},\bm{1})$ multiplet has been proposed 
as an integral over the analytic harmonic superspace \cite{Ivanov:2003nk}. 

Combining the kinetic terms (\ref{n4susy2e7a}) and 
the potential terms (\ref{n4susy2e9}), 
we find the bosonic superconformal actions 
in component fields as \cite{Ivanov:2002pc}
\begin{align}
\label{n4susy2e10a}
S_{\textrm{bosonic}}
&=\mu^{-1}\frac{\alpha^{2}}{1+\alpha}
(S_{\textrm{kin}})_{\textrm{bosonic}}
+\nu(S_{\textrm{pot}})_{\textrm{bosonic}}\nonumber\\
&=\int dt 
\left[
\mu^{-1}\alpha^{2}e^{u}\dot{u}^{2}
+4\mu^{-1}e^{u}
\frac{\dot{\Lambda}\dot{\Lambda}}
{(1+\Lambda\overline{\Lambda})^{2}}
-\frac14\mu\nu^{2}e^{-u}
+i\nu\frac{\overline{\Lambda}\dot{\Lambda}-\Lambda\dot{\overline{\Lambda}}}
{1+\Lambda\overline{\Lambda}}
\right]\nonumber\\
&=
\frac12 \int dt 
\left[
4\alpha^{2}\mu\dot{r}^{2}
+\mu r^{2}\left(
\dot{\vartheta}^{2}+\sin^{2}\vartheta \dot{\varphi}^{2}
\right)
-\frac{\nu^{2}}{\mu r^{2}}
+2\nu \cos\vartheta \dot{\varphi}
\right]\nonumber\\
&=\int dt 
\left[
\mu g_{ij}(X)\dot{X}^{i}\dot{X}^{j}
-\frac{1}{4\mu}\frac{\nu^{2}}{|\bm{X}|^{2}}
+2i\nu\frac{\epsilon^{3ij}X^{i}\dot{X}^{j}}
{\left(X^{2}+|\bm{X}|\right)|\bm{X}|}
\right]
\end{align}
where 
\begin{align}
\label{n4susy2e10a1}
\Lambda&=\tan\frac{\vartheta}{2}e^{i\varphi},&
e^{\frac{u}{2}}&=\frac{1}{\sqrt{2}}\mu r,\\
\label{n4susy2e10a2}
g_{ij}(X)&=\delta_{ij}+(4\alpha^{2}-1)\frac{X_{i}X_{j}}{|\bm{X}|^{2}}.
\end{align}

Observing the two explicit expressions 
(\ref{n4susy2d3}) and (\ref{n4susy2e4}) for the 
two superfields $q^{i}$ and $V^{ij}$ 
in terms of the initial Goldstone superfields, 
we can express the superfields $V^{ij}$ as
\begin{align}
\label{n4susy2e11}
V^{11}&=-i\sqrt{2}q^{1}\overline{q}^{1},&
V^{22}&=-i\sqrt{2}q^{2}\overline{q}^{2},&
V^{12}&=-\frac{i}{\sqrt{2}}
\left(q^{1}\overline{q}^{2}+q^{2}\overline{q}^{1}\right).
\end{align}
Also one can check that if 
the the irreducible constraints (\ref{n4susy2d2}) for 
the $(\bm{4},\bm{4},\bm{0})$ multiplet are 
satisfied by $q^{i},\overline{q}^{i}$, 
then the constraints (\ref{n4susy2e2}) 
for the $(\bm{3},\bm{4},\bm{1})$ multiplet 
are also solved by (\ref{n4susy2e11}) \cite{Ivanov:2003nk}. 
However, it is important to note 
that (\ref{n4susy2e11}) are not general but rather special solutions 
to the$(\bm{3},\bm{4},\bm{1})$ multiplet. 
So the generic $(\bm{3},\bm{4},\bm{1})$ multiplet cannot 
be covered by (\ref{n4susy2e11}).

\subsection{$(2,4,2)$ supermultiplet}
Now we will put $u=0$, $z=0$, $\phi=0$ 
in the supercoset (\ref{d21a3}). 
Then the supercoset contain only 
two bosonic fields $\varphi$, $\overline{\varphi}$ 
or equivalently $\Lambda$, $\overline{\Lambda}$, 
which parametrize the two-sphere $S^{2}\sim SU(2)/U(1)$. 
The condition that 
the spinor covariant derivatives of $\varphi$, $\overline{\varphi}$ 
can be expressed in terms of $\psi$, $\overline{\psi}$ is 
\begin{align}
\label{n4susy2f1}
\omega_{J}|=\omega_{\overline{J}}|=0.
\end{align}
For $\alpha\neq-1$
these the conditions (\ref{n4susy2f1}) are written as
\begin{align}
\label{n4susy2f2}
D^{1}\Lambda&=-\Lambda D^{2}\Lambda,&
\overline{D}_{2}\Lambda&=\Lambda \overline{D}_{1}\Lambda.
\end{align}
Under the constraints (\ref{n4susy2f2}) 
the superfield $\Lambda$, $\overline{\Lambda}$ 
yields the independent component fields
\begin{align}
\label{n4susy2f3}
\begin{cases}
\Lambda,\overline{\Lambda}&\textrm{$2$ physical bosons}\cr
-D^{1}\overline{\Lambda},\overline{D}_{1}\Lambda,
D^{2}\Lambda,-\overline{D}_{2}\overline{\Lambda}
&\textrm{$4$ fermions}\cr
\overline{D}_{1}D^{2}\Lambda,
\overline{D}_{2}D^{1}\overline{\Lambda}&
\textrm{$2$ auxiliary bosons},\cr
\end{cases}
\end{align}
which implies the $(\bm{2},\bm{4},\bm{2})$ supermultiplet. 
This multiplet is called non-linear chiral multiplet 
because the constraints (\ref{n4susy2f2}) can be viewed as 
the modified chirality conditions so that they are also covariant 
with respect to $D(2,1;\alpha)$. 
Note that, apart from the non-linear realization of $D(2,1;\alpha)$, 
the $\mathcal{N}=4$ chiral multiplet $(\bm{2},\bm{4},\bm{2})$ is constructed 
by a complex superfields $\phi$, $\overline{\phi}$ obeying the 
constraints 
\begin{align}
\label{n4susy2f3a}
D^{i}\phi&=0,&
\overline{D}_{j}\overline{\phi}&=0.
\end{align}

It has been discussed that one cannot 
construct superconformal superfield actions 
out of the $(\bm{2},\bm{4},\bm{2})$ multiplet alone 
due to the absence of the dilaton $u$ \cite{Ivanov:2003tm}. 
In order to obtain superconformal superfield actions, 
the coupling to some other $\mathcal{N}=4$ supermultiplets is needed. 

In terms of the hypermultiplet $q$, $\overline{q}$, 
the superfield $\Lambda$, $\overline{\Lambda}$ 
can be written as 
\begin{align}
\label{n4susy2f4}
\Lambda&=-\frac{q^{1}}{q^{2}},&
\overline{\Lambda}&=-\frac{\overline{q}_{1}}{\overline{q}_{2}}.
\end{align}
These are just the special solutions 
to the constraint equations (\ref{n4susy2f2}) for the 
non-linear chiral multiplet.

\subsection{$(1,4,3)$ supermultiplet}
Let us retain the dilaton $u$ alone 
in the supercoset (\ref{d21a3}). 
This corresponds to putting two $SU(2)$ R-symmetry factors 
into the stability subgroup. 
The irreducible condition 
\begin{align}
\label{n4susy2f5}
\omega_{D}|=0
\end{align}
just implies that 
the four spinor derivatives of $u$ is expressed by 
the four fermionic Goldstone superfield $\psi,\overline{\psi}$. 
Therefore the equation 
(\ref{n4susy2f5}) does not impose any constraints on the 
superfield $u$. 
The independent component fields are \cite{Ivanov:1988it}
\begin{align}
\label{n4susy4f6}
\begin{cases}
e^{u}&\textrm{$1$ physical bosons}\cr
D^{i}u, \overline{D}_{i}u&\textrm{$4$ fermions}\cr
[D^{(i},\overline{D}^{j)}]e^{u},
[D^{i},\overline{D}_{i}]e^{u}&\textrm{$3$ auxiliary bosons}\cr
\end{cases}
\end{align}
and this means the $(\bm{1},\bm{4},\bm{3})$ supermultiplet. 
However, as was shown in \cite{Ivanov:1988it}, 
one should impose additional irreducible constraints on 
the dilaton $u$
\begin{align}
\label{n4susy2f7}
D^{i}D_{i}e^{-\alpha u}
=\overline{D}_{i}\overline{D}^{i}e^{-\alpha u}
=[D^{i},\overline{D}_{i}]e^{-\alpha u}=0
\end{align}
for the minimal $(\bm{1},\bm{4},\bm{3})$ multiplet. 
It has been pointed out \cite{Ivanov:2002pc} that 
if we build up the $u$ superfield out of 
the $(\bm{3},\bm{4},\bm{1})$ superfield $V^{ij}$ 
satisfying (\ref{n4susy2e2}) as
\begin{align}
\label{n4susy2f8}
e^{-\alpha u}=\frac{1}{\sqrt{V^{2}}},
\end{align} 
then $u$ automatically obeys the minimal constraints 
(\ref{n4susy2f7}). 
Substituting the relation (\ref{n4susy2f8}) 
into (\ref{n4susy2e9}) and (\ref{n4susy2e8}), 
we obtain the superconformal superfield action 
\footnote{In the original work in \cite{Ivanov:1988it} 
only the $SU(1,1|2)$ invariant action with $\alpha=-1$ was considered.}
\begin{align}
\label{n4susy2g1}
S=\int dtd^{4}\theta\ e^{u}
\end{align}
for $\alpha\neq -1$ and 
\begin{align}
\label{n4susy2g2}
S=\int dtd^{4}\theta\ e^{u}u
\end{align}
for $\alpha=-1$. 
By putting the overall factor, 
we can express the superconformal superfield actions 
for both cases as \cite {Ivanov:2002pc} 
\begin{align}
\label{n4susyg3}
S=\frac{1}{1+\alpha}
\int dtd^{4}\theta\ e^{u}.
\end{align}
Combining (\ref{n4susy2c8}) and (\ref{n4susy2c11}), 
one can check that the superfield action (\ref{n4susyg3}) is 
invariant under the superconformal boost transformations. 
Note that (\ref{n4susyg3}) is not defined at $\alpha=0$ 
because of our choice of the supercoset (\ref{d21a3}) 
and it should be treated separately
\cite{Delduc:2006yp,Delduc:2006pg}. 

Inserting the appropriate set of component fields 
which solve the minimal constraints (\ref{n4susy2f7}) 
into the superfield action (\ref{n4susyg3}),  
integrating over 
the Grassmann coordinates $\theta_{i}, \overline{\theta}^{i}$ 
and integrating out the auxiliary fields, 
one finds the one particle $D(2,1;\alpha)$ superconformal mechanical 
model \cite{Wyllard:1999tm,Fedoruk:2011aa}
\begin{align}
\label{n4susyg4}
S=\frac12 \int 
dt \left[
\dot{x}^{2}+i\left(\overline{\psi}_{i}\dot{\psi}^{i}
-\dot{\overline{\psi}}_{i}\psi^{i}\right)
+\frac{2}{3}(1+2\alpha) \frac{\psi^{i}\overline{\psi}^{j}
\psi_{(i}\overline{\psi}_{j)}}{x^{2}}
\right].
\end{align}
Although the action (\ref{n4susyg4}) does not possess 
bosonic potential at the classical level, 
upon the quantization 
the anti-commutation for the fermions may yield 
a purely bosonic potential term. 
We see that the potential terms just flip the overall sign 
under the transformation $\alpha$ (\ref{d21a2a}). 

As we have already seen (\ref{d212a2c}), 
when $\alpha=-1,0$ the 
$\mathcal{N}=4$ superconformal algebra $D(2,1;\alpha)$ 
is isomorphic to the semi-direct sum of 
$\mathfrak{su}(1,1|2)$ and $\mathfrak{su}(2)$, 
which implies that 
one of the $SU(2)$ symmetry is broken 
and the superalgebra $\mathfrak{su}(1,1|2)$ 
allows for the central charge. 
So the irreducible constraints for the 
bosonic Goldstone superfields can be weakened 
by adding the central charge \cite{Wyllard:1999tm,Fedoruk:2011aa}. 
The constraints (\ref{n4susy2f7}) 
can be modified as
\begin{align}
\label{n4susyg5}
D^{i}D_{i}e^{-\alpha u}
&=0,&
\overline{D}_{i}\overline{D}^{i}e^{-\alpha u}
&=0,&
[D^{i},\overline{D}_{i}]e^{-\alpha u}&=c
\end{align}
or 
\begin{align}
\label{n4susyg6}
D^{i}D_{i}e^{-\alpha u}
&=c,&
\overline{D}_{i}\overline{D}^{i}e^{-\alpha u}
&=c,&
[D^{i},\overline{D}_{i}]e^{-\alpha u}&=0
\end{align}
where $c$ is the central charge of the $\mathfrak{su}(1,1|2)$ 
superalgebra. 
The two constraints correspond 
to the case where the broken $SU(2)$ symmetry 
is taken as the rotation of $\theta$ coordinates 
and $\overline{\theta}$ coordinates respectively \cite{Ivanov:1988it}. 
The solutions to the 
new constraint equations acquire 
the additional term proportional to $\theta\overline{\theta}c$. 
Then one obtains the one particle $SU(1,1|2)$ superconformal mechanical
action \cite{Ivanov:1988it,Wyllard:1999tm,Fedoruk:2011aa}
\begin{align}
\label{n4susyg7}
S=\frac12 \int dt 
\left[
\dot{x}^{2}
+i\left(\overline{\psi}_{i}\dot{\psi}^{i}
-\dot{\overline{\psi}}_{i}\psi^{i}
\right)
-\frac{\left(
c+\overline{\psi}_{i}\psi^{i}\right)^{2}}{x^{2}}
\right].
\end{align}
Note that 
the additional contribution from the central charge $c$ 
yields the inverse square type bosonic potential 
at the classical level.

\subsection{$(0,4,4)$ supermultiplet}
Although we have seen that 
the irreducibility conditions for the supermultiplet 
can be systematically obtained 
by means of the non-linear realization method, 
there is a further possible supermultiplet $(\bm{0},\bm{4},\bm{4})$
\footnote{At least the author does not know the $(\bm{0},\bm{4},\bm{4})$
supermultiplet based on the non-linear realization of the superconformal
group $D(2,1;\alpha)$.}. 
It is described by a fermionic analytic superfield 
in the harmonic superspace (HSS) \cite{Ivanov:2003nk}.

The harmonic superspace (HSS) is the extension of the original
superspace by introducing the new commuting 
harmonic coordinate $u_{i}^{\pm}$, $i=1,2$ 
parametrizing the internal degrees of freedom as 
the two-sphere $S^{2}\sim SU(2)/U(1)$ with $SU(2)$ being the R-symmetry 
\cite{Delduc:1992xp}.  
\begin{align}
\label{n4susyh0}
\mathbb{HR}^{(1+2|4)}
&=(t_{A},\theta^{+},\overline{\theta}^{+},\theta^{-},\overline{\theta}^{-},u_{i}^{+},u_{k}^{-})\nonumber\\
&=(\zeta,u_{i}^{+},u_{k}^{-},\theta^{-},\overline{\theta}^{-})
\end{align}
where 
\begin{align}
\label{n4susyh0a}
&t_{A}:=t-i(\theta^{+}\overline{\theta}^{-}+\theta^{-}\overline{\theta}^{+}),& 
&\theta^{\pm}=\theta^{i}u_{i}^{\pm},&  
&\overline{\theta}^{\pm}=\overline{\theta}^{i}u_{i}^{\pm},\\
\label{n4susyh0b}
&u^{+i}u_{i}^{-}=1,& 
&u_{i}^{+}u_{j}^{-}-u_{j}^{+}u_{i}^{-}=\epsilon_{ij}.
\end{align}
The significant property is the 
existence of an analytic subspace (ASS), 
which is the quotient of $\mathbb{H}^{(1+2|4)}$ 
by $\{\theta^{-},\overline{\theta}^{-}\}$
\begin{align}
\label{n4susyh0c}
\mathbb{AR}^{(1+2|2)}&=(\zeta,u)\nonumber\\
&=(t_{A},\theta^{+},\overline{\theta}^{+},u_{i}^{+},u_{k}^{-}).
\end{align}
The covariant derivatives in the analytic basis of HSS, 
$(\zeta,u,\theta^{-},\overline{\theta}^{-})$ are defined by
\begin{align}
\label{n4susyh0d}
D^{+}&=\frac{\partial}{\partial \theta^{-}},
&
\overline{D}^{+}&=-\frac{\partial}{\partial
 \overline{\theta}^{-}},\\
\label{n4susyh0e}
D^{-}&=
-\frac{\partial}{\partial \theta^{+}}
-2i\overline{\theta}^{-}\frac{\partial}{\partial t_{A}},&
\overline{D}^{-}&=
\frac{\partial}{\partial \overline{\theta}^{+}}
-2i\theta^{-}\frac{\partial}{\partial t_{A}}
\end{align}
and the harmonic covariant derivatives in the analytic basis of HSS are
\begin{align}
\label{n4susyh0f}
D^{\pm\pm}
&=\partial^{\pm\pm}
+2i\theta^{\pm}\overline{\theta}^{\pm}
\frac{\partial}{\partial t_{A}}
+\theta^{\pm}\frac{\partial}{\partial \theta^{\mp}}
+\overline{\theta}^{\pm}\frac{\partial}{\partial \overline{\theta}^{\mp}}.
\end{align}

The constraints for the $(\bm{0},\bm{4},\bm{4})$ superfield 
$\Psi^{+a}(\zeta,u)$, $a=1,2$ 
\footnote{The indices $a=1,2$ denote the doublet of 
the extra $SU(2)$ called the Pauli-G\"{u}rsey group
\cite{Galperin:2001uw}.} 
are givne by \cite{Ivanov:2003nk}
\begin{align}
\label{n4susyh1}
D^{++}\Psi^{+a}=0.
\end{align}
The solution of the constraint (\ref{n4susyh1}) is written as
\begin{align}
\label{n4susyh2}
\Psi^{+a}(\zeta,u)
&=\psi^{ia}u_{i}^{+}
+\theta^{+}\xi^{a}
+\overline{\theta}^{+}\overline{\xi}^{a}
+2i\theta^{+}\overline{\theta}^{+}\dot{\psi}^{ia}u_{i}^{-}
\end{align}
and the independent component fields are
\begin{align}
\label{n4susyh3}
\begin{cases}
\psi^{ia}&\textrm{4 fermions}\cr
\xi^{a},\overline{\xi}^{a}&\textrm{4 auxiliary bosons}.\cr
\end{cases}
\end{align}
The $(\bm{0},\bm{4},\bm{4})$ superfield $\Psi^{+a}$ has been 
discussed in \cite{Delduc:2006yp,Delduc:2006pg,Galajinsky:2014hza}. 
The action takes the form
\begin{align}
\label{n4susyh4}
S&=\frac12 \int dud\zeta^{--}\Psi^{+a}\Psi_{a}^{+}\nonumber\\
&=\int dt \left[
i\psi^{ia}\dot{\psi}_{ia}+\xi^{a}\overline{\xi}_{a}
\right].
\end{align}
Although the action (\ref{n4susyh4}) contains only 
the kinetic term of the free fermions 
and the quadratic term of the bosonic auxiliary fields, 
if we appropriately couple the $(\bm{0},\bm{4},\bm{4})$ multiplet 
to the other $\mathcal{N}=4$ supermultiplets, 
we may produce bosonic potentials 
\cite{Delduc:2006yp,Delduc:2006pg,Galajinsky:2014hza}.

\subsection{Multi-particle model}
\subsubsection{WDVV equation}
\label{wdvvsec1}
We have seen that 
the superspace and superfield formalism 
based on the non-linear realization technique is useful to 
build up $\mathcal{N}=4$ superconformal mechanical models 
possessing $D(2,1;\alpha)$ symmetry. 
However, it is known that 
the direct generalization of the one particle analysis 
does not work well for the construction of 
the $D(2,1;\alpha)$ multi-particle superconformal mechanical systems 
\footnote{In the case of $\alpha=-1,0$ with $SU(1,1|2)$ symmetry, 
the standard $\mathcal{N}=4$ superspace description can be generalized to 
the multi-particle case \cite{Krivonos:2008fx}.}. 
Hence it is insightful to investigate 
the construction for the 
$\mathcal{N}=4$ multi-particle superconformal mechanics 
in the component level.

Let us consider $N$ particles on $\mathbb{R}$ 
with canonical variables $x^{a}$ and their momenta 
$p_{p}$ where $a=1,\cdots, N$ label the particles. 
The $\mathcal{N}=4$ supersymmetry leads to 
two complex fermions $\psi_{i}^{a}, \overline{\psi}^{ai}$, $i=1,2$. 
In addition, 
we also consider a one pair of bosonic isospin variables 
$u^{i}$, $i=1,2$ which parametrize the internal degrees of freedom 
\footnote{It has been discussed \cite{Krivonos:2008fx} 
that isospin variables is needed in order to obtain the multi-particle 
$D(2,1;\alpha)$ superconformal mechanics for $\alpha\neq -1,0$. 
See \cite{Bellucci:2004wn} 
for $\alpha=-1,0$, i.e. $SU(1,1|2)$ superconformal mechanics.}. 

Now we impose the ansatz for the supercharges 
$Q_{i}$ and $\overline{Q}^{i}$ of the form \cite{Krivonos:2008fx}
\begin{align}
\label{wdvv001a}
Q_{i}&=p_{a}\psi^{a}_{i}
+U_{a}(x)K_{ij}\psi^{aj}
+iF_{abc}(x)\psi^{aj}\psi_{j}^{b}\overline{\psi}^{c}_{i},\\
\label{wdvv001aa}
\overline{Q}^{i}&=
p_{a}\overline{\psi}^{ai}
+U_{a}(x)K_{ij}\overline{\psi}^{aj}
-iF_{abc}(x)\overline{\psi}^{aj}\overline{\psi}_{j}^{b}\psi_{i}^{c}
\end{align}
where $U_{a}(x)$ and $F_{abc}(x)$ 
are homogeneous functions of degree $-1$ in $x^{a}$ and  
\begin{align}
\label{wdvv001a1}
K_{ij}=\frac{i}{2}\left(
u_{i}\overline{u}_{j}+u_{j}\overline{u}_{i}
\right).
\end{align}
Let us consider the Dirac brackets 
\begin{align}
\label{wdvv001a2}
\left\{x^{a},p_{b}\right\}&=\delta_{b}^{a},&
\left\{\psi_{i}^{a},\overline{\psi}^{bj}\right\}&=-\frac{i}{2}\delta_{i}^{k}\delta^{ab},&
\left\{u^{i},\overline{u}_{k}\right\}=-i\delta^{i}_{k}.
\end{align}
Then the $\mathcal{N}=4$ superalgebra 
\begin{align}
\label{wdvv001b}
\left\{
Q_{i},\overline{Q}^{j}
\right\}&=2i\delta_{i}^{j}H
\end{align}
implies that \cite{Bellucci:2004wn,Krivonos:2008fx}
\begin{align}
\label{wdvv002a}
\partial_{a}U_{b}
-\partial_{b}U_{a}&=0,\\
\label{wdvv002b}
\partial_{a}F_{bcd}-\partial_{b}F_{acd}&=0
\end{align}
and
\begin{align}
\label{wdvv002c}
F_{cae}F_{ebd}-F_{cbe}F_{ead}&=0,\\
\label{wdvv002d}
\partial_{a}U_{b}
-U_{a}U_{b}
-F_{abc}U_{c}&=0.
\end{align}
The first set of equations (\ref{wdvv002a}) and (\ref{wdvv002b}) 
can be solved by 
\begin{align}
\label{wdvv002e}
U_{a}(x)&=\partial_{a}U(x),&
F_{abc}(x)&=\partial_{a}\partial_{b}\partial_{c}F(x)
\end{align}
where $U(x)$ and $F(x)$ are 
the prepotentials, 
the scalar functions defined up to polynomials of degree $0$ and $2$ 
in $x^{a}$ respectively. 
Therefore we have two non-linear differential equations 
(\ref{wdvv002c}) and (\ref{wdvv002d}) for the prepotential 
$U(x)$ and $F(x)$. 
Quite interestingly the equation (\ref{wdvv002c}) 
is the so-called Witten-Dijkgraaf-Verlinde-Verlinde (WDVV) equation 
\cite{Witten:1989ig,Dijkgraaf:1990dj}. 
It has been established that 
the solution of WDVV equations determines the structure 
of a Frobenius manifold. 
The other equation (\ref{wdvv002d}) 
describes the so-called twisted periods $U_{a}$ of the Frobenius manifold 
\cite{MR2070050,MR2999308} 
\footnote{
Any function $\tilde{p}$ satisfying 
\begin{align}
\frac{\partial \xi_{a}}{\partial p^{b}}
&=\nu G^{cd}\frac{\partial^{3}F_{*}(p)}
{\partial p^{d}\partial p^{a}\partial p^{b}} \xi_{c},&
\xi_{a}&=\frac{\partial \tilde{p}(p;\nu)}{\partial p^{a}}
\end{align}
is called twisted period of the Frobenius manifold 
where $p^{a}$ are periods. 
}. 

Under the conditions (\ref{wdvv002c})  
and (\ref{wdvv002d}) 
the Hamiltonian can be written as
\begin{align}
\label{wdvv002f}
H&=\frac14 p_{a}p_{a}
+\frac18 J^{ij}J_{ij}U_{a}U_{a}\nonumber\\
&-iU_{ab}K_{ij}\psi^{ai}\overline{\psi}^{bj}
-\frac12 F_{abcd}\psi^{ai}\psi_{i}^{b}
\overline{\psi}^{c}_{j}\overline{\psi}^{dj}.
\end{align}
We should note that 
the $\mathcal{N}=4$ superconformal algebra $D(2,1;\alpha)$ 
has not been taken into account so far. 
So the WDVV equation and the twisted period equation  
are just the requirement 
for the conservation of $\mathcal{N}=4$ supersymmetry.

To realize $D(2,1;\alpha)$ superconformal algebra 
let us introduce the conformal generators $D$, $K$, 
superconformal generators $S_{i}$, $\overline{S}^{i}$ 
and the R-symmetry generators 
$J_{ij}$
\begin{align}
\label{wdvv003a}
D&=-\frac14 \left\{x^{p}p_{a}\right\},&
K&=x^{a}x_{a},\\
\label{wdvv003b}
S_{i}&=-2x^{a}\psi_{i}^{a},&
\overline{S}^{i}&=-2x^{a}\overline{\psi}^{ia},\\
\label{wdvv003c}
J_{ij}&=K_{ij}+2i\psi_{(i}^{a}\overline{\psi}_{j)}^{a},\\
\label{wdvv003d}
I_{11}&=i\psi_{i}^{a}\psi^{ia},&
I_{22}&=-i\overline{\psi}^{ia}\overline{\psi}^{a}_{i},&
I_{12}&=i\psi_{i}^{a}\overline{\psi}^{ia}.
\end{align}
From the dilatation invariance 
we require the homogeneity 
\begin{align}
\label{wdvv003e00}
\partial_{b}(x^{a}U_{a})&=(x^{a}\partial_{a}+1)U_{b}=0,\\
\label{wdvv003e01}
\partial_{b}(x^{a}F_{acd})&=(x^{a}\partial_{a}+1)F_{bcd}=0.
\end{align}
The remaining $D(2,1;\alpha)$ superconformal algebra 
(\ref{d21a1a})-(\ref{d21a1f}) then leads to
\begin{align}
\label{wdvv003e}
x^{a}U_{a}&=2\alpha,\\
\label{wdvv003f}
x^{a}F_{abc}&=-(1+2\alpha)\delta_{bc}.
\end{align}
For $\alpha \neq -\frac12$ 
the prepotential $F$ is non-vanishing and 
any two values of $\alpha$ are related by 
a rescaling under the transformation (\ref{d21a2a}). 
In this sense the two conditions (\ref{wdvv003e}) and (\ref{wdvv003f}) 
can be viewed as the normalization conditions. 
In the case of $\alpha$ $=$ $-\frac12$ 
which realizes the $OSp(4|2)$ superconformal mechanics, 
the prepotential $F$ cannot be normalized. 
This corresponds to the fact that 
under the reflection (\ref{d21a2a}) 
$\alpha=-\frac12$ is self-dual 
\footnote{The induced metric defined in (\ref{wdv004c})
is degenerate for $\alpha=-\frac12$.}. 
Therefore we can utilize the families of the solutions $(U,F)$ 
along with the expression (\ref{wdvv002f}) 
to construct $\mathcal{N}=4$ multi-particle superconformal mechanics. 
Since the number of independent equations are 
given by\cite{Fedoruk:2011aa}
\begin{align}
\label{wdvv004a}
\begin{cases}
\frac{1}{12}(N-1)(N-2)^{2}(N-3)&\textrm{for WDVV equation}\cr
\frac12 (N-1)(N-2)&\textrm{for twisted periods},\cr
\end{cases}
\end{align}
when $N\ge4$, i.e. the system contains more than four particles, 
the non-trivial WDVV equation (\ref{wdvv002c}) appears 
and the twisted periods equation (\ref{wdvv002d}) 
gives rise to the non-trivial conditions.

At this stage we with to look for 
the solution $F$ to the WDVV equation (\ref{wdvv002c}) 
and the twisted periods $U_{a}$ defined by (\ref{wdvv002d}). 
However, up to date it is an open mathematical problem to 
list up all the solutions to the WDVV equation 
and only part of the solutions are known 
\cite{MR1669695,Veselov:1999cn,
MR1851578,MR2319765,MR2462356,Lechtenfeld:2009is,Lechtenfeld:2010sk}. 
In \cite{MR1669695} it was shown that 
one can construct the solutions to the WDVV equation (\ref{wdvv002c}) 
by imposing the ansatz 
\begin{align}
\label{wdvv004b1}
F(x)&=\sum_{\bm{\alpha}}f_{\bm{\alpha}}K(\bm{\alpha}\cdot\bm{x})
\end{align}
where
\begin{align}
\label{wdvv004b2}
K(z)=
\begin{cases}
-\frac14 z^{2}\ln z^{2}&\textrm{rational case}\cr
-\frac14 \mathrm{Li}_{3}(e^{2iz})+\frac16 z^{3}&\textrm{trigonometric case}\cr
-\frac14 \mathcal{L}i_{3}(e^{2iz}|\tau)&\textrm{elliptic case}\cr
\end{cases}
\end{align}
with $f_{\bm{\alpha}}\in \mathbb{R}$ and 
$\bm{\alpha}\cdot x=\bm{\alpha}^{a}x_{a}$. 
Here $\left\{\bm{\alpha} \right\}$ are 
the covectors constructing 
a deformed Lie (super)algebra root system \footnote{It is known that 
the root systems of 
some Lie superalgebras give rise to the solutions to the WDVV equation 
\cite{MR2319765,MR2039697}.}. 
$\mathrm{Li}_{3}$ is the trilogarithm and 
$\mathcal{L}i_{3}$ is an elliptic generalization 
\cite{MR1970873,MR2559673,MR2019387,MR2646111}. 
Among the above known solutions to the WDVV equation, 
only the rational case satisfies the normalization conditions 
(\ref{wdvv003e}) and (\ref{wdvv003f}). 
Thus the $D(2,1;\alpha)$ superconformal models may arise for 
\begin{align}
\label{wdvv004b}
F(x)=-\frac14 \sum_{\bm{\alpha}}f_{\bm{\alpha}}(\bm{\alpha}\cdot x)^{2}
\ln \left|\bm{\alpha}\cdot x\right|^{2}
\end{align}
The ansatz (\ref{wdvv004b}) defines the constant metric 
\begin{align}
\label{wdv004c}
g_{ab}=-x^{c}F_{cab}=\sum_{\bm{\alpha}}f_{\bm{\alpha}}\bm{\alpha}\otimes 
\bm{\alpha}.
\end{align}
Then it was established \cite{Veselov:1999cn} 
that certain deformations of root systems 
can solve the WDVV equation (\ref{wdvv002c}) 
and the corresponding collections of covectors $\left\{\bm{\alpha}\right\}$
is called $\vee$-systems \cite{Veselov:2001ic}. 

On the other hand, 
it was observed \cite{MR2999308} that 
the ansatz for the twisted periods $U_{a}$ 
\begin{align}
\label{wdv004d}
U(x)&=\sum_{\beta}u_{\beta}\ln P_{\beta}(x)
\end{align}
can solve the equation (\ref{wdvv002d}) 
where $P_{\beta}(x)$ are homogeneous polynomials 
of degree $n_{\beta}$ in $x$ 
and $u_{\beta}$ is chosen so that
$\sum_{\beta}n_{\beta}u_{\beta}=2\alpha$.

Now let us assume that $\alpha\neq -\frac12$ 
and consider the special solutions to the twisted periods as the form
\begin{align}
\label{wdv004e}
U(x)=\sum_{\bm{\alpha}}u_{\bm{\alpha}}\ln (\bm{\alpha}\cdot x)
\end{align}
where the same covectors $\alpha$ are chosen for $U(x)$ and $F(x)$. 
Then the normalization conditions (\ref{wdvv003e}) and (\ref{wdvv003f}) 
reduce to 
\begin{align}
\label{wdv004f}
\sum_{\bm{\alpha}}u_{\bm{\alpha}}&=2\alpha,\\
\label{wdv004g}
\sum_{\bm{\alpha}}f_{\bm{\alpha}}\bm{\alpha}_{a}\bm{\alpha}_{b}
&=(1+2\alpha)\delta_{ab}
\end{align}
and we get the potential term
\begin{align}
\label{wdv004h}
V(x)&=\frac{K^{ij}K_{ij}}{8}
\sum_{\bm{\alpha},\bm{\beta}}
u_{\bm{\alpha}}u_{\bm{\beta}}
\frac{\bm{\alpha}\cdot \bm{\beta}}
{\left(\bm{\alpha}\cdot x\right)
\left(\bm{\beta}\cdot x\right)}.
\end{align}
By requiring the invariance under permutations 
of the particle labels, 
the WDVV solutions $F$ based on deformed root systems of the Lie algebras 
$A_{n}$, $BCD_{n}$ and $EF_{n}$ and 
the Lie superalgebras have been discussed \cite{Krivonos:2010zy}. 
It is an interesting question 
to reveal the geometrical understanding for 
the relevant WDVV solutions and the relation to the construction of 
the $\mathcal{N}=4$ superconformal mechanical models.

On the contrary, 
we cannot apply the same method to  
the $OSp(4|2)$ superconformal mechanical models for $\alpha=-\frac12$ 
since some formulae become singular. 
One of the illness is the degenerate induced metric 
\begin{align}
\label{wdv005a}
\sum_{\bm{\alpha}}f_{\bm{\alpha}}\bm{\alpha}\otimes \bm{\alpha}=0,
\end{align}
which can be seen from (\ref{wdv004c}) and (\ref{wdv004g}). 
Since this implies the degenerate covectors $\bm{\alpha}$, 
it is natural to consider the degenerate limit 
of the deformed root systems which solve the WDVV equation. 
By observing that 
there exists a degenerate limit in 
the moduli space of the deformed $A_{n}$ root systems, 
the prepotentials for the $OSp(4|2)$ superconformal mechanics have been
proposed as \cite{Krivonos:2010zy}
\begin{align}
\label{wdv005b}
F(x)&=\frac{1}{4N}\sum_{a<b}
(x^{a}-x^{b})^{2}\ln (x^{a}-x^{b})^{2}
-\frac{1}{4N^{2}}
\sum_{a}\left(
Nx^{a}-X
\right)^{2}\ln(Nx^{a}-X)^{2},\\
\label{wdv005c}
U(x)&=-\frac{1}{2N}\sum_{a}\ln \left(
Nx^{a}-X
\right)
\end{align}
where $X=\sum_{a}x^{a}$. 
Correspondingly we get the potential \cite{Krivonos:2010zy,Fedoruk:2011aa}
\begin{align}
\label{wdvv005d}
V(x)&=
\frac{K^{ij}K_{ij}}{8}
\left[
\sum_{a}\frac{1}{(Nx^{a}-X)^{2}}
-\frac{1}{N}\left(
\sum_{a}\frac{1}{Nx^{a}-X}
\right)^{2}
\right]\nonumber\\
&=
\frac{K^{ij}K_{ij}}{8N}
\sum_{a<b}
\left[
\frac{1}{Nx^{a}-X}
-\frac{1}{Nx^{b}-X}
\right].
\end{align}
We should note that the potential 
(\ref{wdvv005d}) does not take the form 
of the Calogero type pairwise interaction 
albeit it is the inverse-square type interaction.

\subsubsection{Sigma-model}
We shall study the $\mathcal{N}=4$ superconformal sigma-model which 
is more general multi-particle $\mathcal{N}=4$ 
superconformal quantum mechanical system \footnote{Note that 
in the $\mathcal{N}=4$ superconformal 
multi-particle mechanical models relevant to the WDVV equation, 
the metric is trivial due to the ansatz (\ref{wdvv001a}), (\ref{wdvv001aa}).}. 

In order to find the condition on the target space geometry, 
we assume that the second, third and fourth supersymmetry transformations 
are expressed as 
\cite{Gibbons:1997iy}
\begin{align}
\label{n4s1a1}
\delta \Phi^{i}&=\epsilon^{r}{(I_{r})^{i}}_{j}D\Phi^{j}
\end{align}
where $\Phi^{i}$ is the $(\bm{1},\bm{2},\bm{1})$ superfields 
and $\epsilon^{r}$, $r=1,2,3$ are the supersymmetry parameters 
and $I_{r}$ are the endomorphisms of the tangent bundle of the 
target space. 
The corresponding $\mathcal{N}=4$ supermultiplet is 
referred to as $\mathcal{N}=4B$ multiplet. 
This is related to the two-dimensional $\mathcal{N}=(4,0)$
supersymmetry. 
Then the $\mathcal{N}=4$ superalgebra imposes the conditions 
\cite{Coles:1990hr,Gibbons:1997iy}
\begin{align}
\label{n4s1a2}
I_{r}I_{s}+I_{s}I_{r}&=-2\delta_{rs},\\
\label{n4s1a3}
N(I_{r},I_{s})&=0
\end{align}
where a $N(F,G)$ is 
Nijenhuis concomitant \cite{MR0082554,MR0074879} 
\begin{align}
\label{n4s1a3}
N(F,G)(X,Y)&=
[FX,GY]-F[X,GY]-F[GX,Y]+FG[X,Y]+G\leftrightarrow F
\end{align}
where $X,Y$ are vector fields on $\mathcal{M}$. 
Thus the target space $\mathcal{M}$ 
possesses three complex sructures $I_{r}$ which 
have vanishing mixed Nijenhuis tensors 
and obey the Clifford algebra (\ref{n4s1a2}). 
Furthermore the three complex structures 
turn out to satisfy the algebra of imaginary unit quaternions
\begin{align}
\label{n4s1a4a}
I_{r}I_{s}&=-\delta_{rs}+\epsilon_{rst}I_{t}
\end{align}
or the $\mathfrak{su}(2)$ R-symmetry algebra
\begin{align}
\label{n4s1a4b}
[I_{r},I_{s}]&=2\epsilon^{rst}I^{t}
\end{align}
since one can construct a third complex structures from other two 
by multiplication. 

Also the supersymmetry invariance 
of the action requires that 
\begin{align}
\label{n4s1a4}
&g_{ij}={(I_{r})^{k}}_{i}{(I_{r})^{l}}_{j}g_{kl},\\
\label{n4s1a5}
&\nabla_{(i}^{(+)}{(I_{r})^{k}}_{j}=0,\\
\label{n4s1a6}
&\partial_{[i}\left({I^{m}}_{j}c_{|m|kl]}
\right)-2{(I_{r})^{m}}_{[i}\partial_{[m}c_{jkl]]}=0.
\end{align}
The first condition(\ref{n4s1a4}) implies that 
the metric $g$ on $\mathcal{M}$ 
is Hermitian with respect to the three complex structures. 
The second condition (\ref{n4s1a5}) is a generalized Yano tensor
condition with torsion and the third condition (\ref{n4s1a6}) 
is imposed on torsion and complex structures.

It has been pointed out \cite{Gibbons:1997iy} 
that the above constraints on the target space $\mathcal{M}$ 
are similar to the defining conditions for 
a weak hyperk\"{a}hler manifold with torsion (HKT) \cite{Howe:1996kj}. 
A weak HKT manifold is a Riemannian manifold 
$\left\{\mathcal{M},g,c\right\}$ with 
a metric $g$, a torsion three-form $c$ and 
three complex structures $I_{r}$, $r=1,2,3$ which 
obey the following conditions \footnote{If $c$ is closed in addition to 
(\ref{n4s1a4a}), (\ref{n4s1a4}), (\ref{n4s1a7}), 
it is called a strong HKT\cite{Howe:1996kj,Gibbons:1997iy}.}
\begin{enumerate}
 \item the three complex structures $I_{r}$ satisfy the algebra 
of imaginary of unit quaternions (\ref{n4s1a4a})
\item the metric is Hermitian with respect to the three complex
      structures; (\ref{n4s1a4})
\item the complex structures are covariant constant 
\begin{align}
\label{n4s1a7}
\nabla_{k}^{(+)}{(I_{r})^{i}}_{j}&=0
\end{align}
with respect to the covariant derivative $\nabla^{(+)}$ with the torsion. 
\end{enumerate}
We see that 
the conditions for the weak HKT geometry are only 
different from the constraints on the target space $\mathcal{M}$ 
in that the covariant constant properties for the 
complex structures (\ref{n4s1a7}) are 
replaced with (\ref{n4s1a5}) and (\ref{n4s1a6}). 
It turns out that the equation (\ref{n4s1a7}) always solves 
the constraints (\ref{n4s1a5}) and (\ref{n4s1a6}). 
Therefore a weak HKT geometry satisfies 
the constraints (\ref{n4s1a3})-(\ref{n4s1a6})  
on the $\mathcal{N}=4B$ supersymmetric sigma-models.

Although it is known that 
the $\mathcal{N}=4$ supermultiplets in one-dimension 
hold the connections to 
the $\mathcal{N}=2$ supersymmetry in two-dimensions as
\begin{align}
\label{n4s1b1}
\textrm{1d $\mathcal{N}=4A$}
&\Leftrightarrow 
\textrm{2d $\mathcal{N}=(2,2)$},\nonumber\\
\textrm{1d $\mathcal{N}=4B$}
&\Leftrightarrow 
\textrm{2d $\mathcal{N}=(4,0)$},
\end{align}
we have seen that 
the target space $\mathcal{M}$ of the $\mathcal{N}=4B$ 
sigma-model is not the HKT geometry in two-dimensions, 
but rather a weak HKT geometry. 
This shows that there are 
one-dimensional supermultiplets 
which cannot be obtained from higher-dimensional supermultiplets.

Furthermore
the $D(2,1;\alpha)$ superconformal algebra 
(\ref{d21a1a})-(\ref{d21a1f}) imposes the additional 
conditions \cite{Michelson:1999zf} \footnote{Here the value
$\alpha=-1,0$ are excluded.}
\begin{align}
\label{n4s1b2}
\mathcal{L}_{D^{r}}{(I_{r})^{i}}_{j}&=-\frac{2}{1+\alpha}
\epsilon^{rst}{(I^{t})_{i}}^{j},&
\mathcal{L}_{D^{r}}g_{ij}&=0
\end{align}
where $D^{r}:=D^{i}{(I^{r})_{i}}^{j}\partial_{j}$.  
These conditions (\ref{n4s1b2}) can be viewed as 
the generalizations of the $\mathcal{N}=2$ superconformal constraints 
(\ref{n2mulc3}).

\subsection{Gauged superconformal mechanics}
Consider the $\mathcal{N}=4$ matrix superfield gauged 
mechanical action in the harmonic superspace 
\cite{Fedoruk:2008hk}
\begin{align}
\label{n4g1a1}
S=&S_{\mathcal{X}}+S_{\mathrm{WZ}}+S_{\mathrm{FI}}
\end{align}
where 
\begin{align}
\label{n4g1a1a}
S_{\mathcal{X}}&=-\frac{1}{4(1+\alpha)}
\int \mu_{H} \mathrm{Tr}\left(
\mathcal{X}^{-\frac{1}{\alpha}}
\right),\\
\label{n4g1a1b}
S_{\mathrm{WZ}}&=\frac{1}{2} \int \mu_{A}^{(-2)} 
\mathcal{V}_{0}\tilde{\mathcal{Z}}^{+}\mathcal{Z}^{+},\\
\label{n4g1a1c}
S_{\mathrm{FI}}&=\frac{i}{2}c \int \mu_{A}^{(-2)} \mathrm{Tr}V^{++}
\end{align}
where the integration measures are defined
\begin{align}
\label{n4g1a2}
\mu_{H}&=dudtd^{4}\theta,&
\mu_{A}^{(-2)}&=dud\zeta^{(-2)}
\end{align}
with harmonic superspace parametrized by the coordinates 
(\ref{n4susyh0})-(\ref{n4susyh0c}). 
The superfields are
\begin{itemize}
 \item the $\mathcal{N}=4$ Grassmann-even Hermitian $n\times n$ 
matrix superfield 
$\mathcal{X}_{a}^{b}(t,\theta^{\pm},\overline{\theta}^{\pm},u^{\pm})$ 
which obeys 
\begin{align}
\label{n4g1a3}
\mathcal{D}^{++}\mathcal{X}&=0,&
\mathcal{D}^{+}\mathcal{D}^{-}\mathcal{X}&=0,&
\left(\mathcal{D}^{+}
\overline{\mathcal{D}}^{-}
+\overline{\mathcal{D}}^{+}
\mathcal{D}^{-}\right)\mathcal{X}&=0,
\end{align}
which is the $(\bm{1},\bm{4},\bm{3})$ supermultiplet 
\item the $\mathcal{N}=4$ Grassmann-even analytic superfield 
$\mathcal{Z}^{+}_{a}(\zeta,u)$ which satisfies 
\begin{align}
\label{n4g1a4}
\mathcal{D}^{++}\mathcal{Z}^{+}&=0,&
D^{+}\mathcal{Z}^{+}&=0,&
\overline{D}^{+}\mathcal{Z}^{+}&=0,
\end{align}
which is the $(\bm{4},\bm{4},\bm{0})$ supermultiplet 
and $\tilde{\mathcal{Z}}^{+}$ being its Hermitian conjugation 
preserving analyticity \cite{Galperin:2001uw,Ivanov:2003nk}
\item the $\mathcal{N}=4$ Grassmann-even $n\times n$ matrix gauge
      superfield $V^{++b}_{a}(\zeta,u)$
\item the unconstrained 
real analytic superfield $\mathcal{V}_{0}(\zeta,u)$ defined by
\begin{align}
\label{n4g1a4a}
\int du \mathcal{V}_{0}(t_{A},\theta^{+},\overline{\theta}^{+},u^{\pm})
|_{\theta^{\pm}=\theta^{i}u_{i}^{\pm},
\overline{\theta}^{\pm}=\overline{\theta}^{i}u_{i}^{\pm}}
&=\mathrm{Tr}\left(\mathcal{X}\right)
\end{align}
\end{itemize}
where the covariant derivative $\mathcal{D}^{++}$ is given by
\begin{align}
\label{n4g1a5}
\mathcal{D}^{++}\mathcal{X}&=D^{++}\mathcal{X}+i[V^{++},\mathcal{X}],\\
\label{n4g1a6}
\mathcal{D}^{++}\mathcal{Z}&=D^{++}\mathcal{Z}^{+}+iV^{++}\mathcal{Z}^{+}. 
\end{align}
The first term $S_{\mathcal{X}}$ of the 
action (\ref{n4g1a1}) is the superconformal action (\ref{n4susyg3}) for 
the $(\bm{1},\bm{4},\bm{3})$ superfield $\mathcal{X}$. 
The second term $S_{\mathrm{WZ}}$ is the Wess-Zumino (WZ) term 
describing $\mathcal{Z}_{a}^{+}$ \cite{Delduc:2006yp}. 
The third term $S_{\mathrm{FI}}$ is the Fayet-Iliopoulos (FI) term 
for the gauge superfield $V^{++}$.

The superconformal boost transformations are 
\cite{Ivanov:2003nk,Delduc:2006yp}
\begin{align}
\label{n4g1b1}
\delta t_{A}&=\alpha^{-1}\Lambda t_{A},&
\delta \theta_{A}&=-\eta^{+}t_{A}+2i(1+\alpha)\eta^{-}\theta^{+}\overline{\theta}^{+},&
\delta u_{i}^{+}&=\Lambda^{++}u_{i}^{-},
\end{align}
\begin{align}
\delta \mu_{H}&=\mu_{H}\left(
2\Lambda -\frac{1+\alpha}{\alpha}\Lambda_{0}
\right),&
\delta \mu_{A}^{(-2)}&=0,
\end{align}
\begin{align}
\delta \mathcal{X}&=-\Lambda_{0}\mathcal{X},&
\delta \mathcal{Z}^{+}&=\Lambda\mathcal{Z}^{+},&
\delta V^{++}&=0
\end{align}
where 
\begin{align}
\label{n4g1b2}
\lambda&=2i\alpha\left(
\overline{\eta}^{-}\theta^{+}-\eta^{-}\overline{\theta}^{+}
\right),\\
\label{n4g1b3}
\Lambda^{++}&=D^{++}\Lambda=2i\alpha\left(
\overline{\eta}^{+}\theta^{+}-\eta^{+}\overline{\theta}^{+}
\right),\\
\label{n4g1b4}
\Lambda_{0}&=2\Lambda-D^{--}\Lambda^{++}.
\end{align}

The action (\ref{n4g1a1}) is invariant under the 
$U(n)$ transformations \cite{Fedoruk:2008hk}
\begin{align}
\label{n4g1c1}
\mathcal{X}&\rightarrow e^{i\Lambda}\mathcal{X}e^{-i\Lambda},\\
\label{n4g1c2}
\mathcal{Z}^{+}&\rightarrow e^{i\Lambda}\mathcal{Z}^{+},\\
\label{n4g1c3}
\tilde{\mathcal{Z}}^{+}&\rightarrow e^{i\Lambda}\mathcal{Z}^{+},\\
\label{n4g1c4}
V^{++}&\rightarrow e^{i\Lambda}V^{++}e^{-i\Lambda}
-ie^{i\Lambda}\left(
D^{++}e^{-i\Lambda}
\right)
\end{align}
where $\Lambda_{a}^{b}(\zeta,u^{\pm})$ 
is the Hermitian analytic matrix gauge parameter. 
From the gauge freedom (\ref{n4g1c1})-(\ref{n4g1c4}) 
let us fix the gauge as 
\begin{align}
\label{n4g1c5}
V^{++}&=-2i\theta^{+}\overline{\theta}^{+}A(t_{A}).
\end{align}
Integrating out 
the auxiliary fields by means of their 
algebraic equations of motion 
and performing the Grassmann integral, 
we obtain the $D(2,1;\alpha)$ superconformal mechanics 
\cite{Fedoruk:2010bt}
\begin{align}
\label{n4g1c6}
S&=\int dt \Biggl[
\dot{x}^{2}
+\frac{i}{2}\left(
\overline{z}_{i}z^{i}-c
\right)
-i\overline{\psi}_{i}\dot{\psi}^{i}
-\dot{\overline{\psi}}_{i}\psi^{i}\nonumber\\
&-\frac{\alpha^{2}(\overline{z}_{i}z^{i})^{2}}{4x^{2}}
+2\alpha \frac{\psi^{i}\overline{\psi}^{j}z_{(i}\overline{z}_{j)}}
{x^{2}}
+\frac23 (1+2\alpha)
\frac{\psi^{i}\overline{\psi}^{j}\psi_{(i}\overline{\psi}_{j)}}
{x^{2}}-A\left(\overline{z}_{i}z^{i}-c\right)
\Biggr].
\end{align}
Using the Noether's method 
the set of generators are evaluated to be \cite{Fedoruk:2010bt}
\begin{align}
\label{n4g1c7}
H&=\frac14
 p^{2}+\alpha^{2}\frac{(\overline{z}_{i}z^{i})^{2}+2\overline{z}_{i}z^{i}}{4x^{2}}
-2\alpha\frac{z^{(i}\overline{z}^{j)}\psi_{(i}\psi_{k)}}
{x^{2}},\nonumber\\
&-(1+2\alpha)\frac{\psi_{i}\psi^{i}\overline{\psi}^{j}\overline{\psi}_{j}}
{2x^{2}}+\frac{(1+2\alpha)^{2}}{16x^{2}},\\
\label{n4g1c8}
D&=tH-\frac14\left\{x,p\right\},\\
\label{n4g1c9}
K&=t^{2}H-\frac12 t\left\{x,p\right\}+x^{2},
\end{align} 
\begin{align}
\label{n4g1d1}
Q^{i}&=p\psi^{i}+2i\alpha\frac{z^{(i}\overline{z}^{j)}\psi_{j}}{x}
+i(1+2\alpha)\frac
{\langle \psi_{j}\psi^{j}\overline{\psi}^{i} \rangle}{x},\\
\label{n4g1d2}
\overline{Q}_{i}&=
p\overline{\psi}_{i}
-2i\alpha\frac{z_{(i}\overline{z}_{j)}\overline{\psi}^{j}}
+i(1+2\alpha)
\frac{\langle
 \overline{\psi}^{j}\overline{\psi}_{j}\psi_{i}\rangle}{x},\\
\label{n4g1d3}
S^{i}&=tQ^{i}-2x\psi^{i}\\
\label{n4g1d4}
\overline{S}_{i}&=t\overline{Q}_{i}-2x\overline{\psi}_{i}
\end{align} 
\begin{align}
\label{n4g1d5}
J^{ij}&=i\left(
z^{(i}\overline{z}^{k)}
+2\psi^{(i}\overline{\psi}^{k)}
\right),\\
\label{n4g1d6}
I^{11}&=-i\psi_{i}\psi^{k},\\
I^{22}&=i\overline{\psi}^{i}\overline{\psi}^{i},\\
\label{n4g1d7}
I^{12}&=-\frac{i}{2}[\psi_{i},\overline{\psi}^{i}]
\end{align}
where $\langle \cdot \rangle$ denotes the Weyl ordering. 
One can show that 
under the canonical relations 
\begin{align}
\label{n4g1e1}
[x,p]&=i,&[z^{i},\overline{z}_{j}]&=\delta^{i}_{j},&
\left\{\psi^{i},\overline{\psi}_{j}\right\}&=-\frac12 \delta_{j}^{i}
\end{align}
the generators form the $D(2,1;\alpha)$ superalgebra
\cite{Fedoruk:2010bt}.

\section{$\mathcal{N}=8$ Superconformal mechanics}
\label{secscqm8}
Up to now much less has been known about higher extended $\mathcal{N}>4$ 
supersymmetric quantum mechanics. 
A study on $\mathcal{N}>4$ supersymmetric quantum mechanics 
was initiated in \cite{deCrombrugghe:1982un} within 
the on-shell Hamiltonian approach.  
As we have discussed in subsection \ref{1dsusysec001a}, 
the $\mathcal{N}=8$ supersymmetry is the maximum case 
in which only the same number of supersymmetry 
is required for the component fields in the minimal supermultiplet 
\footnote{Note that this statement has not been strictly proven 
without the assumptions for 
the particular forms of supersymmetric transformations (\ref{smlt2}), 
(\ref{smlt3}) and the relevant algebras.}. 
In other words, the $\mathcal{N}=8$ supersymmetry is 
the highest supersymmetric case 
in which the superspace and superfield formalism is applicable. 
In fact off-shell actions 
of the $\mathcal{N}=8$ superconformal mechanical models 
are only known for a few cases. 
From the Table \ref{listsup1} 
we see that there are four different possible superconformal group for 
$\mathcal{N}=8$ superconformal mechanics
\footnote{
The relevant $D$-module representations 
for the $d=1$ $\mathcal{N}=2$, $4$ and $8$ superconformal algebras 
have been discussed in \cite{Kuznetsova:2011je}.}:
\begin{enumerate}
\item $SU(1,1|4)$ 
\item $OSp(8|2)$
\item $OSp(4^{*}|4)$ 
\item $F(4)$.
\end{enumerate}
As we will see, 
the $OSp(4^{*}|4)$ superconformal mechanics 
has been constructed from 
the $(\bm{3},\bm{8},\bm{5})$ and the $(\bm{5},\bm{8},\bm{3})$
supermultiplets \cite{Bellucci:2003hn,Bellucci:2004ur}
and $F(4)$ superconformal mechanics has been proposed from 
the $(\bm{1},\bm{8},\bm{7})$ supermultiplet \cite{Delduc:2007dv}.

\subsection{On-shell $SU(1,1|\frac{\mathcal{N}}{2})$ action}
It has been discussed 
\cite{Ivanov:1988it,Akulov:1999bw,Wyllard:1999tm,Fedoruk:2011aa}
that the on-shell one particle component action 
of the $SU(1,1|\frac{\mathcal{N}}{2})$, 
$\mathcal{N}>4$ 
superconformal mechanical models generically take the form
\begin{align}
\label{n8susy0a1}
S=\int dt \left[
\dot{x}^{2}+i\left(
\overline{\psi}_{i}\dot{\psi}^{i}
-\dot{\psi}_{i}\psi^{i}
\right)
-\frac{\left(c+\overline{\psi}_{i}\psi^{i}\right)^{2}}
{x^{2}}
\right]
\end{align}
where the fermionic fields $\psi^{i}$ are the 
spinor representation of the R-symmetry group
$SU(\frac{\mathcal{N}}{2})$. 
It has been pointed out \cite{Wyllard:1999tm} that 
the generators of the superconformal group
$SU(1,1|\frac{\mathcal{N}}{2})$ can be found 
from those of the $SU(1,1|2)$ 
jus by replacing the $SU(2)$ spinor $\psi^{i}$ 
with the $SU(\frac{\mathcal{N}}{2})$ spinors 
and $c$ is a constant parameter. 
Correspondingly the supercharges $Q^{i},\overline{Q}_{i}$ 
and the Hamiltonian $H$ can be expressed as
\begin{align}
\label{n8susy0a2}
Q^{i}&=\psi^{i}\left(
p-2i\frac{c+\overline{\psi}_{i}\psi^{i}}{x}
\right),\\
\label{n8susy0a3}
\overline{Q}_{i}&=
\overline{\psi}_{i}\left(
p+2i\frac{c+\psi_{i}\overline{\psi}^{i}}{x}
\right),\\
\label{n8susy0a4}
H&=\frac{p^{2}}{4}
+\frac{\left[c+\overline{\psi}_{i}\psi^{i}\right]^{2}}
{x^{2}}.
\end{align}
However, 
it has not been completely understood 
how to realize the on-shell action (\ref{n8susy0a1}) 
from the off-shell superspace and superfield formalism.

\subsection{Superspace and supermultiplet}
The $\mathcal{N}=8$ superspace $\mathbb{R}^{(1|8)}$ 
is parametrized by \cite{Bellucci:2003hn,Bellucci:2004ur}
\begin{align}
\label{n8ssz1}
\mathbb{R}^{(1|8)}&=
(t,\theta_{ia},\vartheta_{\alpha A}),&
\overline{(\theta_{ia})}&=\theta^{ia},&
\overline{(\vartheta_{\alpha A})}=\vartheta^{\alpha A}
\end{align}
with $i,a,\alpha,A=1,2$. 
In terms of (\ref{n8ssz1}) 
four commuting $SU(2)$ factors of the R-symmetry will be manifest. 
The covariant derivatives are defined by 
\begin{align}
\label{n8ssz2}
D^{ia}&=\frac{\partial}{\partial \theta_{ia}}
+i\theta^{ia}\frac{\partial}{\partial t},&
\nabla^{\alpha A}&=\frac{\partial}{\partial \vartheta_{\alpha A}}
\end{align}
and they satisfy
\begin{align}
\label{n8ssz3}
\left\{
D^{ia},D^{jb}
\right\}&=2i\epsilon^{ij}\epsilon^{ab}\frac{\partial}{\partial t},&
\left\{
\nabla^{\alpha A},\nabla^{\beta B}
\right\}&=2i\epsilon^{\alpha\beta}\epsilon^{AB}\frac{\partial}{\partial t}.
\end{align}

Although the $\mathcal{N}=8$ superfields 
are useful to find the irreducibility constraints 
and the transformation properties, 
it is hard to reproduce the supersymmetric action 
in terms of the component fields 
because of the large dimension of the integration measure. 
The efficient strategy is to 
split the $\mathcal{N}=8$ supermultiplets 
into the $\mathcal{N}=4$ supermultiplets 
and to deal with the $\mathcal{N}=4$ superspace 
and superfield formalism. 
Such decompositions of the $\mathcal{N}=8$ supermultiplets 
in terms of the $\mathcal{N}=4$ supermultiplets 
can be written as the direct sum \cite{Bellucci:2003hn,Ivanov:2007sh}
\begin{align}
\label{n8ssz4}
(\bm{n},\bm{8},\bm{8-n})
=(\bm{n}_{1},\bm{4},\bm{4}-\bm{n}_{1})\oplus 
(\bm{n}_{2},\bm{4},\bm{4}-\bm{n}_{2})
\end{align}
with $n=n_{1}+n_{2}$.
Here $n$ represents the number of physical bosonic fields in 
the $\mathcal{N}=8$ supermultiplets 
while $n_{1}$ and $n_{2}$ denote the numbers of physical bosons 
in the two $\mathcal{N}=4$ supermultiplets respectively.

\subsubsection{\textrm{$(\bm{0},\bm{8},\bm{8})$} supermultiplet}
The $(\bm{0},\bm{8},\bm{8})$ supermultiplet 
is described by two real fermionic superfields $\Psi^{aA}$,
$\Xi^{i\alpha}$ satisfying the constraints
\begin{align}
\label{n8ssa1}
\nabla^{(\alpha A}\Xi_{i}^{\beta)}&=0,& 
D^{(ia}\Xi^{j}_{\alpha}&=0,\\
\label{n8ssa2}
\nabla^{\alpha(A}\Psi_{a}^{B)}&=0,&
D^{i(a}\Psi_{A}^{b)}&=0,\\
\label{n8ssa3}
\nabla^{\alpha A}\Psi_{A}^{a}&=D^{ia}\Xi^{\alpha}_{i},&
\nabla^{\alpha A}\Xi^{i}_{\alpha}&=-D^{ia}\Psi_{a}^{A}.
\end{align}
(\ref{n8ssa3}) implies that 
the covariant derivative with respect to $\vartheta_{\alpha A}$ 
can be represented by the covariant derivatives with respect to
$\theta_{ia}$.

The $(\bm{0},\bm{8},\bm{8})$ supermultiplet 
possesses a unique splitting 
\begin{align}
\label{n8ssa4}
(\bm{0},\bm{8},\bm{8})&=
(\bm{0},\bm{4},\bm{4})\oplus 
(\bm{0},\bm{4},\bm{4}).
\end{align}
In order to describe the $(\bm{0},\bm{0},\bm{8})$ supermultiplet 
in terms of the $\mathcal{N}=4$ superfields, 
we pick up the appropriate $\mathcal{N}=4$ superspace as
\begin{align}
\label{n8ssa5}
\mathbb{R}^{(1|4)}=(t,\theta_{ia})
\subset \mathbb{R}^{(1|8)}=(t,\theta_{ia},\vartheta_{\alpha A}).
\end{align}
Expanding the superfields in $\vartheta^{iA}$, 
the constraints (\ref{n8ssa3}) 
leave the independent $\mathcal{N}=4$ superfields 
\begin{align}
\label{n8ssa6}
\psi^{aA}&=\Psi^{aA}|_{\vartheta=0},&
\xi^{i\alpha}&=\Xi^{i\alpha}|_{\vartheta=0}.
\end{align}
Then the constraints (\ref{n8ssa1}) and (\ref{n8ssa2}) imply 
that 
\begin{align}
\label{n8ssa7}
D^{a(i}\xi^{j)\alpha}&=0,&
D^{i(a}\psi^{b)A}&=0.
\end{align}
The conditions (\ref{n8ssa7}) correspond to 
the constraints (\ref{n4susyh1}) 
for $(\bm{0},\bm{4},\bm{4})$ supermultiplets 
on the superfields $\xi^{ia}, \psi^{aA}$. 

The $\mathcal{N}=8$ supersymmetric action 
can be written as 
\begin{align}
\label{n8ssa8}
S=\int dtd^{4}\theta 
\left[
\theta^{ia}\theta^{b}_{i}
\psi_{a}^{A}\psi_{bA}
+\theta^{ia}\theta^{j}_{a}\xi_{i}^{\alpha}\xi_{j\alpha}
\right].
\end{align}
Although the action (\ref{n8ssa8}) is not manifestly invariant 
due to the existence of the Grassmann coordinates, 
one can show that it is invariant.

\subsubsection{\textrm{$(\bm{1},\bm{8},\bm{7})$} supermultiplet}
The $(\bm{1},\bm{8},\bm{7})$ supermultiplet 
is described by a single scalar superfield $\mathcal{U}$ 
obeying the conditions
\begin{align}
\label{n8ssb1}
\nabla^{(\alpha i}\nabla^{\beta)j}\mathcal{U}&=0,&
D^{i(a}D^{jb)}\mathcal{U}&=0,\\
\label{n8ssb3}
D^{ia}D^{j}_{a}\mathcal{U}&=-\nabla^{\alpha
 j}\nabla_{\alpha}^{i}\mathcal{U}
\end{align}
The condition (\ref{n8ssb3}) reduce the manifest R-symmetry 
into three $SU(2)$ factors due to the 
identification of the indices $A$ and $i$ of the covariant 
derivatives $\nabla^{\alpha A}$ and $D^{ia}$.

The $(\bm{1},\bm{8},\bm{7})$ has a unique 
decomposition into the $\mathcal{N}=4$ multiplets as
\begin{align}
\label{n8ssb4}
(\bm{1},\bm{8},\bm{7})=
(\bm{1},\bm{4},\bm{3})\oplus 
(\bm{0},\bm{4},\bm{4}).
\end{align}
By choosing the $\mathcal{N}=4$ superspace $\mathbb{R}^{(1|4)}$ 
as in (\ref{n8ssa5}) 
and expanding the superfields in $\vartheta^{iA}$, 
we find the projected $\mathcal{N}=4$ superfields 
\begin{align}
\label{n8ssb5}
u&=\mathcal{U}|_{\vartheta=0},&
\psi^{i\alpha}&=\nabla^{i\alpha}\mathcal{U}|_{\vartheta=0}
\end{align}
obeying
\begin{align}
\label{n8ssb6}
D^{(ia}\psi^{j)\alpha}&=0,&
D^{i(a}D^{jb)}u&=0,
\end{align}
which are viewed as 
the constraint equations (\ref{n4susyh1}) 
and (\ref{n4susy4f6}). 
Thus we can identify 
$\psi^{i\alpha}$ and $u$ 
with the $(\bm{0},\bm{4},\bm{4})$ and 
$(\bm{1},\bm{4},\bm{3})$ superfields respectively.

The general $\mathcal{N}=8$ supersymmetric component action 
of the $(\bm{1},\bm{8},\bm{7})$ supermultiplet 
can be found in \cite{Kuznetsova:2005cd}. 
The harmonic superspace action can be found in 
\cite{Galperin:1984bu,Galperin:1984av,Galperin:2001uw}.

Taking into account the decomposition (\ref{n8ssb4}) 
of the $(\bm{1},\bm{8},\bm{7})$ supermultiplet, 
$\mathcal{N}=8$ superconformal mechanical model has been constructed by 
combining the two supermultiplets 
$(\bm{1},\bm{4},\bm{3})$ and $(\bm{0},\bm{4},\bm{4})$ 
for $D(2,1;\alpha=-\frac13)$ \cite{Delduc:2007dv}. 
Since the possible $\mathcal{N}=8$ superconformal group 
into which one can embed $D(2,1;\alpha=-\frac13)$ is 
only $F(4)$, 
the resulting $\mathcal{N}=8$ superconformal mechanical model 
is identified with $F(4)$ superconformal mechanics.

\subsubsection{\textrm{$(\bm{2},\bm{8},\bm{6})$} supermultiplet}
The $(\bm{2},\bm{8},\bm{6})$ supermultiplet 
contains two scalar bosonic superfields $\mathcal{U},\Phi$ 
which satisfy
\begin{align}
\label{n8ssc1}
\nabla^{(ai}\nabla^{b)j}\mathcal{U}&=0,&
\nabla^{a(i}\nabla^{bj)}\Phi&=0,\\
\label{n8ssc2}
\nabla^{ai}\mathcal{U}&=D^{ia}\Phi,&
\nabla^{ai}\Phi&=-D^{ia}\mathcal{U}
\end{align}
where the indices $i,A$ being identified and 
the indices $a,\alpha$ being identified 
and thus only two $SU(2)$ factors are manifest. 
The $(\bm{2},\bm{8},\bm{6})$ multiplets 
can be regarded as the two $(\bm{1},\bm{8},\bm{7})$ multiplets 
with the additional conditions because 
the two constraints (\ref{n8ssc1}) and (\ref{n8ssc2}) 
lead to 
\begin{align}
\label{n8ssc3}
D^{(ia}D^{j)b}\Phi&=0,&
D^{i(a}D^{jb)}\mathcal{U}&=0,\\
\label{n8ssc4}
D^{ia}D_{a}^{j}\mathcal{U}&=-\nabla^{aj}\nabla^{i}_{a}\mathcal{U},&
D^{ia}D_{i}^{b}\Phi&=-\nabla^{bi}\nabla_{i}^{a}\Phi.
\end{align}

The $(\bm{2},\bm{8},\bm{6})$ multiplet 
has two different decompositions
\begin{align}
\label{n8ssc5}
(\bm{2},\bm{8},\bm{6})
=\begin{cases}
(\bm{1},\bm{4},\bm{3})\oplus 
(\bm{1},\bm{4},\bm{3})\cr
(\bm{2},\bm{4},\bm{2})\oplus
(\bm{0},\bm{4},\bm{4}).\cr
\end{cases}
\end{align}

\begin{enumerate}
 \item \textbf{$(\bm{1},\bm{4},\bm{3})\oplus (\bm{1},\bm{4},\bm{3})$}

Choosing the $\mathcal{N}=4$ superspace (\ref{n8ssa5}) 
and expanding the superfields in $\vartheta^{iA}$, 
we find from (\ref{n8ssc1}) and (\ref{n8ssc2}) 
the independent $\mathcal{N}=4$ superfields 
\begin{align}
\label{n8ssc6}
u&=\mathcal{U}|_{\vartheta=0},&
\phi&=\Phi|_{\vartheta=0}
\end{align}
satisfying 
\begin{align}
\label{n8ssc7}
D^{i(a}D^{jb)}u&=0,&
D^{(ia}D^{j)b}\phi&=0,
\end{align}
which are the constraints equations 
(\ref{n4susy2f7}). 
Therefore the two superfields 
$u$, $\phi$ are regarded as the 
$(\bm{1},\bm{4},\bm{3})$ superfields.

The action can be written as
\begin{align}
\label{n8ssc8}
S=\int dtd^{4}\theta\ 
F(u,\phi)
\end{align}
where the function $F$ satisfies the Laplace equation
\begin{align}
\label{n8ssc8}
\frac{\partial^{2}F}{\partial u^{2}}
+\frac{\partial^{2}F}{\partial \phi^{2}}=0.
\end{align}

\item \textbf{$(\bm{2},\bm{4},\bm{2})\oplus (\bm{0},\bm{4},\bm{4})$}

To realize the decomposition 
$(\bm{2},\bm{8},\bm{6})=$ 
$(\bm{2},\bm{4},\bm{2})\oplus (\bm{0},\bm{4},\bm{4})$ 
we need to modify the choice of the 
$\mathcal{N}=4$ superspace and the superfields. 
Let us introducethe covariant derivatives
\begin{align}
\label{n8ssc9}
\mathcal{D}^{ia}&=
\frac{1}{\sqrt{2}}
\left(D^{ia}-i\nabla^{ai}\right),&
\overline{\mathcal{D}}^{ia}&=
\frac{1}{\sqrt{2}}
\left(D^{ia}+i\nabla^{ai}\right)
\end{align}
and the superfields $\mathcal{V},\overline{\mathcal{V}}$ as
\begin{align}
\label{n8ssc10}
\mathcal{V}&=\mathcal{U}+i\Phi,&
\overline{\mathcal{V}}&=\mathcal{U}-i\Phi.
\end{align}
Then we find a set of constraint equations
\begin{align}
\label{n8ssc11a}
D^{i}\mathcal{V}&=0,&
\nabla^{i}\mathcal{V}&=0,\\
\label{n8ssc11b}
\overline{D}_{i}\overline{\mathcal{V}}&=0,&
\overline{\nabla}_{i}\overline{\mathcal{V}}&=0,\\
\label{n8ssc11c}
D^{i}D_{i}\overline{\mathcal{V}}
&=\overline{\nabla}_{i}\overline{\nabla}^{i}\mathcal{V},&
D^{i}\nabla^{j}\overline{\mathcal{V}}
&=\overline{D}^{i}\overline{\nabla}^{j}\mathcal{V}=0
\end{align}
where we have defined 
\begin{align}
\label{n8ssc12a}
D^{i}&:=\mathcal{D}^{i1},&
\overline{D}^{i}&:=\overline{\mathcal{D}}^{i2},\\
\label{n8ssc12b}
\nabla^{i}&:=\mathcal{D}^{i2},&
\overline{\nabla}^{i}&=-\overline{\mathcal{D}}^{i1}.
\end{align}
Considering a new set of coordinates 
for the $\mathcal{N}=4$ superspace as
\begin{align}
\label{n8ssc12c}
\mathbb{R}^{(1|4)}
&=(t,\theta_{i1}+i\vartheta_{i1},\theta_{i2}-i\vartheta_{i2})
\subset \mathbb{R}^{(1|8)},
\end{align}
we find from the constraints (\ref{n8ssc11a})-(\ref{n8ssc11c}) 
the independent $\mathcal{N}=4$ superfields 
\begin{align}
\label{n8ssc12d}
v&=\mathcal{V}|,&
\overline{v}&=\overline{\mathcal{V}}|,\\
\label{n8ssc12e}
\psi^{i}&=\overline{\nabla}^{i}\mathcal{V}|,&
\overline{\psi}^{i}&=-\nabla^{i}\overline{V}|
\end{align}
satisfying 
\begin{align}
\label{n8ssc12f}
D^{i}v&=0,&
\overline{D}^{i}\overline{v}&=0,\\
\label{n8ssc12g}
D^{i}\psi^{j}&=0,&
\overline{D}^{i}\overline{\psi}^{j}&=0,&
D^{i}\overline{\psi}^{j}&=-\overline{D}^{i}\psi^{j}
\end{align}
Thus we can identify the two sets of the 
superfields, $v,\overline{v}$ and $\psi^{i},\overline{\psi}^{i}$ 
with the $(\bm{2},\bm{4},\bm{2})$ 
and $(\bm{0},\bm{4},\bm{4})$ superfields. 

The invariant action is given by
\begin{align}
\label{n8ssc12g}
S=\int dtd^{4}\theta 
v\overline{v}
-\frac12 \int dtd^{2}\overline{\theta}\psi^{i}\psi_{i}
-\frac12 \int dtd^{2}\theta \overline{\psi}_{i}\overline{\psi}^{i}.
\end{align}
We should note that 
the form of the action (\ref{n8ssc12g}) 
depend on the choice of the $\mathcal{N}=4$ superspace. 
Although the superfield action 
(\ref{n8ssc12g}) looks different from the previous action (\ref{n4susyh4}), 
it turns out to be the same in the component level.

\end{enumerate}

\subsubsection{\textrm{$(\bm{3},\bm{8},\bm{5})$} supermultiplet}
The $(\bm{3},\bm{8},\bm{5})$ supermultiplet includes 
the three bosonic superfields $\mathcal{V}^{ij}=\mathcal{V}^{ji}$ obeying 
\begin{align}
\label{n8ssd1}
D_{a}^{(i}\mathcal{V}^{jk)}&=0,&
{\nabla_{\alpha}}^{(i}\mathcal{V}^{jk)}&=0
\end{align}
and three $SU(2)$ factors are manifest. 
(\ref{n8ssd1}) yield to a further condition
\begin{align}
\label{n8ssd1a}
\partial_{t}
\left(
D_{i}^{a}D_{ja}V^{ij}
+\nabla_{i}^{\alpha}\nabla_{j\alpha}V^{ij}
\right)&=0,
\end{align}
which leads to \cite{Bellucci:2003hn} 
\begin{align}
\label{n8ssd1b}
\nabla_{i}^{\alpha}\nabla_{j\alpha}V^{ij}
&=6m-D_{i}^{a}D_{ja}V^{ij}
\end{align}
where $m$ is a constant parameter.

The $(\bm{3},\bm{8},\bm{5})$ multiplet has 
two decompositions
\begin{align}
\label{n8ssd2}
(\bm{3},\bm{8},\bm{5})
=\begin{cases}
(\bm{3},\bm{4},\bm{1})\oplus
(\bm{0},\bm{4},\bm{4})
\cr
(\bm{1},\bm{4},\bm{3})\oplus
(\bm{2},\bm{4},\bm{2}).\cr
\end{cases}
\end{align}

\begin{enumerate}

\item \textbf{$(\bm{3},\bm{4},\bm{1})\oplus (\bm{0},\bm{4},\bm{4})$}

Let us choose the $\mathcal{N}=4$ superspace (\ref{n8ssa5}) 
and expand the superfields in $\vartheta_{i\alpha}$. 
Then the constraints (\ref{n8ssd1}) leave in $\mathcal{V}^{ij}$ 
the four bosonic and four fermionic $\mathcal{N}=4$ superfields 
\begin{align}
\label{n8ssd2}
v^{ij}&=\mathcal{V}^{ij}|,&
\xi^{i}_{\alpha}&=\nabla_{j\alpha}\mathcal{V}^{ij}|,\\
\label{n8ssd2a}
A&=\nabla_{i}^{\alpha}\nabla_{j\alpha}V^{ij}|
\end{align}
which obey
\begin{align}
\label{n8ssd3}
D_{a}^{(i}v^{jk)}&=0,&
D_{a}^{(i}\xi^{j)}_{\alpha}&=0,\\
\label{n8ssd4}
A&=6m-D_{i}^{a}D_{aj}v^{ij}.
\end{align}
Since (\ref{n8ssd3}) are identified with the constraint equations 
(\ref{n4susy2e2}) and (\ref{n4susyh1}), 
we see that the superfields $v^{ij}$ 
and $\xi^{i}_{\alpha}$ are the $(\bm{3},\bm{4},\bm{1})$ 
and $(\bm{0},\bm{4},\bm{4})$ superfields respectively. 
The remaining equation (\ref{n8ssd4}) is the conservation law type condition 
which gives rise to a constant $m$. 
As observed in \cite{Ivanov:1988it}, 
this is the reminiscent of the $d=4$ $\mathcal{N}=1$ tensor multiplet 
constraints \cite{Gates:1983nr}.

To write down the invariant action 
let us project out the 
$\mathcal{N}=4$ superfields $v^{(ij)}$ and 
$\xi^{i}_{\alpha}$ onto the harmonic superspace as
\begin{align}
\label{n8ssd3a}
v^{++}&=v^{ij}u_{i}^{+}u_{j}^{+},&
v^{+-}&=\frac12 D^{--}v^{++},&
v^{--}&=D^{--}v^{+-},\\
\xi^{+}&=\xi^{i}u_{i}^{+},&
\overline{\xi}^{+}&=\overline{\xi}^{i}u_{i}^{+}.
\end{align}
Then the $OSp(4^{*}|4)$ superconformal action is 
given by \cite{Bellucci:2003hn}
\begin{align}
\label{n8ssd3b}
S&=\int dtd^{4}\theta \sqrt{v^{2}}\nonumber\\
&-\frac{1}{\sqrt{2}}
\int dud\zeta^{--}
\left[
\frac{\xi^{+}\overline{\xi}^{+}}
{(1+c^{--}\hat{v}^{++})^{\frac32}}
+12m\frac{\hat{v}^{++}}
{\sqrt{1+c^{--}\hat{v}^{++}}(1+\sqrt{1+c^{--}\hat{v}^{++}})}
\right]
\end{align}
where 
\begin{align}
\label{n8ssd3c}
dud\zeta^{--}&=dudt_{A}d\theta^{+}d\overline{\theta}^{+},\\
\label{n8ssd3d}
c^{\pm}c^{\pm}&=c^{ik}u_{i}^{\pm}u_{k}^{\pm},&c^{ik}=\mathrm{const.},\\
\label{n8ssd3e}
v^{++}&=\hat{v}^{++}+c^{++}.
\end{align}

\item \textbf{$(\bm{1},\bm{4},\bm{3})\oplus (\bm{2},\bm{4},\bm{2})$}

To obtain the decomposition 
$(\bm{3},\bm{8},\bm{5})=
(\bm{1},\bm{4},\bm{3})\oplus (\bm{2},\bm{4},\bm{2})$, 
we shall introduce the new covariant derivatives 
\begin{align}
\label{n8ssd5}
D^{a}&=\frac{1}{\sqrt{2}}
\left(D^{1a}+i\nabla^{a1}\right),&
\overline{D}_{a}&=\sqrt{1}{\sqrt{2}}\left(
D^{2}_{a}-i\nabla_{a}^{2}
\right),\\
\label{n8ssd6}
\nabla^{a}&=\frac{i}{\sqrt{2}}
\left(D^{2a}+i\nabla^{a2}\right),&
\overline{\nabla}_{a}&=
\frac{i}{\sqrt{2}}\left(
D_{a}^{1}-i\nabla_{a}^{1}
\right)
\end{align}
and the set of coordinates closed under the action of $D^{a},\overline{D}_{a}$
\begin{align}
\label{n8ssd7}
\mathbb{R}^{(1|4)}&
=(t,\theta_{1a}-i\vartheta_{a1},\theta^{1a}+i\vartheta^{a1})\subset 
\mathbb{R}^{(1|8)}.
\end{align}
Defining the $\mathcal{N}=4$ superfields as
\begin{align}
v&=-2i\mathcal{V}^{12},&
\varphi&=\mathcal{V}^{11},&
\overline{\varphi}&=\mathcal{V}^{22},
\end{align}
we find the constraints (\ref{n4susy2f7}) 
for the $(\bm{1},\bm{4},\bm{3})$ supermultiplet 
and the constraints (\ref{n4susy2f3a}) 
for the $(\bm{2},\bm{4},\bm{2})$ chiral supermultiplet
\begin{align}
\label{n8ssd8}
D^{a}D_{a}v&=0,&
\overline{D}_{a}\overline{D}^{a}v&=0,\\
\label{n8ssd9}
D^{a}\varphi&=0,&
\overline{D}_{a}\overline{\varphi}.
\end{align}
from the constraints (\ref{n8ssd1}). 
Therefore the superfields $v$ can be viewed as 
the $(\bm{1},\bm{4},\bm{3})$ superfield 
and $\varphi$ as the $(\bm{2},\bm{4},\bm{2})$ superfield. 
From the constraints (\ref{n8ssd8}) and (\ref{n8ssd9}) 
it follows that 
\begin{align}
\label{n8ssd90a}
\frac{\partial}{\partial t}\left[
D^{a},\overline{D}_{a}
\right]v=0.
\end{align}
Combining (\ref{n8ssd1b}) and (\ref{n8ssd90a}), 
we obtain the constant $m$ \cite{Ivanov:1988it}
\begin{align}
\label{n8ssd90b}
[D^{a},\overline{D}_{a}]v=-2m.
\end{align}

In this case the $\mathcal{N}=8$ supersymmetric free action 
takes the form \cite{Bellucci:2003hn}
\begin{align}
\label{n8ss9a00}
S=-\frac14 \int dtd^{4}\theta 
\left(v^{2}-2\varphi\overline{\varphi}\right).
\end{align}
However, the action (\ref{n8ss9a00}) is not 
invariant under the superconformal transformations. 
Following the strategy of \cite{Galperin:1985tn,Ivanov:2002pc}, 
the $OSp(4^{*}|4)$ superconformal action is given by \cite{Bellucci:2003hn}
\begin{align}
\label{n8ssd9a}
S=-\frac14 \int dtd^{4}\theta 
\left[
v\ln \left(v+\sqrt{v^{2}+\varphi\overline{\varphi}}\right)
-\sqrt{v^{2}+4\varphi\overline{\varphi}}
\right]
\end{align}
whose bosonic part is
\begin{align}
\label{n8ssd9b}
S_{\textrm{bosonic}}
&=\int dt \frac{1}{\sqrt{v^{2}+4\varphi\overline{\varphi}}}
\left[
\dot{v}^{2}+4\dot{\varphi}\dot{\overline{\varphi}}
-m^{2}-2im\dot{v}
-\frac{4im\varphi\dot{\overline{\varphi}}}
{v+\sqrt{v^{2}+\varphi\overline{\varphi}}}
\right].
\end{align}

\end{enumerate}

Therefore the $(\bm{3},\bm{8},\bm{5})$ supermultiplet 
can describe the $OSp(4^{*}|4)$ superconformal mechanics 
\cite{Bellucci:2003hn}. 
By means of the non-linear realization method 
parametrize a coset of the supergroup $OSp(4^{*}|4)$ such that 
$SO(5)\subset OSp(4^{*}|4)$ belongs to the 
stability subgroup 
while one out of three Goldstone bosons is the coset parameter 
associated with the dilatation, the dilaton 
and the remaining two Goldstone bosons parametrize the R-symmetry 
coset $SU(2)_{R}/U(1)_{R}$.
Although the action (\ref{n8ssd3b}) and (\ref{n8ssd9b})
have different manifest $\mathcal{N}=4$ superconformal
symmetries $OSp(4^{*}|2)$ and $SU(1,1|2)$ respectively, 
both of them form $OSp(4^{*}|4)$ superconformal group 
together with the hidden symmetries. 
Hence the two superfield actions 
(\ref{n8ssd3b}) and (\ref{n8ssd9b}) 
exhibit different symmetry aspects of the 
same $\mathcal{N}=8$ superconformal mechanics. 
Note that the two actions 
(\ref{n8ssd3b}) and (\ref{n8ssd9b}) 
produce the same actions (\ref{n8ssd9b}) 
in terms of the component fields 
as they can be obtained from the single $\mathcal{N}=8$ superfield
formulation.

\subsubsection{\textrm{$(\bm{4},\bm{8},\bm{4})$} supermultiplet}
The $(\bm{4},\bm{8},\bm{4})$ supermultiplet 
includes a four superfields $\mathcal{Q}^{a\alpha}$ 
which obeys
\begin{align}
\label{n8sse1}
D_{i}^{(a}\mathcal{Q}^{b)\alpha}&=0,&
\nabla_{i}^{(\alpha}\mathcal{Q}_{a}^{\beta)}&=0.
\end{align} 
The constraints (\ref{n8sse1}) are manifestly covariant 
with respect to the three $SU(2)$ factors for the indices 
$i,a$ and $\alpha$.

There are three different decompositions 
of the $(\bm{4},\bm{8},\bm{4})$ supermultiplet
\begin{align}
\label{n8sse2}
(\bm{4},\bm{8},\bm{4})
=\begin{cases}
(\bm{4},\bm{4},\bm{0})\oplus 
(\bm{0},\bm{4},\bm{4})\cr
(\bm{3},\bm{4},\bm{1})\oplus 
(\bm{1},\bm{4},\bm{3})\cr
(\bm{2},\bm{4},\bm{2})\oplus 
(\bm{2},\bm{4},\bm{2}).\cr
\end{cases}
\end{align}

\begin{enumerate}
 \item \textbf{$(\bm{4},\bm{4},\bm{0})\oplus (\bm{0},\bm{4},\bm{4})$}

Making the choice of the $\mathcal{N}=4$ superspace 
(\ref{n8ssa5}) and expanding the superfields in $\vartheta_{i\alpha}$, 
the constraints (\ref{n8sse1}) yield the independent $\mathcal{N}=4$ 
superfields 
\begin{align}
\label{n8sse2a}
q^{a\alpha}&=\mathcal{Q}^{a\alpha}|,&
\psi^{ia}&=\nabla^{i}_{\alpha}\mathcal{Q}^{a\alpha}|
\end{align}
satisfying the constraint conditions (\ref{n4susy2d2}) for 
the $(\bm{4},\bm{4},\bm{0})$ supermultiplet 
and (\ref{n4susyh2}) for 
the $(\bm{0},\bm{4},\bm{4})$ supermultiplet
\begin{align}
\label{n8sse2b}
D^{i(a}q^{b)\alpha}&=0,&
D^{i(a}\psi^{b)i}&=0.
\end{align}
Thus the superfields $q^{i\alpha}$ 
and $\psi^{ia}$ are the $(\bm{4},\bm{4},\bm{0})$ 
and $(\bm{0},\bm{4},\bm{4})$ superfields respectively.

\item \textbf{$(\bm{3},\bm{4},\bm{1})\oplus (\bm{1},\bm{4},\bm{3})$}

Let us introduce the $\mathcal{N}=8$ superfields 
$\mathcal{V}^{ab}$, $\mathcal{V}$ as
\begin{align}
\label{n8sse2c}
\mathcal{Q}^{a\alpha}&=\delta_{b}^{\alpha}\mathcal{V}^{ab}
-\epsilon^{a\alpha}\mathcal{V},&
\mathcal{V}^{ab}&=\mathcal{V}^{ba}
\end{align}
and pick up the $\mathcal{N}=4$ superspace 
\begin{align}
\label{n8sse2d}
\mathbb{R}^{(1|4)}=
(t,\theta_{1a}+i\vartheta_{1a},\theta_{2a}-i\vartheta_{2a})
\subset \mathbb{R}^{(1|8)}.
\end{align}
Correspondingly we will consider the covariant derivatives 
$D^{a},\overline{D}^{a}$ and 
$\overline{\nabla}^{a},\nabla^{a}$ as
\begin{align}
\label{n8sse2e}
(D^{a},\overline{D}^{a})&=\left(
\mathcal{D}^{1a},\overline{\mathcal{D}}^{2a}
\right),&
(\overline{\nabla}^{a},\nabla^{a})&=
\left(\mathcal{D}^{2a},\overline{\mathcal{D}}^{1a}\right)
\end{align}
where $\mathcal{D}^{ia},\overline{\mathcal{D}}^{ia}$ are defined in 
(\ref{n8ssc9}). 
Then the constraints (\ref{n8sse1}) lead to 
the independent $\mathcal{N}=4$ superfields 
\begin{align}
\label{n8sse2f}
v^{ab}&=\mathcal{V}^{ab},&
v&=\mathcal{V}
\end{align}
which are subjected to 
\begin{align}
\label{n8sse2g}
D^{(a}v^{bc)}&=0,&
\overline{D}^{(a}v^{bc)}&=0,\\
\label{n8sse2h}
D^{(a}\overline{D}^{b)}v&=0.
\end{align}
Thus the superfields $v^{ab}$ and $v$ 
are the $(\bm{3},\bm{4},\bm{1})$ and $(\bm{1},\bm{4},\bm{3})$ 
superfields respectively. 

The supersymmetric invariant free action is given by
\begin{align}
\label{n8sse2i}
S=\int dtd^{4}\theta 
\left[
v^{2}-\frac38 v^{ab}v_{ab}
\right].
\end{align}

\item \textbf{$(\bm{2},\bm{4},\bm{2})\oplus (\bm{2},\bm{4},\bm{2})$}

We shall define the new set of $\mathcal{N}=8$ superfields 
$\mathcal{W}$, $\Phi$ in terms of 
$\mathcal{V}$, $\mathcal{V}^{ab}$ introduced in 
(\ref{n8sse2c}) as
\begin{align}
\label{n8sse2j}
\mathcal{W}&=\mathcal{V}^{11},
&\overline{\mathcal{W}}&=\mathcal{V}^{22},\\
\label{n8sse2k}
\Phi&=\frac23 \left(
\mathcal{V}+\frac32 \mathcal{V}^{12}
\right),&
\overline{\Phi}&=\frac23 \left(
\mathcal{V}-\frac32 \mathcal{V}^{12}
\right)
\end{align}
and the new set of the $\mathcal{N}=4$ covariant derivatives 
$D^{i}$, $\nabla^{i}$ as
\begin{align}
\label{n8sse2l}
D^{i}&=\frac{1}{\sqrt{2}}
\left(\mathcal{D}^{i1}+\overline{\mathcal{D}}^{i1}\right),&
\overline{D}^{i}&=
\frac{1}{\sqrt{2}}
\left(\mathcal{D}^{i2}+\overline{\mathcal{D}}^{i2}\right),\\
\label{n8sse2m}
\nabla^{i}&=\frac{1}{\sqrt{2}}
\left(
\mathcal{D}^{i1}-\overline{\mathcal{D}}^{i1}
\right),&
\overline{\nabla}^{i}&=
-\frac{1}{\sqrt{2}}
\left(
\mathcal{D}^{i2}-\overline{\mathcal{D}}^{i2}
\right)
\end{align}
where $\mathcal{D}^{ia}, \overline{\mathcal{D}}^{ia}$ are introduced in 
(\ref{n8ssc9}). 
Then the constraints (\ref{n8sse1}) provides us 
with the two independent $(\bm{2},\bm{4},\bm{2})$ superfields 
\begin{align}
\label{n8sse2n}
w&=\mathcal{W}|,&\phi&=\Phi|.
\end{align}

The free supersymmetric action can be written as
\begin{align}
\label{n8sse2o}
S=\int dtd^{4}\theta \left[w\overline{w}-\phi\overline{\phi}\right].
\end{align}

\end{enumerate}

The $(\bm{4},\bm{8},\bm{4})$ supermultiplet can be 
constructed by reducing two-dimensional 
$\mathcal{N}=(4,4)$ or heterotic $\mathcal{N}=(8,0)$ sigma model 
\cite{Gates:1994yn}.

\subsubsection{\textrm{$(\bm{5},\bm{8},\bm{3})$} supermultiplet}
The $(\bm{5},\bm{8},\bm{3})$ supermultiplet 
is described by the five bosonic superfields 
$\mathcal{V}_{\alpha a}, \mathcal{U}$ which satisfy 
\begin{align}
\label{n8ssf1}
D^{ib}\mathcal{V}_{\alpha a}
&=-\delta^{b}_{a}\nabla^{i}_{\alpha}\mathcal{U},&
\nabla^{\beta i}\mathcal{V}_{\alpha a}
&=-\delta^{\beta}_{\alpha}D^{i}_{\alpha}\mathcal{U}.
\end{align}
The constraints (\ref{n8ssf1}) 
are covariant not only with respect to 
three $SU(2)$ factors for the indices 
$i,a,\alpha$ but also with respect to 
the $SO(5)$ R-symmetry. 
The $SO(5)$ R-symmetry transformations 
mix the spinor derivatives

The $(\bm{5},\bm{8},\bm{3})$ supermultiplet 
may have two decompositions
\begin{align}
\label{n8ssf1a1}
(\bm{5},\bm{8},\bm{3})
=\begin{cases}
(\bm{1},\bm{4},\bm{3})\oplus 
(\bm{4},\bm{4},\bm{0})\cr
(\bm{3},\bm{4},\bm{1})\oplus 
(\bm{2},\bm{4},\bm{2})\cr
\end{cases}
\end{align}

\begin{enumerate}
 \item \textbf{$(\bm{1},\bm{4},\bm{3})\oplus (\bm{4},\bm{4},\bm{0})$}

Using the $\mathcal{N}=4$ superspace (\ref{n8ssa5}) 
and carrying out the expansion of the superfields in
       $\vartheta_{i\alpha}$, 
we find the independent $\mathcal{N}=4$ superfields 
\begin{align}
\label{n8ssf1a2}
v_{\alpha a}&=\mathcal{V}_{\alpha a}|,&
u&=\mathcal{U}|
\end{align}
which satisfy 
\begin{align}
\label{n8ssf1a3}
D^{i(a}v^{b)\alpha}&=0,&
D^{i(a}D_{i}^{b)}u&=0.
\end{align}
Hence we obtain the $(\bm{4},\bm{4},\bm{0})$ superfield $v_{a\alpha}$ 
and the $(\bm{1},\bm{4},\bm{3})$ superfield $u$.

\item \textbf{$(\bm{3},\bm{4},\bm{1})\oplus (\bm{2},\bm{4},\bm{2})$}

In order to present the decomposition 
$(\bm{5},\bm{8},\bm{3})=$ 
$(\bm{2},\bm{4},\bm{2})\oplus (\bm{2},\bm{4},\bm{2})$, 
we introduce the new set of superfields $\mathcal{W}$, 
$\overline{\mathcal{W}}$ and $\mathcal{W}^{\alpha\beta}$ as
\begin{align}
\label{n8ssf1a4}
\mathcal{W}^{\alpha\beta}&=\frac12 
\left(\mathcal{V}^{\alpha\beta}+\mathcal{V}^{\beta\alpha}\right),&
\mathcal{W}&=-\epsilon_{\alpha a}\mathcal{V}^{\alpha a}+i\mathcal{U}
\end{align}
and the new $\mathcal{N}=4$ superspace 
\begin{align}
\label{n8ssf1a5}
\mathbb{R}^{(1|4)}
&=(t,\theta_{i\alpha}+\vartheta_{\alpha
 i},\theta^{i\alpha}-i\vartheta^{\alpha i}) 
\subset \mathbb{R}^{(1|8)}.
\end{align}
Then the constraints (\ref{n8ssf1}) leave us with 
the independent $\mathcal{N}=4$ superfields 
\begin{align}
\label{n8ssf1a6}
\phi&=\mathcal{W},&
e^{\alpha\beta}&=\mathcal{W}^{\alpha\beta}
\end{align}
which obey 
\begin{align}
\label{n8ssf1a7}
\mathcal{D}^{\alpha}\overline{\phi}&=0,&
\overline{\mathcal{D}}_{\alpha}\phi&=0,\\
\mathcal{D}^{(\alpha}w^{\beta\gamma)}&=0,&
\overline{\mathcal{D}}^{(\alpha}w^{\beta\gamma)}&=0.
\end{align}
Here the $\mathcal{N}=4$ 
covariant derivatives $\mathcal{D}^{\alpha}$, 
$\overline{\mathcal{D}}_{\alpha}$ are defined as
\begin{align}
\label{n8ssf1a8}
\mathcal{D}^{\alpha}&=\mathcal{D}^{1\alpha}&
\overline{\mathcal{D}}_{\alpha}&=
\overline{\mathcal{D}}_{1\alpha}
\end{align}
in terms of the covariant derivatives introduced  
in (\ref{n8ssc9}). 
Therefore the $\mathcal{N}=4$ superfields 
$\phi$ and $w^{\alpha\beta}$ are 
the $(\bm{2},\bm{4},\bm{2})$ superfield and 
the $(\bm{3},\bm{4},\bm{1})$ superfield respectively.

The $\mathcal{N}=4$ supersymmetric free action is given by 
\cite{Bellucci:2003hn}
\begin{align}
\label{n8ssf1b1}
S=\int dtd^{4}\theta 
\left[
w^{2}-\frac34 \phi\overline{\phi}
\right].
\end{align}

The $OSp(4^{*}|4)$ superconformal action can be written as 
\cite{Bellucci:2003hn}
\begin{align}
\label{n8ssf1b2}
S=2\int dtd^{4}\theta 
\frac{\ln \left(
\sqrt{w^{2}}+\sqrt{w^{2}+\frac12 \phi\overline{\phi}}
\right)}
{\sqrt{w^{2}}}
\end{align}
whose bosonic part has the form
\begin{align}
\label{n8ssf1b3}
S_{\mathrm{bosonic}}
=\int dt \frac{\dot{w}_{\alpha}\dot{w}_{\alpha\beta}
+\frac12 \dot{\phi}\dot{\overline{\phi}}}
{\left(w^{2}+\frac12\phi\overline{\phi}\right)^{\frac32}}.
\end{align}
The action (\ref{n8ssf1b3}) can be regarded as 
a conformal invariant type of the $SO(5)$ invariant sigma-model action 
of \cite{Diaconescu:1997ut}.

\end{enumerate}

The $(\bm{5},\bm{8},\bm{3})$ supermultiplet 
can be obtained by the dimensional reduction 
of the $d=4$ $\mathcal{N}=2$ Abelian multiplet \cite{Diaconescu:1997ut}. 
The three extra physical scalar fields 
originate from the spatial component fields of 
the $d=4$ gauge vector potential. 

Using the non-linear realization technique, 
it has been shown \cite{Bellucci:2003hn} 
that the $(\bm{5},\bm{8},\bm{3})$ supermultiplet 
can parametrize a coset of $OSp(4^{*}|4)$ such that 
the four out of five Goldstone bosons parametrize the 
$SO(5)/SO(4)$ coset 
while the remaining one Goldstone boson is the dilaton.

\subsubsection{\textrm{$(\bm{6},\bm{8},\bm{2})$} supermultiplet}
The $(\bm{6},\bm{8},\bm{2})$ supermultiplet 
has two tensor superfields $\mathcal{V}^{ij}, \mathcal{W}^{ab}$ 
subjected to the conditions
\begin{align}
\label{n8susyg1}
D_{a}^{(i}\mathcal{V}^{jk)}&=0,&
\nabla_{a}^{(i}\mathcal{V}^{jk)}&=0,\\
\label{n8susyg2}
D_{i}^{(a}\mathcal{W}^{bc)}&=0,&
\nabla_{i}^{(a}\mathcal{W}^{bc)}&=0,\\
\label{n8susyg3}
D_{j}^{a}\mathcal{V}^{ij}&=\nabla^{bi}\mathcal{W}^{a}_{b},&
\nabla^{a}_{j}\mathcal{V}^{ij}&=-D_{b}^{i}\mathcal{W}^{ab}.
\end{align}
The conditions (\ref{n8susyg3}) 
identify the eight fermions in $\mathcal{V}^{ij}$ 
with those in $\mathcal{W}^{ab}$ 
and also reduce the number of the auxiliary fields to two. 
 
The $(\bm{6},\bm{8},\bm{2})$ supermultiplet can be
decomposed as
\begin{align}
\label{n8susyg4}
(\bm{6},\bm{8},\bm{2})
=\begin{cases}
(\bm{3},\bm{4},\bm{1})\oplus 
(\bm{3},\bm{4},\bm{1})\cr
(\bm{4},\bm{4},\bm{0})\oplus 
(\bm{2},\bm{4},\bm{2}).\cr
\end{cases}
\end{align}

\begin{enumerate}
 \item \textbf{$(\bm{3},\bm{4},\bm{1})\oplus (\bm{3},\bm{4},\bm{1})$}

Using the $\mathcal{N}=4$ superspace (\ref{n8ssa5}) 
and expanding the superfields in $\vartheta$, 
we can project out the $\mathcal{N}=4$ superfields 
\begin{align}
\label{n8susyg4a}
v^{ij}&=\mathcal{V}^{ij},&
w^{ab}&=\mathcal{W}^{ab}
\end{align}
obeying 
\begin{align}
\label{n8susyg4b}
D^{a(i}v^{jk)}&=0,&
D^{i(a}w^{bc)}&=0.
\end{align}
Thus we obtain the two $(\bm{3},\bm{4},\bm{1})$ superfields 
$v^{ij}$ and $w^{ab}$. 

The supersymmetric free action reads 
\begin{align}
\label{n8susyg4c}
S=\int dtd^{4}\theta 
\left(v^{2}-w^{2}\right).
\end{align}

\item \textbf{$(\bm{4},\bm{4},\bm{0})\oplus (\bm{2},\bm{4},\bm{2})$}

This decomposition can be realized by 
combining the 
$(\bm{2},\bm{4},\bm{2})$ chiral multiplet $\phi,\overline{\phi}$ 
and the 
$(\bm{4},\bm{4},\bm{0})$ hypermultiplet $q^{ia}$. 

The invariant free action takes the form
\begin{align}
\label{n8susyg4d}
S=\int dtd^{4}\theta 
\left(q^{2}-4\phi\overline{\phi}\right).
\end{align}

\end{enumerate}

\subsubsection{\textrm{$(\bm{7},\bm{8},\bm{1})$} supermultiplet}
The $(\bm{7},\bm{8},\bm{1})$ supermultiplet 
contains two different types of superfields
$\mathcal{V}^{ij},\mathcal{Q}^{a\alpha}$ 
which obey 
\begin{align}
\label{n8susyh1}
D^{(ia}\mathcal{V}^{jk)}
&=0,&
\nabla^{\alpha(i}\mathcal{V}^{jk)}&=0,\\
\label{n8susyh2}
D^{i(a}\mathcal{Q}^{\alpha b)}&=0,&
\nabla_{i}^{(\alpha}\mathcal{Q}_{a}^{\beta)}&=0,
\\
\label{n8susyh3}
D_{j}^{a}\mathcal{V}^{ij}
&=i\nabla^{i}_{\alpha}\mathcal{Q}^{a\alpha}
,&
\nabla^{\alpha}_{j}\mathcal{V}^{ij}&=
-iD_{a}^{i}\mathcal{Q}^{a\alpha}.
\end{align} 
The constraints 
(\ref{n8susyh1}) extract the  
$(\bm{3},\bm{8},\bm{5})$ 
and $(\bm{4},\bm{8},\bm{4})$ supermultiplets 
from the superfields $\mathcal{V}^{ij}$ and 
$\mathcal{Q}^{a\alpha}$ respectively. 
The constraints (\ref{n8susyh2}) identify 
the fermions in the superfields $\mathcal{V}^{ij}$ 
and $\mathcal{Q}^{a\alpha}$ 
and reduce the number of the auxiliary fields to one.

The $(\bm{7},\bm{8},\bm{1})$ supermultiplet 
has a unique splitting 
\begin{align}
\label{n8susyh4}
(\bm{7},\bm{8},\bm{1})
=(\bm{3},\bm{4},\bm{1})\oplus 
(\bm{4},\bm{4},\bm{0}).
\end{align}

By using the $\mathcal{N}=4$ superspace 
(\ref{n8ssa5}) and 
expanding the superspace in $\vartheta$, 
we find the independent $\mathcal{N}=4$ superfields 
\begin{align}
\label{n8susyh5}
v^{ij}&=\mathcal{V}^{ij}|,&
q^{a\alpha}&=\mathcal{Q}^{a\alpha}|
\end{align}
which satisfy the constraints 
\begin{align}
\label{n8susyh6}
D^{a(i}v^{jk)}&=0,&
D^{i(a}q^{b)\alpha}&=0.
\end{align}
We thus obtain 
the $(\bm{3},\bm{4},\bm{1})$ superfield $v^{ij}$ 
and the $(\bm{4},\bm{4},\bm{0})$ superfield $q^{a\alpha}$.

The invariant free action is given by \cite{Bellucci:2004ur}
\begin{align}
\label{n8susyh7}
S=\int dtd^{4}\theta 
\left[
v^{2}-\frac43 q^{2}
\right].
\end{align}

\subsubsection{\textrm{$(\bm{8},\bm{8},\bm{0})$} supermultiplet}
The $(\bm{8},\bm{8},\bm{0})$ supermultiplet 
possesses two real bosonic superfields $\mathcal{Q}^{aA},
\Phi^{i\alpha}$ 
which obey 
\begin{align}
\label{n8susyi1}
D^{(ia}\Phi^{j)\alpha}&=0,&
\nabla^{(\alpha A}\Phi_{i}^{\beta)}&=0,\\
\label{n8susyi2}
D^{i(a}\mathcal{Q}^{b)A}&=0,&
\nabla^{\alpha(A}\mathcal{Q}^{aB)}&=0,\\
\label{n8susyi3}
\nabla^{\alpha A}\Phi^{i}_{\alpha}&=D^{ia}\mathcal{Q}_{a}^{A},&
\nabla^{\alpha A}\mathcal{Q}_{A}^{a}&=-D^{ia}\Phi_{i}^{\alpha}.
\end{align}
Similar to the $(\bm{0},\bm{8},\bm{8})$ supermultiplet, 
the two conditions (\ref{n8susyi2}) and (\ref{n8susyi3}) means that 
the covariant derivatives with respect to $\vartheta_{\alpha A}$ 
can be written in terms of the covariant derivatives 
with respect to $\theta_{ia}$.

The $(\bm{0},\bm{8},\bm{8})$ supermultiplet 
has a unique decomposition 
\begin{align}
\label{n8susyi4}
(\bm{8},\bm{8},\bm{0})
=(\bm{4},\bm{4},\bm{0})\oplus
(\bm{4},\bm{4},\bm{0}).
\end{align}

Choosing the $\mathcal{N}=4$ superspace 
(\ref{n8ssa5}) and expanding the superfields in $\vartheta$, 
one find the independent $\mathcal{N}=4$ superfields 
\begin{align}
\label{n8susyi4a}
q^{aA}&=\mathcal{Q}^{aA}|,&
\phi^{i\alpha}&=\Phi^{i\alpha}|
\end{align}
satisfying the constraints for $(\bm{4},\bm{4},\bm{0})$ supermultiplet
\begin{align}
\label{n8susyi4b}
D^{a(i}\phi^{j)\alpha}&=0,&
D^{i(a}q^{b)A}&=0.
\end{align}
This implies that 
the $(\bm{8},\bm{8},\bm{0})$ supermultiplet 
can be decomposed as the sum of the 
two $(\bm{4},\bm{4},\bm{0})$ supermultiplets 
as in (\ref{n8susyi4}). 

The invariant free action can be written as \cite{Bellucci:2004ur}
\begin{align}
\label{n8susyi4c}
S=\int dtd^{4}\theta \left[q^{2}-\phi^{2}\right].
\end{align}

\subsection{Multi-particle model}
Let us consider the 
$\mathcal{N}=8$ supersymmetric sigma-model 
\footnote{The $\mathcal{N}=8$ superconformal sigma-model 
has not been well understood. We will only discuss the 
$\mathcal{N}=8$ supersymmetric sigma-model in this thesis.}. 

Suppose we have the extended supersymmetry transformations 
as the form 
\begin{align}
\label{n8susy5a1}
\delta \Phi^{i}&=\epsilon^{A}{(I_{A})^{i}}_{j}D\Phi^{j}
\end{align}
where $\Phi^{i}$ is the $(\bm{1},\bm{2},\bm{1})$ superfields 
and $\epsilon^{A}$, $A=1,\cdots,7$ are the supersymmetry parameters 
and $I_{A}$ are the endomorphism of the tangent bundle of the 
target space. 
This $\mathcal{N}=8$ supermultiplet is called $\mathcal{N}=8B$ multiplet. 
This is related to the two-dimensional $\mathcal{N}=(4,0)$
supersymmetry. 
The closure of the $\mathcal{N}=8$ superalgebra requires that 
\cite{Gibbons:1997iy}
\begin{align}
\label{n8s1a2}
I_{A}I_{B}+I_{B}I_{A}&=-2\delta_{AB},\\
\label{n8s1a3}
N(I_{A},I_{B})&=0
\end{align}
where a $N(F,G)$ is 
Nijenhuis concomitant defined in (\ref{n4s1a3}). 
Thus the target space $\mathcal{M}$ has 
seven complex structures $I_{r}$ which 
have vanishing mixed Nijenhuis tensors 
and the underlying algebraic structure is associated with that of 
octonions. 

The invariance of the action 
under the $\mathcal{N}=8B$ supersymmetry leads to 
\begin{align}
\label{n8s1a4}
&g_{ij}={(I_{A})^{k}}_{i}{(I_{A})^{l}}_{j}g_{kl},\\
\label{n8s1a5}
&\nabla_{(i}^{(+)}{(I_{A})^{k}}_{j}=0,\\
\label{n8s1a6}
&\partial_{[i}\left({I^{m}}_{j}c_{|m|kl]}
\right)-2{(I_{A})^{m}}_{[i}\partial_{[m}c_{jkl]]}=0.
\end{align}
The first condition(\ref{n4s1a4}) implies that 
the metric $g$ on $\mathcal{M}$ 
is Hermitian with respect to the seven complex structures. 
The second condition (\ref{n4s1a5}) is a generalized Yano tensor
condition with torsion and the third condition (\ref{n4s1a6}) 
is imposed on torsion and complex structures.

The Riemannian manifold 
$\left\{\mathcal{M},g,c\right\}$ with 
a metric $g$, a torsion three-form $c$ and 
three complex structures $I_{A}$, $A=1,\cdots,7$ which 
obey the conditions (\ref{n8s1a2})-(\ref{n8s1a6}) is called 
Octonionic K\"{a}hler with torsion manifold (OKT)
\cite{Gibbons:1997iy}.

\part{M2-branes}
\chapter{BLG-model}
\label{blgch1}
The dominant theme of this chapter and the next chapter is 
the world-volume theories of the multiple planar M2-branes 
\footnote{See \cite{Bagger:2012jb} for the excellent review on the
world-volume theories of the multiple planar M2-branes.}. 
We will begin in this chapter with the BLG-model 
\cite{Bagger:2006sk,Bagger:2007jr,Bagger:2007vi,
Gustavsson:2007vu,Gustavsson:2008dy},  
which is one of the candidate descriptions of the 
low-energy dynamics of the multiple planar M2-branes. 
In section \ref{blgsec1} we will set our notations and conventions 
and review the basic properties. 
In section \ref{blgsec2} we will focus on the study 
of the $\mathcal{A}_{4}$ BLG-model that is the non-trivial 
finite dimensional Lie 3-algebra with positive definite metric, 
which may describe two membranes.

\section{Construction}
\label{blgsec1}
The BLG-model is a three-dimensional $\mathcal{N}=8$ 
supersymmetric Chern-Simons matter theory found by Bagger, Lambert 
\cite{Bagger:2006sk,Bagger:2007jr, Bagger:2007vi} 
and Gustavsson \cite{Gustavsson:2007vu,Gustavsson:2008dy}. 
It is characterized by a Lie 3-algebra $\mathcal{A}$, 
which is a generalization of a Lie algebra. 
The action has a manifest $\mathcal{N}=8$
supersymmetry and the $SO(8)_{R}$ R-symmetry. 
It has been shown \cite{Bandres:2008vf} 
that the $SO(4)$ BLG theory has an $OSp(4|8)$ superconformal
symmetry at the classical level.

The field content is
\begin{itemize}
 \item 8 real scalar fields $X^{I}=X_{a}^{I}T^{a}$
 \item 16 (8 on-shell) real fermionic fields $\Psi_{\dot{A}}=\Psi_{\dot{A}a}T^{a}$ 
 \item gauge fields $A_{\mu}=A_{\mu ab}\mathcal{T}^{ab}$.
\end{itemize}
Here $T^{a}$, $a=1,\cdots,\mathrm{dim}\mathcal{A}$ 
is a basis of the Lie 3-algebra $\mathcal{A}$ and 
$\mathcal{T}^{ab}$ is the fundamental object in 
$\mathcal{A}$ which will be introduced in (\ref{fundob}). 
Under the $SO(8)_{R}$ R-symmetry 
the bosonic scalar fields $X^{I}$, $I=1,\cdots,8$ 
are the vector representations $\bm{8}_{v}$ 
while the fermionic fields $\Psi_{\dot{A}}$, $\dot{A}=1,\cdots,8$ 
are the conjugate spinor representations $\bm{8}_{c}$ respectively. 

They also carry the  
$(\dim \mathcal{A})$-dimensional representations of the Lie 3-algebra. 
The gauge fields $A_{\mu ab}$ are 3-algebra $\mathcal{A}$ valued
world-volume vector fields. 
They are antisymmetric under two indices $a,b$ 
of the Lie 3-algebra; $A_{\mu ab}=-A_{\mu ba}$ 
\footnote{For the $\mathcal{A}_{4}$ algebra 
we have a one-to-one correspondence between the fundamental object 
$\mathcal{T}$ and the element $T^{ab}$ of the associated Lie
algebra $\mathfrak{so}(4)$. 
Hence $A_{\mu ab}$ is Lie $\mathfrak{so}(4)$-valued.
Moreover matter fields $X_{a}^{I},\Psi_{\dot{A}a}$ are interpreted as
the fundamental representations $\bm{4}$ of $\mathfrak{so}(4)$.
}.
The mass dimensions of the field content 
and the supersymmetry parameter $\epsilon$ are given by
\begin{equation}
\label{blgdim}
[X_{a}^{I}]=\frac12,\ \ \ [\Psi_{a}]=1,\ \ \ [A_{\mu}]=1,\ \ \ [\epsilon]=-\frac12
\end{equation}

$\Psi_{\dot{A}a}$ is defined as an $SO(1,10)$ Majorana fermion and its
conjugate is given by
\begin{equation}
 \overline{\Psi}:=\Psi^{T}\mathcal{C},
\end{equation}
where $\mathcal{C}$ is a $SO(1,10)$ charge conjugation matrix satisfying
\begin{align}
 \mathcal{C}^{T}&=-\mathcal{C},&
\mathcal{C}\Gamma^{M}\mathcal{C}^{-1}&=-(\Gamma^{M})^{T}.
\end{align}
Gamma matrix $\Gamma^{M}$ is the representation of eleven-dimensional
Clifford algebra
\begin{equation}
 \{\Gamma^{M},\Gamma^{N}\}=2g^{MN}
\end{equation}
\begin{equation}
\label{10gamma}
 \Gamma^{10}:=\Gamma^{0\cdots 9},
\end{equation}
where $g^{MN}=\eta^{MN}=\textrm{diag}(-1,+1,+1,\cdots,+1)$. 
$\Gamma^{M}$ can be decomposed as
\begin{equation}
\label{11gamma}
\begin{cases}
 \Gamma^{\mu}=\gamma^{\mu}\otimes \tilde{\Gamma}^{9}&\mu=0,1,2 \cr
 \Gamma^{I}=\mathbb{I}_{2}\otimes \tilde{\Gamma}^{I-2}&I=3,\cdots,10\cr
\end{cases}
\end{equation}
where 
\begin{equation}
\label{3dgamma}
 \gamma^{0}=\left(
\begin{array}{cc}
0&1\\
-1&0\\
\end{array}
\right)=i\sigma_{2},\ \ \ \gamma^{1}=\left(
\begin{array}{cc}
0&1\\
1&0\\
\end{array}
\right)=\sigma_{1},\ \ \ 
\gamma^{2}=\left(
\begin{array}{cc}
1&0\\
0&-1\\
\end{array}
\right)=\sigma_{3}
\end{equation}
and $\tilde{\Gamma}^{I}$ is an $SO(8)$ $16\times 16$ gamma matrix whose
chirality matrix is $\tilde{\Gamma}^{9}:=\tilde{\Gamma}^{1\cdots 8}$. 
The fermionic field $\Psi$ is a real $\frac12 \cdot
2^{[\frac{11}{2}]}=32$-component Majorana spinor of eleven-dimensional
space-time, obeying the chirality condition\footnote{
32 supercharges in M-theory is broken to 16 due to the existence of
M2-branes and $\Psi$ is identified with the Goldstino corresponding to
the broken supersymmetry. 
Therefore the chirality condition on $\Psi$ is opposite to that of 
supersymmetry parameters $\epsilon$.
}
\begin{equation}
\label{blgchiral}
 \Gamma^{012}\Psi=-\Psi.
\end{equation}
Although at this stage $\Psi$ contains 16 independent real components, 
the number is reduced to 8 when we treat it on-shell. 
From (\ref{11gamma}) it follows that 
\begin{equation}
 \Gamma^{012}=\Gamma^{34\cdots 10}=\mathbb{I}_{2}\otimes \tilde{\Gamma}^{9}
\end{equation}
and 
\begin{equation}
 \Gamma^{34\cdots 10}\Psi=-\Psi.
\end{equation}
Thus $\Psi$ is the conjugate spinor representation 
$\bm{8}_{c}$ of the $SO(8)_{R}$ R-symmetry group.

\subsection{Lie 3-algebra}
The construction of the BLG model is based on the Lie 3-algebra $\mathcal{A}$.
The Lie 3-algebra is an $N$-dimensional vector space
endowed with the totally antisymmetric multi-linear triple product $[A,B,C]$
satisfying the fundamental identity
\begin{equation}
\label{fundid}
 [A,B[C,D,E]]
=[[A,B,C],D,E]
+[C,[A,B,D],E]
+[C,D,[A,B,E]],
\end{equation}
which is a generalization of the Jacobi identity in Lie algebra and
requires that the gauge symmetry $\delta_{AB}X=[A,B,X]$ acts as the
derivation\footnote{Jacobi identity $[A,[B,C]]=[[A,B],C]+[B,[A,C]]$
ensures that the transformation $\delta_{A}X=[A,X]$ behaves as
derivation $\delta_{A}[B,C]=[\delta_{A}B,C]+[B,\delta_{A}C]$.}
\begin{equation}
 \delta_{AB}([C,D,E])
=[\delta_{AB}C,D,E]
+[C,\delta_{AB}D,E]
+[C,D,\delta_{AB}E].
\end{equation}
The supersymmetry algebra of 
the BLG model is closed on-shell when the fundamental
identity (\ref{fundid}) is satisfied \cite{Bagger:2007jr}. 
Let us introduce the basis $\{T^{a}\}_{1\le a\le N}$ of 3-algebra. 
Then the 3-algebra is specified by the metric $h^{ab}$ and 
the structure constant ${f^{abc}}_{d}$
\begin{align}
\label{3algmet}
 h^{ab}&=(T^{a},T^{b}),\\
 [T^{a},T^{b},T^{c}]&={f^{abc}}_{d}T^{d}.
\end{align}
In terms of the structure constant, the fundamental identity
(\ref{fundid}) can be expressed as
\begin{align}
\label{fundid2}
{f^{abc}}_{g}{f^{deg}}_{f}
&={f^{dea}}_{g}{f^{bcg}}_{f}
+{f^{deb}}_{g}{f^{cag}}_{f}
+{f^{dec}}_{g}{f^{abg}}_{f} \\
&=3{f^{de[a}}_{g}{f^{bc]g}}_{f},
\end{align}
which turns out to be equivalent to the relation \cite{Gran:2008vi}
\begin{equation}
\label{fundid3}
 {f^{[abc}}_{g}{f^{d]eg}}_{f}=0.
\end{equation}

Here we will 
define the fundamental object $\mathcal{T}=\mathcal{T}^{ab}$ as
\begin{equation}
 \label{fundob}
\mathcal{T}\cdot X:=[T^{a},T^{b},X],\ \ \ \ \forall X\in \mathcal{A}.
\end{equation}
The fundamental object 
induces derivation and gives the adjoint map
\begin{align}
\label{fundob1}
\textrm{ad}_{T^{a}T^{b}}:X\mapsto [T^{a},T^{b},X],\ \ \ \ 
\forall X\in\mathcal{A}.
\end{align} 
If we require that 
the action of the derivation on the scalar product is invariant
\begin{equation}
 \mathcal{T}\cdot (T^{c},T^{d})
=(\mathcal{T}\cdot T^{c},T^{d})+(T^{c},\mathcal{T}\cdot T^{d})=0, 
\end{equation}
then we obtain the relation
\begin{equation}
\label{metric}
 (T^{a},[T^{b},T^{c},T^{d}])
=-([T^{a},T^{b},T^{c}],T^{d}).
\end{equation}
A Lie 3-algebra is called ``metric'' if it satisfies the relation 
(\ref{metric}). 
This metric property is assumed for all of the BLG theories. 
In terms of the structure constant, (\ref{metric}) is rewritten as
\begin{equation}
\label{metric2}
 f^{abcd}=f^{[abcd]}.
\end{equation}
This antisymmetry of $f^{abcd}$ indicates that the symmetry algebra is contained in $\mathfrak{so}(N)$.
To be more precise, we rewrite the fundamental identity (\ref{fundid}) as
\begin{equation}
\label{fundid4}
\textrm{ad}_{AB}(\textrm{ad}_{CD}X)
-\textrm{ad}_{CD}(\textrm{ad}_{AB}X)
=\textrm{ad}_{([A,B,C],D)+(C,[A,B,D])}X
\end{equation}
or equivalently
\begin{equation}
\label{fundid5}
 \textrm{ad}_{\mathcal{T}}(\textrm{ad}_{\mathcal{S}}X)
-\textrm{ad}_{\mathcal{S}}(\textrm{ad}_{\mathcal{T}}X)
=\textrm{ad}_{\mathcal{TS}}X,\ \ \ \forall \mathcal{T},\mathcal{S}\in
\wedge^{2}\mathcal{A}, \ \ X\in \mathcal{A}.
\end{equation}
Introducing the coordinates of the
$(\dim\mathcal{A}\times \dim\mathcal{A})$ matrices $[T^{a_{1}},T^{a_{2}},\
]=:\mathcal{T}^{a_{1}a_{2}}=:\textrm{ad}_{a_{1}a_{2}}\in
\textrm{End}\mathcal{A}$ as
\begin{align}
{\textrm{ad}_{\mathcal{T}^{a_{1}a_{2}}}}^{l}_{k}
&=(\mathcal{T}^{a_{1}a_{2}})^{l}_{k}
:={f^{a_{1}a_{2}l}}_{k} \\
\mathcal{T}^{a_{1}a_{2}}\cdot T^{k}&=
[T^{a_{1}},T^{a_{2}},T^{k}]={f^{a_{1}a_{2}k}}_{l}T^{l},
\end{align}
then the equations (\ref{fundid4}) and (\ref{fundid5}) may be written in
the form
\begin{equation}
 [(\mathcal{T}^{a_{1}a_{2}}), (\mathcal{T}^{b_{1}b_{2}})]^{s}_{k}
=-f^{a_{1}a_{2}[b_{1}l}f^{b_{2}]}\ {_{lk}}^{s},
\end{equation}
which means that 
\begin{align}
\label{lie1}
 [(\mathcal{T}^{a_{1}a_{2}}),(\mathcal{T}^{b_{1}b_{2}})]^{s}_{k}
&=\frac12
 {C^{a_{1}a_{2}b_{1}b_{2}}}_{c_{1}c_{2}}(\mathcal{T}^{c_{1}c_{2}})^{s}_{k}
\end{align}
where
\begin{equation}
\label{stcon1}
 {C^{a_{1}a_{2}b_{1}b_{2}}}_{c_{1}c_{2}}
={f^{a_{1}a_{2}[b_{1}}}_{[c_{1}}\delta^{b_{2}]}_{c_{2}]}.
\end{equation}
Although (\ref{lie1}) is the same form of the commutator 
in the Lie algebra, 
this does not mean 
${C^{a_{1}a_{2}b_{1}b_{2}}}_{c_{1}c_{2}}$ are the
structure constants of the Lie algebra $\mathfrak{g}$ because
\begin{enumerate}
 \item ${C^{a_{1}a_{2}b_{1}b_{2}}}_{c_{1}c_{2}}$ may not be antisymmetric
       under $(a_{1},a_{2})\leftrightarrow (b_{1},b_{2})$

\item $\mathcal{T}^{c_{1}c_{2}}$ may not be the basis 
of the Lie algebra $\mathfrak{g}$.
\end{enumerate}
However, it has been shown \cite{2010JPhA...43C3001D} 
that when the Lie 3-algebra is simple,
${C^{a_{1}a_{2}b_{1}b_{2}}}_{c_{1}c_{2}}$ are antisymmetric in the upper
indices 
\begin{equation}
 {f^{a_{1}a_{2}[b_{1}}}_{[c_{1}}\delta_{c_{2}]}^{b_{2}]}
=-{f^{b_{1}b_{2}[a_{1}}}_{[c_{1}}\delta_{c_{2}]}^{a_{2}]}
\end{equation}
and define the structure constants of Lie algebra $\mathfrak{g}$. 
Moreover one can find the cases where $\mathcal{T}^{c_{1}c_{2}}$ can be
viewd as the basis of $\mathfrak{g}$.

\subsection{Lagrangian}
The BLG-model Lagrangian is
\begin{align}
\label{blglagrangian}
 \mathcal{L}=
&-\frac12 D^{\mu}X^{Ia}D_{\mu}X_{a}^{I}
+\frac{i}{2}\overline{\Psi}_{\dot{A}}^{a}\Gamma^{\mu}_{\dot{A}\dot{B}}D_{\mu}\Psi_{\dot{B}a}\nonumber\\
&+\frac{i}{4}\overline{\Psi}_{\dot{A}b}
\Gamma^{IJ}_{\dot{A}\dot{B}}X_{c}^{I}X_{d}^{J}\Psi_{\dot{B}a}f^{abcd}
-V(X)+\mathcal{L}_{TCS}
\end{align}
where 
\begin{align}
 V(X)=&\frac{1}{12}f^{abcd}{f^{efg}}_{d}X_{a}^{I}X_{b}^{J}X_{c}^{K}
X_{e}^{I}X_{f}^{J}X_{g}^{K},\\
 \mathcal{L}_{TCS}
=&\frac12 \epsilon^{\mu\nu\lambda}
\left(
f^{abcd}A_{\mu ab}\partial_{\nu}A_{\lambda cd}
+\frac23 {f^{cda}}_{g}f^{efgb}A_{\mu ab}A_{\nu cd}A_{\lambda ef}
\right).
\end{align}
The covariant derivative is defined as
\begin{align}
\label{blgderiv}
 D_{\mu}X_{a}
&:=
\partial_{\mu}X_{a}-A_{\mu cd}[T^{c},T^{d},X]_{a}\nonumber\\
&=\partial_{\mu}X_{a}-\tilde{A}_{\mu a}^{b}X_{b}
\end{align}
where $\tilde{A}_{\mu b}^{a}:={f^{cda}}_{b}A_{\mu cd}$.
Alternatively we can express the Lagrangian in terms of the trace and
the triple product of Lie 3-algebra:
\begin{align}
\label{blglagrangian2}
\mathcal{L}
=&-\frac12 (D_{\mu}X^{I},D^{\mu}X^{I})
+\frac{i}{2}(\overline{\Psi},\Gamma^{\mu}D_{\mu}\Psi)\nonumber\\
&+\frac{i}{4}\left(
\overline{\Psi}\Gamma^{IJ}[X^{I},X^{J},\Psi]
\right)-\frac{1}{12}
\left([X^{I},X^{J},X^{K}],[X^{I},X^{J},X^{K}]\right)\nonumber\\
&+\frac12 \epsilon^{\mu\nu\lambda}
\left[
\textrm{Tr}
\left(A_{\mu ab}\partial_{\nu}\tilde{A}_{\lambda}^{ab}\right)
+\frac23\textrm{Tr}\left(
A_{\mu ab}\tilde{A}^{a}_{\nu g}\tilde{A}^{b}_{\lambda g}
\right)
\right]
\end{align}
Although the kinetic term of the gauge fields is similar to the conventional
Chern-Simons term, it is twisted by the structure constant of the 3-algebra. 
Notice that the gauge fields are non-propagating since it has at most
first order derivative terms. This is consistent with the degrees
of freedom required from supersymmetry. 

From (\ref{blglagrangian}), we obtain the equations of motion
\begin{align}
\label{eom1}
 D^{\mu}D_{\mu}X_{a}^{I}
-\frac{i}{2}\overline{\Psi}_{c}\Gamma^{IJ}X_{d}^{J}\Psi_{b}
{f^{cdb}}_{a}
+\frac12 {f^{bcd}}_{a}{f^{efg}}_{d}
X_{b}^{J}X_{c}^{K}X_{e}^{I}X_{f}^{J}X_{g}^{K}
=&0,\\
\label{eom2}
 \Gamma^{\mu}D_{\mu}\Psi_{a}
+\frac12 \Gamma^{IJ}X_{c}^{I}X_{d}^{J}\Psi_{b}{f^{cdb}}_{a}=&0,\\
\label{eom3}
 \tilde{F}^{b}_{\mu\nu a}+\epsilon_{\mu\nu\lambda}
(X_{c}^{J}D^{\lambda}X_{d}^{J}+\frac{i}{2}\overline{\Psi}_{c}\Gamma^{\lambda}\Psi_{d}){f^{cdb}}_{a}=&0.
\end{align}
Here the field strength of the gauge field is defined as
\begin{equation}
 \tilde{F}^{b}_{\mu\nu a}X_{b}
:=[D_{\mu},D_{\nu}]X_{a}.
\end{equation}
Combining the definition (\ref{blgderiv}) of the covariant derivative,
we can express it as
\begin{equation}
 \tilde{F}^{b}_{\mu\nu a}=
\partial_{\nu}\tilde{A}^{b}_{\mu a}-\partial_{\mu}\tilde{A}_{\nu a}^{b}
-\tilde{A}^{b}_{\mu c}\tilde{A}_{\nu a}^{c}
+\tilde{A}_{\nu c}^{b}\tilde{A}^{c}_{\mu a}.
\end{equation}
The field strength satisfies Bianchi identity
\begin{equation}
\label{bianchi}
 \epsilon^{\mu\nu\lambda}D_{\mu}\tilde{F}_{\nu\lambda b}^{a}=0.
\end{equation}
The stress-energy tensor can be computed as
\begin{equation}
\label{blgemtensor}
 T_{\mu\nu}=D_{\mu}X_{a}^{I}D_{\nu}X^{Ia}
-\eta_{\mu\nu}\left(
\frac12 D_{\lambda}X^{Ia}D^{\lambda}X_{a}^{I}
+V(X)
\right)
\end{equation}
where we set fermionic fields to zero. 
Thus bosonic part of the Hamiltonian density is 
\begin{equation}
\label{blghamiltonian}
 \mathcal{H}=T_{00}=\frac12 D_{0}X^{Ia}X_{0}X_{a}^{I}
+\frac12 D_{\alpha}X^{Ia}D^{\alpha}X_{a}^{I}
+V(X)
\end{equation}
and the momentum density is
\begin{equation}
\label{blgmomentum}
 p_{\alpha}=T_{0\alpha}
=D_{0}X^{Ia}D_{\alpha}X^{I}_{a}.
\end{equation}

\subsection{Gauge transformation}
The gauge transformations of the BLG-model are given by
\begin{align}
\label{gauge1}
 \delta_{\Lambda}X_{a}^{I}=&\Lambda_{cd}[T^{c},T^{d},X^{I}]_{a}\nonumber
\\
=&\Lambda_{cd}{f^{cdb}}_{a}X_{b}^{I}
=\tilde{\Lambda}^{b}_{a}X_{b}^{I},\\
\label{gauge2}
 \delta_{\Lambda}\Psi_{a}=&\Lambda_{cd}[T^{c},T^{d},\Psi]_{a} \nonumber
\\
=&\Lambda_{cd}{f^{cdb}}_{a}\Psi_{b}
=\tilde{\Lambda}^{b}_{a}\Psi_{b},\\
\label{gauge3}
 \delta_{\Lambda}\tilde{A}_{\mu a}^{b} 
=&\partial_{\mu}\tilde{\Lambda}_{\mu a}^{b}-\tilde{\Lambda}^{b}_{c}\tilde{A}^{c}_{\mu a}
+\tilde{A}_{\mu c}^{b}\tilde{\Lambda}_{a}^{c} \nonumber \\
=&D_{\mu}\tilde{\Lambda}^{b}_{a},\\
\label{gauge4}
\delta\tilde{F}_{\mu\nu a}^{b}
=&-\tilde{\Lambda}_{c}^{b}\tilde{F}^{c}_{\mu\nu a}
+\tilde{F}^{b}_{\mu\nu c}\tilde{\Lambda}_{a}^{c},
\end{align}
where $\tilde{\Lambda}^{b}_{a}:={f^{cdb}}_{a}\Lambda_{cd}$ is a gauge
parameter. 
Lagrangian (\ref{blglagrangian}) is invariant up to a total derivative terms
under the above gauge transformations.

\subsection{Supersymmetry transformation}
The $\mathcal{N}=8$ supersymmetry transformations 
of the BLG-model are 
\begin{align}
\label{blgsusy1}
\delta
 X_{a}^{I}&=i\overline{\epsilon}_{A}\Gamma^{I}_{A\dot{B}}\Psi_{\dot{B}a},\\ 
\label{blgsusy2}
\delta\Psi_{\dot{A}a}&=
D_{\mu}X_{a}^{I}\Gamma^{\mu}\Gamma^{I}_{\dot{A}B}\epsilon_{B}
-\frac16 X_{b}^{I}X_{c}^{J}X_{d}^{K}
{f^{bcd}}_{a}\Gamma^{IJK}_{\dot{A}B}\epsilon_{B},\\
\label{blgsusy3}
\delta\tilde{A}_{\mu a}^{b}
&=i\overline{\epsilon}_{A}\Gamma_{\mu}\Gamma^{I}_{A\dot{B}}
X_{c}^{I}\Psi_{\dot{B}d}{f^{cdb}}_{a}.
\end{align}
Here $\epsilon_{A}$, $A=1,\cdots,8$ 
is the unbroken supersymmetry parameter obeying the chirality condition
\begin{equation}
\label{blgchiralmtx}
 \Gamma^{012}\epsilon=\Gamma^{34\cdots 10}\epsilon=\epsilon.
\end{equation}
This implies that $\epsilon_{A}$ 
is a two component three-dimensional Majorana spinor 
and transforms as the spinor representation 
$\bm{8}_{s}$ of the $SO(8)_{R}$ R-symmetry. 
Lagrangian (\ref{blglagrangian}) is invariant 
under the supersymmetric transformations up to a total derivative. 

Using the equations of motion (\ref{eom1}), (\ref{eom2}) and
(\ref{eom3}), we find the following relations
from (\ref{blgsusy1}), (\ref{blgsusy2}) and  (\ref{blgsusy3}):
\begin{align}
\label{off1}
[\delta_{1},\delta_{2}]X_{a}^{I}
&=v^{\lambda}D_{\lambda}X_{a}^{I}+\tilde{\Lambda}^{b}_{a}X_{b}^{I},\\
\label{off2}
[\delta_{1},\delta_{2}]\Psi_{a}
&=v^{\lambda}D_{\lambda}\Psi_{a}
+\tilde{\Lambda}^{b}_{a}\Psi_{b},\\
\label{off3}
[\delta_{1},\delta_{2}]\tilde{A}^{b}_{\mu a}
&=v^{\lambda}\tilde{F}_{\mu\lambda a}^{b}
+D_{\mu}\tilde{\Lambda}^{b}_{a}
\end{align}
where
$v^{\lambda}=-2i\overline{\epsilon}_{2}\Gamma^{\lambda}\epsilon_{1}$
 and
 $\tilde{\Lambda}^{b}_{a}=-2i\overline{\epsilon}_{2}\Gamma^{JK}\epsilon_{1}X_{c}^{J}X_{d}^{K}{f^{cdb}}_{a}$
 are identified 
with a translation parameter and a gauge parameter respectively. 
Thus the supersymmetry transformations close into 
a translation (the first term) 
and a gauge transformation (the second term) on-shell 
and the theory is invariant under 16 supersymmetries and $SO(8)_{R}$
R-symmetry at the classical level.

Allowing the supersymmetry parameter 
$\epsilon$ to has $x$ dependence 
and taking supersymmetry variations of the action, we obtain
\begin{equation}
 \label{noether1}
\delta S=
-i\int d^{3}x D_{\mu}\overline{\epsilon}\left(
D_{\nu}X^{I}_{a}\Gamma^{\nu}\Gamma^{I}\Gamma^{\mu}\Psi^{a}
+\frac16 X_{a}^{I}X_{b}^{J}X_{c}^{K}
f^{abcd}\Gamma^{IJK}\Gamma^{\mu}\Psi_{d}
\right).
\end{equation}
This gives 
\begin{equation}
\label{blgsupercurrent}
 J^{\mu}=-D_{\nu}X^{I}_{a}\Gamma^{\nu}\Gamma^{I}\Gamma^{\mu}\Psi^{a}
-\frac16 X_{a}^{I}X_{b}^{J}X_{c}^{K}
f^{abcd}\Gamma^{IJK}\Gamma^{\mu}\Psi_{d}.
\end{equation}
Then the supercharge is 
\begin{equation}
\label{blgsupercharge}
 Q=\int dx^{2}x J^{0}
=-\int d^{2}x 
\left(
D_{\nu}X^{I}_{a}\Gamma^{\nu}\Gamma^{I}\Gamma^{0}\Psi^{a}
+\frac16 X_{a}^{I}X_{b}^{J}X_{c}^{K}f^{abcd}\Gamma^{IJK}\Gamma^{0}\Psi_{d}
\right).
\end{equation}
From (\ref{blgdim}) one can check that $Q$ has the correct mass
dimension $[Q]=\frac12$ and $J^{0}$ has $[J^{0}]=\frac52$. 
The supercharge $Q$ is the SUSY generator in the sense that\footnote{
The symbol $[A,B\}$ means $AB-(-1)^{AB}BA$ in a $\mathbb{Z}_{2}$-graded algebra.
}
\begin{equation}
 \delta_{\epsilon}\Phi
=i[\overline{\epsilon}Q,\Phi\}
=\begin{cases}
i\overline{\epsilon}_{\dot{P}}[Q^{\dot{P}},\Phi_{B}]&(\textrm{bosonic
  field})\cr
i\overline{\epsilon}_{\dot{P}}\{Q^{\dot{P}},\Phi_{F}^{\dot{Q}}\}&(\textrm{fermionic
  field})\cr
\end{cases}
\end{equation}
where $\dot{P},\dot{Q},\cdots$ are $11$-dimensional spinor indices. 
As an example, we can generate the SUSY transformation for the scalar
fields $X^{I}$
\begin{align}
\delta X^{I}
=&i\overline{\epsilon}[Q,X^{I}] \nonumber \\
=&i\overline{\epsilon}
\left[
-\int d^{2}x \partial_{\nu}X^{J}(x)\Gamma^{\nu}\Gamma^{J}\Gamma^{0}\Psi(x), X^{I}(x')
\right] \nonumber \\
=&-i\overline{\epsilon}\Gamma^{0}\Gamma^{J}\Gamma^{0}\int d^{2}x \Psi(x)
\left[
\partial_{0}X^{J}(x),X^{I}(x')
\right] \nonumber \\
=&i\overline{\epsilon}\Gamma^{J}
\int d^{2}x \Psi(x)\delta^{IJ}\delta(x-x')
=i\overline{\epsilon}\Gamma^{I}\Psi.
\end{align}

\subsection{M2-brane algebra}
Now we want to discuss the algebraic structure 
of the M2-brane by studying the BLG-model. 
Noting that
\footnote{The central charges are proportional to the
world-volume of M2-branes and can be infinite for infinitely extended
M2-branes. Focusing on the charge density, we can avoid the infinities.}
\begin{align}
i\overline{\epsilon}_{\dot{P}}
\{Q^{\dot{P}},Q^{\dot{Q}}\}
&=\int d^{2}x \overline{\epsilon}_{\dot{P}}
\left\{
Q^{\dot{P}},J^{0\dot{Q}}(x)
\right\}\nonumber\\
&=\int d^{2}x (\delta_{\epsilon}J^{0\dot{Q}}(x)),
\end{align}
we obtain \cite{Passerini:2008qt, Furuuchi:2008ki}
\begin{align}
\label{m2alg}
 \left\{Q^{\dot{P}},Q^{\dot{Q}}\right\}
=&-2P_{\mu}(\Gamma^{\mu}\Gamma^{0})^{\dot{P}\dot{Q}}
+Z_{IJ}\Gamma^{IJ}\Gamma^{0} \nonumber \\
&+Z_{\alpha IJKL}(\Gamma^{IJKL}\Gamma^{\alpha}\Gamma^{0})^{\dot{P}\dot{Q}}
+Z_{IJKL}(\Gamma^{IJKL})^{\dot{P}\dot{Q}}
\end{align}
where $\alpha$ is the two-dimensional spatial indice of the M2-brane
world-volume and $P^{\mu}$ is the energy momnetum vector 
$P^{\mu}:=\int d^{2}x T^{0\mu}$. 
The central charges are given by
\begin{align}
\label{blgcc1}
 Z_{IJ}
&=-\int d^{2}x \textrm{Tr}\left(
D_{\alpha}X^{I}D_{\beta}X^{J}\epsilon^{\alpha\beta}
-D_{0}X^{K}[X^{I},X^{J},X^{K}]
\right),\\
\label{blgcc2}
 Z_{\alpha IJKL}
&=\frac13 \int d^{2}x \textrm{Tr}
\left(
D_{\beta}X^{[I} [X^{J},X^{K},X^{L]}]\epsilon^{\alpha\beta}
\right),\\
\label{blgcc3}
 Z_{IJKL}
&=\frac14 \int d^{2}x \textrm{Tr}
\left(
[X^{M},X^{[I},X^{J}],[X^{M},X^{K},X^{L]}]
\right).
\end{align}
Introducing the expression
\begin{equation}
 \tilde{\Gamma}^{I}=
\left(
\begin{array}{cc}
0&\overline{\Gamma}^{I}_{\dot{A}A}\\
\overline{\Gamma}^{I}_{B\dot{B}}&0\\
\end{array}
\right)
\end{equation}
where $(\overline{\Gamma}^{I}_{A\dot{A}})^{T}
=\overline{\Gamma}^{I}_{\dot{A}A}$ are $8\times 8$ real gamma matrices
satisfying 
\begin{align}
\overline{\Gamma}^{I}_{A\dot{A}}\overline{\Gamma}^{J}_{\dot{A}B}
+\overline{\Gamma}^{J}_{A\dot{A}}\overline{\Gamma}_{\dot{A}B}^{I}
=2\delta^{IJ}\delta_{AB},\nonumber\\
\overline{\Gamma}_{\dot{A}A}^{I}\overline{\Gamma}_{A\dot{B}}^{J}
+\overline{\Gamma}^{J}_{\dot{A}A}
\overline{\Gamma}^{I}_{A\dot{B}}
=2\delta^{IJ}\delta_{\dot{A}\dot{B}}
\end{align}
we can rewrite (\ref{blgcc1}) and (\ref{blgcc2}) as surface integrals
\cite{Passerini:2008qt}
\begin{align}
Z^{[AB]}&=-\int d^{2}x \partial_{\alpha} \textrm{Tr}
\left(X^{I},D_{\beta}X^{J}\right)\epsilon^{\alpha\beta}(\overline{\Gamma}^{IJ})^{AB},\\
Z_{\mu}^{(AB)}&=-\frac{1}{12}
\int d^{2}x \partial_{\alpha} \textrm{Tr}
\left(
X^{I},[X^{J},X^{K},X^{L}]
\right){\epsilon^{0\alpha}}_{\mu}
(\overline{\Gamma}^{IJKL})^{AB}
\end{align}
where the symmetric central charge is traceless
$\delta_{AB}Z_{\mu}^{(AB)}=0$ and $A,B,\cdots=1,\cdots,8$ are 
the $SO(8)$ indices. 
(\ref{m2alg}) and (\ref{blgcc1})-(\ref{blgcc3}) are the field
realization of the M2-brane algebra and the central charges \cite{Bergshoeff:1997bh}.
These are useful tools to investigate five constitutes in M-theory, that
is M-wave, M2-brane, M5-brane, M-KK monopole, M9-brane.

\begin{enumerate}
 \item $Z^{[AB]}$

$Z^{[AB]}$ is a world-volume 0-form transforming $\bm{28}$ of $SO(8)$. 
0-form corresponds to a 0-brane (point) on the M2-brane. 
$\bm{28}$ defines a 2-form or 6-form in the transverse space to the M2-brane. 
In the case of 2-form, 0-brane is the result of the intersection with
       two another M2-brane over a point and defines the 2-plane along
       which the second M2-brane is aligned
       \cite{Papadopoulos:1996uq}.
\begin{equation}
\label{m2m2}
 \begin{array}{cccccccccccc}
&0&1&2&3&4&5&6&7&8&9&10\\
\textrm{M2}&\circ&\circ&\circ&\times&\times&\times&\times&\times&\times&\times&\times\\
\textrm{M2}&\circ&\times&\times&\circ&\circ&\times&\times&\times&\times&\times&\times\\
\end{array}
\end{equation}

When choosing 6-form, 0-brane acquires the interpretation as the intersection of M2-brane  with M-KK
       monopole over a point \cite{Bergshoeff:1997tt}.
\begin{equation}
\label{m2kk}
 \begin{array}{cccccccccccc}
&0&1&2&3&4&5&6&7&8&9&10\\
\textrm{M2}&\circ&\circ&\circ&\times&\times&\times&\times&\times&\times&\times&\times\\
\textrm{MKK}&\circ&\times&\times&\circ&\circ&\circ&\circ&\circ&\circ&\times&\times\\
\end{array}
\end{equation}

\item $Z^{(AB)}_{\mu}$

$Z_{\mu}^{(AB)}$ is a world-volume 1-form and $\bm{35}^{+}$ of $SO(8)$. 
1-form corresponds to a 1-brane (string) on the M2-brane. 
$\bm{35}^{+}$ defines a 4-form in the 8-dimensional transverse space. 
1-brane is determined by 4-plane along which four of the spatial
      spaces of the M5-brane are aligned. 
Thus 1-brane has the interpretation as the intersection of M2-brane with
      M5-brane.
\begin{equation}
\label{m2m5}
 \begin{array}{cccccccccccc}
&0&1&2&3&4&5&6&7&8&9&10\\
\textrm{M2}&\circ&\circ&\circ&\times&\times&\times&\times&\times&\times&\times&\times\\
\textrm{M5}&\circ&\circ&\times&\circ&\circ&\circ&\circ&\times&\times&\times&\times\\
\end{array}
\end{equation}

\item $Z_{IJKL}$

Due to the total antisymmetry and the fundamental
      identity, 
$Z_{IJKL}=Z_{[IJKL]}$ vanishes when we consider trace elements. 

However, it is discussed  \cite{Furuuchi:2008ki} 
that if we take into account constant background
      configurations of $X^{I}$ that take values in non-trace elements\footnote{Configurations with non-trace elements are discussed in
      the matrix theory conjecture for M-theory in the light-cone
      quantization \cite{Banks:1996vh}.},
      such configurations may give rise to BPS charges although
      non-abelian fields are infinite dimensional and have an infinite
      norm\footnote{By the novel Higgs mechanism, we can reduce $Z_{IJKL}$ to the form
      $\textrm{Tr}[X^{I},X^{J}][X^{I},X^{J}]$ which is similar to
      D4-brane charge in the D0-brane action in the matrix model. It is
      natural to think that $Z_{IJKL}$ is identified with D6-brane
      charge because the BLG theory action reduces to the D2-brane action rather than
      D0-brane action. Furthermore D6-brane is uplifted to M-KK
      monopole, so $Z_{IJKL}$ is expected to produce the energy bound of the
      configuration of M2-brane and M-KK monopole.}.

\item $P_{\mu}$

$P_{\mu}$ is a 1-form on a world-volume and a singlet $\bm{1}$ of
      $SO(8)$. 
1-form corresponds to a 1-brane (string) on the M2-brane. 
$\bm{1}$ defines a 0-form or 8-form in the transverse space. 
In the case of 0-form, 1-brane can be viewed as the intersection of M2-brane with an M-wave over
      a 1-dimensional string.
\begin{equation}
\label{m2mw}
 \begin{array}{cccccccccccc}
&0&1&2&3&4&5&6&7&8&9&10\\
\textrm{M2}&\circ&\circ&\circ&\times&\times&\times&\times&\times&\times&\times&\times\\
\textrm{MW}&\circ&\circ&\times&\times&\times&\times&\times&\times&\times&\times&\times\\
\end{array}
\end{equation}
In the case of 8-form, 1-brane is the intersection of the M2-brane
      with M9-brane over a string.
\begin{equation}
\label{m2m9}
 \begin{array}{cccccccccccc}
&0&1&2&3&4&5&6&7&8&9&10\\
\textrm{M2}&\circ&\circ&\circ&\times&\times&\times&\times&\times&\times&\times&\times\\
\textrm{M9}&\circ&\circ&\times&\circ&\circ&\circ&\circ&\circ&\circ&\circ&\circ\\
\end{array}
\end{equation}
\end{enumerate}

\section{$\mathcal{A}_{4}$ BLG-theory}
\label{blgsec2}
If we assume that
\begin{enumerate}
 \item the metric $h^{ab}$ of the 3-algebra $\mathcal{A}$ 
is positive definite so that the kinetic
       term and the potential term are all positive,
\item the dimension $N$ of 3-algebra $\mathcal{A}$ is finite,
\end{enumerate}
then the 3-algebra $\mathcal{A}$ is uniquely determined by
\cite{Papadopoulos:2008sk,Gauntlett:2008uf} 
\begin{align}
\label{a4f001}
f^{abcd}&=\frac{2\pi}{k}\epsilon^{abcd}=f\epsilon^{abcd},\\
\label{a4f002}
h^{ab}&=\delta^{ab}
\end{align}
with $a,b=1,\cdots,4$. 
Here $\epsilon^{abcd}$ is an antisymmetric tensor and  
$k$ is the integer  determined by the quantization of the
Chern-Simons level for a non-simply connected gauge group $SO(4)$ 
\cite{Distler:2008mk}. 
The correct normalization can be checked by
using the expression (\ref{a4lagrangian}) and noting that the
coefficient of the Chern-Simons term is $\frac{k}{4\pi}$. 

The 3-algebra characterized by (\ref{a4f001}) and (\ref{a4f002}) 
is called the $\mathcal{A}_{4}$ algebra. 
For the $\mathcal{A}_{4}$ algebra 
we do not distinguish superscripts and subscripts 
since gauge indices $a,b,\cdots$ are 
raised and lowered with Kronecker delta. 
However, $A$ and $\tilde{A}$ should be distinguished 
because of the existence of $f$. 
The corresponding BLG theory has no continuous coupling constant 
but admit a discrete coupling $k$.
The uniqueness up to the Chern-Simons level $k$ makes it difficult to
describe an arbitrary number of coincident M2-branes because the rank of
the gauge algebra is expected to be related to the number of M2-branes
in analogy with D-branes. 

In terms of the antisymmetric tensor $\epsilon_{abcd}$ let us 
introduce the dual generators
\begin{align}
 M_{a_{1}a_{2}}:=\frac12 \epsilon_{a_{1}a_{2}b_{1}b_{2}}\mathcal{T}^{b_{1}b_{2}}
\end{align}
for the fundamental object $\mathcal{T}$. 
Then from the relation
\begin{equation}
 \epsilon^{i_{1}\cdots i_{n}}_{j_{1}\cdots j_{n}}
=\sum_{k=1}^{n}(-1)^{k+1}\delta^{i_{1}}_{j_{k}}
\epsilon^{i_{2}\cdots i_{n}}_{j_{1}\cdots \hat{j}_{k}\cdots j_{n}}
=\sum_{k=1}^{n}(-1)^{k+n}
\epsilon_{j_{1}\cdots \hat{j}_{k}\cdots j_{n}}^{i_{1}\cdots i_{n-1}}
\delta_{j_{k}}^{i_{n}},
\end{equation}
we obtain the commutation relations
\begin{align}
\label{a4f003}
 [M_{a_{1}a_{2}},M_{b_{1}b_{2}}]
=-\delta_{a_{1}b_{2}}M_{a_{2}b_{1}}
-\delta_{a_{2}b_{1}}M_{a_{1}b_{2}}
+\delta_{a_{1}b_{1}}M_{a_{2}b_{2}}
+\delta_{a_{2}b_{2}}M_{a_{1}b_{1}}
\end{align}
The algebraic relation (\ref{a4f003}) 
is recognized as commutators 
of semisimple $\mathfrak{so}(4)$ algebra. 
Thus from the ordinary Lie algebra point of view, 
the $\mathcal{A}_{4}$ BLG theory is based on the $\mathfrak{so}(4)$
gauge algebra. 
It has been discussed \cite{Lambert:2010ji} that 
for the $\mathcal{A}_{4}$ BLG-model there are two 
possible inequivalent gauge groups $G$; 
\begin{align}
\label{a4f004}
G=
\begin{cases}
SO(4)&\cong (SU(2)\times SU(2))/\mathbb{Z}_{2}\cr
Spin(4)&\cong SU(2)\times SU(2).\cr
\end{cases}
\end{align}

\subsection{Quiver gauge structure}
Now we want to discuss the connection between 
the BLG-model based on the Lie 3-algebras 
and the ordinary gauge theories based on the Lie algebras. 
This has been accomplished 
by the remarkable observation \cite{VanRaamsdonk:2008ft} 
that the $\mathcal{A}_{4}$ BLG-model 
can be rewritten as an ordinary gauge 
theory with quiver type gauge group  
and matters in the bifundamental representation 
\footnote{This alternative expression of the $\mathcal{A}_{4}$
BLG-model triggered 
the discovery of the ABJM-model.}.

Since in the $\mathcal{A}_{4}$ theory 
the Higgs fields $X^{I}$ and $\Psi$ are the fundamental 
representation $\bm{4}$ of the $\mathfrak{so}(4)$ 
we can denote them by the four-vectors 
\begin{align}
\label{a4quiv1}
 X^{I}&=\left(
\begin{array}{c}
x_{1}^{I}\\
x_{2}^{I}\\
x_{3}^{I}\\
x_{4}^{I}\\
\end{array}
\right),&
\Psi&=\left(
\begin{array}{c}
\Psi_{1}\\
\Psi_{2}\\
\Psi_{3}\\
\Psi_{4}
\end{array}
\right).
\end{align}
In terms of the Pauli matrices $\sigma_{i}$\footnote{Pauli matrices
$\sigma_{i}$ are given in (\ref{3dgamma}) and normalized such that
$\textrm{Tr}(\sigma_{i}\sigma_{j})=2\delta_{ij}$.}, 
one may express these in the 
bi-fundamental representation $(\bm{2},\bm{2})$ of the 
$\mathfrak{su}(2)\oplus \mathfrak{su}(2)$ gauge algebra as
\begin{align}
\label{a41}
X^{I}=&\frac12 (x_{4}^{I}\mathbb{I}_{2}+ix^{I}_{i}\sigma^{i})
=\frac12\left(
\begin{array}{cc}
x_{4}^{I}+ix_{3}^{I}&x_{2}^{I}+ix_{1}^{I}\\
-x_{2}^{I}+ix_{1}^{I}&x_{4}^{I}-ix_{3}^{I}\\
\end{array}
\right), \nonumber \\
\Psi=&\frac12 (\Psi_{4}\mathbb{I}_{2}+i\Psi_{i}\sigma^{i})
=\frac12\left(
\begin{array}{cc}
\Psi_{4}+i\Psi_{3}&\Psi_{2}+i\Psi_{1}\\
-\Psi_{2}+i\Psi_{1}&\Psi_{4}-i\Psi_{3}\\
\end{array}
\right).
\end{align}
They obey the reality conditions
\begin{align}
X^{I}_{\alpha\dot{\beta}}
=&\epsilon_{\alpha\beta}\epsilon_{\dot{\beta}\dot{\alpha}}(X^{I\dag})^{\dot{\alpha}\beta}
,\nonumber \\
\Psi_{\alpha\dot{\beta}}
=&\epsilon_{\alpha\beta}\epsilon_{\dot{\beta}\dot{\alpha}}
(\Psi^{\dag})^{\dot{\alpha}\beta},
\end{align}
where $\alpha,\beta=1,2$ and $\dot{\alpha},\dot{\beta}=1,2$ 
denote bi-fundamental representation $(\bm{2},\bm{2})$ 
of the $\mathfrak{su}(2)\times \mathfrak{su}(2)$ gauge algebra.

In order to find the adjoint gauge field for each $\mathfrak{su}(2)$ 
gauge symmetry factor, 
we decompose gauge fields $A_{\mu ab}$
 into the sum of the selfdual and anti-selfdual parts
\begin{equation}
 A_{\mu ab}:=-\frac{1}{2f}(A_{\mu ab}^{+}+A_{\mu ab}^{-})
\end{equation}
where
\begin{align}
*A_{\mu ab}^{+}=&\frac12 {\epsilon_{ab}}^{cd}A_{\mu cd}^{+}=A_{\mu
 ab}^{+},\nonumber \\
*A_{\mu ab}^{-}=&\frac12 {\epsilon_{ab}}^{cd}A_{\mu
 cd}^{-}
=-A_{\mu ab}^{-}
\end{align}
and $*$ is the Hodge star acting on the gauge indices and satisfying $*^{2}=1$. 
Noting that $\tilde{A}_{\mu}^{ab}=f^{cdab}A_{\mu cd}$, we also have
\begin{equation}
 \tilde{A}_{\mu}^{cd}=-(A_{\mu}^{+cd}-A_{\mu}^{-cd}).
\end{equation}
Then we define
\begin{align}
\label{a42}
A_{\mu}:=&A_{\mu4i}^{+}\sigma_{i},\nonumber \\
\hat{A}_{\mu}:=&A_{\mu4i}^{-}\sigma_{i}.
\end{align}
Using the expressions (\ref{a41}) and (\ref{a42}), we rewrite the
original BLG-theory Lagrangian (\ref{blglagrangian}) as
\begin{align}
\label{a4lagrangian}
\mathcal{L}
=&-\textrm{Tr}\left(D^{\mu}X^{I\dag}D_{\mu}X^{I}\right)
+i\textrm{Tr}\left(\overline{\Psi}^{\dag}\Gamma^{\mu}D_{\mu}\Psi\right)
 \nonumber \\
&-\frac23 if\textrm{Tr}
\left(
\overline{\Psi}^{\dag}\Gamma^{IJ}X^{I}X^{J\dag}\Psi
+\overline{\Psi}^{\dag}\Gamma^{IJ}X^{J}\Psi^{\dag}X^{I}
+\overline{\Psi}^{\dag}\Gamma^{IJ}\Psi X^{I\dag}X^{J}
\right)
\nonumber \\
&-\frac83
 f^{2}\textrm{Tr}\left(X^{[I}X^{J\dag}X^{k]}X^{K\dag}X^{J}X^{I\dag}\right)
\nonumber \\
&+\frac{1}{2f}\epsilon^{\mu\nu\lambda}
\textrm{Tr}\left(
A_{\mu}\partial_{\nu}A_{\lambda}
+\frac{2}{3}iA_{\mu}A_{\nu}A_{\lambda}\right)
-\frac{1}{2f}\epsilon^{\mu\nu\lambda}
\textrm{Tr}\left(\hat{A}_{\mu}\partial_{\nu}\tilde{A}_{\lambda}
+\frac23 i\hat{A}_{\mu}\hat{A}_{\nu}\hat{A}_{\lambda}
\right).
\end{align}
Here the covariant derivative is defined by
\begin{align}
\label{a4lagrangian1}
 D_{\mu}X^{I}=\partial_{\mu}X^{I}+iA_{\mu}X^{I}
-iX^{I}\hat{A}_{\mu}.
\end{align}
Notice that now the twisted Chern-Simons terms in the original BLG-model
is decomposed into two ordinary Chern-Simons terms for $A$ and $\hat{A}$ with
opposite signs. 
This observation was crucial for the 
discovery of the ABJM-model 
as it opens up the highly extended supersymmetric 
Chern-Simons matter theories with quiver type gauge group. 
We see that 
the theory becomes weakly coupled in the large $k$ limit 
since after rescaling $A\rightarrow \sqrt{f}A$, 
all interaction terms are 
proportional to positive power of $f=\frac{2\pi}{k}$.

The Lagrangian (\ref{a4lagrangian}) is invariant under a new set of
supersymmetry transformations
\begin{align}
\label{a4susy1}
\delta X^{I}=&i\overline{\epsilon}\Gamma^{I}\Psi,\\
\label{a4susy2}
\delta\Psi=&D_{\mu}X^{I}\Gamma^{\mu}\Gamma^{I}\epsilon
+\frac{4\pi}{3}X^{I}X^{J\dag}X^{K}\Gamma^{IJK}\epsilon,\\
\label{a4susy3}
\delta A_{\mu}=&f\overline{\epsilon}\Gamma^{I}
(X^{I}\Psi^{\dag}-\Psi X^{I\dag}),\\
\label{a4susy4}
\delta\hat{A}_{\mu}
=&f\overline{\epsilon}
\Gamma_{\mu}\Gamma^{I}
(\Psi^{\dag}X^{I}-X^{I\dag}\Psi).
\end{align}

\subsection{Superconformal symmetry}
It has been proven \cite{Bandres:2008vf} that 
the $\mathcal{A}_{4}$ BLG theory has $OSp(8|4)$ superconformal symmetry 
that contains the $SO(8)_{R}$ R-symmetry group 
and the three-dimensional 
$Sp(4)\cong Spin(2,3)$ conformal symmetry group as bosonic factor groups 
at the classical level.  
To see the superconformal symmetry explicitly, 
we replace supersymmetry parameter $\epsilon_{A}$ 
by $\Gamma^{\mu}x_{\mu}\eta_{A}$ where $\eta_{A}$ is 
a superconformal symmetry parameter 
and add a term $-\Gamma^{I}X_{a}^{I}\eta$ to
$\delta \Psi_{a}$ in the supersymmetry transformations of the 
BLG-model. 
Then the superconformal symmetry is given by
\begin{align}
\label{blgscy1}
\delta
 X^{I}_{a}=&i\overline{\eta}\Gamma^{\mu}x_{\mu}\Gamma^{I}\Psi_{a},\\
\label{blgscy2}
\delta\Psi_{a}=&D_{\mu}X_{a}^{I}\Gamma^{\mu}\Gamma^{I}\Gamma^{\nu}x_{\nu}\eta
-\frac16 X_{b}^{I}X_{c}^{J}X_{d}^{K}{f^{bcd}}_{a}
\Gamma^{IJK}\Gamma^{\nu}x_{\nu}\eta
-\Gamma^{I}X_{a}^{I}\eta,\\
\label{blgscy3}
\delta\tilde{A}^{b}_{\mu a}
=&i\overline{\eta}x_{\nu}\Gamma^{\nu}\Gamma_{\mu}\Gamma^{I}X_{c}^{I}
\Psi_{d}{f^{cdb}}_{a}
\end{align}
and one can check that 
the action (\ref{blglagrangian}) is invariant 
under the superconformal transformations (\ref{blgscy1})-(\ref{blgscy3}) 
up to total derivative terms.

\subsection{Parity invariance}
Although Chern-Simons theories are parity violating, we can make the
$\mathcal{A}_{4}$ BLG Lagrangian (\ref{a4lagrangian}) parity invariant
by defining parity transformation as a spatial reflection together with
interchange of two $SU(2)$ gauge groups \cite{Bandres:2008vf,
Bagger:2007jr, VanRaamsdonk:2008ft}. 
This implies that we assign an odd parity to $f^{abcd}$. 
In particular, under the reflection $x^{2}\rightarrow -x^{2}$ we require
that
\begin{align}
\label{blgparity}
X_{a}^{I}&\rightarrow X_{a}^{I},&
\tilde{A}_{2b}^{a}&\rightarrow -\tilde{A}_{2b}^{a},\\
\tilde{A}_{0b}^{a}&\rightarrow \tilde{A}_{0b}^{a},&
f^{abcd}&\rightarrow -f^{abcd},\\
\tilde{A}_{1b}^{a}&\rightarrow \tilde{A}_{1b}^{a},&
\Psi_{a}&\rightarrow \Gamma_{2}\Psi_{a}.
\end{align}
Then (\ref{a4lagrangian}) turns out ot be parity conserving.

\subsection{Moduli space}
\begin{figure}
\begin{center}
\includegraphics[width=5.5cm]{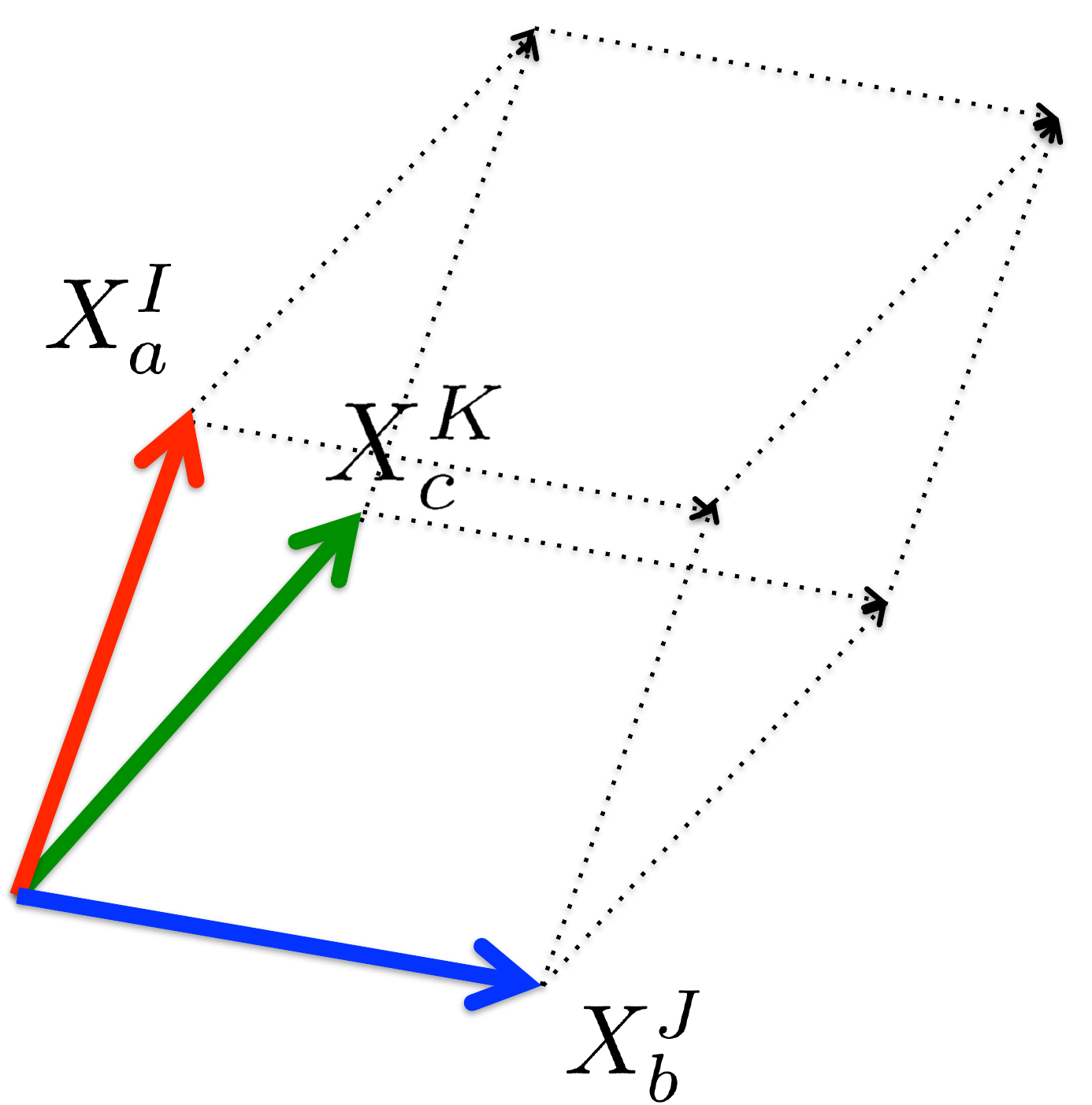}
\caption{The parallelepiped spanned by the three vectors $X_{a}^{I}$,
 $X_{b}^{J}$ and $X_{c}^{K}$. A new vector
 produced by the triple product has the length as the signed volume of
 the parallelepiped. The triple product is zero if and only if all the
 vectors lie in the same plane.}
\label{3algprod}
\end{center}
\end{figure}

The vacuum moduli space of the theory is the configuration space that
minimise the potential modulo gauge transformations. 
For the $\mathcal{A}_{4}$ BLG-model 
it was initially investigated in 
\cite{VanRaamsdonk:2008ft,Lambert:2008et,Distler:2008mk}. 
Since $\mathcal{A}_{4}$ BLG theory has the Euclidean inner product, 
the potential is positive definite and the potential is minimal when
\begin{equation}
 [X^{I},X^{J},X^{K}]=0.
\end{equation}
From the fact that the bosonic scalar fields $X^{I}_{a}$ are eight
vectors in an $\mathbb{R}^{4}$ rotated by the gauge symmetry $SO(4)$, 
the triple product $X_{a}^{I}X_{b}^{J}X_{c}^{K}$ produces a new vector
perpendicular to the three vectors $X_{a}^{I}$, $X_{b}^{J}$ and
$X_{c}^{K}$ whose length is the signed volume of the parallelepiped
spanned by the three vectors in $\mathbb{R}^{4}$ (see Figure \ref{3algprod}).

The bosonic potential is proportional to the square of this volume
summed over each possible triple of vectors. 
Therefore the bosonic potential vanishes if and only if all the three
vectors lie in the same plane. 
This space is labeled by ordered sets of eight vectors in the same
plane. 
One can assume that all vectors lie in the $x_{1}$-$x_{2}$ plane without
losing generality where $x_{a}$ are the coordinates of $T^{a}$. 
Then eight $x_{1}$ coordinates $r^{I}_{1}$ and the eight $x_{2}$
coordinates $r^{I}_{2}$ form two octuplets which are rotated into each
other by the residual $O(2)$ symmetry. 
Thus, up to gauge transformation, the vacuum moduli space is
parametrized by
\begin{equation}
\label{blgmoduli1}
 X_{a}^{I}=r_{1}^{I}T^{1}+r_{2}^{I}T^{2}=\left(
\begin{array}{c}
r_{1}^{I}\\
r_{2}^{I}\\
0\\
0\\
\end{array}
\right),\ \ \ r_{1}^{I},r_{2}^{I}\in \mathbb{R}^{8}.
\end{equation}
In the bi-fundamental notation (\ref{a41}), 
(\ref{blgmoduli1}) is expressed as
\begin{equation}
\label{blgmoduli2}
 X^{I}=\frac{1}{\sqrt{2}}
\left(
\begin{array}{cc}
z^{I}&0\\
0&\overline{z}^{I}\\
\end{array}
\right)
=\frac{1}{\sqrt{2}}
\left(
\begin{array}{cc}
r_{1}+ir_{2}^{I}&0\\
0&r_{1}^{I}-ir_{2}^{I}\\
\end{array}
\right).
\end{equation}
Then one can see that the residual gauge symmetries $g\in SO(4)$ that
preserve the form $X^{I}$ is the block diagonal form
\begin{equation}
 g=\left(
\begin{array}{cc}
g_{1}&0\\
0&g_{2}\\
\end{array}
\right)
\end{equation}
where $g_{1}, g_{2}\in O(2)$ act on 
$(x_{1}, x_{2})$ and $(x_{3},x_{4})$ respectively, with $\det g_{1}=\det g_{2}$.
Since $g_{2}$ acts trivially on (\ref{blgmoduli1}), 
we can ignore it and simply look at $g_{1}\in O(2)$.

Let us discuss the residual gauge symmetry 
in the diagonal configurations (\ref{blgmoduli2}). 
The residual $O(2)$ gauge symmetry in which $g_{1}$ is contained
consists of two types of symmetries:
\begin{enumerate}
 \item simultaneous rotation on $z^{I}$ (continuous symmetry)
\begin{equation}
\label{continblg4}
 U(1)_{12}:z^{I}\rightarrow e^{i\theta}z^{I},\ \ \ \theta \in [0,2\pi)
\end{equation}
\item simultaneous complex conjugation (discrete symmetry)
\begin{equation}
\label{discblg4}
 z^{I}\rightarrow \overline{z}^{I}
\end{equation}
\end{enumerate}
However, the continuous symmetry $U_{12}$ is generically 
broken down for the diagonal configuration (\ref{blgmoduli1}). 
Therefore the remaining component of gauge field become massive 
by the Higgs mechanism. 
To see this we shall write down the effective action. 
Let us firstly define the gauge field $B_{\mu}$ associated with the 
broken $U(1)_{12}$ that rotate $z^{I}$ 
and the preserved gauge field $C_{\mu}$ associated with the 
preserved $U(1)$ by 
\begin{align}
\label{a4modug1}
B_{\mu}:=&\frac{4\pi}{k}A_{\mu}^{34},\\
\label{a4modug2}
C_{\mu}:=&\frac{4\pi}{k}A_{\mu}^{12}.
\end{align}
Note that because of
$\epsilon^{abcd}$ in the covariant derivative (\ref{blgderiv}) 
the broken $U(1)_{12}$ gauge field is associated with $A_{\mu}^{34}$ not 
$A_{\mu}^{12}$. 
Substituting the configurations (\ref{blgmoduli1}), 
(\ref{a4modug1}) and (\ref{a4modug2}) into the 
BLG Lagrangian (\ref{blglagrangian}), 
one can write the kinetic terms on the moduli space and the twisted
Chern-Simons terms as \cite{Lambert:2008et,Distler:2008mk}
\begin{equation}
\label{kintcs}
 \mathcal{L}_{kin}+\mathcal{L}_{TCS}
=-\frac12 |\mathcal{D}_{\mu}z^{I}|^{2}
+\frac{k}{2\pi}\epsilon^{\mu\nu\lambda}B_{\mu}\partial_{\nu}C_{\lambda}
\end{equation}
where $\mathcal{D}_{\mu}z^{I}=\partial_{\mu}z^{I}+iB_{\mu}z^{I}$.

Moreover we can replace the unbroken gauge field 
$C$ with its dual photon 
$\sigma$ that plays a role of a Lagrange multiplier to impose the
      Bianchi identity 
$\epsilon^{\mu\nu\lambda}\partial_{\mu}G_{\nu\lambda}$ 
on the field strength
$G_{\mu\nu}:=\partial_{\mu}C_{\nu}-\partial_{\nu}C_{\mu}$ 
by introducing the additional term 
\footnote{In the original work of \cite{Lambert:2008et,Distler:2008mk} 
the normalization is chosen as 
$\mathcal{L}_{dual}=\frac{1}{8\pi}\sigma\epsilon^{\mu\nu\lambda}\partial_{\mu}G_{\nu\lambda}$
so that $\sigma\in [0,2\pi)$. 
However, this does depend on the two choice of the 
gauge group; $SU(2)\times SU(2)$ and $(SU(2)\times
SU(2))/\mathbb{Z}_{2}$ as pointed in \cite{Lambert:2010ji}. 
}
\begin{equation}
\label{dualphotonblg}
 \mathcal{L}_{dual}=\frac{1}{4\pi}\sigma\epsilon^{\mu\nu\lambda}\partial_{\mu}G_{\nu\lambda}.
\end{equation}

Combining (\ref{kintcs}) and (\ref{dualphotonblg}), 
we can write the low-energy effective action as 
\cite{Lambert:2008et,Distler:2008mk}
\begin{equation}
 \label{modefflag1}
\mathcal{L}_{kin}+\mathcal{L}_{TCS}+\mathcal{L}_{dual}
=-\frac12 |\mathcal{D}_{\mu}z^{I}|^{2}
+\frac{1}{4\pi}\epsilon^{\mu\nu\lambda}
(kB_{\mu}-\partial_{\mu}\sigma)G_{\nu\lambda}.
\end{equation}
The action (\ref{modefflag1}) is 
invariant under the $U(1)_{12}$ gauge symmetry transformations
\begin{align}
\label{gauge1}
z^{I}&\rightarrow e^{i\theta}z^{I},&
\sigma&\rightarrow\sigma+k\theta,&
B_{\mu}&\rightarrow B_{\mu}+\partial_{\mu}\theta.
\end{align}
Using the equation of motion for $G_{\mu\nu}$
\begin{align}
\label{modgauge1b}
B_{\mu}=\frac{\partial_{\mu}\sigma}{k},
\end{align}
the action (\ref{modefflag1}) further reduces to
\begin{align}
\label{modefflag2}
\mathcal{L}
=-\frac12|\partial_{\mu}z^{I}-\frac{i}{k}z^{I}\partial_{\mu}\sigma|^{2}.
\end{align}
By defining the fields
\begin{align}
\label{modefflag3}
w^{I}:=e^{-\frac{i\sigma}{k}}z^{I},
\end{align}
we can absorb the Lagrange multiplier $\sigma$ 
and the action (\ref{modefflag2}) finally becomes
\begin{align}
\label{modefflag4}
\mathcal{L}&=-\frac12 \partial_{\mu}w^{I}\partial^{\mu}\overline{w}^{I}.
\end{align}

As a next step 
we need to determine the periodicity of $\sigma$ which 
yields the gauge symmetry of the moduli parameter $z^{I}$ 
as seen from the redefinition (\ref{modefflag3}). 
The periodicity of $\sigma$ occurs from 
the Dirac quantization of the flux of the field strength. 
Let us consider the case where 
some field $\phi$ couples to a $U(1)$ gauge field $A_{\mu}$ 
as $D_{\mu}\phi=\partial_{\mu}\phi+iA_{\mu}\phi$. 
If we go around a closed path $\gamma$, 
then $\phi$ is parallel transported into 
$\phi_{\gamma}$ $=$ $e^{i\oint_{\gamma}A}\phi$ $=$ $e^{i\int_{\Sigma}F}\phi$ 
where $D$ is a two-dimensional surface whose boundary is $\gamma$ 
and $F$ is the field strength of $A$. 
Since the choice of the surface $\Sigma$ is not unique, 
we require that $q:=\int_{\Sigma}F=2\pi \mathbb{Z}$. 
This is the Dirac quantization for the charge $q$. 
Now we are interested in the Dirac quantization of the 
field strength $G=dC$ of the preserved gauge field $C$ 
since it yields the periodicity for $\sigma$ as we see from 
the action (\ref{modefflag1}). 
However, in our case the charge 
of the field strength $G=dC$ turns out to be different as the Dirac value. 
The result is given by \cite{Distler:2008mk,Lambert:2010ji}
\begin{align}
\label{modefflag5}
\int_{\Sigma}G \in 
\begin{cases}
4\pi \mathbb{Z}&\textrm{for $Spin(4)=SU(2)\times SU(2)$}\cr
2\pi \mathbb{Z}&\textrm{for $SO(4)=(SU(2)\times SU(2))/\mathbb{Z}_{2}$}.\cr
\end{cases}
\end{align}
This is because 
at the generic point of the moduli space 
the $U(1)$ gauge field $C$ sits inside the diagonal 
$SO(3)$ $\in$ $(SU(2)\times SU(2))/\mathbb{Z}_{2}$ 
or $SU(2)\times SU(2)$ 
and the Higgs fields does not transform as the adjoint representations 
of the $U(1)$ but that of the $SO(3)$.  
This situation is similar to the 't Hooft-Polyakov monopoles 
\cite{Montonen:1977sn} where 
all the fields transform in the adjoint representation of $SU(2)\cong
SO(3)$. 
For the $SU(2)\times SU(2)$ group $G$ is thus essentially the sum of two 
independent field strengths and 
we need the additional factor $2$ as $\int_{\Sigma}G\in 4\pi \mathbb{Z}$. 
For the $(SU(2)\times SU(2))/\mathbb{Z}_{2}$ gauge group 
the phase is equal to one only up to a $\mathbb{Z}_{2}$ action 
and we require that $\int_{\Sigma}G\in 2\pi \mathbb{Z}$. 
Noting that 
$dG=\frac12 \epsilon^{\mu\nu\lambda}\partial_{\mu}G_{\nu\lambda}$ 
and lifting the relation (\ref{modefflag5}) to 
the integral of $dG$ over the three-manifold, 
we get 
\begin{align}
\label{modefflag6}
\frac{1}{4\pi}\int \epsilon^{\mu\nu\lambda}\partial_{\mu}G_{\nu\lambda}
\in 
\begin{cases}
2\mathbb{Z}&\textrm{for $Spin(4)=SU(2)\times SU(2)$}\cr
\mathbb{Z}&\textrm{for $SO(4)=(SU(2)\times SU(2))/\mathbb{Z}_{2}$}.\cr
\end{cases}
\end{align}
Since $\sigma$ appears in the action (\ref{modefflag1}) 
as the coupling to 
$\frac{1}{4\pi}\epsilon^{\mu\nu\lambda}\partial_{\mu}G_{\nu\lambda}$,  
which takes the discrete value in (\ref{modefflag5}), 
$\sigma$ must be periodic as
\begin{align}
\label{modefflag7}
\sigma\sim
\begin{cases}
\sigma+\pi&\textrm{for $Spin(4)=SU(2)\times SU(2)$}\cr
\sigma+2\pi&\textrm{for $SO(4)=(SU(2)\times SU(2))/\mathbb{Z}_{2}$}.\cr
\end{cases}
\end{align}
Combining the periodicity (\ref{modefflag7}) and 
the expression (\ref{modefflag3}), 
we can read the gauge identification of $z^{I}$ 
from the continuous transformation (\ref{continblg4}) as
\begin{align}
\label{modefflag8}
z^{I}\cong 
\begin{cases}
e^{\frac{\pi i}{k}}z^{I}&\textrm{for $Spin(4)=SU(2)\times SU(2)$}\cr
e^{\frac{2\pi i}{k}}z^{I}&\textrm{for $SO(4)=(SU(2)\times SU(2))/\mathbb{Z}_{2}$}.\cr
\end{cases}
\end{align}
At this stage we have two types of the gauge equivalences; 
one is (\ref{modefflag8}) from the continuous symmetry (\ref{modefflag8}) 
yielding $\mathbb{Z}_{2k}$ or $\mathbb{Z}_{k}$ 
and the other is from the discrete one (\ref{discblg4}) 
corresponding to $\mathbb{Z}_{2}$. 
Since both of them do not commute, 
we finally obtain the moduli space $\mathcal{M}_{k}$ 
of the $\mathcal{A}_{4}$ BLG-model with the Chern-Simons level $k$ as 
\cite{Lambert:2010ji}
\begin{align}
\label{modefflag9}
\mathcal{M}_{k}=
\begin{cases}
\frac{\mathbb{R}^{8}\times \mathbb{R}^{8}}{D_{4k}}
&\textrm{for $Spin(4)=SU(2)\times SU(2)$}\cr
\frac{\mathbb{R}^{8}\times \mathbb{R}^{8}}{D_{2k}}
&\textrm{for $SO(4)=(SU(2)\times SU(2))/\mathbb{Z}_{2}$}.\cr
\end{cases}
\end{align}
For generic $k$ 
we do not know whether 
these moduli spaces can have a geometrical interpretation of the M2-branes. 
However, for $k=1,2,4$ there is a conjectural space-time 
interpretation of the M2-branes.

\chapter{ABJM-model}
\label{abjmch1}
In this chapter we will review the ABJM-model \cite{Aharony:2008ug} 
which may describe an arbitrary number of M2-branes. 
We will introduce the notations and conventions in 
section \ref{abjmsec1}. 
We will turn to the analysis 
of the moduli space in section \ref{abjmsec2}. 
Then we will discuss the conjectural duality 
between the BLG-model and the ABJM-model 
in section \ref{abjmsec3}.

\section{Construction}
\label{abjmsec1}
The ABJM-model is a three-dimensional $\mathcal{N}=6$ superconformal 
$U(N)_{k}\times \hat{U}(N)_{-k}$ Chern-Simons-matter theory 
proposed as a generalization of the BLG-model in that 
it may describe the dynamics of an arbitrary number of coincident M2-branes 
\cite{Aharony:2008ug}. 
The theory has manifestly only $\mathcal{N}=6$ 
supersymmetry and the corresponding $SU(4)_{R}$ R-symmetry 
at the classical level. 
It has been discussed that
\cite{Aharony:2008ug,Gustavsson:2009pm,Kwon:2009ar} 
at $k=1$ and $k=2$ 
these symmetries are enhanced to $\mathcal{N}=8$ supersymmetry 
and $SO(8)_{R}$ R-symmetry as a quantum effect. 
The theory contains 
\begin{itemize}
 \item 4 complex scalar fields $Y^{A}$
\item 4 Weyl spinors $\psi_{A}$
\item 2 types of gauge fields $A_{\mu}, \hat{A}_{\mu}$.
\end{itemize}
Here the upper and lower indices $A,B,\cdots=1,2,3,4$ denote 
$\bm{4}$ and $\overline{\bm{4}}$ of the $SU(4)_{R}$ respectively. 
The matter fields are $N\times N$ matrices so that 
$Y^{A}$ and $\psi_{A}$ transform as
$(\bm{N},\overline{\bm{N}})$ bi-fundamental representations 
of $U(N)_{k}\times \hat{U}(N)_{-k}$ gauge group, 
while $Y_{A}^{\dag}$ and $\psi^{\dag A}$ do as
$(\overline{\bm{N}},\bm{N})$. 
$A_{\mu}$ is a Chern-Simons $U(N)$ gauge field 
of level $+k$ and $\hat{A}_{\mu}$ is that of level $-k$. 
Also in the theory 
there is a $U(1)_{B}$ flavor symmetry 
and the corresponding baryonic charges 
are assigned $+1$ for bi-fundamental fields, 
$-1$ for anti-bi-fundamental fields 
and $0$ for gauge fields. 
The symmetries in the ABJM-model 
are summarized in Table \ref{abjmcontent}.
\begin{table}
\begin{center}
\begin{tabular}{|c|c|c|c|c|} \hline
 &$U(N)$ &$\hat{U}(N)$ &$SU(4)_{R}$ &$U(1)_{B}$ \\ \hline
$Y^{A}$ &$\bm{N}$ &$\overline{\bm{N}}$ &$\bm{4}$ &$+1$ \\ \hline
$Y_{A}^{\dag}$ &$\overline{\bm{N}}$ &$\bm{N}$ &$\overline{\bm{4}}$ &$-1$
		 \\ \hline
$\psi_{A}$ &$\bm{N}$ &$\overline{\bm{N}}$ &$\overline{\bm{4}}$ &$+1$ \\ \hline
$\psi^{\dag A}$ &$\bm{\overline{N}}$ &$\bm{N}$ &$\bm{4}$
	     &$-1$ \\ \hline
$A_{\mu}$ &$\bm{N}^{2}$ &$\bm{1}$ &$\bm{1}$ &$0$ \\ \hline
$\hat{A}_{\mu}$ &$\bm{1}$ &$\bm{N}^{2}$ &$\bm{1}$ &$0$ \\ \hline
\end{tabular}
\caption{The symmetries and their representations for fields in the
 ABJM-model. 
The bold letters for 
$U(N)$, $\hat{U}(N)$ and $SU(4)_{R}$ symmetries denote 
the representations for the symmetry groups 
and the quantities for $U(1)_{B}$ symmetry are the corresponding charges.}
\label{abjmcontent}
\end{center}
\end{table}

\subsection{Lagrangian}
The Lagrangian of the ABJM-model is given by 
\cite{
Benna:2008zy}
\begin{align}
\label{abjmlag1}
 \mathcal{L}_{\textrm{ABJM}}
=&-\textrm{Tr}
(D_{\mu}Y_{A}^{\dag}D^{\mu}Y^{A})
-i\textrm{Tr}(\psi^{\dag A}
\gamma^{\mu}D_{\mu}\psi_{A})
-V_{\textrm{ferm}}-V_{\textrm{bos}}\nonumber\\
&+\frac{k}{4\pi}\epsilon^{\mu\nu\lambda}\textrm{Tr}\left[
A_{\mu}\partial_{\nu}A_{\lambda}+\frac{2i}{3}
A_{\mu}A_{\nu}A_{\lambda}
-\hat{A}_{\mu}\partial_{\nu}\hat{A}_{\lambda}
-\frac{2i}{3}\hat{A}_{\mu}\hat{A}_{\nu}\hat{A}_{\lambda}\right]
\end{align}
where
\begin{align}
\label{abjmv01}
 V_{\textrm{ferm}}
=&-\frac{2\pi i}{k}
\textrm{Tr}
\Bigl(Y_{A}^{\dag}Y^{A}\psi^{\dag B}\psi_{B}
-\psi^{\dag B}Y^{A}Y_{A}^{\dag}\psi_{B}\nonumber\\
&-2Y_{A}^{\dag}Y^{B}\psi^{\dag A}\psi_{B}
+2Y^{A}Y_{B}^{\dag}\psi_{A}\psi^{\dag B}\nonumber\\
&-\epsilon^{ABCD}Y_{A}^{\dag}
\psi_{B}Y_{C}^{\dag}\psi_{D}
+\epsilon_{ABCD}Y^{A}\psi^{\dag B}Y^{C}\psi^{\dag D}\Bigr),\\
\label{abjmv02}
 V_{\textrm{bos}}
=&-\frac{4\pi^{2}}{3k^{2}}
\textrm{Tr}
\Bigl( Y^{A}Y_{A}^{\dag}Y^{B}Y_{B}^{\dag}Y^{C}Y_{C}^{\dag}
+Y_{A}^{\dag}Y^{A}Y_{B}^{\dag}Y^{B}Y_{C}^{\dag}Y^{C}
\nonumber\\
&+4Y^{A}Y_{B}^{\dag}Y^{C}Y_{A}^{\dag}Y^{B}Y_{C}^{\dag}
-6Y^{A}Y_{B}^{\dag}Y^{B}Y_{A}^{\dag}Y^{C}Y_{C}^{\dag}\Bigr).
\end{align}
Here we use the Dirac matrix  
${(\gamma^{\mu})_{\alpha}}^{\beta}=(i\sigma_{2},\sigma_{1},\sigma_{3})$. 
The spinor indices are raised, 
$\theta^{\alpha}=\epsilon^{\alpha\beta}\theta_{\beta}$, 
and lowered, $\theta_{\alpha}=\epsilon_{\alpha\beta}\theta^{\beta}$ 
with $\epsilon^{12}=-\epsilon_{12}=1$. 
Note that this makes the Dirac matrix 
$\gamma^{\mu}_{\alpha\beta}:=
{(\gamma^{\mu})_{\alpha}}^{\gamma}\epsilon_{\beta\gamma}
=(-\mathbb{I}_{2},-\sigma_{3},\sigma_{1})$ 
symmetric and guarantees the Hermiticity 
of the fermionic kinetic term. 
The covariant derivatives are defined by
\begin{align}
\label{abjmcov1}
 D_{\mu}Y^{A}&=\partial_{\mu}Y^{A}+iA_{\mu}Y^{A}-iY^{A}\hat{A}_{\mu},&
D_{\mu}\psi_{A}&=\partial_{\mu}\psi_{A}+iA_{\mu}\psi_{A}-i\psi_{A}\hat{A}_{\mu}
,\nonumber\\
 D_{\mu}Y^{\dag}_{A}&=\partial_{\mu}Y^{\dag}_{A}-iA_{\mu}Y^{\dag}_{A}
+iY_{A}^{\dag}\hat{A}_{\mu},&
D_{\mu}\psi^{\dag A}&=\partial_{\mu}\psi^{\dag A}-iA_{\mu}\psi^{\dag A}
+i\psi^{\dag A}\hat{A}_{\mu}.
\end{align}

\subsection{Supersymmetry transformation}
The supersymmetry transformation laws are
\begin{align}
\label{abjms01}
 \delta Y^{A}
&=i\omega^{AB}\psi_{B},\\
\label{abjms02}
\delta Y^{\dag}_{A}
&=i\psi^{\dag B}\omega_{AB},\\
\label{abjms03}
 \delta \psi_{A}
&=-\gamma^{\mu}\omega_{AB}D_{\mu}Y^{B}
+\frac{2\pi}{k}\left[
-\omega_{AB}(Y^{C}Y_{C}^{\dag}Y^{B}-Y^{B}Y_{C}^{\dag}Y^{C})
+2\omega_{CD}Y^{C}Y_{A}^{\dag}Y^{D}
\right],\\
\label{abjms04}
 \delta\psi^{\dag A}
&=D_{\mu}Y_{B}^{\dag}\omega^{AB}\gamma^{\mu}
+\frac{2\pi}{k}\left[
-(Y_{B}Y^{C}Y_{C}^{\dag}-Y_{C}^{\dag}Y^{C}Y_{B}^{\dag})\omega^{AB}
+2Y_{D}^{\dag}Y^{A}Y_{C}^{\dag}\omega^{CD}
\right],\\
\label{abjms05}
\delta A_{\mu}
&=\frac{\pi}{k}
\left(
-Y^{A}\psi^{\dag B}\gamma_{\mu}\omega_{AB}
+\omega^{AB}\gamma_{\mu}\psi_{A}Y_{B}^{\dag}
\right),\\
\label{abjms06}
\delta\hat{A}_{\mu}
&=\frac{\pi}{k}
\left(
-\psi^{\dag A}Y^{B}\gamma_{\mu}\omega_{AB}
+\omega^{AB}\gamma_{\mu}Y_{A}^{\dag}\psi_{B}
\right).
\end{align}
The parameter $\omega_{AB}$ is defined by
\begin{align}
\label{abjmsusypara1}
\omega_{AB}:=\epsilon_{i}(\Gamma^{i})_{AB},\ \ \ \ 
\omega^{AB}:=\epsilon_{i}(\Gamma^{i*})^{AB}
\end{align}
where the $SL(2,\mathbb{R})$ spinor $\epsilon^{i}$, $i=1,\cdots,6$ 
transforms as the representation $\bm{6}$ under the $SU(4)_{R}$ and 
$\Gamma^{i}$ is the six-dimensional $4\times 4$ matrix satisfying
\begin{align}
\label{abjmsp1}
(\Gamma^{i})_{AB}&=-(\Gamma^{i})_{BA},\\
\frac12 \epsilon^{ABCD}(\Gamma^{i})_{CD}&=-(\Gamma^{i\dag})^{AB}
=(\Gamma^{i*})^{AB},\\
\left\{\Gamma^{i},\Gamma^{j}\right\}&=2\delta_{ij}.
\end{align}
Note that the supersymmetry parameter $\omega_{AB}$ obeys
\begin{align}
\omega^{AB}&=\omega_{AB}^{*}=\frac12 \epsilon^{ABCD}\omega_{CD}.
\end{align}

\section{Moduli space}
\label{abjmsec2}
In order to determine the vacuum moduli space of the 
$U(N)_{k}\times \hat{U}(N)_{-k}$ ABJM-model, 
we need to consider the minimum of the scalar potential. 
Since the potential turns out to be a perfect square, 
the potential is minimal when the potential vanishes. 
The vanishing condition 
of the bosonic potential is given by 
\begin{align}
\label{abjmv000}
Y^{C}Y_{C}^{\dag}Y^{B}&=0,\\
\label{abjmv001}
Y^{C}Y_{A}^{\dag}Y^{D}&=0.
\end{align}
The generic solution is given by diagonal configurations
\begin{align}
\label{abjmv1}
Y^{A}=\mathrm{diag}(y_{1}^{A}, \cdots, y_{N}^{A})
\end{align}
up to gauge equivalences. 
The configurations (\ref{abjmv1}) are the full moduli space 
because for generic diagonal elements 
one obtains positive definite mass matrix for the off-diagonal 
elements and all off-diagonal elements turn out to be massive. 
The solutions (\ref{abjmv1}) break the gauge group 
$U(N)\times \hat{U}(N)$ to $U(1)^{N}\times U(1)^{N}\times S_{N}$ 
where $S_{N}$ is the Weyl group of $U(N)$ 
that permutes the diagonal elements of all matrices. 
At a generic point of the moduli space, 
only a $U(1)^{N}$ subgroup that does not act 
on the eigenvalues remains unbroken 
and its gauge transformations keep 
There are gauge transformations that 
$Y^{A}$ diagonal. 
Quotienting by such gauge symmetries, 
one finds 
The moduli space of the 
$U(N)_{k}\times \hat{U}(N)_{-k}$ ABJM-model is \cite{Aharony:2008ug} 
\begin{align}
\label{abjmmoduli0}
\mathcal{M}_{N,k}
=\frac{(\mathbb{C}^{4}/\mathbb{Z}_{k})^{N}}{S_{N}}
=\mathrm{Sym}^{N}(\mathbb{C}^{4}/\mathbb{Z}_{k}).
\end{align}
This can be identified with the moduli space 
of $N$ indistinguishable M2-branes moving 
in $\mathbb{C}^{4}/\mathbb{Z}_{k}$ transverse space. 
Therefore the ABJM-model is expected 
to describe the low-energy world-volume theory 
of $N$ coincident M2-branes probing an orbifold 
$\mathbb{C}^{4}/\mathbb{Z}_{k}$. 
The four complex scalar fields $Y^{A}$ 
represent the positions of the membranes in $\mathbb{C}^{4}$.

The orbifold $\mathbb{Z}_{k}$ acts on 
the four complex coordinates $y^{A}$ as
\begin{align}
\label{abjmzk1a}
y^{A}\rightarrow e^{\frac{2\pi i}{k}}y^{A}.
\end{align}
This preserves $SU(4)$ rotational symmetry, which is realized as the
R-symmetry in the ABJM theory. 
The action of the $\mathbb{Z}_{k}$ on the fermionic fields is
\begin{align}
\label{abjmzk1b}
\psi\rightarrow e^{\frac{2\pi (s_{1}+s_{2}+s_{3}+s_{4})}{k}}\psi
\end{align}
where $s_{i}=\pm\frac12$ are the spinor weights. 
The chirality projection implies 
that the sum of all $s_{i}$ must be even, which produces an
eight-dimensional representation. 
The spinors that are left invariant by the orbifold have
$\sum_{i=1}^{4}s_{i}=0, \mod k$. 
This selects six out of eight spinors, 
so the M2-brane theory has 12 supercharges. 
This agrees with ABJM theory. 
Therefore this is consistent 
to the conjecture that the ABJM theory is dual to M-theory on
$AdS_{4}\times S^{7}/\mathbb{Z}_{k}$ with $N$ units of flux \cite{Aharony:2008ug}.

\section{Duality between BLG and ABJ(M)}
\label{abjmsec3}
In \cite{Lambert:2010ji} 
it has been discussed that 
if $N$ and $k$ are co-prime, 
then the vacuum moduli space of the $U(N)_{k}\times \hat{U}(N)_{-k}$ theory
is equivalent to that of the $SU(N)\times SU(N)/\mathbb{Z}_{N}$ theory. 
Consequently there are conjectural dualities 
between the ABJ(M) theory and the BLG theory
\begin{align}
\label{abjmblg}
U(2)_{1}\times \hat{U}(2)_{-1}\ 
\textrm{ABJM theory}
&\Leftrightarrow 
SO(4)\ 
\textrm{BLG theory with}\ 
k=1,\\
U(2)_{2}\times \hat{U}(2)_{-2}\ 
\textrm{ABJM theory}
&\Leftrightarrow
Spin(4)\ 
\textrm{BLG theory with}\ 
k=2,\\
U(3)_{2}\times \hat{U}(2)_{-2}\ 
\textrm{ABJ theory}
&\Leftrightarrow
SO(4)\ 
\textrm{BLG theory with}\ 
k=4. 
\end{align} 
These proposed dualities have been tested 
by the computations of the superconformal indices
\cite{Bashkirov:2011pt}. 
Hence we may regard the $SO(4)$ BLG-model with $k=1$ 
as the world-volume theory of two planar M2-branes 
propagating in a flat space.

\part{SCQM from M2-branes}
\chapter{$\mathcal{N}=16$ Superconformal Mechanics}
\label{secflat1}
Let us turn to the most important part of this thesis 
in which we will see how 
the two subjects discussed so far 
are connected with each other. 
We will initiate our study in this chapter 
by considering the BLG-model wrapped on a torus 
and derive the IR quantum mechanics by shrinking the torus.  
We will see that 
the IR quantum mechanics is the $\mathcal{N}=16$ superconformal 
gauged quantum mechanics 
and also find the $OSp(16|2)$ superconformal quantum mechanics 
from the reduced systems. 

\section{$\mathcal{N}=16$ gauged quantum mechanics}
We shall start our analysis of the wrapped M2-branes 
with the case where 
the two M2-branes wrap a torus $T^{2}$ 
and propagate in a flat transverse space. 
For a torus there is no non-trivial spin connection 
and the world-volume theory of M2-branes is given by 
the BLG action (\ref{blglagrangian}) defined 
on $M_{3}=\mathbb{R}\times T^{2}$. 

\begin{figure}
\begin{center}
\includegraphics[width=5.5cm]{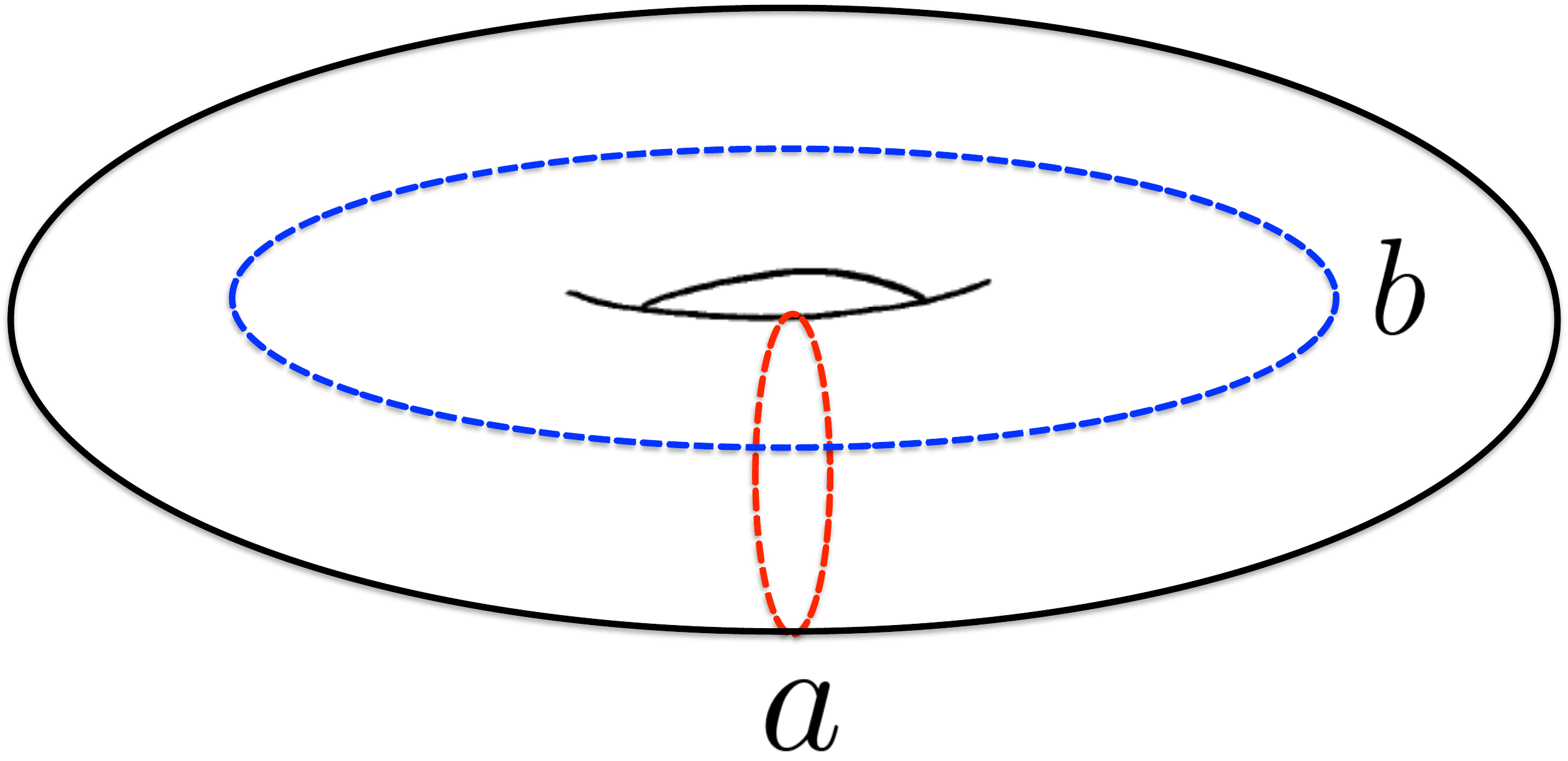}
\caption{A torus with two canonical homology basis $a$ and $b$.}
\label{torfig1}
\end{center}
\end{figure}

A torus is a compact Riemann surface of genus one 
and it is characterized by two periods 
which are defined as the integration 
of a holomorphic differential $\omega$ 
along two canonical homology basis $a$, $b$ of a torus 
(see Figure \ref{torfig1}). 
We will define the periods by
\begin{align}
\label{period0}
\int_{a}\omega=1,\ \ \ \ \ 
\int_{b}\omega=\tau
\end{align}
where $\tau$ is the moduli of the torus and 
it does not take real but complex value. 

Let us consider the limit in which the volume of $T^{2}$ has 
vanishingly small and derive the low-energy effective 
one-dimensional field theory, 
i.e. quantum mechanics on $\mathbb{R}$. 
In order to get such a theory 
we firstly specify the configurations with the lowest energy. 
Since the theories that we are considering are supersymmetric, 
the low-energy configurations can be fixed by the BPS equations. 
Furthermore, since we are now interested in bosonic BPS configurations, 
we require the background values of the fermionic fields to vanish. 
The bosonic fields then turn out to be invariant under 
their supersymmetry transformations. 
Thus the BPS equations are the vanishing conditions of the 
supersymmetry transformations (\ref{blgsusy2}) for fermionic variables. 
We discard the terms which contain the covariant derivatives 
with respect to time since we are interested in 
the low energy dynamics as a fluctuation around 
gauge invariant static configurations. 
We then find the BPS equations
\begin{align}
\label{bpst2a}
&D_{z}X_{a}^{I}=0,\ \ \ 
 D_{\overline{z}}X_{a}^{I}=0,\\
\label{bpst2b}
&[X^{I},X^{J},X^{K}]=0.
\end{align}

Let us now consider 
the $SO(4)$ BLG-model describing two M2-branes. 
For this case the bosonic 
scalar fields $X_{a}^{I}$ and fermionic fields $\Psi_{a}$ 
transform as the fundamental representations 
of the $SO(4)$ gauge symmetry group. 
Let us assume that these Higgs fields 
take non-zero values. 
Then one can express the generic solution to (\ref{bpst2b}) as 
$X_{a}^{I}=\left(X^{I}_{1},X^{I}_{2},0,0\right)^{T}$. 
Inserting these solutions into 
the remaining BPS equations (\ref{bpst2a}), 
we find the set of equations
\begin{align}
\label{bps12a1}
\partial_{z}X_{1}^{I}+\tilde{A}_{z2}^{1}X_{2}^{I}&=0,& 
\partial_{z}X_{2}^{I}-\tilde{A}_{z2}^{1}X_{1}^{I}&=0,\\
\label{bps12b1}
\tilde{A}_{z3}^{1}X_{1}^{I}+\tilde{A}_{z3}^{2}X_{2}^{I}&=0,&
\tilde{A}_{z4}^{1}X_{1}^{I}+\tilde{A}_{z4}^{2}X_{2}^{I}&=0,
\end{align}
and their complex conjugates. 
We should firstly note from the equations (\ref{bps12a1}) that 
the sum of the squares 
$(X_{1}^{I})^{2}+(X_{2}^{I})^{2}$ for 
$I=1,\cdots,8$ has no dependence 
on the locus of the Riemann surface.  
This allows us to write 
\begin{align}
\label{bosconf1a}
X_{1}^{I+2}+iX_{2}^{I+2}
=r^{I}
e^{i(\theta^{I}+\varphi(z,\overline{z}))}.
\end{align}
Here $r^{I}, \theta^{I} \in \mathbb{R}$ can be 
treated as constant on the torus. 
They describe the configuration of the two membranes in the $I$-th
direction. 
On the other hand, 
$\varphi(z,\overline{z})$ may depend on $z$ and $\overline{z}$. 
From the equations (\ref{bps12a1}) 
we can write $\tilde{A}_{z2}^{1}=\partial_{z}\varphi$. 
Furthermore the second set of equations (\ref{bps12b1}) 
requires us to turn off four of six gauge fields; 
$\tilde{A}_{z3}^{1}=\tilde{A}_{z3}^{2}=
\tilde{A}_{z4}^{1}=\tilde{A}_{z4}^{2}=0$. 
These components of the gauge field become massive by the Higgs mechanism. 
Notice that the above set of solutions automatically obeys 
the integrability condition for (\ref{bpst2a}) 
since the Abelian gauge field $\tilde{A}_{z2}^{1}$ is flat.

One can find further restrictions by noting that 
the flat Abelian gauge fields 
$\tilde{A}_{z2}^{1}$ on a torus can 
only take particular forms. 
Let us cut a torus along the canonical basis $a$ and $b$. 
Then the sections of a flat bundle are 
characterized by their transition functions, 
i.e. constant phases around $a$ and $b$. 
Thus the sections of the flat bundle 
can be completely classified by their twists 
$e^{2\pi i\xi}$, $e^{-2\pi i\zeta}$ 
on the homology along cycles $a$, $b$ 
with $\xi$ and $\zeta$ being real parameters. 
This space is the torus $\mathbb{C}/L_{\tau}$ 
where $L_{\tau}$ is the lattice generated by 
$\mathbb{Z}+\tau\mathbb{Z}$. 
It is called the Jacobi variety of $T^{2}$ denoted
by $\mathrm{Jac}(T^{2})$. 
The twists on the homology can be described as a point 
on the Jacobi variety. 
Thus one can express the flat gauge field as
\cite{AlvarezGaume:1986es}
\begin{align}
\label{az12}
\tilde{A}_{z2}^{1}
=-2\pi\frac{\Theta}{\tau-\overline{\tau}}\omega,\ \ \ \ \ 
\tilde{A}_{\overline{z}2}^{1}
=2\pi\frac{\overline{\Theta}}{\tau-\overline{\tau}}
\overline{\omega}
\end{align}
where $\Theta:=\zeta+\overline{\tau}\xi$ is   
the complex parameter describing the twists on 
the homology along two cycles.  
Consequently one can write 
\begin{align}
\label{phi}
\varphi(z,\overline{z})
=2\pi\frac{\overline{\Theta}}{\tau-\overline{\tau}}\overline{z}
-2\pi\frac{\Theta}{\tau-\overline{\tau}}z. 
\end{align}
Due to the fact that the angular variable $\varphi(z,\overline{z})$ in the 
$X^{I}_{1} X_{2}^{I}$-plane 
characterizes the ratio of two bosonic degrees of freedom 
for the two M2-branes, 
it must take same values modulo $2\pi \mathbb{Z}$ 
around two cycles, namely 
under the shifts of the coordinates $z$; 
$z\rightarrow z+1$ and $z\rightarrow z+\tau$ . 
Hence both the coordinates $\xi$ and $\zeta$ 
must be integer values. 
Along with the expression (\ref{az12}),  
the discretization of these coordinates leads to the quantization of the 
gauge fields  
$\tilde{A}_{z2}^{1}$ and $\tilde{A}_{\overline{z}2}^{1}$. 
Thus the BPS configurations are given by
\begin{align}
\label{0bos2}
X^{I+2}
&=
\left(
\begin{array}{c}
X_{A}^{I}\\
X_{B}^{I}\\
0\\
0\\
\end{array}
\right)
=\left(
\begin{array}{c}
\cos(\theta^{I}+\varphi(z,\overline{z}))\\
\sin(\theta^{I}+\varphi(z,\overline{z}))\\
0\\
0\\
\end{array}
\right)r^{I},
\nonumber\\
\tilde{A}_{z}
&=\left(
\begin{array}{cccc}
0&-2\pi\frac{\Theta}{\tau-\overline{\tau}}\omega_{z}&0&0\\
2\pi\frac{\Theta}{\tau-\overline{\tau}}\omega_{z}&0&0&0\\
0&0&0&\tilde{A}_{z4}^{3}(z,\overline{z})\\
0&0&-\tilde{A}_{z4}^{3}(z,\overline{z})&0\\
\end{array}
\right).
\end{align}
It should be noted that the Abelian gauge fields 
$\tilde{A}_{z4}^{3}$ and $\tilde{A}_{\overline{z}4}^{3}$ 
associated with the preserved $U(1)$ symmetry 
do not receive any constraints from the BPS conditions. 
Due to the bosonic configurations (\ref{0bos2}) 
and the supersymmetry transformations (\ref{blgsusy1}), 
the fermionic partners can be written as
\begin{align}
\label{0fer1}
\Psi_{\pm}
=\left(
\begin{array}{c}
\Psi_{\pm A}\\
\Psi_{\pm B}\\
0\\
0\\
\end{array}
\right),\ \ \ 
\overline{\Psi}^{\pm}=\left(
\begin{array}{c}
\overline{\Psi}^{\pm}_{A}\\
\overline{\Psi}^{\pm}_{B}\\
0\\
0\\
\end{array}
\right)
\end{align}
where 
$\overline{\Psi}$ is the conjugate spinor that is defined by 
$\overline{\Psi}:=\Psi^{T}\tilde{C}$ by using  
the $SO(8)$ charge conjugation matrix $\tilde{C}$. 
$\Psi_{+}^{a}$ and $\overline{\Psi}^{+a}$ 
are the $SO(2)_{E}$ spinors with the positive chiralities 
while $\Psi_{-}^{a}$ and $\overline{\Psi}^{-a}$ 
possess the negative chiralities.  
Both fermionic fields transform 
as the representation $\bm{8}_{c}$ of the $SO(8)_{R}$. 
The subscripts $A,B$ are the renewed labels of 
the original gauge indices $1$ and $2$ 
\footnote{In order to avoid the confusion 
coming from the various possible explicit numerical subscripts, 
we here relabel the gauge indices $a=1,2$ as $a=A,B$.}. 

Keeping the above static BPS configurations (\ref{0bos2}) and (\ref{0fer1}), 
we now want to consider the evolution of time for the remaining degrees
of freedom. 
Let us compactify the system on $T^{2}$. 
Plugging the configurations (\ref{0bos2}) and (\ref{0fer1}) 
into the action (\ref{blglagrangian}), we find
\begin{align}
\label{effs0}
S=
\int_{\mathbb{R}}dt
\int_{T^{2}}d^{2}z 
&
\Biggl[
\frac12 D_{0}X^{Ia}D_{0}X_{a}^{I}
-\frac{i}{2}
\overline{\Psi}^{\alpha a}D_{0}\Psi_{\alpha a}\nonumber\\
&-\frac{k}{2\pi}\tilde{A}_{02}^{1}\tilde{F}_{z\overline{z}4}^{3}
-\frac{k}{4\pi}
\left(
\tilde{A}_{z2}^{1}\dot{\tilde{A}}_{\overline{z}4}^{3}
-\tilde{A}_{\overline{z}2}^{1}\dot{\tilde{A}}_{z4}^{3}
\right)
\Biggr]
\end{align}
where the Greek letters $\alpha=+,-$ 
denote the $SO(2)_{E}$ spinor indices. 
The terms in the first line of the action (\ref{effs0}) 
deduce from the kinetic terms of the BLG action 
while the terms in the second line arise from 
the twisted topological Chern-Simons terms. 

As we have discussed, since the gauge fields $\tilde{A}_{z2}^{1}$ and 
$\tilde{A}_{\overline{z}2}^{1}$ are quantized and 
their time derivatives do not appear in the action (\ref{effs0}), 
these fields can be considered as auxiliary fields. 
By making use of the equations of motion one can exclude them 
and find the constraints 
$\dot{\tilde{A}}_{z4}^{3}=\dot{\tilde{A}}_{\overline{z}4}^{3}=0$. 
Hence we see that the corresponding field strength 
$\tilde{F}_{z\overline{z}4}^{3}$ does not depend on time. 
To proceed further we carry out the dimensional reduction on $T^{2}$ 
by rescaling the fields as
\begin{align}
\label{resc1}
X^{I'}=R^{2}X^{I}, \ \ \ \ \ 
\Psi_{\alpha a}'=R^{2}\Psi_{\alpha a}, \ \ \ \ \ 
{\overline{\Psi}^{\alpha a}}'=R^{2}\overline{\Psi}^{\alpha a}
\end{align}
where $R$ is the circumference of the torus. 
At this stage the fields get the canonical dimensions in the reduced
theory.  
The bosonic field $X^{I'}$ has 
mass dimension $-1/2$ and 
the fermionic filed $\Psi'$ acquires mass dimension zero.

Let us carry out the integration on the torus with respect to 
the coordinates $z$, $\overline{z}$ 
by applying the Kaluza-Klein ansatz for the 
Abelian gauge field $\tilde{A}_{0}^{12}$ 
and omit the unimportant primes on the fields. 
Then we find the effective action 
\begin{align}
\label{effs1}
S=
\int_{\mathbb{R}}dt
\Biggl[
\frac12 D_{0}X^{Ia}D_{0}X^{I}_{a}
-\frac{i}{2}
\overline{\Psi}^{\alpha a}D_{0}\Psi_{\alpha a}
-kC_{1}(E)\tilde{A}_{02}^{1}
\Biggr]
\end{align}
where
\begin{align}
\label{ch01}
C_{1}(E)=\int_{T^{2}}c_{1}(E)
:=\frac{1}{2\pi}\int_{T^{2}}
d^{2}z \tilde{F}_{z\overline{z}4}^{3}
\end{align}
is the Chern number that results from the integration 
of the first Chern class $c_{1}(E)$ of the 
$U(1)$ principal bundle $E\rightarrow T^{2}$ over the torus.  
It is associated with the preserved $U(1)$ gauge field
$\tilde{A}_{z4}^{3}$. 
Thus the last term in the action (\ref{effs1}) 
can be recognized as a Fayet-Iliopoulos (FI) term 
as in (\ref{confg2}).

The action (\ref{effs1}) is invariant 
under the following one-dimensional conformal transformations;
\begin{align}
\label{tconf1}
\delta t&=f(t)=a+bt+ct^{2}, 
&\delta\partial_{0}&=-\dot{f}\partial_{0},\\
\delta X_{a}^{I}&=\frac12 \dot{f} X_{a}^{I}, 
&\delta \tilde{A}_{02}^{1}&=-\dot{f}\tilde{A}_{02}^{1},\\
\delta \Psi_{\alpha a}&=0,&
\delta \overline{\Psi}^{\alpha a}&=0
\end{align}
where $f(t)$ is a quadratic function of time 
that containts real infinitesimal parameters $a$, $b$ and $c$. 
The action (\ref{effs1}) is also invariant 
under the $\mathcal{N}=16$ supersymmetry transformations
\begin{align}
\label{tsusy01}
\delta X^{I}_{a}
&=i\overline{\epsilon}^{+}\tilde{\Gamma}^{I}\Psi_{-a}
-i\overline{\epsilon}_{-}\tilde{\Gamma}^{I}\Psi_{+a}, 
&
\delta \tilde{A}_{02}^{1}&=0,
\\
\label{tsusy02}
\delta \Psi_{+a}
&=-D_{0}X_{a}^{I}\tilde{\Gamma}^{I}\epsilon_{-}, 
&\delta \Psi_{-a}
&=D_{0}X_{a}^{I}\tilde{\Gamma}^{I}\epsilon_{+}. 
\end{align}
Therefore the low-energy effective theory (\ref{effs1}) 
is the $\mathcal{N}=16$ superconformal gauged quantum mechanics with 
the FI term.

%
%
%
%
%
%
%
%

\section{Reduction}
\label{secflat2}
As we have already argued, 
gauged conformal mechanics and the Calogero model reduce to 
conformal mechanical models with inverse-square type 
potentials after integrating out the auxiliary gauge fields. 
In fact our gauged mechanical action (\ref{effs1}) 
is quadratic in the $U(1)$ gauge field $\tilde{A}_{02}^{1}$ 
and contain no time derivative of the Abelian gauge field. 
Hence $\tilde{A}_{02}^{1}$ 
can be regarded as an auxiliary field 
and it does not contribute to the Hamiltonian. 
In other words, the Hamiltonian is invariant under the action 
of the $U(1)$ gauge group on the phase space $\mathcal{M}$.  
This means that 
the corresponding moment map $\mu:\mathcal{M}\rightarrow \mathfrak{u}(1)^{*}$ 
is the constant of motion \cite{MR0690288} and therefore 
one can reduce the original phase space $\mathcal{M}$ 
to a smaller one $\mathcal{M}_{c}=\mu^{-1}(c)$ 
with decreased degrees of freedom by choosing the 
specific inverse of 
the moment map at a point $c\in \mathfrak{u}(1)^{*}$ 
\footnote{
The components of the moment map form a system 
being in involution since the gauge group is Abelian. 
So we do not need to divide by the 
non-trivial coadjoint isotropy subgroup to obtain the reduced phase space.}.

In order to get our reduced system, 
let us eliminate the auxiliary field $\tilde{A}_{02}^{1}$ in two steps. 
Firstly we fix a specific gauge 
and then extract and impose the Gauss law constraint 
to ensure the consistency of the gauge choice. 
We will choose the temporal gauge $\tilde{A}_{0}=0$. 
We see that the solutions 
to the equations of motion for $\tilde{A}_{0}$ are 
\begin{align}
\label{0a0}
\tilde{A}_{02}^{1}
&=\frac{kC_{1}(E)+\sum_{I}(r^{I})^{2}\dot{\theta}^{I}
+i\overline{\Psi}^{\alpha}_{A}\Psi_{\alpha B}
}
{\sum_{I}(r^{I})^{2}},\\
\tilde{A}_{03}^{1}
&=\tilde{A}_{04}^{1}
=\tilde{A}_{03}^{2}
=\tilde{A}_{04}^{2}=0.
\end{align}
We therefore can read the Gauss law constraint
\begin{align}
\label{1a0}
\phi_{0}:=kC_{1}(E)+
\sum_{I}(r^{I})^{2}\dot{\theta}^{I}
+i\overline{\Psi}^{\alpha}_{A}\Psi_{\alpha B}=0.
\end{align}
This constraint equation is nothing but a moment map condition. 
To see the physical meaning of this constraint, 
we note that $(r^{I})^{2}\dot{\theta}^{I}$ 
represents the ``angular momentum'', the $SO(2)$-charge corresponding to
the rotation in the $X^{I}_{1}X^{I}_{2}$-plane 
while the fermionic bilinear term 
$i\overline{\Psi}_{A}^{\alpha}\Psi_{\alpha B}$ 
provides the charge of the $SO(2)$ symmetry group of 
the two different types of fermionic variables $\Psi_{A}$ and $\Psi_{B}$.  
In other words, the equation (\ref{1a0}) tells us that 
the total $SO(2)$ charge rotating the internal degrees of freedom 
for the two membranes is specified by the Chern-Simons level $k$ 
and the Chern number $C_{1}(E)$.

Under the constraint $\phi_{0}=0$, 
one can write a new Lagrangian by adding $\lambda \phi_{0}$ 
where $\lambda$ is the Lagrange multiplier. 
The result is given by
\begin{align}
\label{effs2}
S=\int_{\mathbb{R}}
dt &\Biggl[
\frac12 \sum_{I}(\dot{r}^{I})^{2}
+\frac12 \sum_{I} (r^{I}\dot{\theta}^{I})^{2}
-\frac{i}{2}\overline{\Psi}^{\alpha a}\dot{\Psi}_{\alpha a}\nonumber\\
&+\lambda\left(
kC_{1}(E)+
\sum_{I}(r^{I})^{2}\dot{\theta}^{I}
+i\overline{\Psi}_{A}^{\alpha}\Psi_{\alpha B}
\right)
\Biggr].
\end{align}
Note that the variables $\theta^{I}$'s are absent in the action
(\ref{effs2}). 
This means that they are cyclic coordinates and 
their canonical momenta $p_{\theta^{I}}=(r^{I})^{2}\dot{\theta}^{I}$ 
are just the constant of motion. 

Now we can eliminate cyclic coordinates 
from the Lagrangian by introducing the Routhian. 
As we have discussed in 
section \ref{gcqmsec1}, the Routhian is a hybrid 
between the Lagrangian and the Hamiltonian, 
defined by performing a Legendre transformation 
on the cyclic coordinates
\begin{align}
\label{routh1}
R(r^{I},\dot{r}^{I},h^{I},\Psi):=
L(r^{I},\dot{r}^{I},\dot{\theta}^{I},\Psi)
-\sum_{I}\dot{\theta}^{I}p_{\theta^{I}}.
\end{align}
According to the partial Legendre transformation, 
the canonical variables $r^{I}$ and $\Psi$ still 
obey the Euler-Lagrange equations 
whereas the cyclic coordinates $\theta^{I}$ and their momenta 
$h^{I}:=p_{\theta^{I}}$ 
satisfy the Hamilton equations. 
However, one can see that 
the latter set of equations are trivial statements. 
They correspond to 
the constant property of $h^{I}$ (i.e. $\dot{h}^{I}=0$) 
and the definition of $h^{I}$ 
(i.e. $\dot{\theta}^{I}=\frac{h^{I}}{(r^{I})^{2}}$). 
In other words, classically the Routhian is not  
$R(r^{I},\dot{r}^{I}, h^{I},\Psi)$ 
but $R(r^{I},\dot{r}^{I},\Psi)$ endowed 
with the constant of motion $h^{I}$'s. 
Thus the action (\ref{effs2}) can be rewritten as
\begin{align}
\label{effs3}
S=\int_{\mathbb{R}}
dt &\Biggl[ 
\frac12 \sum_{I}(\dot{r}^{I})^{2}
-\frac12\sum_{I}\frac{(h^{I})^{2}}{(r^{I})^{2}}
-\frac{i}{2}\overline{\Psi}^{\alpha a}\dot{\Psi}_{\alpha a}
+\lambda\left(
kC_{1}(E)+\sum_{I}h^{I}
+i\overline{\Psi}^{\alpha}_{A}\Psi_{\alpha B}
\right)
\Biggr].
\end{align}
By integrating out $\lambda$, 
one finds the reduced effective action 
\begin{align}
\label{0eff0}
S=\frac12\ \int_{\mathbb{R}}
dt 
\Biggl[
&\dot{q}^{2}
+\sum_{I\neq K}(\dot{r}^{I})^{2}
-i\Psi^{\alpha a}\dot{\Psi}_{\alpha a}\nonumber\\
&-\frac{\left[
kC_{1}(E)+\sum_{I\neq K}h^{I}
+i\Psi^{\alpha}_{A}\Psi_{\alpha B}\right]^{2}}{q^{2}}
-\sum_{I\neq K}\frac{(h^{I})^{2}}{(r^{I})^{2}}
\Biggr]
\end{align}
where we have taken the $SO(8)$ charge conjugation matrix $\tilde{C}$ 
as an identity matrix for simplicity 
\footnote{For the symmetric charge conjugation matrix 
one can reduce it to an identity matrix by an appropriate unitary
transformation. }. 
Here we have defined the quantity $q:=r^{K}$ 
where $K$ represents the direction 
in which $h^{K}$ is fixed  by other conserved
quantities $h^{I}$'s. 
We remark that the terms in the numerator of the inverse-square type potential 
are the constant of motion, 
which commutes with the Hamiltonian.

Now we want to study the classical properties of the theory (\ref{0eff0}). 
From the action (\ref{0eff0}) 
we can read the classical equations of motion
\begin{align}
\label{eomqm1}
\ddot{q}&=\frac{[kC_{1}(E)+\sum_{I\neq K}h^{I}
+i\Psi^{\alpha}_{A}\Psi_{\alpha B}]^{2}}{q^{3}},\\
\label{eomqm1a}
\ddot{r}^{I}&=\frac{(h^{I})^{2}}{(r^{I})^{3}},\\
\label{eomqm2a}
\dot{\Psi}_{\alpha A}
&=-\frac{[kC_{1}(E)+\sum_{I\neq K}h^{I}
+i\Psi^{\alpha}_{A}\Psi_{\alpha B}]}{q^{2}}
\Psi_{\alpha B},\\
\label{eomqm3a}
\dot{\Psi}_{\alpha B}
&=\frac{[kC_{1}(E)+\sum_{I\neq K}h^{I}
+i\Psi_{A}^{\alpha}\Psi_{\alpha B}]}{q^{2}}
\Psi_{\alpha A}. 
\end{align}
Using the equations of motion 
(\ref{eomqm2a}) and (\ref{eomqm3a}), 
we can check that the Gauss law constraint 
(\ref{1a0}) has no time dependence. 
Namely, $\phi_{0}$ is the integral of motion.

The canonical momenta are given by
\begin{align}
\label{mom01}
p&:=\frac{\partial L}{\partial \dot{q}}=\dot{q},&
p_{I}&:=\frac{\partial L}{\partial \dot{r}^{I}}=\dot{r}^{I},\\
\label{mom02}
\pi^{\alpha a}
&:=\frac{\vec{\partial}L}
{\partial \dot{\Psi}_{\alpha a}}
=\frac{i}{2}\Psi^{\alpha a}.
\end{align}
As usual the fermionic momenta $\pi^{\alpha a}$ 
do not depend on the velocities 
but on the fermionic variables themselves. 
Thus we imposes the second-class constraints
\begin{align}
\label{prcons1}
\phi_{1}^{\alpha a}
&:=\pi^{\alpha a}-\frac{i}{2}\Psi^{\alpha a}=0.
\end{align}
Under the constraints, 
one finds the Dirac brackets
\begin{align}
\label{dirac1}
\left[q,p\right]_{DB}&=1,& 
\left[r^{I},p_{J}\right]_{DB}&={\delta^{I}}_{J},\\ 
\label{dirac2}
\left[
\Psi_{\alpha a\dot{A}},\pi^{\beta b\dot{B}}
\right]_{DB}&=
\frac12 \delta_{\alpha\beta}\delta_{ab}\delta_{\dot{A}\dot{B}},&
\left[
\Psi_{\alpha a\dot{A}},\Psi^{\beta b\dot{B}}
\right]_{DB}
&=-i\delta_{\alpha\beta}\delta_{ab}\delta_{\dot{A}\dot{B}}.
\end{align}

The action (\ref{0eff0}) 
is invariant under the 
one-dimensional conformal transformations
\begin{align}
\label{conf0}
\delta t&=f(t)=a+bt+ct^{2},
&\delta\partial_{0}&=-\dot{f}\partial_{0},\\
\label{conf1} 
\delta q&=\frac12 \dot{f}q,
&\delta r^{I}&=\frac12 \dot{f}r^{I},\\
\label{conf2}
\delta \Psi_{\alpha a}&=0.
\end{align}
Here the constant parameters $a$, $b$ and $c$ are infinitesimal
parameters, which correspond to   
translation, dilatation and conformal boost respectively. 
We find the corresponding Noether charges, 
the Hamiltonian $H$, the dilatation operator $D$ and 
the conformal boost operator $K$ as
\begin{align}
\label{hkdconf1}
H&=\frac12 \left[
p^{2}+\frac{
\left(
kC_{1}(E)+\sum_{I\neq K}h^{I}
+i\Psi^{\alpha}_{A}\Psi_{\alpha B}
\right)
^{2}}
{q^{2}}
+\sum_{I\neq K}\left(
p_{I}^{2}+\frac{(h^{I})^{2}}{(r^{I})^{2}}
\right)
\right]
,\\
\label{hkdconf2}
D&=tH
-\frac14 
\left[\left(qp+pq\right)+\sum_{I\neq K}
\left( r^{I}p_{I}+p_{I}r^{I} \right)
\right]
,\\
\label{hkdconf3}
K&=t^{2}H
-\frac{1}{2}t\left[
\left(qp+pq\right)
+\sum_{I\neq K}\left(
r^{I}p_{I}+p_{I}r^{I}
\right)
\right]
+\frac12 
\left[
q^{2}+\sum_{I\neq K}(r^{I})^{2}
\right].
\end{align}


The action (\ref{0eff0}) is invariant 
under the fermionic transformations
\begin{align}
\label{susy001a}
\delta q&=
\frac{i}{\sqrt{2}}
\left(\epsilon^{-}\Psi_{-A}
-\epsilon^{+}\Psi_{+A}
\right)
+\frac{i}{\sqrt{2}}\left(
\epsilon^{-}\Psi_{-B}
-\epsilon^{+}\Psi_{+B}
\right)
,\\
\label{susy001}
\delta r^{I}&=
i\cos\theta^{I}\left(
\epsilon^{+}\tilde{\Gamma}^{I}\Psi_{-A}
-\epsilon^{-}\tilde{\Gamma}^{I}\Psi_{+A}
\right)
+i\sin\theta^{I}\left(
\epsilon^{+}\tilde{\Gamma}^{I}\Psi_{-B}
-\epsilon^{-}\tilde{\Gamma}^{I}\Psi_{+B}
\right)
,\\
\label{susy002}
\delta\Psi_{+A\dot{A}}
&=
-\frac{1}{\sqrt{2}}\left(\dot{q}-\frac{h^{K}}{q}\right)
\epsilon_{+\dot{A}}
-\frac{i}{\sqrt{2}}\frac{l}{q}\Psi_{+B\dot{A}}
-\sum_{I\neq K}\left(
\dot{r}^{I}\cos\theta^{I}-\sin\theta^{I}\frac{h^{I}}{r^{I}}
\right)\tilde{\Gamma}^{I}\epsilon_{-\dot{A}},\\
\label{susy003}
\delta\Psi_{-A\dot{A}}
&=
\frac{1}{\sqrt{2}}\left(\dot{q}-\frac{h^{K}}{q}\right)
\epsilon_{-\dot{A}}
-\frac{i}{\sqrt{2}}\frac{l}{q}\Psi_{-B\dot{A}}
+\sum_{I\neq K}\left(
\dot{r}^{I}\cos\theta^{I}-\sin\theta^{I}\frac{h^{I}}{r^{I}}
\right)\tilde{\Gamma}^{I}\epsilon_{+\dot{A}},\\
\label{susy004}
\delta\Psi_{+B\dot{A}}
&=
-\frac{1}{\sqrt{2}}\left(\dot{q}+\frac{h^{K}}{q}\right)\epsilon_{+\dot{A}}
+\frac{i}{\sqrt{2}}\frac{l}{q}\Psi_{+A\dot{A}}
-\sum_{I\neq K}\left(
\dot{r}^{I}\sin\theta^{I}+\cos\theta^{I}\frac{h^{I}}{r^{I}}
\right)\tilde{\Gamma}^{I}\epsilon_{-\dot{A}},\\
\label{susy005}
\delta\Psi_{-B\dot{A}}
&=
\frac{1}{\sqrt{2}}\left(\dot{q}+\frac{h^{K}}{q}\right)\epsilon_{-\dot{A}}
+\frac{i}{\sqrt{2}}\frac{l}{q}\Psi_{-A\dot{A}}
+\sum_{I\neq K}\left(
\dot{r}^{I}\sin\theta^{I}+\cos\theta^{I}\frac{h^{I}}{r^{I}}
\right)\tilde{\Gamma}^{I}\epsilon_{+\dot{A}}.
\end{align}
Here we have defined the quantities
\begin{align}
\label{theta01}
\theta^{I}(t)
&:=h^{I}\int^{t} \frac{dt'}{(r^{I}(t'))^{2}},\\
\label{l01}
l&:=
\left(
\Psi^{+A}\epsilon_{+}
-\Psi^{-A}\epsilon_{-}
\right)
-
\left(
\Psi^{+B}\epsilon_{+}
-\Psi^{-B}\epsilon_{-}
\right). 
\end{align}
One can see that the supersymmetry is in general non-local 
in that the transformations involve the integrals of the function of 
the non-local variables. 
The origin of the non-locality is comes from the Routh reduction. 
Thus there may exist the infinite number of 
the associated conserved charges 
and things may happen to be much more exotic 
\footnote{The action (\ref{0eff0}) is invariant 
under the fermionic transformations (\ref{susy001a})-(\ref{susy005}), 
however, the Gauss constraint (\ref{prcons1}) may not be invariant under those
transformations. Although in that case the original system may be
modified, 
we here just want to study the reduced system 
without the auxiliary gauge field. 
}. 
But, as seen from the (\ref{hkdconf1}), 
one can focus on 
the motion in the $K$-th direction associated with  
the local supersymmetry since it is essentially decoupled 
from others with non-local ones
as their Hamiltonians commute with each other. 
This leads us to treat them separately. 
and also indicates that the theory holds 
the local conserved supercurrents 
and the non-local supercurrents which are in involution.

\section{$OSp(16|2)$ superconformal mechanics}
\label{secflat3}
We shall focus on the study 
of the motion in the $K$-th direction 
which is associated with the local charges 
and investigate the algebraic 
structure of the symmetry group in the quantum mechanics. 
From now on we will consider the case where 
the all independent conserved charges $h^{I}$'s are zeros. 
This is realized when the internal degrees of freedom 
for two M2-branes are unbiased.  
We note that for the purpose of the exploration 
of the algebraic structure for the $K$-th motion, 
this particular charge assignment does not 
affect the following discussion 
because non-vanishing $h^{I}$'s can only yield   
a constant shift of the coupling constant in the inverse-square type potential. 
From (\ref{0eff0}) we can read the effective action for the 
dynamics in the $K$-th direction 
\begin{align}
\label{effk1}
S=\frac12\ \int_{\mathbb{R}}
dt 
\Biggl[ 
\dot{q}^{2}
-i\Psi^{\alpha a}\dot{\Psi}_{\alpha a}
-\frac{\left(
kC_{1}(E)
+i\Psi^{\alpha}_{A}\Psi_{\alpha B}
\right)^{2}}
{q^{2}}
\Biggr].
\end{align}
One can see that 
the reduced action (\ref{effk1}) contains the 
inverse-square type potential that is similar 
to the known $\mathcal{N}>4$ superconformal mechanical potentials  
discussed in (\ref{n8susy0a1}) (also see 
\cite{Ivanov:1988it,Akulov:1999bw,Wyllard:1999tm,Fedoruk:2011aa}). 
%

We shall study the existing symmetry in the effective action 
(\ref{effk1}). 
The action (\ref{effk1}) may be rewritten as $SU(1,1|16)$ 
superconformal quantum mechanics in the form of (\ref{n8susy0a1}). 
However, 
we should note that the same form of the Lagrangian does not necessarily 
lead to the same symmetry in the theory 
if we have additional constraints or symmetries. 
In fact in our setup 
the bilinear terms for fermions are treated as 
conserved quantities due to the Gauss constraint (\ref{1a0}). 
This implies that 
the gauge indices $a,b,\cdots=A,B$ 
should be distinguished from other indices $\alpha,\beta,\cdots$ 
and $\dot{A},\dot{B},\cdots$ and 
prevents us from forming 32 supercharges. 
Put it another way, 
our theory describes the radial motion of the wrapped membranes 
and thus we have at most 16 supercharges on the branes by the
projection. 
So we will only focus on the remaining $\mathcal{N}=16$
supersymmetry due to the constraint. 
The simplest way to read the consistent supersymmetry 
for the wrapped branes is just look at the 
supersymmetry transformations for the original BLG-model. 
From (\ref{blgsusy1})-(\ref{blgsusy3}) 
we see that the action (\ref{effk1}) is invariant under 
the following $\mathcal{N}=16$ supersymmetry transformation laws
\begin{align}
\label{tsusy001}
\delta q&=
\frac{i}{\sqrt{2}}\left(
\epsilon^{-}\Psi_{-A}
-\epsilon^{+}\Psi_{+A}
\right)
+\frac{i}{\sqrt{2}}\left(
\epsilon^{-}\Psi_{-B}
-\epsilon^{+}\Psi_{+B}
\right)
,\\
\label{tsusy002}
\delta \Psi_{+A\dot{A}}
&=
-\frac{1}{\sqrt{2}}\left(
\dot{q}+\frac{g}{q}
\right)\epsilon_{+\dot{A}}
-\frac{i}{\sqrt{2}}\frac{l}{q}\Psi_{+B\dot{A}}
,\\
\label{tsusy003}
\delta \Psi_{-A\dot{A}}
&=
\frac{1}{\sqrt{2}}\left(\dot{q}+\frac{g}{q}\right)
\epsilon_{-\dot{A}}
-\frac{i}{\sqrt{2}}\frac{l}{q}\Psi_{-B\dot{A}}
,\\
\label{tsusy004}
\delta\Psi_{+B\dot{A}}
&=
-\frac{1}{\sqrt{2}}\left(
\dot{q}-\frac{g}{q}
\right)\epsilon_{+\dot{A}}
+\frac{i}{\sqrt{2}}\frac{l}{q}\Psi_{+A\dot{A}}
,\\
\label{tsusy005}
\delta \Psi_{-B\dot{A}}
&=
\frac{1}{\sqrt{2}}
\left(
\dot{q}-\frac{g}{q}
\right)\epsilon_{-\dot{A}}
+\frac{i}{\sqrt{2}}\frac{l}{g}\Psi_{-A\dot{A}}
\end{align}
where we have defined 
\begin{align}
\label{g01}
g&:=kC_{1}(E)
+i\Psi_{A}^{\alpha}\Psi_{\alpha B}.
\end{align}
Unlike the set of transformations 
(\ref{susy001a})-(\ref{susy005}), 
the supersymmetry transformations (\ref{tsusy001})-(\ref{tsusy005}) 
are local.  
Therefore the conventional Noether's procedure can be applied. 
By using the Noether's method, 
the corresponding supercharges are found to be
\begin{align}
\label{tsch1}
Q_{+\dot{A}}
&=\frac{1}{\sqrt{2}}\left(p+\frac{g}{q}\right)\Psi_{+A\dot{A}}
+\frac{1}{\sqrt{2}}\left(p-\frac{g}{q}\right)\Psi_{+B\dot{A}}
,\\
Q_{-\dot{A}}
&=
-\frac{1}{\sqrt{2}}\left(p+\frac{g}{q}\right)\Psi_{-A\dot{A}}
-\frac{1}{\sqrt{2}}\left(p-\frac{g}{q}\right)\Psi_{-B\dot{A}}.
\end{align}
The action (\ref{effk1}) is invariant 
under the conformal transformations 
$\delta t=f(t)$, $\delta q=\frac12 \dot{f}q$ and  
$\delta\Psi_{\alpha a}=0$. 
By applying the Noether's method, 
we find three generators, the Hamiltonian $H$, 
the dilatation generator $D$ and the conformal boost generator $K$; 
\begin{align}
\label{hkdconf01}
H&=\frac12 p^{2}+\frac{\left[
kC_{1}(E)+i\Psi^{\alpha}_{A}\Psi_{\alpha B}
\right]^{2}}{2q^{2}},\\
\label{hkdconf02}
D&=-\frac14 \{q,p\},\\
\label{hkdconf03}
K&=\frac12 q^{2}
\end{align}
where $\{ , \}$ stands for an anti-commutator.

To quantize the theory, 
we impose the (anti)commutation relations 
for the canonical variables from 
the classical Dirac brackets (\ref{dirac1}) and (\ref{dirac2}) 
\begin{align}
\label{canonical0}
[q,p]&=i,&
\left\{
\Psi_{\alpha a\dot{A}},\Psi_{\beta b\dot{B}}
\right\}
&=\delta_{\alpha\beta}\delta_{ab}\delta_{\dot{A}\dot{B}}.
\end{align}

The conformal symmetry and the supersymmetry combine 
into the superconformal symmetry. 
We will define the superconformal boost generators by
\begin{align}
\label{scqmboost01}
S_{+\dot{A}}&=\frac{1}{\sqrt{2}}q\left(
\Psi_{+A\dot{A}}+\Psi_{+B\dot{A}}
\right),\\
S_{-\dot{A}}&=-\frac{1}{\sqrt{2}}q\left(
\Psi_{-A\dot{A}}+\Psi_{-B\dot{A}}
\right).
\end{align}

According to the extended supersymmetry 
the theory holds the internal R-symmetry rotating  
the fermionic charges. 
Let us introduce the R-symmetry generators as
\begin{align}
\label{j01}
(J_{\alpha\beta})_{\dot{A}\dot{B}}
=i\Psi_{\alpha a \dot{A}}\Psi_{\beta \dot{B}}^{a}.
\end{align}
One can check that the R-symmetry generators satisfy the relations
\begin{align}
\label{bmtx1}
(J_{++})_{\dot{A}\dot{B}}&=-(J_{++})_{\dot{B}\dot{A}},\\
\label{bmtx2}
(J_{--})_{\dot{A}\dot{B}}&=-(J_{--})_{\dot{B}\dot{A}},\\
\label{bmtx3}
(J_{+-})_{\dot{A}\dot{B}}&=-(J_{-+})_{\dot{B}\dot{A}}
\end{align}
and therefore the matrices $J_{++}$, $J_{--}$ and $J_{-+}$ involve 
$28$, $28$ and $64$ independent entries respectively 
while $J_{-+}$ yields no independent ones 
because of the relations (\ref{bmtx3}). 
Hence the R-symmetry matrix totally includes 
$28+28+64=120$ elements.

Making use of the canonical (anti)commutation relations (\ref{canonical0}) 
and taking account into the Weyl ordering for the fermionic bilinear terms, 
we find the (anti)commutation relations among the generators
\begin{align}
\label{osp01a}
\begin{array}{ccc}
[H,D]=iH,
&[K,D]=-iK,
&[H,K]=2iD,
\cr
\end{array}
\end{align}
\begin{align}
\label{osp01b}
\begin{array}{ccc}
[(J_{\alpha\beta})_{\dot{A}\dot{B}},H]=0,
&[(J_{\alpha\beta})_{\dot{A}\dot{B}},D]=0,
&[(J_{\alpha\beta})_{\dot{A}\dot{B}},K]=0,
\end{array}
\end{align}
\begin{align}
\label{osp01c}
[(J_{\alpha\beta})_{\dot{A}\dot{B}},
(J_{\gamma\delta})_{\dot{C}\dot{D}}]
&=i(J_{\gamma\beta})_{\dot{C}\dot{B}}
\delta_{\alpha\delta}\delta_{\dot{A}\dot{D}}
-i(J_{\alpha\delta})_{\dot{A}\dot{D}}
\delta_{\beta\gamma}\delta_{\dot{B}\dot{C}},
\nonumber\\
&+i(J_{\delta\beta})_{\dot{D}\dot{B}}
\delta_{\alpha\gamma}\delta_{\dot{A}\dot{C}}
-i(J_{\alpha\gamma})_{\dot{A}\dot{C}}
\delta_{\beta\delta}\delta_{\dot{B}\dot{D}},
\end{align}
\begin{align}
\label{osp01d}
\begin{array}{ccc}
[H,Q_{\alpha\dot{A}}]=0,
&[D,Q_{\alpha\dot{A}}]=-\frac{i}{2} Q_{\alpha\dot{A}},
&[K,Q_{\alpha\dot{A}}]=iS_{\alpha\dot{A}},
\cr
\end{array}
\end{align}
\begin{align}
\label{osp01e}
\begin{array}{ccc}
[H,S_{\alpha\dot{A}}]=-iQ_{\alpha\dot{A}},
&[D,S_{\alpha\dot{A}}]=\frac{i}{2}S_{\alpha\dot{A}},
&[K,S_{\alpha\dot{A}}]=0,
\cr
\end{array}
\end{align}
\begin{align}
\label{osp01f}
\{Q_{\alpha\dot{A}},Q_{\beta\dot{B}}\}
&=2H\delta_{\alpha\beta}\delta_{\dot{A}\dot{B}},\nonumber\\
\{S_{\alpha\dot{A}},S_{\beta\dot{B}}\}
&=2K\delta_{\alpha\beta}\delta_{\dot{A}\dot{B}},\nonumber\\
\{Q_{\alpha\dot{A}},
S_{\beta\dot{B}}\}
&=-2D\delta_{\alpha\beta}\delta_{\dot{A}\dot{B}}
+{(J_{\alpha\beta})}_{\dot{A}\dot{B}},
%
\end{align}
\begin{align}
\label{osp01g}
[(J_{\alpha\beta})_{\dot{A}\dot{B}},Q_{\gamma\dot{C}}]
&=i\left(
Q_{\alpha\dot{A}}\delta_{\beta\gamma}\delta_{\dot{B}\dot{C}}
-Q_{\beta\dot{B}}\delta_{\alpha\gamma}\delta_{\dot{A}\dot{C}}
\right),
\nonumber\\
[(J_{\alpha\beta})_{\dot{A}\dot{B}},S_{\gamma\dot{C}}]
&=i\left(
S_{\alpha\dot{A}}\delta_{\beta\gamma}\delta_{\dot{B}\dot{C}}
-S_{\beta\dot{B}}\delta_{\alpha\gamma}\delta_{\dot{A}\dot{C}}
\right).
%
%
%
%
%
%
%
%
\end{align}

The Hamiltonian $H$, the dilatation generator $D$ and 
the conformal boost generator $K$ obey  
the $\mathfrak{sl}(2,\mathbb{R})$ one-dimensional conformal algebra
(\ref{osp01a}). 

As the fermionic generators in the superconformal algebra, 
there are sixteen supercharges 
$Q_{\alpha\dot{A}}$ and as many 
superconformal generators $S_{\alpha\dot{A}}$. 
One can see from (\ref{osp01b}) and (\ref{osp01g}) 
that the R-symmetry generators $(J_{\alpha\beta})_{\dot{A}\dot{B}}$ 
commute with the bosonic generators 
$H$, $D$ and $K$ 
while they rotate of the fermionic generators $Q_{\alpha\dot{A}}$ 
and $S_{\alpha\dot{A}}$. 
The commutation relation (\ref{osp01c}) indicates that 
120 components of the R-symmetry generator 
$(J_{\alpha\beta})_{\dot{A}\dot{B}}$ 
form the $\mathfrak{so}(16)$ algebra. 
Hence it can be concluded that 
the theory (\ref{effk1}) is the $OSp(16|2)$ superconformal mechanics. 
We see that this fits in the list 
of the one-dimensional superconformal group 
that we have already shown in Table \ref{listsup1}.

It is true that the quantum mechanics (\ref{effk1}) has  
the $\mathcal{N}=16$ superconformal symmetry, 
but, it is not clear that 
the theory (\ref{effk1}) actually captures 
the dynamics of the wrapped membranes around a torus since 
it is not totally same as the superconformal gauged quantum mechanics 
(\ref{effs1}) due to the reduction process.

Note that 
the original $SO(8)$ R-symmetry is now enhanced to $SO(16)$ 
in our quantum mechanics. 
It is not so strange 
as a similar phenomenon has been already observed in 
$d=11$ supergravity. 
In $d=11$ supergravity 
the original Lorentzian symmetry group $SO(1,10)$ 
can break down into the subgroup $SO(1,2)\times SO(8)$ 
via a partial choice of gauge for the elfbein. 
As pointed out in 
\cite{deWit:1985iy,Nicolai:1986jk,deWit:1986mz} 
the enhanced $SO(1,2)\times SO(16)$ tangent space symmetry 
has been discovered  
by introducing new gauge degrees of freedom. 
It would be an interesting to understand 
such enhanced R-symmetry by investigating from the gravity dual
description.

\chapter{$\mathcal{N}=12$ Superconformal Mechanics}
\label{abjmscqm}
Similar to the previous chapter, 
we will consider the ABJM-model wrapped on a torus 
and derive the IR quantum mechanics by shrinking the torus in this chapter.  
We will derive 
the IR $\mathcal{N}=12$ superconformal gauged quantum mechanics　
and extract the corresponding $SU(1,1|6)$ superconformal quantum mechanics 
from the reduced systems. 

\section{$\mathcal{N}=12$ gauged quantum mechanics}
\label{abjmscqmsec1}
We now want to consider an arbitrary number of 
M2-branes wrapped around a torus, 
which may be described by the $U(N)_{k}\times \hat{U}(N)_{-k}$ ABJM-model on
$\mathbb{R}\times T^{2}$. 
From the discussion in the chapter \ref{abjmch1}, 
this theory is expected to describe the dynamics of $N$ coincident M2-branes 
with the world-volume $M_{3}=\mathbb{R}\times T^{2}$ 
propagating in a transverse space which holds an $SU(4)$ holonomy. 
The crucial point is now that 
the volume of the torus yields a typical energy scale in the theory 
and we can take a further limit where the 
energy is lower than the inverse of the size of the torus. 
Such low-effective theory describes the fluctuations 
around static BPS configurations obeying the BPS equations. 
From the supersymmetry transformations 
(\ref{abjms03}), (\ref{abjms04}) for fermionic fields,  
we obtain the set of the BPS equations: 
\begin{align}
\label{bpsabjm01}
&D_{z}Y^{A}=0,\ \ \ D_{\overline{z}}Y^{A}=0,\\
\label{bpsabjm02}
&Y^{C}Y_{C}^{\dag}Y^{B}-Y^{B}Y_{C}^{\dag}Y^{C}=0,\\
\label{bpsabjm03}
&Y^{C}Y_{A}^{\dag}Y^{D}=0.
\end{align}
In order for 
the algebraic equations 
(\ref{bpsabjm02}) and (\ref{bpsabjm03}) to be satisfied, 
the bosonic matter fields $Y^{A}$ and $Y_{A}^{\dag}$ should be in the
diagonal form
\begin{align}
\label{ydiag1}
Y^{A}&=\mathrm{diag}(y_{1}^{A}, \cdots, y_{N}^{A}),& 
Y^{\dag}_{A}&=\mathrm{diag}(\overline{y}_{A1}, \cdots,
 \overline{y}_{AN}). 
\end{align}
Here $y_{a}^{A}$ is a complex scalar field 
where $a=1,\cdots, N$ are gauge indices characterizing 
the internal degrees of freedom of the multiple M2-branes 
and $A=1,\cdots, 4$ are $SU(4)_{R}$ indices. 
In the above diagonal configurations, 
all the off-diagonal elements of the gauge fields become massive and 
the gauge group $U(N)\times \hat{U}(N)$ is spontaneously broken down to 
$U(1)^{N}$ \cite{Aharony:2008ug}. 
We now define 
\begin{align}
\label{qgauge1}
\mathcal{A}^{+}_{\mu a}&:=A_{\mu aa}+\hat{A}_{\mu aa},&
\mathcal{A}^{-}_{\mu a}&:=A_{\mu aa}-\hat{A}_{\mu aa}
\end{align}
We see that 
all the couplings are associated 
with the gauge fields $\mathcal{A}_{\mu a}^{-}$ 
while the other gauge fields $\mathcal{A}_{\mu a}^{+}$ have no 
couplings to matter fields as 
the preserved $U(1)$ gauge group. 
Making use of the
expressions (\ref{ydiag1}) and (\ref{qgauge1}), 
one can rewrite the equations (\ref{bpsabjm01}) as 
\begin{align}
\label{bpsabjm011}
\partial_{z}y_{a}^{A}+i\mathcal{A}_{za}^{-}y_{a}^{A}&=0,&
\partial_{z}\overline{y}_{Aa}-i\mathcal{A}_{za}^{-}\overline{y}_{Aa}&=0,
\\
\label{bpsabjm012}
\partial_{\overline{z}}y_{a}^{A}+i\mathcal{A}_{\overline{z}a}^{-}y_{a}^{A}&=0,& 
\partial_{\overline{z}}\overline{y}_{Aa}-i\mathcal{A}_{\overline{z}a}^{-}\overline{y}_{Aa}&=0,\\
\label{bpsabjm013}
A_{z ab}=\hat{A}_{z ab}
=A_{\overline{z}ab}&=\hat{A}_{\overline{z}ab}=0
&\mathrm{for\ }\ a\neq b &.
\end{align}
The first line and the second line are the constraint equations 
for diagonal elements and last one imposes the condition 
on the off-diagonal elements. 
The generic solutions to 
the equations (\ref{bpsabjm011}) and (\ref{bpsabjm012}) are
\begin{align}
\label{bpsabjm14}
y_{a}^{A}&=r_{a}^{A}e^{i
\left(
\varphi_{a}(z,\overline{z})+\theta^{A}_{a}
\right)},
\\
\label{bpsabjm15}
\mathcal{A}_{za}^{-}&=-\partial_{z}\varphi_{a}(z,\overline{z})
\end{align}
where $r_{a}^{A}$, $\theta_{a}^{A}$ $\in \mathbb{R}$ 
have no dependence on the coordinate $z$ and $\overline{z}$ 
while $\varphi_{a}(z,\overline{z}) \in \mathbb{R}$ 
is a function of $z$ and $\overline{z}$. 
The above expression (\ref{bpsabjm15}) ensures that 
the $U(1)$ gauge field $\mathcal{A}_{z}^{-}$ is flat.  
Thus $\varphi_{a}$, $\mathcal{A}_{za}^{-}$ and
$\mathcal{A}_{\overline{z}a}^{-}$ can be expressed as
\cite{AlvarezGaume:1986es}
\begin{align}
\label{phiabjm1}
&\varphi_{a}(z,\overline{z})
=-2\pi\frac{\Theta_{a}}{\tau-\overline{\tau}}z+
2\pi\frac{\overline{\Theta}_{a}}{\tau-\overline{\tau}}\overline{z},\\
\label{azabjm1}
&\mathcal{A}_{za}^{-}
=2\pi\frac{\Theta_{a}}{\tau-\overline{\tau}}\omega, \ \ \ \ \ \ \ 
\mathcal{A}_{\overline{z}a}^{-}
=-2\pi\frac{\overline{\Theta}_{a}}{\tau-\overline{\tau}}\overline{\omega}
\end{align}
where $\tau$ is the moduli of the torus 
as in (\ref{period0}) and 
the complex quantities 
$\Theta_{a}:=\zeta_{a}+\overline{\tau}\xi_{a}$, 
$a=1, \cdots, N$ are the coordinates of the product space 
of the $N$ Jacobi varieties which characterize the $N$ $U(1)$ flat bundles. 
In order for the bosonic matter fields to 
describe the transverse locations of the M2-branes, 
One needs to require the single-valuedness of the variables $y_{a}^{A}$ as
\begin{align}
\label{abjmsing1}
y_{a}^{A}(z+1,\overline{z}+1)
&=y_{a}^{A}(z,\overline{z}),
\nonumber\\
y_{a}^{A}(z+\tau,\overline{z}+\overline{\tau})
&=y_{a}^{A}(z,\overline{z}).
\end{align}  
These conditions imply that the coordinates 
$\xi_{a}$ and $\zeta_{a}$ take integer values.  
This results in the quantization of the variables $\varphi_{a}$,
$\mathcal{A}_{za}^{-}$ and $\mathcal{A}_{\overline{z}a}^{-}$, 
as discussed for the BLG case. 
The resulting static BPS configurations then turn out to be 
\begin{align}
\label{1bps1}
Y^{A}&=\mathrm{diag}(y_{1}^{A}, \cdots, y_{N}^{A})
=\textrm{diag}
\left(
r_{1}^{A}e^{i(\varphi_{1}(z,\overline{z})+\theta_{1}^{A})},
\cdots,
r_{N}^{A}e^{i(\varphi_{N}(z,\overline{z})+\theta_{N}^{A})}
\right),\nonumber\\
Y_{A}^{\dag}&=\mathrm{diag}(\overline{y}_{A1}, \cdots,
 \overline{y}_{AN})
=\textrm{diag}
\left(
r_{1}^{A}e^{-i(\varphi_{1}(z,\overline{z})+\theta_{1}^{A})},
\cdots,
r_{N}^{A}e^{-i(\varphi_{N}(z,\overline{z})+\theta_{N}^{A})}
\right),
\nonumber\\
A_{z}&=\textrm{diag}
\left(
A_{z 11},\cdots,A_{z NN}
\right),\nonumber\\
\hat{A}_{z}&=A_{z}+\partial_{z}\varphi
=\textrm{diag}\left(
A_{z 11}+\partial_{z}\varphi_{1},
\cdots,A_{z NN}+\partial_{z}\varphi_{N}
\right).
\end{align}
According to the supersymmetry the above bosonic configurations 
should be paired with the fermionic fields
\begin{align}
\label{1bps4}
\psi_{\pm A}&=
\mathrm{diag}\left(
\psi_{\pm A 1}, \cdots, \psi_{\pm A N}
\right),&
\psi^{\dag A}_{\pm}&=
\mathrm{diag}\left(
\psi_{\pm}^{\dag A 1}, \cdots, \psi_{\pm}^{\dag A N} 
\right)
\end{align}
where the subscripts $\pm$ on the fermionic fields 
label the $SO(2)_{E}$ spinor representation.

By substituting the set of BPS configurations 
(\ref{1bps1}) and (\ref{1bps4}) 
into the ABJM action (\ref{abjmlag1}) 
we find
\begin{align}
\label{abjmeff1a}
S=
\int_{\mathbb{R}} dt\int_{T^{2}}d^{2}z 
\sum_{A}
\sum_{a=1}^{N}\Biggl[
&D_{0}\overline{y}_{A}^{a}D_{0}y_{a}^{A}
-i\psi_{+}^{\dag Aa}D_{0}\psi_{+Aa}
-i\psi_{-}^{\dag Aa}D_{0}\psi_{-Aa}
\nonumber\\
&
+\frac{k}{4\pi}
\left(
\mathcal{A}^{-}_{0a}\mathcal{F}_{z\overline{z}a}^{+}
+\frac12 \mathcal{A}_{\overline{z}a}^{-}\dot{\mathcal{A}}_{za}^{+}
-\frac12 \mathcal{A}_{za}^{-}\dot{\mathcal{A}}_{\overline{z}a}^{+}
\right)
\Biggr].
\end{align}
Since $\mathcal{A}_{z}^{-}$ and $\mathcal{A}_{\overline{z}}^{-}$ 
are quantized and their time derivative terms do not show up in the
action, 
they can be regarded as auxiliary fields and we can integrate then out. 
As a result, we get constraints 
$\dot{\mathcal{A}}_{za}^{+}=\dot{\mathcal{A}}_{\overline{z}a}^{+}=0$. 
They imply that 
the gauge fields $\mathcal{A}_{za}^{+}$ and $\mathcal{A}_{\overline{z}a}^{+}$ 
on the Riemann surface do not depend on time.  

Under the above constraints we can carry out the integration over the
torus 
and can find the low-energy effective action 
\begin{align}
\label{effss2}
S=\int_{\mathbb{R}}dt 
\left[
D_{0}\overline{y}^{a}_{A}D_{0}y_{a}^{A}
-i\psi^{\dag \alpha Aa}D_{0}\psi_{\alpha Aa}
+kC_{1}(E_{a})\mathcal{A}_{0a}^{-}
\right].
\end{align}
Here the repeated indices implicated as a sum and 
the indices $\alpha,\beta,\cdots=+,-$ represent the $SO(2)_{E}$ spinor indices. 
We have defined the covariant derivatives by
\begin{align}
\label{abjmgcov1}
D_{0}y^{A}_{a}
&=\dot{y}_{a}^{A}+i\mathcal{A}^{-}_{0a}y_{a}^{A},&
D_{0}\overline{y}_{Aa}
&=\dot{\overline{y}}_{Aa}-i\mathcal{A}_{0a}^{-}\overline{y}_{Aa}
,\nonumber\\
D_{0}\psi_{\alpha Aa}
&=\dot{\psi}_{\alpha Aa}+i\mathcal{A}^{-}_{0a}\psi_{\alpha Aa},& 
D_{0}\psi_{\alpha a}^{\dag A}&=\dot{\psi}^{\dag A}_{\alpha a}
-i\mathcal{A}^{-}_{0a}\psi^{\dag A}_{\alpha a}.
\end{align}
We have also introduced the Chern number of the $a$-th $U(1)$ principal bundle 
$E_{a}\rightarrow T^{2}$ over the torus 
associated with 
the preserved $U(1)$ gauge fields $A_{z aa}$ as
\begin{align}
\label{ch02}
C_{1}(E_{a}):=
\frac{1}{2\pi}\int_{T^{2}}
F_{z\overline{z}aa}
=\frac{1}{4\pi}\int_{T^{2}}
\mathcal{F}^{+}_{z\overline{z} a}.
\end{align}

The action (\ref{effss2}) is invariant 
under the one-dimensional conformal transformations
\begin{align}
\label{tconf2}
\delta t&=f(t)=a+bt+ct^{2}, 
&\delta \partial_{0}&=-\dot{f}\partial_{0},\\
\delta y_{a}^{A}&=\frac12 \dot{f}y_{a}^{A}, 
&\delta \overline{y}_{Aa}&=\frac12 \dot{f}\overline{y}_{Aa},\\
\delta\psi_{\alpha Aa}&=0, 
&\delta\psi^{\dag A}_{\alpha a}&=0,\\
\delta \mathcal{A}_{0a}^{-}&=-\dot{f}\mathcal{A}_{0a}^{-}.
\end{align}
It is invariant 
under the $\mathcal{N}=12$ supersymmetry transformations 
\begin{align}
\label{tsusy03}
\delta
 y_{a}^{A}&=i\omega^{\alpha AB}\psi_{\alpha Ba},&
\delta
\overline{y}_{Aa}&=i\psi^{\dag \alpha B}_{a}\omega_{\alpha AB},\\
\delta\psi_{\alpha Aa}
&=\omega_{\alpha AB}D_{0}y_{a}^{B},&
\delta\psi_{\alpha a}^{\dag A}
&=-D_{0}\overline{y}_{Ba}\omega_{\alpha}^{AB},\\
\delta\mathcal{A}^{-}_{0a}
&=0.
\end{align}
Here the supersymmetry parameters 
$\omega_{+AB}:=\epsilon_{+i}(\Gamma^{i})_{AB}$ 
and $\omega_{-AB}:=\epsilon_{-i}(\Gamma^{i})_{AB}$ transform as 
$\bm{6}_{+}$ and $\bm{6}_{-}$ under $SU(4)\times SO(2)_{E}$ respectively. 
Accordingly the low-energy effective theory is described by 
the $\mathcal{N}=12$ superconformal gauged quantum mechanics (\ref{effss2}).

\section{Reduction}
\label{abjmscqm1}
We see that the effective action (\ref{effss2}) 
is quadratic in $\mathcal{A}_{0a}^{-}$ 
and contains no time derivatives of $\mathcal{A}_{0a}^{-}$. 
It means that they are auxiliary fields and we want to integrate them
out as in the BLG case. 
Let us choose the temporal gauge; $\mathcal{A}_{0a}^{-}=0$.  
Then the algebraic equations of motion for  
the auxiliary gauge fields $\mathcal{A}^{-}_{0a}$ turn out to yield 
the Gauss law constraints, which are the moment map conditions
\begin{align}
\label{1a1}
\phi_{0a}:=kC_{1}(E_{a})
+2\sum_{A}(r_{a}^{A})^{2}\dot{\theta}_{a}^{A}
+\sum_{A}
\psi^{\dag \alpha A a}\psi_{\alpha A a}
=0
\end{align}
for $a=1, \cdots, N$. 
Note that 
although the set of constraint equations (\ref{1a1}) has the same form
as (\ref{1a0}), there is a difference in 
the physical implications for these constraint equations. 
The angular variable $\theta_{a}^{A}$'s  
are defined not in the two-dimensional space of the internal degrees of freedom 
as in (\ref{1a0}), 
but in the transverse configuration space of the $a$-th M2-brane 
in the $A$-th complex plane. 

By defining the conserved charges 
$h_{a}^{A}:=2(r_{a}^{A})^{2}\dot{\theta}_{a}^{A}$, 
using the above constraints (\ref{1a1}) 
and proceeding the reduction procedure as in the derivation of 
(\ref{0eff0}), 
one can integrate out the auxiliary gauge fields $\mathcal{A}_{0a}^{-}$ 
and find the reduced effective action 
endowed with the inverse-square type potential
\begin{align}
\label{abjmeff1b}
S=\int_{\mathbb{R}}dt 
&\sum_{a=1}^{N}
\Biggl[
\dot{x}^{2}_{a}
-\frac{i}{2}\sum_{A\neq B}
\left(
\psi^{\dag\alpha A a}\dot{\psi}_{\alpha A a}
-\dot{\psi}^{\dag Aa}\psi_{\alpha Aa}
\right)
\nonumber\\
&
+\sum_{A\neq B}(\dot{r}_{a}^{A})^{2}
-\frac{i}{2}
\left(
\lambda^{\dag \alpha a}\dot{\lambda}_{\alpha a}
-\dot{\lambda}^{\dag\alpha a}\lambda_{\alpha a}
\right)
\nonumber\\
&
-\frac{\left[
kC_{1}(E_{a})
+\sum_{A\neq B}h_{a}^{A}
+\sum_{A\neq B}
\psi^{\dag \alpha A a}\psi_{\alpha A a}
+\lambda^{\dag \alpha a}\lambda_{\alpha a}
\right]^{2}}
{4x_{a}^{2}}
-\sum_{A\neq B}\frac{(h_{a}^{A})^{2}}{4(r_{a}^{A})^{2}}
\Biggr].
\end{align}
Here we have defined the variable $x_{a}:=r_{a}^{B}$ 
which describes the motion of 
the $a$-th M2-brane in the $B$-th complex plane 
in which the corresponding ``angular momentum'' $h_{a}^{B}$ is fixed  
by the other preserved charges.  
We have also introduced the fermion
$\lambda_{\alpha a}:=\psi_{\alpha B a}$ with $A=B$, 
which is the superpartner of 
$r_{a}^{C}$, $C=1,2,3$, 
as we will see from the supersymmetry transformations 
(\ref{abjmqmsusy05}) and (\ref{abjmqmsusy06}).

From the action (\ref{abjmeff1b}) one can read the following 
equations of motion
\begin{align}
\label{eomabjm00}
\ddot{x}_{a}
&=\frac{\left[
kC_{1}(E_{a})
+\sum_{A\neq B}h_{a}^{A}
+\sum_{A\neq B}\psi^{\dag\alpha Aa}\psi_{\alpha Aa}
+\lambda^{\dag\alpha a}\lambda_{\alpha a}
\right]^{2}}
{4x_{a}^{3}},\\
\label{eomabjm01}
\ddot{r}_{a}^{A}
&=\frac{(h_{a}^{A})^{2}}
{4(r_{a}^{A})^{3}},\\
\label{eomabjm02}
\dot{\psi}_{\alpha Aa}
&=i\frac{
kC_{1}(E_{a})
+\sum_{A\neq B}h_{a}^{A}
+\sum_{A\neq B}\psi^{\dag\alpha Aa}\psi_{\alpha Aa}
+\lambda^{\dag\alpha a}\lambda_{\alpha a}}
{2x_{a}}\psi_{\alpha A a},
\\
\label{eomabjm03}
\dot{\psi}^{\dag \alpha Aa}
&=-i
\frac{kC_{1}(E_{a})
+\sum_{A\neq B}h_{a}^{A}
+\sum_{A\neq B}\psi^{\dag\alpha Aa}\psi_{\alpha Aa}
+\lambda^{\dag\alpha a}\lambda_{\alpha a}}
{2x_{a}}\psi^{\dag\alpha Aa},
\\
\label{eomabjm04}
\dot{\lambda}_{\alpha a}
&=i\frac{
kC_{1}(E_{a})
+\sum_{A\neq B}h_{a}^{A}
+\sum_{A\neq B}\psi^{\dag\alpha Aa}\psi_{\alpha Aa}
+\lambda^{\dag\alpha a}\lambda_{\alpha a}}
{2x_{a}}\lambda_{\alpha a},\\
\label{eomabjm05}
\dot{\lambda}^{\dag \alpha a}
&=-i
\frac{kC_{1}(E_{a})
+\sum_{A\neq B}h_{a}^{A}
+\sum_{A\neq B}\psi^{\dag\alpha Aa}\psi_{\alpha Aa}
+\lambda^{\dag\alpha a}\lambda_{\alpha a}}
{2x_{a}}\lambda^{\dag\alpha a}.
\end{align}
Making use of the fermionic equations of motion 
(\ref{eomabjm02})-(\ref{eomabjm05}), 
one can see that the Gauss law constraint (\ref{1a1}) 
does not depend on time, i.e. $\dot{\phi}_{0a}=0$.

The canonical momenta are
\begin{align}
\label{mom11}
p^{a}&:=\frac{\partial L}{\partial \dot{x}_{a}}
=2\dot{x}^{a},& 
P^{a}_{A}&:=\frac{\partial L}{\partial \dot{r}_{a}^{A}}
=2\dot{r}^{a}_{A},\\
\label{mom12}
\pi^{\alpha Aa}
&:=\frac{\vec{\partial}L}{\partial \dot{\psi}_{\alpha Aa}}
=\frac{i}{2}\psi^{\dag \alpha A a},&
\tilde{\pi}_{\alpha Aa}
&:=\frac{\vec{\partial}L}{\partial \dot{\psi}^{\dag\alpha Aa}}
=\frac{i}{2}\psi_{\alpha Aa},
\\
\label{mom13}
\Pi^{\alpha a}
&:=\frac{\vec{\partial}L}{\partial \dot{\lambda}_{\alpha a}}
=\frac{i}{2}\lambda^{\dag \alpha a},&
\tilde{\Pi}_{\alpha a}
&:=\frac{\vec{\partial}L}{\partial \dot{\lambda}^{\dag \alpha a}}
=\frac{i}{2}\lambda_{\alpha a}.
\end{align}
The fermionic canonical momenta lead to 
the second class constraints
\begin{align}
\label{prcons2}
\phi_{1}^{\alpha Aa}
&:=\pi^{\alpha Aa}-\frac{i}{2}\psi^{\dag Aa}=0,&
\phi_{2\alpha Aa}
&:=\tilde{\pi}_{\alpha Aa}-\frac{i}{2}\psi_{\alpha Aa}=0,\\
\label{prcons3}
\phi_{3}^{\alpha a}
&:=\Pi^{\alpha a}-\frac{i}{2}\lambda^{\dag \alpha a}=0,&
\phi_{4\alpha a}
&:=\tilde{\Pi}_{\alpha a}-\frac{i}{2}\lambda_{\alpha a}=0,
\end{align}
from which 
we can obtain the Dirac brackets
\begin{align}
\label{dirac3}
[x_{a},p^{b}]_{DB}&=\delta_{ab},&
[r_{a}^{A},P_{B}^{b}]_{DB}=\delta_{AB}\delta_{ab},\\
\label{dirac4}
\left[
\psi_{\alpha Aa},\psi^{\dag \beta Bb}
\right]_{DB}
&=i\delta_{\alpha\beta}\delta_{AB}\delta_{ab},&
\left[
\lambda_{\alpha a},\lambda^{\dag\beta b}
\right]_{DB}
=i\delta_{\alpha\beta}\delta_{ab}.
\end{align}

The action (\ref{abjmeff1b}) is invariant under 
the one-dimensional conformal transformations
\begin{align}
\label{abjmconf00}
\delta t&=f(t)=a+bt+ct^{2}, 
&\delta \partial_{0}&=-\dot{f}\partial_{0},\\
\delta x_{a}&=\frac12 \dot{f}x_{a}, 
&\delta r_{a}^{A}&=\frac12 \dot{f}r_{a}^{A},\\
\delta\psi_{\alpha Aa}&=0, 
&\delta\psi^{\dag \alpha A}_{a}&=0,\\
\delta\lambda_{\alpha a}&=0, 
&\delta\lambda^{\dag\alpha}_{a}&=0.
\end{align}
Appealing the Noether's method,  
one finds the $SL(2,\mathbb{R})$ generators
\begin{align}
\label{abjmconf01}
H&=\sum_{a=1}^{N}
\Biggl[
\frac{p_{a}^{2}}{4}
+\frac{\left(
kC_{1}(E_{a})
+\sum_{A\neq B}h_{a}^{A}
+\sum_{A}
\psi^{\dag \alpha A a}\psi_{\alpha A a}
+\lambda^{\dag \alpha a}\lambda_{\alpha a}
\right)
^{2}}{4x_{a}^{2}}\nonumber\\
&+\sum_{A\neq B}\left(
\frac{(P_{a}^{A})^{2}}{4}
+\frac{(h_{a}^{A})^{2}}{4(r_{a}^{A})^{2}}
\right)
\Biggr],\\
\label{abjmconf02}
D&=-\frac14 \sum_{a=1}^{N}
\left[
\left\{x_{a},p_{a}\right\}
+\sum_{A\neq B}
\left\{r_{a}^{A},P_{a}^{A}\right\}
\right],\\
\label{abjmconf03}
K&=\sum_{a=1}^{N}\left[
x_{a}^{2}+\sum_{A\neq B}(r_{a}^{A})^{2}
\right].
\end{align}
Here we have absorbed the explicit time dependent part 
of $D$ and $K$ by taking the similarity transformations (\ref{simdk0}). 

The action (\ref{abjmeff1b}) is also invariant 
under the following fermionic transformations 
\begin{align}
\label{abjmqmsusy01}
\delta x_{a}&=
\frac{i}{\sqrt{2}}
\left(
\epsilon^{\alpha C}\psi_{\alpha Ca}
+\epsilon^{\dag}_{\alpha C}\psi^{\dag\alpha C}_{a}
\right)
,\\
\label{abjmqmsusy02}
\delta r_{a}^{C}&=
\frac{i}{2}
\left[
\left(
\omega^{\alpha CD}\psi_{\alpha Da}
\right)
e^{-i\theta_{a}^{C}}
+
\left(
\psi^{\dag \alpha D}_{a}\omega_{\alpha CD}
\right)e^{i\theta_{a}^{C}}
-
\left(\epsilon^{\alpha C}\lambda_{\alpha a}
\right)e^{-i\theta_{a}^{C}}
-
\left(\epsilon_{\alpha C}^{\dag}\lambda^{\dag\alpha}_{a}
\right)e^{i\theta_{a}^{C}}
\right]
,\\
\label{abjmqmsusy03}
\delta \psi_{\alpha Ca}&=
\left(
\dot{r}_{a}^{D}+i\frac{h_{a}^{D}}{2r_{a}^{D}}
\right)e^{i\theta_{a}^{D}}\omega_{\alpha CD}\nonumber\\
&+
\sqrt{2}\left(
\dot{x}_{a}
-i\frac{kC_{1}(E_{a})
+\sum_{D\neq B}h_{a}^{D}
+\psi^{\dag \alpha Da}\psi_{\alpha Da}
+\lambda^{\dag \alpha a}\lambda_{\alpha a}}{2x_{a}}
\right)\epsilon_{\alpha C}^{\dag}
-\frac{i}{\sqrt{2}}
\frac{l_{a}}{x_{a}}\psi_{\alpha Ca}
,\\
\label{abjmqmsusy04}
\delta \psi^{\dag\alpha C}_{a}&=
-\left(
\dot{r}_{a}^{D}-i\frac{h_{a}^{D}}{2r_{a}^{D}}
\right)e^{-i\theta_{a}^{D}}\omega_{\alpha}^{CD}\nonumber\\
&+
\sqrt{2}\left(
\dot{x}_{a}
+i\frac{
kC_{1}(E_{a})
+\sum_{D\neq B}h_{a}^{D}
+\psi^{\dag \alpha Da}\psi_{\alpha Da}
+\lambda^{\dag\alpha a}\lambda_{\alpha a}
}
{2x_{a}}
\right)\epsilon^{\alpha C}
+\frac{i}{\sqrt{2}}
\frac{l_{a}}{x_{a}}\psi_{a}^{\dag\alpha C}
,\\
\label{abjmqmsusy05}
\delta \lambda_{\alpha a}
&=
-\epsilon_{\alpha C}^{\dag}
\left(\dot{r}_{a}^{C}+i\frac{h_{a}^{C}}
{2r_{a}^{C}}
\right)e^{i\theta^{C}_{a}},\\
\label{abjmqmsusy06}
\delta \lambda^{\dag\alpha}_{a}
&=-\left(
\dot{r}_{a}^{C}-i\frac{h_{a}^{C}}
{2r_{a}^{C}}
\right)e^{-i\theta_{a}^{C}}\epsilon^{\alpha C}
\end{align}
with $C,D=1,2,3$ labeling the R-symmetry. 
Here $\epsilon^{\alpha C}$ and 
their Hermitian conjugate $\epsilon_{\alpha C}^{\dag}$ 
are infinitesimal fermionic parameters and we have defined the quantities
\begin{align}
\label{abjmpa1}
\theta_{a}^{C}(t)&=h^{C}_{a}\int^{t} \frac{dt'}{(r^{C}_{a}(t'))^{2}},\\
\label{abjmpa2}
l_{a}&=\epsilon\psi_{a}-\epsilon^{\dag}\psi_{a}^{\dag}.
\end{align}

\section{$SU(1,1|6)$ superconformal mechanics}
\label{abjmscqm2}
Since the non-local quantities are included in 
the fermionic transformations
(\ref{abjmqmsusy01})-(\ref{abjmqmsusy06}), 
we may have infinitely many conserved non-local charges. 
Similar to the BLG case, from (\ref{abjmconf01}), 
we see that the Hamiltonian associated with local charges can be
decoupled. 
The Hamiltonian describing 
the motion in the $B$-th complex plane associated with 
the variable $x_{a}$ possesses the local charges 
while the others associated with the variables $r_{a}^{C}$'s 
have non-local charges. 
Since they are decoupled 
and one can analyze the dynamics in the $B$-th direction
separately. 
For simplicity, 
as in the section \ref{secflat3}, 
it is useful to assign the conserved charges $h_{a}^{A}$ and 
$\lambda^{\dag\alpha a}\lambda_{\alpha a}$ to be zeros. 
The low-energy dynamics 
in the $B$-th complex plane is then given by the action 
\begin{align}
\label{abjmscqm0}
S&=\int_{\mathbb{R}}dt \sum_{a=1}^{N}
\Biggl[
\dot{x}_{a}^{2}-i\psi^{\dag\alpha Aa}\dot{\psi}_{\alpha Aa}
-\frac{\left(
kC_{1}(E_{a})+\psi^{\dag\alpha Aa}\psi_{\alpha Aa}
\right)^{2}}
{4x_{a}^{2}}
\Biggr]
\end{align}
where $A=1,2,3$ represent the R-symmetry indices. 
We note that the action (\ref{abjmscqm0}) 
takes the same structure as (\ref{n8susy0a1}) 
\cite{Ivanov:1988it,Wyllard:1999tm,Fedoruk:2011aa} 
for $SU(1,1|\frac{\mathcal{N}}{2})$, $\mathcal{N}>4$ 
superconformal quantum mechanics. 
 
The action (\ref{abjmscqm0}) is invariant under 
the $\mathcal{N}=12$ supersymmetry transformation laws
\begin{align}
\label{su116susy1}
\delta x_{a}&=
\frac{i}{\sqrt{2}}
\left(
\epsilon^{\alpha A}\psi_{Aa}^{\alpha}
+\epsilon^{\dag}_{\alpha A}\psi_{a}^{\dag\alpha A}
\right),
\\
\label{su116susy2}
\delta\psi_{\alpha Aa}
&=
\sqrt{2}
\left(
\dot{x}_{a}-i\frac{g_{a}}{2x_{a}}
\right)\epsilon_{\alpha A}^{\dag}
-\frac{i}{\sqrt{2}}\frac{l_{a}}{x_{a}}\psi_{\alpha Aa},
\\
\label{su116susy3}
\delta\psi^{\dag \alpha A}_{a}
&=
\sqrt{2}
\left(
\dot{x}_{a}+i\frac{g_{a}}{2x_{a}}
\right)\epsilon^{\alpha A}
+\frac{i}{\sqrt{2}}\frac{l_{a}}{x_{a}}
\psi_{a}^{\dag \alpha A}
\end{align}
where 
\begin{align}
\label{g02}
g_{a}=kC_{1}(E_{a})
+\psi^{\dag \alpha Aa}\psi_{\alpha Aa}.
\end{align}
The supersymmetry transformations 
(\ref{su116susy1})-(\ref{su116susy3}) 
are generated by the supercharges
\begin{align}
\label{tsch2}
Q_{\alpha A}&=
\frac{i}{\sqrt{2}}
\left(
p^{a}-\frac{g_{a}}{x_{a}}
\right)
\psi_{\alpha Aa},\\
\tilde{Q}^{\alpha A}&=
\frac{i}{\sqrt{2}}
\left(
p^{a}+\frac{g_{a}}{x_{a}}
\right)
\psi^{\dag \alpha A}.
\end{align}

The action (\ref{abjmscqm0}) also 
has the one-dimensional conformal invariance. 
The Noether charges are found to be
\begin{align}
\label{hkdconf11}
H&=\sum_{a=1}^{N}
\left[
\frac{p_{a}^{2}}{4}
+\frac{
\left(
kC_{1}(E_{a})
+\psi^{\dag \alpha Aa}\psi_{\alpha A a}
\right)^{2}}
{4x_{a}^{2}}
\right],\\
\label{hkdcon12}
D&=-\frac14\sum_{a=1}^{N}\left\{x_{a},p^{a}\right\},\\
\label{hkdcon13}
K&=\sum_{a=1}^{N}x_{a}^{2}.
\end{align}

Due to the Dirac brackets (\ref{dirac3}) and (\ref{dirac4}), 
quantum operators of the canonical coordinates 
and momenta satisfy the (anti)commutation relations
\begin{align}
\label{canonical1}
[x_{a},p^{b}]&=i\delta_{ab},&
\left\{
\psi_{\alpha Aa},\psi^{\dag \beta Bb}
\right\}=-\delta_{\alpha\beta}\delta_{AB}\delta_{ab}.
\end{align}

The combination of the supercharges and 
the conformal generators leads to 
the superconformal boost generators
\begin{align}
\label{scboost2}
S_{\alpha A}&=
\sqrt{2}i\sum_{a}x_{a}\psi_{\alpha Aa},\\
\tilde{S}^{\alpha A}
&=\sqrt{2}i\sum_{a}x_{a}\psi^{\dag \alpha A}_{a}.
\end{align}
We will define the R-symmetry generator by
\begin{align}
\label{j2}
(J_{\alpha\beta})_{AB}
=i\sum_{a}\psi^{\dag\beta B}_{a}\psi_{\alpha A a}.
\end{align}
We see that (\ref{j2}) is a complex $6\times 6$ matrix 
with $\alpha,\beta=+,-$ and 
$A,B=1,2,3$ and it contains $36$ complex valued elements.

Taking into account the canonical relations (\ref{canonical1}) 
and the Weyl ordering \footnote{One needs to pick up constant shifts 
as a quantum effect.}, we find that 
the set of generators form the following algebra
\begin{align}
\label{su116a}
\begin{array}{ccc}
[H,D]=iH,
&[K,D]=-iK,
&[H,K]=2iD,
\cr
\end{array}
\end{align}
\begin{align}
\label{su116b}
\begin{array}{ccc}
[(J_{\alpha\beta})_{AB},H]=0,
&[(J_{\alpha\beta})_{AB},D]=0,
&[(J_{\alpha\beta})_{AB},K]=0,
\end{array}
\end{align}
\begin{align}
\label{su116c}
[(J_{\alpha\beta})_{AB},
(J_{\gamma\delta})_{CD}]
&=i(J_{\alpha\delta})_{AD}
\delta_{\beta\gamma}\delta_{BC}
-i(J_{\gamma\beta})_{CB}
\delta_{\alpha\delta}\delta_{AD},
\end{align}
\begin{align}
\label{su116d}
\begin{array}{ccc}
[H,Q_{\alpha A}]=0,
&[D,Q_{\alpha A}]=-\frac{i}{2} Q_{\alpha A},
&[K,Q_{\alpha A}]=iS_{\alpha A},
\cr
[H,\tilde{Q}^{\alpha A}]=0,
&[D,\tilde{Q}^{\alpha A}]
=-\frac{i}{2}\tilde{Q}^{\alpha A},
&[K,\tilde{Q}^{\alpha A}]
=i\tilde{S}^{\alpha A},
\end{array}
\end{align}
\begin{align}
\label{su1161e}
\begin{array}{ccc}
[H,S_{\alpha A}]=-iQ_{\alpha A},
&[D,S_{\alpha A}]=\frac{i}{2}S_{\alpha A},
&[K,S_{\alpha A}]=0,
\cr
[H,\tilde{S}^{\alpha A}]
=-i\tilde{Q}^{\alpha A},
&[D,\tilde{S}^{\alpha A}]
=\frac{i}{2}\tilde{S}^{\alpha A},
&[K,\tilde{S}^{\alpha A}]=0,
\cr
\end{array}
\end{align}
\begin{align}
\label{su116f}
\{Q_{\alpha A},\tilde{Q}^{\beta B}\}
&=2H\delta_{\alpha\beta}\delta_{AB},\nonumber\\
\{S_{\alpha A},\tilde{S}^{\beta B}\}
&=2K\delta_{\alpha\beta}\delta_{AB},\nonumber\\
\{Q_{\alpha A},\tilde{S}^{\beta B}\}
&=-2D\delta_{\alpha\beta}\delta_{AB}
-2(J_{\alpha\beta})_{AB},
\nonumber\\
\{\tilde{Q}^{\alpha A},S_{\beta B}\}
&=-2D\delta_{\alpha\beta}\delta_{AB}
-2(J_{\alpha\beta}^{\dag})_{AB},
\end{align}
\begin{align}
\label{su116g}
[(J_{\alpha\beta})_{AB},Q_{\gamma C}]
&=iQ_{\alpha A}\delta_{\beta\gamma}\delta_{BC},&
[(J_{\alpha\beta})_{AB},S_{\gamma,C}]
&=iS_{\alpha A}\delta_{\beta\gamma}\delta_{BC},
\nonumber\\
[(J_{\alpha\beta})_{AB},\tilde{Q}^{\gamma C}]
&=-i\tilde{Q}^{\alpha A}\delta_{\beta\gamma}\delta_{BC},&
[(J_{\alpha\beta})_{AB},\tilde{S}^{\gamma,C}]
&=-i\tilde{S}^{\alpha A}\delta_{\beta\gamma}\delta_{BC}.
\end{align}

We see that 
the Hamiltonian $H$, 
the dilatation generator $D$ and 
the conformal boost generator form the 
$\mathfrak{so}(1,2)=\mathfrak{sl}(2,\mathbb{R})=\mathfrak{su}(1,1)$ 
one-dimensional conformal algebra. 
Note that 
there are twelve supercharges 
since each of the supercharges $Q_{\alpha A}$ 
and $\tilde{Q}^{\alpha A}=-(Q_{\alpha A})^{\dag}$ 
contains six real components. 
The supercharges can be viewed as the square roots of the Hamiltonian $H$. 
In addition to the 
supercharges, there are as many superconformal charges 
$S_{\alpha A}$ and $\tilde{S}^{\alpha A}$, 
which can be recognized 
as the square roots of the conformal boost generator $K$. 
The anti-commutators of the fermionic charges generate 
an extra bosonic R-symmetry generators $(J_{\alpha\beta})_{AB}$. 
They rotate the fermionic generators 
and form the $\mathfrak{u}(6)$ algebra (\ref{su116c}). 
Hence the effective action (\ref{abjmscqm0}) describes 
the $SU(1,1|6)$ superconformal mechanics.  
We see that this belongs to the list of the 
one-dimensional superconformal group 
which we have argued in Table \ref{listsup1}.

Following the AdS$_{2}$/CFT$_{1}$ correspondence, 
we expect that the superconformal quantum mechanical models 
(\ref{effs1}) and (\ref{effss2}) 
may be related to $\textrm{AdS}_{2}\times T^{2}$ solutions, 
the so-called magnetic brane solutions 
\cite{Almheiri:2011cb,Donos:2011pn}. 
It may be interesting to check those correspondences.

\chapter{Curved Branes and Topological Twisting}
\label{curvedtwistingcg01}
In this chapter 
we will investigate the topological twisting 
and its relevant application as 
the world-volume description of curved branes 
in string theory and M-theory, 
which was firstly pointed out in \cite{Bershadsky:1995qy}. 
In section \ref{twistsec01} 
we will discuss various topological twisting procedures. 
In section \ref{twistsec02} 
we will explain 
that the topologically twisted theories may yield the 
world-volume theories of the curved branes.

\section{Topological twisting}
\label{twistsec01}
Topological twisting is a modification of the Euclidean rotational
group of a supersymmetric theory through an embedding into a global
symmetry of the theory. 
The resulting theory will be topological if the twisted supersymmetry
generators include at least one space-time scalar. 
Equivalently one can regard the twisting procedure as a gauging of an
internal symmetry group in which a global symmetry is promoted to a
space-time symmetry. In many cases, gauging can be performed by coupling
of the internal symmetry current to the spin connection of the
underlying manifold to the Lagrangian. 
We will give many examples of the topological twisting in the following.

\subsection{$d=4$, $\mathcal{N}=2$ SYM theories}
\label{d4n2twistsymsubsec1}
Let us consider topological twisting of 
$d=4$, $\mathcal{N}=2$ super Yang-Mills (SYM) theories 
\cite{Witten:1988ze}. 
We take $M_{4}=\mathbb{R}^{4}$ whose rotational symmetry group is
$Spin(4)_{E}\cong SU(2)_{l}\times SU(2)_{r}$. 
The global symmetry of the theory is $U(2)_{R}\simeq SU(2)\times U(1)$ R-symmetry. 
The field content is
\begin{itemize}
\item complex scalar field $\phi$
\item 2 complex fermionic fields $\lambda_{\alpha}^{i},
      \overline{\lambda}_{\dot{\alpha} i}$
\item gauge field $A_{\alpha\dot{\alpha}}$
\end{itemize}
where $\alpha$ are indices of the fundamental representation of
$SU(2)_{l}$ and $\dot{\alpha}$ are indices of the fundamental
representation of $SU(2)_{r}$.
$i$ denotes the fundamental representation of $SU(2)_{R}$. 
 These indices are raised and lowered with
the antisymmetric tensor $\epsilon_{\alpha\beta},
\epsilon_{\dot{\alpha}\dot{\beta}}, \epsilon_{ij}$ such that 
$\epsilon_{12}=\epsilon^{12}=1$. 
All fields are the adjoint representation of compact group $G$. 
The scaling dimensions are
\begin{equation}
 [\phi]=1,\ \ \ [\psi]=[\lambda]=\frac32,\ \ \ [A]=1,\ \ \
 \ \ \ [\epsilon]=-\frac12
\end{equation}
where $\epsilon$ is a supersymmetry parameter.

The supersymmetry transformations are 
\begin{align}
&\delta
A_{\mu}=-i\overline{\lambda}^{\dot{\alpha}}_{i}\sigma_{\mu\alpha\dot{\alpha}}
\epsilon^{\alpha
 i}+i\overline{\epsilon}^{\dot{\alpha}}_{i}\sigma_{\mu\alpha\dot{\alpha}}\lambda^{\alpha
 i},\\
&\delta\lambda_{\alpha}^{i}
=\sigma^{\mu\nu}_{\alpha\beta}\epsilon^{\beta i}F_{\mu\nu}
+i\epsilon_{\alpha}^{i}[\phi,\overline{\phi}]
+i\sqrt{2}\sigma^{\mu}_{\alpha\dot{\alpha}}D_{\mu}\phi\epsilon^{ij}\overline{\epsilon}_{j}^{\dot{\alpha}},\\
&\delta\overline{\lambda}_{\dot{\alpha}i}
=\sigma^{\mu\nu}_{\dot{\alpha}\dot{\beta}}
\overline{\epsilon}^{\dot{\beta}}_{i}F_{\mu\nu}
-i\overline{\epsilon}_{\dot{\alpha}i}[\phi,\overline{\phi}]
+i\sqrt{2}D_{\mu}\sigma^{\mu}_{\dot{\alpha}\alpha}\overline{\phi}\epsilon_{ij}\epsilon^{\alpha j},\\
&\delta \phi=\sqrt{2}\epsilon^{\alpha i}\lambda_{\alpha i},\\
&\delta \overline{\phi}=\sqrt{2}\overline{\epsilon}^{\dot{\alpha}}_{i}
\overline{\lambda}^{i}_{\dot{\alpha}}
\end{align}
where $\epsilon^{i}_{\alpha}$ and $\overline{\epsilon}_{\dot{\alpha}i}$
are supersymmetry parameters that transform as $(\bm{2},\bm{1},\bm{2})$
and $(\bm{1},\bm{2},\bm{2})$ respectively.

The Lorentzian action is given by
\begin{align}
\mathcal{L}
=&\frac{1}{e^{2}}
\int_{M}d^{4}x \textrm{Tr} 
\Biggl(
-\frac14 F_{\mu\nu}F^{\mu\nu}
-i\overline{\lambda}^{\dot{\alpha}}_{i}
\sigma_{\alpha\dot{\alpha}}^{\mu\nu}D_{\mu}\lambda^{\alpha i}
-D_{\mu}\overline{\phi}D^{\mu}\phi
\nonumber \\
&-\frac12[\overline{\phi},\phi]^{2}
-\frac{1}{\sqrt{2}}\overline{\varphi}\epsilon_{ij}
[\lambda^{\alpha i},\lambda_{\alpha}^{j}]
+\frac{i}{\sqrt{2}}\phi \epsilon^{ij}
[\overline{\lambda}_{\dot{\alpha}i}, \overline{\lambda}^{\dot{\alpha}}_{i}]
\Biggr).
\end{align}
Here $\textrm{Tr}$ is an invariant quadratic form on the Lie algebra.

The classical $\mathcal{N}=2$ theory has a $U(2)$ symmetry acting on the
two fermion $(\lambda, \overline{\lambda})$. 
The center $U(1)_{R}\subset U(2)$ is anomalous. 
On a given 4-manifold $M_{4}$ and for a given instanton number$k$, 
the total violation $\Delta U$ of the $U(1)_{R}$ charge is given by the
dimension of the Yang-Mills instanton moduli space \cite{Witten:1988ze}.
For $SU(2)$ this is 
\begin{align}
\Delta U=\dim \mathcal{M}=8k-\frac32 (\chi+\sigma)
\end{align}
where $\chi$ and $\sigma$ are the Euler
characteristic and signature of $M_{4}$\footnote{The quantity
$\frac{\chi+\sigma}{2}$ is always integer.}. 
This was first discussed in \cite{Witten:1988ze}.

The fields and supersymmetry parameters transform under
$SO(4)_{E}\times U(2)_{R}\simeq SU(2)_{l}\times SU(2)_{r}\times
SU(2)_{R}\times U(1)_{R}$ as
\begin{align}
 \phi:& (\bm{1},\bm{1},\bm{1})_{2}\oplus (\bm{1},\bm{1},\bm{1})_{-2} \\
\psi,\lambda:& (\bm{2},\bm{1},\bm{2})_{1}\oplus (\bm{1},\bm{2},\bm{2})_{-1}\\
A_{\mu}:&(\bm{2},\bm{2},\bm{1})_{0}\\
\epsilon:&(\bm{2},\bm{1},\bm{2})_{1}\oplus (\bm{1},\bm{2},\bm{2})_{-1}.
\end{align}
To perform the topological twisting, we leave $SU(2)_{l}$ undisturbed
and pick a homomorphism
\begin{equation}
\label{homsym2}
\pi:SU(2)_{r}\rightarrow SU(2)_{R},
\end{equation}
and replace $SU(2)_{r}$ by a
diagonal subgroup $SU(2)'_{r}=(1+\pi)(SU(2))\subset SU(2)_{r}\times
SU(2)_{R}$. Then under the new rotational symmetry $SO(4)'_{E}\simeq
SU(2)_{l}\times SU(2)_{r}'$, the fields and supersymmetry parameters transform as
\begin{align}
\label{4dn2symtwist1}
 \phi\rightarrow& (\bm{1},\bm{1})_{2}\oplus (\bm{1},\bm{1})_{-2} \\
\label{4dn2symtwist2}
\psi,\lambda\rightarrow& (\bm{2},\bm{2})_{1}\oplus
 (\bm{1},\bm{1})_{-1}\oplus (\bm{1},\bm{3})_{-1}\\
\label{4dn2symtwist3}
A_{\mu}\rightarrow&(\bm{2},\bm{2})_{0}\\
\label{4dn2symtwist5}
\epsilon\rightarrow& (\bm{2},\bm{2})_{1}\oplus
 (\bm{1},\bm{1})_{-1}\oplus (\bm{1},\bm{3})_{-1}.
\end{align}
Thus the bosonic field content is
\begin{itemize}
 \item complex scalar field $\phi$: $(\bm{1},\bm{1})_{2}\oplus
       (\bm{1},\bm{1})_{-2}$
\item gauge field $A_{\mu}$: $(\bm{2},\bm{2})_{0}$
\end{itemize}
and the fermionic field content is
\begin{itemize}
 \item scalar field $\eta$: $(\bm{1},\bm{1})_{-1}$
\item 1-form $\psi_{\mu}$: $(\bm{2},\bm{2})_{1}$
\item 2-form (self-dual antisymmetric 2-tensor) $\chi_{\mu\nu}^{+}$: $(\bm{1},\bm{3})_{-1}$.
\end{itemize}
From (\ref{4dn2symtwist5}), 
one can see that there exists one BRST charge.

In $d=4$, $\mathcal{N}=2$ SYM theories, the possible anomalies are related
to the global $SU(2)$ anomaly \cite{Witten:1982fp}, which only appear
when the corresponding moduli space is not orientable
\cite{Witten:1988ze}. 
In Donaldson-Witten theory the moduli space is given by anti-self-dual
connections, which is orientable \cite{MR1079726}. 
Thus the twisted theory is anomaly free.

The twisted Lagrangian is
\begin{align}
\mathcal{L}
&=\textrm{Tr}\Biggl(
\frac14 F_{\alpha\beta}F^{\alpha\beta}
-\frac12D_{\alpha}\phi D^{\alpha}\sigma
-i\eta D_{\alpha}\psi^{\alpha}+iD_{\alpha}\psi_{\beta}\cdot
 \chi^{\alpha\beta}\nonumber \\
&-\frac{i}{8}\phi[\chi_{\alpha\beta},\chi^{\alpha\beta}]
-\frac{i}{2}\sigma[\psi_{\alpha},\psi^{\alpha}]
-\frac{i}{2}\phi[\eta,\eta]-\frac18[\phi,\sigma]^{2}
\Biggr).
\end{align}

For the closure of supersymmetry algebra, it it necessary to introduce
an auxiliary field $T_{ij}=T_{ji}$. 
It has scaling dimension $[T]=2$ and transform
$(\bm{1},\bm{1},\bm{3})_{0}$ under $SU(2)_{l}\times SU(2)_{r}\times
SU(2)_{R}\times U(1)_{R}$. After twisting they transform
$(\bm{1},\bm{3})_{0}$ and identified with a 2-form. 

Twisted $\mathcal{N}=2$ supersymmetric gauge theories have an off-shell
formulation such that the action can be expressed as a $Q$-exact
expression up to a $\theta$-term\footnote{
Because of the chiral anomaly inherent to the R-symmetry of
$\mathcal{N}=2$ SYM theories, observables are independent of
$\theta$-term up to rescaling. Thus one can ignore $\theta$-term. 
}, where $Q$ is the BRST charge.

\subsection{$d=4$, $\mathcal{N}=2$ SCFT on $C\times \Sigma$}
\label{d4n2scftsubsec1}
We now consider a four-dimensional  $\mathcal{N}=2$ superconformal field 
theory (SCFT) on $M_{4}=C\times \Sigma$ whose holonomy group is reduced to
$U(1)_{C}\times U(1)_{\Sigma}$, where $C$ and $\Sigma$ are Riemann
surfaces. This has been discussed in \cite{Kapustin:2006hi}. 
The global symmetry of the theory is $SU(2)_{R}\times U(1)_{R}$
R-symmetry and $U(1)_{B}$ symmetry. 
The field content is
\begin{itemize}
 \item complex scalar field in the adjoint representation: $\varphi, \overline{\varphi}$
\item 2 complex scalar fields in representation $R, R^{\vee}$ of $G$
      (squarks): $q,\tilde{q}$
\item gauge field $A_{\mu}$ 
\item 2 gauginos:
      $\psi,\lambda$
\item 2 complex left-handed quarks in representation
      $R^{\vee}, R$ of $G$: $\psi_{q},\psi_{\tilde{q}}$
\item 2 complex right-handed quarks in representation $R,
      R^{\vee}$ of $G$: $\overline{\psi_{q}},\overline{\psi_{\tilde{q}}}$.
\end{itemize}

Before topological twisting, fields transform under $U(1)_{C}\times
U(1)_{\Sigma}\times SU(2)_{R}\times U(1)_{R}\times U(1)_{B}$ as
\begin{align}
&\varphi,\overline{\varphi}:
\bm{1}_{0020}\oplus \bm{1}_{00-20}\\
&q,\tilde{q}:\bm{2}_{000-}\oplus \bm{2}_{000+}\\
&A_{\mu}:\bm{1}_{2000}\oplus \bm{1}_{-2000}
\oplus \bm{1}_{0200}\oplus \bm{1}_{0-200}\\
&\psi,\lambda:\bm{2}_{+-+0}\oplus \bm{2}_{-++0}\\
&\overline{\psi},\overline{\lambda}:\bm{2}_{---0}\oplus \bm{2}_{++-0}\\
&\psi_{q},\psi_{\tilde{q}}:\bm{1}_{+---}\oplus \bm{1}_{+--+}
\oplus \bm{1}_{-+--}\oplus \bm{1}_{-+-+}\\
&\overline{\psi}_{q},\overline{\psi}_{\tilde{q}}:
\bm{1}_{--++}\oplus \bm{1}_{++++}\oplus \bm{1}_{--+-}
\oplus \bm{1}_{+++-}
\end{align}
. To perform
the topological twisting, we pick a homomorphism
$\pi:U(1)_{E}\rightarrow SU(2)_{R}\times U(1)_{R}\times U(1)_{B}$ and
replace $U(1)_{E}$ by $U(1)'_{E}=(1+\pi)(U(1)_{E})\subset U(1)_{E}\times
SU(2)_{R}\times U(1)_{R}\times U(1)_{B}$.

To pick a homomorphism, we consider the maximal torus $U(1)_{R}'$ of $SU(2)_{R}$
\begin{equation}
 SU(2)_{R}\supset U(1)_{R}'.
\end{equation}
We assign $U(1)_{R}'$ charge for each field as in Table
\ref{scftfig1}\footnote{Our assignment is different from that in
\cite{Kapustin:2006hi} where the both charges for $q,\tilde{q}$ are $-$.}.
\begin{table}
\begin{center}
\begin{tabular}{|c|c|} \hline
&$U(1)_{R}'$\\ \hline\hline
$\varphi$&$0$ \\ 
$\overline{\varphi}$&$0$\\
$\psi$&$+$ \\
$\lambda$&$-$ \\
$q$&$+$ \\ 
$\tilde{q}$&$-$ \\
$\psi_{q}$&$0$ \\
$\psi_{\tilde{q}}$&$0$\\ \hline
\end{tabular}
\caption{The $U(1)_{R}'\subset SU(2)_{R}$ charge assignments for $d=4$, 
$\mathcal{N}=2$ SCFT field content.}
\label{scftfig1}
\end{center}
\end{table}
Then all of the $U(1)$ charges are summarized in Table \ref{scftfig2}. 
\begin{table}
\begin{center}
\begin{tabular}{|c|c|c|c|c|c|c|} \hline
&$U(1)_{C}$&$U(1)_{\Sigma}$&$U(1)_{R}'$&$U(1)_{R}$&$U(1)_{B}$&section \\ \hline\hline
$\varphi$&$0$&$0$&$0$&$2$&$0$&$\mathcal{O}_{C}\otimes
			 \mathcal{O}_{\Sigma}$ \\
$\overline{\varphi}$&$0$&$0$&$0$&$-2$&$0$&$\mathcal{O}_{C}\otimes \mathcal{O}_{\Sigma}$ \\
$\psi_{+}$&$+$&$-$&$+$&$+$&$0$&$K_{C}^{\frac12}\otimes K_{\Sigma}^{-\frac12}$ \\
$\psi_{-}$&$-$&$+$&$+$&$+$&$0$&$K_{C}^{-\frac12}\otimes
			 K_{\Sigma}^{\frac12}$\\
$\lambda_{+}$&$+$&$-$&$-$&$+$&$0$&$K_{C}^{\frac12}\otimes K_{\Sigma}^{-\frac12}$ \\
$\lambda_{-}$&$-$&$+$&$-$&$+$&$0$&$K_{C}^{-\frac12}\otimes
			 K_{\Sigma}^{\frac12}$ \\
$\overline{\psi}_{+}$&$-$&$-$&$-$&$-$&$0$&$K_{C}^{-\frac12}\otimes K_{\Sigma}^{-\frac12}$ \\
$\overline{\psi}_{-}$&$+$&$+$&$-$&$-$&$0$&$K^{\frac12}\otimes
			 K_{\Sigma}^{\frac12}$ \\ 
$\overline{\lambda}_{+}$&$-$&$-$&$+$&$-$&$0$&$K_{C}^{-\frac12}\otimes
			 K_{\Sigma}^{-\frac12}$\\
$\overline{\lambda}_{-}$&$+$&$+$&$+$&$-$&$0$&$K_{C}^{\frac12}\otimes
			 K_{\Sigma}^{\frac12}$\\ 
$q$&$0$&$0$&$+$&$0$&$-$&$\mathcal{O}_{C}\otimes \mathcal{O}_{\Sigma}$ \\
$\tilde{q}$&$0$&$0$&$-$&$0$&$+$&$\mathcal{O}_{C}\otimes
			 \mathcal{O}_{\Sigma}$ \\ 
$\psi_{q+}$&$+$&$-$&$0$&$-$&$-$&$K_{C}^{\frac12}\otimes K_{\Sigma}^{-\frac12}$ \\
$\psi_{q-}$&$-$&$+$&$0$&$-$&$-$&$K_{C}^{-\frac12}\otimes
			 K_{\Sigma}^{\frac12}$ \\ 
$\psi_{\tilde{q}+}$&$+$&$-$&$0$&$-$&$+$&$K_{C}^{\frac12}\otimes
			 K_{\Sigma}^{-\frac12}$ \\ 
$\psi_{\tilde{q}-}$&$-$&$+$&$0$&$-$&$+$&$K_{C}^{-\frac12}\otimes
			 K_{\Sigma}^{\frac12}$ \\ 
$\overline{\psi}_{q+}$&$-$&$-$&$0$&$+$&$+$&$K_{C}^{-\frac12}\otimes K_{\Sigma}^{-\frac12}$ \\
$\overline{\psi}_{q-}$&$+$&$+$&$0$&$+$&$+$&$K_{C}^{\frac12}\otimes
			 K_{\Sigma}^{\frac12}$ \\ 
$\overline{\psi}_{\tilde{q}+}$&$-$&$-$&$0$&$+$&$-$&$K_{C}^{-\frac12}\otimes
			 K^{-\frac12}_{\Sigma}$\\
$\overline{\psi}_{\tilde{q}-}$&$+$&$+$&$0$&$+$&$-$&$K_{C}^{\frac12}\otimes
			 K_{\Sigma}^{\frac12}$ \\ \hline
\end{tabular}
\caption{$U(1)_{R}$ charge assignments for $d=4$, $\mathcal{N}=2$ SCFT
 field content. The subscripts $\pm$ indicate the upper and lower
 components of spinors. We denote the trivial bundle as $\mathcal{O}$
 and the canonical bundle as $K$.}
\label{scftfig2}
\end{center}
\end{table}
In Table \ref{scftfig2}, the subscripts $\pm$ indicate the upper and lower
components of spinors. 
If $\Sigma$ is flat, we should twist only $U(1)_{C}$ and there are two
types of twisting
\begin{align}
\textrm{A-twist}: & U(1)_{C}\rightarrow U(1)_{R}'\\
\textrm{B-twist}: & U(2)_{C}\rightarrow U(1)_{R}.
\end{align}
The field content of A-twist and B-twist listed in Table \ref{4scftab}.
\begin{table}
\begin{center}
\begin{tabular}{|c|c|c|c|c|} \hline
 &\multicolumn{2}{|c|}{A-twist}&\multicolumn{2}{|c|}{B-twist}\\ \hline
fields&$U(1)_{C}'$&$U(1)_{\Sigma}'$&$U(1)_{C}'$&$U(1)_{\Sigma}$ \\ \hline\hline
$\varphi$&$0$&$0$&$2$&$0$ \\ 
$\overline{\varphi}$&$0$&$0$&$-2$&$0$ \\ 
$\psi_{+}$&$2$&$-$&$2$&$-$ \\
$\psi_{-}$&$0$&$+$&$0$&$+$ \\
$\lambda_{+}$&$0$&$-$&$2$&$-$ \\
$\lambda_{-}$&$-2$&$+$&$0$&$+$ \\
$\overline{\psi}_{+}$&$-2$&$-$&$-2$&$-$ \\
$\overline{\psi}_{-}$&$0$&$+$&$0$&$+$ \\ 
$\overline{\lambda}_{+}$&$0$&$-$&$-2$&$-$ \\
$\overline{\lambda}_{-}$&$2$&$+$&$0$&$+$ \\ 
$q$&$+$&$0$&$0$&$0$ \\
$\tilde{q}$&$-$&$0$&$0$&$0$ \\ 
$\psi_{q+}$&$+$&$-$&$0$&$-$ \\
$\psi_{q-}$&$-$&$+$&$-2$&$+$ \\ 
$\psi_{\tilde{q}+}$&$+$&$-$&$0$&$-$ \\ 
$\psi_{\tilde{q}-}$&$-$&$+$&$2$&$+$ \\ 
$\overline{\psi}_{q+}$&$-$&$-$&$0$&$-$ \\
$\overline{\psi}_{q-}$&$+$&$+$&$2$&$+$ \\ 
$\overline{\psi}_{\tilde{q}+}$&$-$&$-$&$0$&$-$ \\
$\overline{\psi}_{\tilde{q}-}$&$+$&$+$&$2$&$+$ \\ \hline
\end{tabular}
\caption{The spin of the fields for A-twisted and B-twisted 
$d=4$, $\mathcal{N}=2$ SCFT on $C\times \Sigma$.}
\label{4scftab}
\end{center}
\end{table}
If both $C$ and $\Sigma$ are curved, we should also twist
$U(1)_{\Sigma}$. 
Although there are many possibilities for twisting, we consider the
following cases
\begin{align}
\textrm{AA-twist}&: U(1)_{C}\rightarrow U(1)_{R}',
 U(1)_{\Sigma}\rightarrow U(1)_{R}'\\
\textrm{BA-twist}&: U(1)_{C}\rightarrow U(1)_{R},
 U(1)_{\Sigma}\rightarrow U(1)_{R}'\\
\textrm{BB-twist}&: U(1)_{C}\rightarrow U(1)_{R}, U(1)_{\Sigma}\rightarrow U(1)_{R}\\
\textrm{BA+-twist}&: U(1)_{C}\rightarrow U(1)_{R},
 U(1)_{\Sigma}\rightarrow U(1)_{R}'\times U(1)_{B}.
\end{align}
The results of twisting are given in Table \ref{4scftab2}.
\begin{table}
\begin{center}
\begin{tabular}{|c|c|c|c|c|c|c|c|c|} \hline
 &\multicolumn{2}{|c|}{AA-twist}&\multicolumn{2}{|c|}{BA-twist}&
\multicolumn{2}{|c|}{BB-twist}&\multicolumn{2}{|c|}{BA+-twist} \\ \hline
fields&$U(1)_{C}'$&$U(1)_{\Sigma}'$&$U(1)_{C}'$&$U(1)_{\Sigma}$&$U(1)_{C}'$&$U(1)_{\Sigma}'$&$U(1)_{C}'$&$U(1)_{\Sigma}'$ \\ \hline\hline
$\varphi$&$0$&$0$&$2$&$0$&$2$&$2$&$2$&$0$ \\ 
$\overline{\varphi}$&$0$&$0$&$-2$&$0$&$-2$&$-2$&$-2$&$0$ \\ 
$\psi_{+}$&$2$&$0$&$2$&$0$&$2$&$0$&$2$&$0$ \\
$\psi_{-}$&$0$&$2$&$0$&$2$&$0$&$2$&$0$&$2$ \\
$\lambda_{+}$&$0$&$-2$&$2$&$-2$&$2$&$0$&$2$&$-2$ \\
$\lambda_{-}$&$-2$&$0$&$0$&$0$&$0$&$2$&$0$&$0$ \\
$\overline{\psi}_{+}$&$-2$&$-2$&$-2$&$-2$&$-2$&$-2$&$-2$&$-2$ \\
$\overline{\psi}_{-}$&$0$&$0$&$0$&$0$&$0$&$0$&$0$&$0$ \\ 
$\overline{\lambda}_{+}$&$0$&$0$&$-2$&$0$&$-2$&$-2$&$-2$&$0$ \\
$\overline{\lambda}_{-}$&$2$&$2$&$0$&$2$&$0$&$0$&$0$&$2$ \\ 
$q$&$-$&$-$&$0$&$-$&$0$&$0$&$0$&$0$ \\
$\tilde{q}$&$-$&$-$&$0$&$-$&$0$&$0$&$0$&$0$ \\ 
$\psi_{q+}$&$+$&$-$&$0$&$-$&$0$&$-2$&$0$&$-2$ \\
$\psi_{q-}$&$-$&$+$&$-2$&$+$&$-2$&$0$&$-2$&$0$ \\ 
$\psi_{\tilde{q}+}$&$+$&$-$&$0$&$-$&$0$&$-2$&$0$&$0$ \\ 
$\psi_{\tilde{q}-}$&$-$&$+$&$-2$&$+$&$-2$&$0$&$-2$&$2$ \\ 
$\overline{\psi}_{q+}$&$-$&$-$&$0$&$-$&$0$&$0$&$0$&$0$ \\
$\overline{\psi}_{q-}$&$+$&$+$&$2$&$+$&$2$&$2$&$2$&$2$ \\ 
$\overline{\psi}_{\tilde{q}+}$&$-$&$-$&$0$&$-$&$0$&$0$&$0$&$-2$ \\
$\overline{\psi}_{\tilde{q}-}$&$+$&$+$&$2$&$+$&$2$&$2$&$2$&$0$ \\ \hline
\end{tabular}
\caption{The spin of the fields for AA, BA, BB-twisted 4d $\mathcal{N}=2$ SCFT on $C\times \Sigma$.}
\label{4scftab2}
\end{center}
\end{table}

\subsubsection{A-twist}
After twisting, the bosonic scalar fields
$\varphi,\overline{\varphi},q,\tilde{q}$ remain scalars and the
left-handed quarks $\psi_{q},\psi_{\tilde{q}}$ remain sections of 
\begin{align}
K_{C}^{\frac12}\otimes K_{\Sigma}^{-\frac12}
+K_{C}^{-\frac12}\otimes K_{\Sigma}^{\frac12}
\end{align}
and the right-handed quarks
$\overline{\psi}_{q},\overline{\psi}_{\tilde{q}}$ are sections of 
\begin{align}
K_{C}^{-\frac12}\otimes K_{\Sigma}^{-\frac12}
+K_{C}^{\frac12}\otimes K_{\Sigma}^{\frac12}.
\end{align}
On the other hand squarks $q,\tilde{q}$ become sections of
$K_{C}^{\frac12}$ and $K_{C}^{-\frac12}$. 
The gauginos $\psi$ reduces to sections of 
\begin{align}
\label{4scfta}
K_{C}\otimes K_{\Sigma}^{-\frac12}+\mathcal{O}_{C}\otimes K_{\Sigma}^{\frac12}
\end{align}
and $\lambda$ become sections of
\begin{align}
\label{4scftb}
\mathcal{O}_{C}\otimes K_{\Sigma}^{-\frac12}
+K_{C}^{-1}\otimes K_{\Sigma}^{\frac12}.
\end{align}
Their right-handed partners are sections of
\begin{align}
\label{4scftc}
&K_{C}^{-1}\otimes K_{\Sigma}^{-\frac12}
+\mathcal{O}_{C}\otimes K_{\Sigma}^{\frac12}\\
\label{4scftd}
&\mathcal{O}_{C}\otimes K_{\Sigma}^{-\frac12}
+K_{C}\otimes K_{\Sigma}^{\frac12}.
\end{align}

In the original theory, we have eight supercharges. 
Since the transformations of supercharges under R-symmetry are identical
to those of gauginos and only scalars on $C$ survive in the twisted
theory, (\ref{4scfta})-(\ref{4scftd}) shows that
there remains four supercharges
\begin{align}
&\mathcal{O}_{C}\otimes K_{\Sigma}^{\frac12},\ \ \ 
\mathcal{O}_{C}\otimes K_{\Sigma}^{-\frac12},\nonumber \\
&\mathcal{O}_{C}\otimes K_{\Sigma}^{\frac12},\ \ \ 
\mathcal{O}_{C}\otimes K_{\Sigma}^{-\frac12}.
\end{align}
Two of them transform as spinors of positive chirality on $\Sigma$ and
the other two transform as those of negative chirality on $\Sigma$. 
Therefore if one takes into account the dimensional reduction to
$\Sigma$, the theory on $\Sigma$ has $(2,2)$ supersymmetry.

\subsubsection{B-twist}
After the twisting the bosonic scalars $\varphi$ becomes section of $K_{C}$
and squarks $q,\tilde{q}$ are unchanged. 
The quarks $\psi_{q}$ and $\psi_{\tilde{q}}$ become sections of 
\begin{align}
&\mathcal{O}_{C}\otimes K_{\Sigma}^{-\frac12}
+K^{-1}_{C}\otimes K_{\Sigma}^{\frac12}\\
&\mathcal{O}_{C}\otimes K_{\Sigma}^{-\frac12}
+K_{C}\otimes K_{\Sigma}^{\frac12}.
\end{align} 
The gauginos $\psi$ and $\lambda$ are sections of
\begin{align}
\label{4scfte}
&K_{C}\otimes K_{\Sigma}^{-\frac12}
+\mathcal{O}_{C}\otimes K_{\Sigma}^{\frac12}\\
\label{4scftf}
&K_{C}\otimes K_{\Sigma}^{-\frac12}
+\mathcal{O}_{C}\otimes K_{\Sigma}^{\frac12}
\end{align}
and $\tilde{\psi}$ and $\tilde{\lambda}$ are sections of
\begin{align}
\label{4scftg}
&K_{C}^{-1}\otimes K_{\Sigma}^{-\frac12}
+\mathcal{O}_{C}\otimes K_{\Sigma}^{\frac12}\\
\label{4scfth}
&K_{C}^{-1}\otimes K_{\Sigma}^{-\frac12}
+\mathcal{O}_{C}\otimes K_{\Sigma}^{\frac12}.
\end{align}
From (\ref{4scfte})-(\ref{4scfth}), 
we see that there are four supercharges
\begin{align}
&\mathcal{O}_{C}\otimes K_{\Sigma}^{\frac12},\ \ \ 
\mathcal{O}_{C}\otimes K_{\Sigma}^{\frac12}\\
&\mathcal{O}_{C}\otimes K_{\Sigma}^{\frac12},\ \ \ 
\mathcal{O}_{C}\otimes K_{\Sigma}^{\frac12},
\end{align}
which transform as spinors of the positive chirality on $\Sigma$. 
Thus the theory on $\Sigma$ can have $(4,0)$ supersymmetry.

\subsubsection{AA-twist}
After twisting, we have
\begin{align}
&\varphi\in \Gamma(\mathcal{O}_{C}\otimes \mathcal{O}_{\Sigma})\\
&q\in \Gamma(K_{C}^{\frac12}\otimes K_{\Sigma}^{-\frac12}),\ \ \ 
\tilde{q}\in \Gamma(K_{C}^{-\frac12}\otimes K_{\Sigma}^{-\frac12})\\
&\psi_{q}\in \Gamma(K_{C}^{\frac12}\otimes K_{\Sigma}^{-\frac12}
+K_{C}^{-\frac12}\otimes K_{\Sigma}^{\frac12}),\ \ \ \psi_{\tilde{q}}\in (K_{C}^{\frac12}\otimes K_{\Sigma}^{-\frac12}
+K_{C}^{-\frac12}\otimes K_{\Sigma}^{\frac12})\\
&\overline{\psi}_{q}\in \Gamma(K_{C}^{-\frac12}\otimes K_{\Sigma}^{-\frac12}
+K_{C}^{\frac12}\otimes K_{\Sigma}^{\frac12}),\ \ \ \overline{\psi}_{\tilde{q}}\in \Gamma(K_{C}^{-\frac12}\otimes
 K_{\Sigma}^{-\frac12}
+K_{C}^{\frac12}\otimes K_{\Sigma}^{\frac12})\\
&\psi\in \Gamma(K_{C}\otimes \mathcal{O}_{\Sigma}+\mathcal{O}_{C}\otimes
 K_{\Sigma}),\ \ \ \lambda\in \Gamma(\mathcal{O}_{C}\otimes K_{\Sigma}^{-1}
+K_{C}^{-1}\otimes \mathcal{O}_{\Sigma})\\
&\overline{\psi}\in \Gamma(K_{C}^{-1}\otimes K_{\Sigma}^{-1}
+\mathcal{O}_{C}\otimes \mathcal{O}_{\Sigma}),\ \ \ \overline{\lambda}\in \Gamma(\mathcal{O}_{C}\otimes \mathcal{O}_{\Sigma}
+K_{C}\otimes K_{\Sigma}).
\end{align}
In other word, the fields transform under $U(1)_{C}'\times
U(1)_{\Sigma}'$ as
\begin{align}
&\varphi,\overline{\varphi}\rightarrow \bm{1}_{00}\oplus \bm{1}_{00}\\
&\psi,\lambda\rightarrow \bm{1}_{20}\oplus \bm{1}_{02}\oplus
 \bm{1}_{0-2}\oplus \bm{1}_{-20}\\
&\overline{\psi},\overline{\lambda}\rightarrow 
\bm{1}_{-2-2}\oplus \bm{1}_{00}\oplus \bm{1}_{00}\oplus \bm{1}_{22}\\
&q,\tilde{q}\rightarrow \bm{1}_{++}\oplus \bm{1}_{--}\\
&\psi_{q},\psi_{\tilde{q}}\rightarrow \bm{1}_{+-}\oplus \bm{1}_{-+}
\oplus \bm{1}_{+-}\oplus \bm{1}_{-+}\\
&\overline{\psi}_{q},\overline{\psi}_{\tilde{q}}
\rightarrow \bm{1}_{--}\oplus \bm{1}_{++}\oplus \bm{1}_{--}\oplus \bm{1}_{++}.
\end{align}
Therefore the bosonic field content is 
\begin{itemize}
 \item 2 scalar fields $\phi,\sigma$: $\bm{1}_{00}\oplus \bm{1}_{00}$
\item gauge fields $A_{z},A_{w}$: 
$\bm{1}_{20}\oplus \bm{1}_{-20}\oplus
      \bm{1}_{02}\oplus \bm{1}_{0-2}$
\item spinor fields $\tilde{q},\tilde{q}$: $\bm{1}_{++}\oplus \bm{1}_{--}$
\end{itemize}
and the fermionic field content is 
\begin{itemize}
 \item scalar field $\eta$: $\bm{1}_{00}$
\item 1-forms $\psi_{z},\psi_{w}$: $\bm{1}_{20}\oplus \bm{1}_{02}\oplus
      \bm{1}_{-02}\oplus \bm{1}_{-20}$
\item 2-form $\chi$: $\bm{1}_{-2-2}\oplus \bm{1}_{22}$
\item spinor fields $\psi_{q},\psi_{\tilde{q}}, \overline{\psi}_{q},
      \overline{\psi}_{\tilde{q}}$: $2 \left(\bm{1}_{+-}\oplus
      \bm{1}_{-+}\oplus \bm{1}_{--}\oplus \bm{1}_{++}\right)$
\end{itemize}
Focusing on the vector multiplet, one can check that the field content
is same as that of $\mathcal{N}=2$ twisted Donaldson-Witten theory as
expected. 

In the twisted theory we have two supercharges. 
Both of them are right-handed in 4-dimensions. 
Noting the spin of $U(1)_{C}\times U(1)_{\Sigma}$ for
$\overline{\psi}_{-}, \overline{\lambda}_{+}$
\begin{align}
&\overline{\psi}_{-}:(+,+)\xrightarrow{\textrm{AA-twist}}(0,0)\\
&\overline{\lambda}_{+}:(-,-)\xrightarrow{\textrm{AA-twist}}(0,0),
\end{align}
one can see that the two supercharges have the opposite chiralities on
both $C$ and $\Sigma$.

\subsubsection{BA-twist}
After twisting we obtain 
\begin{align}
&\varphi\in \Gamma(K_{C}\otimes \mathcal{O}_{\Sigma})\\
&q\in\Gamma(\mathcal{O}_{C}\otimes K_{\Sigma}^{-1}),\ \ \ \tilde{q}\in \Gamma(\mathcal{O}_{C}\otimes K_{\Sigma}^{-1})\\
&\psi_{q},\psi_{\tilde{q}}\in \Gamma(\mathcal{O}_{C}\otimes K_{\Sigma}^{-\frac12}
+K_{C}^{-1}\otimes K_{\Sigma}^{\frac12})\\
&\overline{\psi}_{q}, \overline{\psi}_{\tilde{q}}\in 
\Gamma(\mathcal{O}_{C}\otimes K_{\Sigma}^{-\frac12}
+K_{C}\otimes K_{\Sigma}^{\frac12})\\
&\psi\in \Gamma(K_{C}\otimes \mathcal{O}_{\Sigma}+\mathcal{O}_{C}\otimes
 K_{\Sigma}), \ \ \ \lambda\in \Gamma(K_{C}\otimes K_{\Sigma}^{-1}
+\mathcal{O}_{C}\otimes \mathcal{O}_{\Sigma})\\
&\overline{\psi}\in \Gamma(K_{C}^{-1}\otimes K_{\Sigma}^{-1}
+\mathcal{O}_{C}\otimes \mathcal{O}_{\Sigma}), \ \ \ \overline{\lambda}\in \Gamma(K^{-1}_{C}\otimes \mathcal{O}_{\Sigma}
+\mathcal{O}_{C}\otimes K_{\Sigma}).
\end{align}
Therefore there exists two supercharges. 
One is left-handed and the other is right-handed in 4-dimensions. 
From the fact
\begin{align}
&\lambda_{-}:(-,+)\rightarrow (0,0)\\
&\overline{\psi}_{-}:(+,+)\rightarrow (0,0),
\end{align}
it turns out that two supercharges have the same chirality on $\Sigma$
and the opposite chirality on $C$.

\subsection{$d=4$, $\mathcal{N}=4$ SYM theories}
\label{4dn4symsubsec1}
Let us start with $d=10$, $\mathcal{N}=1$ SYM theory. 
$d=4$, $\mathcal{N}=4$ SYM theory is most easily derived by dimensional
reduction from ten dimensions \footnote{
From Nahm's theorem \cite{Nahm:1977tg}, ten-dimensional is the maximum
possible dimension for SYM theory.
}.

The field content of $d=10$, $\mathcal{N}=1$ SYM theory is
\begin{itemize}
 \item gauge field $A$
\item 16 fermionic fields (gauginos) $\Psi$
\end{itemize}
The gauge field $A$ is a connection on a $G$-bundle $E$. The fermionic
field $\Psi$ is a positive chirality spinor field with values in the
adjoint representation of $G$, that is a section of $S^{+}\otimes
\textrm{ad}(E)$. We should note that the ten-dimensional spin representations
\begin{align}
\textrm{$\bm{16}_{s}$ and $\bm{16}_{c}$ are \ } 
\begin{cases}
\textrm{real and dual to each other}:&\textrm{Lorentz signature}\cr
\textrm{complex conjugate}:&\textrm{Euclidean signature}\cr
\end{cases}.
\end{align}

The gaugino $\Psi$ is a ten-dimensional positive chirality spinor field
\begin{align}
\label{n4chmx1}
\Gamma^{11}\Psi=\Psi
\end{align}
where 
\begin{align}
\Gamma^{11}:=i\Gamma^{12\cdots 10}.
\end{align}
The conjugate is given by
\begin{align}
\overline{\Psi}:=\Psi^{T}\mathcal{C}
\end{align}
where $\mathcal{C}$ is a ten-dimensional charge conjugation matrix
satisfying\footnote{Although there is another definition given by
\begin{align}
\label{10symchmtx}
\mathcal{C}^{T}=-\mathcal{C},\ \ \ \mathcal{C}\Gamma^{M}\mathcal{C}^{-1}=-\Gamma^{M}
\end{align}
which corresponds to $\mathcal{C}_{+}$, we choose $\mathcal{C}_{-}$ in (\ref{10symchmtx}). } 
\begin{align}
\mathcal{C}^{T}=\mathcal{C},\ \ \ \mathcal{C}\Gamma^{M}\mathcal{C}^{-1}=\Gamma^{M}.
\end{align}

The Lagrangian of Euclidean $d=10$, $\mathcal{N}=1$ SYM theory is given by
\begin{align}
\label{10dn1sym1}
\mathcal{L}=
\frac{1}{e^{2}}\textrm{Tr}
\left(\frac14 F_{MN}F^{MN}+\frac12
  \overline{\Psi}\Gamma^{M}D_{M}\Psi\right)
\end{align}
where $M,N,\cdots=1,2,\cdots, 10$ are indices of ten-dimensional
space-time and we define\footnote{
This is preferred convention in physics (see Table \ref{conn}).
}
\begin{align}
\label{10dn1sym2}
&D_{M}\Psi:=\partial_{M}\Psi-i[A_{M},\Psi],\\
&F_{MN}:=\partial_{M} A_{N}-\partial_{N}A_{M}
-i[A_{M},A_{N}].
\end{align}
\begin{table}
\begin{center}
\begin{tabular}{|c|c|} \hline
physics convention&mathmatics convention\\ \hline\hline
$D=d-iA'$ ($A'$:Hermitian)&$D=d+A$ ($A$:anti-Hermitian),\\
$F=i(D)^{2}=dA'-iA'\wedge A'$&$F=D^{2}=dA+A\wedge A$\\ \hline
\end{tabular}
\caption{The physical and mathematical definitions of connection and
 field strength. The relations are given by $-iF'=F, -iA'=A$. Although the anti-Hermitian fields $A$ are unnatural for
 $G=U(1)$, they may
 avoid unnatural factors $i$.}
\label{conn}
\end{center}
\end{table}

The 16 supersymmetries are
\begin{align}
&\delta A_{M}=\overline{\Psi}\Gamma_{M}\epsilon
=-\overline{\epsilon}\Gamma_{M}\Psi\\
&\delta\Psi=\frac12 F_{MN}\Gamma^{MN}\epsilon.
\end{align}

To consider the dimensional reduction of $d=10$, $\mathcal{N}=1$ SYM
theories to four dimensions, 
we decompose the gamma matrices under $SO(10)\supset SO(4)_{E}\times
SO(6)$ as 
\begin{align}
\begin{cases}
\Gamma^{\mu}=\gamma^{\mu}\otimes \hat{\Gamma}^{7}\\
\Gamma^{I}=\mathbb{I}_{4}\otimes \hat{\Gamma}^{I}\\
\end{cases}
\end{align}
where $\mu=1,2,3,4$ and $I=5,6,7,8,9,10$. 
$\hat{\Gamma}^{I}$ are six-dimensional gamma matrices satisfying
\begin{align}
\{\hat{\Gamma}^{I},\hat{\Gamma}^{J}\}=2\delta^{IJ},\ \ \ 
(\hat{\Gamma}^{I})^{\dag}=\Gamma^{I}
\end{align}
\begin{equation}
\hat{\Gamma}^{7}
=i\hat{\Gamma}^{12\cdots 6}=\left(
\begin{array}{cc}
\mathbb{I}_{4}&0\\
0&-\mathbb{I}_{4}\\
\end{array}
\right)
\end{equation}
and $\gamma^{\mu}$ are four-dimensional gamma matrices
\begin{align}
\label{4dgamma1}
\left\{\gamma^{\mu},\gamma^{\nu}\right\}=2\delta^{\mu\nu},\ \ \
 (\gamma^{\mu})^{\dag}=\gamma^{\mu},\ \ \ 
\gamma^{5}:=\gamma^{1\cdots 4}.
\end{align}
The charge conjugation matrix $\mathcal{C}$ 
and the chiral matrix are decomposed as
\begin{align}
\label{10dsymcdecom1}
&\mathcal{C}=C\otimes \hat{C},\\
\label{10dsymchdecom}
&\Gamma^{11}=\gamma^{5}\otimes \hat{\Gamma}^{7}
\end{align}
where $C$ is the four-dimensional charge conjugation matrix satisfying
\footnote{Here we choose $C$ as $C_{-}$.}
\begin{align}
C^{T}=-C,\ \ \ C\gamma^{\mu}C^{-1}=(\gamma^{\mu})^{T},\ \ \ C\gamma^{5}C^{-1}=(\gamma^{5})^{T}
\end{align}
and $\hat{C}$ is the six-dimensional charge conjugation matrix
\begin{align}
\hat{C}^{T}=-\hat{C},\ \ \ \hat{C}\hat{\Gamma}^{I}\hat{C}^{-1}=(\hat{\Gamma}^{I})^{T},\ \ \ 
\ \ \ 
C\hat{\Gamma}^{7}C^{-1}=-(\hat{\Gamma}^{7})^{T}
\end{align}

The global symmetry of the theory is $SU(4)_{R}$ R-symmetry. 
The field content is
\begin{itemize}
\item 6 real scalar fields $\phi^{I} (I=5,6,\cdots,10)$
\item 16 fermionic fields (gauginos) $\psi^{A} (A=1,2,3,4)$
\item gauge field $A_{\mu}$
\end{itemize}
where indices $I,J,\cdots$ and $A,B,\cdots$ are $\bm{6}$ and $\bm{4}$ of
$SU(4)_{R}$ R-symmetry.

Performing the dimensional reduction of (\ref{10dn1sym1}), 
we obtain the Lagrangian of
$d=4$, $\mathcal{N}=4$ SYM theories
\begin{align}
\label{4dn4symlag1}
\mathcal{L}
=\frac{1}{e^{2}}
\textrm{Tr}
\Biggl(
&\frac14 F_{\mu\nu}F^{\mu\nu}+\frac12 D_{\mu}\phi_{I}D^{\mu}\phi^{I}
-\frac14 [\phi_{I},\phi_{I}][\phi^{I},\phi^{I}]
&+\frac12 \overline{\psi}\Gamma^{\mu}D_{\mu}\psi
-\frac{i}{2}\overline{\psi}\Gamma^{I}[\phi_{I},\psi]
\Biggr)
\end{align}
where $\phi_{I}:=A_{I} (5\le I\le 10)$ and $\mu,\nu=1,2,3,4$. 
If $G$ is simple and if we require that Lagrangian is quadratic in
derivatives, the above Lagrangian is unique except for the change of
parameter $e$. 
However, we may have $\theta$-term that measures the topology of the
$G$-bundle $E$
\begin{align}
\mathcal{L}_{\theta}
=\frac{i\theta}{16\pi^{2}}
\textrm{Tr}(F\wedge F)
=\frac{i\theta}{16\pi^{2}}
\textrm{Tr}(*F_{\mu\nu}F^{\mu\nu}),
\end{align}
which is $\theta$ times the second Chern class or instanton number of
the bundle\footnote{In $\mathcal{N}=4$ SYM theories $\theta$-terms are
observable because there is no chiral anomaly and we cannot shift them. This situation is
different from $\mathcal{N}=2$ SYM.}. 
The parameter $e$ and $\theta$ combine into a complex parameter
\begin{align}
\tau=\frac{\theta}{2\pi}+\frac{4\pi i}{e^{2}}.
\end{align}

The Lagrangian (\ref{4dn4symlag1}) is invariant under 16 supersymmetries
\begin{align}
\label{4dn4symsusy1}
&\delta\phi_{I}=\overline{\psi}\Gamma_{I}\epsilon
=-\overline{\epsilon}\Gamma_{I}\psi,\\
\label{4dn4symsusy2}
&\delta A_{\mu}=\overline{\psi}\Gamma_{\mu}\epsilon
=-\overline{\epsilon}
\Gamma_{\mu}\psi,\\
\label{4dn4symsusy3}
&\delta\psi =\frac12 F_{\mu\nu}F^{\mu\nu}\epsilon
+D_{\mu}\Phi_{I}\Gamma^{\mu}\Gamma^{I}\epsilon
-\frac{i}{2}[\phi_{I},\phi_{J}]\Gamma^{IJ}\epsilon.
\end{align}

Before topological twisting, fields transform under
$SO(4)_{E}\times SU(4)_{R}\simeq SU(2)_{l}\times SU(2)_{r}\times SU(4)_{R}$ as
\begin{align}
 \phi:& (\bm{1},\bm{1},\bm{6}) \\
\psi:& (\bm{2},\bm{1},\overline{\bm{4}})\oplus (\bm{1},\bm{2},\bm{4})\\
A_{\mu}:&(\bm{2},\bm{2},\bm{1}).
\end{align}
To perform the fully topological twisting, we pick a homomorphism
$\pi:SO(4)_{E}\rightarrow SU(4)_{R}$ and replace $SO(4)_{E}$ by $SO(4)'_{E}=(1+\pi)(SO(4)_{E})\subset SO(4)_{E}\times
SO(6)_{R}$.

The choice of $\pi$ amounts to embedding $SO(4)_{E}\simeq SU(2)_{l}\times SU(2)_{r}$ in
$SU(4)_{R}$ as
\begin{equation}
\pi:SU(2)_{l}\times SU(2)_{r}\rightarrow \left(
\begin{array}{cc}
SU(2)_{l}&0\\
0&SU(2)_{r}\\
\end{array}
\right),
\end{equation}
which leads us to consider the decomposition
\begin{equation}
 \label{sym4decom}
SU(4)\supset SU(2)\times SU(2)\times U(1).
\end{equation}
Under (\ref{sym4decom}), we still have several possible embedding 
determined by telling how the $\bm{4}$ of $SU(4)_{R}$ transforms under
$SU(2)_{l}\times SU(2)_{r}$. 
Up to an exchange of left and right, there are three inequivalent
transformations of $\bm{4}$ of $SU(4)_{R}$ under (\ref{sym4decom}).
\begin{align}
\label{4dn4twist}
\begin{array}{ccl}
(\textrm{i})&\textrm{GL twist}&\bm{4}=(\bm{2},\bm{1})_{1}\oplus (\bm{1},\bm{2})_{-1}\\
(\textrm{ii})&\textrm{VW twist}&\bm{4}=(\bm{1},\bm{2})_{1}\oplus (\bm{1},\bm{2})_{-1}\\
(\textrm{iii})&\textrm{DW twist}&\bm{4}=(\bm{1},\bm{2})_{0}\oplus
  (\bm{1},\bm{1})_{1}\oplus (\bm{1},\bm{1})_{-1}\\
\end{array}
\end{align}

\subsubsection{Geometric Langlands (GL) twist}
GL twist has the branching
\begin{align}
 \bm{4}&=(\bm{2},\bm{1})_{1}\oplus (\bm{1},\bm{2})_{-1}\nonumber \\
\overline{\bm{4}}&=(\bm{2},\bm{1})_{-1}\oplus (\bm{1},\bm{2})_{1}.
\end{align}

\begin{enumerate}
 \item \textbf{fermionic fields}

Noting that
\begin{align}
(\bm{2},\bm{1})_{0}\times
\left((\bm{2},\bm{1})_{-1}\oplus (\bm{1},\bm{2})_{1}\right)=(\bm{1},\bm{1})_{-1}\oplus (\bm{3},\bm{1})_{-1}\oplus (\bm{2},\bm{2})_{1}
\end{align}
and 
\begin{align}
(\bm{1},\bm{2})_{0}
\times \left(
(\bm{2},\bm{1})_{1}\oplus (\bm{1},\bm{2}_{-1})
\right)
=(\bm{2},\bm{2})_{1}\oplus (\bm{1},\bm{1})_{-1}\oplus (\bm{1},\bm{3})_{-1},
\end{align}
one can see that the fermionic fields transform under $SU(2)_{l}'\times SU(2)_{r}'\times U(1)$ as 
\begin{equation}
\label{glfermion}
 (\bm{1},\bm{1})_{-1}\oplus (\bm{3},\bm{1})_{-1}
\oplus (\bm{2},\bm{2})_{1}
\oplus (\bm{2},\bm{2})_{1}
\oplus (\bm{1},\bm{1})_{-1}
\oplus (\bm{1},\bm{3})_{-1}.
\end{equation}
Similarly one can obtain the transformations of supersymmetries. 
Thus, from (\ref{glfermion}) we see that GL twist leads to two unbroken
       BRST charges which have the same $U(1)$ charge.

\item \textbf{bosonic fields}

The bosonic scalar field $\bm{6}_{v}$ of $SO(6)_{R}$ is produced by the
      product of $SO(6)$ spinor $\bm{8}=\bm{4}+\overline{\bm{4}}$ as
\begin{align}
\bm{8}\times \bm{8}
=&(\bm{4}+\overline{\bm{4}})\times (\bm{4}+\overline{\bm{4}})\nonumber \\
=&\bm{4}\times \bm{4}+\bm{4}\times \overline{\bm{4}}
+\overline{\bm{4}}\times \bm{4}+\overline{\bm{4}}\times
 \overline{\bm{4}}\nonumber \\
=&([1]+[3])+([0]+[2])+([0]+[2])+([1]+[3])\nonumber \\
=&(\bm{6}+\bm{10})
+(\bm{1}+\bm{15})
+(\bm{1}+\bm{15})
+(\bm{6}+\bm{10}).
\end{align}
Note that $\bm{6}_{v}$ is the antisymmetric product of $\bm{4}$
\begin{align}
\label{glboson}
 \bm{6}_{v}=&(\bm{4}+\bm{4})_{a}\nonumber \\
=&\left(
\left((\bm{2},\bm{1})_{1}\oplus (\bm{1},\bm{2})_{-1}\right)
\otimes \left(
(\bm{2},\bm{1})_{1}\oplus (\bm{1},\bm{2})_{-1}
\right)
\right)_{a}\nonumber \\
=&\left(
\left(
(\bm{1},\bm{1})_{2}\oplus (\bm{1},\bm{1})_{-2}
\oplus (\bm{2},\bm{2})_{0}
\right)
\oplus \left(
(\bm{3},\bm{1})_{2}\oplus (\bm{2},\bm{2})_{0}
\oplus (\bm{1},\bm{3})_{-2}
\right)
\right)_{a}\nonumber \\
=&(\bm{1},\bm{1})_{2}\oplus (\bm{1},\bm{1})_{-2}
\oplus (\bm{2},\bm{2})_{0}.
\end{align}
Thus the bosonic field content consists of 
\begin{itemize}
 \item 2 scalar fields : $(\bm{1},\bm{1})_{2}\oplus (\bm{1},\bm{1})_{-2}$
\item 1-form : $(\bm{2},\bm{2})_{0}$
\item gauge field : $(\bm{2},\bm{2})_{0}$
\end{itemize}
and the fermionic field content is 
\begin{itemize}
 \item 2 scalar fields : $(\bm{1},\bm{1})_{-1}\oplus
       (\bm{1},\bm{1})_{-1}$
\item 2 1-forms : $(\bm{2},\bm{2})_{1}\oplus (\bm{2},\bm{2})_{1}$
\item 2 2-forms \footnote{Note that this is not self-dual.} : $(\bm{3},\bm{1})_{-1}\oplus (\bm{1},\bm{3})_{-1}$.
\end{itemize}

\end{enumerate}

Under decomposition $SO(10)\supset SO(4)_{E}\times SO(4)\times
SO(2)$, the decomposition of the ten-dimensional gamma matrices is given by
\begin{align}
\label{4dsymmtx1}
\begin{cases}
\Gamma^{\mu}=\gamma^{\mu}\otimes \gamma^{5}\otimes -\sigma_{3}\cr
\Gamma^{\mu+4}=\mathbb{I}_{4}\otimes \gamma^{\mu}\otimes \mathbb{I}_{2}\cr
\Gamma^{i+8}=\mathbb{I}_{4}\otimes \gamma^{5}\otimes \sigma_{i}\cr
\end{cases}
\end{align}
where $\mu=1,2,3,4$ and $i=1,2$. $\sigma_{i}$ are Pauli matrices and
$\gamma^{\mu}$ are four-dimensional gamma matrices defined by
(\ref{4dgamma1}). 
The charge conjugation matrix $\mathcal{C}$ is decomposed as
\begin{align}
\label{4dsymcdecom1}
\mathcal{C}=C\otimes C\otimes \sigma_{1}
\end{align}

Under the decomposition (\ref{4dsymmtx1}), ten-dimensional chirality
matrix is expressed as
\begin{align}
\label{4dn4chiralmtxdecom}
\Gamma^{11}=-\left(
\gamma^{5}\otimes \gamma^{5}\otimes \sigma_{3}\right).
\end{align}

\begin{enumerate}
 \item \textbf{bosonic fields}

For the bosonic fields we redefine
\begin{align}
&\Phi_{\mu}:=\phi_{4+\mu},\ \ \ (\mu=1,2,3,4),\\
&A_{\mu}^{\pm}:=\frac{1}{\sqrt{2}}(A_{\mu}\pm i\Phi_{\mu}),\\
&\varphi:=\frac{1}{\sqrt{2}}(\phi_{9}+i\phi_{10}),\ \ \ \overline{\varphi}:=\frac{1}{\sqrt{2}}
(\phi_{9}-i\phi_{10}).
\end{align}

\item \textbf{fermionic fields}

Noting the fermionic field content of GL twist, one can decompose the
      10-dimensional fermionic field $\Psi$ under the decomposition 
$SO(10)\supset SO(4)_{E}\times SO(4)\times SO(2)$ as
\footnote{The inverse of charge conjugation matrices $C^{-1}$ is
      included just because of the convenience of the calculation.}
\begin{align}
\Psi_{pq\alpha}
=\frac{1}{\sqrt{2}}
\left(
\eta_{\alpha}\mathbb{I}_{4 pq}
+\psi_{\mu\alpha}\gamma^{\mu}_{pq}
+\frac12 \chi_{\mu\nu\alpha}\gamma^{\mu\nu}_{pq}
+\omega_{\mu \alpha}\gamma^{\mu5}_{pq}
+\zeta_{\alpha}\gamma^{5}_{pq}
\right)C^{-1}
\end{align}
where $p,q$ and $\alpha$ are indices of $SO(4)_{E}, SO(4)$ and $SO(2)$
respectively. 
$\eta_{\alpha},\zeta_{\alpha}$ are scalars $(\bm{1},\bm{1})_{-}\oplus (\bm{1},\bm{1})_{-}$ and $\psi_{\mu\alpha},
\omega_{\mu\alpha}$ are 1-forms $(\bm{2},\bm{2})_{+}\oplus (\bm{2},\bm{2})_{+}$ and $\chi_{\mu\nu}=-\chi_{\nu\mu}$ is a
2-form $(\bm{3},\bm{1})_{-}\oplus (\bm{1},\bm{3})_{-}$.

From the decompositions (\ref{4dsymmtx1}) and the
chirality condition (\ref{n4chmx1}) for the femionic fields, we see that
\begin{align}
&\sigma_{3}\eta=\eta,\ \ \ \sigma_{3}\psi_{\mu}=\psi_{\mu},\\
&\sigma_{3}\chi_{\mu\nu}=-\chi_{\mu\nu},\ \ \ \sigma_{3}\omega_{\mu}=\omega_{\mu},\\
&\sigma_{3}\zeta=-\zeta
\end{align}
Therefore $\psi_{\mu}, \omega_{\mu}$ have $U(1)$ ghost charge $+1$ and 
$\eta, \chi_{\mu\nu}, \zeta$ have $U(1)$ charge $-1$. This is consistent
      to the results of the fermionic field content for GL twist.

\item \textbf{supersymmetries}

For the sypersymmetries we can also expand as
\begin{align}
\epsilon_{pq\alpha}
=\frac{1}{\sqrt{2}}\left(
\epsilon_{\alpha}\mathbb{I}_{4 pq}
+\epsilon_{\mu\alpha}\gamma^{\mu}_{pq}
+\frac12 \epsilon_{\mu\nu\alpha}\gamma^{\mu\nu}_{pq}
+\tilde{\epsilon}_{\mu\alpha}\gamma^{\mu5}
+\tilde{\epsilon}_{\alpha}\gamma^{5}_{pq}
\right)C^{-1}.
\end{align}

From the decompositions (\ref{4dsymmtx1}) and the
chirality condition for the supersymmetries, we see that
\begin{align}
&\sigma_{3}\epsilon=-\epsilon,\ \ \ \sigma_{3}\tilde{\epsilon}=-\tilde{\epsilon}
\end{align}
Therefore both BRST charges $\epsilon$ and $\tilde{\epsilon}$ have
      $U(1)$ charge $-1$ as expected.

\end{enumerate}

The BRST transformation is given by
\begin{align}
\label{4d4a}
&\delta
A_{\mu}=-2(\overline{\epsilon}\omega^{\mu})-2(\overline{\tilde{\epsilon}}\psi^{\mu}),\\
\label{4d4b}
&\delta\Phi_{\mu}
=-2i(\overline{\epsilon}\psi^{\mu})
-2i(\overline{\tilde{\epsilon}}\omega^{\mu}),\\
\label{4d4c}
&\delta\varphi=2i\overline{\epsilon}\sigma_{+}\zeta
+2i\overline{\tilde{\epsilon}}\sigma_{+}\eta,\\
\label{4d4d}
&\delta\tilde{\varphi}=2i\overline{\epsilon}\sigma_{-}\zeta
+2i\overline{\tilde{\epsilon}}\sigma_{-}\eta,\\
\label{4d4e}
&\delta\eta=iD_{\mu}\Phi^{\mu}\tilde{\epsilon}
-\frac12[\phi^{i},\phi^{j}](\epsilon_{ij}\epsilon),\\
\label{4d4f}
&\delta\psi_{\mu}=-iD_{\mu}\phi_{i}(\sigma_{i}\epsilon)
-i[\Phi_{\mu},\phi_{i}](\sigma_{i}\tilde{\epsilon}),\\
\label{4d4g}
&\delta\chi_{\mu\nu}=-F_{\mu\nu}\epsilon
+\epsilon_{\mu\nu\rho\sigma}F^{\rho\sigma}\tilde{\epsilon}
+2i\epsilon_{\mu\nu\rho\sigma}D^{\rho}\Phi^{\sigma}\epsilon
+2iD_{\mu}\Phi_{\nu}\tilde{\epsilon} \nonumber \\
&\ \ \ \ \ \ 
+i[\Phi_{\mu},\Phi_{\nu}]\epsilon-i\epsilon_{\mu\nu\rho\sigma}[\Phi^{\rho},\Phi^{\sigma}]\tilde{\epsilon},\\
\label{4d4h}
&\delta\omega_{\mu}
=-iD_{\mu}\phi_{i}(\sigma_{i}\tilde{\epsilon})
+i[\Phi_{\mu},\phi_{i}](\sigma_{i}\epsilon),\\
\label{4d4i}
&\delta\zeta=-iD_{\mu}\Phi^{\mu}\epsilon
-\frac12 (\epsilon_{ij}\tilde{\epsilon})[\phi_{i},\phi_{j}]
\end{align}
where we introduce
\begin{align}
&\sigma_{+}:=\frac{1}{\sqrt{2}}(\sigma_{1}+i\sigma_{2}),\\
&\sigma_{-}:=\frac{1}{\sqrt{2}}(\sigma_{1}-i\sigma_{2}).
\end{align}

\subsubsection{Vafa-Witten (VW) twist}
VW twist corresponds to the following branching:
\begin{align}
\label{vwtwist001}
\bm{4}&=(\bm{1},\bm{2})_{1}\oplus (\bm{1},\bm{2})_{-1}\nonumber \\
\overline{\bm{4}}&=(\bm{1},\bm{2})_{-1}\oplus (\bm{1},\bm{2})_{1}.
\end{align}

\begin{enumerate}
 \item \textbf{fermionic fields}

Noting that
\begin{equation}
 (\bm{2},\bm{1})_{0}\times \left((\bm{1},\bm{2})_{-1}\oplus
  (\bm{1},\bm{2})_{1}\right)
=(\bm{2},\bm{2})_{-1}\oplus (\bm{2},\bm{2})_{1}
\end{equation}
and 
\begin{equation}
 (\bm{1},\bm{2})_{0}\times \left(
(\bm{1},\bm{2})_{1}\oplus (\bm{1},\bm{2})_{-1}
\right)
=(\bm{1},\bm{1})_{1}
\oplus (\bm{1},\bm{3})_{1}
\oplus (\bm{1},\bm{1})_{-1}
\oplus (\bm{1},\bm{3})_{-1},
\end{equation}
it turns out that the fermionic fields transform under $SU(2)_{l}'\times SU(2)_{r}'\times
U(1)$ as
\begin{equation}
 \label{vwfermion}
(\bm{2},\bm{2})_{-1}\oplus (\bm{2},\bm{2})_{1}
\oplus (\bm{1},\bm{1})_{1}
\oplus (\bm{1},\bm{3})_{1}
\oplus (\bm{1},\bm{1})_{-1}
\oplus (\bm{1},\bm{3})_{-1}.
\end{equation}
Therefore VW twist gives rise to two unbroken BRST charges which have
       the opposite $U(1)$ charge.

\item \textbf{bosonic fields}

The bosonic scalar field $\bm{6}_{v}$ of $SO(6)_{R}$ is given by the
      antisymmetric product of $\bm{4}$
\begin{align}
\label{vwboson}
\bm{6}_{v}
=&(\bm{4}\times \bm{4})_{a}\nonumber \\
=&\left(\left(
(\bm{1},\bm{2})_{1}\oplus (\bm{1},\bm{2})_{-1}
\right)
\times \left(
(\bm{1},\bm{2})_{1}\oplus (\bm{1},\bm{2})_{-1}
\right)\right)_{a}\nonumber \\
=&\left(
\left(
(\bm{1},\bm{1})_{0}
\oplus (\bm{1},\bm{3})_{0}
\oplus (\bm{1},\bm{1})_{2}
\oplus (\bm{1},\bm{1})_{-2}
\right)\oplus 
\left(
(\bm{1},\bm{3})_{2}
\oplus (\bm{1},\bm{1})_{0}
\oplus (\bm{1},\bm{3})_{0}
\oplus (\bm{1},\bm{3})_{-2}
\right)
\right)_{a}\nonumber \\
=&(\bm{1},\bm{1})_{0}
\oplus (\bm{1},\bm{3})_{0}
\oplus (\bm{1},\bm{1})_{2}
\oplus (\bm{1},\bm{1})_{-2}.
\end{align}
Thus the bosonic field content consists of 
\begin{itemize}
 \item 3 scalar fields : $(\bm{1},\bm{1})_{-2}\oplus (\bm{1},\bm{1})_{0}\oplus (\bm{1},\bm{1})_{2}$
\item 2-form : $(\bm{1},\bm{3})_{0}$
\item gauge field : $(\bm{2},\bm{2})_{0}$
\end{itemize}
and the fermionic one is
\begin{itemize}
 \item 2 scalar fields : $(\bm{1},\bm{1})_{+}\oplus (\bm{1},\bm{1})_{-}$
\item 1-form : $(\bm{2},\bm{2})_{-}\oplus (\bm{2},\bm{2})_{+}$
\item 2 2-from : $(\bm{1},\bm{3})_{-}\oplus (\bm{1},\bm{3})_{+}$.
\end{itemize}

\end{enumerate}

\subsubsection{Donaldson-Witten (DW) twist}
DW twist has the branching
\begin{align}
\bm{4}&=(\bm{1},\bm{2})_{0}\oplus (\bm{1},\bm{1})_{1}\oplus (\bm{1},\bm{1})_{-1}\nonumber \\
\overline{\bm{4}}&=(\bm{1},\bm{2})_{0}\oplus (\bm{1},\bm{1})_{-1}\oplus (\bm{1},\bm{1})_{1}.
\end{align}

\begin{enumerate}
 \item \textbf{fermionic fields}

Noticing that
\begin{equation}
 (\bm{2},\bm{1})_{0}\times 
\left(
(\bm{1},\bm{2})_{0}\oplus (\bm{1},\bm{1})_{-1}
\oplus (\bm{1},\bm{1})_{1}
\right)
=(\bm{2},\bm{2})_{0}
\oplus (\bm{2},\bm{1})_{-1}
\oplus (\bm{2},\bm{1})_{1}
\end{equation}
and
\begin{equation}
 (\bm{1},\bm{2})_{0}
\times \left(
(\bm{1},\bm{2})_{0}\oplus (\bm{1},\bm{1})_{1}
\oplus (\bm{1},\bm{1})_{-1}
\right)
=(\bm{1},\bm{1})_{0}
\oplus (\bm{1},\bm{3})_{0}
\oplus (\bm{1},\bm{2})_{1}
\oplus (\bm{1},\bm{2})_{-1},
\end{equation}
one can see that fermionic fields transform under 
$SU(2)_{l}'\times SU(2)_{r}'\times U(1)$ as
\begin{equation}
\label{dwfermion}
 (\bm{2},\bm{2})_{0}
\oplus (\bm{2},\bm{1})_{-1}
\oplus (\bm{2},\bm{1})_{1}
\oplus (\bm{1},\bm{1})_{0}
\oplus (\bm{2},\bm{3})_{0}
\oplus (\bm{1},\bm{2})_{1}
\oplus (\bm{1},\bm{2})_{-1},
\end{equation}
which implies that DW twist allows one unbroken BRST charge.

\item \textbf{bosonic fields}

The bosonic scalar field $\bm{6}_{v}$ of $SO(6)_{R}$ is given by the
      antisymmetric product of $\bm{4}$
\begin{align}
\label{dwboson}
\bm{6}_{v}
=&(\bm{4}\times \bm{4})_{a}\nonumber \\
=&\left(\left(
(\bm{1},\bm{2})_{0}\oplus (\bm{1},\bm{1})_{1}
\oplus (\bm{1},\bm{1})_{-1}
\right)
\times \left(
(\bm{1},\bm{2})_{0}\oplus (\bm{1},\bm{1})_{1}
\oplus (\bm{1},\bm{1})_{-1}
\right)\right)_{a}\nonumber \\
=&\left(
\left(
2(\bm{1},\bm{1})_{0}
\oplus (\bm{1},\bm{2})_{1}
\oplus (\bm{1},\bm{2})_{-1}
\right)\oplus 
\left(
(\bm{1},\bm{1})_{0}
\oplus (\bm{1},\bm{3})_{0}
\oplus (\bm{1},\bm{2})_{1}
\oplus (\bm{1},\bm{2})_{-1}
\oplus (\bm{1},\bm{1})_{2}
\oplus (\bm{1},\bm{1})_{-2}
\right)
\right)_{a}\nonumber \\
=&2(\bm{1},\bm{1})_{0}
\oplus (\bm{1},\bm{2})_{1}
\oplus (\bm{1},\bm{2})_{-1}.
\end{align}
Thus the bosonic field content consists of 
\begin{itemize}
 \item 2 scalar fields : $(\bm{1},\bm{1})_{0}\oplus (\bm{1},\bm{1})_{0}$

\item 2 spinor fields : $(\bm{1},\bm{2})_{1}\oplus (\bm{1},\bm{2})_{-1}$

\item gauge field : $(\bm{2},\bm{2})_{0}$
\end{itemize}

\end{enumerate}

\subsection{$d=4$, $\mathcal{N}=4$ SYM theories on $C\times \Sigma$}
Now we discuss a $d=4$, $\mathcal{N}=4$ SYM theory on $M_{4}=C\times
\Sigma$. 
We consider the twisting by using the embedding
\begin{align}
U(1)_{C}\rightarrow U(1)_{R}.
\end{align}

The assignment of $U(1)$ charges are given in Table \ref{vwp1} where
$+,-$ signs denote upper and lower components of spinors and
right-handed fermions indicated with bars. 
\begin{table}
\begin{center}
\begin{tabular}{|c|c|c|c|c|c|} \hline
&$U(1)_{C}$&$U(1)_{\Sigma}$&$U(1)_{R}$&$U(1)_{C}'$&$U(1)_{\Sigma}'$\\ \hline\hline
$\phi^{1}$&$0$&$0$&$0$&$0$&$0$\\ 
$\phi^{2}$&$0$&$0$&$0$&$0$&$0$\\
$\phi^{3}$&$0$&$0$&$0$&$0$&$0$\\
$\phi^{4}$&$0$&$0$&$0$&$0$&$0$\\
$\phi^{5}$&$0$&$0$&$0$&$2$&$2$\\
$\phi^{6}$&$0$&$0$&$0$&$-2$&$-2$\\ \hline
$\psi^{1}_{+}$&$+$&$-$&$+$&$2$&$-$\\
$\psi^{2}_{+}$&$+$&$-$&$+$&$2$&$-$\\
$\psi^{3}_{+}$&$+$&$-$&$+$&$0$&$-$\\
$\psi^{4}_{+}$&$+$&$-$&$+$&$0$&$-$\\
$\psi^{1}_{-}$&$-$&$+$&$+$&$0$&$+$\\
$\psi^{2}_{-}$&$-$&$+$&$+$&$0$&$+$\\
$\psi^{3}_{-}$&$-$&$+$&$-$&$-2$&$+$\\
$\psi^{4}_{-}$&$-$&$+$&$-$&$-2$&$+$\\ \hline
$\overline{\psi}^{1}_{+}$&$-$&$-$&$-$&$0$&$-$\\
$\overline{\psi}^{2}_{+}$&$-$&$-$&$-$&$0$&$-$\\
$\overline{\psi}^{3}_{+}$&$-$&$-$&$-$&$-2$&$-$\\
$\overline{\psi}^{4}_{+}$&$-$&$-$&$-$&$-2$&$-$\\
$\overline{\psi}^{1}_{-}$&$+$&$+$&$+$&$2$&$+$\\
$\overline{\psi}^{2}_{-}$&$+$&$+$&$+$&$2$&$+$\\
$\overline{\psi}^{3}_{-}$&$+$&$-$&$+$&$0$&$+$\\
$\overline{\psi}^{4}_{-}$&$+$&$-$&$+$&$0$&$+$\\ \hline
\end{tabular}
\caption{$U(1)$ charges for VW partial twisting.}
\label{vwp1}
\end{center}
\end{table}
Under $U(1)_{C}'\times U(1)_{\Sigma}'\times SU(2)_{2}$ the fields
transform as
\begin{align}
&\phi\rightarrow \bm{1}_{00}\oplus \bm{3}_{00}\oplus \bm{1}_{20}\oplus \bm{1}_{-20}\\
&\psi\rightarrow \bm{2}_{2-}\oplus \bm{2}_{0-}\oplus \bm{2}_{0+}
\oplus \bm{2}_{-2+}\\
&\overline{\psi}\rightarrow \bm{2}_{0-}\oplus \bm{2}_{-2-}
\oplus \bm{2}_{2+}\oplus \bm{2}_{0+}\\
&A_{\mu}\rightarrow \bm{1}_{20}\oplus \bm{1}_{02}\oplus \bm{1}_{0-2}.
\end{align}
The bosonic field content is
\begin{itemize}
 \item 2 complex scalar fields $\phi,\overline{\phi}$: $\bm{1}_{00}\oplus\bm{3}_{00}$
\item 1-form $\Phi_{w},\Phi_{\overline{w}}$ : $\bm{1}_{20}\oplus
      \bm{1}_{-20}$
\item gauge field $A_{z},A_{\overline{z}},A_{w},A_{\overline{w}}$ :
      $\bm{1}_{20}\oplus \bm{1}_{-20}\oplus \bm{1}_{02}\oplus \bm{1}_{0-2}$
\end{itemize}
and the fermionic field content is
\begin{itemize}
 \item 8 scalar fields $\eta_{-}^{a}, \chi_{+}^{a},
       \overline{\eta}_{-}^{a}, \overline{\chi}_{+}^{a}$ :
       $\bm{2}_{0-}\oplus \bm{2}_{0+}\oplus \bm{2}_{0-}\oplus
       \bm{2}_{0+}$
\item 1-form
      $\lambda_{w-}^{a},\chi_{\overline{w}+}^{a},\overline{\lambda}_{\overline{w}}^{a},\overline{\chi}_{w+}^{a}$
      : $\bm{2}_{2-}\oplus \bm{2}_{-2+}\oplus\bm{2}_{-2-}\oplus \bm{2}_{2+}$ 
\end{itemize}
where $a=1,2$ are the indices of the fundamental representations of the
unbroken $SU(2)_{2}$ symmetry.

One can see that there are eight BRST charges. 
Four of them transform as spinors with positive chirality on $\Sigma$
corresponding to $\chi_{+}^{a}$ and $\overline{\chi}_{+}^{a}$ and the
others, that is $\eta^{a}_{-}$ and $\overline{\eta}^{a}_{-}$ transform
as those with negative chirality on $\Sigma$. 
Therefore we can regard the theory on $\Sigma$ has $(4,4)$
supersymmetry.

\subsection{$d=3$, $\mathcal{N}=4$ SYM theories}
The global symmetry of the theory is $SO(4)\simeq SU(2)_{1}\times SU(2)_{2}$ R-symmetry. 
In order to understand the symmetry, it is convenient to construct
$d=3$, $\mathcal{N}=4$ theories by dimensional reduction 
from $d=6$, $\mathcal{N}=1$ supersymmetric gauge theories. 
$SU(2)_{1}$ is the double cover of rotational symmetry $SO(3)$ in the
three reduced coordinates and $SU(2)_{2}$ is the R-symmetry in
six-dimensional $\mathcal{N}=1$ SYM theories \cite{Seiberg:1996nz,
Hanany:1996ie, deBoer:1996mp, Porrati:1996xi, Chalmers:1996xh}.

The field content is
\begin{itemize}
 \item 3 scalar fields $\phi^{i}$
\item fermionic field $\psi$
\item gauge fields $A_{\mu}$
\end{itemize}

Before topological twisting, fields transform under $SU(2)_{E}\times
SU(2)_{1}\times SU(2)_{2}$ as
\begin{align}
 \phi:& (\bm{1},\bm{3},\bm{1}) \\
\psi:& (\bm{2},\bm{2},\bm{2})\\
A_{\mu}:&(\bm{3},\bm{1},\bm{1}).
\end{align}
To perform the fully topological twisting, we pick a homomorphism
$\pi:SU(2)_{E}\rightarrow SU(2)_{1}\times SU(2)_{2}$ and replace $SU(2)_{E}$ by $SU(2)'_{E}=(1+\pi)(SU(2)_{E})\subset SU(2)_{E}\times
SU(2)_{1}\times SU(2)_{2}$.

We many have two choices of $\pi$ as
\[
 \begin{cases}
(\textrm{i}) \textrm{A-twist} &SU(2)_{E}\rightarrow SU(2)_{2}\cr
(\textrm{ii}) \textrm{B-twist} &SU(2)_{E}\rightarrow SU(2)_{1}\cr
\end{cases}.
\]

\subsubsection{A-twist}
After twisting, the fields transform under the new symmetry
$SU(2)_{E}'\times SU(2)_{1}$ as
\begin{align}
\label{3dn4symtwist1}
\phi\rightarrow& (\bm{1},\bm{3}) \\
\label{3dn4symtwist2}
 \psi\rightarrow& (\bm{1},\bm{2})\oplus (\bm{3},\bm{2}) \\
\label{3dn4symtwist3}
A_{\mu}\rightarrow&(\bm{3},\bm{1}).
\end{align}
There are two BRST charges and this is just the dimensional reduction of Donaldson-Witten theory
(twisted $d=4$, $\mathcal{N}=2$ theory).
The field content is same as $d=3$ super BF model \cite{Witten:1989sx,
Birmingham:1989is, Birmingham:1991ty, Blau:1991bn} associated with the
Casson invariant.

\subsubsection{B-twist}
After twisting, the fields transform under the new symmetry
$SU(2)_{E}'\times SU(2)_{2}$ as
\begin{align}
\label{3dn4symtwist1}
\phi\rightarrow& (\bm{3},\bm{1}) \\
\label{3dn4symtwist2}
 \psi\rightarrow& (\bm{1},\bm{2})\oplus (\bm{3},\bm{2}) \\
\label{3dn4symtwist3}
A_{\mu}\rightarrow&(\bm{3},\bm{1}).
\end{align}
In B-twist we also have two BRST charges. 
As B-twist is related A-twist under the exchange of $SU(2)_{1}$ and
$SU(2)_{2}$, it is expected that B-twist may be regarded as a mirror
discription of the Casson invariant because $d=3$, $\mathcal{N}=4$ mirror
symmetry has mirror pair under this exchange.

\subsection{$d=3$, $\mathcal{N}=8$ SYM theories}
\label{3dn8symsubsec1}
The global symmetry of the theory is $Spin(7)_{R}$ R-symmetry. 
If we construct the theories by the dimensional reduction of $d=10$, 
$\mathcal{N}=1$ SYM theories, this is recognized as double cover of
rotational symmetry $SO(7)$ in the seven reduced coordinates. 
The field content is
\begin{itemize}
 \item 7 scalar fields $\phi^{i}$
\item fermionic field $\psi$
\item gauge fields $A_{\mu}$
\end{itemize}

Before topological twisting, fields transform under $SU(2)_{E}\times
Spin(7)_{R}$ as 
\begin{align}
 \phi:& (\bm{1},\bm{7}) \\
\psi:& (\bm{2},\bm{8})\\
A_{\mu}:&(\bm{3},\bm{1}).
\end{align}
To perform the fully topological twisting, we pick a homomorphism
$\pi:SU(2)_{E}\rightarrow Spin(7)_{R}$ and replace $SU(2)_{E}$ by
$SU(2)'_{E}=(1+\pi)(SU(2)_{E})\subset SU(2)_{E}\times Spin(7)_{R}$.

The homomorphism $\pi$ is determined by the decomposition of $Spin(7)$
under $SU(2)$ and the embedding of $SU(2)$ in $Spin(7)$. 
Although there are many possible decompositions, we consider the
following branchings
\begin{align}
\label{3dn8decom}
 Spin(7)\supset SU(2)_{1}\times SU(2)_{2}\times
 SU(2)_{3}.
\end{align}
Under (\ref{3dn8decom}), $\bm{7}$ and $\bm{8}$ of $Spin(7)$ decomposed
       as \cite{Slansky:1981yr}
\begin{align}
\bm{7}&=(\bm{2},\bm{2},\bm{1})\oplus (\bm{1},\bm{1},\bm{3})\\
\bm{8}&=(\bm{2},\bm{1},\bm{2})\oplus (\bm{1},\bm{2},\bm{2}).
\end{align}
Then one can consider two different types of embedding with the residual
       global symmetry \footnote{Note that
       $SU(2)_{E}\rightarrow SU(2)_{2}$ is same as B twist.} $SU(2)\times SU(2)$
\begin{align}
&\textrm{A-twist}: SU(2)_{E}\rightarrow SU(2)_{3}\nonumber \\
&\textrm{B-twist}: SU(2)_{E}\rightarrow SU(2)_{1}.
\end{align}

\subsubsection{A-twist}

After twisting the fields transform under $SU(2)_{E}'\times
       SU(2)_{1}\times SU(2)_{2}$ as
\begin{align}
\label{3dn8symtwist1}
\phi&\rightarrow (\bm{1},\bm{2},\bm{2})\oplus (\bm{3},\bm{1},\bm{1})\\
\label{3dn8symtwist2}
\psi&\rightarrow (\bm{1},\bm{2},\bm{1})\oplus (\bm{3},\bm{2},\bm{1})
\oplus (\bm{1},\bm{1},\bm{2})\oplus (\bm{3},\bm{1},\bm{2}).
\end{align}
The bosonic field content consists of
\begin{itemize}
 \item 4 scalar fields: $(\bm{1},\bm{2},\bm{2})$
\item 1-form: $(\bm{3},\bm{1},\bm{1})$
\item gauge field: $(\bm{3},\bm{1},\bm{1})$
\end{itemize}
and fermionic fields are
\begin{itemize}
 \item 4 scalar fields: $(\bm{1},\bm{2},\bm{1})\oplus (\bm{1},\bm{1},\bm{2})$
\item 4 vector fields: $(\bm{3},\bm{2},\bm{1})\oplus (\bm{3},\bm{1},\bm{2})$.
\end{itemize}
Thus in A-twist, there are four BRST charges transforming as two $SU(2)$
       doublets. 
It turns out that A-twist topological theories is the dimensional
       reduction of twisted $d=4$, $\mathcal{N}=4$ theories with GL twist
       and VW twist.

\subsubsection{B-twist}
After twisting, the fields transform under $SU(2)_{E}'\times
      SU(2)_{2}\times SU(2)_{3}$ as
\begin{align}
\label{3dn8symtwist3}
\phi\rightarrow&(\bm{2},\bm{2},\bm{1})\oplus (\bm{1},\bm{1},\bm{3})\\
\label{3dn8symtwist4}
 \psi\rightarrow&(\bm{1},\bm{1},\bm{2})\oplus (\bm{3},\bm{1},\bm{2})
\oplus (\bm{2},\bm{2},\bm{2}).
\end{align}
The bosonic field content is 
\begin{itemize}
 \item 3 scalar fields: $(\bm{1},\bm{1},\bm{3})$
\item 2 spinor fields: $(\bm{2},\bm{2},\bm{1})$
\item gauge fields: $(\bm{3},\bm{1},\bm{1})$
\end{itemize}
and the fermionic field content is
\begin{itemize}
 \item 2 scalar fields: $(\bm{1},\bm{1},\bm{2})$
\item 2 vector fields: $(\bm{3},\bm{1},\bm{2})$
\item 4 spinor fields: $(\bm{2},\bm{2},\bm{2})$.
\end{itemize}
In B-twist, we have two BRST charges transforming as a $SU(2)$ doublet. 
B-twist topological theories are the dimensional reduction of twisted
$d=4$, $\mathcal{N}=4$ theories with DW twist.

\subsection{$d=3$, $\mathcal{N}=8$ SYM theories on $\mathbb{R}\times \Sigma$}
Now consider a three-dimensional $\mathcal{N}=8$ SYM theories on
$M_{3}=\mathbb{R}\times \Sigma$.

Before twisting, fields transform under $SO(2)_{E}\times Spin(7)_{R}$ as
\begin{align}
\phi:& \bm{7}_{0}\\
\psi:& \bm{8}_{+}\oplus \bm{8}_{-}\\
A_{\mu}:& \bm{1}_{-2}\oplus \bm{1}_{0}\oplus \bm{1}_{2}.
\end{align}
To determine the homomorphism, we consider the decomposition
of $Spin(7)$ under $SO(2)$ as
\begin{align}
\label{3dn8t1}
&\textrm{A-twist}: Spin(7)\supset SO(5)\times SO(2)\\
\label{3dn8t2}
&\textrm{B-twist}: Spin(7)\supset SO(3)\times SO(4)\supset SO(3)\times
 SO(2)_{1}\times SO(2)_{2}\\
\label{3dn8t3}
&\textrm{C-twist}: Spin(7)\supset SO(6)\supset SO(2)_{1}\times SO(2)_{2}\times SO(2)_{3}
\end{align}

\subsubsection{A-twist}
Under (\ref{3dn8t1}), $\bm{7}$ and $\bm{8}$ of $Spin(7)_{R}$ decomposed
as
\begin{align}
\bm{7}&=\bm{5}_{0}\oplus \bm{1}_{-2}\oplus \bm{1}_{2}\\
\bm{8}&=\bm{4}_{+}\oplus \overline{\bm{4}}_{-}.
\end{align}
Then after twisting, fields transform under $SO(2)_{E}\times SO(5)_{R}$
as
\begin{align}
&\bm{7}_{0}\rightarrow \bm{5}_{0}\oplus \bm{1}_{-2}\oplus \bm{1}_{2}\\
&\bm{8}_{+}\oplus \bm{8}_{-}
\rightarrow \bm{4}_{2}\oplus \overline{\bm{4}}_{0}\oplus
 \bm{4}_{0}\oplus \overline{\bm{4}}_{-2}\\
&\bm{1}_{-2}\oplus \bm{1}_{0}\oplus \bm{1}_{2}
\rightarrow \bm{1}_{-2}\oplus \bm{1}_{0}\oplus \bm{1}_{2}.
\end{align}
Thus there are eight BRST charges in A-twisted $d=3$, $\mathcal{N}=8$ SYM
theory on $\mathbb{R}\times \Sigma$.

\subsubsection{B-twist}
Under (\ref{3dn8t2}), $\bm{7}$ and $\bm{8}$ of $Spin(7)_{R}$ decomposed
as
\begin{align}
\bm{7}&=\bm{3}_{00}\oplus \bm{1}_{0-2}\oplus \bm{1}_{02}
\oplus \bm{1}_{20}\oplus \bm{1}_{-20}
\\
\bm{8}&=\bm{2}_{++}\oplus \bm{2}_{+-}
\oplus \bm{2}_{-+}\oplus \bm{2}_{--}.
\end{align}
We normalize $SO(2)_{1}, SO(2)_{2}$ charges by dividing by two and
simply take a sum of all of the charges including the original rotational  $SO(2)_{E}$ charges.   
Performing this twisting, fields transform under $SO(2)_{E}'\times SO(3)_{R}$
as
\begin{align}
&\bm{7}_{0}\rightarrow \bm{3}_{0}\oplus 2 (\bm{1}_{+})\oplus 2 (\bm{1}_{-})\\
&\bm{8}_{+}\oplus \bm{8}_{-}
\rightarrow \bm{2}_{2}\oplus 2(\bm{2}_{+})\oplus 2(\bm{2}_{0})
\oplus 2(\bm{2}_{-})\oplus \bm{2}_{-2}
 \\
&\bm{1}_{-2}\oplus \bm{1}_{0}\oplus \bm{1}_{2}
\rightarrow \bm{1}_{-2}\oplus \bm{1}_{0}\oplus \bm{1}_{2}.
\end{align}
Thus there are four BRST charges in B-twisted $d=3$, $\mathcal{N}=8$ SYM
theory on $\mathbb{R}\times \Sigma$.

\subsubsection{C-twist}
Under (\ref{3dn8t3}), $\bm{7}$ and $\bm{8}$ of $Spin(7)_{R}$ decomposed
as
\begin{align}
\bm{7}&=\bm{1}_{200}\oplus \bm{1}_{-200}\oplus \bm{1}_{020}
\oplus \bm{1}_{0-20}\oplus \bm{1}_{002}\oplus \bm{1}_{00-2}\oplus \bm{1}_{000}
\\
\bm{8}&=\bm{1}_{+++}\oplus \bm{1}_{++-}
\oplus \bm{1}_{+-+}\oplus \bm{1}_{+--}
\oplus \bm{1}_{-++}\oplus \bm{1}_{-+-}
\oplus \bm{1}_{--+}\oplus \bm{1}_{---}.
\end{align}
We normalize $SO(2)_{1}, SO(2)_{2}, SO(2)_{3}$ charges by dividing by three and
simply take a sum of all of the charges including the original rotational  $SO(2)_{E}$ charges.   
Performing this twisting, fields transform under $SO(2)_{E}'$
as
\begin{align}
&\bm{7}_{0}\rightarrow 3(\bm{1}_{\frac23})\oplus
 3(\bm{1}_{-\frac23})\oplus \bm{1}_{0}\\
&\bm{8}_{+}\oplus \bm{8}_{-}
\rightarrow 2 (\bm{1}_{0})\oplus 2 (\bm{1}_{\frac23})\oplus 
2(\bm{1}_{-\frac23})
\oplus \bm{1}_{\frac43}\oplus \bm{1}_{-\frac43}
 \\
&\bm{1}_{-2}\oplus \bm{1}_{0}\oplus \bm{1}_{2}
\rightarrow \bm{1}_{-2}\oplus \bm{1}_{0}\oplus \bm{1}_{2}.
\end{align}
Thus there are two BRST charges in B-twisted $d=3$, $\mathcal{N}=8$ SYM
theory on $\mathbb{R}\times \Sigma$.

\subsection{$d=2$, $\mathcal{N}=8$ SYM theories}
The global symmetry of the theory is $Spin(8)_{R}$ R-symmetry. 
The field content is 
\begin{itemize}
 \item 8 scalar fields $\phi^{i}$
\item fermionic fields $\psi$
\item gauge field $A_{\mu}$.
\end{itemize}

Before topological twisting, fields transform under $SO(2)_{E}\times
Spin(8)_{R}$ as
\begin{align}
\phi&:\bm{8}_{v0}\\
\psi&:\bm{8}_{c+}\oplus \bm{8}_{s-}\\
A_{\mu}&:\bm{1}_{-2}\oplus \bm{1}_{2}.
\end{align}
To perform the topological twisting, 
we pick a homomorphism $\pi:SO(2)_{E}\rightarrow Spin(8)_{R}$ and
replace $SO(2)_{E}$ by $SO(2)_{E}'=(1+\pi)(SO(2)_{E})\subset
SO(2)_{E}\times Spin(8)_{R}$. 

The homomorphism $\pi$ is determined by the decomposition of $Spin(8)$
under $SO(2)$ and the embedding of $SO(2)$ in $Spin(8)$. 
Although there are many possible decompositions, we consider the
following two types of branching
\begin{align}
\label{2dn8decom1}
\textrm{A-twist}:\ Spin(8)_{R}&\supset SO(6)_{R}\times SO(2)_{1}\\
\label{2dn8decom2}
\textrm{B-twist}:\ Spin(8)_{R}&\supset 
SO(4)_{1}\times SO(4)_{2}
\supset SO(4)_{1}\times SU(2)_{1}\times
 SU(2)_{2}\nonumber\\
&\supset SO(4)_{1}\times
      SU(2)_{1}\times SO(2)_{2}
\end{align}

\subsubsection{A-twist}
Under (\ref{2dn8decom1}), $\bm{8}_{v}$, $\bm{8}_{s}$ and $\bm{8}_{c}$ of
       $Spin(8)_{R}$ decomposed as
\begin{align}
\bm{8}_{v}=&\bm{6}_{0}\oplus\bm{1}_{2}\oplus\bm{1}_{-2}\\
\bm{8}_{s}=&\bm{4}_{+}\oplus \overline{\bm{4}}_{-}\\
\bm{8}_{c}=&\bm{4}_{-}\oplus \overline{\bm{4}}_{+}.
\end{align}
Then the choice of $\pi$ amounts to 
$SO(2)_{E}'\rightarrow SO(2)_{1}$ and the fields transform under
$SO(2)_{E}\times SO(6)_{R}$ as 
\begin{align}
\bm{8}_{v}\rightarrow&\bm{6}_{0}\oplus \bm{1}_{2}\oplus \bm{1}_{-2}\\
\bm{8}_{s}\rightarrow&\bm{4}_{0}\oplus \overline{\bm{4}}_{-2}\\
\bm{8}_{c}\rightarrow&\bm{4}_{0}\oplus \overline{\bm{4}}_{2}
\end{align}
There are eight supercharges transforming $\bm{4}$ of $SO(6)_{R}$. 
This is just the dimensional reduction of A-twisted $d=3$, $\mathcal{N}=8$
SYM theory. 

To see this, let us consider the further decomposition
\begin{equation}
\label{su4dec1}
 SO(6)_{R}\simeq SU(4)_{R}
\supset SU(2)_{1}\times SU(2)_{2}\times U(1).
\end{equation}
Under (\ref{su4dec1}), 
$\bm{6}$ and $\bm{4}$ decomposed as \footnote{As in (\ref{4dn4twist}),
we have other possible decompositions. (\ref{su4dec1}) is same as GL twist.}
\begin{align}
\label{su4dec2}
\bm{6}=&(\bm{2},\bm{2})_{0}
\oplus (\bm{1},\bm{1})_{2}\oplus (\bm{1},\bm{1})_{-2}\\
\label{su4dec3}
\bm{4}=&(\bm{2},\bm{1})_{+}
\oplus (\bm{1},\bm{2})_{-}\\
\label{su4dec4}
\overline{\bm{4}}=&(\bm{2},\bm{1})_{-}\oplus (\bm{1},\bm{2})_{+}.
\end{align}
Then the fields transform under $SO(2)_{E}'\times SU(2)_{1}\times SU(2)_{2}$ as
\begin{align}
\bm{8}_{v}&\rightarrow
(\bm{2},\bm{2})_{0}\oplus (\bm{1},\bm{1})_{0}
\oplus(\bm{1},\bm{1})_{0}\oplus (\bm{1},\bm{1})_{2}
\oplus (\bm{1},\bm{1})_{-2}\\
\bm{8}_{s}&\rightarrow
(\bm{2},\bm{1})_{0}
\oplus (\bm{1},\bm{2})_{0}
\oplus (\bm{2},\bm{1})_{-2}
\oplus (\bm{1},\bm{2})_{-2}\\
\bm{8}_{c}&\rightarrow 
(\bm{2},\bm{1})_{0}\oplus (\bm{1},\bm{2})_{0}
\oplus (\bm{2},\bm{1})_{2}\oplus (\bm{1},\bm{2})_{2}
\end{align}
Thus bosonic field content is 
\begin{itemize}
 \item 6 scalar fields: $2(\bm{1},\bm{1})_{0}\oplus (\bm{2},\bm{2})_{0}$
\item 1-form: $(\bm{1},\bm{1})_{2}\oplus (\bm{1},\bm{1})_{-2}$
\end{itemize}
and the fermionic field content is 
\begin{itemize}
 \item 8 scalar fields: $2(\bm{2},\bm{1})_{0}\oplus
       2(\bm{1},\bm{2})_{0}$
\item 4 1-forms: $(\bm{2},\bm{1})_{2}\oplus (\bm{2},\bm{1})_{-2}\oplus
      (\bm{1},\bm{2})_{2}\oplus (\bm{1},\bm{2})_{-2}$.
\end{itemize}
The two bosonic scalars $2(\bm{1},\bm{1})_{0}$ correspond to the third
components of the gauge field and 1-form in three dimensions. 
Also the four fermionic scalars $(\bm{2},\bm{1})_{0}\oplus
(\bm{1},\bm{2})_{0}$ are the third components of the 1-form in 
three dimensions.

\subsubsection{B-twist}
Under (\ref{2dn8decom2}), $\bm{8}_{v}$, $\bm{8}_{s}$ and $\bm{8}_{c}$ of
$Spin(8)$ decomposed as
\begin{align}
\bm{8}_{v}=&(\bm{4},\bm{1})\oplus (\bm{1},\bm{4})\nonumber \\
=&(\bm{4},\bm{1},\bm{1})\oplus
 (\bm{1},\bm{2},\bm{2})
=(\bm{4},\bm{1})_{0}\oplus (\bm{1},\bm{2})_{+}\oplus (\bm{1},\bm{2})_{-}\\
\bm{8}_{s}=&(\bm{2},\bm{2})\oplus (\bm{2}',\bm{2}')\nonumber \\
=&(\bm{2},\bm{2},\bm{1})\oplus (\bm{2}',\bm{1},\bm{2})
=(\bm{2},\bm{2})_{0}\oplus (\bm{2}',\bm{1})_{+}
\oplus (\bm{2}',\bm{1})_{-}\\
\bm{8}_{c}=&(\bm{2},\bm{2}')\oplus (\bm{2}',\bm{2})\nonumber \\
=&(\bm{2},\bm{1},\bm{2})\oplus (\bm{2}',\bm{2},\bm{1})
=(\bm{2},\bm{1})_{+}\oplus (\bm{2},\bm{1})_{-}\oplus (\bm{2}',\bm{2})_{0}.
\end{align}
Then the choice of $\pi$ amounts to $SO(2)_{E}\rightarrow SO(2)_{2}$ and
the fields transform under $SO(2)_{E}'\times SO(4)_{R}\times SO(2)_{1}$
as
\begin{align}
\bm{8}_{v}&\rightarrow (\bm{4},\bm{1})_{0}
\oplus (\bm{1},\bm{2})_{+}\oplus (\bm{1},\bm{2})_{-}\\
\bm{8}_{s}&\rightarrow (\bm{2},\bm{2})_{-}\oplus (\bm{2}',\bm{1})_{0}
\oplus (\bm{2}',\bm{1})_{-2}\\
\bm{8}_{c}&\rightarrow (\bm{2},\bm{1})_{2}
\oplus (\bm{2},\bm{1})_{0}
\oplus (\bm{2}',\bm{2})_{+}.
\end{align}
The bosonic field contnt is
\begin{itemize}
 \item 4 scalar fields: $(\bm{4},\bm{1})_{0}$
\item 4 spinors: $(\bm{1},\bm{2})_{+}\oplus (\bm{1},\bm{2})_{-}$
\end{itemize}
and the fermionic field content is
\begin{itemize}
 \item 4 scalar fields: $(\bm{2}',\bm{1})_{0}\oplus (\bm{2},\bm{1})_{0}$
\item 8 spinors: $(\bm{2},\bm{2})_{-}\oplus (\bm{2}',\bm{2})_{+}$
\item 2 1-forms: $(\bm{2}',\bm{1})_{-2}\oplus (\bm{2},\bm{1})_{2}$.
\end{itemize}
There are four BRST charges in B-twist.

\subsection{$d=3$, $\mathcal{N}=4$ Chern-Simons matter theories}
Gaiotto and Witten gave a general prescription for coupling Chern-Simons
theory to hypermultiplets, which allows for a new large class of 
three-dimensional $\mathcal{N}=4$ supersymmetric
gauge theories \cite{Gaiotto:2008sd}.
Gaiotto-Witten theory can be regarded as a three-dimensional $\mathcal{N}=4$
gauged sigma-model with a hyperk\"{a}hler target space $X$.

The field content is 
\begin{itemize}
\item gauge field $A^{m}_{\mu}$
 \item hypermultiplet boson $q^{A}_{\alpha}$
\item hypermultiplet fermion $\psi^{A}_{\dot{\alpha}}$
\item twisted hypermultiplet $\tilde{q}^{A}_{\dot{\alpha}}$
\item twisted hypermultiplet $\tilde{\psi}^{A}_{\alpha}$
\end{itemize}
where $m$ is the adjoint indices raised by invariant quadratic form
$k^{mn}$ of the gauge group. 
The gauge group is a subgroup of $Sp(2n)$ and we denote the
anti-Hermitian generators of the gauge group by $(t^{m})^{A}_{B}
(A,B,\cdots=1,\cdots,2n)$, which satisfy
\begin{align}
[t^{m},t^{n}]={f^{mn}}_{p}t^{p},\ \ \ t_{AB}=\omega_{AC}t^{C}_{B}
\end{align}
where $\omega_{AB}$ are the anti-symmetric invariant tensor.

The hyper-multiplet fields satisfy the reality condition
\begin{align}
(q_{\alpha}^{A})^{*}=\epsilon^{\alpha\beta}\omega_{AB}q_{\beta}^{B},\\
(\psi_{\alpha}^{A})^{*}=\epsilon^{\dot{\alpha}\dot{\beta}}\omega_{AB}\psi_{\dot{\beta}}^{B}
\end{align}
where $(\alpha,\beta;\dot{\alpha},\dot{\beta})$ are the indices of 
$SU(2)\times SU(2)$ R-symmetry.

For $\mathcal{N}=4$ supersymmetry, 
$t_{AB}^{m}$ satisfy the fundamental identity
\begin{align}
\label{3dn4csmid1}
k_{mn}t^{m}_{(AB}t^{n}_{C)D}=0
\end{align}
where $A,B,C,\cdots$ are symmetrized. 
(\ref{3dn4csmid1}) is nothing but the Jacobi identity for three
fermionic generators of a Lie superalgebra
\begin{align}
&[M^{m},M^{n}]={f^{mn}}_{p}M^{p},\\
&[M^{m},Q_{A}]=Q_{B}(t^{m})^{B}_{A},\\
&\{Q_{A},Q_{B}\}=t^{m}_{AB}M_{m}.
\end{align}

Lagrangian of the Gaiotto-Witten theory is given by
\footnote{An overall coefficients of the Lagrangian should satisfy an
integrality condition to make the quantum theory well-defined. But here
we suppress them.}
\begin{align}
\label{gwlag}
\mathcal{L}=
&\frac12 \epsilon^{\mu\nu\lambda}
\left(
k_{mn}A_{\mu}^{m}\partial_{\nu}A_{\lambda}^{n}
+\frac13 f_{mnp}A_{\mu}^{m}A_{\nu}^{n}A_{\lambda}^{p}
\right)\nonumber \\
&+\omega_{AB}
\left(
-\epsilon^{\alpha\beta}Dq_{\alpha}^{A}Dq_{\beta}^{B}
+i\epsilon^{\dot{\alpha}\dot{\beta}}\psi_{\dot{\alpha}}^{A}
D_{\mu}\gamma^{\mu}\psi_{\dot{\beta}}^{B}
\right)\nonumber \\
&-ik_{mn}\epsilon^{\alpha\beta}
\epsilon^{\dot{\gamma}\dot{\delta}}
j^{m}_{\alpha\dot{\gamma}}j^{n}_{\beta\dot{\delta}}
-\frac{1}{12}f_{mnp}(\mu^{m})^{\alpha}_{\beta}
(\mu^{n})^{\beta}_{\gamma}
(\mu^{p})^{\gamma}_{\alpha}
\end{align}
where 
\begin{align}
\mu_{\alpha\beta}^{m}:=t^{m}_{AB}q_{\alpha}^{A}q_{\beta}^{B},\\
j_{\alpha\dot{B}}^{m}:=t_{AB}^{m}q_{\alpha}^{A}\psi_{\dot{\beta}}^{B},\\
\rho_{\dot{\alpha}\dot{\beta}}^{m}
:=t_{AB}^{m}\psi_{\dot{\alpha}}^{A}\psi_{\dot{\beta}}^{B}
\end{align}
are the momentum map multiplet.

The Euclidean Lagrangian is
\begin{align}
\label{egwlag}
\mathcal{L}=
&-\frac{i}{2} \epsilon^{\mu\nu\lambda}
\left(
k_{mn}A_{\mu}^{m}\partial_{\nu}A_{\lambda}^{n}
+\frac13 f_{mnp}A_{\mu}^{m}A_{\nu}^{n}A_{\lambda}^{p}
\right)\nonumber \\
&-\omega_{AB}
\left(
-\epsilon^{\alpha\beta}Dq_{\alpha}^{A}Dq_{\beta}^{B}
+i\epsilon^{\dot{\alpha}\dot{\beta}}\psi_{\dot{\alpha}}^{A}
D_{\mu}\gamma^{\mu}\psi_{\dot{\beta}}^{B}
\right)\nonumber \\
&+ik_{mn}\epsilon^{\alpha\beta}
\epsilon^{\dot{\gamma}\dot{\delta}}
j^{m}_{\alpha\dot{\gamma}}j^{n}_{\beta\dot{\delta}}
+\frac{1}{12}f_{mnp}(\mu^{m})^{\alpha}_{\beta}
(\mu^{n})^{\beta}_{\gamma}
(\mu^{p})^{\gamma}_{\alpha}.
\end{align}
Note that it differs from the Lorentzian Lagrangian (\ref{gwlag}) 
by the factor $(-i)$ for Chern-Simons term and an overall sign for the
matter terms. 
Now the fermionic fields do not obey the reality conditions.

The supersymmetry transformations are
\begin{align}
&\delta
 q^{A}_{\alpha}=i\epsilon^{\dot{\alpha}}_{\alpha}\psi_{\dot{\alpha}}^{A},\\
&\delta\psi_{\dot{\alpha}}^{A}
=\left(
D^{\mu}\gamma_{\mu}q_{\alpha}^{A}
+\frac13 k_{mn}(t^{m})^{A}_{B}
q^{B}_{\beta}(\mu^{n})^{\beta}_{\alpha}
\right)\epsilon^{\alpha}_{\dot{\alpha}},\\
&\delta A_{\mu}^{m}
=i\epsilon^{\alpha\dot{\alpha}}\gamma_{\mu}j^{m}_{\alpha\dot{\alpha}}.
\end{align}
The supersymmetry parameter $\epsilon$ transforms as 
$(\bm{2},\bm{2})$ in $SU(2)\times SU(2)$ R-symmetry and satisfies the
reality condition
\begin{equation}
 (\epsilon_{\alpha}^{\dot{\alpha}})^{*}
=-\epsilon^{\alpha\beta}\epsilon_{\dot{\alpha}\dot{\beta}}\epsilon_{\beta}^{\dot{\beta}}.
\end{equation}
The supersymmetry transformations are same as in the Lorentzian case.

Furthermore we can add twisted hyper-multiplets
$(\tilde{q}_{\dot{\alpha}}^{A}, \tilde{\psi}^{A}_{\alpha})$ to
Gaiotto-Witten theory. This is regarded as 
a non-linear sigma model \cite{Gaiotto:2008sd}. 
The Lagrangian is 
\begin{align}
\label{3dn4gwlag1}
\mathcal{L}
&=\frac12 \epsilon^{\mu\nu\lambda}
\left(
k_{mn}A_{\mu}^{m}\partial_{\nu}A_{\lambda}^{n}
+\frac13 f_{mnp}A_{\mu}^{m}A_{\nu}^{n}A_{\lambda}^{p}
\right)\nonumber \\
&+\omega_{AB}
\left(
-\epsilon^{\alpha\beta}Dq_{\alpha}^{A}Dq_{\beta}^{B}
+i\epsilon^{\dot{\alpha}\dot{\beta}}\psi_{\dot{\alpha}}^{A}
D_{\mu}\gamma^{\mu}\psi_{\dot{\beta}}^{B}\right)
+\omega_{AB}
\left(
-\epsilon^{\dot{\alpha}\dot{\beta}}D\tilde{q}_{\dot{\alpha}}^{B}
+i\epsilon^{\alpha\beta}\tilde{\psi}_{\alpha}^{A}D_{\mu}\gamma^{\mu}\tilde{\psi}^{B}_{\beta}
\right)
\nonumber \\
&-ik_{mn}\left(\epsilon^{\alpha\beta}
\epsilon^{\dot{\gamma}\dot{\delta}}
j^{m}_{\alpha\dot{\gamma}}j^{n}_{\beta\dot{\delta}}
+\epsilon^{\dot{\alpha}\dot{\beta}}\epsilon^{\gamma\delta}
\tilde{j}^{m}_{\dot{\alpha}\gamma}\tilde{j}^{n}_{\dot{\beta}\delta}
+4\epsilon^{\alpha\gamma}\epsilon^{\dot{\beta}\dot{\delta}}
j^{m}_{\alpha\dot{\beta}}\tilde{j}^{n}_{\dot{\delta}\gamma}
-\epsilon^{\dot{\alpha}\dot{\gamma}}\epsilon^{\dot{\beta}\dot{\delta}}
\tilde{\mu}^{m}_{\dot{\alpha}\dot{\beta}}
\rho^{n}_{\dot{\gamma}\dot{\delta}}
-\epsilon^{\alpha\gamma}\epsilon^{\beta\delta}
\mu^{m}_{\alpha\beta}\tilde{\rho}_{\gamma\delta}
\right)\nonumber \\
&-\frac{1}{12}f_{mnp}(\mu^{m})^{\alpha}_{\beta}
(\mu^{n})^{\beta}_{\gamma}
(\mu^{p})^{\gamma}_{\alpha}
+(\tilde{\mu}^{m})^{\dot{\alpha}}_{\dot{\beta}}
(\tilde{\mu}^{n})^{\dot{\beta}}_{\dot{\gamma}}
(\tilde{\mu}^{\rho})^{\dot{\gamma}}_{\dot{\alpha}}
\nonumber \\
&-\frac12\tilde{\mu}^{mn}
(\mu_{m})^{\alpha}_{\beta}
(\mu_{n})^{\beta}_{\alpha}
-\frac12 \mu^{mn}
(\tilde{\mu}_{m})^{\dot{\alpha}}_{\dot{\beta}}
(\tilde{\mu}_{n})^{\dot{\beta}}_{\dot{\alpha}}
\end{align}
where the twisted moment map is defined by
\begin{align}
&\mu^{mn}
=\epsilon^{\alpha\beta}
(t^{m}t^{n})_{AB}q_{\alpha}^{A}q_{\beta}^{B},\\
&\tilde{\mu}^{mn}
=\epsilon^{\dot{\alpha}\dot{\beta}}
(\overline{t}^{m}\overline{t}^{n})_{AB}
\tilde{q}_{\dot{\alpha}}^{A}
\tilde{q}_{\dot{\beta}}^{B}.
\end{align}

The supersymmetry transformation is
\begin{align}
&\delta q_{\alpha}^{A}
=i\epsilon_{\alpha}^{\dot{\alpha}}\psi_{\dot{\alpha}}^{A},\\
&\delta\psi_{\dot{\alpha}}^{A}
=\left(
D_{\mu}\gamma^{\mu}q^{A}_{\alpha}
+\frac13 (t_{m})^{A}_{B}q_{\beta}^{B}(\mu^{m})^{\beta}_{\alpha}
\right)\epsilon^{\alpha}_{\dot{\alpha}},\\
&\delta\tilde{\psi}_{\alpha}^{A}
=\left(
D_{\mu}\gamma^{\mu}q^{A}_{\dot{\alpha}}
+\frac13 (\tilde{t}_{m})\tilde{q}^{B}_{\dot{\beta}}
(\tilde{\mu}^{m})^{\dot{\beta}}_{\dot{\alpha}}
\right)\epsilon_{\alpha}^{\dot{\alpha}}
-(\tilde{t}_{m})^{A}_{B}
\tilde{q}^{B}_{\dot{\beta}}(\mu^{m})^{\beta}_{\alpha}
\epsilon^{\beta\dot{\beta}},\\
&\delta A_{\mu}^{m}
=i\epsilon^{\alpha\dot{\alpha}}\gamma_{\mu}
(j_{\alpha\dot{\alpha}}^{m}+\tilde{j}^{m}_{\dot{\alpha}\alpha}).
\end{align}

The topological twisting for Gaiotto-Witten theory was
discussed in \cite{Kapustin:2009cd, Koh:2009um}. 
\begin{enumerate}
 \item flat target space $X$

If the target $X$ is flat, Gaiotto-Witten theory has $SU(2)\times SU(2)$
       R-symmetry. The topologically twisted theory is equivalent to the
       pure Chern-Simons theory whose gauge group is a supergroup
       \cite{Kapustin:2009cd}. In other words, the topologically twisted
       Gaiotto-Witten theory is obtained from the supergroup
       Chern-Simons theory by gauge fixing the odd part of the
       supergroup and the even part of the supergroup gives rise to
       gauge group $G$. 

\item general target space $X$

For general target $X$, Gaiotto-Witten theory has $SU(2)$ R-symmetry. 
     The topologically twisted Gaiotto-Witten theory can be interpreted as
      a gauged Rozansky-Witten theory \cite{Rozansky:1996bq}, that is a
      hybrid of Chern-Simons and Rozansky-Witten theory \cite{Kapustin:2009cd}. 
It is associated to a quadruple:
\begin{enumerate}
 \item $G$: a compact Lie group
\item $\kappa$: invariant metric on the Lie algebra
\item $X$: hyperk\"{a}hler manifold with a tri-holomorphic action of $G$
\item $I$: complex structure on $X$ such that the complex moment map
      with respect to the complex symplectic form $\Omega_{I}$ is
      isotropic with respect to $\kappa$ 
\end{enumerate}

\end{enumerate}

Before twisting, fields and supercharges transform under
$SU(2)_{E}\times SU(2)_{1}\times SU(2)_{2}$ as
\begin{align}
&q: (\bm{1},\bm{2},\bm{1})\\
&\psi: (\bm{2},\bm{1},\bm{2})\\
&Q: (\bm{2},\bm{2},\bm{2}).
\end{align}
We may also have fields of twisted hypermultiplet, which transform as
\begin{align}
&\tilde{q}: (\bm{1},\bm{1},\bm{2})\\
&\tilde{\psi}: (\bm{2},\bm{2},\bm{1}).
\end{align}
Depending on which $SU(2)$ factor we use, 
we may think two types of twisting
\begin{align}
&\textrm{A-twist}: SU(2)_{E}\rightarrow SU(2)_{2}\\
&\textrm{B-twist}: SU(2)_{E}\rightarrow SU(2)_{1}.
\end{align}
However, if both hypermultiplet and twisted hypermultiplet are present,
A-twist and B-twist are the same. 
We call it AB-twist.

\subsubsection{A-twist}
After A-twisting, the fields and supercharges transform under
$SU(2)_{E}'\times SU(2)$ as
\begin{align}
&q\rightarrow (\bm{1},\bm{2})\\
&\psi\rightarrow (\bm{1},\bm{1})\oplus (\bm{3},\bm{1})\\
&Q\rightarrow (\bm{1},\bm{2})\oplus (\bm{3},\bm{1}).
\end{align}
Thus in the bosonic field content we have
\begin{itemize}
 \item 2 scalar fields: $(\bm{1},\bm{2})$
\end{itemize}
and in the fermionic field content we include
\begin{itemize}
 \item scalar field $(\bm{1},\bm{1})$
\item 1-form $(\bm{3},\bm{1})$.
\end{itemize}
There are two BRST charges in A-twisted theory.

\begin{enumerate}
 \item \textbf{bosonic fields}

In A-twist all of the hypermultiplet scalars remain scalars.

\item \textbf{fermionic fields}

We decompose the fermionic fields under $SU(2)_{E}\times SU(2)_{1}\times
      SU(2)_{2}$ as
\begin{align}
(\psi_{\alpha'})_{\dot{\alpha}}^{A}
=\frac{1}{\sqrt{2}}\left(
\psi^{A}\mathbb{I}_{\alpha'\dot{\alpha}}+\Psi_{i}^{A}(\sigma_{i})_{\alpha'\dot{\alpha}}
\right)\sigma_{2}^{-1}
\end{align}
where $\alpha',\dot{\alpha}$ are the indices of $SU(2)_{E}, SU(2)_{2}$
respectively. $A=1,2,\cdots, 2n$ is again the index of $Sp(2n)$.

\item \textbf{supersymmetries}

For the supersymmetries we expand as
\begin{align}
(\epsilon^{\alpha'})^{\dot{\alpha}}_{\alpha}
=\frac{1}{\sqrt{2}}\left(
\epsilon_{\alpha}\mathbb{I}^{\alpha'\dot{\alpha}}
+\epsilon_{i\alpha}(\sigma_{i})^{\alpha'\dot{\alpha}}
\right)(\sigma_{2})^{-1}.
\end{align}

\end{enumerate}

\subsubsection{B-twist}
After B-twisting, the fields tranform under $SU(2)_{E}'\times SU(2)$ as
\begin{align}
&q\rightarrow (\bm{2},\bm{1})\\
&\psi\rightarrow (\bm{2},\bm{2})\\
&Q\rightarrow (\bm{1},\bm{2})\oplus (\bm{3},\bm{2}).
\end{align}
Thus the bosonic field content is
\begin{itemize}
 \item spinor field: $(\bm{2},\bm{1})$
\end{itemize}
and the fermionic field content is
\begin{itemize}
 \item 2 spinor fields $(\bm{2},\bm{2})$.
\end{itemize}
We have two BRST charges in B-twisted theory.

\subsubsection{AB-twist}
After twisting, the fields transform under $SU(2)_{E}'\times SU(2)_{2}$
as 
\begin{align}
&q\rightarrow (\bm{1},\bm{2})\\
&\psi\rightarrow (\bm{1},\bm{1})\oplus (\bm{3},\bm{1})\\
&\tilde{q}\rightarrow (\bm{2},\bm{1})\\
&\tilde{\psi}\rightarrow (\bm{2},\bm{2})\\
&Q\rightarrow (\bm{1},\bm{2})\oplus (\bm{3},\bm{2}).
\end{align}
Thus the bosonic field content is
\begin{itemize}
 \item 2 scalar fields : $(\bm{1},\bm{2})$
\item spinor field: $(\bm{2},\bm{1})$
\end{itemize}
and the fermionic field content is 
\begin{itemize}
 \item scalar fields : $(\bm{1},\bm{1})$
\item 1-form : $(\bm{3},\bm{1})$
\item 2 spinor fields: $(\bm{2},\bm{2})$.
\end{itemize}
Again we have two BRST charges in AB-twisted theory.

\subsection{$d=3$, $\mathcal{N}=5$ Chern-Simons matter theories}
Three-dimensional $\mathcal{N}\ge 5$ theories can be
understood in the Gaiotto-Witten framework by adding twisted
hypermultiplets \cite{Hosomichi:2008jd}. The target spaces of
$\mathcal{N}\ge 5$ theories are only flat spaces and their orbifolds.

We may consider the decomposition of $SO(5)_{R}$ R-symmetry under
$SO(3)\simeq SU(2)$ as
\begin{align}
\label{3dn51}
&SO(5)\supset SO(2)\times SO(3)\\
\label{3dn52}
&SO(5)\supset SO(4)\supset SU(2)_{1}\times SU(2)_{2}.
\end{align}
Under (\ref{3dn51}), $\bm{5}$ of $SO(5)_{R}$ decomposed as
\begin{align}
\bm{5}=\bm{3}_{0}\oplus \bm{1}_{-2}\oplus \bm{1}_{2}.
\end{align}
Noting that
\begin{align}
\bm{2}_{0}\times \left(
\bm{3}_{0}\oplus \bm{1}_{-2}\oplus \bm{1}_{2}
\right)=\bm{2}_{0}\oplus \bm{4}_{0}\oplus \bm{2}_{2}\oplus \bm{2}_{-2},
\end{align}
we see that there are no BRST charges. 

On the other hand, under (\ref{3dn52}), $\bm{5}$ and $\bm{4}$ of $SO(5)_{R}$
decomposed as
\begin{align}
&\bm{5}=\bm{4}+\bm{1}=(\bm{2},\bm{2})\oplus (\bm{1},\bm{1})\\
&\bm{4}=\bm{4}=(\bm{2},\bm{1})\oplus (\bm{1},\bm{2}).
\end{align}
This is nothing but the AB-twist in $d=3$, $\mathcal{N}=4$ Chern-Simons matter
theory.

\subsection{$d=3$, $\mathcal{N}=6$ Chern-Simons matter theories}
We may consider the decomposition of $SO(6)_{R}$ R-symmetry under
$SO(3)\simeq SU(2)$ as
\begin{align}
\label{3dn61}
&SO(6)\supset SO(3)\times SO(3)\\
\label{3dn62}
&SO(6)\supset SO(2)\times SO(4)\supset SO(2)\times SU(2)_{1}\times SU(2)_{2}.
\end{align}
Under (\ref{3dn61}), $\bm{5}$ of $SO(5)_{R}$ decomposed as
\begin{align}
\bm{6}=(\bm{3},\bm{1})\oplus (\bm{1},\bm{3}).
\end{align}
Noting that
\begin{align}
\bm{2}\times \bm{3}=\bm{2}\oplus \bm{4}\oplus \bm{2}\oplus \bm{2},
\end{align}
we see that there are no BRST charges. 

On the other hand, under (\ref{3dn62}), $\bm{6}$ of $SO(6)_{R}$
decomposed as
\begin{align}
\bm{6}=\bm{4}_{0}\oplus \bm{1}_{-2}\oplus \bm{1}_{2}
=(\bm{2},\bm{2})_{0}\oplus (\bm{1},\bm{1})_{-2}\oplus (\bm{1},\bm{1})_{2}.
\end{align}
As seen from the appearance of $(\bm{2},\bm{2})$, this is nothing but
the AB-twist in $d=3$, $\mathcal{N}=4$ Chern-Simons matter theory.

\subsection{$d=3$, $\mathcal{N}=8$ Chern-Simons matter theories}
\label{3dn8symsubsec}
To perfom the topological twisting, 
we put the BLG theory on a three-dimensional Euclidean space. 
The fermionic fields and supersymmetry parameters are defined as
eleven-dimensional fermions and their conjugate are given by
\begin{align}
\overline{\Psi}:=\Psi^{T}\mathcal{C}
\end{align}
where $\mathcal{C}$ is a eleven-dimensional matrix satisfying
\begin{align}
\mathcal{C}^{T}=-\mathcal{C},\ \ \ \mathcal{C}\Gamma^{M}\mathcal{C}^{-1}
=-(\Gamma^{M})^{T}.
\end{align}
Gamma matrix $\Gamma^{M}$ is the representation 
of eleven-dimensional Clifford algebra
\begin{align}
\left\{
\Gamma^{M},\Gamma^{N}
\right\}=2g^{MN},\ \ \ \Gamma^{11}:=i\Gamma^{12\cdots 10}.
\end{align}
$\Gamma^{M}$ can be decomposed under $SO(11)\supset SO(3)\times SO(8)$
as
\begin{align}
\begin{cases}
\Gamma^{i}=\sigma_{i}\otimes \tilde{\Gamma}^{9}\cr
\Gamma^{I+3}=\mathbb{I}\otimes \tilde{\Gamma}^{I}\cr
\end{cases}
\end{align}
where $\tilde{\Gamma}^{9}:=\tilde{\Gamma}^{1\cdots 8}$. 
Note that 
\begin{align}
\Gamma^{123}=i\Gamma^{45\cdots 11}
=i(\mathbb{I}\otimes \tilde{\Gamma}^{9}).
\end{align} 
The fermionic fields $\Psi$ are $\bm{8}_{c}$ of $SO(8)_{R}$, so 
they satisfy the chirality condition
\begin{align}
\Gamma^{45\cdots 11}\Psi=-\Psi,\\
\Gamma^{123}\Psi=-i\Psi.
\end{align}

The Euclidean BLG Lagrangian is given by
\begin{align}
\label{eblglag}
\mathcal{L}
&=\frac12 (D_{\mu}X^{I},D^{\mu}X^{I})
-\frac{i}{2}(\overline{\Psi},\Gamma^{\mu}D_{\mu}\Psi)
-\frac{i}{4}\left(
\overline{\Psi}\Gamma^{IJ}[X^{I},X^{J},\Psi]
\right) \nonumber \\ 
&+\frac{1}{12}
\left([X^{I},X^{J},X^{K}],[X^{I},X^{J},X^{K}]\right)
-\frac{i}{2} \epsilon^{\mu\nu\lambda}
\left[
\textrm{Tr}
\left(A_{\mu ab}\partial_{\nu}\tilde{A}_{\lambda}^{ab}\right)
+\frac23\textrm{Tr}\left(
A_{\mu ab}\tilde{A}^{a}_{\nu g}\tilde{A}^{b}_{\lambda g}
\right)
\right],
\end{align}
which differs from Lorentzian case by the factor $(-i)$ for the
Chern-Simons terms and a overall sign factors for matter terms.

The supersymmetry transformations are
\begin{align}
&\delta X_{a}^{I}
=i\overline{\epsilon}\Gamma^{I}\Psi_{a},\\
&\delta\Psi_{a}
=D_{\mu}X_{a}^{I}\Gamma^{\mu}\Gamma^{I}\epsilon
-\frac16 X_{b}^{I}X_{c}^{J}X_{d}^{K}
{f^{bcd}}_{a}\Gamma^{IJK}\epsilon,\\
&\delta \tilde{A}_{\mu b}^{a}
=i\overline{\epsilon}
\Gamma_{\mu}\Gamma^{I}X_{c}^{I}\Psi_{d}{f^{cda}}_{b}.
\end{align}
This is exactly same as in the Lorentzian transformations
(\ref{blgsusy1})-(\ref{blgsusy3}). 
$\epsilon$ is the unbroken supersymmetry parameters obeying\footnote{
This convention is different from that in \cite{Koh:2009um}. There are
two choices for $\tilde{\Gamma}^{9}=\pm \tilde{\Gamma}^{12\cdots 8}$. We
take $\tilde{\Gamma}^{9}=+\tilde{\Gamma}^{12\cdots 8}$.
}
\begin{align}
&\Gamma^{34\cdots 11}\epsilon=\epsilon,\\
&\Gamma^{123}\epsilon=i\epsilon
\end{align}

Before topological twisting, 
fields and supersymmetry parameter transform under $SU(2)_{E}\times
SO(8)_{R}$ as 
\begin{align}
&X: (\bm{1},\bm{8}_{v})\\
&\Psi: (\bm{2},\bm{8}_{c})\\
&\epsilon: (\bm{2},\bm{8}_{s}).
\end{align}
Although there are many possible ways of the twisting, 
we consider the following decomposition $SO(8)_{R}$ under $SO(3)$ \cite{Slansky:1981yr}
\begin{align}
\label{eblga}
\textrm{A-twist}:& SO(8)\supset SO(5)\times SO(3)\\
\label{eblgb}
\textrm{B-twist}:& SO(8)\supset G_{2}\supset SU(2)_{1}\times SU(2)_{2}.
\end{align}

\subsubsection{A-twist}
Under (\ref{eblga}), we can obtain the decomposition of $\bm{8}_{v}$, $\bm{8}_{s}$ and $\bm{8}_{c}$ as
\begin{align}
&\bm{8}_{v}=(\bm{5},\bm{1})\oplus (\bm{1},\bm{3})\\
&\bm{8}_{s}=(\bm{4},\bm{2})\\
&\bm{8}_{c}=(\bm{4},\bm{2}).
\end{align}
Then the transformations under $SO(3)'_{E}\times SO(5)_{R}$ of the fields are
\begin{align}
&X\rightarrow (\bm{1},\bm{5})\oplus (\bm{3},\bm{1})\\
&\Psi\rightarrow (\bm{1},\bm{4})\oplus (\bm{3},\bm{4})\\
&\epsilon\rightarrow (\bm{1},\bm{4})\oplus (\bm{3},\bm{4}).
\end{align}
The bosonic field content is
\begin{itemize}
 \item 5 scalar fields : $(\bm{1},\bm{5})$
\item 1-form : $(\bm{3},\bm{1})$
\end{itemize}
and the fermionic field content is
\begin{itemize}
 \item 4 scalar fields : $(\bm{1},\bm{4})$
\item 4 1-form : $(\bm{3},\bm{4})$.
\end{itemize}
Therefore there exists four supercharges.

We decompose the gamma matrices under $SO(11)\supset SO(3)_{E}\times
SO(5)\times SO(3)$ as
\begin{align}
\begin{cases}
\Gamma^{i}=\sigma_{i}\otimes \mathbb{I}_{4}\otimes \mathbb{I}_{2}\otimes
 \sigma_{1}\\
\Gamma^{\mu+3}=\mathbb{I}_{2}\otimes \gamma^{\mu}\otimes \mathbb{I}_{2}
\otimes \sigma_{2}\\
\Gamma^{\mu+9}=\mathbb{I}_{2}\otimes \mathbb{I}_{4}\otimes \sigma_{i}
\otimes \sigma_{3}\\
\end{cases}
\end{align}
where $\sigma_{i}$ are Pauli matrices and $\gamma^{\mu}$ are 
five-dimensional gamma matrices satisfying 
\begin{align}
\left\{\gamma^{\mu}, \gamma^{\nu}\right\}=2\delta^{\mu\nu}&\gamma^{5}:=\gamma^{1234}.
\end{align}
The charge conjugation matrix can be expressed as
\begin{align}
\mathcal{C}=\sigma_{2}\otimes C\otimes \sigma_{2}\otimes \mathbb{I}_{2}
\end{align}
where $\sigma_{2}$ is a three-dimensional charge conjugation matrix
\begin{align}
(\sigma_{2})^{T}
=-\sigma_{2}&\sigma_{2}\sigma_{i}\sigma_{2}^{-1}
=-(\sigma_{2})^{T}
\end{align}
and $C$ is a five-dimensional charge conjugation matrix
\begin{align}
(C)^{T}=-C&C\gamma^{\mu}C^{-1}=(\gamma^{\mu})^{T}.
\end{align}
The chirality matrix is given by
\begin{align}
\Gamma^{123}=
=-i\Gamma^{45\cdots 11}
i(\mathbb{I}\otimes \mathbb{I}\otimes \mathbb{I}\otimes \sigma_{1}).
\end{align}

\begin{enumerate}
 \item \textbf{bosonic fields}

For the bosonic fields we redefine
\begin{align}
&\phi^{I}:=X^{I} (I=4,5,6,7,8),\nonumber \\
&\Phi^{\mu}:=X^{\mu+8} (\mu=1,2,3)
\end{align}

\item \textbf{fermionic fields}

We expand the elven-dimensional fermionic fields 
$\Psi$ under the decomposition $SO(11)\supset SO(3)_{E}\times
      SO(5)\times SO(3)$ as
\begin{align}
\Psi_{p\alpha q}
=\frac{1}{\sqrt{2}}
\left(
\psi_{\alpha}\mathbb{I}_{pq}+\Psi_{i\alpha}{\sigma_{ipq}}
\right)\sigma_{2}^{-1}
\end{align}
where $p,q,\alpha$ are indices of $SO(3)_{E}, SO(5), SO(3)$
      respectively. 
$\psi_{\alpha}$ and $\Psi_{i\alpha}$ are scalars $(\bm{1},\bm{4})$ and a
      1-form $(\bm{3},\bm{4})$.

\end{enumerate}

\subsubsection{B-twist}
Under (\ref{eblgb}), we can obtain the decomposition of $\bm{8}_{v}$, $\bm{8}_{s}$ and $\bm{8}_{c}$ as
\begin{align}
&\bm{8}_{v}=\bm{7}+\bm{1}=(\bm{1},\bm{3})\oplus (\bm{2},\bm{2})\oplus (\bm{1},\bm{1})\\
&\bm{8}_{s}=\bm{7}+\bm{1}=(\bm{1},\bm{3})\oplus (\bm{2},\bm{2})\oplus
 (\bm{1},\bm{1}) \\
&\bm{8}_{c}=\bm{7}+\bm{1}=(\bm{1},\bm{3})\oplus (\bm{2},\bm{2})\oplus (\bm{1},\bm{1}).
\end{align}
Choosing the homomorphism as $SU(2)_{E}\rightarrow SU(2)_{1}$, the transformations under $SO(3)'_{E}\times SU(2)_{2}'$ of the fields are
\begin{align}
&X\rightarrow (\bm{1},\bm{3})\oplus (\bm{2},\bm{2})\oplus (\bm{1,\bm{1}})\\
&\Psi\rightarrow (\bm{2},\bm{3})\oplus (\bm{1},\bm{2})\oplus
 (\bm{3},2)\oplus (\bm{2},\bm{1})\\
&\epsilon\rightarrow 
(\bm{2},\bm{3})\oplus (\bm{1},\bm{2})\oplus
 (\bm{3},2)\oplus (\bm{2},\bm{1}).
\end{align}
The bosonic field content is
\begin{itemize}
 \item 4 scalar fields : $(\bm{1},\bm{3})\oplus (\bm{1},\bm{1})$
\item 2 spinor fields : $(\bm{2},\bm{2})$
\end{itemize}
and the fermionic field content is
\begin{itemize}
 \item 2 scalar fields : $(\bm{1},\bm{2})$
\item 4 spinor fields : $(\bm{2},\bm{3})\oplus (\bm{2},\bm{1})$.
\item 2 1-form : $(\bm{3},\bm{2})$.
\end{itemize}
Therefore there exists two supercharges.

\subsection{$d=2$, $\mathcal{N}=(2,2)$ non-linear sigma-model}
There are two types global symmetry in the theory, which are called
$U(1)_{V}$ vector R-symmetry and $U(1)_{A}$ axial R-symmetry 
\footnote{Although $SO(2)_{E}$ rotational symmetry acts on the variables
$z,\overline{z},\theta^{\pm}$ and $\overline{\theta}$ simultaneously, 
one can construct two $U(1)$ groups that act only on a subset of variables
and leave the measure invariant and keep the chiral fields to be chiral
\begin{align}
\label{vecrsym}
U(1)_{V}:&\begin{cases}(\theta^{+},\overline{\theta}^{+})&\rightarrow
 (e^{-i\alpha}\theta^{+}, e^{i\alpha}\overline{\theta}^{+})\cr
(\theta^{-},\overline{\theta}^{-})&\rightarrow 
(e^{-i\alpha}\theta^{+},e^{i\alpha}\overline{\theta}^{+}) \cr
\end{cases}
 \\
\label{axialrsym}
U(1)_{A}:&\begin{cases}(\theta^{+},\overline{\theta}^{+})&\rightarrow
 (e^{-i\alpha}\theta^{+},e^{i\alpha}\overline{\theta}^{+})\cr 
(\theta^{-},\overline{\theta}^{-})&\rightarrow
	   (e^{i\alpha}\theta^{-},e^{-i\alpha}\overline{\theta}^{-})\cr
\end{cases}.
\end{align}
}. 

The field content is
\begin{itemize}
 \item scalar fields $\phi^{I}(z,\overline{z})$
\item fermionic fields $\psi_{+}^{I}(z,\overline{z}),\psi_{-}^{I}(z,\overline{z})$
\end{itemize}
The bosonic field $\phi^{I}(z,\overline{z})$ is a map from 2-dimensional
genus $g$ Riemann surface $\Sigma$ to a target space $X$ of metric
$g$
\begin{align}
\phi^{I}(z,\overline{z}):\Sigma\rightarrow X
\end{align}
where $z,\overline{z}$ are the local coordinates on $\Sigma$.
The fermionic field $\psi_{+}^{I}$ is a section of $K^{\frac12}\otimes
\phi^{*}(TX)$ and $\psi_{-}^{I}$ is a section of $K^{-\frac{1}{2}}\otimes
\phi^{*}(TX)$
\begin{equation}
\label{2d22psi1}
 \psi_{\pm}^{I}(z,\overline{z})\in 
\Gamma(K^{\pm\frac12}\otimes \phi^{*}(TX))
\end{equation}
where  $TX$ is the holomorphic tangent bundle to $X$. 
$K$ and $K^{-1}$ are the canonical and anti-canonical bundle on $\Sigma$
(i.e. the bundle of $(1,0)$ and $(0,1)$ forms)
and $K^{\frac12}$ and $K^{-\frac12}$ are square roots of these. 

Here we consider the case where we have $\mathcal{N}=(2,2)$
supersymmetry, which require that $X$ is K\"{a}hler. 
We denote the local complex coordinates by $\phi^{i}$ and their complex
conjugate by $\phi^{\overline{i}}=\overline{\phi}^{\overline{i}}$. 
As the complexified tangent bundle $TX$ has a decomposition as
$TX=T^{1,0}X\oplus T^{0,1}X$, (\ref{2d22psi1}) becomes
\begin{align}
\label{2d22psi2}
\psi_{+}^{i}&\in \Gamma(K^{\frac12}\otimes \phi^{*}(T^{1,0}X)),\nonumber \\
\psi_{+}^{\overline{i}}&\in \Gamma(K^{\frac12}\otimes
 \phi^{*}(T^{0,1}X)),\nonumber \\
\psi_{-}^{i}&\in \Gamma(K^{-\frac12}\otimes
 \phi^{*}(T^{1,0}X)),\nonumber \\
\psi_{-}^{\overline{i}}&\in \Gamma(K^{-\frac12}\otimes \phi^{*}(T^{0,1}X))
\end{align}
and $\psi_{\pm}^{i}$ and $\psi_{\pm}^{\overline{i}}$ are left- and
right-moving fermionic fields respectively.

The action is
\begin{align}
\label{2d22action}
S=2t\int_{\Sigma}
d^{2}z\left(
\frac{1}{2}g_{IJ}\partial_{z}\phi^{I}\partial_{\overline{z}}\phi^{J}
+ig_{i\overline{j}}\psi_{-}^{\overline{j}}D_{z}\psi_{-}^{i}
+ig_{i\overline{j}}\psi_{+}^{\overline{j}}D_{\overline{z}}\psi_{+}^{i}
+R_{i\overline{j}k\overline{l}}\psi_{+}^{i}\psi_{+}^{\overline{j}}\psi_{-}^{k}\psi_{-}^{\overline{l}}
\right)
\end{align}
where $t$ is a coupling constant or a string tension depending on the
overall volume of $X$ and $R_{i\overline{j}k\overline{l}}$ is the
Riemann tensor of target space $X$.

Originally fields transform under $SO(2)_{E}$
$\times$ $U(1)_{V}\times U(1)_{A}$ as in Table \ref{2d22sigma1} 
\footnote{
From (\ref{vecrsym}) and (\ref{axialrsym}), the vector R-rotations and
the axial R-rotations of superfield are given by
\begin{align}
&e^{i\alpha F_{V}}:\Phi(x^{\mu},\theta^{\pm},\overline{\theta}^{\pm})
\mapsto e^{i\alpha
 q_{V}}\Phi(x^{\mu},e^{-i\alpha}\theta^{\pm},e^{i\alpha}\overline{\theta}^{\pm})\\
&e^{i\beta F_{A}}:\Phi(x^{\mu},\theta^{\pm},\overline{\theta}^{\pm})
\mapsto e^{i\beta q_{A}}\Phi(x^{\mu},e^{\mp i\beta}\theta^{\pm},e^{\pm i\beta}\overline{\theta}^{\pm})
\end{align}
where $F_{V}, F_{A}$ are the generators of the vector and the axial
R-symmetry and $q_{V}, q_{A}$ are the vector and the axial
R-charges respectively. 
Therefore we see that
\begin{align}
&\psi_{+new}^{i}=e^{i\alpha(1-q_{V})}\psi_{+old}^{i},   
&\psi_{+new}^{i}=e^{i\beta(1-q_{A})}\psi_{+old}^{i} 
\\
&\psi_{-new}^{i}=e^{i\alpha(1-q_{V})}\psi_{-old}^{i},   
&\psi_{-new}^{i}=e^{i\beta(-1-q_{A})}\psi_{-old}^{i} 
\\
&\psi_{+new}^{\overline{i}}=e^{i\alpha(-1+q_{V})}\psi_{+old}^{\overline{i}},
&\psi_{+new}^{\overline{i}}=e^{i\beta(-1-q_{A})}\psi_{+old}^{\overline{i}} 
\\
&\psi_{-new}^{\overline{i}}=e^{i\alpha(-1+q_{V})}\psi_{+old}^{\overline{i}},  
&\psi_{-new}^{\overline{i}}=e^{i\alpha(1-q_{A})}\psi_{-old}^{\overline{i}} .
\end{align} 
Setting $q_{V}=q_{A}=0$, we obtain the $U(1)_{V}$ and the $U(1)_{A}$
charges in Table \ref{2d22sigma1}.
}.
\begin{table}
\begin{center}
\begin{tabular}{|c|c|c|c|c|} \hline
&$U(1)_{E}$&$U(1)_{V}$&$U(1)_{A}$&$\mathcal{L}$\\ \hline\hline
$\phi$&$0$&$0$&$0$&$\mathcal{O}$ \\
$\psi_{+}^{i}$&$+$&$-$&$+$&$K^{\frac12}$ \\
$\psi_{-}^{i}$&$-$&$-$&$-$&$K^{-\frac12}$ \\
$\psi_{+}^{\overline{i}}$&$+$&$+$&$-$&$K^{\frac12}$ \\ 
$\psi_{-}^{\overline{i}}$&$-$&$+$&$+$&$K^{-\frac12}$ \\ \hline
\end{tabular}
\caption{$U(1)$ charges of the $d=2$, $\mathcal{N}=(2,2)$ sigma model
 fields. $\mathcal{L}$ is the complex line bundle on $\Sigma$ in which
 the fields take values. $\mathcal{O}$ is the trivial bundle and
 $K$ is the canonical bundle.}
\label{2d22sigma1}
\end{center}
\end{table} 
Likewise supersymmetry generators transform as Table \ref{2d22sigma11}.
\begin{table}
\begin{center}
\begin{tabular}{|c|c|c|c|c|} \hline
&$U(1)_{E}$&$U(1)_{V}$&$U(1)_{A}$&$\mathcal{L}$\\ \hline\hline
$Q_{+}$&$+$&$-$&$+$&$K^{\frac12}$ \\
$Q_{-}$&$-$&$-$&$-$&$K^{-\frac12}$ \\
$\overline{Q}_{+}$&$+$&$+$&$-$&$K^{\frac12}$ \\ 
$\overline{Q}_{-}$&$-$&$+$&$+$&$K^{-\frac12}$ \\ \hline
\end{tabular}
\caption{$U(1)$ charges of the $d=2$, $\mathcal{N}=(2,2)$ sigma model
 supersymmetry generators.}
\label{2d22sigma11}
\end{center}
\end{table} 
Depending on which R-symmetry we use, there are two homomorphisms for the twisting
\begin{align}
\textrm{A-twist}:&U(1)_{E}\rightarrow U(1)_{V}\\
\textrm{B-twist}:&U(1)_{E}\rightarrow U(1)_{A}.
\end{align}
The results of the twisting are summarized in Table \ref{2d22sigma2} and
Table \ref{2d22sigma22}.
\begin{table}
\begin{center}
\begin{tabular}{|c|cc|cc|} \hline
 &\multicolumn{2}{|c|}{A-twist}&\multicolumn{2}{|c|}{B-twist}\\
fields&$U(1)'_{E}$&$\mathcal{L}$&$U(1)_{E}'$&$\mathcal{L}$ \\ \hline\hline
$\phi$&$0$&$\mathcal{O}$&$0$&$\mathcal{O}$ \\
$\psi_{+}^{i}$&$2$&$K$&$2$&$K$ \\
$\psi_{-}^{i}$&$0$&$\mathcal{O}$&$-2$&$K^{-1}$ \\
$\psi_{+}^{\overline{i}}$&$0$&$\mathcal{O}$&$0$&$\mathcal{O}$ \\ 
$\psi_{-}^{\overline{i}}$&$-2$&$K^{-1}$&$0$&$\mathcal{O}$ \\ \hline
\end{tabular}
\caption{The spin of field for A-twisted and B-twisted $d=2$, $\mathcal{N}=(2,2)$ sigma model.}
\label{2d22sigma2}
\end{center}
\end{table}
\begin{table}
\begin{center}
\begin{tabular}{|c|cc|cc|} \hline
 &\multicolumn{2}{|c|}{A-twist}&\multicolumn{2}{|c|}{B-twist}\\
fields&$U(1)'_{E}$&$\mathcal{L}$&$U(1)_{E}'$&$\mathcal{L}$ \\ \hline\hline
$Q_{+}$&$2$&$K$&$2$&$K$ \\
$Q_{-}$&$0$&$\mathcal{O}$&$-2$&$K^{-1}$ \\
$\overline{Q}_{+}$&$0$&$\mathcal{O}$&$0$&$\mathcal{O}$ \\ 
$\overline{Q}_{-}$&$-2$&$K^{-1}$&$0$&$\mathcal{O}$ \\ \hline
\end{tabular}
\caption{The spin of the supersymmetry generators 
for A-twisted and B-twisted $d=2$, $\mathcal{N}=(2,2)$ sigma model.}
\label{2d22sigma22}
\end{center}
\end{table}

\subsubsection{A-model}
After performing A-twist, the bundles in which the fermionic fields take
values are modified as
\begin{align}
\label{amodelbdl}
\psi_{z}^{i}&\in \Gamma(K\otimes \phi^{*}(T^{1,0}X)),\nonumber \\
\psi^{i}&\in \Gamma(\phi^{*}(T^{0,1}X)),\nonumber \\
\psi^{\overline{i}}&\in \Gamma(\phi^{*}(T^{1,0}X)),\nonumber \\
\psi^{\overline{i}}_{\overline{z}}&\in\Gamma(K^{-1}\otimes \phi^{*}(T^{0,1}X)).
\end{align}

\subsubsection{B-model}
B-twist changes the bundles in which fermionic fields take values as
\begin{align}
\label{bmodelbdl}
\psi_{z}^{i}&\in \Gamma(K\otimes \phi^{*}(T^{1,0}X)),\nonumber \\
\psi^{\overline{i}}&\in \Gamma(\phi^{*}(T^{0,1}X)),\nonumber \\
\psi_{\overline{z}}^{i}&\in \Gamma(K^{-1}\otimes
 \phi^{*}(T^{1,0}X)),\nonumber \\
\psi^{\overline{i}}&\in \Gamma(\phi^{*}(T^{0,1}X)).
\end{align}

These topological twisted theories 
are known as A-model and B-model topological sigma-models, or 
topological string theories
\cite{Witten:1988xj,Witten:1989ig,Witten:1990hr,Witten:1991zz} 
\footnote{See \cite{Hori:2003ic} for the detailed review on the
topological string theory.}.

\section{Curved branes and twisted theories}
\label{twistsec02}
Let us consider the gauge theories arising from the dimensional
reduction of ten-dimensional $\mathcal{N}=1$ SYM theory to $(p+1)$
dimensions. 
It is known that these theories describe the low energy world-volume dynamics of
flat D$p$-branes \cite{Witten:1995im}. 

On the other hand, 
when one consider curved branes wrapping around a non-trivial cycle $\mathcal{C}$ in the ambient
manifold $X$, the cycle has to be identified with a calibrated
submanifold and satisfy some stringent conditions to preserve some fraction of the supersymmetries.

As discussed in \cite{Bershadsky:1995qy}, curved world-brane theories are
obtained by topological twisting along the directions where the
world-volume is curved. 
To see this, we remember that 
the bosonic scalar fields are associated with translations of the
D-brane. 
Thus when D-brane wrap around curved cycle $\mathcal{C}$ in $X$, there are only $(10-\dim X)$ actual scalar fields and 
the other translational modes are identified with the section of the
normal bundle $N_{\mathcal{C}}$ to $\mathcal{C}$ in $X$. 
Therefore these modes should be twisted if the normal bundle is
non-trivial and so are their superpartners. 

From the above observations, for given supersymmetric cycles
$\mathcal{C}$ and their ambient manifolds $X$, one can determine
\begin{enumerate}
 \item the bosonic field content
\item the number of scalar supercharges
\end{enumerate}
of the world-volume topological gauge theories of D-branes. 
On the contrary, one can check whether there exists supersymmetric
cycles with the required properties for given topological gauge
theories. 

Noting that there is a global invariance under the rotational 
$SO(10-\dim X)$ symmetry \footnote{
When one considers Euclidean D-branes, 
the invariant rotational symmetry
is $SO(1,9-\dim X)$ because curved D-branes do not wrap the time
direction.} 
of the uncompactified dimensions, the original 
R-symmetry $SO(9-p)$ should be decomposed as
\begin{equation}
\label{branching1}
 SO(9-p)\subset SO(10-\dim X)\times SO(\dim X-p-1).
\end{equation}
Then, under the branching (\ref{branching1}) of R-symmetry, we try to perform
topological twisting by using the second factor $SO(\dim X-p-1)$
corresponding to the normal bundle $N_{\mathcal{C}}$ in $X$. 
A relevant information is given by Table \ref{ambmfd}.

\begin{table}
\begin{center}
\begin{tabular}{|c|c|c|c|} \hline
ambient manifolds (dimensions)&holonomy&submanifolds&SUSY\\ \hline \hline
Calabi-Yau 2-fold (4)&$SU(2)\subset SO(4)$&holomorphic curve
	 (2)&$\frac12$\\ 
Calabi-Yau 3-fold (6)&$SU(3)\subset SO(6)$&Lagrangian (3)&$\frac14$\\
$G_{2}$ manifold (7)&$G_{2}\subset SO(7)$&coassociative (4)&$\frac18$\\
&&associative (3)&$\frac18$ \\
$Spin(7)$ manifold (8)&$Spin(7)\subset SO(8)$&Cayley
	 (4)&$\frac{1}{16}$\\
Calabi-Yau 4-fold (8)&$SU(4)\subset SO(8)$&Lagrangian (4)&$\frac18$\\
Hyperk\"{a}hler manifold (8)&$Sp(2)\subset SO(8)$& &$\frac{3}{16}$\\
$CY_{2}\times CY_{2}$ (8)&$SU(2)\times SU(2)\subset SO(8)$& &$\frac{1}{4}$\\
Calabi-Yau 5-fold (10)&$SU(5)\subset SO(10)$&Lagrangian (5)&$\frac{1}{16}$\\ \hline
\end{tabular}
\caption{The ambient manifolds and the examples of calibrated submanifolds that preserve
 the fraction of supersymmetry. Note that all of the Calabi-Yau manifolds
 include holomorphic submanifolds as calibrated
 submanifolds.}
\label{ambmfd}
\end{center}
\end{table}

The preserved fraction of supersymmetries are derived as follows. 
The holonomy group of K3 surface is $SU(2)$, so the spinor of $SO(4)$ is
 decomposed under $SO(4)\supset SU(2)_{H}\times SU(2)\times U(1)\supset
 SU(2)_{H}$ as
\begin{align}
\bm{4}=(\bm{2},\bm{1})_{-}\oplus (\bm{1},\bm{2})_{+}
=\bm{2}\oplus \bm{1}\oplus \bm{1}.
\end{align}
Thus 2 of 4 supercharges are constant spinors and 
we have $\frac12$ BPS background.

The holonomy group of Calabi-Yau 3-fold is $SU(3)$, so the spinor of $SO(6)$ is
 decomposed under $SO(6)\supset SU(3)$ as
\begin{align}
\bm{4}=\bm{3}\oplus \bm{1}.
\end{align}
Thus 1 of 4 supercharges is constant spinor and 
we have $\frac14$ BPS background.

The holonomy group of Calabi-Yau 4-fold is $SU(4)$, so the spinor of $SO(8)$ is
 decomposed under $SO(8)\supset SU(4)$ as
\begin{align}
\bm{8}_{s}\oplus \bm{8}_{c}=
\bm{6}\oplus \bm{1}\oplus \bm{1}\oplus \bm{4}\oplus \overline{\bm{4}}
\end{align}
Thus 2 of 16 supercharges is constant spinor and 
we have $\frac18$ BPS background.

The holonomy group of $G_{2}$ manifold is $G_{2}$, so the spinor of $SO(7)$ is
 decomposed under $SO(7)\supset G_{2}$ as
\begin{align}
\bm{8}=\bm{1}\oplus \bm{7}
\end{align}
Thus 1 of 8 supercharges is constant spinor and 
we have $\frac18$ BPS background.

The holonomy group of $Spin(7)$ manifold is $Spin(7)$, so the spinor of $SO(8)$ is
 decomposed under $SO(8)\supset Spin(7)$ as
\begin{align}
\bm{8}_{s}\oplus \bm{8}_{c}=
\bm{7}\oplus \bm{1}\oplus \bm{8}
\end{align}
Thus 1 of 16 supercharges is constant spinor and 
we have $\frac{1}{16}$ BPS background.

The holonomy group of Calabi-Yau 5-fold is $SU(5)$, so the spinor of $SO(10)$ is
 decomposed under $SO(10)\supset SU(5)$ as
\begin{align}
\bm{16}\oplus \bm{16}'
=\bm{1}_{-5}\oplus \overline{\bm{5}}_{3}\oplus \bm{10}_{-}
\oplus \bm{1}_{5}\oplus \bm{5}_{-3}\oplus \bm{10}_{+}
\end{align}
Thus 2 of 32 supercharges is constant spinor and 
we have $\frac{1}{16}$ BPS background.

\subsection{D3-branes and twisted $d=4$, $\mathcal{N}=4$ SYM theories}
The first example is the 
low-energy effective field theories of 
the D3-branes wrapped on curved four-manifold. 
A set of these descriptions can be obtained by 
the three distinct topologically twisted $d=4$, $\mathcal{N}=4$ SYM theories. 
\begin{enumerate}

\item \textbf{GL twist D-brane}

The fact that $\dim\mathcal{C}=4$ and that there are two scalar fields
means that the theory describes 4-cycle $\mathcal{C}$ in
$4+(6-2)=8$-dimensional manifold $X$. 
The existence of two preserved BRST charges indicates that 8-manifold
preserve $\frac{2}{16}=\frac{1}{8}$ of the supersymmetry. 
From the above facts and Table \ref{ambmfd}, 
$X$ is a Calabi-Yau 4-fold and $\mathcal{C}$ is a special Lagrangian
submanifold\footnote{Special Lagrangian submanifold is a submanifold
for which the real part of the holomorphic form restricts to the
volume form on the submanifold.}.

Moreover it it known that in the case where special Lagrangian
submanifold is embedded in Calabi-Yau 4-fold, the normal bundle
$N_{\mathcal{C}}$ can be identified with the cotangent bundle
      $T_{\mathcal{C}}^{*}$ \cite{MR1664890}.
This is consistent to the fact that the remaining four scalar fields
combine to form one 1-form on $\mathcal{C}$.

Note that a global $U(1)$ ghost number symmetry corresponds to the
      rotational symmetry of the two uncompactified dimensions. 
The two scalars having opposite $U(1)$ charges are identified
      with the 2-dimensional vector and the 1-form is a $U(1)$-singlet.

\item \textbf{VW twist D-brane}

The fact that $\dim\mathcal{C}=4$ and that there are three scalar fields
means that the theory describes 4-cycle $\mathcal{C}$ in
$4+(6-3)=7$-dimensional manifold $X$. 
The existence of two preserved BRST charges indicates that 7-manifold
preserve $\frac{2}{16}=\frac{1}{8}$ of the supersymmetry. 
From the above facts and Table \ref{ambmfd}, 
$X$ is a $G_{2}$ manifold and $\mathcal{C}$ is a coassociative
submanifold. 

It is known that for a coassociative 4-submanifold in
      $G_{2}$ manifold, the normal bundle is $(\bm{1},\bm{3})_{0}$
      \cite{MR1664890}. This is consistent to the results obtained by
      the twisting.

\item \textbf{DW twist D-brane}

The fact that $\dim\mathcal{C}=4$ and that there are two scalar fields
means that the theory describes 4-cycle $\mathcal{C}$ in
$4+(6-2)=8$-dimensional manifold $X$. 
The existence of one preserved BRST charge indicates that 8-manifold
preserve $\frac{1}{16}=\frac{1}{16}$ of the supersymmetry. 
From the above facts and Table \ref{ambmfd}, 
$X$ is a $Spin(7)$ manifold and $\mathcal{C}$ is a Cayley submanifold.

It it known that for the $Spin(7)$ manifold the normal bundle is
      $S_{+}\oplus V$ where $S_{+}$ is a spin bundle of a given
      chirality and $V$ is a 2-dimensional bundle
      \cite{Bershadsky:1995qy}. When $V$ is trivial, this becomes
      $S_{+}\oplus S_{+}$, that is $(\bm{1},\bm{2})_{+}\oplus
      (\bm{1},\bm{2})_{-}$.

\end{enumerate}
These results are summarized in Table \ref{d34dsym}

\begin{table}
\begin{center}
\begin{tabular}{|c|c|c|c|} \hline
twist&submanifold (dimension)&ambient manifold (dimension)&SUSY\\ \hline \hline
GL twist&Lagrangian (4)&Calabi-Yau 4-fold (8)&$\frac{16}{8}=2$\\ 
VW twist&coassociative (4)&$G_{2}$ manifold (7)&$\frac{16}{8}=2$\\
DW twist&Cayley (4)&$Spin(7)$ manifold (8)&$\frac{16}{16}=1$\\ \hline
\end{tabular}
\caption{Three types of topological twists for $d=4$, $\mathcal{N}=4$ SYM
 theories, curved D3-branes (submanifolds) and
 ambient manifolds.}
\label{d34dsym}
\end{center}
\end{table}

\subsection{D2-branes and twisted $d=3$, $\mathcal{N}=8$ SYM theories}
The D2-branes wrapped on three-manifold are 
given by the topologically twisted $d=3$, $\mathcal{N}=8$ SYM theories. 
\begin{enumerate}
 \item \textbf{A-twist}

The fact that $\dim\mathcal{C}=3$ and that there are four scalar fields
       means that the theory describes 3-cycle $\mathcal{C}$ in
       $3+(7-4)=6$-dimensional manifold $X$. The existence of the four
       preserved BRST charges indicates that 6-manifold preserve
       $\frac{4}{16}=\frac{1}{4}$ of the supersymmetry. From the above
       facts and Table \ref{ambmfd}, $X$ is a Calabi-Yau 3-fold and
       $\mathcal{C}$ is a special Lagrangian submanifold. 

Also it is
       known that the normal bundle $N_{\mathcal{C}}$ can be identified
       with the cotangent bundle $T^{*}_{\mathcal{C}}$
       \cite{MR1664890}. This is consistent to the fact that the
       remaining three scalar fields combine to form one 1-form on
       $\mathcal{C}$. 

A global $SU(2)_{1}\times SU(2)_{2}\simeq SU(4)$
       ghost number symmetry corresponds to the rotational symmetry of
       the four uncompactified dimensions. The four scalars transform as
       a $\bm{4}_{v}$ of $SO(4)$ and the 1-form is an $SO(4)$-singlet.

\item \textbf{B-twist}

The fact that $\dim\mathcal{C}=3$ and that there are three scalar fields
       means that the theory describes 3-cycle $\mathcal{C}$ in
       $3+(7-3)=7$-dimensional manifold $X$. The existence of the two
       preserved BRST charges indicates that 7-manifold preserve
       $\frac{2}{16}=\frac{1}{8}$ of the supersymmetry. From the above
       facts and Table \ref{ambmfd}, $X$ is a $G_{2}$ manifold and
       $\mathcal{C}$ is an associative submanifold. 

Also it is
       known that for an associative 3-submanifold in $G_{2}$ manifold,
       the normal bundle is $N_{\mathcal{C}}=S\otimes V$ where $S$ is a
       spinor bundle of $\mathcal{C}$ and $V$ is a rank two
       $SU(2)$-bundle. This is consistent to the fact that the twisted
       bosonic spinors $(\bm{2},\bm{2},\bm{1})$ are an $SU(2)$-doublet
       of spinors on $\mathcal{C}$. 

Again a global $SU(2)_{3}\simeq SO(3)$ symmetry corresponds to the
       rotational symmetry of the four uncompactified dimensions. The
       three scalars transform as a $\bm{3}_{v}$ of $SO(3)$ and the
      twisted bosonic spinors $(\bm{2},\bm{2},\bm{1})$ are $SO(3)$-singlet. 

\end{enumerate}

These results are summarized in Table \ref{d23dsym}

\begin{table}
\begin{center}
\begin{tabular}{|c|c|c|c|} \hline
twist&submanifold (dimension)&ambient manifold (dimension)&SUSY\\ \hline \hline
A-twist&Lagrangian (3)&Calabi-Yau 3-fold (6)&$\frac{16}{4}=4$\\ 
B-twist&associative (3)&$G_{2}$ manifold (7)&$\frac{16}{8}=2$\\ \hline
\end{tabular}
\caption{Two types of topological twists for $d=3$, $\mathcal{N}=8$ SYM
 theories, curved D2-branes (submanifolds) and
 ambient manifolds.}
\label{d23dsym}
\end{center}
\end{table}

\subsection{D2-branes and twisted $d=3$, $\mathcal{N}=8$ SYM theories on
  $\mathbb{R}\times \Sigma$}
\label{d2curvesubsec1}
The low-energy effective theories of 
the D2-branes wrapping on the holomorphic Riemann surface $\Sigma$ 
are the partially twisted $d=3$, $\mathcal{N}=8$ SYM theories: 
\begin{enumerate}
 \item \textbf{A-twist}

The fact that $\dim\Sigma=2$ and that there are five scalar fields
       means that the theory describes 2-cycle $\Sigma$ in
       $2+(7-5)=4$-dimensional manifold $X$. The existence of the eight
       preserved BRST charges indicates that 4-manifold preserve
       $\frac{8}{16}=\frac{1}{2}$ of the supersymmetry. From the above
       facts and Table (\ref{ambmfd}), $X$ is a K3 surface and
       $\Sigma$ is a holomorphic curve.

\item \textbf{B-twist}

The fact that $\dim\Sigma=2$ and that there are three scalar fields
       means that the theory describes 2-cycle $\Sigma$ in
       $2+(7-3)=6$-dimensional manifold $X$. The existence of the four
       preserved BRST charges indicates that 6-manifold preserve
       $\frac{4}{16}=\frac{1}{4}$ of the supersymmetry. From the above
       facts and Table (\ref{ambmfd}), $X$ is a Calabi-Yau 3-fold and
       $\Sigma$ is a holomorphic curve.

\item \textbf{C-twist}

The fact that $\dim\Sigma=2$ and that there are one scalar field
       means that the theory describes 2-cycle $\Sigma$ in
       $2+(7-1)=8$-dimensional manifold $X$. The existence of the two
       preserved BRST charges indicates that 6-manifold preserve
       $\frac{2}{16}=\frac{1}{8}$ of the supersymmetry. From the above
       facts and Table (\ref{ambmfd}), $X$ is a Calabi-Yau 4-fold and
       $\Sigma$ is a holomorphic curve. 

\end{enumerate}

These results are summarized in Table \ref{d2sy3}

\begin{table}
\begin{center}
\begin{tabular}{|c|c|c|c|} \hline
twist&submanifold (dimension)&ambient manifold (dimension)&SUSY\\ \hline \hline
A-twist&holomorphic (2)&K3 surface (4)&$\frac{16}{2}=8$\\ 
B-twist&holomorphic (2)&Calabi-Yau 3-fold (6)&$\frac{16}{4}=4$\\
C-twist&holomorphic (2)&Calabi-Yau 4-fold (8)&$\frac{16}{8}=2$\\ \hline
\end{tabular}
\caption{Three types of topological twists for $d=3$, $\mathcal{N}=8$ SYM
 theories on $\mathbb{R}\times \Sigma$, curved D2-branes (submanifolds) and
 ambient manifolds.}
\label{d2sy3}
\end{center}
\end{table}

\subsection{Relationship between $d=4$ and $d=3$ twists}
\label{4d3dtwistsubsec1}
The $d=4$ twisting and the $d=3$ twisting are connected 
via dimensional reduction. 
\begin{enumerate}
 \item \textbf{DW twist and B-twist}

Let us define Cayley 4-form in local coordinates $\mathbb{R}^{8}$ as
       \cite{MR916718, MR1664890}
\begin{align}
\Omega_{Cayley}:=dx^{0123}
+&\left(dx^{01}-dx^{23}\right)
\wedge \left(
dx^{45}-dx^{67}
\right)\nonumber \\
+&\left(dx^{02}+dx^{13}\right)
\wedge \left(dx^{46}+dx^{57}\right)\nonumber \\
+&\left(
dx^{03}-dx^{12}
\right)\wedge \left(
dx^{47}-dx^{56}
\right)+dx^{4567}
\end{align}
where $dx^{ijk}:=dx^{i}\wedge dx^{j}\wedge dx^{k}$, etc. 
Then one can define $Spin(7)$ manifold to be the subgroup of $GL(8)$ that
       preserve $\Omega_{Cayley}$. Integrating over the fibre $x^{0}$,
       we obtain
\begin{align}
\pi_{*}\Omega_{Cayley}
=dx^{123}+&dx^{1}\wedge \left(dx^{45}-dx^{67}\right)\nonumber \\
+&dx^{2}\wedge \left(
dx^{46}+dx^{57}
\right)\nonumber \\
+&dx^{3}\wedge \left(
dx^{47}-dx^{56}
\right)
\end{align}
On the other hand, the associative 3-form $\Omega_{ass}$ characterizing
       associative 3-manifolds of $G_{2}$ manifolds is defined as \cite{MR916718}
\begin{align}
\Omega_{ass}:=
dx^{456}+&dx^{4}\wedge \left(dx^{01}-dx^{23}\right)\nonumber \\
+&dx^{5}\wedge \left(
dx^{02}+dx^{13}
\right)\nonumber \\
+&dx^{6}\wedge \left(
dx^{03}-dx^{12}
\right)
\end{align}
Thus 
\begin{equation}
 \pi_{*}\Omega_{Cayley}=\Omega_{ass}.
\end{equation} 
Therefore DW twist is related to B twist by the dimensional reduction.

\item \textbf{VW twist and A-twist}

VW twist theory corresponds to coassociative
      submanifolds of $G_{2}$ manifolds characterized by the Hodge dual
      4-form $\Omega_{coass}=*\Omega_{ass}$, which is expressed as
      \cite{MR1664890}
\begin{align}
\Omega_{coass}
=dx^{0123}
&-dx^{56}
\wedge \left(
dx^{01}-dx^{23}
\right)\nonumber \\
&+dx^{46}
\wedge \left(
dx^{02}+dx^{13}
\right) \nonumber \\
&-dx^{45}\wedge
\left(dx^{03}-dx^{12}\right).
\end{align}
Integrating this over $x^{0}$, one obtains
\begin{equation}
 \pi_{*}\Omega_{coass}
=dx^{123}-dx^{156}+dx^{246}
-dx^{345}.
\end{equation}
On the other hand, the holomorphic volume form of a
      Calabi-Yau 3-fold characterizing special Lagrangian submanifolds
      is 
\begin{align}
\Omega_{slag}
=&dz^{1}\wedge dz^{2}\wedge dz^{3}\nonumber \\
=&\left(
dx^{123}-dx^{453}-dx^{156}-dx^{426}
\right)\nonumber+i\left(
dx^{423}+dx^{513}+dx^{612}-dx^{456}
\right).
\end{align}
Thus
\begin{equation}
 \pi_{*}\Omega_{coass}=\textrm{Re}\Omega_{slag}.
\end{equation}
Therefore VW twist is associated with A twist by the dimensional reduction.

\item \textbf{GL twist and A-twist}

Suppose that the Calabi-Yau 4-fold is locally of the form
\begin{equation}
 CY_{4}=CY_{3}\times T^{2}.
\end{equation}
Then special Lagrangian of $CY_{4}$ wrapping around one of the circles
      reduce to special Lagrangian submanifolds of $CY_{3}$ by the
      double dimensional reduction. Therefore GL twist is related to
      A-twist by the double dimensional reduction.
\end{enumerate}

\subsection{D1-branes and twisted  $d=2$, $\mathcal{N}=8$ SYM theories}
\label{d1curvesubsec1}
The world-volume theories 
of the D1-branes wrapped on holomorphic Riemann surfaces 
are topologically twisted $d=2$, $\mathcal{N}=8$ SYM theories. 
\begin{enumerate}
 \item \textbf{A-twist}

The fact that $\dim\mathcal{C}=2$ and that there are six scalar fields
       means that the theory describes 2-cycle $\mathcal{C}$ in
       $2+(8-6)=4$-dimensional manifold $X$. The existence of the eight
       preserved BRST charges indicates that 4-manifold preserve
       $\frac{8}{16}=\frac{1}{2}$ of the supersymmetry. From the above
       facts and Table \ref{ambmfd}, $X$ is a K3 surface and
       $\mathcal{C}$ is a holomorphic curve. 

Let us consider the normal bundle $N_{\mathcal{C}}$. 
Noting that 
\begin{align}
&T_{X}=T_{\mathcal{C}}\oplus N_{\mathcal{C}},\\
&c_{1}(T_{X})=0
\end{align}
for holomorphic genus $g$ curve $\mathcal{C}$ in Calabi-Yau $n$-folds
       $X$, we see that
\begin{equation}
 c_{1}(N_{\mathcal{C}})=-c_{1}(T_{\mathcal{C}})=2g-2.
\end{equation}
Alternatively as $\wedge^{n}T_{X}$ is trivial, 
one has
\begin{equation}
 \wedge^{n}T_{X}=T_{\mathcal{C}}\wedge^{n-1}N_{\mathcal{C}}=1,
\end{equation}
which gives the condition of the canonical bundle $K_{\mathcal{C}}$ on
       $\mathcal{C}$
\begin{align}
\label{normalbdl1}
\wedge^{n-1}N_{\mathcal{C}}=K_{\mathcal{C}}
\end{align}
because $T_{\mathcal{C}}=K_{\mathcal{C}}^{-1}$. 

If $X$ is a K3 surface and $\mathcal{C}$ is a holomorphic curve, 
then $n=2$ and $N_{\mathcal{C}}$ has rank one and (\ref{normalbdl1})
       becomes
\begin{equation}
\label{normalbdl2}
 N_{\mathcal{C}}=K_{\mathcal{C}}.
\end{equation}
This is consistent to the fact that remaining two scalar fields combine
       to form a single one-form on $\mathcal{C}$.

A global $SO(6)_{R}$ ghost number symmetry corresponds to the rotational symmetry of
       the six uncompactified dimensions. The six scalars transform as
       a $\bm{6}_{v}$ of $SO(6)$ and the one-form is an $SO(6)$-singlet.

\item \textbf{B-twist}

The fact that $\dim\mathcal{C}=2$ and that there are four scalar fields
       means that the theory describes 2-cycle $\mathcal{C}$ in
       $2+(8-4)=6$-dimensional manifold $X$. The existence of the four
       preserved BRST charges indicates that 6-manifold preserve
       $\frac{4}{16}=\frac{1}{4}$ of the supersymmetry. From the above
       facts and Table \ref{ambmfd}, $X$ is a Calabi-Yau 3-fold and
       $\mathcal{C}$ is an holomorphic curve. 

In this case (\ref{normalbdl1}) becomes
\begin{equation}
\label{normalbdl3}
 \wedge^{2}N_{\mathcal{C}}=K_{\mathcal{C}}
\end{equation}
and generally this is solved by
\begin{equation}
 N_{\mathcal{C}}=K^{\frac12}_{\mathcal{C}}\otimes V
\end{equation}
where $V$ is a rank two bundle with trivial determinant.

\end{enumerate}

These results are summarized in Table \ref{d12dsym}

\begin{table}
\begin{center}
\begin{tabular}{|c|c|c|c|} \hline
twist&submanifold (dimension)&ambient manifold (dimension)&SUSY\\ \hline \hline
A-twist&holomorphic curve (2)&K3 surface (4)&$\frac{16}{2}=8$\\ 
B-twist&holomorphic curve (2)&Calabi-Yau 3-fold (6)&$\frac{16}{4}=4$\\ \hline
\end{tabular}
\caption{Two types of topological twists for $d=2$, $\mathcal{N}=8$ SYM
 theories, curved D1-branes (submanifolds) and
 ambient manifolds.}
\label{d12dsym}
\end{center}
\end{table}

\subsection{M2-branes and twisted BLG theory}
\label{m2euclid1}
The low-energy description of the two M2-branes wrapping curved three-fold 
are as follows: 
\begin{enumerate}
 \item \textbf{A-twist}

The fact that $\dim\mathcal{C}=3$ and that there are five scalar fields
       means that the theory describes 3-cycle $\mathcal{C}$ in
       $3+(8-5)=6$-dimensional manifold $X$. The existence of the four
       preserved BRST charges indicates that 6-manifold preserve
       $\frac{4}{16}=\frac{1}{4}$ of the supersymmetry. From the above
       facts and Table \ref{ambmfd}, $X$ is a Calabi-Yau 3-fold and
       $\mathcal{C}$ is a special Lagrangian submanifold. 

Also it is
       known that the normal bundle $N_{\mathcal{C}}$ can be identified
       with the cotangent bundle $T^{*}_{\mathcal{C}}$
       \cite{MR1664890}. This is consistent to the fact that the
       remaining three scalar fields combine to form one 1-form on
       $\mathcal{C}$. 

A global $SO(5)$
       ghost number symmetry corresponds to the rotational symmetry of
       the four uncompactified dimensions. The five scalars transform as
       a $\bm{5}_{v}$ of $SO(5)$ and the 1-form is an $SO(5)$-singlet.

\item \textbf{B-twist}

The fact that $\dim\mathcal{C}=3$ and that there are three scalar fields
       means that the theory describes 3-cycle $\mathcal{C}$ in
       $3+(8-4)=7$-dimensional manifold $X$. The existence of the two
       preserved BRST charges indicates that 7-manifold preserve
       $\frac{2}{16}=\frac{1}{8}$ of the supersymmetry. From the above
       facts and Table \ref{ambmfd}, $X$ is a $G_{2}$ manifold and
       $\mathcal{C}$ is an associative submanifold. 



\end{enumerate}

These results are summarized in Table \ref{m2instanton1} and same as
that of D2-brane instantons (Table \ref{d23dsym}).

\begin{table}
\begin{center}
\begin{tabular}{|c|c|c|c|} \hline
twist&submanifold (dimension)&ambient manifold (dimension)&SUSY\\ \hline \hline
A-twist&Lagrangian (3)&Calabi-Yau 3-fold (6)&$\frac{16}{4}=4$\\ 
B-twist&associative (3)&$G_{2}$ manifold (7)&$\frac{16}{8}=2$\\ \hline
\end{tabular}
\caption{Two types of topological twists for BLG model, curved M2-branes (submanifolds) and
 ambient manifolds.}
\label{m2instanton1}
\end{center}
\end{table}

\chapter{Curved M2-branes and Topological Twisting}
\label{seccurv}
In this chapter 
we will return to the study of the M2-branes 
and discuss that the topologically twisted $\mathcal{A}_{4}$ BLG-model 
may describe the two wrapped M2-branes around a holomorphic Riemann
surface in Calabi-Yau manifold based on the work of \cite{Okazaki:2014sga}. 
We will study the preserved supersymmetry on the wrapped branes around 
a holomorphic Riemann surface inside a Calabi-Yau manifold in 
section \ref{seccurv1}, \ref{seccurv1a} and \ref{seccurv1b}. 
In section \ref{seccurv2} we will specify 
the appropriate twisting procedures for our wrapped M2-branes.

\section{M2-branes wrapping a holomorphic curve}
\label{seccurv1}
Now we are ready to discuss the M2-branes wrapping curved Riemann surface. 
Recall that the BLG action (\ref{blglagrangian}) 
and the ABJM action (\ref{abjmlag1}) are conjectured to  
describe the dynamics 
of probe multiple M2-branes moving in 
a fixed background geometry characterized by an $SO(8)$ 
and an $SU(4)$ holonomy respectively.  
In these cases, 
the world-volume $M_{3}$ is taken as 
a flat space-time, 
that is $\mathbb{R}^{1,2}$ or $\mathbb{R}\times T^{2}$.   
In the following we will consider more general situations 
where curved M2-branes reside in some fixed curved background
geometries, as discussed in the previous chapter. 
When one naively puts the theory on a generic three-dimensional manifold, 
one cannot preserve supersymmetries.  
In order to retain the partial supersymmetry, 
we shall wrap the M2-branes around 
a Riemann surface $\Sigma_{g}$ of genus $g$ 
(i.e. supersymmetric two-cycles) as the form
\begin{equation}
\label{m2geo}
 M_{3}=\mathbb{R}\times (\Sigma_{g}\subset X)
\end{equation}
where $\mathbb{R}$ is a time direction 
and $X$ is a real $2(n+1)$-dimensional space 
such that it preserves supersymmetry 
and contain no non-trivial three-form gauge field. 
In this setup 
holomorphic two-cycles in Calabi-Yau spaces are
the only known supersymmetric two-cycles, 
i.e. calibrated two-cycles, in special holonomy manifolds. 
The calibrations for them are K\"{a}hler calibrations. 
Therefore we will consider the ambient space $X$ as an $(n+1)$-dimensional 
Calabi-Yau manifold and take the other transverse space as flat Euclidean. 
Thus the M-theory geometry we are considering takes the following form; 
\begin{align} 
\label{m2geo1}
\mathbb{R}^{1,8-2n}\times  CY_{n+1}.
\end{align}

\section{Supersymmetry in Calabi-Yau space}
\label{seccurv1a}
As a first step to count the number of preserved supersymmetries 
in our setup, 
one needs to count the dimension of the vector space 
formed by the corresponding Killing spinor $\epsilon$, 
which yields the amount of supersymmetries in the background geometry. 
For the eleven-dimensional background geometries 
without non-trivial four-form flux, 
the Killing spinor equation is given by
\begin{align}
\label{killspinor1}
\nabla_{M}\epsilon=
\left(\partial_{M}+\frac14 \omega_{MPQ}\Gamma^{PQ}\right)\epsilon=0.
\end{align}
Here $\omega_{MPQ}$, 
$M,N,P,Q=0,1,\cdots,10$ denotes an eleven-dimensional 
Levi-Civita spin connection. 
The integrability condition reads
\begin{align}
\label{killspinor2}
[\nabla_{M},\nabla_{N}]\epsilon
=\frac14 R_{MNPQ}\Gamma^{PQ}\epsilon=0, 
\end{align}
This implies that 
a Killing spinor $\epsilon$ transforms as a trivial representation, 
i.e. singlet under the 
restricted holonomy group $H\subset Spin(1,10)$ 
generated by the generator $R_{MNPQ}\Gamma^{PQ}$. 
Thus one can see that the number of preserved supersymmetries in the 
background geometries with the special holonomy  
is counted as the number of singlets 
in the decomposition 
of the spinor representation $\bm{32}$ of $Spin(1,10)$ into 
some representation of the holonomy group $H$. 
For our case the special holonomy manifolds are taken as 
Calabi-Yau $(n+1)$-folds with the holonomy $H=SU(n+1)$, $n=1,2,3,4$. 
The decompositions of the spinor representation are as follows.
\begin{enumerate}
\item $CY_{5}$

In this case the geometry takes the form $\mathbb{R}\times CY_{5}$. 
This decomposes the $Spin(10)$ into $SU(5)$ and 
the corresponding splitting of the spinor representation is
\begin{align}
\label{cy5susy}
\bm{16}&=\bm{10}_{-}\oplus \overline{\bm{5}}_{3}\oplus \bm{1}_{-5}
\nonumber\\
\bm{16}'&=\bm{10}_{+}\oplus \bm{5}_{-3}\oplus \bm{1}_{5}
\end{align}
where the capital letters denote the representations 
of the $SU(5)$ and the subscripts represent the $U(1)$ charges 
under the decomposition $Spin(10)\rightarrow SU(5)\times U(1)$. 
The appearance of two singlets tells us that 
the space $\mathbb{R}\times CY_{5}$ preserves 
$\frac{2}{32}=\frac{1}{16}$ supersymmetries. 

Now we will define an explicit set of projections 
one the Killing spinors. 
Let us consider the situations 
where the Calabi-Yau spaces fill in the order 
$(x^{1},x^{2})$, $(x^{9},x^{10})$, $(x^{7},x^{8})$, 
$(x^{5},x^{6})$ and $(x^{3},x^{4})$. 
Then we can define the Killig spinors 
by the eigenvalues $\pm 1$ 
for the following set of commuting matrices 
\begin{align}
\label{projcy}
\Gamma^{12910},\ \ \ \ \ 
\Gamma^{91078},\ \ \ \ \ 
\Gamma^{7856},\ \ \ \ \ 
\Gamma^{5634}.
\end{align}
We will define the corresponding Killing spinors for $CY_{5}$ 
by the projection
\begin{align}
\label{cy5susy2}
\Gamma^{12910}\epsilon=\Gamma^{91078}\epsilon
=\Gamma^{7856}\epsilon=\Gamma^{5634}\epsilon=-\epsilon.
\end{align}
Note that this implies that $\Gamma^{012}\epsilon=\epsilon$. 
\item $CY_{4}$

In this case the geometry is 
the product space $\mathbb{R}^{1,2}\times CY_{4}$. 
Correspondingly the 
$Spin(8)$ decomposes into $SU(4)$ and 
the branching rule of the spinor representation is given by
\begin{align}
\label{cy4susy}
\bm{8}_{s}&=\bm{6}_{0}\oplus \bm{1}_{2}\oplus \bm{1}_{-2}\nonumber\\
\bm{8}_{c}&=\bm{4}_{-}\oplus \bm{4}_{+}.
\end{align}
One can see that the decomposition produces  
two singlets out of sixteen components. 
Hence the geometry $\mathbb{R}^{1,2}\times CY_{4}$ 
may preserve $\frac{2}{16}=\frac{1}{8}$ supersymmetries. 
For this case the projection on the Killing spinor is represented as
\begin{align}
\label{cy4susy2}
\Gamma^{12910}\epsilon=\Gamma^{91078}\epsilon
=\Gamma^{7856}\epsilon=-\epsilon.
\end{align}
\item $CY_{3}$

In this case the geometry is of the form $\mathbb{R}^{1,4}\times CY_{3}$. 
As a consequence, 
the $Spin(6)$ decomposes into $SU(3)$ 
and the decomposition of the spinor representation is given by
\begin{align}
\label{cy3susy}
\bm{4}&=\bm{3}_{-}\oplus \bm{1}_{3}\nonumber\\
\overline{\bm{4}}&=\bm{3}_{+}\oplus \bm{1}_{-3}.
\end{align}
The presence of two singlets from eight components  
implies that $\frac{2}{8}=\frac{1}{4}$ supersymmetries 
can be preserved for the 
product space $\mathbb{R}^{1,4}\times CY_{3}$. 
The Killing spinor can be defined by the following projection
\begin{align}
\label{cy3susy2}
\Gamma^{12910}\epsilon=\Gamma^{91078}\epsilon=-\epsilon.
\end{align}

\item $CY_{2}$

In this case the geometry takes the product form 
$\mathbb{R}^{1,6}\times CY_{2}$. 
This leads to the decomposition of $Spin(4)$ into $SU(2)\times SU(2)$ 
and it gives rise to the decomposition of the spinor representation as
\begin{align}
\label{cy2susy}
\bm{2}&=(\bm{2},\bm{1})\nonumber\\
\bm{2}'&=(\bm{1},\bm{2}).
\end{align}
We see that there are 
two singlets under one part of the $SU(2)$. 
This implies that 
there are $\frac{2}{4}=\frac{1}{2}$ supersymmetries in the 
background geometry $\mathbb{R}^{1,6}\times CY_{2}$. 
We can choose the corresponding  Killing spinors 
such that they satisfy the following projection 
\begin{align}
\label{cy2susy2}
\Gamma^{12910}\epsilon=-\epsilon.
\end{align}
\end{enumerate}

\section{Calibration and supersymmetric cycle}
\label{seccurv1b}
As a next step we shall consider the situation where 
the M2-branes wrapping a Riemann surface $\Sigma_{g}$ 
propagate in a Calabi-Yau space without back reaction. 
In order for supersymmetry to be preserved on the world-volume, 
$\Sigma_{g}$ needs to be a calibrated two-cycle, 
i.e. holomorphic two-cycle of a Calabi-Yau manifold. 
Let us briefly recall and review the 
mathematical concepts on a calibration. 
In general a calibration on a special holonomy manifold $X$ 
is defined by a differential $p$-form $\varphi$ satisfying \cite{MR666108}
\begin{align}
\label{calibdef1}
d\varphi&=0,\\
\label{calibdef2}
\varphi|_{\mathcal{C}_{p}}&\le \mathrm{Vol}|_{\mathcal{C}_{p}}, \ \ \
 \forall \mathcal{C}_{p}
\end{align}
where $\mathcal{C}_{p}$ is any $p$-cycle in 
the special holonomy manifold $X$ 
and $\mathrm{Vol}$ is the volume form on the cycle induced from the
metric on $X$. 
Note that we have defined the inequality locally, 
namely $\varphi|_{\mathcal{C}_{p}}=a \cdot
\mathrm{Vol}|_{\mathcal{C}_{p}}$ 
for some $a\in \mathbb{R}$, and 
$\varphi|_{\mathcal{C}_{p}}\le \mathrm{Vol}|_{\mathcal{C}_{p}}$ 
if $a\le 1$. 
Then a $p$-cycle $\Sigma$ is said to be calibrated by $\varphi$ 
if it satisfies 
\begin{align}
\label{calibdef3}
\varphi|_{\Sigma}=\mathrm{Vol}|_{\Sigma}.
\end{align}
An important property is that a calibrated submanifold 
can be regarded as a minimal surface 
in their homology class 
\begin{align}
\mathrm{Vol}(\Sigma)=
\int_{\Sigma}\varphi=\int_{M_{p+1}}d\varphi 
+\int_{\Sigma'}\varphi=\int_{\Sigma'}\varphi
\le \mathrm{Vol}(\Sigma').
\end{align}
Here we denote another $p$-cycle in the same homology class 
by $\Sigma'$ so that $\partial M_{p+1}=\Sigma-\Sigma'$.

Calabi-Yau $(n+1)$-folds are known to allow for two different types 
of calibrations. 
One is the K\"{a}hler form $J$ and the other is the real part of 
holomorphic $(n+1,0)$-form $\Omega$.
Such calibrations can be constructed as bilinear forms of spinors 
\cite{MR1208563,MR1045637}
\begin{align}
\label{calib4}
J_{MN}&=i\epsilon^{\dag}\Gamma_{MN}\epsilon,\\
\Omega_{M_{1}\cdots M_{n+1}}
&=\epsilon^{T}\Gamma_{M_{1}\cdots M_{2(n+1)}}\epsilon.
\end{align}

Let us now consider the condition 
such that a bosonic configuration of the M2-branes is supersymmetric. 
As we can always introduce a second probe brane 
without breaking supersymmetry if it is wrapped around  
the supersymmetric cycle which the original probe brane is wrapped, 
an easy way to find such condition is to start with an effective
world-volume action of a single M2-brane \cite{Becker:1995kb}. 
The action for a single M2-brane coupled to $d=11$ supergravity is given
by \cite{Bergshoeff:1987cm}
\begin{align}
\label{memac01}
S=\int d^{3}x\Biggl[&
\frac12 \sqrt{-h}h^{\mu\nu}
\partial_{\mu}X^{M}\partial_{\nu}X^{N}g_{MN}
-\frac12 \sqrt{-h}\nonumber\\
&-i\sqrt{-h}h^{\mu\nu}
\overline{\Theta}\Gamma_{\mu}\nabla_{\nu}\Theta
+\frac16 \epsilon^{\mu\nu\lambda}C_{MNP}
\partial_{\mu}X^{M}\partial_{\nu}X^{N}\partial_{\lambda}X^{P}+\cdots
\Biggr]
\end{align}
where $h_{\mu\nu},\mu,\nu=0,1,2$ is the metric of the world-volume, 
$h=\mathrm{det}(h_{\mu\nu})$, 
$g_{MN},M=0,1,\cdots,10$ is the $d=11$ space-time metric.  
$X^{M}$ is an eleven-dimensional space-time coordinate 
and $\Theta$ is a fermionic coordinate. 
$C_{MNP}$ is a three-form gauge field, 
which is now vanishing in our background geometries.  
The action (\ref{memac01}) possesses the invariance under 
the rigid supersymmetry transformations
\begin{align}
\label{memsusy01}
\delta_{\epsilon}X^{M}&=i\overline{\epsilon}\Gamma^{M}\Theta,\\
\label{memsusy02}
\delta_{\epsilon}\Theta&=\epsilon
\end{align}
where $\epsilon$ is a constant anti-commuting spinor. 
Additionally the action (\ref{memac01}) has a local fermionic symmetry, 
the so-called $\kappa$-symmetry whose transformation is given by
\begin{align}
\label{kappasym1}
\delta_{\kappa} X^{M}&=
2i\overline{\Theta}\Gamma^{M}P_{+}\kappa(x),\\
\label{kappasym2}
\delta_{\kappa} \Theta&=
2P_{+}\kappa(x)
\end{align}
where $\kappa(x)$ is a non-constant spinor and 
we have defined the matrices 
\begin{align}
\label{pro11a}
P_{\pm}=
\frac12 
\left(
1\pm\frac{1}{6\sqrt{-h}}\epsilon^{\mu\nu\lambda}
\partial_{\mu}X^{M}
\partial_{\nu}X^{N}
\partial_{\lambda}X^{P}
\Gamma_{MNP}
\right)
\end{align}
as projection operators obeying the relations
\begin{align}
\label{pro11b}
P_{\pm}^{2}=1,\ \ \ \ \ 
P_{+}P_{-}=0,\ \ \ \ \ 
P_{+}+P_{-}=1.
\end{align}

To extract the physical degrees of freedom, 
one needs to fix the suitable gauge 
that excludes the local world-volume reparametrization 
and the local $\kappa$-symmetry. 
First of all, we will fix the reparametrization by choosing $x^{0}=X^{0}$. 
Then one can express the projection operator (\ref{pro11a}) as
\begin{align}
\label{pro11c}
P_{\pm}&=\frac12 
\left(
1\pm \Gamma
\right)
\end{align}
where 
\begin{align}
\Gamma:=\frac{1}{2\sqrt{\mathrm{det} (h_{\Sigma ij})}}\Gamma^{0}\epsilon^{ij}
\partial_{i}X^{M}\partial_{j}X^{N}\Gamma_{MN}.
\end{align}
Here we have introduced $h_{\Sigma ij}, i,j=1,2$ 
as the metric of the Riemann surface wrapped 
by the M2-brane and 
the quantity $\sqrt{\mathrm{det}(h_{\Sigma ij})}$ as the area 
of the wrapped surface. 
The next step we should do is to fix the local $\kappa$-symmetry on the
world-volume. 
In order to make a bosonic world-volume configuration supersymmetric, 
the global supersymmetry transformations (\ref{memsusy02}) 
have to be canceled by the $\kappa$-symmetry transformations 
(\ref{kappasym2}) 
\begin{align}
\label{memsusy03}
\left(
\delta_{\epsilon}+\delta_{\kappa}
\right)\Theta=\epsilon+2P_{+}\kappa(x)=0.
\end{align}
From the action of $P_{-}$ on both sides we find that 
\begin{align}
\label{memsusy04}
P_{-}\epsilon=
\frac{1-\Gamma}{2}\epsilon
=0.
\end{align}
Thus the supersymmetry preserved by the M2-branes 
is given by the Killing spinor $\epsilon$ that satisfies the projection 
(\ref{memsusy03}). 
By noting that $\Gamma^{2}=1$ and $\Gamma^{\dag}=\Gamma$, 
one can find that
\begin{align}
\label{memsusy04}
\epsilon^{\dag}
\frac{1-\Gamma}{2}
\epsilon
=\epsilon^{\dag}\frac{(1-\Gamma)(1-\Gamma)}{4}\epsilon
=\left|\frac{1-\Gamma}{2}\epsilon\right|^{2}\ge 0. 
\end{align}
By utilizing the normalization 
of the Killing spinors so that $\epsilon^{\dag}\epsilon=1$, 
we can rewrite the inequality (\ref{memsusy04}) as
\begin{align}
\label{memsusy05}
\mathrm{Vol}(\Sigma_{g})\ge \varphi
\end{align}
Here $\mathrm{Vol}(\Sigma_{g})=\sqrt{\mathrm{det}(h_{\Sigma ij})}$ is 
the area of the wrapped Riemann surface and $\varphi$ is the differential
two-form given by
\begin{align}
\label{memsusy06}
\varphi=-\frac12
\left( 
\overline{\epsilon}\Gamma_{MN}\epsilon 
\right)
dX^{M}\wedge dX^{N}.
\end{align}
Thus the two-form (\ref{memsusy06}) 
obeys the second condition (\ref{calibdef2}) for the calibration 
and enjoys the bilinear expression for K\"{a}hler calibration $J$ 
(see (\ref{calib4})). 
Also one can show that 
the two-form (\ref{memsusy06}) satisfies the first condition
(\ref{calibdef1}) for the calibration 
by observing the explicit expression (\ref{memsusy06})
\footnote{It can also be checked by using the supersymmetry algebra
\cite{Gutowski:1999tu}.}.  
Then one can conclude that the two-form (\ref{memsusy06}) is 
identified with a K\"{a}hler calibration and that 
the supersymmetric two-cycle $\Sigma_{g}$ 
wrapped by the M2-branes is recognized as calibrated two-cycle, 
i.e. a holomorphic two-cycle. 
We note that 
(\ref{memsusy03}) is equivalent to 
the chirality condition $\Gamma^{012}\epsilon=\epsilon$ 
for the supersymmetry parameters in the BLG-model (see
(\ref{blgchiralmtx})).

Now we are ready to count the amount of preserved
supersymmetries for the wrapped M2-brane configurations 
by taking account into the two different types of projections. 
One is the set of the projections (\ref{cy5susy2}), (\ref{cy4susy2}), (\ref{cy3susy2}) and
(\ref{cy2susy2}) 
for the background Calabi-Yau manifolds. 
The other is the projection (\ref{memsusy03}) (or (\ref{blgchiralmtx})) 
for the M2-branes wrapped 
around a holomorphic two-cycle $\Sigma_{g}$. 
For most of the cases, 
one can see that wrapped branes break half of the preserved
supersymmetries in the special holonomy manifolds due to the additional
projection for the wrapped branes around calibrated submanifolds. 
But we should note that 
for the Calabi-Yau 5-fold 
the projection condition (\ref{memsusy03}) 
for the wrapped M2-branes leads to no 
further constraint on the conserved two Killing spinors. 
This corresponds to the fact that 
the M2-branes can wrap a holomorphic two-cycle 
in a Calabi-Yau 5-fold without losing the supersymmetry. 
The numbers of preserved supersymmetries by the wrapped M2-branes 
around holomorphic two-cycles $\Sigma_{g}$ in Calabi-Yau manifolds   
are shown 
in Table \ref{calib}. 
\begin{table}
\begin{center}
\begin{tabular}{|c|c|} \hline
\ \ $M_{3}$ \ \ 
&supersymmetry\\ \hline\hline
$\mathbb{R}\times (\Sigma_{g}\subset \mathrm{K3})$&8\\ \hline
$\mathbb{R}\times (\Sigma_{g}\subset CY_{3})$&4 \\ \hline
$\mathbb{R}\times (\Sigma_{g}\subset CY_{4})$&2 \\ \hline
$\mathbb{R}\times (\Sigma_{g}\subset CY_{5})$&2 \\ \hline
\end{tabular}
\caption{
The amounts of the preserved supersymmetries 
for the M2-branes wrapping holomorphic curves $\Sigma_{g}$ 
in Calabi-Yau spaces. 
Note that the M2-branes can wrap a holomorphic curve 
in a $CY_{5}$ without loss of the supersymmetries.}
\label{calib}
\end{center}
\end{table}
Performing the dimensional reduction on $\Sigma_{g}$ 
we can get the low-energy effective quantum mechanics on $\mathbb{R}$, 
as we will discuss in the following. 
Such quantum mechanics on $\mathbb{R}$ will possess 
the same number of supersymmetries.

\section{Topological twisting}
\label{seccurv2}
It is true that a quantum field theory on the curved M2-branes 
may interact with gravity, but 
at the low-energy one can obtain 
a supersymmetric quantum field theory 
on $\mathbb{R}\times\Sigma_{g}$ by taking the appropriate decoupling limit 
$l_{p}\rightarrow 0$ where $l_{p}$ is the Planck length 
 while keeping the size of $\Sigma_{g}$ and that
of $X$ fixed. 
As a first step to derive such low-energy effective field theories 
on the curved branes, 
let us recall why the BLG-model is conjectured to describe the dynamics 
of the planar M2-branes. 
The BLG-model contains the fields and supercharges 
which transform under 
$SO(2)_{E}\times SO(8)_{R}$ as
\begin{align}
\label{blgfieldrep1}
X_{a}^{I}& : {\bm{8}_{v}}_{0}\nonumber \\
\Psi_{a}& : \bm{8}_{c+}\oplus \bm{8}_{c-}\nonumber \\
\epsilon& : \bm{8}_{s+}\oplus \bm{8}_{s-}.
\end{align}
The eight bosonic scalar fields $X^{I}$'s are the vector
representations of the R-symmetry $SO(8)_{R}$ which 
corresponds to the rotational symmetry group 
of the normal directions of the multiple M2-branes. 
They can be regarded as the sections of the trivial normal bundle. 
If we consider the geometry given in (\ref{m2geo}), 
the tangent bundle $T_{X}$ of the Calabi-Yau space $X$ 
has the decomposition as
\begin{equation}
 T_{X}=T_{\Sigma}\oplus N_{\Sigma}.
\end{equation}
Here $T_{\Sigma}$ denotes the tangent bundle over the Riemann surface
$\Sigma_{g}$ and $N_{\Sigma}$ stands for the
normal bundle over the Riemann surface. 
Hence we must be careful to treat with the existence of 
the non-trivial normal bundle of calibrated cycles. 
This situation naturally leads us 
to introduce new bosonic dynamical variables 
instead of the original bosonic scalar fields. 
Such transitions from trivial representations, i.e. scalars, 
to the representations describing the non-trivial 
non-trivial normal bundles are closely related to 
the way in which the field theory on the geometry 
$\mathbb{R}\times \Sigma_{g}$ 
can preserve supersymmetry. 
The coupling to the curvature on the Riemann surface 
enforces us to introduce  
the coupling to an external $SO(2n)$ gauge group, 
the R-symmetry background. 
Therefore we can preserve supersymmetry on the holomorphic two-cycle 
by making an appropriate choice of the $SO(2)$ Abelian factors 
from the $SO(2n)$ R-symmetry group.

As we have already discussed in the previous chapter, 
there has been an important observation that 
such an effective field theories on curved branes can be realized  
by topological twisting \cite{Bershadsky:1995qy}. 
Now we want to consider the topologically twisted BLG-model 
that yields the low-energy description for the curved M2-branes 
\footnote{
For the ABJM-model 
the geometric meaning of the topological twisting is less clear 
because the classical $SU(4)_{R}$ R-symmetry reflects the orbifolds. 
In this paper we will focus on the BLG-model.
}.
Recall that topological twisting procedure can be achieved 
by combining the original Euclidean rotational group $SO(2)_{E}$ on the
Riemann surface with 
a different subgroup $SO(2)'_{E}$ of $SO(2)_{E}\times SO(8)_{R}$. 
Although there exist various possible ways to choose such subgroups,  
we now consider the following decomposition
\begin{align}
\label{splitso8}
 SO(8)\supset
& SO(8-2n)\times SO(2n) \nonumber \\
\supset
& SO(8-2n)\times SO(2)_{1}\times \cdots \times SO(2)_{n}.
\end{align}
Note that $SO(8-2n)$ corresponds to 
the rotational group of the Euclidean flat space
normal to the Riemann surface, 
whereas the $SO(2)_{i}$ represent the 
diagonal subgroups of the $SO(2n)$ R-symmetry group. 
This decomposition implies that 
the Calabi-Yau manifold $X$ is constructed 
with the decomposable line bundles over the Riemann surface as the
form 
\begin{align}
\label{cyconst1}
X=\mathcal{L}_{1}\oplus
\cdots \oplus
\mathcal{L}_{n}
\rightarrow \Sigma_{g}.
\end{align} 
The decomposition (\ref{splitso8}) leads to 
the corresponding branching rule of  
the R-charges for $\bm{8}_{v}$, $\bm{8}_{s}$
and $\bm{8}_{c}$ are determined as follows:

\begin{enumerate}
 \item $SO(8)\supset SO(6)\times SO(2)_{1}$
\begin{align}
\label{k3twist}
\bm{8}_{v}=&\bm{6}_{0}\oplus\bm{1}_{2}\oplus\bm{1}_{-2} \nonumber \\
\bm{8}_{s}=&\bm{4}_{+}\oplus\overline{\bm{4}}_{-} \nonumber \\
\bm{8}_{c}=&\bm{4}_{-}\oplus\overline{\bm{4}}_{+}.
\end{align}

\item $SO(8)
\supset SO(4)\times SO(2)_{1}\times SO(2)_{2}$
\begin{align}
\label{cy3twist1}
\bm{8}_{v}
=&\bm{4}_{00}\oplus \bm{1}_{02}\oplus \bm{1}_{0-2}\oplus
 \bm{1}_{20}\oplus \bm{1}_{-20}\nonumber \\
\bm{8}_{s}
=&\bm{2}_{++}\oplus \bm{2}'_{+-}\oplus \bm{2}_{--}\oplus
 \bm{2}_{-+}'\nonumber \\
\bm{8}_{c}
=&\bm{2}_{-+}\oplus \bm{2}'_{--}\oplus \bm{2}_{+-}\oplus \bm{2}_{++}'.
\end{align}

\item $SO(8)
\supset SO(2)\times SO(2)_{1}\times SO(2)_{2}\times
      SO(2)_{3}$
\begin{align}
\label{cy4twist1}
\bm{8}_{v}
=&
\bm{2}_{000}\oplus\bm{1}_{002}\oplus\bm{1}_{00-2}
\oplus\bm{1}_{020}\oplus\bm{1}_{0-20}
\oplus\bm{1}_{200}\oplus\bm{1}_{-200}
\nonumber \\
\bm{8}_{s}
=&
\bm{1}_{+++}\oplus\bm{1}_{++-}
\oplus\bm{1}_{+--}\oplus\bm{1}_{+-+}
\oplus\bm{1}_{--+}\oplus\bm{1}_{---}
\oplus\bm{1}_{-+-}\oplus\bm{1}_{-++}
\nonumber \\
\bm{8}_{c}
=&
\bm{1}_{-++}\oplus\bm{1}_{-+-}
\oplus\bm{1}_{---}\oplus\bm{1}_{--+}
\oplus\bm{1}_{+-+}\oplus\bm{1}_{+--}
\oplus\bm{1}_{++-}\oplus\bm{1}_{+++}.
\end{align}

\item $SO(8)
\supset SO(2)_{1}\times SO(2)_{2}\times
      SO(2)_{3}\times SO(2)_{4}$
\begin{align}
\label{cy5twist1}
\bm{8}_{v}
=&
\bm{1}_{0002}\oplus\bm{1}_{000-2}\oplus\bm{1}_{0020}\oplus\bm{1}_{00-20}
\oplus\bm{1}_{0200}\oplus\bm{1}_{0-200}
\oplus\bm{1}_{2000}\oplus\bm{1}_{-2000}
\nonumber \\
\bm{8}_{s}
=&
\bm{1}_{++++}\oplus\bm{1}_{++--}
\oplus\bm{1}_{+--+}\oplus\bm{1}_{+-+-}
\oplus\bm{1}_{--++}\oplus\bm{1}_{----}
\oplus\bm{1}_{-+-+}\oplus\bm{1}_{-++-}
\nonumber \\
\bm{8}_{c}
=&
\bm{1}_{-+++}\oplus\bm{1}_{-+--}
\oplus\bm{1}_{---+}\oplus\bm{1}_{--+-}
\oplus\bm{1}_{+-++}\oplus\bm{1}_{+---}
\oplus\bm{1}_{++-+}\oplus\bm{1}_{+++-}.
\end{align}

\end{enumerate}
Making use of one of the decompositions (\ref{k3twist})-(\ref{cy5twist1}), 
one can define a new generator $s'$ as the $SO(2)_{E}'$ charge by
\begin{align}
\label{twistcharge1}
s':=s-\sum_{i=1}^{n}a_{i}T_{i}
\end{align}
where $s$ is a generator of the original rotational group
$SO(2)_{E}$ on the Euclidean Riemann surface, 
$T_{i}$ stands for a 
generator of the diagonal subgroup $SO(2)_{i}$ of the R-symmetry group $SO(2n)$ 
and $a_{i}$'s denote the constant parameters 
which characterize distinct twisting procedures. 
Let us normalize these $SO(2)$ charges $s'$, $s$ and $T_{i}$ 
so that they are twice as the usual spin on the Riemann
surface. 
By noting that the parameters $a_{i}$'s are related to 
the degrees of the line bundles $\mathcal{L}_{i}$'s as 
\begin{align}
\label{bdldeg}
\deg (\mathcal{L}_{i})=
\begin{cases}
2|g-1|a_{i}& \textrm{for} \ \ g\neq 1\cr
a_{i}& \textrm{for} \ \ g=1\cr
\end{cases}
\end{align}
and that the degrees determine the first Chern classes, 
we can find the conditions that $X$ is Calabi-Yau as follows:
\begin{align}
\label{cycond1}
\sum_{i=1}^{n}a_{i}=
\begin{cases}
-1& \textrm{for} \ \ g=0\cr
0& \textrm{for} \ \ g=1\cr
1& \textrm{for} \ \ g>1\cr 
\end{cases}.
\end{align}
The above Calabi-Yau conditions (\ref{cycond1}) 
guarantee that 
there exist covariant constant spinors 
in the twisted theories. 
It can be easily checked that 
the topological twisting underlying the decompositions 
(\ref{k3twist}), (\ref{cy3twist1}), (\ref{cy4twist1}) and 
(\ref{cy5twist1}) preserve eight, four, two and two  
supersymmetries as expected for $CY_{2}$, $CY_{3}$, $CY_{4}$ and $CY_{5}$ 
respectively.

Hence when we have the decomposable line bundle structures 
of the Calabi-Yau manifolds (\ref{cyconst1}), 
we can specify the topological twist by considering 
the two conditions (\ref{bdldeg}) and (\ref{cycond1}). 
For instance, if we take a $CY_{2}$, i.e. for 
$a_{2}=a_{3}=a_{4}=0$, 
the local geometry can be viewed as 
the cotangent bundle $T^{*}\Sigma_{g}$ over the 
Riemann surface and 
a single twisting parameter $a_{1}$ can be determined uniquely 
from the Calabi-Yau condition.  
For other Calabi-Yau manifolds 
the Calabi-Yau conditions are not enough to fix the 
parameters $a_{i}$'s and 
there may exist infinitely many methods of the topological twists, 
which depends on the degrees of the line bundles.

\chapter{SCQM from M2-branes in a K3 surface}
\label{cy2sec}
In this chapter we will give further detailed investigation 
on the wrapped M2-branes 
on the holomorphic Riemann surface of genus $g>1$ in a K3 surface. 
Firstly we will discuss the field content and the supersymmetry 
in the twisted theory and their consistency in section \ref{cysec0}. 
Then we will derive the twisted theory in section \ref{cy2sec1}. 
Finally we will compactify the twisted theory 
on the Riemann surface and find the IR quantum mechanics in section
\ref{cy2sec2}. 
The theory turns out to be 
the $\mathcal{N}=8$ superconformal gauged quantum mechanics. 

\section{K3 twisting}
\label{cysec0}
In order to obtain the world-volume description for 
the membranes wrapping a holomorphic Riemann surface of genus $g>1$ 
in a K3 surface, we should perform the 
topological twisting by using the decomposition (\ref{k3twist}). 
As we have mentioned earlier, in this case 
the existence of covariant constant spinors 
fixes the twisting procedure because 
the external gauge field used for the twist 
is nothing but an $SO(2)$ Abelian background itself in this case. 
Note that the twisting for $\Sigma_{g}=\mathbb{P}^{1}$ can be 
realized just by the orientation reversal.

For the twisted field theory with $g>1$, 
the decomposition 
$SO(2)_{E}\times SO(8)_{R}\rightarrow SO(2)_{E}'\times SO(6)_{R}$ 
yields the new field content and the supersymmetry parameters 
characterized by the following representations: 
\begin{align}
\label{k3content}
&X^{I}:\bm{8}_{v0}\rightarrow\bm{6}_{0}\oplus \bm{1}_{2}\oplus
 \bm{1}_{-2} \nonumber \\
&\epsilon:\bm{8}_{s+}\oplus \bm{8}_{s-}\rightarrow\bm{4}_{0}\oplus\overline{\bm{4}}_{2}
\oplus\bm{4}_{-2}\oplus \overline{\bm{4}}_{0}\nonumber \\
&\Psi:\bm{8}_{c+}\oplus\bm{8}_{c-}\rightarrow
\bm{4}_{2}\oplus \overline{\bm{4}}_{0}
\oplus \bm{4}_{0}\oplus \overline{\bm{4}}_{-2}.
\end{align}
The results of the topological twisting for the components of fields 
and supersymmetry parameters are shown 
in Table \ref{k3twist1} and \ref{k3twist2} respectively. 
In the twisted theory the bosonic field content involves  
six scalar fields 
$\phi^{I}$ as the representation $\bm{6}_{0}$
and one-forms 
$\Phi_{z}$, $\Phi_{\overline{z}}$ which transform as 
$\bm{1}_{2}\oplus \bm{1}_{-2}$. 
The fermionic field content contains  
eight scalar fields 
$\psi,\tilde{\lambda}$ 
as the representations $\bm{4}_{0}\oplus\overline{\bm{4}}_{0}$
and 
one-forms $\Psi_{z}, \tilde{\Psi}_{\overline{z}}$ 
as the representations $\bm{4}_{2}\oplus \overline{\bm{4}}_{-2}$. 
The supersymmetry parameters split into  
eight scalars 
$\epsilon,\tilde{\epsilon}$ 
as $\bm{4}_{0}\oplus \overline{\bm{4}}_{0}$ 
and 
one-forms $\tilde{\epsilon}_{z}, \epsilon_{\overline{z}}$ 
as $\overline{\bm{4}}_{2}\oplus \bm{4}_{-2}$. 
\begin{table}
\begin{center}
\begin{tabular}{@{\vrule width 1pt}c|c|c|c|c@{\ \vrule width 1pt}} \hline
&$SO(2)_{E}$ &$SO(2)_{1}$ &$SO(2)_{E}'$&$\mathcal{L}$  \\ \hline \hline
$\phi^{1}$ &$0$&$0$&$0$&$\mathcal{O}$  \\
$\phi^{2}$ &$0$&$0$&$0$&$\mathcal{O}$  \\
$\phi^{3}$ &$0$&$0$&$0$&$\mathcal{O}$  \\
$\phi^{4}$ &$0$&$0$&$0$&$\mathcal{O}$   \\
$\phi^{5}$ &$0$&$0$&$0$&$\mathcal{O}$  \\
$\phi^{6}$ &$0$&$0$&$0$&$\mathcal{O}$   \\
$\Phi_{\overline{z}}$ &$0$&$2$&$-2$&$K^{-1}$   \\
$\Phi_{z}$ &$0$&$-2$&$2$&$K$  \\ \hline
$\Psi_{z1}$ &$1$&$-1$&$2$&$K$   \\
$\Psi_{z2}$ &$1$&$-1$&$2$&$K$  \\
$\Psi_{z3}$ &$1$&$-1$&$2$&$K$  \\
$\Psi_{z4}$ &$1$&$-1$&$2$&$K$  \\
$\tilde{\lambda}_{1}$ &$1$&$1$&$0$&$\mathcal{O}$   \\
$\tilde{\lambda}_{2}$ &$1$&$1$&$0$&$\mathcal{O}$  \\
$\tilde{\lambda}_{3}$ &$1$&$1$&$0$&$\mathcal{O}$   \\
$\tilde{\lambda}_{4}$ &$1$&$1$&$0$&$\mathcal{O}$  \\ 
$\tilde{\Psi}_{\overline{z}1}$ &$-1$&$1$&$-2$&$K^{-1}$   \\
$\tilde{\Psi}_{\overline{z}2}$ &$-1$&$1$&$-2$&$K^{-1}$  \\
$\tilde{\Psi}_{\overline{z}3}$ &$-1$&$1$&$-2$&$K^{-1}$  \\
$\tilde{\Psi}_{\overline{z}4}$ &$-1$&$1$&$-2$&$K^{-1}$  \\
$\psi_{1}$ &$-1$&$-1$&$0$&$\mathcal{O}$   \\
$\psi_{2}$ &$-1$&$-1$&$0$&$\mathcal{O}$  \\
$\psi_{3}$ &$-1$&$-1$&$0$&$\mathcal{O}$   \\
$\psi_{4}$ &$-1$&$-1$&$0$&$\mathcal{O}$  \\ \hline
\end{tabular}
\caption{The twisting for bosonic scalar fields $X^{I}$'s 
and fermionic fields $\Psi$'s of the BLG-model 
when the Riemann surface of genus $g>1$ 
is embedded in a K3 surface. 
$\mathcal{L}$ is the complex line bundle over 
$\Sigma_{g}$ in which the fields take values. $\mathcal{O}$ and $K$ 
are the trivial bundle and the canonical bundle respectively.}
\label{k3twist1}
\end{center}
\end{table}
\begin{table}
\begin{center}
\begin{tabular}{@{\vrule width 1pt}c|c|c|c|c@{\ \vrule width 1pt}} \hline
&$SO(2)_{E}$ &$SO(2)_{1}$ &$SO(2)_{E}'$&$\mathcal{L}$ \\ \hline \hline
$\epsilon_{1}$ &$1$&$1$ &$0$&$\mathcal{O}$ \\
$\epsilon_{2}$ &$1$&$1$ &$0$&$\mathcal{O}$ \\
$\epsilon_{3}$ &$1$&$1$ &$0$&$\mathcal{O}$ \\
$\epsilon_{4}$ &$1$&$1$ &$0$&$\mathcal{O}$ \\
$\tilde{\epsilon}_{z1}$ &$1$&$-1$ &$2$&$K$ \\
$\tilde{\epsilon}_{z2}$ &$1$&$-1$ &$2$&$K$ \\
$\tilde{\epsilon}_{z3}$ &$1$&$-1$ &$2$&$K$ \\
$\tilde{\epsilon}_{z4}$ &$1$&$-1$ &$2$&$K$ \\ 
$\tilde{\epsilon}_{1}$ &$-1$&$-1$ &$0$&$\mathcal{O}$ \\
$\tilde{\epsilon}_{2}$ &$-1$&$-1$ &$0$&$\mathcal{O}$ \\
$\tilde{\epsilon}_{3}$ &$-1$&$-1$ &$0$&$\mathcal{O}$ \\
$\tilde{\epsilon}_{4}$ &$-1$&$-1$ &$0$&$\mathcal{O}$ \\
$\epsilon_{\overline{z}1}$ &$-1$&$1$ &$-2$&$K^{-1}$ \\
$\epsilon_{\overline{z}2}$ &$-1$&$1$ &$-2$&$K^{-1}$  \\
$\epsilon_{\overline{z}3}$ &$-1$&$1$ &$-2$&$K^{-1}$ \\
$\epsilon_{\overline{z}4}$ &$-1$&$1$ &$-2$&$K^{-1}$ \\ \hline
\end{tabular}
\caption{The twisted supersymmetry parameters 
of the BLG-model probing a K3 surface. 
The eight covariant constant spinors 
play the role of BRST generators in the twisted theory. 
The result is consistent to the fact that 
a holomorphic curve inside a K3 surface can preserve a
 half of the supersymmetries (see Table \ref{calib}).}
\label{k3twist2}
\end{center}
\end{table}
%
%
%
%
%
%
%
%
%
%
%
%
%
%
%
In the following we will distinguish $\bm{4}$ and $\overline{\bm{4}}$ 
by putting tildes over the fermions.

It is instructive to comment 
on the geometric implications 
of the above field content. 
The significant is that 
we have six bosonic scalar fields 
and eight fermionic scalar charges in the twisted theory. 
The fact that a Riemann surface is 
a real two-dimensional manifold 
and that there are six scalar fields in the twisted theory 
implies that the theory should describe the case where the 
two-cycle lives in a $2+(8-6)=4$-dimensional curved manifold $X$. 
In addition, the presence of eight scalar supercharges says that 
the four-manifold preserves $\frac{8}{16}=\frac12$ of the supersymmetries. 
This is realized when a holomorphic curve $\Sigma_{g}$ is
embedded in a K3 surface.

Remember that the K3 geometry is the cotangent bundle $T^{*}\Sigma_{g}$
locally. 
The two scalar fields combine to yield  
one-forms on the Riemann surface. 
They represent the motion of the M2-branes 
along the non-trivial normal bundle $N_{\Sigma}$ 
over the Riemann surface inside the K3 surface. 
Under the $SO(6)$ rotational group 
of the six uncompactified dimensions, 
the six scalars transform as vector representations $\bm{6}_{v}$ 
and the one-forms are just singlets. 
We take the eleven-dimensional space-time configuration as
\begin{equation}
\label{k3}
 \begin{array}{cccccccccccc}
&0&1&2&3&4&5&6&7&8&9&10\\
\textrm{K3}&\times&\circ&\circ&\times&\times&\times&\times&\times&\times&\circ&\circ\\
\textrm{M2}&\circ&\circ&\circ&\times&\times&\times&\times&\times&\times&\times&\times \\
\Sigma_{g}&\times&\circ&\circ&\times&\times&\times&\times&\times&\times&\times&\times\\
\end{array}
\end{equation}
where $\circ$ stands for the direction 
to which the geometrical objects stretch, 
while  $\times$ represents the direction in which they localize. 
One can see that the projection defined in (\ref{cy2susy2}) 
for the K3 surface corresponds to the configuration (\ref{k3}). 
The world-volume of the M2-branes is extended to 
a time direction $x^{0}$ and spacial directions $x^{1}, x^{2}$.  
The two spacial directions $x^{1}$, $x^{2}$ 
are taken as tangent to the Riemann surface in the K3 surface. 
On the other hand, the transverse directions 
of the M2-branes is now split into two parts; 
one is the non-trivial normal bundle $N_{\Sigma}$ in the K3 surface, 
extending to two directions $x^{9}$, $x^{10}$ and 
the other is the flat Euclidean space labeled
by $x^{3},\cdots, x^{8}$.

\section{Twisted theory}
\label{cy2sec1}
The space-time configuration 
(\ref{k3}) breaks down the space-time symmetry group 
$SO(1,10)$ to $SO(2)_{E}\times SO(6)_{R}\times SO(2)_{1}$. 
Then one can decompose the $SO(1,10)$ gamma matrix as
\begin{equation}
\label{mtx1}
\begin{cases}
 \Gamma^{\mu}=\gamma^{\mu}\otimes
  \hat{\Gamma}^{7}
\otimes \sigma_{2}
&\mu=0,1,2 \cr
 \Gamma^{I+2}=\mathbb{I}_{2}\otimes
 \hat{\Gamma}^{I} \otimes 
\sigma_{2}&I=1,\cdots, 6\cr
 \Gamma^{i+8}=\mathbb{I}_{2}\otimes
 \mathbb{I}_{8}\otimes
 \gamma^{i}&i=1,2 \cr
\end{cases}
\end{equation}
where
\footnote{$(d+1)$-th component
of $d=t+s$ dimensional gamma matrices can be defined by \cite{Kugo:1982bn}
\[
 \Gamma^{d+1}:=\sqrt{(-1)^{\frac{s-t}{2}}} \Gamma^{12\cdots d}
\]
where $s$ and $t$ are correspond to the dimension of space 
and time respectively. In the above case $s=6$ and $t=0$
. Note that minus sign should be included in
(\ref{gamma7}) since we are now considering the decomposition of (\ref{11gamma}).}
 $\hat{\Gamma}^{I}$ is the $SO(6)$ gamma matrix satisfying
\begin{align}
\{\hat{\Gamma}^{I},\hat{\Gamma}^{J}\}=2\delta^{IJ},\ \ \ 
(\hat{\Gamma}^{I})^{\dag}=\Gamma^{I}
\end{align}
\begin{equation}
\label{gamma7}
\hat{\Gamma}^{7}
=-i\hat{\Gamma}^{12\cdots 6}=\left(
\begin{array}{cc}
\mathbb{I}_{4}&0\\
0&-\mathbb{I}_{4}\\
\end{array}
\right).
\end{equation} 
Correspondingly the $SO(1,10)$ charge conjugation matrix $\mathcal{C}$ 
can be expressed as
\begin{equation}
\label{ccmtx}
 \mathcal{C}=\epsilon \otimes \hat{C} \otimes \epsilon
\end{equation}
where $\epsilon:=i\sigma_{2}$ is introduced as the charge conjugation
matrix obeying the relations
\begin{equation}
\epsilon^{T}=-\epsilon,\ \ \
\epsilon\gamma^{\mu}\epsilon^{-1}=-(\gamma^{\mu})^{T} 
\end{equation}
and $\hat{C}$ is the $SO(6)$ charge conjugation matrix obeying
\footnote{
In even dimensional space-time, a charge conjugation matrix can be
defined in two ways. 
Instead of (\ref{ccmtx1}), 
we may define
\begin{equation}
 \hat{C}^{T}=\hat{C},\ \ \ \hat{C}\hat{\Gamma}^{I}\hat{C}^{-1}=-(\hat{\Gamma}^{I})^{T}.
\end{equation}
However, Majorana spinors are only allowed for (\ref{ccmtx1}).
}
\begin{align}
\label{ccmtx1}
\hat{C}^{T}=-\hat{C},\ \ \ \ \ \ \hat{C}\hat{\Gamma}^{I}\hat{C}^{-1}=(\hat{\Gamma}^{I})^{T},\ \ \ \ \ \ 
\hat{C}\hat{\Gamma}^{7}\hat{C}^{-1}=-(\hat{\Gamma}^{7})^{T}.
\end{align}
Note that the decomposition (\ref{mtx1}) 
leads us to write the $SO(8)$ chiral matrix as
\begin{equation}
\label{chiralmtx}
\Gamma^{012}=
\Gamma^{34\cdots 10}=
\mathbb{I}_{2}\otimes \hat{\Gamma}^{7}\otimes \sigma_{2}.
\end{equation}

We will define the twisted bosonic fields by
\begin{align}
\label{new1}
\phi^{I}&:=X^{I+2},\\
\label{new2}
\Phi_{z}&:=\frac{1}{\sqrt{2}}(X^{9}-iX^{10}), 
&\Phi_{\overline{z}}&:=\frac{1}{\sqrt{2}}(X^{9}+iX^{10}),\\
\label{new3}
A_{z}&:=\frac{1}{\sqrt{2}}(A_{1}-iA_{2}), 
&A_{\overline{z}}&:=\frac{1}{\sqrt{2}}(A_{1}+iA_{2}).
\end{align}
Here the bosonic scalar fields $\phi^{I}$'s are 
the vector representations $\bm{6}_{v}$ of the $SO(6)$ global
symmetry and the indices $I=1,\cdots,6$ correspond to 
the flat transverse directions. 
The bosonic one-froms, $\Phi_{z}$ and $\Phi_{\overline{z}}$ are 
the trivial representation of the $SO(6)$  
and they correspond to the motion in the non-trivial 
normal geometry $N_{\Sigma}$ of the holomorphic 
Riemann surface in the K3 surface. 
These bosonic matter fields $\phi^{I}, \Phi_{z}$ and $\Phi_{\overline{z}}$ 
are the 3-algebra valued.

Let us turn to the twisted fermionic objects. 
In the original BLG-model the fermions $\Psi$ are 
$SL(2,\mathbb{R})$ spinors transforming as the spinor 
representations $\bm{8}_{c}$ of the $SO(8)_{R}$ R-symmetry. 
Under the splitting 
$Spin(1,10)$ $\rightarrow$ $Spin(2)\times Spin(6)\times Spin(2)$,  
the fermionic fields $\Psi$ decompose into the four distinct representations 
$\bm{4}_{2}$, $\overline{\bm{4}}_{0}$, 
$\bm{4}_{0}$ and $\overline{\bm{4}}_{-2}$, 
which we will denote by 
$\Psi_{z}$, $\tilde{\lambda}$, $\psi$ and
$\tilde{\Psi}_{\overline{z}}$ respectively. 
From the above definitions and notations 
we can be expand fermionic field $\Psi$ as
\begin{align}
\label{new4}
\Psi_{A}^{\alpha\beta}
=\frac{i}{\sqrt{2}}
\psi_{A}(\gamma_{+}\epsilon^{-1})^{\alpha\beta}
+i\tilde{\Psi}_{\overline{z}A}
(\gamma^{\overline{z}}\epsilon^{-1})^{\alpha\beta}
-\frac{i}{\sqrt{2}}\tilde{\lambda}_{A}
(\gamma_{-}\epsilon^{-1})^{\alpha\beta}
-i\Psi_{zA}(\gamma^{z}\epsilon^{-1})^{\alpha\beta}.
\end{align}
Here the three indices $\alpha,A$ and $\beta$ represent 
the $SO(2)_{E}$ spinor,
the $SO(6)_{R}$ spinor 
and the $SO(2)_{1}$ spinor respectively. 
Also we have defined the matrices $\gamma_{\pm}, \gamma^{z}$ and
$\gamma^{\overline{z}}$ by 
\begin{align}
\label{gammapm}
&\gamma_{+}:=\frac{1}{\sqrt{2}}(\mathbb{I}_{2}+\sigma_{2}),\ \ \ \ \ 
\gamma_{-}:=\frac{1}{\sqrt{2}}(\mathbb{I}_{2}-\sigma_{2}),\\
\label{gammaz1}
&\gamma^{z}
:=\frac{1}{\sqrt{2}}
(\gamma^{1}+i\gamma^{2})=\frac{1}{\sqrt{2}}\left(
\begin{array}{cc}
i&1\\
1&-i\\
\end{array}
\right),\\
\label{gammaz2}
&\gamma^{\overline{z}}
:=\frac{1}{\sqrt{2}}(\gamma^{1}-i\gamma^{2})
=\frac{1}{\sqrt{2}}\left(
\begin{array}{cc}
-i&1\\
1&i\\
\end{array}
\right).
\end{align}
These matrices allow us to perform the topological twisting, 
as we identify the index $\alpha$ 
with the index $\beta$. 
The two matrices $\gamma_{+}$ and $\gamma^{\overline{z}}$ are associated with 
the representations $\bm{8}_{c-}$ 
and give rise to $\bm{4}_{0}$ and $\overline{\bm{4}}_{-2}$. 
On the other hand, 
the remaining two matrices $\gamma_{-}$ and $\gamma^{z}$ are 
associated with $\bm{8}_{c+}$ 
and yield $\bm{4}_{2}$ and $\overline{\bm{4}}_{0}$. 
From the decomposition (\ref{chiralmtx}) 
and the chirality condition (\ref{blgchiral}) for $\Psi$, 
we can easily check that the expansion (\ref{new4}) 
of the fermionic field yields the relations; 
$\hat{\Gamma}^{7}\psi=\psi$, 
$\hat{\Gamma}^{7}\tilde{\Psi}_{\overline{z}}=-\tilde{\Psi}_{\overline{z}}$, 
$\hat{\Gamma}^{7}\tilde{\lambda}=-\tilde{\lambda}$ and
$\hat{\Gamma}^{7}\Psi_{z}=\Psi_{z}$. 
For the $\mathcal{A}_{4}$ algebra  
these fermions are the fundamental representations 
of the $SO(4)$ gauge group. 
Let us define the conjugate of the $SO(6)$ spinors by
\begin{align}
&\overline{\psi}:=\psi^{T}\hat{C},\ \ \
 \overline{\tilde{\lambda}}:=\tilde{\lambda}^{T}\hat{C},\ \ \ 
\overline{\Psi}_{z}:=\Psi_{z}^{T}\hat{C},\ \ \ 
\overline{\tilde{\Psi}}_{\overline{z}}:=\tilde{\Psi}_{\overline{z}}^{T}\hat{C}.
\end{align}

Similarly we can also expand 
the supersymmetry parameters. 
They originally transform as the 
$SL(2,\mathbb{R})$ spinor representations of the rotational group 
of the world-volume and transform 
as the spinor representation $\bm{8}_{s}$ of the $SO(8)$ R-symmetry. 
In the twisted theory, as we have already argued, 
they reduce to the four distinct representations 
$\bm{4}_{0}$, $\overline{\bm{4}}_{2}$, 
$\bm{4}_{-2}$ and $\overline{\bm{4}}_{0}$.  
Therefore one can expand supersymmetry parameters as
\begin{align}
\label{new5}
\epsilon_{A}^{\alpha\beta}
=\frac{i}{\sqrt{2}}\tilde{\epsilon}_{A}
(\gamma_{+}\epsilon^{-1})^{\alpha\beta}
+i\epsilon_{\overline{z}A}
(\gamma^{\overline{z}}\epsilon^{-1})^{\alpha\beta}
-\frac{i}{\sqrt{2}}\epsilon_{A}
(\gamma_{-}\epsilon^{-1})^{\alpha\beta}
-i\tilde{\epsilon}_{zA}
(\gamma^{z}\epsilon^{-1})^{\alpha\beta}
\end{align}
where the indices $\alpha$, $A$ and $\beta$ 
denote $SO(2)_{E}$, $SO(6)_{R}$
and $SO(2)_{1}$ respectively. 
Note that $\epsilon$ and $\tilde{\epsilon}$ 
are covariant constant on an arbitrary Riemann surface 
and therefore they can be identified with preserved supercharges. 
This fact implies that the effective theory will possess 
the corresponding eight supercharges.

By inserting the expressions (\ref{mtx1}), (\ref{new1}), (\ref{new2}),
(\ref{new3}) and (\ref{new4}) into 
the original BLG Lagrangian (\ref{blglagrangian}), 
we arrive at the topologically twisted BLG Lagrangian
\begin{align}
\label{twistedlagrangian}
 \mathcal{L}
&=\frac12 (D_{0}\phi^{I},D_{0}\phi^{I})
-(D_{z}\phi^{I},D_{\overline{z}}\phi^{I})
+(D_{0}\Phi_{z},D_{0}\Phi_{\overline{z}})
-2(D_{z}\Phi_{\overline{w}},D_{\overline{z}}\Phi_{w})
\nonumber\\
&
+(\overline{\tilde{\lambda}},D_{0}\psi)
+(\overline{\Psi}_{z},D_{0}\tilde{\Psi}_{\overline{z}})
-(\overline{\tilde{\Psi}}_{\overline{z}},D_{0}\Psi_{z})
-2i(\overline{\tilde{\Psi}}_{\overline{z}},D_{z}\psi)
+2i(\overline{\tilde{\lambda}},D_{\overline{z}}\Psi_{z})
\nonumber \\
&
+\frac{i}{2}(\overline{\tilde{\lambda}}\hat{\Gamma}^{IJ},
[\phi^{I},\phi^{J},\psi])
-i(\overline{\tilde{\Psi}}_{\overline{z}}\hat{\Gamma}^{IJ},
[\phi^{I},\phi^{J},\Psi_{z}])
\nonumber \\
&
+2i(\overline{\psi}\hat{\Gamma}^{I},
[\Phi_{\overline{z}},\phi^{I},\Psi_{z}])
-2i(\overline{\tilde{\lambda}}\hat{\Gamma}^{I},
[\Phi_{z},\phi^{I},\tilde{\Psi}_{\overline{z}}])
\nonumber \\
&
+i(\overline{\tilde{\lambda}},
[\Phi_{z},\Phi_{\overline{z}},\psi])
-2i(\overline{\tilde{\Psi}}_{\overline{w}},
[\Phi_{z},\Phi_{\overline{z}},\Psi_{w}])
\nonumber \\
&
-\frac{1}{12}
\left(
[\phi^{I},\phi^{J},\phi^{K}],
[\phi^{I},\phi^{J},\phi^{K}]
\right)
-\frac12 \left(
[\Phi_{z},\phi^{I},\phi^{J}],
[\Phi_{\overline{z}},\phi^{I},\phi^{J}]
\right)
\nonumber \\
&
-\frac12 \left(
[\Phi_{z},\Phi_{w},\phi^{I}],
[\Phi_{\overline{z}},\Phi_{\overline{w}},\phi^{I}]
\right)
-\frac12 \left(
[\Phi_{z},\Phi_{\overline{w}},\phi^{I}],
[\Phi_{\overline{z}},\Phi_{w},\phi^{I}]
\right)
\nonumber \\
&
+\frac16 \left(
[\Phi_{z},\Phi_{w},\Phi_{v}],
[\Phi_{\overline{z}},\Phi_{\overline{w}},\Phi_{\overline{v}}]
\right)
+\frac12 \left(
[\Phi_{z},\Phi_{w},\Phi_{\overline{v}}],
[\Phi_{\overline{z}},\Phi_{\overline{w}},\Phi_{v}]
\right)
+\mathcal{L}_{\textrm{TCS}}.
\end{align}
Here the parentheses 
$(\ ,\ )$ is implicated as the trace form on the 
3-algebra introduced in (\ref{3algmet}).  
The covariant derivatives are defined by 
$ D_{z}:=\frac{1}{\sqrt{2}}(D_{1}-iD_{2})$ and 
$D_{\overline{z}}:=\frac{1}{\sqrt{2}}(D_{1}+iD_{2})$.

By plugging the expressions (\ref{mtx1}), (\ref{new1}), (\ref{new2}),
(\ref{new3}), (\ref{new4}) and (\ref{new5}) into the supersymmetry
transformations (\ref{blgsusy1})-(\ref{blgsusy3}) for the original BLG theory, 
we find the following BRST transformations
\begin{align}
\label{brstbos1}
\delta\phi^{I}_{a}
&=i\overline{\tilde{\epsilon}}\hat{\Gamma}^{I}\tilde{\lambda}_{a}
-i\overline{\epsilon}\hat{\Gamma}^{I}\psi_{a},\\
\label{brstbos2}
\delta\Phi_{za}
&=-i\overline{\tilde{\epsilon}}\Psi_{za}
,\\
\label{brstbos3}
\delta\Phi_{\overline{z}a}
&=-i\overline{\epsilon}\tilde{\Psi}_{\overline{z}a}
,\\
\label{brstfermi1}
\delta\psi_{a}
&=
iD_{0}\phi^{I}_{a}\hat{\Gamma}\tilde{\epsilon}
-2D_{\overline{z}}\Phi_{za}\epsilon
+\frac16
 [\phi^{I},\phi^{J},\phi^{K}]_{a}\hat{\Gamma}^{IJK}\tilde{\epsilon}
+[\Phi_{z},\Phi_{\overline{z}},\phi^{I}]_{a}
\hat{\Gamma}^{I}\tilde{\epsilon}
,\\
\label{brstfermi2}
\delta\tilde{\lambda}_{a}
&=
iD_{0}\phi^{I}_{a}\hat{\Gamma}^{I}\epsilon
-2D_{z}\Phi_{\overline{z}a}\tilde{\epsilon}
-\frac16 
[\phi^{I},\phi^{J},\phi^{K}]_{a}\hat{\Gamma}^{IJK}\epsilon
+[\Phi_{z},\Phi_{\overline{z}},\phi^{I}]_{a}
\hat{\Gamma}^{I}\epsilon
,\\
\label{brstfermi3}
\delta\Psi_{za}
&=
-D_{z}\phi^{I}\hat{\Gamma}^{I}\tilde{\epsilon}
-iD_{0}\Phi_{z}\epsilon
+\frac12 [\Phi_{z},\phi^{I},\phi^{J}]_{a}
\hat{\Gamma}^{IJ}\epsilon
+\frac13 [\Phi_{w},\Phi_{\overline{w}},\Phi_{z}]_{a}
\epsilon,\\
\label{brstfermi4}
\delta\tilde{\Psi}_{\overline{z}a}
&=
D_{\overline{z}}\phi^{I}_{a}\hat{\Gamma}^{I}\epsilon
+iD_{0}\Phi_{\overline{z}a}\tilde{\epsilon}
+\frac12
 [\Phi_{\overline{z}},\phi^{I},\phi^{J}]_{a}
\hat{\Gamma}^{IJ}\tilde{\epsilon}
+\frac13 
[\Phi_{\overline{w}},\Phi_{w},\Phi_{\overline{z}}]_{a}
\tilde{\epsilon},\\
\label{brstbos4}
\delta\tilde{A}_{0 a}^{b}
&=-\overline{\epsilon}\hat{\Gamma}^{I}\phi_{c}^{I}\psi_{d}{f^{cdb}}_{a}
-\overline{\tilde{\epsilon}}\hat{\Gamma}^{I}\phi_{c}^{I}
\tilde{\lambda}_{d}{f^{cdb}}_{a}
-2\overline{\epsilon}\Phi_{zc}\tilde{\Psi}_{\overline{z}d}{f^{cdb}}_{a}
+2\overline{\tilde{\epsilon}}\Phi_{\overline{z}c}\Psi_{zd}{f^{cdb}}_{a},\\
\label{brstbos5}
\delta\tilde{A}^{b}_{za}
&=2i\overline{\epsilon}\hat{\Gamma}^{I}\phi_{c}^{I}\Psi_{zd}{f^{cdb}}_{a}
+2i\overline{\epsilon}\Phi_{zc}\tilde{\lambda}_{d}{f^{cdb}}_{a},\\
\label{brstbos6}
\delta\tilde{A}_{\overline{z}a}^{b}
&=
-2i\overline{\tilde{\epsilon}}
\hat{\Gamma}^{I}\phi_{c}^{I}\tilde{\Psi}_{\overline{z}d}{f^{cdb}}_{a}
+2i\overline{\tilde{\epsilon}}\Phi_{\overline{z}c}
\psi_{d}{f^{cdb}}_{a}.
\end{align}

\section{Derivation of quantum mechanics}
\label{cy2sec2}
In the previous section 
we have derived the topologically twisted BLG-model 
as the low-energy effective field theories 
on the curved M2-branes. 
Now we attempt to reduce the theory further 
to a low-energy effective one-dimensional field theory on $\mathbb{R}$. 
As mentioned in the analysis for 
the M2-branes wrapped on a torus, 
when the size of the Riemann surface shrinks, 
only the light degrees of freedom are relevant. 
To keep track of them  
we have to find the static configurations that minimize the energy, 
that is the zero-energy conditions. 
One can take the zero-energy conditions 
as a set of the BPS equations. 
In addition, we set all the fermionic fields to zero 
because we are interested in bosonic BPS configurations. 
The BPS equations, derived from the vanishing of the 
BRST transformations 
(\ref{brstfermi1})-(\ref{brstfermi4}) 
for the fermions, are
\begin{align}
\label{bps1}
&D_{z}\phi^{I}=0,\ \ \ 
D_{\overline{z}}\phi^{I}=0,\\
\label{bps2}
&D_{z}\Phi_{\overline{z}}=0,\ \ \ D_{\overline{z}}\Phi_{z}=0,\\
\label{bps3}
&[\phi^{I},\phi^{J},\phi^{K}]=0,\\
\label{bps4}
&[\Phi_{z},\Phi_{\overline{z}},\phi^{I}]=0,\ \ \ 
[\Phi_{z},\phi^{I},\phi^{J}]=0,\ \ \ 
[\Phi_{\overline{z}},\phi^{I},\phi^{J}]=0,\\
\label{bps5}
&[\Phi_{w},\Phi_{\overline{w}},\Phi_{z}]=0,\ \ \ 
[\Phi_{\overline{w}},\Phi_{w},\Phi_{\overline{z}}]=0.
\end{align}

According to the algebraic equations 
(\ref{bps3}), (\ref{bps4}) and (\ref{bps5}), 
all the bosonic matter fields are required to reside in the same plane 
in the $SO(4)$ gauge group. 
Hence one can write those bosonic fields as
\begin{align}
\label{bpsk3a}
\phi^{I}=(\phi^{I1},\phi^{I2},0,0)^{T},\ \ \ \ \ 
\Phi_{z}=(\Phi_{z}^{1},\Phi_{z}^{2},0,0)^{T},\ \ \ \ \ 
\Phi_{\overline{z}}
=(\Phi_{\overline{z}}^{1},\Phi_{\overline{z}}^{2},0,0)^{T}.
\end{align} 
Due to the supersymmetry the 
corresponding fermionic partners can be written as
\begin{align}
\label{bpsk3b}
\psi&=(\psi^{1},\psi^{2},0,0)^{T},& 
\tilde{\lambda}&=(\tilde{\lambda}^{1},\tilde{\lambda}^{2},0,0)^{T},\\
\label{bpsk3b1}
\Psi_{z}&=(\Psi_{z}^{1},\Psi_{z}^{2},0,0)^{T},&
\tilde{\Psi}_{\overline{z}}&=(\tilde{\Psi}_{\overline{z}}^{1},\tilde{\Psi}_{\overline{z}}^{2},0,0)^{T}.
\end{align} 
We note that 
under the BPS configurations (\ref{bpsk3a})-(\ref{bpsk3b1}) 
the original $SO(4)$ gauge group is broken down to $U(1)\times U(1)$. 
Combining these solutions with 
the BPS equations (\ref{bps1}), (\ref{bps2}) 
one finds that $\tilde{A}_{z3}^{1}=\tilde{A}_{z3}^{2}=
\tilde{A}_{z4}^{1}=\tilde{A}_{z4}^{2}=0$. 
This means that 
these vanishing components of the gauge field now become massive 
by the Higgs mechanism. 
Thus we will consider the time evolution for 
the surviving degrees of freedom in the low-energy effective theory.

In order to achieve this consistently, 
we should impose the Gauss law constraint. 
This turns out to require the flatness of the gauge field; 
$\tilde{F}_{z\overline{z}}=0$. 
At this stage we should remember that 
we are now interested in the case where 
the genus of the Riemann surface is larger than one. 
In such a case 
it is natural to assume that 
the flat connections are irreducible.  
When one considers irreducible flat connections, 
the Laplacian on the Riemann surface has no zero modes. 
Thus one can set the scalar fields to be zero $\phi^{I}=0$ 
\footnote{
Such BPS solutions with the irreducible connections 
have been considered in 
the four-dimensional topologically twisted Yang-Mills theories 
defined on the product of two Riemann surfaces  
\cite{Bershadsky:1995vm,Kapustin:2006pk,Kapustin:2006hi} and  
the corresponding decoupling limit for the brane description 
has been argued in \cite{Maldacena:2000mw}.}. 
Therefore the above set of equations over the compact Riemann surface 
of genus $g>1$ now becomes 
\begin{align}
\label{bps01}
&\tilde{F}_{z\overline{z}2}^{1}=0,\\
\label{bps02}
&\partial_{\overline{z}}\Phi_{z1}
+\tilde{A}_{\overline{z}2}^{1}\Phi_{z2}=0,\\
\label{bps03}
&\partial_{\overline{z}}\Phi_{z2}
-\tilde{A}_{\overline{z}2}^{1}\Phi_{z1}=0.
\end{align}

\begin{figure}
\begin{center}
\includegraphics[width=12.5cm]{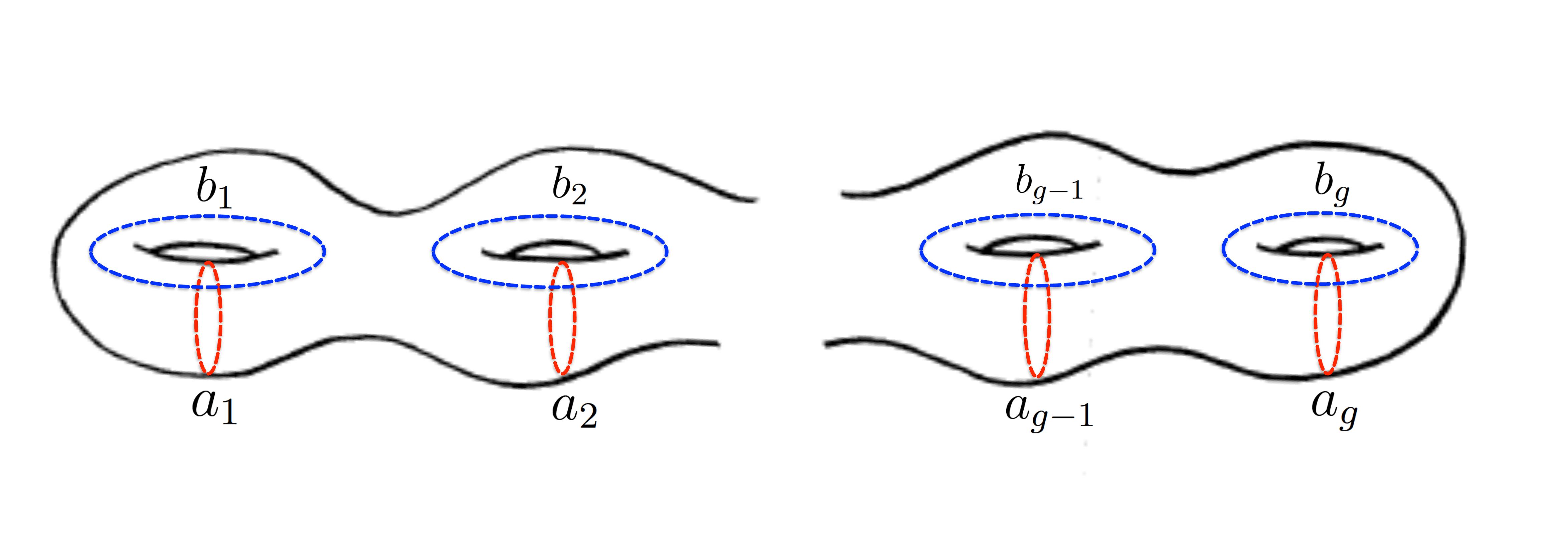}
\caption{A Riemann surface $\Sigma_{g}$ of genus $g$. 
$a_{i}$ and $b_{i}$ generate $H_{1}(\Sigma_{g})$.}
\label{rsu01}
\end{center}
\end{figure}

Let us determine the generic BPS configuration satisfying  
(\ref{bps01})-(\ref{bps03}). 
As we are now considering a compact Riemann surface 
of genus $g>1$,  
there exist $g$ holomorphic $(1,0)$-forms $\omega_{i}$, $i=1,\cdots,g$ 
and $g$ anti-holomorphic
$(0,1)$-forms $\overline{\omega}_{i}$. 
We will normalize them as
\begin{align}
\label{period1}
\int_{a_{i}}\omega_{j}=\delta_{ij},\ \ \ 
\int_{b_{i}}\omega_{j}=\Omega_{ij}
\end{align}
where $a_{i}$, $b_{i}$ are the canonical homology basis 
for $H_{1}(\Sigma_{g})$ (see Figure \ref{rsu01}). 
The $(g\times g)$ matrix $\Omega$ is the so-called period matrix 
of the Riemann surface. 
It is a complex symmetric matrix with imaginary part being positive. 
Note that the equation (\ref{bps01}) requires that 
the $U(1)$ gauge field $\tilde{A}_{z2}^{1}$ is flat. 
As we have argued earlier, 
the space of the $U(1)$ flat connection on a compact Riemann surface 
is the Jacobi variety denoted by 
$\mathrm{Jac}(\Sigma_{g})$. 
We can express the flat gauge fields as 
\cite{AlvarezGaume:1986es}
\begin{align}
\label{az1}
\tilde{A}_{z2}^{1}
&=-2\pi \sum_{i,j=1}^{g}
\left(
\Omega-\overline{\Omega}
\right)^{-1}_{ij}
\Theta^{i}\omega_{j}, \ \ \ \ \ 
\tilde{A}_{\overline{z}2}^{1}
=2\pi \sum_{i,j=1}^{g}
\left(
\Omega-\overline{\Omega}
\right)^{-1}_{ij}
\overline{\Theta}^{i}\overline{\omega}_{j}
\end{align}
where $\Theta^{i}:=\zeta^{i}+\overline{\Omega}_{ij}\xi^{j}$ 
stands for the complex coordinate of the Jacobi variety
$\mathrm{Jac}(\Sigma_{g})$. 
It characterizes the twists  
$e^{2\pi i\xi^{i}}$ and 
$e^{-2\pi i\zeta^{i}}$ 
around the $i$-th pair of homology cycles $a_{i}$ and $b_{i}$. 
We note that $\xi^{i}\rightarrow \xi^{i}+m^{i}$, 
$\zeta^{i}\rightarrow \zeta^{i}+n^{i}$ 
for $n^{i}, m^{i}\in \mathbb{Z}$ 
yield the identical point on $\mathrm{Jac}(\Sigma_{g})$. 
This means that 
$\mathrm{Jac}(\Sigma_{g})=\mathbb{C}^{g}/L_{\Omega}$ 
with $L_{\Omega}$ being the lattice generated 
by $\mathbb{Z}^{g}+\Omega\mathbb{Z}^{g}$. 
Let us define a function
\begin{align}
\label{phifct1}
\varphi:=-2\pi\sum_{i,j=1}^{g}\left(
\Omega-\overline{\Omega}
\right)^{-1}_{ij}
\left(
\Theta^{i}f_{j}(z)
-\overline{\Theta}^{i}\overline{f}_{j}(\overline{z})
\right)
\end{align}
where $f_{i}(z):=\int^{z}\omega_{i}$ is the holomorphic 
function of $z$ that satisfying the relations 
$f_{i}|_{a_{j}}=\delta_{ij}$ and $f_{i}|_{b_{j}}=\Omega_{ij}$. 
Then the flat gauge fields can be expressed as 
\begin{align}
\label{az2}
\tilde{A}_{z2}^{1}=\partial_{z}\varphi, \ \ \ \ \ \ \ \ \ \ 
\tilde{A}_{\overline{z}2}^{1}=\partial_{\overline{z}}\varphi. 
\end{align}
In terms of the above expressions (\ref{az1}) for the $U(1)$ flat connection, 
we can write the generic solutions 
to the equation (\ref{bps02}) and (\ref{bps03}) as
\begin{align}
\label{bosconf1b}
\Phi_{z1}(z,\overline{z})-i\Phi_{z2}(z,\overline{z})
=&e^{-i\varphi(z,\overline{z})}\sum_{i=1}^{g}x_{A}^{i}\omega_{i},
\nonumber\\
\Phi_{z1}(z,\overline{z})+i\Phi_{z2}(z,\overline{z})
=&e^{i\varphi(z,\overline{z})}\sum_{i=1}^{g}x_{B}^{i}\omega_{i}.
\end{align}
Here the variables $x_{A}^{i}$, $x_{B}^{i} \in \mathbb{C}$ 
are constant on the Riemann surface. 
As the result of the limit 
where the size of the Riemann surface $\Sigma_{g}$ goes to zero, 
the space-time configurations of the M2-branes 
need to be single-valued functions 
of $z$ and $\overline{z}$ 
in the low-energy effective quantum mechanics. 
Thus the coordinates $\xi^{i}$ and $\zeta^{i}$ can only be integer
values and the $U(1)$ flat gauge fields 
$\tilde{A}_{z2}^{1}$ and $\tilde{A}_{\overline{z}2}^{1}$ are quantized. 
This corresponds to fixing the point of the $\mathrm{Jac}(\Sigma_{g})$.

To sum up, the generic bosonic BPS 
configurations are
\begin{align}
\label{bpsconf1}
&\phi^{I}=0\nonumber\\
&\Phi_{z}=\sum_{i=1}^{g}
\left(\begin{array}{c}
\frac12\left(
e^{-i\varphi}x_{A}^{i}
+e^{i\varphi}x_{B}^{i}
\right)\\
\frac{i}{2}
\left(e^{-i\varphi}x_{A}^{i}-e^{i\varphi}x_{B}^{i}
\right)
\\
0\\
0\\
\end{array}
\right)\omega_{i},\ \ \ 
\Phi_{\overline{z}}=
\sum_{i=1}^{g}
\left(\begin{array}{c}
\frac12\left(
e^{i\varphi}\overline{x}_{A}^{i}
+e^{-i\varphi}\overline{x}_{B}^{i}
\right)\\
-\frac{i}{2}
\left(e^{i\varphi}\overline{x}_{A}^{i}
-e^{-i\varphi}\overline{x}_{B}^{i}
\right)
\\
0\\
0\\
\end{array}
\right)
\overline{\omega}_{i}
,\nonumber\\
&\tilde{A}_{z}=\left(
\begin{array}{cccc}
0&\partial_{z}\varphi(z,\overline{z})&0&0\\
-\partial_{z}\varphi(z,\overline{z})&0&0&0\\
0&0&0&\tilde{A}_{z4}^{3}(z,\overline{z})\\
0&0&-\tilde{A}_{z4}^{3}(z,\overline{z})&0\\
\end{array}
\right).
\end{align}
Here $\tilde{A}_{z4}^{3}$ and 
$\tilde{A}^{3}_{\overline{z}4}$ are the Abelian gauge fields 
associated with the preserved $U(1)$ symmetry. 
Note that they do not receive 
any constraints from the BPS conditions.

Due to the supersymmetry 
the corresponding fermionic fields 
can be written as
\begin{align}
\label{bpsconf2}
\psi&=0,&
\tilde{\lambda}&=0,\nonumber\\
\Psi_{z}
&=\sum_{i=1}^{g}
\left(
\begin{array}{c}
\frac12\left(
\Psi_{A}^{i}+\Psi_{B}^{i}
\right)\\
\frac{i}{2}\left(
\Psi_{A}^{i}-\Psi_{B}^{i}
\right)\\
0\\
0\\
\end{array}
\right)
\omega_{i}
,& 
\tilde{\Psi}_{\overline{z}}
&=\sum_{i=1}^{g}
\left(
\begin{array}{c}
\frac12 \left(
\tilde{\Psi}_{A}^{i}+\tilde{\Psi}_{B}^{i}
\right)\\
-\frac{i}{2} \left(
\tilde{\Psi}_{A}^{i}-\tilde{\Psi}_{B}^{i}\right)\\
0\\
0\\
\end{array}
\right)\overline{\omega}_{i}.
\end{align}

Substituting the BPS configuration (\ref{bpsconf1}) 
and (\ref{bpsconf2}) 
into the twisted action (\ref{twistedlagrangian}), one finds
\begin{align}
\label{twistbpsaction1}
S=\int_{\mathbb{R}}dt\int_{\Sigma_{g}}d^{2}z
\Biggl[
\left(
D_{0}\Phi_{z}^{a},D_{0}\Phi_{\overline{z}a}
\right)
+
\left(
\overline{\Psi}_{z}^{a},D_{0}\tilde{\Psi}_{\overline{z}a}
\right)
-
\left(
\overline{\tilde{\Psi}}_{\overline{z}}^{a},D_{0}\Psi_{za}
\right)
\nonumber\\
-\frac{k}{2\pi}\tilde{A}_{02}^{1}\tilde{F}_{z\overline{z}4}^{3}
-\frac{k}{4\pi}
\left(
\tilde{A}_{z2}^{1}\dot{\tilde{A}}_{\overline{z}4}^{3}
-\tilde{A}_{\overline{z}2}^{1}\dot{\tilde{A}}_{z4}^{3}
\right)
\Biggr].
\end{align}
Now that the gauge fields $\tilde{A}_{z2}^{1}$, $\tilde{A}_{\overline{z}2}^{1}$ 
are quantized and 
there are no their time derivatives in the effective action, 
one can integrate out them as the auxiliary fields. 
Then we are left with the constraints 
$\dot{\tilde{A}}_{z4}^{3}=\dot{\tilde{A}}_{\overline{z}4}^{3}=0$. 

To proceed the integration over the Riemann surface, 
we should note the Riemann bilinear relation \cite{MR1139765} 
\begin{align}
\label{bil01}
\int_{\Sigma_{g}}\omega \wedge \eta
=\sum_{i=1}^{g}
\left[
\int_{a_{i}}\omega\int_{b_{i}}\eta
-\int_{b_{i}}\omega\int_{a_{i}}\eta
\right]. 
\end{align}
After the integration over the Riemann surface $\Sigma_{g}$ 
one finds the gauged quantum mechanical action
\begin{align}
\label{eff001}
S=\int_{\mathbb{R}}dt 
\Biggl[
&
\sum_{i,j}
\left(
\mathrm{Im}\ \Omega
\right)_{ij}
\left(
D_{0}x^{ia}D_{0}\overline{x}_{a}^{j}
+\overline{\Psi}^{ia}D_{0}\tilde{\Psi}_{a}^{j}
-\overline{\tilde{\Psi}}^{ia}D_{0}\Psi_{a}^{j}
\right)
-kC_{1}(E)\tilde{A}_{02}^{1}
\Biggr]
\end{align}
where the indices $a=A,B$ represent the 
two internal degrees of freedom for the two M2-branes. 
We have defined the covariant derivatives by
\begin{align}
D_{0}x_{A}^{i}&=\dot{x}_{A}^{i}+i\tilde{A}^{1}_{02}x_{A}^{i}, 
&D_{0}x_{B}^{i}&=\dot{x}_{B}^{i}-i\tilde{A}^{1}_{02}x_{B}^{i},\\
D_{0}\Psi_{A}^{i}&=\dot{\Psi}_{A}^{i}+i\tilde{A}_{02}^{1}\Psi_{A}^{i}, 
&D_{0}\Psi_{B}^{i}&=\dot{\Psi}_{B}^{i}-i\tilde{A}_{02}^{1}\Psi_{B}^{i},\\
D_{0}\tilde{\Psi}_{A}^{i}&=\dot{\tilde{\Psi}}_{A}^{i}
-i\tilde{A}_{02}^{1}\tilde{\Psi}_{A}^{i}, 
&D_{0}\Psi_{B}^{i}&=\dot{\tilde{\Psi}}_{B}^{i}
+i\tilde{A}_{02}^{1}\tilde{\Psi}_{B}^{i}.
\end{align}
We have also introduced the Chern number $C_{1}(E) \in \mathbb{Z}$ 
associated to the $U(1)$ principal bundle $E\rightarrow \Sigma_{g}$
over the Riemann surface
\begin{align}
\label{ch03}
C_{1}(E)=\int_{\Sigma_{g}}c_{1}(E)
=\frac{1}{2\pi}\int_{\Sigma_{g}}d^{2}z \tilde{F}_{z\overline{z}4}^{3}.
\end{align}
The action (\ref{eff001}) is invariant 
under the one-dimensional $SL(2,\mathbb{R})$ conformal transformations
\begin{align}
\label{k3qmconf1}
\delta t&=f(t)=a+bt+ct^{2}, 
&\delta\partial_{0}&=-\dot{f}\partial_{0},\\
\delta x_{a}^{i}&=\frac12 \dot{f} x_{a}^{i}, 
&\delta \tilde{A}_{02}^{1}&=-\dot{f}\tilde{A}_{02}^{1},\\
\delta \Psi^{i}_{a}&=0, 
&\delta \tilde{\Psi}^{i}_{a}&=0.
\end{align}
The action (\ref{eff001}) possesses the invariance 
under the $\mathcal{N}=8$ supersymmetry transformation laws
\begin{align}
\label{k3qmsusy1}
\delta x^{i}_{a}&=2i\overline{\tilde{\epsilon}}\Psi_{a}^{i}, 
&\delta
 \overline{x}^{i}_{a}&=2i\overline{\epsilon}\tilde{\Psi}_{a}^{i},\\
\delta\Psi_{a}^{i}&=-iD_{0}x_{a}^{i}\epsilon, 
&\delta\tilde{\Psi}_{a}^{i}&=iD_{0}\overline{x}_{a}^{i}\tilde{\epsilon},\\
\delta\tilde{A}_{02}^{1}&=0.
\end{align}

Therefore as the consequence of the topological reduction 
of the twisted BLG model (\ref{twistedlagrangian}), 
we obtain the $\mathcal{N}=8$ superconformal gauged quantum mechanics 
(\ref{eff001}) which may describe the low-energy dynamics  
of the two wrapped M2-branes around $\Sigma_{g}$ 
probing a K3 surface.

One can see from the action (\ref{eff001}) 
that the $U(1)$ gauge field $\tilde{A}_{02}^{1}$ 
can be regarded as an auxiliary field 
because its kinetic term is absent. 
Thus the gauge field does not contribute to the Hamiltonian. 
As argued earlier for the similar gauged quantum mechanical models, 
the corresponding gauge symmetry gives rise to an integral of motion as a
moment map $\mu:\mathcal{M}\rightarrow \mathfrak{u}(1)^{*}$ 
and it allows us to reduce the phase space $\mathcal{M}$ 
to smaller one $\mathcal{M}_{c}=\mu^{-1}(c)$ by choosing the inverse of 
the moment map at a point $c\in \mathfrak{u}(1)^{*}$. 
Let us fix the gauge as a temporal gauge; $\tilde{A}_{02}^{1}=0$. 
Then one finds the action
\begin{align}
\label{eff0001}
S=\int_{\mathbb{R}} dt 
\sum_{i,j}\left(
\mathrm{Im}\Omega
\right)_{ij}\left(
\dot{x}^{ia}\dot{\overline{x}}^{j}_{a}
+\overline{\Psi}^{ia}\dot{\tilde{\Psi}}^{j}_{a}
-\overline{\tilde{\Psi}}^{ia}\dot{\Psi}^{j}_{a}
\right)
\end{align}
and we are left with the Gauss law constraint
\begin{align}
\label{k3gauss1}
\phi_{0}:=
kC_{1}(E)
+i\sum_{i,j}\left(\textrm{Im}\Omega\right)_{ij}
\left[
K_{ij}
+2\left(
\overline{\Psi}^{i}_{A}\tilde{\Psi}_{A}^{j}
-\overline{\Psi}_{B}^{i}\tilde{\Psi}_{B}^{j}
\right)
\right]
=0
\end{align}
where 
\begin{align}
\label{kconst}
K_{ij}:=
\left(
\dot{x}^{i}_{A}\overline{x}^{j}_{A}
-x_{A}^{i}\dot{\overline{x}}_{A}^{j}
\right)
-
\left(
\dot{x}^{i}_{B}\overline{x}^{j}_{B}
-x^{i}_{B}\dot{\overline{x}}_{B}^{j}
\right).
\end{align} 
The constraint equation (\ref{k3gauss1}) 
is the moment map condition and 
it tells that all states in the Hilbert space should be gauge
invariant. 
Although we have obtained the reduced Lagrangian 
by making use of the conserved quantities via Routh reduction 
for the previous gauged mechanical systems, 
in this case the symmetry of 
the system seems not so large enough to get the reduced Lagrangian. 
It is still open to know whether 
we can obtain the reduced Lagrangian description from 
the superconformal gauged quantum mechanics 
(\ref{eff001}) (or (\ref{eff0001}) together with (\ref{k3gauss1})). 
However, this particular form of the gauged mechanical action may 
suggest that the obstructed construction of SCQM models 
can be extended to by gauging procedure 
as in \cite{MR0478225,Polychronakos:1991bx,Fedoruk:2008hk}.

We finally comment on 
the corresponding supermultiplet for 
our $\mathcal{N}=8$ superconformal quantum mechanics (\ref{eff001}). 
Although the superspace and superfield formalism 
is quite useful, we do not know whether our $\mathcal{N}=8$
superconformal quantum mechanics can be derived  
via superspace and superfield formalism 
since our derivation is not based on the superfield formulation and 
the reduced quantum mechanical description is not available so far. 
If it exists, 
the corresponding supermultiplet may be inferred as 
the $g$ sets of $(\bm{2}, \bm{8}, \bm{6})$ multiplet 
by observing the representations (\ref{k3content}) 
of the fields under the remaining R-symmetry $SO(6)$. 
However, after the integration of  
the auxiliary gauge field $\tilde{A}_{02}^{1}$,  
the physical degrees of freedom may be reduced 
and thus the supermultiplets needs to be modified.

\chapter{Conclusion and Discussion}
\label{secconc}
\section{Conclusion}
In this thesis we have established the new connection between two subjects; 
the superconformal quantum mechanics and the M2-branes 
by examining the IR superconformal quantum mechanics 
resulting from the multiple M2-branes 
wrapped around a compact Riemann surface $\Sigma_{g}$ 
after shrinking the size of the Riemann surface. 

We have seen that 
conformal symmetry and supersymmetry 
in quantum mechanics, i.e. one-dimensional field theory 
are rather out of the way in that 
they contain numerous unfamiliar features 
which are not observed in higher dimensional field theories.

Instead of the morbid Hamiltonian, 
one can label the state in terms of 
the eigenstate of the compact operator 
$L_{0}=\frac12 (H+K)$ and the second Casimir operator of the
$SL(2,\mathbb{R})$ conformal symmetry group. 
Although one cannot assume the existence of 
both normalizable conformally invariant states 
and invariant primary operators 
due to the fact that the quantum mechanics is based on the Hilbert space 
not on the Fock space, 
the 2-point, 3-point and 4-point functions which satisfy the conformal
constraints can be constructed by using those two defects 
\cite{Chamon:2011xk,Jackiw:2012ur}. 
We have also discussed 
the interesting observations \cite{Claus:1998ts,Gibbons:1998fa} 
that the motion of the particle 
near the horizon of the extreme Reissner-Nordstr\"{o}m 
black hole is described by the (super)conformal mechanics.  
This indicates that 
(super)conformal quantum mechanics may caputure 
the information of the dual AdS$_{2}$ gravity. 
Obviously further surveys are needed to understand 
AdS$_{2}$/CFT$_{1}$ correspondence.

Due to the reduced Poincar\'{e} symmetry, 
one-dimensional supersymmetry has the special properties    
that (i) the number of the component fields in the supermultiplet 
is larger than the number $\mathcal{N}$ 
of supersymmetry if $\mathcal{N}$ is greater thatn eight and  
that (ii) the number $n$ of physical bosonic component fields 
is not necessarily same as that of the fermions. 
These facts allow us to construct various supermultiplets 
$(\bm{n},\mathcal{N},\mathcal{N}-\bm{n})$ 
only for $\mathcal{N}=1,2,4$ and $8$ supersymmetric quantum mechanics. 
Indeed we have argued that 
for such supersymmetric quantum mechanics 
there have been continuous attempts 
to construct superconformal mechanical models 
by appealing the superspace and superfield formalism.

We have shown that 
the IR quantum mechanics arising 
from the BLG-model and the ABJM-model wrapped on a torus are
the $\mathcal{N}=16$ and $\mathcal{N}=12$ 
superconformal gauged quantum mechanical models respectively. 
Furthermore after the integration of the auxiliary gauge fields, 
we found that 
the $OSp(16|2)$ quantum mechanics (\ref{effk1}) 
and $SU(1,1|6)$ quantum mechanics (\ref{abjmscqm0}) emerge 
from the reduced theories. 
Both of them are $\mathcal{N}>8$ superconformal quantum mechanical
models which have not been available by the superspace and superfield
formalism so far. 
It is interesting to investigate their spectrums, 
wavefunctions and correlation functions for those new superconformal 
mechanical models.

We have also surveyed the membranes wrapped around a genus 
$g\neq 1$ Riemann surface. 
In this case the surface is singled out as 
a calibrated holomorphic curve in a Calabi-Yau manifold  
to preserve supersymmetry. 
We have found that 
the IR quantum mechanical models have 
$\mathcal{N}=8$, $4$, $2$ and $2$ supersymmetries for 
K3, $CY_{3}$, $CY_{4}$ and $CY_{5}$ respectively. 
Especially when the Calabi-Yau manifolds are 
constructed via decomposable line bundles 
over the Riemann surface,  
the K3 surface essentially allows for a unique topological twist  
while for the other Calabi-Yau manifolds 
there are infinitely many topological twists 
which are specified by the degrees of the line bundles.

We have especially analyzed the 
two membranes wrapping a holomorphic genus $g>1$ curve 
embedded in a K3 surface based on the topologically twisted BLG-model. 
We have found 
the new $\mathcal{N}=8$ superconformal gauged quantum mechanics
(\ref{eff001}) that 
may describe the low-energy dynamics of the wrapped M2-branes 
in a K3 surface. 
It is known that \cite{MR0478225,Polychronakos:1991bx,Fedoruk:2008hk} 
there are the connections of the gauged quantum mechanics 
to the conformal mechanical models, the Calogero model and their
generalizations. 
An interesting question is what type of interaction potential, 
if it exists, 
may characterize our superconformal ``gauged'' quantum mehcanics 
(\ref{eff001}). 
The structure of the resulting theory 
may indicate that generic SCQM takes the form of 
superconformal gauged quantum mechanics along with 
auxiliary gauge fields.

\section{Future directions}
There may be a number of future aspects of the present work. 
In the following 
we will briefly discuss the possible three applications.

\subsection{AdS$_{2}/$CFT$_{1}$ correspondence}
AdS$_{d+1}$/CFT$_{d}$ correspondence \cite{Maldacena:1997re} 
is an important example of the holographic principle \cite{Susskind:1994vu}. 

For $d=2$ 
it has been shown \cite{Brown:1986nw} that 
the Hilbert space of the any quantum gravity 
on an asymptotically AdS$_{3}$ space-time 
is a representation of the two-dimensional conformal group 
and that the central charge of the $d=2$ CFT is given by
\begin{align}
\label{ads3cft2a}
c&=\frac{3l}{2G}
\end{align}
where $l$ is the AdS$_{3}$ radius and 
$G$ is Newton constant. 
The relationship between the BTZ black hole and the 
state in the two-dimensional CFT indicates that 
the entropy of the black may be defined as 
the logarism of the degeneracy of the corresponding states in the CFT. 
In this perspective 
the entropy of the $d=3$ Ba\~{a}dos-Teitelboim-Zanelli (BTZ) 
black hole is computed by counting the states of the 
$d=2$ conformal field theory on the boudnary of AdS$_{3}$ 
\cite{Strominger:1997eq}
\begin{align}
\label{ads3cft2b}
S=2\pi\sqrt{\frac{cn_{R}}{6}}
+2\pi\sqrt{\frac{cn_{L}}{6}}
\end{align}
where $n_{R}$ and $n_{L}$ are the eigenvalues of 
the Virasoro generators 
$L_{0}$ and $\overline{L}_{0}$ respectively. 
For large $L_{0}$ one can use the Cardy formula to 
evaluate the degeneracy of the states 
and it has been shown 
\cite{Saida:1999ec,Kraus:2005vz,Sahoo:2006vz,Nampuri:2007gw} 
that the result agrees with the one obtained by Wald's formula 
\cite{Wald:1993nt,Jacobson:1993vj,Iyer:1994ys,Jacobson:1994qe}.

 
The case of $d=1$, i.e. AdS$_{2}$/CFT$_{1}$ correspondence 
\cite{Strominger:1998yg,Maldacena:1998uz,Nakatsu:1998st,Townsend:1998qp,
Spradlin:1999bn,Cadoni:1999ja,Blum:1999pc,NavarroSalas:1999up,
Caldarelli:2000xk,Cadoni:2000gm,
Bellucci:2002va,
Strominger:2003tm,Leiva:2003kd,
Giveon:2004zz,
Azeyanagi:2007bj,
Gupta:2008ki,Hartman:2008dq,Galajinsky:2008ce,Sen:2008vm,Sen:2008yk,
Sen:2011cn} 
is less understood, 
however, it is extremely significant case of 
$\textrm{AdS}_{d+1}/\textrm{CFT}_{d}$ correspondence 
in that all known extremal black holes contain 
the AdS$_{2}$ factor in their near horizon geometries 
\cite{Kunduri:2007vf,Figueras:2008qh}. 
The two candidates for the CFT$_{1}$ have been proposed
\begin{align}
&\textrm{(i) conformal quantum mechanics}\nonumber\\ 
&\textrm{(ii) a chiral half of a $d=2$ CFT}\nonumber. 
\end{align}
For the former 
only the global $SL(2,\mathbb{R})$ acts nontrivially 
on the Hilbert space, while 
in the in latter case 
one copy of the Virasoro generators acts nontrivially on the Hilbert
space. 
In \cite{Hartman:2008dq} 
the central charge for the CFT$_{1}$ which 
corresponds to the quantum gravity with a $U(1)$ gauge field on
AdS$_{2}$ has been given by 
\begin{align}
\label{ads2cft1a}
c&=\frac{3kE^{2}l^{4}}{4}
\end{align}
where $l$ is the AdS$_{2}$ radius, 
$E$ is the electric field and $k$ is the level of the 
$U(1)$ current. 
The expression is similar to (\ref{ads3cft2a}) for 
AdS$_{3}$/CFT$_{2}$ correspondence.  
It has been discussed \cite{Strominger:1998yg,Hartman:2008dq} 
that the latter idea of the non-trivial action of the Virasoro 
could be consistent and 
AdS$_{2}$/CFT$_{1}$ correspondence reduces to the 
CFT$_{2}$/CFT$_{2}$ duality on the strip. 
As discussed in \cite{Strominger:1998yg,Hartman:2008dq}, 
this idea could be true 
when AdS$_{2}$ is generated as a $S^{1}$ compactification 
of AdS$_{3}$, 
however, there may be other types of the AdS$_{2}$ which 
do not arise as a $S^{1}$ compactification of AdS$_{3}$ 
and therefore the former possibility could still 
be a good candidate of the CFT$_{1}$. 
In the former perspective, 
it has been proposed \cite{Sen:2008yk} that 
the logarithm of the ground state degeneracy in a 
conformal quantum mechanics living on the boundary of AdS$_{2}$ 
yields the definition of the entropy of the extremal black hole in the
quantum theory. 
Furthermore it has been pointed out in \cite{Chamon:2011xk,Jackiw:2012ur} 
that the correlation functions of 
the conformal quantum mechanics \cite{deAlfaro:1976je} 
have the expected scaling behaviors from AdS$_{2}$/CFT$_{1}$
correpondence  
although one cannot assume the existence of the normalized and conformal
invariant vacuum states in conformal quantum mechanics 
as in other higher dimensional conformal field theories. 
It is interesting to investigate 
whether our superconformal quantum mechanics 
resulting from the wrapped M2-branes 
around a compact Riemann surface in M-theory could provide 
some examples of the AdS$_{2}$/CFT$_{1}$ correspondence 
in the former perspective.

\subsection{Indices and the reduced Gromov-Witten invariants}

Another topic is the computation of the indices and their applications. 
For instance, 
the BPS partition function which gives rise to the 
counding of the BPS states 
may be related to the number of the 
supersymmetric two-cycles of genus $g$ in our setup. 
Indeed, in the setup where 
the curved D3-branes wrapping supersymmetric two-cycles 
embedded in K3 surface, 
the formula for the numbers of rational curves with $g$ double points 
on a K3 surface, 
the so-called reduced Gromov-Witten invariants \cite{MR1750955} 
has been conjectured by Yau and Zaslow \cite{Yau:1995mv} in the 
computation of the BPS partition function by 
appealing the string duality. 
Closely related to their setup, 
our $\mathcal{N}=8$ superconformal gauged quantum mechanics 
(\ref{eff001}) appears from the wrapped M2-branes instead of the D3-brane. 
It would be interesting to compute the indices and 
to extract enumerative information and structure from our model. 

In order to compute the indices 
we take a trace over the eigenstates. 
As discussed in section \ref{secana0}, 
it is difficult to calculate 
a trace over the eigenstate of the Hamiltonian $H$ 
for the superconformal quantum mechanics 
because there is no normalizable ground state 
and its spectrum is continuous. 
As proposed in \cite{BrittoPacumio:2000sv}, 
the indices in superconformal quantum mechanics can be defined by 
taking a trace over the eigenstates of the compact operator 
$L_{0}=\frac12 (H+K)$ 
which has a normalizable ground state 
and the discrete eigenvalues with equal spacing as
\begin{align}
\label{ind1a}
\mathcal{I}(\mathcal{O})&=\mathrm{Tr}_{L_{0}}(-1)^{2J}\mathcal{O}
e^{-\beta (L_{0}-J)}
\end{align} 
where $J$ is the R-symmetry generator 
and $\mathcal{O}$ is some operator in the theory. 
It is an open problem to evaluate indices 
and understand their physical and mathematical implication 
for our superconformal quantum mechanics.

\subsection{1d-2d relation}
Finally we want to comment on the ``1d-2d relation'', 
which is motivated by the fascinating stories 
arising from the compactification of M5-branes, 
for example, the AGT-relation \cite{Alday:2009aq}, 
the DGG-relation \cite{Dimofte:2011ju} 
and the 2d-4d relation \cite{Gadde:2013sca}. 
It has been argued that 
the world-volume theories of multiple M5-branes can be described by 
the six-dimensional superconformal field theories 
labeled by a simply-laced Lie algebra $\mathfrak{g}$, 
the so-called $(2,0)$ theories.  
Via compactification, such theories 
leads to a family $T[M_{6-d},\mathfrak{g}]$ of 
$d$-dimensional superconformal field theories which can be labeled by 
a choice of a specific manifold $M_{6-d}$ 
and a Lie algebra $\mathfrak{g}$. 
From this perspective 
the AGT-relation, the DGG-relation and 2d-4d relation 
are regarded as the decomposition of the six-dimensional world-volume 
of M5-branes as $6=4+2$, $3+3$ and $2+4$ respectively.

On the other hand, 
the world-volume theories of multiple M2-branes can be described 
by the three-dimensional superconformal field theories. 
Unlike the M5-branes 
we know the explicit Lagrangian for such world-volume theories as 
the BLG-model and the ABJ(M) model. 
It would be attractive 
to find the new relationship between 
the superconformal field theories and 
the geometries or relevant dualities from M2-branes, 
i.e. ``1d-2d relation'' arising from 
the decomposition of the three-dimensional world-volume of M2-branes 
as $3=1+2$. 
As an exchange of the order
we may have two ways of the compactification
\begin{align}
\label{1d2d00a}
\begin{array}{ccccc}
& &\textrm{3d SCFT on $\mathbb{R}\times \Sigma_{g}$}& &\cr
 &\swarrow & &\searrow& \cr
\textrm{1d SCQM on $\mathbb{R}$}& & & &
\textrm{2d TQFT on $\Sigma_{g}$},\cr
\end{array}
\end{align}
which suggests a new set of dualities 
in the sense that the partition functions or indices on both sides 
yield the same result. 
As we discussed in section \ref{secscqm4}, 
the WDVV equation \cite{Witten:1989ig,Dijkgraaf:1990dj} 
and the twisted periods \cite{MR2070050,MR2999308} 
which are relevant to two-dimensional geometries 
and topological field theories appear 
from the constraint conditions 
for the constructions of $\mathcal{N}=4$ superconformal mechanics. 
It would be interesting to investigate 
whether our M-theoretical construction of 
superconformal quantum mechanics 
could help to understand and generalize such relations 
as the ``1d-2d relation''.


\bibliographystyle{utphys}
\bibliography{ref}

\end{document}